\documentclass[a4paper, 12pt]{book}

\usepackage[english]{babel}
\usepackage[nosort,compress]{cite}
\usepackage{graphicx,epsfig,subfigure,color}
\usepackage[dvipsnames]{xcolor}
\usepackage[linktoc=page,bookmarks=false,colorlinks=false,
linkbordercolor=RoyalBlue,citebordercolor=ForestGreen,urlbordercolor=CornflowerBlue]{hyperref}
\usepackage[normal,font=small,labelfont=bf,labelsep=period]{caption}
\usepackage{amsmath,amssymb,array,arydshln,enumerate,slashed,multirow,fancybox,marvosym,wasysym}
\usepackage{fancyhdr}

\definecolor{darkblue}{rgb}{0,0.2,0.6}
\definecolor{darkgreen}{rgb}{0,0.4,0}

\font\tenrsfs=rsfs10 at 12pt

\font\sevenrsfs=rsfs7
\font\fiversfs=rsfs5
\newfam\rsfsfam
\textfont\rsfsfam=\tenrsfs
\scriptfont\rsfsfam=\sevenrsfs
\scriptscriptfont\rsfsfam=\fiversfs
\def\mathscr#1{{\fam\rsfsfam\relax#1}}

\font\teneusm=eusm10 at 12pt
\newfam\eusmfam
\textfont\eusmfam=\teneusm
\def\EuScript#1{{\fam\eusmfam\relax#1}}

\font\tenbbm=bbm10 at 12pt
\newfam\bbmfam
\textfont\bbmfam=\tenbbm
\def\mathbbm#1{{\fam\bbmfam\relax#1}}

\def\L{\mathscr{L}}
\def\H{\mathscr{H}}
\def\O{\mathcal{O}}
\def\Q{\mathcal{Q}}
\def\A{\mathcal{A}}
\def\B{\mathcal{B}}
\def\E{\mathcal{E}}
\def\M{\mathcal{M}}
\def\Vcal{\mathcal{V}}
\def\I{\mathcal{I}}

\def\Ord{\EuScript{O}}

\def\d{{\rm d}}
\def\tr{\mathop{\rm tr}}
\def\diag{\mathop{\rm diag}}

\renewcommand{\Re}{\mathop{\rm Re}}
\renewcommand{\Im}{\mathop{\rm Im}}

\newcommand{\mub}{\bar\mu}

\newcommand{\MS}{\overline{\mbox{\sc ms}}}
\newcommand{\os}{{\rm OS}}
\def\SU{{SU}}
\def\U{{U}}
\def\SO{{SO}}
\def\CP{CP}

\def\ss0{{\scriptscriptstyle 0}}
\newcommand\mpl{M_{\rm Pl}}

\def\one{\boldsymbol{1}}
\def\two{\boldsymbol{2}}
\def\twobar{\boldsymbol{\bar{2}}}
\def\three{\boldsymbol{3}}
\def\threebar{\boldsymbol{\bar{3}}}
\def\four{\boldsymbol{4}}
\def\five{\boldsymbol{5}}
\def\eight{\boldsymbol{8}}
\def\ten{\boldsymbol{10}}

\newcommand{\UtreLC}{$U(3)^3_\mathrm{LC}$}
\newcommand{\UtreRC}{$U(3)^3_\mathrm{RC}$}
\newcommand{\UdueLC}{$U(2)^3_\mathrm{LC}$}
\newcommand{\UdueRC}{$U(2)^3_\mathrm{RC}$}
\newcommand{\Utre}{$U(3)^3$}
\newcommand{\Udue}{$U(2)^3$}

\newcommand{\qL}{\boldsymbol{q}_{\boldsymbol{L}}}
\newcommand{\uR}{\boldsymbol{u}_{\boldsymbol{R}}}
\newcommand{\dR}{\boldsymbol{d}_{\boldsymbol{R}}}

\newcommand{\qLbar}{\boldsymbol{\bar q}_{\boldsymbol{L}}}
\newcommand{\uRbar}{\boldsymbol{\bar u}_{\boldsymbol{R}}}
\newcommand{\dRbar}{\boldsymbol{\bar d}_{\boldsymbol{R}}}

\newcommand{\ELqL}{\boldsymbol{q}_{\boldsymbol{L}}}
\newcommand{\ELuR}{\boldsymbol{u}_{\boldsymbol{R}}}
\newcommand{\ELdR}{\boldsymbol{d}_{\boldsymbol{R}}}
\newcommand{\EHqL}{q_{3L}}
\newcommand{\EHuR}{t_{R}}
\newcommand{\EHdR}{b_{R}}
\newcommand{\ELqLbar}{\boldsymbol{\bar q}_{\boldsymbol{L}}}

\newcommand{\EHqLbar}{\bar q_{3L}}

\newcommand{\CLquR}{\boldsymbol{Q_R^u}}

\newcommand{\CHquR}{Q_{3R}^u}

\newcommand{\CLuLbar}{\boldsymbol{\bar U_L}}

\newcommand{\CHuLbar}{\bar T_{L}}

\newcommand{\ELeR}{\boldsymbol{e}_{\boldsymbol{R}}}

\newcommand{\ELlLbar}{\boldsymbol{\bar\ell}_{\boldsymbol{L}}}

\newcommand{\V}{\boldsymbol{V}}
\newcommand{\Ve}{\boldsymbol{V_e}}
\newcommand{\Vu}{\boldsymbol{V_u}}
\newcommand{\Vd}{\boldsymbol{V_d}}
\newcommand{\Vudag}{\boldsymbol{V}_{\!\!\boldsymbol{u}}^{\dag}}
\newcommand{\Vddag}{\boldsymbol{V}_{\!\!\!\boldsymbol{d}}^{\dag}}

\newcommand{\Mtexp}{173.10}
\newcommand{\Mterr}{0.59}
\newcommand{\Mhexp}{125.66}
\newcommand{\Mherr}{0.34}
\newcommand{\MWexp}{80.384}
\newcommand{\MWerr}{0.014}
\newcommand{\Mhexperr}{\Mhexp\pm\Mherr~\GeV}
\newcommand{\Mtdiff}{ \bigg(\frac{M_t}{\GeV}-\Mtexp\bigg)}
\newcommand{\asdiff}{\, \frac{\alpha_3(M_Z)-0.1184}{0.0007} }
\newcommand{\Mhdiff}{\bigg(\frac{M_h}{\GeV}-\Mhexp\bigg)}
\newcommand{\MWdiff}{\frac{M_W - \MWexp\GeV}{\MWerr\GeV}  }

\newcommand{\GeV}{\,{\rm GeV}}

\newcommand{\eq}[1]{~{\rm(\ref{eq:#1})}}
\def\eqg#1{eq.~(\ref{#1})}

\def\bea{\begin{eqnarray}}
\def\eea{\end{eqnarray}}

\def\psl{\hbox{\hbox{${p}$}}\kern-1.9mm{\hbox{${/}$}}}

\newcommand*\xbar[1]{%
  \hbox{%
    \vbox{%
      \hrule height 0.5pt 
      \kern0.4ex
      \hbox{%
        \kern-0.22em
        \ensuremath{#1}%
        \kern-0.08em
}}}}

\makeatletter
\def\cleardoublepage{\clearpage\if@twoside
\ifodd\c@page
\else\hbox{}\thispagestyle{empty}\newpage
\if@twocolumn\hbox{}\newpage\fi\fi\fi}
\makeatother

\usepackage{titlesec}
\titleformat{\chapter}[hang]{\Huge\bf}{\thechapter\quad}{0 pt}{}[]

\makeatletter
\renewcommand\tableofcontents{%
    \chapter*{\contentsname
        \@mkboth{Contents}{Contents}}%
    \@starttoc{toc}%
    \if@restonecol\twocolumn\fi
    }
\makeatother

\title{\Huge\bf Implications of the discovery of a Higgs boson with a mass of 125 GeV}
\author{\Large Dario Buttazzo}

\begin{document}

\frontmatter

\setcounter{tocdepth}{2}

\begin{titlepage}

\begin{center}

{\LARGE\sc Scuola Normale Superiore di Pisa}

\bigskip

{\Large\sc Classe di Scienze}

\bigskip

{\Large Corso di Perfezionamento in Fisica}

\vfill

\includegraphics[width=.2\textwidth]{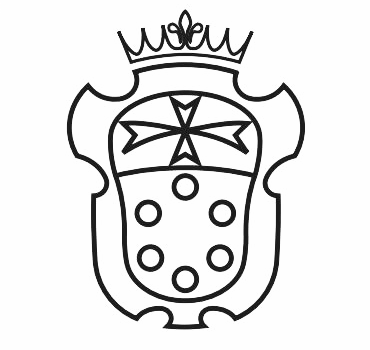}

\vfill

\centerline{\scalebox{1.13}
{\vbox{\LARGE\bf Implications of the discovery of a Higgs\\ boson with a mass of 125 GeV}}
}

\vfill

{\Large Ph.D. Thesis}

\vfill\vfill\vfill

\begin{minipage}{0.4\textwidth}
\begin{flushleft}

{\large\it Candidate}

\medskip

{\large Dario Buttazzo}

\end{flushleft}
\end{minipage}\hfill%
\begin{minipage}{0.4\textwidth}
\begin{flushright}

{\large\it Supervisor}

\medskip

{\large Prof. Riccardo Barbieri}

\end{flushright}
\end{minipage}

\vfill

\end{center}

\end{titlepage}

\chapter*{}
\thispagestyle{empty}
\hfill{\it\large To my family}
\vfill
\eject

\chapter*{}
\centerline{\Large\bf Abstract}
\bigskip

\noindent{The discovery of a Higgs-like particle by the ATLAS and CMS experiments at the LHC has been a major event for particle physics.
The rather precise knowledge of the mass of the Higgs boson and of its couplings to the other Standard Model fields has important consequences for the physical phenomena taking place at the Fermi scale of electroweak symmetry breaking. We will analyze some of these implications in the most motivated frameworks for physics at that scale -- supersymmetry, models of a composite Higgs boson, and the Standard Model itself.}

\noindent{At the same time, precision experiments in flavour physics require a highly non-generic structure of flavour and \CP\ transitions. This is relevant to any model of electroweak symmetry breaking with a relatively low scale of new phenomena, motivated by naturalness, where some mechanism has to be found in order to keep unwanted flavour effects under control. We will discuss in particular the consequences of the approximate $U(2)^3$ symmetry exhibited by the quarks of the Standard Model.}

\noindent{The combined analysis of the indirect constraints from flavour, Higgs and electroweak physics will allow us to outline a picture of some most natural models of physics at the Fermi scale. This is particularly interesting in view of the forthcoming improvements in the direct experimental investigation of the phenomena at that energies.
Although non trivially, a few models emerge that look capable of accommodating a 125 GeV Higgs boson, consistently with all the other constraints, with new particles in an interesting mass range for discovery at the LHC, as well as associated flavour signals.}

\noindent{Finally, the measurement of the last parameter of the Standard Model -- the Higgs quartic coupling -- has important consequences even if no new physics is present close to the Fermi scale: its near-critical value, which puts the electroweak vacuum in a metastable state close to a phase transition, may have an interesting connection with Planck-scale physics. We derive the bound for vacuum stability with full two-loop precision and use it to explore some possible scenarios of near-criticality.}

\noindent{This thesis is mainly based on the work published in the papers \cite{Barbieri:2012uh,Barbieri:2012bh,Barbieri:2012tu,Barbieri:2013hxa,Barbieri:2013nka,Buttazzo:2013uya}.}

\tableofcontents

\mainmatter

\chapter{Introduction}\label{Introduction}

In the last three decades the Standard Model of elementary particles (SM) \cite{Glashow:1961tr,Weinberg:1967tq,Salam:1968rm} has passed an impressive number of experimental tests.
The precision experiments at LEP have explored the nature of the gauge interactions with an incredibly high precision\cite{ALEPH:2005ab}, leaving little room for theories that differ significantly from the SM at energies below the TeV scale.
With the advent of the Large Hadron Collider a whole new energy range is opening up to experimental particle physics. For the first time the Fermi scale, at which electroweak symmetry breaking (EWSB) occurs, is being thoroughly explored.
After the first run of the LHC the Standard Model has been proven to be essentially at work even in the multi-TeV domain: a Higgs boson -- the last missing particle predicted by the model -- has been discovered \cite{Chatrchyan:2012ufa,Aad:2012tfa} and is showing standard-like properties through the measurement of its interactions \cite{Giardino:2013bma,CMS:ril,CMS:xwa,CMS:bxa,CMS:utj,Chatrchyan:2013vaa,CMS:zwa,CMS:gya,ATLAS:2013rma,ATLAS:2013qma,ATLAS:2013pma,ATLAS:2013oma,ATLAS:2013nma,ATLAS:2013mma,ATLAS:2013wla}, while no other significative signs of new phenomena have been observed. Needless to say, the discovery of the new particle exactly in the range of mass allowed by the precision tests of LEP is certainly another striking success of the theory.

Despite this enormous success, the SM as it is comes with two major theoretical issues, which seem to point to the existence of some new physics phenomena beyond the Standard Model (BSM) at not-too-short distances.

The observed Higgs mass of about 125 GeV fits well to the indirect prediction from the electroweak precision tests (EWPT), but leaves open a big question: why is it so light? For very other particle in the SM there is a symmetry protecting it to acquire a big mass by radiative corrections, but there is no such mechanism for scalar particles, which get a mass contribution from every scale they are coupled to, allowing them in principle to be as heavy as the Planck mass. This very precise fine-tuning of the Higgs mass to the Fermi scale is known as the {\it hierarchy} or {\it naturalness} problem of the Standard Model \cite{Wilson:1970ag,Gildener:1976ai,Gildener:1976ih,tHooft:1979bh,Maiani:1979cx}.
A light Higgs implies a large accidental cancellation between different, in principle unrelated physical quantities, which one wants to avoid, or at least explain by means of symmetries in a natural theory. In this view, suitable new phenomena should then appear at a low enough scale in order to suppress the large radiative corrections to the scalar masses.

A Higgs boson is thus a very special particle, naturally tied to the high-energy cut-off of the theory it lives in. The measurements of its mass and its couplings allow to get indirect informations about the short-distance physics where the Standard Model is supposed to break down. One of the aims of this work is to try to give a picture of some implications of the 125 GeV mass of the newly discovered Higgs boson, in the few most motivated models of particle physics at the Fermi scale.

The most popular solution to the hierarchy problem, enforced by the hints of a weakly interacting BSM physics coming from EWPT, is supersymmetry (SUSY).\footnote{See chapter~\ref{SUSY} for a list of references.}
In supersymmetric extensions of the Standard Model the Higgs mass is not renormalized in the high-energy limit where SUSY is unbroken, and is thus protected from ultraviolet (UV) radiative corrections. At low energies, it gets contributions proportional to the soft SUSY-breaking mass of each particle it couples to -- in particular from the stops and the gluinos, which have the strongest coupling to the Higgs sector. In a natural supersymmetric model these superparners can thus not have a too large mass. 
At the same time, the supersymmetric Higgs mass is a well-predicted quantity at tree-level, related to the electroweak scale in a very precise and model dependent way. The observed value of 125 GeV is a bit too high for a natural Minimal Supersymmetric Standard Model (MSSM), since it requires a largish radiative contribution from the stops, and thus a sizable amount of fine-tuning.
A more complex scalar sector with at least an extra singlet (NMSSM), where the tree-level value of the Higgs mass is larger, can instead accommodate lighter superpartners (compatibly with direct searches) and a lower SUSY-breaking scale.

Another possible solution to the hierarchy problem is that some new strong interaction sets in at a nearby energy scale, making the Higgs boson a composite object at that scale, and protecting its mass from short-distance radiative corrections. In a generic strongly interacting theory the compositeness scale would be close to the Higgs mass, being thus in conflict with EWPT and direct searches of heavy resonances at colliders. A natural and appealing way to describe a light composite Higgs is to consider it as a pseudo-Nambu-Goldstone boson of some spontaneously broken global symmetry of the strong sector \cite{Kaplan:1983fs,Georgi:1984af,Contino:2003ve, Agashe:2004rs,Giudice:2007fh}. In such a scenario the Higgs boson is significantly lighter than every other typical composite state, in a very similar fashion to what happens for the pions in QCD, allowing the threshold of the strong dynamics to be safely above the Fermi scale. Here the shift symmetry of the Goldstone field is the feature that really protects its mass from short-distance contributions.
Even in that case, however, the gap between the two scales cannot be too large in a natural theory, since radiative corrections below the compositeness threshold still tend to push the Higgs mass towards the cut-off of the theory. In addition to this, the measured value of 125 GeV lies close to the lower edge of the allowed spectrum in strongly interacting models, requiring a coupling to fermion partners which is well within the perturbative regime. This has important consequences in particular for the flavour structure of the theory.

The knowledge of the value of the Higgs mass already tells much about the form of new physics at the TeV scale and about its fine-tunings. More quantitative constraints come with the measurement of the Higgs couplings, which are entering quickly in the precision regime at the LHC. In strongly interacting models their close-to-standard values put lower bounds on the compositeness scale \cite{Contino:2006qr}, similarly to what EWPT do. In supersymmetry, where more than one Higgs boson is predicted, they can be used to constrain masses and mixings of the extra Higgs states \cite{Barbieri:2013hxa,Barbieri:2013nka,Djouadi:2013vqa}.

\bigskip

The second major puzzle of the Standard Model has to do with the physics of flavour and \CP\ transitions. The experimental progress in the last decade has shown in a rather spectacular way that the Cabibbo-Kobayashi-Maskawa (CKM) picture of flavour physics in the quark sector is fundamentally at work \cite{Charles:2011va,Derkach:2013da}. High-precision experiments in the physics of Kaon mixing and decays, as well as in B physics, set a lower bound of several thousands of TeV on the scale at which generic departures from the SM can appear \cite{Isidori:2010kg}.
While this is not a physical inconsistency, one wants to avoid to push the origin of flavour into the very short-distance range for two reasons. On the one hand, for the reasons described above, such a high scale of new physics could not be identified with a natural scale of EWSB. On the other hand, if the origin of flavour were completely decoupled from electroweak physics, this would leave little hope to understand the origin of the large hierarchies exhibited by the Yukawa couplings of the Standard Model. On this basis we are lead to consider extensions of the SM with highly non-generic flavour structures.

A possibility is to invoke a suitable flavour symmetry in order to imprint some specific structure into the new BSM physics. Probably the best known example of this kind is the so-called Minimal Flavour Violation (MFV) paradigm \cite{Chivukula:1987py,Hall:1990ac,DAmbrosio:2002ex}, where one assumes any kind of new physics to be formally invariant under the full $\U(3)_q\times \U(3)_u\times \U(3)_d$ flavour symmetry of the quark sector, apart from breaking terms which are taken proportional to the SM Yukawa couplings. An important shortcoming of MFV is that the $\U(3)^3$ group does not reflect the large hierarchies of the quark masses and mixings. A better choice is to consider a smaller $\U(2)_q\times \U(2)_u\times \U(2)_d$ symmetry acting on the first two families of quarks \cite{Barbieri:2012uh}. This is a good approximate symmetry of the SM Lagrangian, broken at most by an amount of a few times $10^{-2}$, which is the size of $V_{cb}$. A first objective of this work is to study in detail the consequences of such a flavour symmetry in a completely model-independent way, i.e. in an Effective Field Theory (EFT) approach. It turns out that a theory of flavour based on a $\U(2)^3$ symmetry naturally incorporates the pattern of Yukawa couplings of the SM, and makes predictions of new flavour and \CP\ phenomena which are fully consistent with the present experimental bounds and are potentially observable at the LHC.

Remarkably, some of the most stringent constraints on models of a composite Higgs boson, especially in view of its light mass, come from flavour physics \cite{Barbieri:2012tu}. An efficient way to couple the SM fermions to a composite Higgs boson, and thus to the strongly interacting sector, is to introduce a whole plethora of heavy fermionic resonances which mix with the known elementary fermions, in what goes under the name of {\it partial compositeness} \cite{Kaplan:1991dc,Contino:2006nn}. In this framework it is possible to suppress flavour and \CP\ violating effects through the small couplings of the standard fermions to the strong sector. This is a very efficient tool to avoid dangerous effects even without relying on some particular flavour structure, but, as we will see, even in this case there is still some strong constraint coming from experimental bounds, especially from \CP\ violation in the Kaon system.
It will be interesting to study the possibility to embed a flavour symmetry in specific composite Higgs models.
We will analyze both the case of $\U(3)^3$ and of $\U(2)^3$, and will compare them to the case of an anarchic generation of flavour. A detailed analysis of all the relevant, often competing bounds coming from flavour, from collider data, and from EWPT is carried out, in order to get an accurate picture of which are the models of flavour consistent with a low scale of strongly interacting new physics and a Higgs mass of 125 GeV.

\bigskip

It is worth to notice that the observed value of the Higgs mass does not lie in the most favorable range for either of the known natural solutions to the hierarchy problem -- supersymmetry and compositeness -- requiring in both cases some additional ingredient with respect to the simplest models in order to be in agreement with the predictions. It must also be said that the discovered resonance looks SM-like to a good extent. Furthermore, the direct searches of new particles expected both in supersymmetry or in Higgs compositeness performed in the first phase of the LHC has given negative results.
Still, both options are not excluded with the present experimental data. The second run of the LHC at center-of-mass energy of 14 TeV will probably have a relevant word about the question of whether the naturalness paradigm, in one of its declinations, is able to explain the Fermi scale.
To anticipate a possible negative outcome, one may find it interesting to investigate the consequences of a picture where the Standard Model is assumed to hold up to high energies, imagining that some unknown mechanism different from naturalness explains the lightness of the Higgs. This is even more the case given the very special value of the Higgs mass in the context of the SM itself: the observed value of 125 GeV lies indeed remarkably close to the critical point where the quartic scalar coupling turns negative at some high energy scale and electroweak vacuum becomes unstable \cite{Degrassi:2012ry,Buttazzo:2013uya}. This near-criticality, which may be an intriguing hint of some physics at very high scales, motivates an accurate study of the higher-order corrections to the lower bound for vacuum stability.

\bigskip

We will analyze the implications of a 125 GeV Higgs with standard-like couplings in the three outlined scenarios in different parts of this thesis. In the first part we analyze the flavour problem of BSM physics in a model-independent way, using an Effective Field Theory approach, and we show how a $\U(2)^3$ flavour symmetry may be a suitable framework for models of new physics at the Fermi scale. In the second part we discuss composite Higgs models, and we use the tools introduced in the previous part to investigate the relation between Higgs properties, flavour physics and EWPT. In the third part we consider the Higgs sector of supersymmetric theories, focussing the attention in particular on the search of the extra Higgs states which are predicted in these models. In the last part we compute the vacuum stability bound in the Standard Model and explore the near-criticality of a 125 GeV Higgs boson mass in the hypothesis that the SM is valid up to a high energy scale.

\chapter{The Standard Model of elementary particles}\label{chap:SM}

The Standard Model of elementary particles \cite{Glashow:1961tr,Weinberg:1967tq,Salam:1968rm} is the renormalizable quantum field theory which successfully describes all the known particles and their gauge interactions (excluding gravity).\footnote{See \cite{Barbieri:2007gi} for a review.} Let us briefly review its main features, and see how naturalness arguments do require new physics beyond it to appear not too much above the Fermi scale.

\section{The SM Lagrangian}\label{SM/Lagrangian}

The Lagrangian of the Standard Model can be schematically written as
\begin{equation}\label{SM}
\L = \L_g + \L_H + \L_Y + \L_{\nu},
\end{equation}
where $\L_g$ describes the gauge interactions of vector bosons and fermions, $\L_H$ is the Higgs Lagrangian which triggers electroweak symmetry breaking, $\L_Y$ contains the Yukawa interactions between the Higgs field and the fermions, which are responsible for flavour physics in the SM, and $\L_{\nu}$ generates the small neutrino masses.
Let us briefly review the particle content of the different sectors of the theory.

\paragraph{Gauge sector.}
The electromagnetic, weak and strong interactions are described by a gauge theory based on the symmetry group
\begin{equation}\label{gaugegroupSM}
\SU(3)_c\times \SU(2)_L\times \U(1)_Y.
\end{equation}
In terms of particles, this comprises all the spin-1 bosons, namely the eight gluons of the strong interaction and four weakly coupled vector bosons. The interactions among them are completely determined by the gauge symmetry,
\begin{equation}\label{gauge}
\L_g = -\frac{1}{4}G_{\mu\nu}^{\alpha} G^{\mu\nu}_{\alpha} - \frac{1}{4}W_{\mu\nu}^a W^{\mu\nu}_a - \frac{1}{4}B_{\mu\nu}B^{\mu\nu},
\end{equation}
where $G_{\mu\nu}$, $W_{\mu\nu}$ and $B_{\mu\nu}$ are the field strengths associated with the various gauge fields.

Chiral Weyl fermions are minimally coupled to the gauge fields,
\begin{equation}
\L_f = \sum_f i\bar\psi_f \gamma^{\mu}D_{\mu}^{(f)}\psi_f,
\end{equation}
where the covariant derivative is $D_{\mu}^{(f)} = \partial_{\mu} - ig_s G_{\mu}^a \mathcal{T}^{(f)}_a - igW_{\mu}^i T_i^{(f)} - ig'Y^{(f)}B_{\mu}$, $\mathcal{T}_{\alpha}^{(f)}$, $T_a^{(f)}$ are respectively the $\SU(3)_c$ and $\SU(2)_L$ generators in the representation of the gauge group corresponding to $\psi_f$, and $Y^{(f)}$ is the hypercharge of $\psi_f$. Only the left-handed fermions are charged under $\SU(2)_L$, while the right-handed fermions are singlets. In the quark sector there is thus one doublet and two singlets, corresponding to the left- and right-handed up and down quarks; analogously, in the lepton sector there is one left-handed doublet and -- if we include also the sterile right-handed neutrino -- two right-handed singlets. All the fermions come in three identical copies (generation), which differ only in the Yukawa sector, i.e. for their masses. The whole matter content of the Standard Model can be summarized in table \ref{matter}.
\begin{table}[h]
\renewcommand{\arraystretch}{1.3}
\centering%
\begin{tabular}{c|cccccc}
& $q_L$ & $u_R$ & $d_R$ & $\ell_L$ & $e_R$ & $\nu_R$\\
\hline
$\SU(3)_c$ & {\bf 3} & {\bf 3} & {\bf 3} & {\bf 1} & {\bf 1} & {\bf 1}\\
$\SU(2)_L$ & {\bf 2} & {\bf 1} & {\bf 1} & {\bf 2} & {\bf 1} & {\bf 1}\\
$Y$ & 1/6 & 2/3 & -1/3 & -1/2 & -1 & 0
\end{tabular}
\caption{Matter content of the Standard Model.\label{matter}}
\end{table}

\paragraph{Electroweak symmetry breaking sector.} In order to give masses to the $W$ and $Z$ bosons, the full gauge symmetry of \eqref{gauge} has to be spontaneously broken. In the SM this is achieved through the coupling of the gauge bosons to a weakly interacting doublet of scalar fields, the Brout-Englert-Higgs field $H$ \cite{Higgs:1964ia,Englert:1964et,Guralnik:1964eu}. The most general renormalizable gauge-invariant Lagrangian for $H$ is
\begin{equation}\label{Higgslag}
\L_{H} = |D_{\mu}H|^2 - V(H),
\end{equation}
where
\begin{equation}\label{higgspot}
V(H) = -\frac{m_h^2}{2} H^{\dag}H + \lambda (H^{\dag}H)^2.
\end{equation}
If $m_h^2 > 0$ the potential has a vacuum state with a nonzero field configuration which breaks the electroweak gauge symmetry down to the electromagnetic $\U(1)_{\rm em}$. The Higgs field can then be written as
\begin{equation}\label{Higgsfield}
H = \begin{pmatrix}\chi^-\\ (v + h + i\eta)/\sqrt{2},\end{pmatrix}
\end{equation}
where the vacuum expectation value (vev) is $v\simeq 246$ GeV. The three Nambu-Goldstone bosons $\chi^\pm$, $\eta$ of the spontaneous breaking are eaten-up by the combinations of gauge bosons $W^{\pm}$ and $Z$ which get masses
\begin{equation}\label{WZmasses}
m_W = \frac{g\, v}{2} \simeq 80.39 \,{\rm GeV},\qquad m_Z = \frac{m_W}{\cos\theta_w} \simeq 91.19 \,{\rm GeV},
\end{equation}
where $\tan\theta_w = g'/g$ is the weak mixing angle.
The photon remains massless, as required by the residual unbroken gauge symmetry.
The fourth scalar degree of freedom corresponds to the Higgs boson $h$ \cite{Higgs:1964ia}. Its mass $m_h = v\sqrt{2\lambda}$ is a free parameter of the theory and has to be determined experimentally. If the particle which was recently observed at the ATLAS and CMS experiments at the Large Hadron Collider turns out to be the Higgs boson of the SM, its mass $m_h \simeq 125$~GeV would be in perfect agreement with the indirect bounds from the LEP data,\footnote{See next section.} and would imply a tree-level value of the quaric self-coupling $\lambda = m_h^2/2v^2\simeq 0.13$.

\paragraph{Flavour sector.} The $\SU(2)_L\times \U(1)_Y$ symmetry does not allow a gauge invariant mass term for the fermions. They must thus acquire their mass as a consequence of EWSB.

The Yukawa interaction terms between the Higgs field and the fermions can be written as
\begin{equation}\label{yukawa}
\L_Y = \frac{1}{\sqrt{2}}\sum_{i,j}y_u^{ij}\bar q_L^i H^{c} u_R^j + y_d^{ij}\bar q_L^i H d_R^j + y_{\ell}^{ij}\bar\ell_L^i H e_R^j,
\end{equation}
where $H^c = i\sigma^2 H^*$ and $i,j$ index the three generations.
When $H$ gets a vev \eqref{yukawa} generates the masses of the fermions, as well as mixings among them, plus interaction terms with the Higgs boson $h$. Using the global $\SU(2)_L$ invariance of \eqref{gauge} one is allowed to perform rotations and phase redefinitions of the various fields, putting the previous Lagrangian in the following form
\begin{equation}\label{massbasis}
\L_Y = \bar u_L^i V^{\dag}_{ij}m_{u_j}u_R^j + m_{d_i}\bar d_L^i d_R^i + m_{e_i}\bar e_L^ie_R^i,
\end{equation}
where the Cabibbo-Kobayashi-Maskawa (CKM) matrix $V$ is unitary, and a sum over all the indices is understood.

Here we are neglecting the terms involving the sterile right-handed neutrinos $\nu_R$, which give rise to the small neutrino masses.

\section{Electroweak precision observables}\label{SM/EWPT}
Starting from \eqref{SM} one can calculate the renormalized effective action at a given energy scale, which contains all possible operators allowed by the symmetries, and includes quantum corrections from loops.
One interesting subset of these, which have been tightly constrained by the experiments at LEP, is constituted by the transversal terms in the vacuum polarizations of the gauge fields, i.e. the quadratic part of the effective gauge Lagrangian. After EWSB we can parametrize it as follows
\begin{equation}\label{vacpol}
\L_{\rm eff}^{(2)} = -\frac{1}{2}W_{\mu}^3\Pi_{33}(q^2)W_{\mu}^3 - \frac{1}{2}B_{\mu}\Pi_{00}(q^2)B_{\mu} - W_{\mu}^3\Pi_{30}(q^2)B_{\mu} - W_{\mu}^+\Pi_{WW}(q^2)W_{\mu}^-.
\end{equation}

The useful thing in this approach is that, since we are dealing with an effective low-energy action, the form of \eqref{vacpol} is independent of any high-energy completion of the Standard Model, but the functions $\Pi_{IJ}$ get corrections from loops involving both the SM and -- possibly -- the new physics degrees of freedom. For that reason, and since the $\Pi$'s are measured with high precision, they are an excellent tool to constrain any BSM theory, as well as to constrain unknown SM parameters, such as the Higgs mass (before it was measured).

Let us concentrate on the terms in \eqref{vacpol} that are most sensitive to a high energy scale, namely, as one can easily verify by power counting, the first two orders of a derivative expansion, $\Pi(0)$ and $\Pi'(0)$. Some of the effective operators that we are considering were already contained in the bare Lagrangian \eqref{SM}, so their coefficients are nothing else than renormalized SM parameters which have to be determined by experiments. More in detail, $\Pi'_{WW}(0)$ and $\Pi'_{00}(0)$ fix the normalization of the kinetic terms $W_{\mu\nu}^aW^{\mu\nu}_a$ and $B_{\mu\nu}B^{\mu\nu}$, i.e. the value of the coupling constants $g$ and $g'$. $\Pi_{ZZ}(0)$, in turn, is the mass of the $Z$ boson, which is determined in terms of $v$. We get two more conditions requiring that the photon remains massless, namely $\Pi_{\gamma\gamma}(0)$ = $\Pi_{\gamma Z}(0)$ = 0. In the end we are left with three independent physical quantities: two $\Pi'$'s and one $\Pi$, which can be parametrized as follows,
\begin{equation}\label{STU}
\hat T = \frac{\Pi_{33}(0) - \Pi_{WW}(0)}{m_W^2},\quad \hat S = \frac{g}{g'}\Pi'_{30}(0),\quad \hat U = \Pi'_{33}(0) - \Pi'_{WW}(0).
\end{equation}

Four other parameters appear at the next order in $q^2$,
\begin{equation}\begin{aligned}\label{VWXY}
\hat V &= \frac{m_W^2}{2}\left(\Pi''_{33}(0) - \Pi''_{WW}(0)\right), & \hat W &= \frac{m_W^2}{2}\Pi''_{33}(0),\\
\hat X &= \frac{m_W^2}{2}\Pi''_{30}(0), & \hat Y &= \frac{m_W^2}{2}\Pi''_{00}(0),
\end{aligned}\end{equation}
all of which are determined in the SM, but are experimentally less tightly constrained than the previous ones.

\begin{figure}
\centering%
\includegraphics[width=\textwidth]{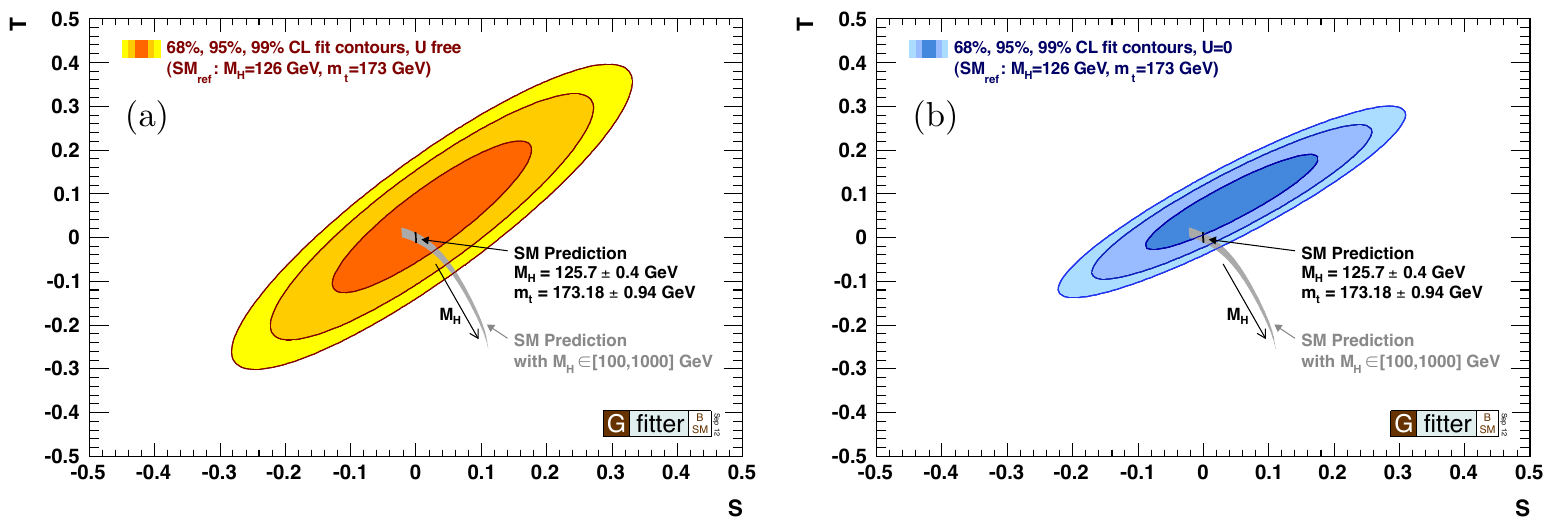}
\caption{Experimental constraints on the $S$ and $T$ parameters with respect to the SM reference
($m_{h,{\rm ref}} = 126$ GeV and $m_{t,{\rm ref}} = 173$ GeV) from \cite{Baak:2012kk}. Shown are the 68\%, 95\% and 99\% C.L. allowed regions, where the third parameter U is left unconstrained (left) or fixed to 0 (right). The prediction in the SM is given by
the black (grey) area when including (excluding) the $m_h$ measurements.\label{obliquefit}}
\end{figure}

One-loop diagrams involving the exchange of a Higgs particle correct the $\hat S$ and $\hat T$ parameters by an amount
\begin{equation}
\delta\hat S \simeq \frac{G_Fm_W^2}{12\sqrt{2}\pi^2}\log\left({m_h}/{\mu}\right),\qquad \delta\hat T\simeq -\frac{3 G_F m_W^2}{4\sqrt{2}\pi^2}\tan^2\theta_w\log\left(m_h/\mu\right).
\end{equation}
Note that the $\hat T$ parameter vanishes in the limit where the {\it custodial symmetry} $\SU(2)_L\times \SU(2)_R$ is exactly respected and the relation $m_W^2 = m_Z^2$ holds.
A best fit to these quantities from all the LEP, Tevatron and LHC measurements, shown in figure~\ref{obliquefit}, yields a prediction for the Higgs mass of \cite{Baak:2012kk}
\begin{equation}
m_h = 94^{+25}_{-22}\,{\rm GeV},\qquad m_h < 163 \,{\rm GeV}\;{\rm at \;95\%\; C.L.}
\end{equation}
which is in good agreement with the measured value of about 125 GeV.

Suppose now that some new physics beyond the Standard Model is present at an energy scale $\Lambda_{\rm NP}$. Again in an effective field theory approach, the exchange of these UV degrees of freedom will generate non-renormalizable corrections to the Lagrangian \eqref{SM},
\begin{equation}\label{SMeff4}
\L = \L_{\rm SM} + \sum_i \frac{c_i}{\Lambda_{\rm NP}^{d_i-4}}{\O}_i,
\end{equation}
where ${\O}_i$ are operators of dimension $d_i$. These new operators, in turn, will contribute to the parameters $\hat S, \hat T, \hat U, \hat V, \hat W, \hat X, \hat Y$, and from the experimental constraints on these contributions we can set a lower bound on the scale $\Lambda_{\rm NP}$ of new physics.
Just as an example, we give here the form of the operators contributing to the parameters $\hat S$ and $\hat T$,
\begin{align}
{\O}_H &= |H^{\dag}D_{\mu}H|^2 & &\Longrightarrow & \Delta\hat T &= -\frac{2}{g^2}\frac{m_W^2}{\Lambda_{NP}^2}c_{H},\label{deltaT}\\
{\O}_{WB} &= g g' (H^{\dag}\sigma^a H)W_{\mu\nu}^a B^{\mu\nu} & &\Longrightarrow & \Delta\hat S &= 4\frac{m_W^2}{\Lambda_{NP}^2}c_{WB}.
\end{align}

\section{The hierarchy problem: is Nature natural?}\label{SM/hierarchy}

In the SM, the mass terms for vectors and fermions are forbidden by the $\SU(2)_L~\times~\U(1)_Y$ gauge symmetry. Since they have to vanish in the limit where the symmetry is unbroken, they are all proportional to the vev. This is true not only at tree level (where $m_W = g v/2$, $m_t = \lambda_t v/\sqrt{2}$, \dots) but at any order in perturbation theory, beacuse the renormalization procedure preserves all the symmetries.  The size of any loop correction is thus controlled by the tree-level masses.
We say that gauge boson masses and fermion masses are {\it protected}, i.e. the point $m = 0$ is stable under radiative corrections. For fermions the same property holds also in absence of a gauge symmetry, because of the chiral symmetry which is broken by the mass term. In general any point of the parameter space with an enhanced symmetry is stable under renormalization group (RG) running.

\begin{figure}[t]
\centering%
\hfill%
\includegraphics{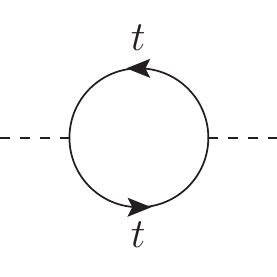}\hfill%
\includegraphics{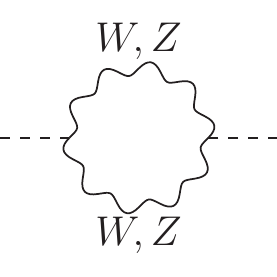}\hfill%
\includegraphics{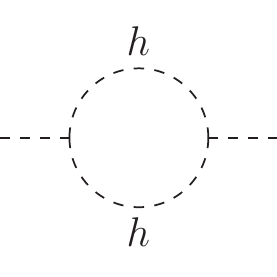}\hfill%
\includegraphics{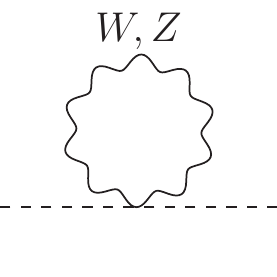}\hfill%
\includegraphics{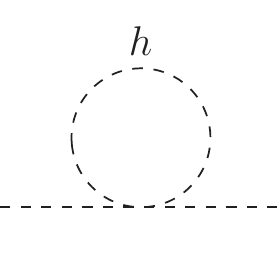}\hfill\hfill
\caption{Feynman diagrams contributing to $m_h$ at one loop in the Standard Model.\label{SMHiggstwoptf}}
\end{figure}

The same property does not hold for scalar particles. The mass of the Higgs boson $m_h$ is an arbitrary parameter of the model, not protected by any approximate symmetry, which is {\it additively} renormalized: it gets radiative corrections proportional to the mass of any particle which couples to it. In that sense the point $m_h = 0$ is UV-unstable. This is easily seen in the Standard Model, where the one-loop corrections to the Higgs mass are generated by the diagrams in figure~\ref{SMHiggstwoptf} and are given in appendix~\ref{app:1loop}.
However, if we compute the beta function for the running mass we get
\begin{equation}
\beta_{m_h^2} = \frac{ d m_h^2}{d\log\mub} = \frac{3 m_h^2}{8\pi^2}\Big(2\lambda + y_t^2 - \frac{3 g^2}{4} - \frac{g'^2}{4}\Big),
\end{equation}
i.e. the running of the mass parameter $m_h^2$ is proportional to itself. This is true in the pure SM because the masses of the particles are all proportional to the EWSB scale $v$.

Suppose now that the SM is modified at some energy $\Lambda_{\rm NP} > \Lambda_{\rm SM}$, where $\Lambda_{\rm SM}\simeq4\pi m_W$ is the typical energy scale of the SM. If the Higgs boson is coupled to the new physics sector, then its mass will get a correction also from loops of the new heavy particles, which will be quadratic in their mass $M \approx \Lambda_{\rm NP}$. If we want a UV completion of the Standard Model in which the Higgs mass is a predictable quantity, this constitutes a problem.

To make the statement more precise, let us calculate explicitly the one-loop correction to the Higgs pole mass arising from a fermion with Dirac mass $M$ and Yukawa coupling $y$. From a diagram analogous to the first one of figure~\ref{SMHiggstwoptf}, using dimensional regularization we get
\begin{align}\label{deltamh}
\delta m_h^2 &= \Re\left. \Pi_{hh}\right|_{p^2 = m_h^2} = \frac{y^2}{2(4\pi)^2}\Re\big[\Delta_{\epsilon} + (m_h^2 - 4M^2)B_0(m_h; M, M) - 2 A_0(M) \big]\notag\\
&= \frac{y^2}{2(4\pi)^2}\Big(\Delta_{\epsilon} + (6M^2 - m_h^2)\log\frac{m_h^2}{\mub^2} + f(m_h, M)\Big),
\end{align}
where $\Delta_{\epsilon}$ is the pole which has to be subtracted by a counterterm, $A_0$ and $B_0$ are the finite parts of the Passarino-Veltman one-loop functions defined in appendix \ref{app:1loop}, $\mub$ is the renormalization scale and $f$ is some function.
Very similar equations hold for scalar and vector particles circulating in the loop (see eq.~\eqref{Higgsmassthreshold1loop} in the appendix).
The term $f(m_h^2,M^2)$ in \eqref{deltamh} is unphysical since it does not depend on $\mub$ and it can be subtracted together with the divergence in a suitable renormalization scheme -- anyway it drops out from mass differences between different scales. The logarithm, on the other hand, contributes to the beta function of the running Higgs mass as
\begin{equation}\label{betamh}
\beta_{m_h^2} = \frac{d\, m_h^2(\mub)}{d\log\mub} = \frac{y^2}{(4\pi)^2}(m_h^2 - 6M^2) + \cdots.
\end{equation}
The renormalization group running thus generates a mass term $m_h\approx M^2$, even if one sets this term to zero at a given scale, if the running is done over a sufficiently large energy range.
Fixing the boundary conditions for the renormalization group equation at the high scale $\Lambda_{\rm NP}$, where one imagines some UV-completion to determine the masses and couplings, the relation between the Higgs mass at the two scales $\Lambda_{\rm NP}$ and $\Lambda_{\rm SM}$ then reads
\begin{equation}\label{mhcorrection}
m_h^2(\Lambda_{\rm SM}) \simeq m_h^2(\Lambda_{\rm NP}) - \#\,\Lambda_{\rm NP}^2\log\frac{\Lambda_{\rm NP}}{\Lambda_{\rm SM}}.
\end{equation}
where \# is a numerical factor which includes also coupling constants.
The hierarchy problem can now be stated in the following way: if the scale $\Lambda_{\rm NP}$ is much higher than $m_h$, then the two contributions in \eqref{mhcorrection} have to balance out with a very high accuracy in order to generate a Higgs boson mass much smaller than $\Lambda_{\rm NP}$.

This can better be formalized in terms of the amount of {\it fine-tuning}
\begin{equation}
\Delta = \frac{d\log m_h^2(\Lambda_{\rm SM})}{d\log m_h^2(\Lambda_{\rm NP})}\propto \frac{\Lambda_{\rm NP}^2}{m_h^2(\Lambda_{\rm SM})},
\end{equation}
which is the precision to which the initial conditions at the high scale have to be given in order to have the Higgs mass at the low scale determined up to a factor of order 1. Let us see some explicit example to get an idea of the numbers we are talking about: if we take $\Lambda_{\rm NP}$ to be, say, of the order of the Planck scale, then we get $\Delta\sim 10^{34}$ for a Higgs mass of about $125$ GeV. If we accept an amount of fine tuning at the percent level, namely an accidental cancellation between the initial conditions $m_h(\Lambda_{\rm NP})$ and the quantum corrections of the order of one percent, then the scale of new physics cannot be much higher than the TeV.

A simple way to reformulate the hierarchy problem is to consider the Standard Model as an effective field theory (EFT), valid up to the maximum energy scale $\Lambda_{\rm NP}$. Its Lagrangian can then be written in the form
\begin{equation}\label{SMEFT}
\L_{\rm SM, eff} = \sum_i C_i(\mub)\Lambda_{\rm NP}^{4-d_i}{\O}_i,
\end{equation}
where the $\O_i$ are operators of dimension $d_i$ and $C_i(\mub)$ are their Wilson coefficients, which in an effective field theory are not predicted, and are usually of order 1 unless some symmetry is operative. The Higgs mass term is an operator of dimension two, and thus comes with a factor $\Lambda_{\rm NP}^2$. If the cut-off scale is very big, the only way to get a small mass is to have a large suppression of the Wilson coefficient at the Fermi scale: a fine-tuning. On the other hand, a large cut-off in \eqref{SMEFT} seems to be preferred by the experimental constraints on the operators of dimension greater than four, as in \eqref{SMeff4}. We will discuss this issue in detail in the next chapters.

There are different approaches one can take to face the hierarchy problem.
If the Standard Model were a complete description of Nature, and no new phenomena appear at any energy, then there would be no hierarchy problem at all. The Higgs mass is not predicted in the SM, as every other parameter in the Lagrangian, and its renormalized value is arbitrary and has to be determined by experiments: there is no reason to set the input for the renormalization group running at some high scale.
However, usually one prefers to view the SM as an effective field theory valid in a limited energy range.
After all, we know that some new physics {\it has to appear} at least at the Planck scale, where quantum gravity effects are expected to become important. Moreover, there are many indirect hints of new physics at high energies coming from dark matter observations, the baryon asymmetry in the Universe, the smallness of neutrino masses, the Yukawa hierarchies in flavour physics, and so on.\footnote{One could include in the list also inflation and the cosmological constant, which are related to two other enormous fine-tuning problems.}
This of course does not automatically mean that the coupling of the Higgs boson to all of those new degrees of freedom must lead to the quadratic corrections that we estimated in \eqref{mhcorrection}. A possible scenario could be that the Higgs mass is protected from quantum gravity effects through some unknown mechanism, while any other phenomenon below the Planck scale is sufficiently decoupled from the Standard Model to make its correction irrelevant\cite{Farina:2013mla}.

Another possibility is to ignore the hierarchy problem and accept a fine tuned Standard Model; after all, fine-tunings do exist elsewhere in Nature.\footnote{The binding energy of the deuteron, $\Delta E_D\simeq 2$ MeV, is fine-tuned to some few percent.} In that case one has to find some guideline other than naturalness to go beyond the SM. The near-criticality of the Higgs boson quartic coupling and of its beta-function may be an intriguing possibility in this direction.

If one insists with naturalness, still viewing the Standard Model as an effective field theory, then one has to conclude that new (BSM) physics has to appear at the TeV scale. The models that describe these new phenomena can be divided into two classes, depending on whether they are strongly coupled or weakly coupled.

\begin{itemize}
\item Among weakly interacting theories beyond the Standard Model the main candidate is certainly supersymmetry. The non-renormalization theorem causes all the quadratic divergences to cancel out exactly above the scale $\Lambda_{\rm soft}$ of supersymmetry-breaking sparticle masses. The main corrections to the Higgs mass thus come from SM loops, cut-off approximately at that scale, which are proportional to $\Lambda_{\rm soft}^2$.

\item In composite models the Higgs boson is a resonance of some new strongly interacting sector. Since it makes no sense to speak about the Higgs particle above its compositeness scale, it is automatically protected from Planck-scale radiative corrections. We will see in the following that, again because of the SM quantum corrections, a natural compositeness scale has to lie somewhere in the TeV range.
\end{itemize}

\part{The flavour problem}

\chapter{Flavour physics beyond the Standard Model}\label{Flavour}

Some of the most precise measurements in elementary particle physics concern flavour changing processes, namely transitions between different generations of quarks and leptons, and \CP\ violating observables. In the Standard Model all the flavour and \CP\ processes in the quark sector are governed by one single unitary matrix -- the Cabibbo-Kobayashi-Maskawa (CKM) matrix \cite{Cabibbo:1963yz,Kobayashi:1973fv}. High-precision experiments at colliders and heavy flavour factories have shown that the CKM picture is correct with a 20--30\% accuracy for a large number of observable processes \cite{Charles:2011va,Derkach:2013da}.

Thus, if one wants to extend the SM at the TeV scale, the new physics cannot have an arbitrary flavour structure, since its effects in low energy flavour and \CP\ processes would spoil the SM predictions. This is true, for instance, both for supersymmetric and strongly interacting models.

Moreover, the Yukawa couplings of the SM -- which are responsible for quark mixings and \CP\ phases -- present a very evident hierarchical structure, which may be generated by some more fundamental underlying mechanism. At present no concrete theory of flavour explaining the dynamical mechanism which generates the Yukawa patterns, being fully consistent with precision bounds on flavour observables, is known. Nevertheless, a handful of useful parametrizations and extensions of the CKM picture exist, which may give some insight on the origins of flavour physics.

In this chapter we adopt an effective theory point of view to parametrize all the possible flavour and \CP\ operators that may arise in a BSM theory. We show also the most relevant experimental bounds on the scale of generic new physics coming from flavour observables. We review the CKM structure of the Standard Model, and the Minimal Flavour Violation paradigm, which is its minimal extension.
We consider, here and in the following, mainly the quark sector, with some minor remark about the lepton sector when useful.

\section{Effective flavour operators}\label{Flavour/EFT}
From the point of view of an effective field theory, any high-energy extension of the Standard Model which preserves its gauge group \eqref{gaugegroupSM} can be described by a set of effective operators of dimension greater than four. New physics contributions to operators of lower dimension are not relevant since they can be reabsorbed in a renormalization of the SM couplings.

Let us consider the generic effective Lagrangian
\begin{equation}\label{Heff}
\L_{\rm eff} = -\H_{\rm eff} = \sum_i \frac{1}{\Lambda_i^2}C_i(\mu)\O_i,
\end{equation}
where $\mu$ is the renormalization scale. Requiring gauge invariance, and the conservation of baryonic and leptonic numbers, the operators which contribute to each possible $q_i\leftrightarrow q_j$ flavour transition can be classified in the following way:
\begin{itemize}
\item $\Delta F = 2$ operators contribute to heavy meson mixings
\begin{align}
\Q_1 &= (\bar q_{Li}\gamma_{\mu}q_{Lj})(\bar q_{Li}\gamma_{\mu} q_{Lj}),& \Q_1' &= (\bar q_{Ri}\gamma_{\mu}q_{Rj})(\bar q_{Ri}\gamma_{\mu} q_{Rj}),\label{Q1}\\
\Q_2 &= (\bar q_{Ri} q_{Lj})(\bar q_{Ri} q_{Lj}),& \Q_2' &= (\bar q_{Li}q_{Rj})(\bar q_{Li}q_{Rj}),\label{Q2}\\
\Q_3 &= (\bar q_{Ri}^{\alpha} q_{Lj}^{\beta})(\bar q_{Ri}^{\beta} q_{Lj}^{\alpha}),& \Q_3' &= (\bar q_{Li}^{\alpha}q_{Rj}^{\beta})(\bar q_{Li}^{\beta}q_{Rj}^{\alpha}),\label{Q3}\\
\Q_4 &= (\bar q_{Ri}q_{Lj})(\bar q_{Li}q_{Rj}),& \Q_5 &= (\bar q_{Ri}^{\alpha}q_{Lj}^{\beta})(\bar q_{Li}^{\beta}q_{Rj}^{\alpha});\label{Q4}
\end{align}
\item four-quark $\Delta F = 1$ operators contribute to hadronic heavy meson decays
\begin{align}
\O_1^k &= (\bar q_{Li}\gamma_{\mu}q_{Lj})(\bar q_{Lk}\gamma_{\mu}q_{Lk}), & \O_1^{\prime k} &= (\bar q_{Ri}\gamma_{\mu}q_{Rj})(\bar q_{Rk}\gamma_{\mu}q_{Rk}),\\
\O_2^{k} &= (\bar q_{Li}^{\alpha}\gamma_{\mu}q_{Lj}^{\beta})(\bar q_{Lk}^{\beta}\gamma_{\mu}q_{Lk}^{\alpha}), &
\O_2^{\prime k} &= (\bar q_{Ri}^{\alpha}\gamma_{\mu}q_{Rj}^{\beta})(\bar q_{Rk}^{\beta}\gamma_{\mu}q_{Rk}^{\alpha}),\\
\O_5^k &= (\bar q_{Li}\gamma_{\mu}q_{Lj})(\bar q_{Rk}\gamma_{\mu}q_{Rk}), & \O_5^{\prime k} &= (\bar q_{Ri}\gamma_{\mu}q_{Rj})(\bar q_{Lk}\gamma_{\mu}q_{Lk}),\\
\O_6^k &= (\bar q_{Li}^{\alpha}\gamma_{\mu}q_{Lj}^{\beta})(\bar q_{Rk}^{\beta}\gamma_{\mu}q_{Rk}^{\alpha}), &
\O_6^{\prime k} &= (\bar q_{Ri}^{\alpha}\gamma_{\mu}q_{Rj}^{\beta})(\bar q_{Lk}^{\beta}\gamma_{\mu}q_{Lk}^{\alpha}),
\end{align}
and in addition to these, in principle there are also operators with scalar and tensor Lorentz structure, such as $(\bar q_{Li}q_{Rj})^2$ or $(\bar q_{Li}\sigma_{\mu\nu}q_{Rj})^2$;
\item magnetic and chromomagnetic $\Delta F =1$ operators contribute to $q_i\to q_j\gamma$ transitions
\begin{align}
\O_{7\gamma} &= m_j (\bar q_{Li}\sigma_{\mu\nu}q_{Rj})eF^{\mu\nu},& \O_{7\gamma}' &= m_i(\bar q_{Ri}\sigma_{\mu\nu}q_{Lj})eF^{\mu\nu},\\
\O_{8g} &= m_j (\bar q_{Li}\sigma_{\mu\nu}q_{Rj})g_sG^{\mu\nu},& \O_{8g}' &= m_i(\bar q_{Ri}\sigma_{\mu\nu}q_{Lj})g_sG^{\mu\nu};
\end{align}
notice that operators proportional to $\partial_{\mu}F^{\mu\nu}$ and $\partial_{\mu}G^{\mu\nu}$ can also be present;
\item the following $\Delta F = 1$ operators contribute to semi-leptonic heavy meson decays
\begin{align}
\O_9^{k} &= (\bar q_{Li}\gamma_{\mu}q_{Lj})(\bar\ell_{Lk}\gamma_{\mu}\ell_{Lk}),& \O_9^{\prime k} &= (\bar q_{Ri}\gamma_{\mu}q_{Rj})(\bar e_{Rk}\gamma_{\mu} e_{Rk}),\\
\O_{10}^{k} &= (\bar q_{Li}\gamma_{\mu}q_{Lj})(\bar e_{Rk}\gamma_{\mu} e_{Rk}),& \O_{10}^{\prime k} &= (\bar q_{Ri}\gamma_{\mu}q_{Rj})(\bar\ell_{Lk}\gamma_{\mu}\ell_{Lk}),
\end{align}
plus scalar and tensor operators as before;
\item the following $\Delta F = 1$ operators contribute to $Z\to d_i d_j$ after EWSB
\begin{align}
\O_H &= (\bar q_{Li}\gamma_{\mu}q_{Lj})(H^{\dag}D_{\mu}H), & \O_H' &= (\bar q_{Ri}\gamma_{\mu}q_{Rj})(H^{\dag}D_{\mu}H).
\end{align}
\end{itemize}
Moreover we will consider also the following flavour-diagonal operators which give rise to (chromo-)electric and (chromo-)magnetic dipole moments of the quarks
\begin{align}
\O_{7\gamma}^{\d} &= m_q (\bar q_{L}\sigma_{\mu\nu}q_{R})eF^{\mu\nu},& \O_{8g}^{\d} &= m_q (\bar q_{L}\sigma_{\mu\nu}q_{R})g_sG^{\mu\nu}.
\end{align}
Lorentz structures different than the previous ones can be obtained through Fiertz identities. We will specify them in the particular cases where a distinction between the various forms has to be made.

Many of these operators are generated also in the SM itself, with a typical scale $\Lambda_{\rm SM}\sim G_F^{-1/2}$ and Wilson coefficients $C_i$ that are fully predicted in terms of the renormalizable SM couplings. Flavour bounds on a given model of new physics are then set in the following way: one has to calculate the SM and BSM contributions to the relevant Wilson coefficients at the scale $\mu_{\rm exp}$ where the measurements are done, and impose that the total contribution to the physical amplitude agree with the experimental data.

As we will see in the next section, it turns out that contributions from the SM alone are consistent with experiments -- with few tensions of significance less that $3 \sigma$ -- in all known cases. Assuming that there is no particular suppression of new physics contributions, there are very stringent lower bounds on the scale $\Lambda$, in some cases of several thousands of TeV. This is known as the {\it flavour problem} of BSM models.

\section{The CKM picture}\label{Flavour/CKM}

In the Standard Model the only sources of flavour violation are the Yukawa couplings between fermions and the Higgs field, described by the Lagrangian \eqref{yukawa}, or \eqref{massbasis} where the CKM matrix is made explicit rotating the down quark fields to the mass basis. One can further rotate also the left-handed up quarks $u_L^i\to V^{\dag}_{ij} u_L^j$ in order to obtain, after EWSB, the Lagrangian for the physical mass eigenstates
\begin{equation}\label{yukawaphysical}
\L = \sum_{i=u,c,t} m_{u_i}\bar u_L^i u_R^i + \sum_{i = d,s,b} m_{d_i}\bar d_L^i d_R^i + g W_{\mu}^+ \bar u_L^i V_{ij}\gamma^{\mu} d_L^j +  {\rm h.c.} + \cdots,
\end{equation}
where the dots indicate other flavour diagonal terms. Now the Yukawa couplings are completely diagonal and proportional to the quark masses, while the only source of flavour violation is the charged-current interaction. Notice that, as flavour violation in the gauge sector comes uniquely from rotations of the quark fields, all the neutral currents remain flavour diagonal. Moreover, in the limit where the small neutrino masses are neglected, there are no flavour transitions in the lepton sector, since there is only one Yukawa matrix which can be diagonalized redefining the fields $\ell_L$ and $e_R$.

The independent parameters of the flavour sector -- apart from the quark masses -- are three angles and one \CP\ phase, and are all contained in the CKM matrix $V$. Indeed $V$ is a $3\times 3$ unitary matrix and thus contains 9 parameters, of which 3 are rotation angles and 6 are phases. Redefining the phases of the six up- and down-type quarks, imposing conservation of the baryon number, one can eliminate 5 of these phases, remaining with one single physical phase. Thus, in the Standard Model one has the following well known results
\begin{itemize}
\item flavour transitions at tree-level are present only in charged-current interactions of quarks; flavour-changing neutral currents (FCNC) are generated only at loop-level;
\item all the flavour effects are proportional to the CKM matrix $V$;
\item all the \CP\ violating effects depend on one single phase, the Jarlskog invariant $J = \Im(V_{ud}V_{us}^*V_{ts}V_{td}^*).$
\end{itemize}

A standard parametrization of the physical quantities of the CKM matrix -- which differs from the one that we choose e.g. in \eqref{U(2)CKM} -- is the Wolfenstein parametrization
\begin{equation}
V_{CKM} = \begin{pmatrix}1 - \frac{\lambda^2}{2} & \lambda & A\lambda^3(\rho - i\eta)\\
-\lambda & 1 - \frac{\lambda^2}{2} & A\lambda^2\\
A\lambda^3(1 - \rho - i\eta) & -A\lambda^2 & 1\end{pmatrix},
\end{equation}
where $\lambda$ is the sine of the Cabibbo angle, and the relative size of the various entries is clear.

\begin{figure}
\centering%
\hfill
\includegraphics[width=0.4\textwidth]{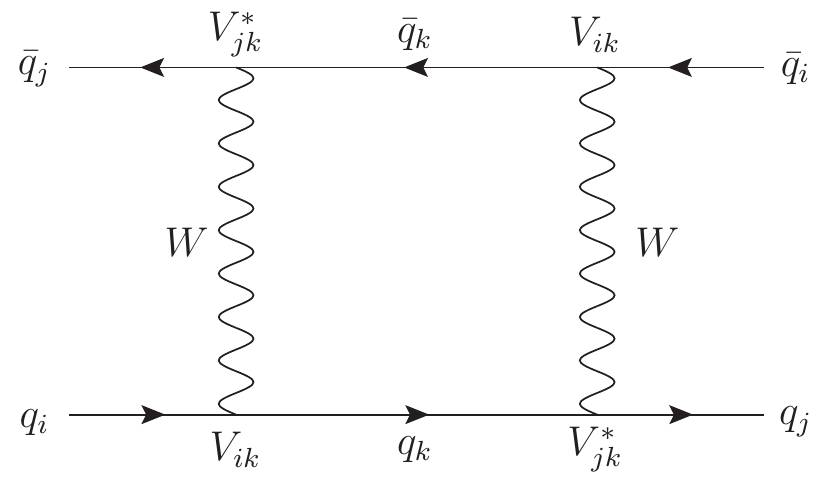}\hfill
\includegraphics[width=0.4\textwidth]{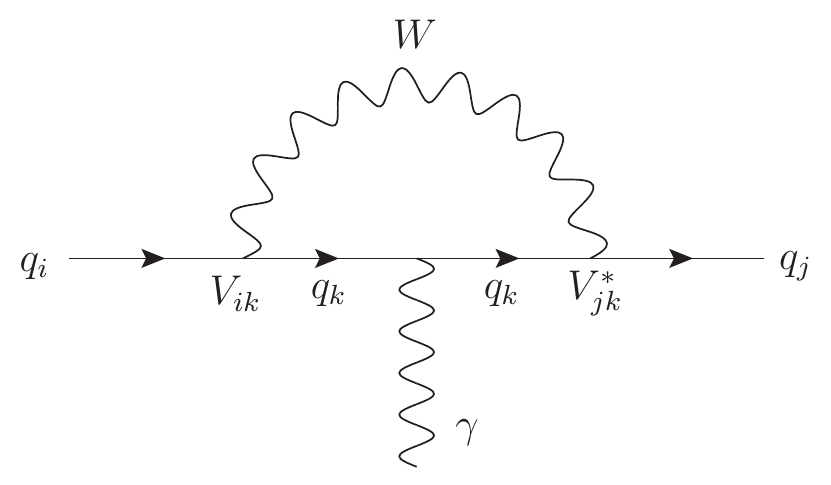}\hfill\hfill
\caption{Two examples of Feynman diagrams contributing respectively to $\Delta F = 2$ (left) and $\Delta F = 1$ (right) FCNC in the Standard Model.\label{flavourSM}}
\end{figure}

The flavour operators $\O_i$ introduced in the previous section are generated at one loop in the SM, and since the exchange of at least one charged $W$ is involved, they are always suppressed by factors of the small off-diagonal entries of the CKM matrix. All the relevant contributions come from box diagrams and penguin diagrams of the type shown in figure~\ref{flavourSM}. The one-loop effective Hamiltonians for the most relevant processes can be written as (see e.g. \cite{Buras:1998raa})
\begin{align}
\Delta\H_{\Delta S = 2}^{\rm box} &= \frac{G_F^2m_W^2}{16\pi^2}\sum_{i,j = c,t}\xi_i\xi_j S_0(\mu_i,\mu_j)\Q_1,\label{DS2}\\
\Delta\H_{s\to d\nu\bar\nu}^{\rm box} &= \frac{G_F}{\sqrt{2}}\frac{\alpha}{2\pi\sin^2\theta_w}\sum_{q = c,t}\left(-4\xi_q B_0(\mu_i)\right) \O_9^{\nu},\label{nunu}\\
\Delta\H_{s\to d\mu^+\mu^-}^{\rm box} &= \frac{G_F}{\sqrt{2}}\frac{\alpha}{2\pi\sin^2\theta_w}\sum_{q = c,t}\xi_q B_0(\mu_i) \O_9^{\mu},\label{mumu}
\end{align}
\begin{align}
\Delta\H_{s\to d Z}^{\rm penguin} &= \frac{G_F}{\sqrt{2}}\frac{e}{2\pi^2}m_Z^2\frac{\cos\theta_w}{\sin\theta_w}\sum_{q = c,t}\xi_q C_0(\mu_i) (\bar s_{L}\gamma_{\mu}d_{L})Z^{\mu},\label{sdZ}\\
\Delta\H_{s\to d\gamma}^{\rm penguin} &= \frac{G_F}{\sqrt{2}}\frac{e}{8\pi^2}\sum_{q = c,t}\xi_q D_0(\mu_i) (\bar s_{L}\gamma_{\nu}d_{L})\partial_{\mu}F^{\mu\nu},\label{sdgamma}\\
\Delta\H_{s\to d g}^{\rm penguin} &= \frac{G_F}{\sqrt{2}}\frac{g_s}{8\pi^2}\sum_{q = c,t}\xi_q E_0(\mu_i)(\bar s^{\alpha}_{L}\gamma_{\nu}T_{\alpha\beta}^a d_{L}^{\beta})\partial_{\mu}G^{\mu\nu}_a,\label{sdg}
\end{align}
where $\xi_q = V_{qs}V_{qd}^*$, $\mu_i = m_i/m_W$, and where analogous expressions for $b\to s, d$ and $c\to u$ transitions hold. In the $b\to s,d$ cases also the operators $\O_{7\gamma}$ and $\O_{8g}$ are generated,
\begin{align}
\Delta\H_{b\to q\gamma}^{\prime{\rm penguin}} &= \frac{G_F}{\sqrt{2}}\frac{1}{8\pi^2} \sum_{i = c,t}\tilde\xi_i D_0'(\mu_i)\O_{7\gamma},&
\Delta\H_{b\to q g}^{\prime{\rm penguin}} &= \frac{G_F}{\sqrt{2}}\frac{1}{8\pi^2} \sum_{i = c,t}\tilde\xi_i E_0'(\mu_i)\O_{8g},\label{btogammag}
\end{align}
where now $\tilde\xi_i = V_{ib}V_{iq}^*$.
The functions $S_0$, $B_0$, $C_0$, $D_0$, $E_0$, $D_0'$, $E_0'$ come from the loop integrals, and their values are given in appendix \ref{app:loops}.

Depending on the process that one actually considers, some of the terms in (\ref{DS2})--(\ref{btogammag}) will dominate over the others because of the hierarchies of the CKM and the quark masses.
Additional suppressions often arise when one considers the imaginary parts of the amplitudes, which lead to \CP\ violation, since the result in this case has always to be proportional to the rephasing-invariant combination $J$.

The effective vertices given above are all in terms of the elementary quark states. Going from these expressions to the physical amplitudes for meson mixings and decays is often a very nontrivial task, since QCD corrections in the non-perturbative regime have to be taken into account. There are nevertheless a few clean processes where these corrections cancel out, or are under control.

\begin{figure}[t]
\centering%
\hfill\includegraphics[width=.45\textwidth]{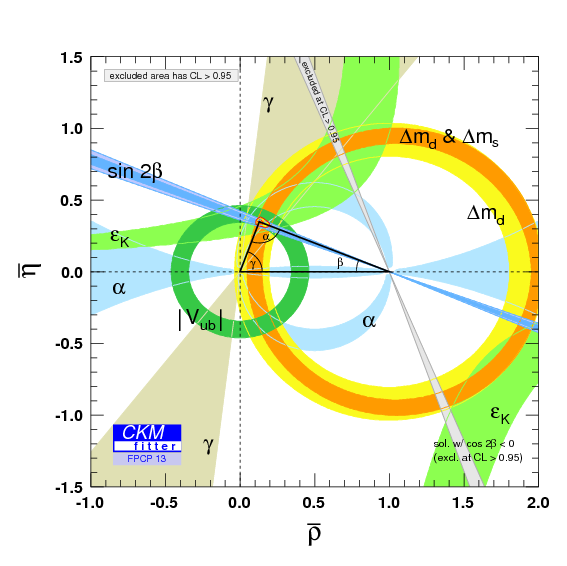}
\hfill\raisebox{.35cm}{\includegraphics[width=.54\textwidth]{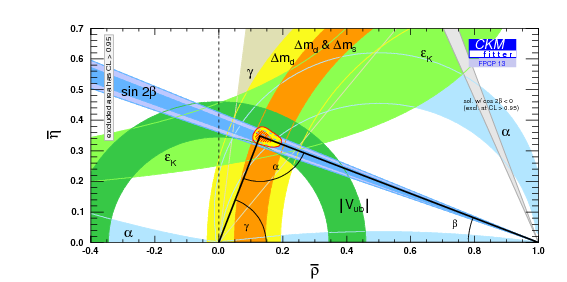}}
\caption{Fit to the parameters $\rho$ and $\eta$ from CKMfitter\cite{Charles:2011va}. The red hashed region of the global combination corresponds to 68\% C.L. The right panel is a zoom on the best fit region.
\label{CKMfit}}
\end{figure}

In figure~\ref{CKMfit}, a global fit to the CKM parameters, using the constraints from all relevant flavour processes, shows the good agreement of the data with the SM predictions. A small tension between the measure of $\sin(2\beta)$ from $B$ decays and the constraint from $\epsilon_K$ is nevertheless present at the level of about $2\sigma$ \cite{Lunghi:2008aa,Buras:2008nn,Altmannshofer:2009ne,Lunghi:2010gv,Bevan:2010gi,Brod:2011ty}.

\section{Bounds from flavour and \CP\ observables}\label{Flavour/bounds}

If one looks at the previous equations, it is clear why the bounds on new physics coming from flavour observables are so strong. If one assumes all the coefficients to be of order one, the ratios $(\Lambda_i^2 G_F)^{-1}$ has to be small enough to reproduce the large suppression factors proportional to the CKM matrix, even if one only requires the new physics contributions to be comparable with the SM ones. Notice that however for many observables -- when the QCD corrections are under control -- the bounds are much more stringent than this naive estimate, since both the theoretical and the experimental uncertainties are very small.

If, on the other hand, one wants to keep the scale $\Lambda$ in the few-TeV range, there must be some mechanism at work which suppresses the coefficients $C_i$ in a similar way to what happens in the SM.

Let us work out some exemplificative bounds from \CP\ violation in $\Delta F =1$, $\Delta F = 2$ and flavour conserving observables.

\subsection{Direct \CP\ violations in $K$ decays: $\epsilon'$}\label{Flavour/bounds/epsilonprime}
Direct \CP\ violation in the $K\to \pi\pi$ decays is described by the $\epsilon'$ parameter, which comes from the interference of two different amplitudes and is defined as
\begin{equation}
\epsilon' = \frac{1}{3}\left(\frac{\A(K_L\to \pi^+\pi^-)}{\A(K_S\to \pi^+\pi^-)} - \frac{\A(K_L\to \pi^0\pi^0)}{\A(K_S\to \pi^0\pi^0)}\right).
\end{equation}
The dominant contribution to $\epsilon'$ reads
\begin{equation}
\left|\frac{\epsilon'}{\epsilon}\right| \simeq \frac{\left| \Im \A_2 \right|}{\sqrt{2} \, |\epsilon| \,\Re \A_0} \,,
\end{equation}
where $\A_I = \A(K \to (\pi \pi)_I)$, $I$ being the isospin of the $\pi\pi$ pair, and $\epsilon$ indicates the indirect \CP\ violation parameter in $K$-$\bar K$ mixing.

The strongest bound in most concrete models is on the operators
 \begin{equation}\label{effectiveoperatorsepsilonprime}
 \Delta{\L}^{\Delta S = 1}_{4f,LR} = \frac{1}{\Lambda^2} (C^d_5{\O}^d_5 + C^u_5{\O}^u_5 + C^d_6{\O}^d_6 + C^u_6{\O}^u_6) + \text{h.c.},
 \end{equation}
 where
 \begin{align}\label{O56}
{\O}_5^q = ( \bar{d}_L \gamma_\mu s_L ) (\bar{q}_R \gamma_\mu q_R), \qquad
{\O}_6^q = ( \bar{d}_L^\alpha \gamma_\mu s_L^\beta) ( \bar{q}_R^\beta \gamma_\mu q_R^\alpha),\qquad q=u,d.
\end{align}

Using $\langle (\pi \pi)_{I=2} |{\O}_i^u +{\O}_i^d | K \rangle \simeq 0$ from isospin conservation and neglecting contributions from other operators, assuming them to be subleading,  we obtain
\begin{align}
\Im \A_2 = \frac{1}{\Lambda^2}\left[ \left( C_5^d-C_5^u\right) \langle (\pi \pi)_{I=2} |{\O}_5^d| K \rangle + \left( C_6^d-C_6^u\right) \langle (\pi \pi)_{I=2} |{\O}_6^d| K \rangle \right].
\end{align}
From \cite{Bosch:1999wr} we have at the scale $\mu = m_c$
\begin{align}
\langle (\pi \pi)_{I=2} |{\O}_5^d | K \rangle & \simeq -\frac{1}{6 \sqrt{3}}\big( m_K^2 \rho^2 - m_K^2 + m_\pi^2\big) f_\pi  B_7^{(3/2)}(m_c), \nonumber \\
\langle (\pi \pi)_{I=2} |{\O}_6^d | K \rangle & \simeq -\frac{1}{2 \sqrt{3}} \big( m_K^2 \rho^2 - \frac{1}{6} ( m_K^2 - m_\pi^2) \big)  f_\pi B_8^{(3/2)}(m_c) ,
\end{align}
where $\rho = m_K/m_s$. In the following we set $B_7^{(3/2)}(m_c) =  B_8^{(3/2)}(m_c) =1$.

The coefficients $C^{(3/2)}_i = C_i^d-C_i^u$ at the low scale $m_c$ read in terms of those at the high scale $\Lambda$\cite{Buchalla:1995vs}
\begin{align}
C^{(3/2)}_5 (m_c) &= \eta_5 C^{(3/2)}_5 (\Lambda), \nonumber \\
C^{(3/2)}_6 (m_c) &= \eta_6 C^{(3/2)}_6(\Lambda) + \frac{1}{3}(\eta_6 - \eta_5) C^{(3/2)}_5(\Lambda),
\end{align}
where
\begin{align}
\eta_5 &= \left(\frac{\alpha_s(\Lambda)}{\alpha_s(m_t)}\right)^{\frac{3}{21}} \left(\frac{\alpha_s(m_t)}{\alpha_s(m_b)}\right)^{\frac{3}{23}}\left(\frac{\alpha_s(m_b)}{\alpha_s(m_c)}\right)^{\frac{3}{25}}  \simeq 0.82\,,\notag \\
\eta_6 &= \left(\frac{\alpha_s(\Lambda)}{\alpha_s(m_t)}\right)^{-\frac{24}{21}} \left(\frac{\alpha_s(m_t)}{\alpha_s(m_b)}\right)^{-\frac{24}{23}} \left(\frac{\alpha_s(m_b)}{\alpha_s(m_c)}\right)^{-\frac{24}{25}} \simeq 4.83
\end{align}
for $\Lambda\simeq 3$ TeV.

Requiring the extra contribution from $\Delta{\L}^{4f, \Delta S = 1}_L$ to respect the experimental bound $|\epsilon^\prime/\epsilon | < |\epsilon^\prime/\epsilon |_\text{exp} \simeq 1.7\times 10^{-3}$, we obtain 
\begin{align}\label{boundepsilonprimeLambda}
C_5^{u,d} &\lesssim 6\times 10^{-6}\, {\Lambda^2}\,{\rm TeV}^{-2}, & C_6^{u,d} &\lesssim 2\times 10^{-6}\,\Lambda^2\,{\rm TeV}^{-2}.
\end{align}
The same analysis can be repeated for the operators with exchanged chiralities $\O_5^{\prime q}$ and $\O_6^{\prime q}$, thus the same bounds hold for their coefficients $C_{5,6}^{\prime q}$. If one assumes all Wilson coefficients to be of order one, \eqref{boundepsilonprimeLambda} can be expressed as a lower bound on the scale $\Lambda$ of the order of
\begin{equation}
\Lambda \gtrsim 10^3\,{\rm TeV}.
\end{equation}

Also the $s\to d$ chromomagnetic dipole
\begin{align}
 \Delta{\L}^{\Delta S = 1}_\text{mag} &= 
 \frac{m_s}{\Lambda^2} {C}_{8g}^K \left( \bar{d}_L \sigma_{\mu\nu} T^a s_R \right) g_s G_{\mu\nu}^a
\end{align}
contributes to $\epsilon'$. Following the analysis in \cite{Mertens:2011ts}, one obtains the bound
\begin{equation}
{C}_{8g}^K\lesssim 4\times 10^{-4}\, \Lambda^2 \,{\rm TeV},
\label{eq:DS1boundeff}
\end{equation}
which is somewhat weaker than the one on four-fermion operators.

Taking into account the uncertainties in the estimate of the SM contribution to $\epsilon^\prime/\epsilon$, which could cancel against a new physics contribution, as well as the uncertainties in the $B_i$ parameters, this bound might perhaps be relaxed by a factor of a few.\footnote{Note that in Supersymmetry the heaviness of the first generation squark circulating in the box loop suppresses the coefficients $c_5^{u,d}$ and $c_6^{u,d}$.}

\subsection{Indirect \CP\ violation in $\Delta F = 2$ observables}\label{sec:DF=2}\label{Flavour/bounds/DF2}
Indirect \CP\ violation in the $K^0$, $B^0_{s,d}$ and $D^0$ systems comes from the superposition of two states of opposite \CP\ in the mixing. 

The operators contributing to $\Delta F = 2$ observables are listed in \eqref{Q1}--\eqref{Q4}. Let us consider the most relevant ones, namely the left-left operators which arise also in the SM, and the left-right operators which are enhanced by a chirality factor. The effective Lagrangian in this case is
\begin{equation}
\Delta{\L}^{\Delta F = 2} = \frac{C_1}{\Lambda_1^2} \Q_1 + \frac{C_4}{\Lambda_4^2} \Q_4 + \frac{C_5}{\Lambda_5^2}\Q_5.
\end{equation}
The mass differences $\Delta m_K$, $\Delta m_B$, $\Delta m_D$ set a bound on the real part of the Wilson coefficients, while the imaginary part is constrained by \CP-violating parameters such as $\epsilon_K$ for the $K^0$ system, or $S_{\psi K}$ for the $B_d$ system.

\begin{table}[t]
\centering%
\renewcommand{\arraystretch}{1.2}
\begin{tabular}{cccccc}
Operator & \multicolumn{2}{c}{Bounds on $\Lambda_i$ [TeV] ($C_i = 1$)} & \multicolumn{2}{c}{Bounds on $C_i$ ($\Lambda = 1\,{\rm TeV})$} & Observables\\
& $\Re$ & $\Im$ & $\Re$ & $\Im$\\
\hline
$\Q_1^K$ & $9.8 \times 10^2$ & $1.6 \times 10^4$ & $9.0 \times 10^{-7}$ & $3.4 \times 10^{-9}$ & $\Delta m_K, \epsilon_K$\\
$\Q_4^K$ & $1.8 \times 10^4$ & $3.2 \times 10^5$ & $6.9 \times 10^{-9}$ & $2.6 \times 10^{-11}$ & $\Delta m_K, \epsilon_K$\\
\hline
$\Q_1^D$ & $1.2 \times 10^3$ & $2.9 \times 10^3$ & $5.6 \times 10^{-7}$ & $1.0 \times 10^{-7}$ & $\Delta m_D, \phi_D$\\
$\Q_4^D$ & $6.2 \times 10^3$ & $1.5 \times 10^4$ & $5.7 \times 10^{-8}$ & $1.1 \times 10^{-8}$ & $\Delta m_D, \phi_D$\\
\hline
$\Q_1^{B_d}$ & $5.1 \times 10^2$ & $9.3 \times 10^2$ & $3.3 \times 10^{-6}$ & $1.0 \times 10^{-6}$ & $\Delta m_{B_d}, S_{\psi K_S}$\\
$\Q_4^{B_d}$ & $1.9 \times 10^3$ & $3.6 \times 10^3$ & $5.6 \times 10^{-7}$ & $1.7 \times 10^{-7}$ & $\Delta m_{B_d}, S_{\psi K_S}$\\
\hline
$\Q_1^{B_s}$ & \multicolumn{2}{c}{$1.1 \times 10^2$} & \multicolumn{2}{c}{$7.6 \times 10^{-5}$} & $\Delta m_{B_s}$\\
$\Q_4^{B_s}$ & \multicolumn{2}{c}{$3.7 \times 10^2$} & \multicolumn{2}{c}{$1.3 \times 10^{-5}$} & $\Delta m_{B_s}$\\
\hline
\end{tabular}
\caption{Main bounds on the coefficients of $\Delta F = 2$ operators (from \cite{Isidori:2010kg}).\label{tableDF=2}}
\end{table}

In table \ref{tableDF=2} we collect the bounds on new physics in $\Delta F=2$ observables as reported in \cite{Isidori:2010kg}.
If one assumes Wilson coefficients of order one, the bounds on the scales $\Lambda_i$ are everywhere greater or of the order of $10^3$~TeV; in the particular case of the chirality enhanced operator $\Q_4^K$ the measure of $\epsilon_K$ requires $\Lambda_4$ to be even greater than $10^5$ TeV.

\subsection{Electric dipole moment of the neutron}\label{EDMsec}\label{Flavour/bounds/EDM}
The presence of phases in flavour-diagonal chirality breaking operators has to be consistent with the limits coming from the neutron electric dipole moment.
The relevant contraints come from the \CP\ violating contributions to the operators
\begin{align}\label{EDM}
\Delta{\L}_\text{mag}^{\Delta F=0} &= \frac{1}{\Lambda^2} \left[
\tilde c_u^g e^{i\tilde \phi_u^g} m_u(\bar u_L\sigma_{\mu\nu}T^a u_R)
+ \tilde c_d^g e^{i\tilde\phi_d^g} m_u(\bar u_L\sigma_{\mu\nu} u_R)
\right] g_s G^{\mu\nu}_a \nonumber \\
 &+
 \frac{1}{\Lambda^2} \left[
\tilde c_u^\gamma e^{i\tilde\phi_u^\gamma} m_d(\bar d_L\sigma_{\mu\nu}T^a d_R) + \tilde c_d^\gamma e^{i\tilde\phi_d^\gamma} m_d(\bar d_L\sigma_{\mu\nu} d_R)
\right] eF^{\mu\nu} +\text{h.c.}\, ,
\end{align}
where we have made all the phases explicit. 
In terms of the coefficients of \eqref{EDM}, the up and down quark electric dipole moments (EDM) and chromoelectric dipole moments (CEDM), defined as in \cite{Pospelov:2000bw}, are
\begin{align}
d_{q} = 2 e \, \frac{m_q}{\Lambda^2}\,\tilde c_{q}^{\gamma}\sin(\tilde\phi_i^{\gamma}),\qquad \tilde d_q = 2 \, \frac{m_q}{\Lambda^2}\, \tilde c_q^g\sin(\tilde \phi_q^g),\qquad q=u,d.
\end{align}
The contribution to the neutron EDM reads \cite{Pospelov:2000bw}
\begin{equation} \label{neutron}
d_n = (1 \pm 0.5) \left( 1.4( d_d - \tfrac{1}{4} d_u) + 1.1 e ( \tilde{d}_d + \tfrac{1}{2} \tilde{d}_u)  \right) \, ,
\end{equation}
where all the coefficients are defined at a hadronic scale of 1~GeV.

Taking into account the renormalization-group evolution between 3~TeV and the hadronic scale, the 90\% C.L. experimental bound
$|d_n| < 2.9 \times 10^{-26} ~e\,\text{cm}$ \cite{Baker:2006ts}
implies for the parameters at the high scale
\begin{align}
\tilde c_{u}^{\gamma}\sin(\tilde\phi_u^{\gamma}) &\lesssim 1.9 \times 10^{-2}\left(\frac{\Lambda}{3~\text{TeV}}\right)^2, & \tilde c_{d}^{\gamma}\sin(\tilde\phi_d^{\gamma}) \lesssim 2.4 \times 10^{-3}\left(\frac{\Lambda}{3~\text{TeV}}\right)^2,\label{EDMboundgamma}\\
\tilde{c}_{u}^g\sin(\tilde\phi_u^{g}) &\lesssim 7.1 \times 10^{-3}\left(\frac{\Lambda}{3~\text{TeV}}\right)^2, & \tilde{c}_{d}^g\sin(\tilde\phi_d^{g}) \lesssim 1.8 \times 10^{-3}\left(\frac{\Lambda}{3~\text{TeV}}\right)^2.\label{EDMboundg}
\end{align}

\section{Flavour symmetries}\label{Flavour/symmetries}

The flavour problem, as discussed in the previous discussion, can be summarized as follows. If one describes possible deviations from the CKM picture by a phenomenological effective Lagrangian of the form \eqref{Heff} with generic flavour structure, in several cases the lower bounds on the scales $\Lambda_i$ are above thousands of TeV. If one interprets this result as due to a very short-distance origin of flavour phenomena, with no expected deviation from the CKM picture close to the Fermi scale, this constitutes a problem for natural theories of EWSB beyond the SM. Moreover, with flavour physics originating at such a high scale, it would probably be very difficult to provide a testable explanation of the pattern of quark masses and mixings.

This is not, however, a necessity. It is for example conceivable that the success of the CKM picture be due to the existence of a suitable flavour symmetry, appropriately broken in some definite direction, keeping under control the coefficients in front of every operator in \eqref{Heff}. An ideal situation would be one such that the effective Lagrangian
\begin{equation}
{\L}_{\rm eff} = \sum_i \frac{c_i \xi_i}{\Lambda_\text{NP}^2}{\O}_i ~+\text{h.c.}
\label{ideal}
\end{equation}
is compatible with current data, where $\xi_i$ are small parameters controlled by the flavour symmetry and otherwise $c_i$ are ${\Ord}(1)$ coefficients. With $\Lambda_\text{NP}$ sufficiently close to the Fermi scale, this might leave room for new observable flavour phenomena. Such effects would indeed be very welcome in most extensions of the SM in the EWSB sector and, if observed, might help to shed light on a possible theory of flavour.

\subsection{Minimal Flavour Violation}\label{Flavour/symmetries/MFV}
An example largely studied in the literature is represented by the hypothesis of Minimal Flavour Violation (MFV), based on the invariance of the quark Lagrangian under the full  $\U(3)_q\times \U(3)_u\times \U(3)_d$ flavour group of the quark sector of the Standard Model\cite{Chivukula:1987py,Hall:1990ac,DAmbrosio:2002ex}. In the limit where this symmetry is exact, no flavour violating terms are allowed at all, whether in the higher dimensional operators of \eqref{Heff}, nor in the SM Yukawa sector.

In order to reproduce the observed quark mixings of the CKM matrix, the flavour symmetry has to be broken. The MFV paradigm consists in the following assumptions:
\begin{itemize}
\item the SM Yukawa couplings $\hat y_u$ and $\hat y_d$ are the only effective sources of flavour violation, even when the short distance contributions from $\L_{\rm eff}$ are taken into account;
\item if the flavour-breaking terms are treated as {\it spurions} transforming in suitable representations of $\U(3)^3$, the effective operators in $\L_{\rm eff}$ are the most generic functions of the fields and the Yukawa couplings that can be constructed in a way compatible with the flavour symmetry.
\end{itemize}
The consequence of this approach is that the Wilson coefficients of the effective operators $\O_i$ are suppressed by suitable powers of the CKM matrix, exactly like in the Standard Model. Of course the exact value of the new physics contributions is not determined in terms of the SM couplings, but their size will in general be comparable with the SM contributions, apart from a loop factor.

In order to reproduce the Yukawa structure \eqref{yukawa} of the SM, it is clear that the spurions must transform under $\U(3)_q\times \U(3)_u\times \U(3)_d$ as
\begin{align}
\hat Y_u&\sim (\three, \threebar, \one), & \hat Y_d&\sim (\three, \one, \threebar).
\end{align}
A drawback of the MFV approach is the fact that the full $\U(3)^3$ flavour group is not a good symmetry of the Standard Model, being badly broken at least by the order 1 top Yukawa coupling. As a consequence, when trying to construct flavour-violating effective operators in terms of $\hat Y_u$, $\hat Y_d$ one can not perform an expansion in the spurions, and one has to consider arbitrary powers of at least the flavour-invariant combination $\hat Y_u \hat Y_u^{\dag}$. In models with more than one Higgs doublet, if the ratio of the vacuum expectation values of the fields coupled to the up- and down-type quarks $\tan\beta$ is large, also the bottom Yukawa coupling $y_b$ becomes a relevant breaking parameter, and powers of $\hat Y_d \hat Y_d^{\dag}$ have to be included in the effective operators as well.

One can choose a basis for the quark fields where the spurions take the form
\begin{align}
\hat Y_u &= V_0^{\dag} \hat Y_u^{\rm diag}, & \hat Y_d &= \hat Y_d^{\rm diag},
\end{align}
where $V_0$ is a unitary matrix dependent on one single phase.\footnote{The matrix $V_0$ does in general not coincide with the CKM matrix because of the $\Ord(1)$ corrections coming from all the terms $(\hat Y_u \hat Y_u^{\dag})^n$. Similarly, the spurions $\hat Y_u$, $\hat Y_d$ can not be identified with the physical Yukawa coupling matrices $\hat y_u$ and $\hat y_d$.}
Therefore the CKM phase is the only source of \CP\ violation if no new phase is born outside of $\hat Y_u$ or $\hat Y_d$.
The powers of $\hat Y_u Y_u^{\dag}$ in this basis reduce to
\begin{align}
(\hat Y_u \hat Y_u^{\dag})^n &= \big(V_0 |\hat Y_u^{\rm diag}|^2 V_0^{\dag}\big)^n\approx V_0 \mathcal{I}_3 V_0^{\dag}, & (\hat Y_d \hat Y_d^{\dag})^n &= |\hat Y_d^{\rm diag}|^{2n},
\end{align}
with $\mathcal{I}_3 = \diag(0,0,1)$. To determine the relevant flavour-violating operators one has to reduce the kinetic terms to canonical form and the mass matrices to real diagonal form. In turn this depends on the value of $\tan\beta$, which determines the need to include or not powers of $\hat Y_d \hat Y_d^{\dag}$ in the effective operators.
For moderate $\tan\beta$ the most generic flavour-changing quark bilinears which are formally $\U(3)^3$-invariant and can be built in terms of the spurions, in the physical basis, have the approximate form ($i, j = d, s, b$):
\begin{align}\label{MFVdbilinear}
&\bar d_L \sigma_{\mu\nu}\big(V^{\dag}\mathcal{I}_3 V \hat Y_d^{\rm diag}\big)d_R\simeq y_b\, \xi_{ij}\bar d_L^i \sigma_{\mu\nu}d_R^j, & &\bar d_L \gamma_{\mu}\big(V^{\dag}\mathcal{I}_3 V\big) d_L\simeq \xi_{ij} \bar d_L^i \gamma_{\mu}d_L^j,
\end{align}
where $\xi_{ij} = V_{ti}^*V_{tj}$, $V$ is the CKM matrix. Similar equations can be derived for the up-type quarks.

The relevant $\Delta S, \Delta B = 2$ operators, which come from the contraction of two bilinears of \eqref{MFVdbilinear} and are all of the form $\mathcal{Q}_1$, are thus suppressed by a factor $\xi_{ij}^2$, which is exactly the suppression of the SM contributions with the exchange of a top quark in the loop. Analogously, $\Delta S,\Delta B = 1$ operators are proportional to $\xi_{ij}$, which again is the suppression factor coming from the top quark contribution in the SM. The same results hold in the up-quark sector for $\Delta C = 1,2$ operators, replacing $\xi_{ij}$ with $\zeta_{ic} = V_{ib}V_{jb}^*$. 

New \CP\ effects can still appear in this picture, since the coefficients of the effective operators, although being similar in size to the SM ones, have not to be aligned in phase with them. Nevertheless the large suppression allows the new effects to appear at a scale much lower than the one allowed in a generic model. The strongest bounds come from $\epsilon_K$ and $b\to s\gamma$ decays, and set a lower bound on $\Lambda_{\rm NP}$ of the order of a few TeV \cite{DAmbrosio:2002ex}.

At large $\tan\beta$ both powers of $\hat Y_u \hat Y_u^{\dag}$ and $\hat Y_d \hat Y_d^{\dag}$ are relevant in effective operators \cite{Feldmann:2008ja,Kagan:2009bn,Colangelo:2008qp}, and \eqref{MFVdbilinear} is not exhaustive. The
fact that $y_t$ and $y_b$ are both of $\Ord(1)$ leads to the the breaking of $\U(3)^3$ down to $\U(2)^3$. In this case, after suitable $\U(3)^3$ transformations, one can write the spurions as
\begin{equation}\label{Ylargetanbeta}
\hat Y_{u,d} = e^{\pm i \hat \chi/2}y_{t,b}\begin{pmatrix}\Delta \hat Y_{u,d} & 0\\ 0 & 1\end{pmatrix},
\end{equation}
where $\Delta \hat Y_{u,d}$ are $2\times 2$ matrices and
\begin{equation}
\hat \chi = \begin{pmatrix}0 & \boldsymbol{\chi}\\ \boldsymbol{\chi}^{\dag} & 0\end{pmatrix},
\end{equation}
with $\boldsymbol{\chi}$ a 2-vector, is a hermitian $3\times 3$ matrix which determines the misalignment of $\hat Y_{u,d}$ in the 13 and 23 directions, which is known to be small (of order $V_{cb}$) from the CKM matrix. Expanding in $\hat\chi$ one has from \eqref{Ylargetanbeta}
\begin{equation}
\hat Y_{u,d} \simeq y_{t,b}\begin{pmatrix}\Delta \hat Y_{u,d} & \pm i \boldsymbol{\chi}/2\\
\pm i \boldsymbol{\chi}^{\dag} \Delta \hat Y_{u,d}/2 & 1\end{pmatrix}.
\end{equation}

A thorough analysis of the effective operators relevan in this case will be the topic of the next chapter, where we will extend the MFV paradigm in order to work with the better approximate $\U(2)^3$ symmetry.

\chapter{An approximate $\U(2)^3$ flavour symmetry}\label{U2}

The quark sector of the Standard Model exhibits an approximate $\U(2)_q\times \U(2)_u\times \U(2)_d$ flavour symmetry acting on the first two generations of quarks of different $\SU(2)\times \U(1)$ quantum numbers, $q_L, u_R$ and $d_R$\cite{Barbieri:2011ci,Barbieri:2011fc,Barbieri:2012uh}. The symmetry is exact in the limit where one neglects the masses of the first two generations and their mixings with the third generation quarks. The smallness of these masses (with respect to the top mass) and mixings ensures that $\U(2)^3$ is indeed a good approximate symmetry of the SM Lagrangian, broken at most by an amount of order a few $\times 10^{-2}$. This is the size of $V_{cb}$, comparable to or bigger than the mass ratios $m_{c, u}/m_t$ or $m_{s, d}/m_b$.

To describe the breaking of $\U(2)^3$ we assume that it is encoded in a few small dimensionless parameters. Their origin is unknown and may be different, for example, in different models of EWSB, but we require that they have  definite transformation properties under  $U(2)^3$ itself, so that the overall Lagrangian, fundamental or effective as it may be, remains formally invariant. This is what we mean by saying that $U(2)^3$ is broken in specific directions. Along these lines, the simplest way to give masses to both the up and down quarks of the first two generations is to introduce two (sets of) parameters $\Delta_u$, $\Delta_d$,  transforming as
\begin{align}\label{DeltaY}
\Delta_u &= (\two, \twobar, \one), & \Delta_d &= (\two, \one, \twobar)
\end{align}
under $\U(2)_q\times \U(2)_u\times \U(2)_d$. If these {\it bi-doublets} were the only breaking parameters, the third generation, made of singlets under $\U(2)^3$, would not   communicate with the first two generations at all. For this to happen one needs single doublets, at least one,  under any  of the three $\U(2)$'s. The only such doublet that can explain the observed small mixing between the third and the first two generations, in terms of a correspondingly small parameter, transforms under $\U(2)_q\times \U(2)_u\times \U(2)_d$ as 
\begin{equation}\label{V}
\V = (\two,\one,\one).
\end{equation}
A single doublet under $\U(2)_u$ or $\U(2)_d$ instead of $\U(2)_q$ would have to be of order unity. This is the minimal set of breaking parameters\footnote{Notice that the spurions $\Delta_{u,d}$ and $\V$ correspond to $\Delta \hat Y_{u,d}$ and $\boldsymbol{\chi}$ in \eqref{Ylargetanbeta}, and thus MFV at large $\tan\beta$ is equivalent, in an effective field theory framework, to $\U(2)^3$.} required for a realistic description of quark masses and mixings; we call this setup Minimal $\U(2)^3$.
One can extend this picture by considering all the possible breaking terms of $\U(2)^3$ entering the quark mass terms, thus including also the two doublets
\begin{align}
\Vu &= (\one, \two, \one), & \Vd &= (\one, \one, \two).\label{VuVd}
\end{align}
We call this situation Generic $\U(2)^3$.

To summarize, we assume that $\U(2)^3$ is an approximate symmetry of the flavour sector of the SM only weakly broken in the directions $\Delta_u$, $\Delta_d$, $\V$, and possibly $\Vu$ and $\Vd$.
A key assumption is that the {\it spurions} transform under $\U(2)^3$ as defined in \eqref{DeltaY}, \eqref{V} and \eqref{VuVd}, so that every term in the quark mass bilinears is formally invariant.

\section{Effective operators in the physical quark basis}\label{U2/EFT}

Given the basic distinction between the third and the first two generations of quarks, we adopt for the left-handed doublets and the right-handed charge $2/3$ and $-1/3$ quark singlets respectively the self-explanatory notation
\begin{equation}
q_L = \begin{pmatrix} \ELqL \\ \EHqL\end{pmatrix},\qquad
u_R = \begin{pmatrix} \ELuR \\ \EHuR\end{pmatrix},\qquad
d_R = \begin{pmatrix} \ELdR \\ \EHdR\end{pmatrix}.
\label{BoldNotation}
\end{equation}

By sole $\U(2)^3$ transformations it is possible and useful to restrict and define the physical parameters appearing in the spurions.  In Minimal $\U(2)^3$ we choose:
\begin{equation}
\V = \begin{pmatrix}0\\ \epsilon_L\end{pmatrix},\qquad \Delta_u = L_{12}^u\,\Delta_u^{\rm diag},\qquad
 \Delta_d = \Phi_L L_{12}^d\,\Delta_d^{\rm diag},
 \end{equation}
 where $\epsilon_L$ is a real parameter, $L_{12}^{u,d}$ are rotation matrices in the space of the first two generations with angles $\theta_L^{u,d}$ and $\Phi_L = {\rm diag}\big(e^{i\phi},1\big)$, i.e. four parameters in total. Incidentally this shows that, if \CP\ violation only resides in $\V, \Delta_u, \Delta_d$, there is a single physical phase, $\phi$, which gives rise to the CKM phase.
 
Similarly in Generic $\U(2)^3$ we set:
\begin{equation}\label{genericV}
\V = \begin{pmatrix}0\\ \epsilon_L\end{pmatrix},\qquad \Vu = \begin{pmatrix}0\\ \epsilon^u_R\end{pmatrix},\qquad \Vd = \begin{pmatrix}0\\ \epsilon_R^d\end{pmatrix},
\end{equation}
\begin{align}\label{genericY}
\Delta_u &= L_{12}^u\,\Delta_u^{\rm diag}\,\Phi_R^u R_{12}^u, & \Delta_d &= \Phi_L L_{12}^d\,\Delta_d^{\rm diag}\,\Phi_R^d R_{12}^d,\\
\Phi_L &= {\rm diag}\big(e^{i\phi},1\big), & \Phi_R^{u, d} &= {\rm diag}\big(e^{i\phi_1^{u,d}}, e^{i\phi_2^{u,d}}\big),
\end{align}
 which adds to the four parameters of Minimal $\U(2)^3$ other four real parameters, $\epsilon_R^{u,d}, \theta_R^{u,d}$ and four phases, $\phi_{1,2}^{u,d}$. For later convenience we define $s^{u,d}_L=\sin{\theta_L^{u,d}}$ and $s^{u,d}_R=\sin{\theta_R^{u,d}}$.
 
For the remaining part of this section we consider only the Generic case, since the Minimal case can be easily obtained from it in the limit where all the right-handed parameters are set to zero.
 
The effective operators are constructed from the most generic quark bilinears which contain the spurions and are formally invariant under $\U(2)^3$.
To a sufficient approximation, the chirality conserving bilinears for the left-handed doublet take the form
\begin{equation}\label{ccL}
\bar q_{L}\gamma^{\mu}X_L^{\alpha} q_{L} = a_L^{\alpha}\bar q_{3L}\gamma^{\mu}q_{3L} + b_L^{\alpha}\qLbar\gamma^{\mu}\qL + c_L^{\alpha}\qLbar\!\V\gamma^{\mu} q_{3L} + d_L^{\alpha}(\qLbar\!\V)\gamma^{\mu}(\V^{\dag}\qL) + {\rm h.c.},
\end{equation}
while for the right-handed singlets
\begin{align}
\bar u_{R}\gamma^{\mu}X_{uR}^{\alpha}u_{R}\! &= a_{uR}^{\alpha}\bar t_R\gamma^{\mu}t_R\! +  b_{uR}^{\alpha}\uRbar\gamma^{\mu}\uR + c_{uR}^{\alpha}\uRbar\Vu \gamma^{\mu}t_R + d_{uR}^{\alpha}(\uRbar \Vu)\gamma^{\mu}(\Vudag\uR)\! + {\rm h.c.},\label{ccu}\\
\bar d_{R}\gamma^{\mu}X_{dR}^{\alpha}d_{R} &= a_{dR}^{\alpha}\bar b_R\gamma^{\mu}b_R + b_{dR}^{\alpha}\dRbar\gamma^{\mu}\dR + c_{dR}^{\alpha}\dRbar\Vd\gamma^{\mu}b_R + d_{dR}^{\alpha}(\dRbar \Vd)\gamma^{\mu}(\Vddag\dR) + {\rm h.c.},\label{ccd}
\end{align}
where all the parameters except the c's are real by hermiticity. These bilinears give rise to four-fermion operators, as well as to the kinetic terms.

Similarly, the chirality breaking bilinears are, to lowest order in the spurions,
\begin{align}
\bar q_{L}(M_u^{\beta}) u_{R} &= \lambda_t\Big(a_u^{\beta}\bar q_{3L}t_R + b_u^{\beta}(\qLbar\V)t_R + c_u^{\beta}\qLbar\Delta_u \uR + d_u^{\beta}\bar q_{3L}(\Vudag\uR)
+ e_u^{\beta}(\qLbar\V)(\Vudag\uR)\Big), \label{cb_u}\\
\bar q_{L}(M_d^{\beta}) d_{R} &= \lambda_b\Big(a_d^{\beta}\bar q_{3L}b_R + b_d^{\beta}(\qLbar\V)b_R + c_d^{\beta}\qLbar\Delta_d \dR + d_d^{\beta}\bar q_{3L}(\Vddag\dR)\nonumber\\
&\quad + e_d^{\beta}(\qLbar\V)(\Vddag\dR)\Big), \label{cb_d}
\end{align}
where now all the parameters are complex. They generate the Yukawa couplings $\hat y_u, \hat y_d$, magnetic and chromomagnetic interaction terms, as well as left-right four-fermion operators. A $\sigma_{\mu\nu}$ is understood in the previous equations when needed. Notice that all the parameters in the kinetic and Yukawa terms, except one, can be made real through rephasings of the fields \cite{Barbieri:2012uh}.

In each of these equations we have neglected higher order terms in the spurions which, unless differently specified, do not affect the following considerations and we have included order 1 coefficients, $a, b, c, \dots$, dependent upon the specific bilinear under consideration. In each  of the chirality breaking bilinears we have factored out the parameters $\lambda_t \approx m_t/v$ and $\lambda_b\approx m_b/v$. While $\lambda_t$ is of order unity, the smallness of $\lambda_b$ may be attributed to an approximate $\U(1)_d$ acting on all the right-handed down quarks in the same way inside and outside $\U(2)^3$. In presence of more than one Higgs doublet, $\lambda_b$ can get bigger values.

We are interested in the expressions for the operators \eqref{ccL}--\eqref{cb_d} in the physical basis where the quark masses are diagonal, and the kinetic terms are canonical.
The kinetic terms are put in the canonical form by real rotations in the $(2,3)$ sector plus wavefunction renormalizations of the fields. One can check that these transformations do not alter, to a sufficient accuracy, the structure of the other operators, but cause only $\Ord(1)$ redefinitions of the parameters.

The diagonalization of the mass terms can be done perturbatively by taking into account the smallness of $\epsilon_L, \epsilon_R^{u,d}$ and  $\Delta_{u, d}^{\rm diag}$. As a consequence, to a sufficient approximation, the unitary transformations that bring these mass matrices to diagonal form are influenced on the left side only by the four parameters of  Minimal $\U(2)^3$, $\epsilon_L, \theta_L^{u,d}, \phi$, whereas those on the right side depend on the extra parameters of Generic $\U(2)^3$, $\epsilon_R^{u,d}, \theta_R^{u,d}, \phi_{1,2}^{u,d}$.
One goes to the physical basis for the quarks by
\begin{align}
u_L&\mapsto L_{23}^u L_{12}^u u_L\equiv L^u u_L, & d_L&\mapsto U_{23}^d U_{12}^d d_L\equiv U^d d_L, \label{U2transformationsreal}\\
u_R&\mapsto R_{23}^u R_{12}^u u_R\equiv R^u u_R, & d_R&\mapsto V_{23}^d V_{12}^d d_R\equiv V^d d_R,\label{U2transformationsrealR}
\end{align}
which diagonalize approximately the mass terms up to transformations of order $\epsilon_L y_{u,d,c,s}$, $\epsilon^u_R y_{u,c}$ and $\epsilon^d_R y_{d,s}$. Here and in the following $U_{ij}$ ($V_{ij}$) stand always for unitary left (right) matrices in the $(i,j)$ sector, while $L_{ij}$ ($R_{ij}$) indicate orthogonal left (right) matrices. In particular $U_{12}^d = \Phi_L L_{12}^d$ and $V_{12}^{u,d} = \Phi_R^{u,d} R_{12}^{u,d}$.

In turn this leads to a unique form of the standard CKM matrix
 \begin{equation}\label{CKM}
V_{CKM}\simeq (L_{12}^u)^T(L_{23}^u)^T U_{23}^d U_{12}^d\equiv (R_{12}^u)^T U_{23}^{\epsilon} U_{12}^d,
\end{equation}
where $U_{23}^{\epsilon}$ is a unitary transformation of order $\epsilon_L$. This expression leads to the following parametrization
\begin{equation}\label{U(2)CKM}
V_{CKM} = \begin{pmatrix}
c^u_L c^d_L & \lambda & s^u_L s\,e^{-i\delta}\\
-\lambda & c^u_L c^d_L & c^u_L s\\
-s^d_L s\,e^{i(\delta - \phi)} & -c^d_L s & 1
\end{pmatrix},
\end{equation}
where $s\sim \Ord(\epsilon_L)$, $c^{u,d}_L=\cos{\theta_L^{u,d}}$ and
$s^u_L c^d_L - s^d_L c^u_L e^{i\phi} = \lambda e^{i\delta}$.
Using this parametrization of the CKM matrix, a direct fit of the tree-level flavour observables, presumably not influenced by new physics,  results in \cite{Barbieri:2011fc}
\begin{align}
s^u_L &= 0.086\pm0.003
\,,&
s^d_L &= -0.22\pm0.01
\,,\\
s &= 0.0411 \pm 0.0005
\,,&
\phi &= (-97\pm 9)^\circ
\,.
\end{align}
At this stage, the extra ``right-handed'' parameters present in Generic $\U(2)^3$ are unconstrained, since they do not enter the CKM matrix.

The explicit results for the coefficients of all the quark bilinears in the physical quark-mass basis, which generate a suppression factors $\xi_i$ as in \eqref{ideal} for each effective operator they contribute to, are derived in appendix~\ref{app:bilinears}.

\section{Flavour and \CP\ observables in Minimal $\U(2)^3$}\label{U2/minimal}

\begin{table}[t]
\renewcommand{\arraystretch}{1.0}
 \begin{center}
\begin{tabular}{llllll}
\hline
$|V_{ud}|$ & $0.97425(22)$ &\cite{Hardy:2008gy}& $f_K$  & $(155.8\pm1.7)$ MeV & \cite{Laiho:2009eu}\\
$|V_{us}|$ & $0.2254(13)$ &\cite{Antonelli:2010yf}& $\hat B_K$ & $0.737\pm0.020$ &\cite{Laiho:2009eu} \\
$|V_{cb}|$ & $(40.6\pm1.3)\times10^{-3}$ &\cite{Nakamura:2010zzi}& $\kappa_\epsilon$ & $0.94\pm0.02$ & \cite{Buras:2010pza}\\
$|V_{ub}|$ & $(3.97\pm0.45)\times10^{-3}$ &\cite{Kowalewski:2011zz}& $f_{B_s}\sqrt{\hat B_s}$  & $(288\pm15)$ MeV &\cite{Lunghi:2011xy}\\
$\gamma_{\rm CKM}$ & $(74\pm11)^\circ$ &\cite{Bevan:2010gi}& $\xi$ & $1.237\pm0.032$ &\cite{Laiho:2009eu}\\
$|\epsilon_K|$ & $(2.229\pm0.010)\times10^{-3}$ &\cite{Nakamura:2010zzi} &$\eta_{tt}$&$0.5765(65)$&\cite{Buras:1990fn}\\ 
$S_{\psi K_S}$ & $0.673\pm0.023$ &\cite{Asner:2010qj} &$\eta_{ct}$&$0.496(47)$&\cite{Brod:2010mj}\\
$\Delta M_d$ & $(0.507\pm0.004)\,\text{ps}^{-1}$ &\cite{Asner:2010qj} &$\eta_{cc}$&$1.38(53)$&\cite{Brod:2011ty}\\
$\Delta M_s/\Delta M_d$ & $(35.05\pm0.42)$ &\cite{Abulencia:2006ze,Asner:2010qj} &&&\\
$\phi_s$ & $-0.002 \pm 0.087$ & \cite{LHCb-TALK-2012-029} &&&\\
\hline
 \end{tabular}
 \end{center}
\caption{Observables and hadronic parameters used as input to the $\Delta F=2$ fits.}
\label{tab:inputs}
\end{table}

The general form of the flavour-changing effective operators can be summarized in
\begin{equation}
{\L}_{\rm eff} = \Delta{\L}_{4f,L}  + \Delta{\L}_\text{mag} + \Delta{\L}_{H},
\label{Leffgeneral}
\end{equation}
where $\Delta{\L}_{4f,L} $ is the set of four-fermion operators with flavour violation in the left-handed sector, $\Delta{\L}_\text{mag}$ contains the chirality-breaking dipole operators, and $\Delta\L_H$ contains the Higgs-dependent operator $\O_H$. Notice that sizeable flavour-violations in the right-handed sector are absent in Minimal $\U(2)^3$.

In what follows, we will write each single term in \eqref{Leffgeneral} as
\begin{equation}
{\L}_{\rm eff} = -\H_{\rm eff} = \frac{1}{\Lambda^2} \sum_i C_i{\O}_i ~+\text{h.c.},
\end{equation}
where the coefficients $C_i$ of the operators relevant for the process under examination can be read off the equations \eqref{firstOcoefficient}--\eqref{lastOcoefficient} and will be specified case by case. The bounds on these coefficients arise from $\Delta F=2$ transitions, $\epsilon_K$, $B_d^0$-$\bar{B}_d^0$ mixing, $B_s^0$-$\bar{B}_s^0$ mixing, from $K\to\pi\pi$ decays, and from $B$ decays -- mostly $b\rightarrow s \gamma$, $b\rightarrow s \ell\bar{\ell}$, $b\to s\nu\bar{\nu}$.

The previous analysis leads to the following exhaustive set of relevant flavour changing effective operators,\footnote{The notation is the same as in section~\ref{Flavour/EFT}. We omit $\Delta F=1$ operators which only enter low-energy observables in a fixed linear combination with the ones considered here and do not lead to qualitatively new effects.} all weighted  by the square of an inverse mass scale $1/\Lambda$:
\begin{itemize}
\item $\Delta F = 2$ operators $\Q_1^{K,B}\subset\Delta\L_{4f,L}$, with a real coefficient required for $\Q_1^K$;
\item chirality-breaking $\Delta B = 1$ operators $\O_{7\gamma}^B, \O_{8g}^B\subset\Delta\L_{\rm mag}$, the analogous $\Delta S = 1$ operators being suppressed as $m_s/m_b$; 
\item chirality conserving $\Delta F = 1$ semileptonic operators $\O_9^{B,K}, \O_{10}^{B,K}\subset\Delta\L_{4f,L}$, plus $\O_H^{K,B}\subset \Delta\L_H$, again with a real coefficient for the operators in the $K$ sector;
\end{itemize}

Analogous operators involving the up quarks are present. However if these operators are weighted by the same scale $\Lambda$ as for the down quarks, they are phenomenologically irrelevant unless some of the relative dimensionless coefficients are at least one order of magnitude bigger than the ones in the down sector. This is in particular the case for operators contributing to $D-\bar{D}$ mixing, to direct \CP\ violation in $D$-decays or to top decays, $t\rightarrow c \gamma$ or $t\rightarrow c Z$.
 
As we now show, the operators in $\Delta{\L}_{4f,L}$, $\Delta{\L}_\text{mag}$ and $\Delta\L_H$, controlled by Minimal $\U(2)^3$  breaking, are broadly consistent with an overall scale at 3 TeV and otherwise model-dependent coefficients $c_i$ in the range $0.2$ to $1$, depending on their phases\cite{Barbieri:2012uh}.

\subsection{$\Delta F=2$ processes}\label{U2/minimal/DF2}

\begin{figure}[t]
\centering
\includegraphics[width=.325\textwidth]{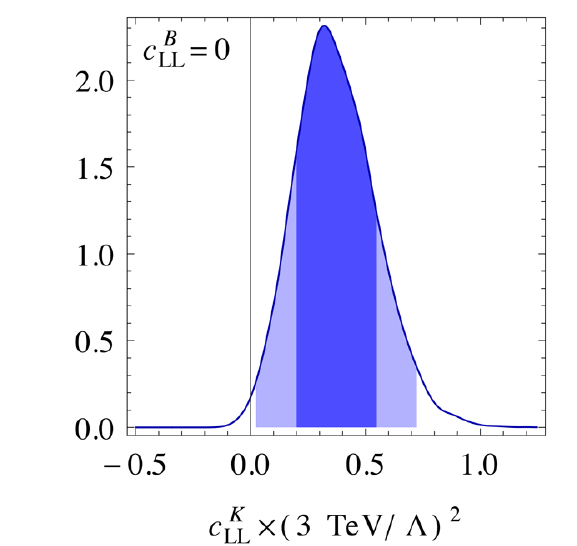}
\includegraphics[width=.325\textwidth]{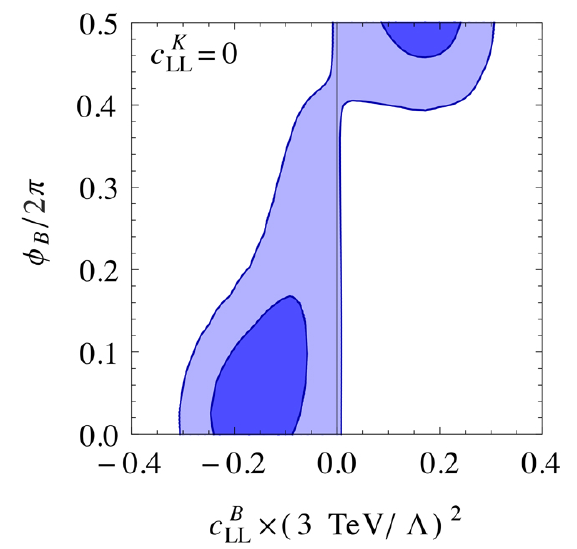}
\raisebox{1.3mm}{\includegraphics[width=.31\textwidth]{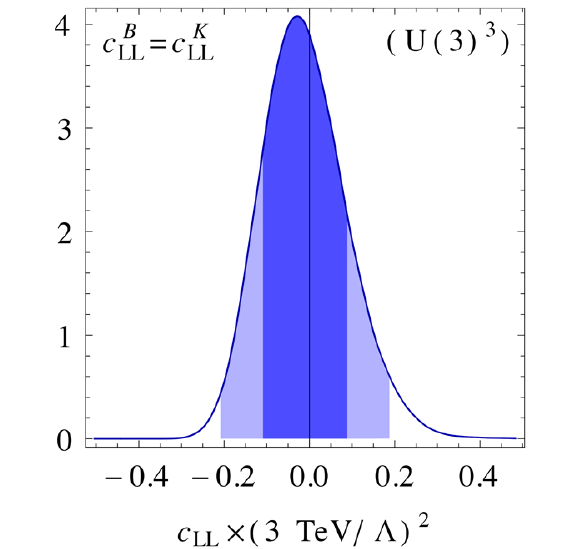}}
\caption{Fit predictions (68 and 95\% Bayesian credible regions) in $\Delta F=2$ fits with $c_{LL}^B=0$ (left), $c_{LL}^K=0$ (centre), and in the $\U(3)^3$ case $c_{LL}^B=c_{LL}^K$, $\phi_B=0$ (right).}
\label{fig:DF2singlefits}
\end{figure}

The relevant $\Delta F=2$ operators generated in the $\U(2)^3$ framework come only from the left-handed sector and read
\begin{equation}
\H_{4f,L}^{\Delta F = 2} =
\frac{c_{LL}^K }{\Lambda^2} \xi_{ds}^2 \frac{1}{2}\left({\bar d}_L \gamma_\mu  s_L \right)^2
+\sum_{i=d,s}
\frac{c_{LL}^B e^{i\phi_B}}{\Lambda^2} \xi_{ib}^2\frac{1}{2}\left({\bar d}_L^i \gamma_\mu  b_L \right)^2
+\text{h.c.}\,,
\label{eq:HeffDF2}
\end{equation}
where $c_{LL}^{K,B}$ are real, model dependent parameters that can be of ${\Ord}(1)$. The $\U(3)^3$ case at low $\tan\beta$ is recovered for $c_{LL}^K=c_{LL}^B$ and $\phi_B=0$. Notice that, as in MFV, the factors of $\xi_{ij}$ multiplying the coefficients are the same as for the top contribution in the SM. As a consequence one can verify that the observables in $K$, $B_d$ and $B_s$ meson mixing are modified in the following way \cite{Barbieri:2011ci} 
\begin{align}
\epsilon_K&=\epsilon_K^\text{SM(tt)}\left(1 - h_K\right) +\epsilon_K^\text{SM(tc+cc)} ,
\label{eq:epsKxF}\\
S_{\psi K_S} &=\sin\left(2\beta + \text{arg}\left(1 - h_B e^{i\phi_B}\right)\right) ,
\label{eq:Spk} \\
S_{\psi\phi} &=\sin\left(2|\beta_s| - \text{arg}\left(1 - h_B e^{i\phi_B}\right)\right) ,
\label{eq:SpsiphixF}\\
\Delta M_d &=\Delta M_d^\text{SM}\,\left|1 - h_B e^{i\phi_B}\right| ,
\label{eq:DMdxF}\\
\frac{\Delta M_d}{\Delta M_s} &= \frac{\Delta M_d^\text{SM}}{\Delta M_s^\text{SM}} \,,
\label{eq:MdMs}
\end{align}
where
\begin{align}
h_{K,B} &= c_{LL}^{K,B}
\frac{4s_w^4}{\alpha_{em}^2S_0(x_t)}
\frac{m_W^2}{\Lambda^2}
\approx
1.08\,c_{LL}^{K,B}
\left[ \frac{3\,\text{TeV}}{\Lambda} \right]^2
\,.
\end{align}

To compare the effective Hamiltonian (\ref{eq:HeffDF2}) with the data, the dependence of the $\Delta F=2$ observables on the CKM matrix elements has to be taken into account. To this end, we performed global fits of the CKM Wolfenstein parameters $A$, $\lambda$, $\bar\rho$ and $\bar\eta$ as well as the coefficients $c_{LL}^{K,B}$ and the phase $\phi_B$ to the set of experimental observables collected in the left-hand column of table~\ref{tab:inputs}, by means of a Markov Chain Monte Carlo, assuming all errors to be Gaussian.\footnote{A similar analysis was recently performed in \cite{Charles:2013aka} with updated values of some hadronic parameters and flavour observables, finding comparable results.}

\begin{figure}[t]
\centering\hfill
\includegraphics[width=.4\textwidth]{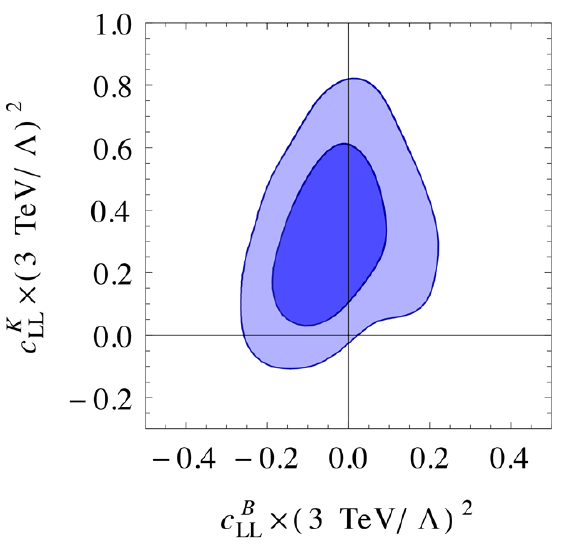}\hfill
\includegraphics[width=.4\textwidth]{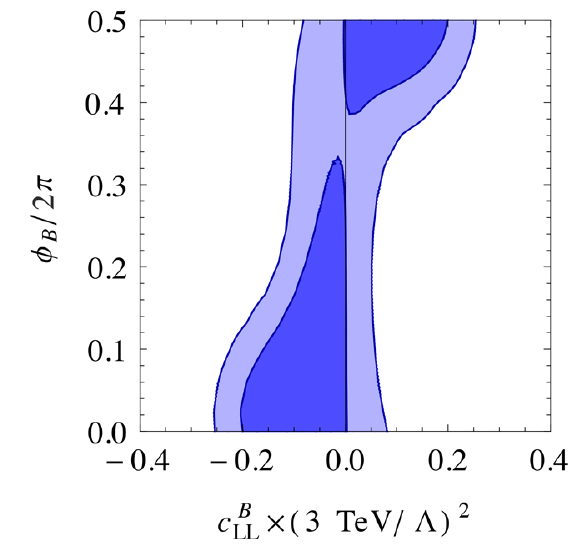}\hfill\hfill
\caption{Fit predictions (68 and 95\% Bayesian credible regions) in $\Delta F=2$ fits for the parameters $c_{LL}^K$, $c_{LL}^B$ and $\phi_B$ in $\U(2)^3$.}
\label{fig:DF2fits}
\end{figure}

The results of four different fits are shown in figures~\ref{fig:DF2singlefits} and \ref{fig:DF2fits}. The left panel of figure~\ref{fig:DF2singlefits} shows the fit prediction for $c_{LL}^K$ in a fit with $c_{LL}^B=0$. The centre panel shows the fit prediction in the $(c_{LL}^B,\phi_B)$ plane in a fit with $c_{LL}^K=0$.
The preference for non-SM values of the parameters in both cases arises from the tension in the SM CKM fit between $\epsilon_K$ (when using the experimental data for $V_{cb}$ and $\sin2\beta$ as inputs) and $S_{\psi K_S}=\sin2\beta$ \cite{Lunghi:2008aa,Buras:2008nn,Altmannshofer:2009ne,Lunghi:2010gv,Bevan:2010gi,Brod:2011ty}. As is well known, this tension can be solved either by increasing $\epsilon_K$ (as in the first case) or by decreasing $S_{\psi K_S}$ by means of a new physics contribution to the $B_d$ mixing phase (as in the second case). In the second case, also a positive contribution to $S_{\psi\phi}=-\sin2\phi_s$ is generated.
The right panel of figure~\ref{fig:DF2singlefits} shows the fit prediction in a fit where $\phi_B=0$ and $c_{LL}^B=c_{LL}^K\equiv c_{LL}$, i.e. the $\U(3)^3$ or MFV limit. In that case, a positive $c_{LL}$ cannot solve the CKM tension, since it would lead to an increase not only in $\epsilon_K$, but also in $\Delta M_{d,s}$.

The two plots of figure~\ref{fig:DF2fits} show the projections onto the $(c_{LL}^K,c_{LL}^B)$ and $(c_{LL}^B,\phi_B)$ planes of the fit with all 3 parameters in (\ref{eq:HeffDF2}) non-zero. Since both solutions to the CKM tension now compete with each other, the individual parameters are less constrained individually.

\subsection{$\Delta S = 1$: $\epsilon'/\epsilon$}\label{U2/minimal/epsilonprime}
An observable that is relevant in other contexts as well, like in MFV \cite{DAmbrosio:2002ex}, is direct \CP\ violation in $K$ decays, as summarized in the parameter $\epsilon^\prime$.
Either in $\U(2)^3$ or in MFV, a contribution to $\epsilon^\prime$ arises from the effective Hamiltonian
 \begin{equation}\label{operatorsepsilonprime}
{\H}^{\Delta S = 1}_{4f,LR} = \frac{1}{\Lambda^2} \xi_{ds} (c^d_5{\O}^d_5 + c^u_5{\O}^u_5 + c^d_6{\O}^d_6 + c^u_6{\O}^u_6) + \text{h.c.}, \qquad\xi_{ds} = V_{td}V_{ts}^*\,.
 \end{equation}
Following the analysis of section \ref{Flavour/bounds/epsilonprime} and imposing the bounds \eqref{boundepsilonprimeLambda} we obtain 
\begin{align}\label{boundepsilonprime}
c_5^{u,d} &\lesssim 0.4 \left(\frac{\Lambda}{3~\text{TeV}}\right)^2, & c_6^{u,d} &\lesssim 0.13 \left(\frac{\Lambda}{3~\text{TeV}}\right)^2.
\end{align}

The limit on $c_6^{u,d}$ is actually one of the strongest bounds among all the $\U(2)^3$ parameters. Nevertheless, as already stated before, the uncertainties in the estimate of the SM contribution to $\epsilon^\prime/\epsilon$ could cancel against the new physics contribution, relaxing the previous bound by a factor of a few.

\subsection{$\Delta B=1$ processes}\label{U2/minimal/DB1}
The $\U(2)^3$ predictions for $\Delta F=1$ processes are more model-dependent because a larger number of operators is relevant. In addition, the main prediction of universality of $b\to s$ and $b\to d$ amplitudes --  but not $s\to d$ amplitudes -- is not well tested. Firstly, current data are better for $b\to s$ decays compared to $b\to d$ decays. Secondly, the only clean $s\to d$ processes are $K\to\pi\nu\bar\nu$ decays, but $b\to q\nu\bar\nu$ processes have not been observed yet.
Thus, in the following we will present the constraints on the effective Hamiltonian
\begin{align}
{\H}^{\Delta B = 1} &= \Delta\H_{4f,L}^{\Delta B = 1} + \Delta\H_{\rm mag} + \Delta\H_H\notag\\
&= \sum_{i=d,s} \xi_{ib}
\bigg[
\frac{c_{7\gamma}e^{i\phi_{7\gamma}}}{\Lambda^2}m_{b} \left( {\bar d^i}_L \sigma_{\mu\nu}  b_R\right) e F^{\mu\nu}
+
\frac{c_{8g}e^{i\phi_{8g}}}{\Lambda^2}m_{b} \left( {\bar d^i}_L \sigma_{\mu\nu} T^a b_R \right) g_s G^{\mu\nu\,a}
\notag\\
&\qquad\qquad +
\frac{c_Le^{i\phi_{L}}}{\Lambda^2} \left( {\bar d}_L^i \gamma_{\mu} b_L \right) \left( {\bar \ell}_L \gamma_{\mu} \ell_L \right)+
\frac{c_R e^{i\phi_{R}}}{\Lambda^2} \left( {\bar d}_L^i \gamma_{\mu} b_L \right) \left( {\bar e}_R \gamma_{\mu} e_R \right)\notag\\
&\qquad\qquad +
\frac{c_He^{i\phi_{H}}}{\Lambda^2} \frac{v^2}{2} \left( {\bar d}_L^i  \gamma_\mu b_L \right)\frac{g}{c_w}Z^{\mu}
\bigg]
+\text{h.c.}\,,
\label{eq:HeffDF1}
\end{align}
using only data from inclusive and exclusive $b\to s$ decays, making use of the results of \cite{Altmannshofer:2011gn,Altmannshofer:2013foa}.
In general all the coefficients in~(\ref{eq:HeffDF1}) can be relevant and of ${\Ord}(1)$.
Since the chromomagnetic penguin operator enters the observables considered in the following only through operator mixing with the electromagnetic one, we will ignore $c_{8g}$ in the following.

\begin{figure}[tb]
\centering
\includegraphics[width=0.8\textwidth]{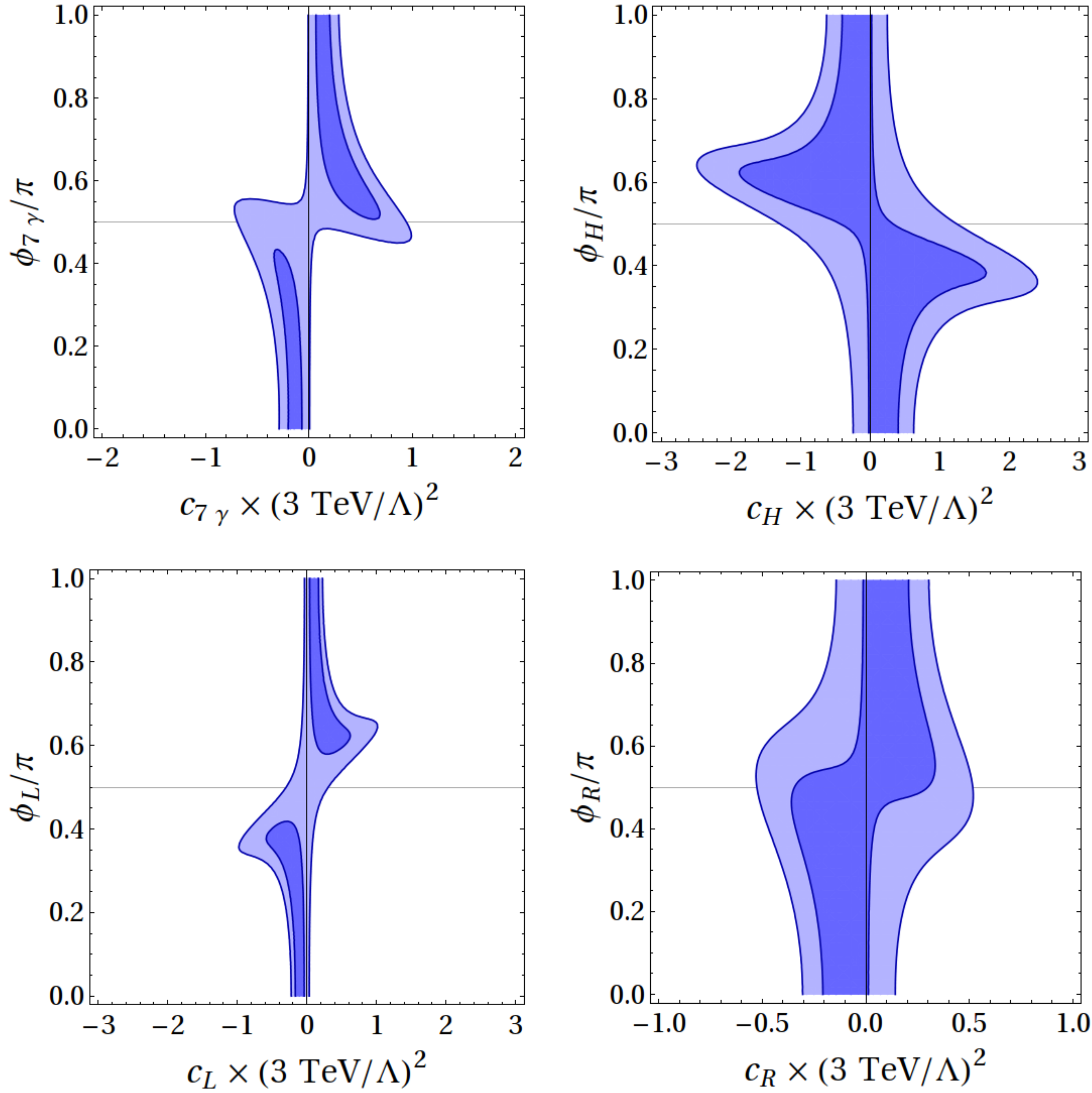}
\caption{68 and 95\% C.L. allowed regions for the $\Delta F=1$ coefficients in $\U(2)^3$, using the results of a global analysis of inclusive and exclusive $b\to s$ decays \cite{Altmannshofer:2011gn, Altmannshofer:2013foa} (courtesy of D.~M.~Straub).}
\label{fig:DF1fits}
\end{figure}

Figure~\ref{fig:DF1fits} shows the constraints on the coefficients of the four operators in (\ref{eq:HeffDF1}) and their phases. The constraints are particularly strong for the magnetic penguin operator and the semi-leptonic left-left vector operator. In the first case, this is due to the $B\to X_s\gamma$ branching ratio, in the second case it is due to the forward-backward asymmetry in the $B\to K^*\mu^+\mu^-$ decay and the branching ratio of the recently observed $B_s\to \mu^+\mu^-$ decay. Interestingly, in both cases, the constraint is much weaker for maximal phases of the new physics contribution, since the interference with the (real) SM contribution is minimized in that case. For the two operators in the right-hand column of figure~\ref{fig:DF1fits}, this effect is less pronounced. The reason is that both coefficients are accidentally small in the SM: in the first case due to the small $Z$ coupling to charged leptons proportional to $(1-4s_w^2)$ and in the second case due to $C_9(\mu_b)\approx-C_{10}(\mu_b)$ (in the convention of  \cite{Altmannshofer:2011gn}).

To ease the interpretation of figures~\ref{fig:DF2fits} and \ref{fig:DF1fits}, all the coefficients are normalized, as explicitly indicated, to a scale $\Lambda = 3$ TeV, which might represent both the scale of a new strong interaction responsible for EWSB, $\Lambda_s \approx 4\pi v$, or the effective scale from loops involving the exchange of some new weakly interacting particle(s) of mass of ${\Ord}(v)$. Interestingly the fits of the current flavour data are generally consistent with coefficients of order unity, at least if there exist sizable non vanishing phases, when they are allowed. Note in this respect that $\U(3)^3$ at low $\tan{\beta}$ does not allow phases in $c_{L, R, H}$ in figure~\ref{fig:DF1fits}, with correspondingly more stringent constraints especially on $c_L$. A possible interpretation of these figures~\ref{fig:DF2fits},~\ref{fig:DF1fits} is that the current flavour data are at the level of probing the $\U(2)^3$ hypothesis in a region of parameter space relevant to several new theories of EWSB.

\subsection{Electric dipole moments}\label{U2/minimal/EDM}

The presence of phases in the flavour-diagonal chirality breaking operators of $\Delta\L_{\rm mag}$ has to be consistent with the limits coming from the neutron electric dipole moment.

The relevant contraints come from the \CP\ violating contributions to the operators $\O_{7\gamma}$ and $\O_{8g}$ of section~\ref{EDMsec}. The bounds on their coefficients are the ones reported in \eqref{EDMboundgamma} and \eqref{EDMboundg}.
Note that the bounds are automatically satisfied if one does not allow for phases outside the spurions. Otherwise, even with generic phases $\tilde \phi_{u,d}^{\gamma,g}$, the smallness of the coefficients can be explained taking the new physics scale related to the first generation decoupled from the scale $\Lambda$ of EWSB, as allowed by the $\U(2)^3$ symmetry.
This can be realized in concrete models if the operators in \eqref{EDM} come from Feynman diagrams involving the exchange of heavy first generation partners.

It should also be noted that the coefficients of the flavour diagonal operators, analogue of the ones in the second line of \eqref{eq:HeffDF1}, are limited by the electroweak precision tests at a similar level to the ones shown in figure~\ref{fig:DF1fits}.

\subsection{Up quark sector within $\U(2)^3$}\label{U2/minimal/up}

Flavour and \CP\ violation in the up quark sector could arise in the $\U(2)^3$ framework as contributions to $D$-$\bar{D}$ mixing, to direct \CP\ violation in $D$-decays,
to the top decays $t\rightarrow c \gamma$ and $t\rightarrow c Z$ and to the top chromo-electric dipole moment.

Within our setup, the relevant effective operators for the above processes are:
\begin{align}
\Delta\H_{4f,L}^{D}  & = \frac{c_{LL}^D }{\Lambda^2}\, \zeta_{uc}^2 \, \dfrac{1}{2}\left({\bar u}_L
\gamma_\mu  c_L \right)^2, \label{eq:Dmixing} \\
\Delta\H_{\rm mag}^{D}  & = \frac{c_{8g}^{D} e^{i \phi_{8g}^{D}}}{\Lambda^2} \,m_c \,\zeta_{uc}
\left( {\bar u}_L \sigma_{\mu \nu} c_R \right) g_s G^{\mu \nu},
\label{eq:Ddecay}\\
\Delta\H_{\rm mag}^{\Delta T = 1} & = \frac{m_t}{\Lambda^2}\zeta_{ct} \left( {\bar c}_L \sigma_{\mu \nu} t_R \right)\left(c_{7\gamma}^{t} e^{i \phi_{7\gamma}^{t}}eF^{\mu\nu} + c_{7Z}^{t} e^{i \phi_{7Z}^{t}}\frac{g}{c_w}Z^{\mu\nu}\right),\label{eq:topcb}\\
\Delta\H_{H}^{\Delta T = 1} & = \frac{c_{H}^{t} e^{i \phi_{H}^{t}}}{\Lambda^2}\frac{v^2}{2} \,\zeta_{ct}
\left({\bar c}_L \gamma_\mu t_L \right) \frac{g}{c_w} Z^{\mu},
\label{eq:topcc}\\
\Delta\H_{\rm mag}^{\Delta T = 0} & = \frac{c_\text{dm} e^{i \phi_\text{dm}}}{\Lambda^2}\,m_t
\left( {\bar t}_L \sigma_{\mu \nu} t_R \right) g_s G_{\mu \nu},
\end{align}
where
$c_w = \cos \theta_w$ and the $c$'s are real parameters, with the phases made explicit wherever present.
All these coefficients are model dependent and, in principle, can be of ${\Ord}(1)$.
Since $\Lambda \gg v$, the requirement of $\SU(2)_L$ invariance correlates $c_{LL}^{D}$, $c_H^{t}$ and
$\phi_H^{t}$ with the analogous parameters in the down sector. One can easily see they have to be equal
within a few percent, and so they have to respect bounds similar to those for $c_{LL}^K$, $c_H$ and
$\phi_H$ (see figures \ref{fig:DF2fits} and \ref{fig:DF1fits}).

In the neutral $D$ meson system the SM short distance contribution to the mixing is orders of magnitudes
below the long distance one, thus complicating the theoretical calculation of the mass and width splittings
$x$ and $y$ (see \cite{Gedalia:2009kh}, also for a discussion of the relevant parameters).
Despite the above uncertainties, many studies (see \cite{Falk:2001hx, Falk:2004wg} and references therein)
indicate that the standard model could naturally account for
the values $x \sim y \sim 1 \%$, thus explaining the measured $95\%$ CL intervals
$x \in [0.10,0.81]\%$, $y \in [0.56,0.92]\%$ \cite{Amhis:2012bh}.
Here, like in \cite{Gedalia:2009kh, Isidori:2011qw}, we take the conservative approach
of using the above data as upper bounds to constrain new physics contributions.
Referring to the analysis carried out in \cite{Gedalia:2009kh, Isidori:2011qw}, within our framework it turns
out that the most effective bound is the one on the coefficient in the operator $\Delta\H_{4f,L}^{D}$.
In our notation it reads
\begin{equation}
(c_{LL}^D)^2 \left(\dfrac{3 \,\text{TeV}}{\Lambda}\right)^2 < 90,
\label{eq:Dmix}
\end{equation}
so to saturate it we would need values of $c_{LL}^D$ that are excluded, since they would imply a too large contribution to
$\Delta F = 2$ observables in the down sector.

Suppose now that the recent evidence for \CP\ violation in $D$ decays measured by LHCb \cite{Aaij:2011in} and CDF \cite{CDF-Note-10784}
is at least in part due to new physics. The quantity of interest is the difference between the time-integrated \CP\ asymmetries in the decays
$D^0 \to K^+ K^-$ and $D^0 \to \pi^+ \pi^-$, for which the world average
$\Delta A_\text{\CP} = A_{KK} - A_{\pi \pi} = -(3.29 \pm 1.21)\times 10^{-3}$ is reported in \cite{Amhis:2012bh}.
In reference \cite{Isidori:2011qw} all the possible effective
operators contributing to the asymmetry are considered, while respecting at the same time the bounds coming from $D$-$\bar{D}$ mixing and
from $\epsilon_K'/\epsilon_K$.
Following that analysis, the only operator that can give a relevant contribution in our setup is
$\Delta\H_{\rm mag}^{D}$, $\Delta A_\text{\CP}$ being proportional to the imaginary part of the relative coefficient.
Referring to the estimations carried out in \cite{Isidori:2011qw} and \cite{Giudice:2012qq} for the hadronic
matrix elements, to reproduce the measured value of $\Delta A_\text{\CP}$ one would need
\begin{equation}
c_{8g}^D \sin ({\rm arg} \xi_{uc} + \phi_{8g}^D) \left( \frac{3\, \text{TeV}}{\Lambda}\right)^2 \simeq 20,
\label{eq:Ddevay}
\end{equation}
a value out of reach if we want to keep the parameter $c_{8g}^D$ to be of order one.

The LHC sensitivity to top-quark FCNC at 14 TeV with $100 \, \text{fb}^{-1}$ of data is expected to be (at $95 \%$ C.L.)
\cite{Carvalho:2007yi}:
$\text{BR}(t \to c\, Z, \,u\, Z) \simeq 5.5 \times 10^{-5}$ and $\text{BR}(t \to c\, \gamma,\, u\,\gamma)
\simeq 1.2 \times 10^{-5}$. Here we concentrate on the charm channels, since both in the SM and in our framework
the up ones are CKM suppressed. In the SM, $\text{BR}(t \to c\, Z,\, c\, \gamma)$ can be estimated to be
of order $ (m_b^2/m_W^2)^2 \, |V_{cb}|^2 \, \alpha^2/s_w^2 \sim 10^{-12}$, so that an experimental observation
will be a clear signal of new physics. To estimate the $\U(2)^3$ effects for these processes, we
follow the analysis carried out in \cite{Fox:2007in}. The dominant contributions are those given by the operator
$\O_{7\gamma}^t$ in $\Delta\H_{\rm mag}^{\Delta T = 1}$ for $t \to c \gamma$, and by both $\Delta\H_{\rm mag}^{\Delta T = 1}$ and
$\Delta\H_H^{\Delta T = 1}$ for $t \to c Z$. We obtain
\begin{align}
\text{BR}(t \to c\, \gamma) &\simeq 1.7 \times 10^{-8} \left(\dfrac{3 \, \text{TeV}}{\Lambda}\right)^{\!\! 4}
{c_{7\gamma}^t}^2,\\
\text{BR}(t \to c\, Z) &\simeq 8.5 \times 10^{-8} \left(\dfrac{3 \, \text{TeV}}{\Lambda}\right)^{\!\! 4} \!\!
\left(0.61 {c_{7Z}^t}^2 + 0.39\, {c_H^t}^2 + 0.83\, c_{7Z}^t c_H^t \cos (\phi_H^t - \phi_{7Z}^t)\right),
\label{eq:topBR}
\end{align}
leading us to conclude that any non-zero evidence for these decays at the LHC could not be explained in our setup,
unless we allow the dimensionless coefficients to take values more than one order of magnitude bigger
than the corresponding ones in the down sector (actually this could be possible only for $c_{7Z}^t$ and $c_{7\gamma}^t$
but not for $c_H^t$, because of its correlation with $c_H$ and of the bounds of figure~\ref{fig:DF1fits}).

Finally, the recent analysis carried out in \cite{Kamenik:2011dk} has improved previous bounds \cite{CorderoCid:2007uc}
on the top CEDM $\tilde{d}_t$
by two orders of magnitude, via previously unnoticed contributions of $\tilde{d}_t$ to the neutron
electric dipole moment.
In deriving this bound, the authors of \cite{Kamenik:2011dk} have assumed
the up and down quark EDMs $d_{u,d}$ and CEDMs $\tilde{d}_{u,d}$ to be negligible.
This is relevant in our context if
\begin{itemize}
 \item we allow for generic phases outside the spurions $\V$,
 $\Delta_u$ and $\Delta_d$,
 \item we assume that some other mechanism is responsible for making $d_{u,d}$ and $\tilde{d}_{u,d}$
 negligible. Notice that this is actually the case in SUSY with heavier first two generations, where
 on the contrary there is no further suppression of $\tilde{d}_t$ with respect to the EFT natural estimate.
\end{itemize}
Then,
the bound given in \cite{Kamenik:2011dk} imposes
\begin{equation}
c_\text{dm} |\sin \phi_\text{dm}|\, \left( \dfrac{3\, \text{TeV}}{\Lambda} \right)^2 < 0.6 \,,
\label{topCEDM}
\end{equation}
so that future experimental improvements in the determination of the neutron EDM
will start to challenge the $\U(2)^3$ scenario with \CP\ violating phases outside the spurions, if the
hypothesis of negligible $d_{u,d}$ and $\tilde{d}_{u,d}$ is realized.

\section{Flavour and \CP\ violation in Generic  $\U(2)^3$}\label{U2/generic}

Generic $\U(2)^3$, introducing physical rotations in the right handed sector as well, gives rise to extra flavour and \CP\ violating contributions in \eqref{eq:genL}. Their general form can be summarized in
\begin{equation}
\Delta{\L} = \Delta{\L}_{4f,L}  + \Delta{\L}_\text{mag} + \Delta\L_H + \Delta{\L}_{4f,R}  + \Delta{\L}_{4f,LR}\, ,
\label{eq:genL}
\end{equation}
where $\Delta{\L}^{4f}_{L, R, LR} $ are the sets of four-fermion operators with flavour violation respectively in the left-handed sector, in the right-handed sector and in both, while $\Delta{\L}_{\rm mag}$ and $\Delta\L_H$ were defined in \eqref{Leffgeneral}. The most significant new effects are contained in $\Delta{\L}_\text{mag}$ and in $\Delta{\L}^{4f}_{LR}$. Notice that sizeable contributions to $\Delta{\L}^{4f}_{R}$ and $\Delta{\L}^{4f}_{LR}$ are absent in Minimal $\U(2)^3$. In the following, we first discuss the relevant new effects with respect to Minimal~$\U(2)^3$, which show up in $\Delta C = 1$, $\Delta S = 1$ and $\Delta S = 2$ observables, as well as in flavour conserving electric dipole moments. We then see how in $B$ and $t$ decays and in $D$-$\bar{D}$ mixing the new effects are at most analogous in magnitude to those of the Minimal breaking case.

\subsection{$\Delta C=1$: $D$ decays}\label{U2/generic/DC1}

\CP\ asymmetries in $D$ decays receive contributions from chromo-magnetic dipole operators with both chiralities,
\begin{equation}
{\L}^{\Delta C=1}_\text{mag} = \frac{1}{\Lambda^2} c^D_{8g} e^{i \phi^D_{8g}}
\zeta_{uc} \left[
e^{-i\phi_2^u} \frac{\epsilon^u_R}{\epsilon_L}\,{\O}_{8g}
+
e^{i\phi_1^u} \frac{s^u_R}{s^u_L} \frac{\epsilon_R^u}{\epsilon_L}\,{\O}_{8g}'
\right] + \text{h.c.}\\
\end{equation}
where
\begin{equation}
{\O}_{8g} = m_t(\bar u_L\sigma_{\mu\nu}T^a c_R)g_s G^{\mu\nu}_a,\qquad{\O}_{8g}' = m_t(\bar u_R\sigma_{\mu\nu}T^a c_L)g_s G^{\mu\nu}_a,
\end{equation}
and with $\phi^D_{8g}$ we account for the possibility of \CP\ violating phases outside the spurions (see appendix~\ref{app:bilinears} for details). 
Most notably, the recently observed \CP\ asymmetry difference between $D\to KK$ and $D\to \pi\pi$ decays, even if roughly consistent with the SM prediction, could in part be due to new physics contributions to the chromo-magnetic operators. Following \cite{Isidori:2011qw,Giudice:2012qq} we write at the scale $\mu = m_c$
\begin{equation}
\Delta A_\text{\CP} \simeq - \frac{2}{\lambda} \Big[\Im(V_{cb}^* V_{ub}) \Im\left(\Delta R^{\text{SM}} \right) + \frac{1}{\Lambda^2}
\Big(
\Im\big( C_8\big) \Im\big( \Delta R^{\text{NP}} \big) +
\Im ( C_8' ) \Im\big( {\Delta R'}^{\text{NP}} \big) 
\Big) \Big]
,
\label{DeltaAcp}
\end{equation}
where the Cabibbo angle $\lambda$ is defined in \eqref{CKM}, $\Delta R^{(\prime)\text{SM},\text{NP}} = R_K^{(\prime)\text{SM},\text{NP}} + R_{\pi}^{(\prime)\text{SM}, \text{NP}}$, $R_{K,\pi}^{\text{SM}}$ are the ratios between the subleading and the dominant SM hadronic matrix elements, and
\begin{align}
\label{Dhadr}
{R_{K}^{(\prime)\text{NP}}} &\simeq V_{cs}^* V_{us} \frac{\langle K^+ K^- |{\O}_8^{(\prime)}| D\rangle}{\langle K^+ K^- |{\L}^{\text{SM}}_{\rm eff}| D \rangle} \sim 0.1 \times \frac{4 \pi^2 m_t}{m_c} \frac{\sqrt{2}}{G_F},\\
R_{\pi}^{(\prime)\text{NP}} &\simeq V_{cd}^* V_{ud} \frac{\langle \pi^+ \pi^- |{\O}_8^{(\prime)}| D\rangle}{\langle \pi^+ \pi^- |{\L}^{\text{SM}}_{\rm eff}| D \rangle} \simeq R_K^{(\prime)\text{NP}}.
\end{align}
In our estimates we will assume maximal strong phases, which imply $|\Im\, \Delta R^{(\prime)\text{NP}}| \simeq 2 R_{K}^{\text(\prime){\rm NP}}$. The SM contribution can be naively estimated to be $\Delta R^{\text{SM}} \sim \alpha_s(m_c)/\pi \sim 0.1$, but larger values from long distance contributions could arise, 
making a precise prediction within the SM still an open issue (see e.g. \cite{Brod:2012ud,Isidori:2012yx} for recent works on this).

Requiring the new physics contribution to $\Delta A_\text{\CP}$ to be less than the central value of the 
world average of experimental results\cite{Amhis:2012bh} $\Delta A_\text{\CP}^\text{exp}=(-3.29\pm1.21)\times 10^{-3}$ implies
\begin{align}
c^D_{8g}\frac{\epsilon^u_R}{\epsilon_L}\frac{\sin\left(\delta -\phi_2^u + \phi^D_{8g}\right)}{\sin \delta}
&\lesssim 0.17
\left( \frac{\Lambda}{3\, {\rm TeV}} \right)^{\! 2}, &
c^D_{8g}\frac{s^u_R}{s^u_L} \frac{\epsilon^u_R}{\epsilon_L}\frac{\sin(\delta + \phi_1^u - \phi^D_{8g})}{\sin \delta}
&\lesssim 0.17
\left( \frac{\Lambda}{3\, {\rm TeV}} \right)^{\! 2}.
\label{eq:DC1bound}
\end{align}
The bound can be saturated without violating indirect constraints on these operators arising from $\epsilon'$ or $D$-$\bar D$ mixing due to weak operator mixing \cite{Isidori:2011qw}. We stress that the bounds in \eqref{eq:DC1bound} carry an order one uncertainty coming from the normalized matrix elements $R_{\pi,K}$.

\subsection{Electric dipole moments}\label{U2/generic/EDM}

In the flavour conserving case, important constraints arise from the up and down quark electric dipole moments (EDMs) and chromo-electric dipole moments (CEDMs). In addition to \eqref{EDM} there are new contributions coming from the \CP\ violating part of the operators
\begin{align}\label{EDMright}
{\L}^{\Delta F=0}_\text{mag} &= \frac{m_t}{\Lambda^2} \xi_{uu}\,e^{-i \phi_1^u} \frac{s^u_R}{s^u_L} \frac{\epsilon^u_R}{\epsilon_L}\left[
c_u^g e^{i\phi_u^g} (\bar u_L\sigma_{\mu\nu}T^a u_R)g_sG^{\mu\nu}_a
+ c_u^\gamma e^{i\phi_u^\gamma} (\bar u_L\sigma_{\mu\nu}u_R)eF^{\mu\nu}
\right] \nonumber \\
 &+
 \frac{m_b}{\Lambda^2} \xi_{dd} \,e^{-i \phi_1^d} \frac{s^d_R}{s^d_L} \frac{\epsilon^d_R}{\epsilon_L}\left[
c_d^g e^{i\phi_d^g} (\bar d_L\sigma_{\mu\nu}T^a d_R)g_sG^{\mu\nu}_a
+ c_d^\gamma e^{i\phi_d^\gamma} (\bar d_L\sigma_{\mu\nu}d_R)eF^{\mu\nu}
\right] + \text{h.c.}\, ,
\end{align}
where we remind that $\phi_1^{u,d}$ are non zero even if there are no \CP\ phases outside the spurions. The new contributions to the quark (C)EDMs are
\begin{equation}
d_u = 2e\frac{m_t}{\Lambda^2}\xi_{uu}\frac{s^u_R}{s^u_L}\frac{\epsilon^u_R}{\epsilon_L}c_u^{\gamma}\sin(\phi_u^{\gamma} - \phi_1^u),\qquad \tilde d_u = 2\frac{m_t}{\Lambda^2}\xi_{uu} \frac{s^u_R}{s^u_L}\frac{\epsilon^u_R}{\epsilon_L}c_u^{g}\sin(\phi_u^{g} - \phi_1^u), \qquad (u\leftrightarrow d).
\end{equation}
From \eqref{neutron}, considering again the running of the Wilson coefficients from 3 TeV down to the hadronic scale of 1 GeV, the experimental bound on the neutron EDM implies for the parameters at the high scale
\begin{align}
c_u^{\gamma}\,|\sin (\phi_u^{\gamma}-\phi_1^u )| \frac{s^u_R}{s^u_L}\,\frac{\epsilon^u_R}{\epsilon_L}\lesssim 1.2 \times 10^{-2}
\left( \frac{\Lambda}{3\, {\rm TeV}} \right)^2
,
\nonumber\\
c_d^{\gamma}\,|\sin (\phi_d^{\gamma}-\phi_1^d )|  \frac{s^d_R}{s^d_L}\,\frac{\epsilon^d_R}{\epsilon_L}\lesssim 3.2 \times 10^{-2}
\left( \frac{\Lambda}{3\, {\rm TeV}} \right)^2
,
\nonumber\\
c_u^{g}\,|\sin (\phi_u^{g}-\phi_1^u )|  \frac{s^u_R}{s^u_L}\,\frac{\epsilon^u_R}{\epsilon_L}\lesssim 4.4 \times 10^{-3}
\left( \frac{\Lambda}{3\, {\rm TeV}} \right)^2
,
\nonumber\\
c_d^{g}\,|\sin (\phi_d^{g}-\phi_1^d )|  \frac{s^d_R}{s^d_L}\,\frac{\epsilon^d_R}{\epsilon_L}\lesssim 2.5 \times 10^{-2}
\left( \frac{\Lambda}{3\, {\rm TeV}} \right)^2
.
\label{eq:DF0bound}
\end{align}

Notice that since the operators of \eqref{EDMright} are generated through the right-handed mixings with the third generation, the coefficients $c_{u,d}^{\gamma, g}$ can no longer be suppressed by the large mass of first-generation-partners as in the Minimal case, and the bounds above will constrain $s^u_R\epsilon^u_R$ and $s^d_R\epsilon^d_R$.

\subsection{$\Delta S=1$: $\epsilon'/\epsilon$}\label{U2/generic/epsilonprime}

The $s\to d$ chromomagnetic dipole
\begin{align}
 \Delta{\L}^{\Delta S = 1}_\text{mag} &= 
 \frac{m_t}{\Lambda^2} {c}^K_{8g} e^{i(\phi^K_{8g}-\phi_2^d)}\lambda_b \xi_{ds} \frac{\epsilon^d_R}{\epsilon_L} \left( \bar{d}_L \sigma_{\mu\nu} T^a s_R \right) g_s G_{\mu\nu}^a
\end{align}
contributes to $\epsilon'$. Following the analysis in \cite{Mertens:2011ts}, one obtains the bound
\begin{equation}
{c}^K_{8g} \frac{\sin (\beta +\phi^K_{8g} - \phi_2^d)}{\sin \beta}\,\frac{\epsilon^d_R}{\epsilon_L}
\lesssim 0.7 \left(\frac{\Lambda}{3 \,{\rm TeV}} \right)^2 \,.
\label{eq:DS1bound}
\end{equation}
Furthermore, in addition to the LR four fermion operators in \eqref{operatorsepsilonprime}, there is a contribution to $\epsilon'$ also from the operators with exchanged chiralities
\begin{equation}
\Delta{\L}^{\Delta S = 1}_{4f,LR} = \frac{1}{\Lambda^2} \xi_{ds} \frac{s^d_R}{s^d_L} \left(\frac{\epsilon^d_R}{\epsilon_L}\right)^2 e^{i(\phi_1^d-\phi_2^d)}(c^{\prime d}_5{\O}^{\prime d}_5 + c^{\prime u}_5{\O}^{\prime u}_5 + c^{\prime d}_6{\O}^{\prime d}_6 + c^{\prime u}_6{\O}^{\prime u}_6) + \text{h.c.}\, ,
\end{equation}
where
\begin{align}
{\O}_5^{\prime q} = ( \bar{d}_R \gamma_\mu s_R ) (\bar{q}_L \gamma_\mu q_L), \qquad
{\O}_6^{\prime q} = ( \bar{d}_R^\alpha \gamma_\mu s_R^\beta) ( \bar{q}_L^\beta \gamma_\mu q_L^\alpha),\qquad q=u,d.
\end{align}
From \eqref{boundepsilonprimeLambda} one gets
\begin{align}
c_5^{\prime u,d} \frac{\sin (\beta +\phi_1^d - \phi_2^d)}{\sin \beta}\, \frac{s^d_R}{s^d_L} \left(\frac{\epsilon^d_R}{\epsilon_L}\right)^2
&\lesssim 0.4\left(\frac{\Lambda}{3~\text{TeV}}\right)^2,\\
c_6^{\prime u,d} \frac{\sin (\beta +\phi_1^d - \phi_2^d)}{\sin \beta}\, \frac{s^d_R}{s^d_L} \left(\frac{\epsilon^d_R}{\epsilon_L}\right)^2
&\lesssim 0.13\left(\frac{\Lambda}{3~\text{TeV}}\right)^2,
\end{align}
which is not particularly relevant since a stronger bound on the combination $(s^d_R/s^d_L)(\epsilon^d_R/\epsilon_L)^2$ comes from $\epsilon_K$.

\subsection{$\Delta S=2$: $\epsilon_K$}\label{U2/generic/epsilonK}
   
Finally, the only relevant new effect contained in $\Delta{\L}_{4f,LR}$ arises from the $\Delta S=2$ operators $\Q_4$ and $\Q_5$, contributing to $\epsilon_K$, which are enhanced by a chiral factor and by renormalization-group effects. The relevant operators are
\begin{equation}
\Delta{\L}^{\Delta S=2}_{4f,LR} = \frac{1}{\Lambda^2}\frac{s^d_R}{s^d_L} \left(\frac{\epsilon^d_R}{\epsilon_L}\right)^2\xi_{ds}^2 e^{i(\phi_1^d-\phi_2^d)}\left[ c^K_4\lambda_b^2  \left(\bar{d}_L s_R\right) \left(\bar{d}_R s_L \right) + c^K_5 \left(\bar{d}_L \gamma_\mu s_L \right)\left(\bar{d}_R \gamma_\mu s_R \right)\right]
.
\end{equation}
Using bounds from \cite{Isidori:2010kg}, one gets
\begin{equation}
 c^K_5 \frac{\sin (2 \beta +\phi_1^d - \phi_2^d)}{\sin 2 \beta}\, \frac{s^d_R}{s^d_L} \left(\frac{\epsilon^d_R}{\epsilon_L}\right)^2
\lesssim 6 \times 10^{-3} \left(\frac{\Lambda}{3 \,{\rm TeV}} \right)^2\,.
\label{eq:DS2bound}
\end{equation}

\subsection{$D$ mixing, $B$ and top FCNCs}\label{U2/generic/other}

In the $D$ and $B$ systems there are no enhancements of the matrix elements of the operators in $\Delta{\L}_{LR}^{4f}$ and $\Delta{\L}_{R}^{4f}$, unlike what happens for $K$ mesons. Moreover the new contributions to these operators are all suppressed by some powers of $\epsilon^{u,d}_R/\epsilon_L$ (see appendix~\ref{app:bilinears}). Therefore they are all subleading with respect to those of Minimal $\U(2)^3$, once we take into account the bounds from the other observables that we have discussed. An analogous suppression holds also for the operators that contain chirality breaking bilinears involving one third generation quark, relevant for $B$ and top FCNCs. 
A four fermion operator of the form $(\bar{u}_
L c_R)(\bar{u}_R c_L)$
might in principle be relevant for $D$-$\bar{D}$ mixing. However, taking into account the bounds on $\epsilon^u_R/\epsilon_L$,  this new contribution gives effects of the same size of those already present in Minimal $\U(2)^3$ and far from the current sensitivity.
Consequently the phenomenology of $B$ decays is the same for Minimal and Generic $\U(2)^3$. The only difference  in the latter is that \CP\ violating effects are generated also if we set to zero the phases outside the spurions, though suppressed by at least one power of $\epsilon^d_R/\epsilon_L$.

Concerning the up quark sector, 
given the future expected sensitivities for top FCNC \cite{Carvalho:2007yi} and \CP\ violation in $D$-$\bar{D}$ mixing \cite{Aushev:2010bq, Merk:2011zz}, within the $\U(2)^3$ framework we continue to expect no significant effects in these processes (see \cite{Barbieri:2012uh} for the size of the largest contributions). We stress that, while an observation of a flavour changing top decay at LHC would generically put the $\U(2)^3$ framework into trouble, a hypothetical observation of \CP\ violation in $D$-$\bar{D}$ mixing would call for a careful discussion of the long distance contribution.

\section{Lepton flavour violation}

If one tries to extend the considerations developed so far to the lepton sector, one faces two problems. First, while the hierarchy of charged lepton masses is comparable to the mass hierarchies in the quark sector, the leptonic charged-current mixing matrix  does not exhibit the hierarchical pattern of the CKM matrix. Secondly, in the likely possibility that the observed neutrinos are of Majorana type and are light because of a small mixing to heavy right-handed neutrinos, the relevant parameters in the Yukawa couplings of the  lepton sector,  $H \bar{\ell}_L \hat y_e e_R$ and $\tilde{H} \bar{\ell}_L \hat y_\nu \nu_R$,  are augmented, relative to the ones in the quark sector, by the presence of the right-handed neutrino mass matrix, $\nu_R^T M \nu_R$, which is unknown. To overcome these problems we make the following two hypotheses: \begin{itemize} \item we suppose that the charged leptons, thorough  $\hat y_e$, behave in a similar way to the quarks, whereas $\hat y_\nu$ and $M$ are responsible for the anomalously large neutrino mixing angles; \item we assume that $\hat y_\nu$ has no significant influence on flavour physics near the Fermi scale in spite of the presence at this scale of new degrees of freedom carrying flavour indices, like sneutrinos or heavy composite leptons. One can imagine many reasons for this to be the case, like e.g. in the discussion of the next section.\end{itemize}

Extending $\U(3)^3$ and $\U(2)^3$ to the leptonic sector, we consider respectively a $\U(3)_\ell\times \U(3)_e$ and a $\U(2)_\ell\times \U(2)_e$ symmetry. Here comes another significant difference between the two cases: in the $\U(3)_\ell\times \U(3)_e$ case, with $\hat y_e$ transforming as a $(\three, \threebar)$, there is no new flavor changing phenomenon at the Fermi scale other than the  leptonic charged-current mixing matrix, since $\hat y_e$  can be diagonalized by a $\U(3)_\ell\times \U(3)_e$  transformation.
On the contrary, let us assume that the  $\U(2)_\ell\times \U(2)_e$ symmetry be  broken {\em weakly} by the spurions
\begin{align}
\Delta_e &\sim (\two,\twobar) \,, & 
\Ve &\sim (\two,\one) \,.
\label{lept_spur}
\end{align}
By proceeding in the same way as for the quarks in section~\ref{U2/EFT}, one can set
\begin{align}
\Ve &= (0, \eta)^T, & \Delta_e &= R_{12}^e \Delta_e^{\rm diag}
\end{align}
and see the occurrence of flavour changing bilinears with two important differences relative to the quark case. Firstly, one cannot relate the size of $\eta$ or of the angle $\theta_e$ in $R_{12}^e$ to the mixing matrix in the leptonic charged current. Secondly, due to the importance of $\mu \rightarrow e \gamma$, one has to include in the expansion of the 
chirality breaking bilinears, analogue of eq. (\ref{cb_d}), subleading terms like $(\ELlLbar \Ve) (\Ve^\dagger \Delta_e \ELeR)$, and perform the rotation to the physical basis to the corresponding order in the spurions.

The flavour changing dimension six effective operators in the lepton sector can then be written as
\begin{itemize}
\item Chirality breaking:
\begin{align}
&c_\tau \zeta_{i\tau} m_\tau \left(\bar{e}_{Li}\sigma_{\nu\rho} \tau_R\right) e F_{\nu\rho}, &
&c_\mu \zeta_{e\mu} m_\mu \left(\bar{e}_{L}\sigma_{\nu\rho} \mu_R\right) e F_{\nu\rho},\label{leptoncb}
\end{align}
\item Chirality conserving:
\begin{align}
&c^\beta_\tau \zeta_{i\tau} \left(\bar{e}_{Li}\gamma_\nu \tau_L\right) O^\beta_\nu, &
&c^\beta_\mu  \zeta_{e\mu}\left(\bar{e}_{L}\gamma_\nu \mu_L\right)O^\beta_\nu,\label{leptoncc}
\end{align}
\end{itemize}
where
\begin{equation}
O^\beta_\mu = \left(\bar{\ell}_L\gamma_\mu \ell_L\right),~\left(\bar{e}_R\gamma_\mu e_R\right),
~\left(\bar{q}_L\gamma_\mu q_L\right),~\left(\bar{u}_R\gamma_\mu u_R\right),~\left(\bar{d}_R\gamma_\mu d_R\right),
~\left(H^\dagger D_\mu H\right),
\end{equation}
and $\zeta_{ij}= U_{eL}^{3i*}U_{eL}^{3j}$ and $|U_{eL}|\simeq R_{12}^eR_{23}^e$ is the left-handed charged lepton Yukawa diagonalization matrix.
All these coefficients are  model dependent and, in principle, of similar order.

\begin{table}[tbp]
\begin{center}
\renewcommand{\arraystretch}{1.5}
\renewcommand\tabcolsep{5pt}
\begin{tabular}{ccccccccc}
\hline
$\mu\to e\gamma$ & $2.4\times10^{-12}$ &\cite{Adam:2011ch}& $\mu\to 3e$ & $1.0\times10^{-12}$ &\cite{Bellgardt:1987du}& $\mu\to e \text{ (Ti)}$ & $6.1\times10^{-13}$ &\cite{Wintz:1998rp} \\
$\tau\to e\gamma$ & $3.3\times10^{-8}$ &\cite{Aubert:2009ag}& $\tau\to 3e$ &$2.7\times10^{-8}$ &\cite{Hayasaka:2010np}&&\\
$\tau\to \mu\gamma$ & $4.3\times10^{-8}$ &\cite{Aubert:2009ag}& $\tau\to 3\mu$ & $2.1\times10^{-8}$ &\cite{Hayasaka:2010np}&&\\
\hline
\end{tabular}
\end{center}
\caption{90\% C.L. experimental upper bounds on the branching ratios of 6 LFV decays and on the $\mu\to e$ conversion rate in Titanium.}
\label{tab:lfvexp}
\end{table}

The current bounds on lepton-flavour-violating (LFV) processes are collected in table~\ref{tab:lfvexp}. Using them, bounds can be set on the above coefficients, making assumptions about the mass scale of new physics and the size of the mixing angles $\zeta_{ij}$. The relevant contributions to LFV observables are induced by the operators
\begin{align}\label{LFVeff}
\H_\text{LFV} =
&\sum_{j>i}\frac{\zeta_{ij}}{\Lambda^2}\bigg[
c_j m_{e_j} (\bar{e}_{Li}\sigma_{\mu \nu} e_{Rj}) e F_{\mu\nu}+
c_j^{l}(\bar{e}_{Li} \gamma_\mu e_{Lj})(\bar e_{Li} \gamma^\mu e_{Li})
+
c_j^{e}(\bar{e}_{Li} \gamma_\mu e_{Lj})(\bar e_{Ri} \gamma^\mu e_{Ri})\notag\\
&+
c_j^{u}(\bar{e}_{Li} \gamma_\mu e_{Lj})(\bar u \gamma^\mu u)+
c_j^{d}(\bar{e}_{Li} \gamma_\mu e_{Lj})(\bar d \gamma^\mu d)
\bigg]
+\text{h.c.}
\end{align}
From \eqref{LFVeff} one derives the predicted branching ratios for LFV decays.
The $\ell^j\to \ell^i\gamma$ branching ratio is given by
\begin{equation}
 \text{BR}( \ell^j \rightarrow  \ell^i \gamma)
= \frac{192\pi^3 \alpha}{G_F^2} \frac{|\zeta_{ij} c_j|^2}{\Lambda^4}
\;b^{ij}\,,
\end{equation}
where $b^{ij}=\text{BR}( \ell^j\rightarrow \ell^i\nu\bar\nu)$.
The $ \ell^j \rightarrow  \ell^i\bar \ell^i \ell^i$ branching ratio reads \cite{Hisano:1995cp,Arganda:2005ji}
\begin{multline}
\text{BR}( \ell^j \rightarrow 3 \ell^i)=\frac{1}{2G_F^2}b^{ij}\frac{|\zeta_{ij}|^2}{\Lambda^4}
\bigg[
e^4|c_j|^2\Big(16\log\frac{m_{ \ell_j}}{m_{ \ell_i}}-22\Big)
+\frac{1}{2}|c^l_j|^2
+\frac{1}{4}|c^e_j|^2
\\
+e^2\left(2c_j c^{l*}_j+c_j c^{e*}_j+\text{h.c.}\right)
\bigg] .
\end{multline}
For $\mu$-$e$ conversion, instead, one obtains \cite{Barbieri:1995tw,Hisano:1995cp}
\begin{equation}
\Gamma(\mu\to e)=
\frac{\alpha^3}{4\pi^2}
\frac{Z_\text{eff}^4}{Z} |F(q)|^2 m_\mu^5
\frac{|\zeta_{ij}|^2}{\Lambda^4}
\left|
(2Z+N)c_j^u+(2N+Z)c_j^d
-
2Ze^2 c_j
\right|^2 \,.
\end{equation}
In the case of ${}^{48}_{22}$Ti, one has $Z_\text{eff}=17.6$ and $|F(q^2)|=0.54$ and the conversion rate is defined as 
\begin{equation}
\text{CR}(\mu\text{ Ti}\to e\text{ Ti}) =
\frac{\Gamma(\mu\text{ Ti}\to e\text{ Ti})}{\Gamma(\mu\text{ Ti}\to \text{capture})},
\end{equation}
where the capture rate is $\Gamma(\mu\text{ Ti}\to \text{capture})=(2.590 \pm 0.012)\times10^6 ~\text{s}^{-1}$.

\begin{table}[tbp]
\begin{center}
\renewcommand{\arraystretch}{1.5}
\begin{tabular}{lcccl}
\hline
Operator & $|\widetilde c_{e\mu}|$ & $|\widetilde c_ {e\tau}|$ & $|\widetilde c_ {\mu\tau}|$ & constrained from\\
\hline
$m_{e_j} (\bar{e}_{Li}\sigma_{\mu \nu} e_{Rj}) e F_{\mu\nu}$    & 0.07 & 0.79 & 0.2 & $\ell_j\to \ell_i\gamma$\\
$(\bar{e}_{Li} \gamma_\mu e_{Lj})(\bar e_{Li} \gamma^\mu e_{Li})$ & 0.6 & 9.4 & 1.8 & $\ell_j\to 3\ell_i$ \\
$(\bar{e}_{Li} \gamma_\mu e_{Lj})(\bar e_{Ri} \gamma^\mu e_{Ri})$ & 0.9 & 13 & 2.6 & $\ell_j\to 3\ell_i$ \\
$(\bar{e}_{Li} \gamma_\mu e_{Lj})(\bar u \gamma^\mu u)$  & 0.03 &--&-- & $\mu\to e \text{ (Ti)}$\\
$(\bar{e}_{Li} \gamma_\mu e_{Lj})(\bar d \gamma^\mu d)$  & 0.03 &--&-- & $\mu\to e \text{ (Ti)}$\\
\hline
\end{tabular}
\renewcommand{\arraystretch}{1.0}
\end{center}
\caption{90\% C.L. upper bounds on the reduced coefficients defined in (\ref{eq:ctilde}). The last column lists the processes giving the strongest constraint on the respective operators.}
\label{tab:lfvbounds}
\end{table}

In table~\ref{tab:lfvbounds}, we show the bounds on the dimensionless reduced coefficients
\begin{equation}
\widetilde c_{ij} = c_{j} \times \left[\frac{3\,\text{TeV}}{\Lambda}\right]^2\left[\frac{\zeta_{ij}}{V_{ti}V_{tj}^*}\right],
\label{eq:ctilde}
\end{equation}
where the indices $i,j = e, \mu, \tau$ refer to the specific flavour transitions.
For $\Lambda = 3$ TeV some of these bounds are significant, especially from $\mu$-decay processes. Note however that the normalization of  the $\zeta_{ij}$ to the corresponding products of the  CKM matrix elements should only be taken as indicative. Note furthermore that the operators contributing to $\mu\to e$ conversion are suppressed in specific models by explicit gauge coupling factors (e.g. supersymmetry) or by small mass mixing terms (e.g. composite Higgs models).

\section{Summary and comparison of bounds}

The motivation for the occurrence of new physics in the TeV range related to EWSB  also calls for the existence of new flavour physics phenomena at similar energies.
For this to be compatible with the numerous experimental constraints, which do not allow large deviations from the CKM picture of flavour and \CP\ violation, some physical mechanism must be operative in order to keep the size of flavour transitions under control.

We have analyzed the case where this mechanism is due to an approximate $\U(2)^3$ symmetry acting on the first two generations of quarks, motivated by the pattern of masses and mixing exhibited by the quark sector of the Standard Model. In what we called Minimal $\U(2)^3$ we have crucially assumed that the breaking of  $\U(2)^3$ takes place along definite directions motivated by minimality. 
The first  conclusion that we draw from general effective field theory considerations is apparent from figures~\ref{fig:DF2fits} and \ref{fig:DF1fits}: current experimental data still allow for new flavour physics phenomena hiding at an interesting scale of about 3~TeV -- a clearly relevant energy range in BSM theories of EWSB, either perturbative or strongly interacting.

\begin{figure}[t]
\centering
\includegraphics[width=\textwidth]{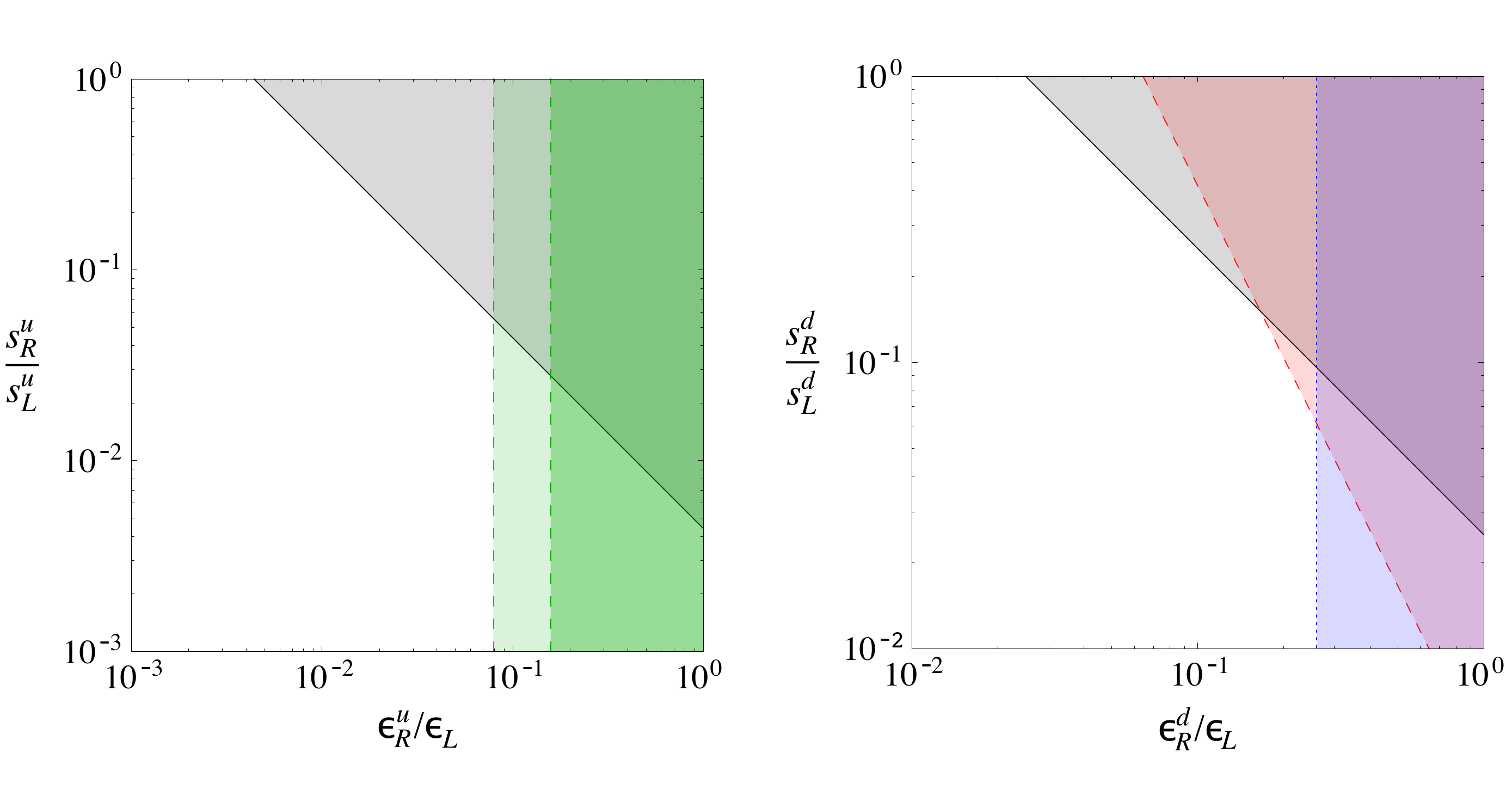}
\caption{Bounds on the free parameters of Generic $\U(2)^3$ breaking, normalized to the parameters present in Minimal $\U(2)^3$ breaking (determined by the CKM), with maximal phases.
The black solid line in both plots shows the bound from the neutron EDM (the shaded region is disfavoured at 90\% C.L.). In the left-hand plot, the green dashed lines correspond to a new physics contribution to $\Delta A_\text{\CP}$ of 50\% and 100\% of the experimental central value. The darker shaded region is disfavoured, while in the lighter region, new physics could account for the large experimental value. In the right-hand plot, the red dashed line shows the bound from $\epsilon_K$ and the blue dotted line the one from $\epsilon'$.
}
\label{fig:bounds}
\end{figure}

Generic $\U(2)^3$ introduces new parameters which do not have a correspondence with the ones of standard CKM. 
As such, even insisting on an effective scale at 3 TeV, one cannot predict the size of the extra effects introduced in Generic $\U(2)^3$. We have seen, however,  where the main constraints on the new parameters come from: in the up sector from  \CP\ asymmetries in $D$ decays and from the neutron EDM and, in the down sector, also from the neutron EDM and from \CP\ violation in the Kaon system. 
Assuming all the real model-dependent parameters to be unity and all the phases to be such as to maximize the corresponding bounds on the $\U(2)^3$ breaking parameters, to be conservative, figure~\ref{fig:bounds} compares the strength of the bounds from the different observables. Always with an effective scale at 3 TeV and barring cancellations among  phases, the size of the new breaking terms included in Generic $\U(2)^3$ have to be smaller than the corresponding ones in Minimal $\U(2)^3$.
Note that $\Delta A_\text{\CP}$ could be due to new physics compatible with $\U(2)^3$ if $\epsilon^u_R\sim0.1\epsilon_L$. However, with phases that maximize all the constraints, the bound on the up-quark CEDM then requires the angle $s^u_R$ to be more than one order of magnitude smaller than the corresponding ``left-handed'' angle $s^u_L$, whose size is determined by the CKM matrix.
If $\epsilon^d_R\lesssim0.1\epsilon_L$, bounds from the kaon system and the down quark (C)EDM are satisfied even without a considerable alignment of the $\Delta_d$ spurion.
Needless to say, in concrete models the relative strength of these bounds could vary by factors of a few.

Summarizing, it seems possible to collect flavour effects from new physics in an effective Lagrangian of the form
 \begin{equation}
\Delta{\L} = \sum_i \frac{c_i \xi_i}{(4\pi v)^2}{\O}_i ~+\text{h.c.}
\label{ideal2}
\end{equation}
where the $\xi_i$ are suitable combinations of the standard CKM matrix elements and $|c_i| = 0.2\div 1$ consistently with current experimental constraints.

The main observables that deserve attention, in view of conceivable experimental progress, are \CP\ violation in the mixing of the $B_s$ system, rates and/or asymmetries in $B$ decays, like $b\rightarrow s(d)\gamma$, $b\rightarrow s(d) \ell\bar{\ell}$, $b\to s(d)\nu\bar{\nu}$  and in $K\rightarrow \pi \nu\bar{\nu}$ decays, either charged or neutral.
In the Generic case, sizable effects are predicted in $K$ and $D$ decays. It is for example possible to attribute the recently measured \CP\ violation in $D~\rightarrow~\pi\pi, K K$ to one such breaking term consistently with any other constraint. However, no big improvement is expected here in the near future, mainly because of the large theoretical uncertainties in the SM amplitudes coming from QCD corrections at low energies.
Both Minimal and Generic $\U(2)^3$ are unlikely to give rise to any sizeable effect neither in top FCNC nor in \CP\ violation in $D$-$\bar{D}$ mixing at forseen experiments.

\begin{table}[t]
\renewcommand{\arraystretch}{1.5}
 \begin{center}
\begin{tabular}{lcccccc}
&\multicolumn{2}{c}{Chirality conserving} & \multicolumn{2}{c}{Chirality breaking}\\
\hline
& $\Delta B = 1,2$ & $\Delta S = 1,2$ & $\Delta B = 1$ & $\Delta C = 1$ 
\\\hline
$\U(3)^3$ moderate $t_\beta$ & \multicolumn{2}{l}{\hspace{.61cm}\ovalbox{$\mathbbm R\qquad\qquad\;\; \mathbbm R$}} & $\mathbbm C$ & 0 
\\
Minimal $\U(2)^3$, $\U(3)^3$ large $t_\beta$  & $\mathbbm C$ & $\mathbbm R$ & $\mathbbm C$ & 0 
\\
Generic $\U(2)^3$ & $\mathbbm C$ & $\mathbbm C$ & $\mathbbm C$ & $\mathbbm C$ 
\\\hline
 \end{tabular}
 \end{center}
\caption{Expected new physics effects in $\U(3)^3$ and both Minimal and Generic $\U(2)^3$, for chirality conserving and chirality breaking $\Delta F=1,2$ FCNC operators in the $B$, $K$, $D$ systems. $\mathbbm R$ denotes possible effects, but aligned in phase with the SM, $\mathbbm C$ denotes possible effects with a new phase, and 0 means no or negligible effects. In $\U(3)^3$ with moderate $\tan\beta$ an additional feature is that the effects in $b\to q$ ($q = d, s$) and $s\to d$ transitions are perfectly correlated.
In two Higgs doublet models at large $\tan\beta$, the possible effects in $\U(3)^3$ correspond to the ones in $\U(2)^3$ and the correlation can be broken.}
\label{tableEFT}
\end{table} 
 
A synthetic description of new physics effects in Minimal $\U(2)^3$ and in Generic $\U(2)^3$ is given in table~\ref{tableEFT}
and compared with MFV (i.e. $\U(3)^3$ at moderate or large $\tan{\beta}$). On top of the qualitative differences shown in table~\ref{tableEFT}, the size of the possible effects is significantly more constrained in $\U(3)^3$ at moderate  $\tan{\beta}$ than in all other cases.

\bigskip

A complete analysis of flavour should include also leptons. The peculiar properties of the neutrino mixing matrix, quite different from the quark one, and -- perhaps not unrelated -- the weaker information available in the lepton sector relative to the quark sector, due to the possible role of the right-handed neutrino mass matrix, make the extension to the lepton sector not straightforward. 
It is however conceivable that the observed neutrino masses and mixings have a very high-energy origin, with little impact on Fermi scale physics. Taking this view, we have assumed that the charged leptons have flavour properties which are very similar to the ones of the quarks, with a natural extension of $\U(2)^3$ to an $\U(2)_\ell\times \U(2)_e$ symmetry. One interesting feature characteristic of this symmetry, suitably broken as in \eqref{lept_spur}, is the comparison between the chirality-breaking operators responsible for the $\mu\to e$ and $\tau\to \mu$ transitions, which are suppressed respectively as $\zeta_{\mu\tau}m_{\tau}$ and $\zeta_{e\mu}m_{\mu}$.
While this can only be a qualitative picture, since at present we lack any direct information on the mixing matrix of the lepton sector, it gives nevertheless a coherent description of lepton flavour violation at the TeV scale.

\part{A composite Higgs boson}

\chapter{Composite Higgs models}\label{CHM}

The discovery of a Higgs-like scalar particle by the ATLAS and CMS experiments at the Large Hadron Collider brings new focus on the longstanding issue of electroweak symmetry breaking (EWSB). Higgsless Technicolor models, where the symmetry is broken dynamically by a fermion condensate like in QCD -- a situation already disfavored by EWPTs -- are now ruled out by direct observation.
It is now clear that the $\SU(2)_L\times \U(1)_Y$ symmetry is broken by the vacuum expectation value of a weakly interacting scalar field. Still, the exact dynamical mechanism which gives rise to the scalar potential and to its vacuum expectation value remains unknown. In particular, a relevant question is if this mechanism is strongly or weakly interacting.

Here we are concerned with the possibility that the newly discovered particle is a composite state coming from a new strongly interacting sector at the TeV scale, and in particular with the view that tries to explain a natural Fermi scale in terms of the Higgs particle as a composite pseudo-Nambu-Goldstone boson (PNGB) \cite{Giudice:2007fh,Kaplan:1983fs,Georgi:1984af,Contino:2003ve,Agashe:2004rs}.

In this chapter, following mainly the discussion of ref. \cite{Contino:2010rs}, we introduce the tools that are needed for the description of a strongly interacting scalar field with non-standard couplings stemming from the composite dynamics. After a model-independent review of the mechanism of electroweak symmetry breaking, in section \ref{CHM/PNGB} we introduce the basic concepts of a light PNGB Higgs and show how naturalness arguments require a low scale of compositeness. The vector and fermion resonances of the strongly interacting sector are described in section \ref{CHM/partialcompositeness} using the formalism of partial compositeness.

The phenomenological implications of the compositeness paradigm following from experimental data, and in particular from the measurement of the mass of the Higgs boson, are explored in the next chapter.

\section{The electroweak chiral Lagrangian}\label{CHM/chiral}

Let us make a step backward, and consider in more detail the spontaneous breaking of the electroweak symmetry. To understand exactly what the role of the Higgs boson is, and to what extent it can be non-standard, we want to describe EWSB in a model-independent way, i.e. without specifying any concrete mechanism for it.

The most general gauge-invariant Lagrangian that we can write in terms of the Goldstone bosons $\pi^a$ of the breaking $\SU(2)_L\times \U(1)_Y\to \U(1)_{\rm em}$ and the SM fields, the {\it electroweak chiral Lagrangian}, reads
\begin{equation}\begin{aligned}\label{chiral}
\L_{\pi} &= \frac{v^2}{4}\tr\left[(D_{\mu}\Sigma)^{\dag}D^{\mu}\Sigma\right] + a_0v^2\tr\left[\Sigma^{\dag}D_{\mu}\Sigma\,\sigma^3\right]^2\\
&\quad + v\sum_{i,j}\left(Y_{ij}^u \bar u_L^i \Sigma u_R^j + Y_{ij}^d \bar d_L^i\Sigma d_R^j\right) + {\rm h.c.} + \;\cdots
\end{aligned}\end{equation}
where the dots indicate terms of higher order in the derivatives, and the chiral field is
\begin{equation}\label{Sigma}
\Sigma = \exp\left(\frac{i\sigma^a\pi^a}{v}\right),\qquad D_{\mu}\Sigma = \partial_{\mu}\Sigma - igW_{\mu}^a\frac{\sigma^a}{2}\Sigma + ig'B_{\mu}\Sigma\frac{\sigma^3}{2}.
\end{equation}
$\Sigma$ transforms under the local $\SU(2)_L\times \U(1)_Y$ as $\Sigma\mapsto U_L(x)\Sigma \,U_Y^{\dag}(x)$, where
\begin{equation}\label{Sigmatransformations}
U_L(x) = \exp\left(i\epsilon^a_L(x)\frac{\sigma^a}{2}\right),\qquad U_Y(x) = \exp\left(i\epsilon_Y(x)\frac{\sigma^3}{2}\right).
\end{equation}
It is clear from \eqref{Sigma} and \eqref{Sigmatransformations} that these transformation properties can be embedded in the larger global $\SU(2)_L\times \SU(2)_R$ group, under which $\Sigma\mapsto U_L\Sigma \,U_R^{\dag}$.
Since the second term in \eqref{chiral} breaks explicitly the custodial symmetry and contributes in a sizable way to the $T$ parameter, we neglect it in the following.

Notice that in the unitary gauge where $\Sigma\equiv 1$, the chiral Lagrangian reproduces exactly the mass term for the gauge fields and the fermions.
Thus EWSB happens as a consequence of the spontaneous breaking of the global symmetry $\SU(2)_L\times \SU(2)_R\to \SU(2)_V$, as $\langle\Sigma\rangle = 1$. Here the exact mechanism through which that happens is not specified: in the Standard Model at the minimum of the Higgs potential $\langle\pi^a\rangle=0$, while in Technicolor models the symmetry is broken dynamically by some strong interaction.

We can now ask ourselves how \eqref{chiral} is modified if we introduce one additional scalar degree of freedom -- the Higgs boson. The most general, gauge invariant, Lagrangian reads
\begin{equation}\begin{aligned}\label{scalar}
\L_{\pi,h} &= \frac{1}{2}(\partial_{\mu} h)^2 - V(h) + \frac{v^2}{4}\tr\left[(D_{\mu}\Sigma)^{\dag}D^{\mu}\Sigma\right]\Big(1 + 2a\frac{h}{v} + b\frac{h^2}{v^2} + \cdots\Big)\\
&\quad + v\sum_{i,j}\left(Y_{ij}^u \bar u_L^i \Sigma u_R^j + Y_{ij}^d \bar d_L^i\Sigma d_R^j\right)\Big(1 + c\frac{h}{v} + \cdots\Big) + {\rm h.c.}\,,
\end{aligned}\end{equation}
where $a,b,c$ are free parameters and we neglected terms of higher order in $h/v$ which have dimension greater than 4.

The Standard Model Higgs boson has $a=b=c=1$. In this case $h$ and the Goldstone bosons $\pi^a$ form a linear representation of $\SU(2)_L\times \SU(2)_R$ and can be viewed as the components of the Higgs doublet $H$. In a generic composite Higgs model these couplings will deviate from their standard value.

\subsection{$WW$ scattering and unitarity}\label{CHM/chiral/unitarity}
The chiral Lagrangian \eqref{chiral} is not renormalizable. This is easily seen as the kinetic term $(D_{\mu}\Sigma)^2$ contains derivative interactions with an arbitrary number of Goldstone bosons, suppressed by a suitable power of $v$, which are operators of dimension greater than 4.

A consequence of this fact is the well known non-unitarity of the longitudinal $WW$ scattering amplitude. Indeed, the equivalence theorem tells us that at sufficiently high energies $E\gg m_W$ the amplitude for the longitudinal polarization of $W$ bosons equals the one for Goldstones,
\begin{equation}
\A(W^+_LW^-_L\to W^+_LW^-_L) = \A(\pi^+\pi^-\to\pi^+\pi^-) = \frac{1}{v^2}(s+t),
\end{equation}
hence the unitarity bound is saturated at an energy $\sqrt{s}\simeq 4\pi v\simeq 3\,{\rm TeV}$ for the $s$-wave scattering.

\begin{figure}[t]
\centering%
\hfill%
\includegraphics{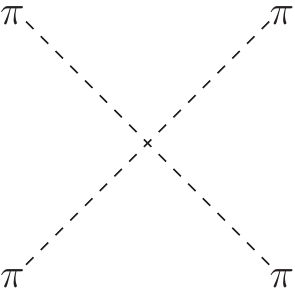}\hfill%
\includegraphics{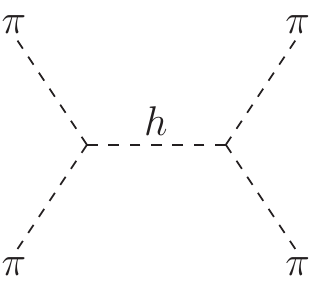}\hfill%
\includegraphics{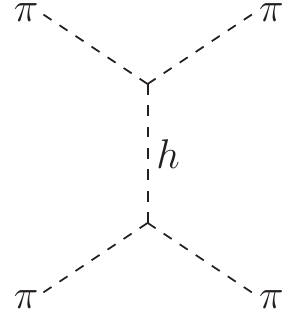}\hfill\hfill
\caption{Feynman diagrams contributing to $\pi\pi\to\pi\pi$ scattering at energies below the compositeness scale.\label{higgsunitarity}}
\end{figure}
In the Standard Model unitarity is restored at tree-level by the exchange of a Higgs particle. Using the Lagrangian \eqref{scalar} for a generic scalar particle, from the diagrams of figure~\ref{higgsunitarity} we get the amplitude for $\pi\pi$ scattering at tree-level
\begin{equation}\label{pipi}
\A(\pi\pi\to\pi\pi) = \frac{s+t}{v^2} - \frac{a^2}{v^2}\left(\frac{s^2}{s - m_h^2} + \frac{t^2}{t - m_h^2}\right) = \frac{s+t}{v^2}(1-a^2) + \Ord\Big(\frac{m_h^2}{E^2}\Big).
\end{equation}
\begin{figure}[b]
\centering%
\hfill%
\includegraphics{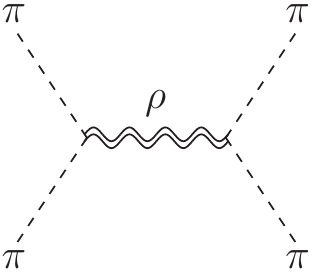}\hfill%
\includegraphics{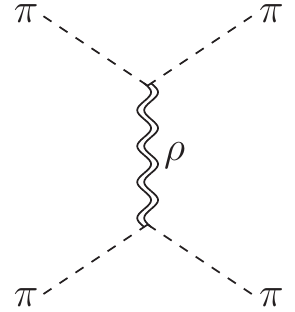}\hfill\hfill
\caption{Feynman diagrams unitarizing $\pi\pi\to\pi\pi$ scattering at the compositeness scale.\label{fig2comp}}
\end{figure}
If the Higgs coupling has its SM value $a = 1$, the part of the amplitude that grows with the energy is exactly canceled, and the theory remains unitary at every scale.

On the contrary, in Technicolor models of strong EWSB unitarity is restored by a new strong interaction that sets in at the scale $\Lambda \simeq 4\pi v$. Composite resonances of the strong sector are excited at that scale; the exchange of these resonances contributes to the scattering amplitudes, and helps to preserve unitarity. Exact unitarity is recovered if the whole tower of resonances of the strong sector is taken into account. This is exactly what happens in QCD, where the exchange of the $\rho$ meson -- and of its resonances, at higher energies -- unitarizes pion scattering.

If the Higgs boson is a composite object, it can be considered as a bound state of a strongly interacting sector. This situation, from the point of view of the unitarity bounds, is an intermediate between the standard Higgs model, and a theory of strong EWSB. In general, if the Higgs is composite, the parameters of \eqref{scalar} will differ from 1, so that the scattering amplitude \eqref{pipi} will still grow as $s+t$, eventually saturating the unitarity bound. But this happens now at a higher scale with respect to the higgsless case, namely at $\sqrt{s}\simeq 4\pi v / \sqrt{1-a^2}$. At this point the strongly interacting sector provides for exact unitarity, as before.

For completeness, one can consider also the scattering amplitudes involving the Higgs boson and fermions. Again from \eqref{scalar} we get
\begin{align}
\A(\pi\pi\to h h) &= \frac{s}{v^2}(b - a^2),\label{pih}\\
\A(\pi\pi\to\bar\psi_i\psi_i) &= m_{\psi}\frac{\sqrt{s}}{v}(1 - ac),\label{pifermions}
\end{align}
and we see that the amplitudes are unitarized only for $a = b = c = 1$, which are the SM Higgs couplings. For any departure of these values, new degrees of freedom are needed at some higher energy scale, in order to have a consistent theory.

\section{The Higgs as a pseudo-Nambu-Goldstone boson}\label{CHM/PNGB}

From now on we assume that the Higgs boson is part of a composite sector. In other words, this means that its couplings are not the ones of a standard $\SU(2)_L$ doublet and above some energy scale $\Lambda$ the unitarity bound is saturated. There the theory becomes strongly interacting, with a coupling constant which can be assumed to be $1\lesssim g_{\rho}\lesssim 4\pi$, and effective operators of this strong dynamics excite a tower of resonances.

These composite resonances, if not too heavy, should in principle be observable at collider experiments \cite{Barbieri:2009tx}. 
In general, unless some mechanism is at work which generates a separation of scales, one expects the compositeness scale to roughly coincide with the Fermi scale, $\Lambda\sim 4\pi v$, and the vector and fermion resonances, which have to appear below this energy scale, to have masses which are comparable with the Higgs mass. The direct searches at the LHC are currently sensitive to masses in the 500--700 GeV range for composite fermions \cite{CMS:2012ab,Chatrchyan:2012vu,Chatrchyan:2012af,ATLAS:2012hpa}, and up to a few TeV for composite vector states \cite{Chatrchyan:2013qha,Chatrchyan:2013lga,ATLAS:2012qjz,ATLAS:2013zzj}, and have seen no evidence for new particles up to now.

It is therefore clear that the Higgs boson has to be by far the lightest state of the composite sector. This situation reminds what happens in QCD, where one has a typical scale of strong interactions $\Lambda_{\rm QCD}$, a set of scalar states -- the pions -- which are lighter than this scale, and all the other resonances -- first of all the rho mesons -- with a much higher mass. This is explained by the fact that the pions are the Nambu-Goldstone bosons associated with the spontaneous breaking of the approximate $\SU(2)_L\times \SU(2)_R$ chiral symmetry of QCD. In a very similar fashion, one can assume that also the Higgs is a (pseudo-)Nambu-Goldstone boson associated to some global symmetry breaking of the strong sector.

Consider a global symmetry group $G$ which is spontaneously broken down to a subgroup $H$ at some scale $f > v$. The Nambu-Goldstone bosons of this breaking live in the coset space $G/H$ and -- if the coset is a symmetric space -- transform under $G$ linearly as
\begin{equation}\label{shift}
\pi^{\hat a}\mapsto \pi^{\hat a} + \epsilon^{\hat a},\qquad\qquad \exp\left(i\epsilon^{\tilde a} T^{\tilde a} + i\epsilon^{\hat a} T^{\hat a}\right)\in G,
\end{equation}
where $T^{\tilde a}$ are the unbroken generators of $H$, and $T^{\hat a}$ are the broken generators. Since the Lagrangian has to be invariant under the full group $G$, the shift symmetry \eqref{shift} forbids any non-derivative interaction among the $\pi$'s, including a mass term $m_{\pi}^2 \pi^2$, and the Nambu-Goldstone bosons are thus massless \cite{Nambu:1960tm,Goldstone:1961eq}.
If a gauge symmetry is involved, the number of massless degrees of freedom that eventually appear in the breaking corresponds to the total number of broken generators minus the number of broken gauge generators, since the Nambu-Goldstone bosons associated with the latter are eaten-up by the gauge fields.
For simplicity, let us assume that the only gauge fields that are relevant are the ones of the Standard Model. Since the $\SU(2)_L\times \U(1)_Y$ generators remain unbroken at this stage, we have $n = \dim(G) - \dim(H)$ Nambu-Goldstone bosons associated with the breaking $G\to H$.

\begin{figure}[t]
\centering%
\includegraphics[width=.4\textwidth]{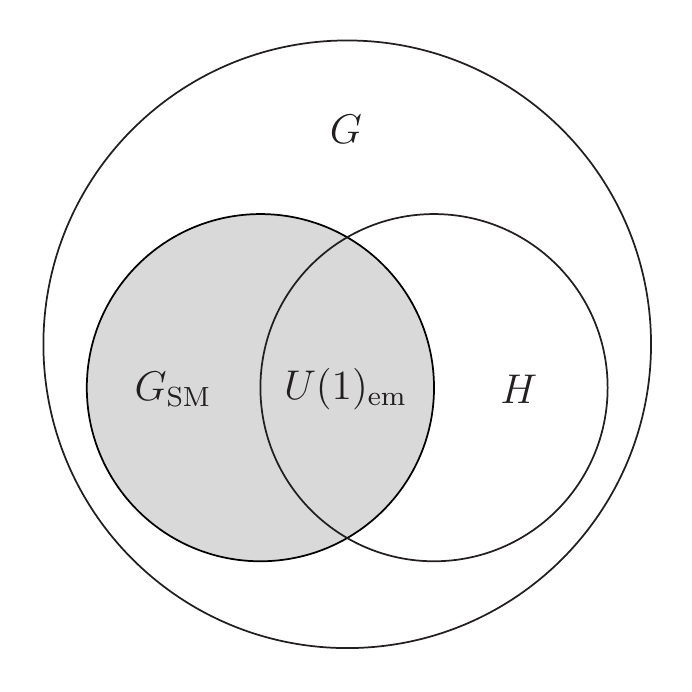}
\caption{Structure of the symmetry breakings in a generic composite Higgs model. The misalignment between the gauged $G_{\rm  SM}$ and $H$ causes the global breaking $G\to H$ to trigger also EWSB.}
\end{figure}

Let us make some observations:
\begin{itemize}
\item At least four Goldstone bosons are needed to reproduce the $\SU(2)_L$ Higgs doublet of the Standard Model.
\item In analogy with the chiral Lagrangian \eqref{chiral} we have to impose that the strong sector respects a custodial $\SU(2)_L\times \SU(2)_R\simeq \SO(4)$ symmetry. Without this assumption, one would have large contributions -- in principle of order 1 -- to the $T$ parameter. Here one comment can be made: if the Higgs appears in a complete representation of the custodial group, even if the strong sector is not symmetric under it, the contributions to $T$ are suppressed by a factor of $v^2/f^2$ as in \eqref{deltaT}. To avoid a strong lower bound on the scale $f$ from EWPT we will assume a custodial invariant strong sector in the following.
\item The global group $G$ must contain the Standard Model gauge group $\SU(2)_L\times \U(1)_Y$ as an embedding, since the Higgs must couple to the gauge fields. The gauging of only a subgroup of $G$ breaks explicitly the global symmetry. If the SM group is contained in $G$ but not in $H$, i.e. there is a misalignment between $G_{\rm SM}$ and $H$, the $G\to H$ breaking triggers also EWSB.
\end{itemize}

\subsection{Minimal composite Higgs models}\label{CHM/PNGB/MCHM}

The minimal group which satisfies the above properties is $SO(5)\times \U(1)_X$ \cite{Agashe:2004rs} (see \cite{Contino:2010rs} for a review). In these minimal models there are exactly four Goldstone bosons -- i.e. only one Higgs doublet -- living in the coset $SO(5)/SO(4)$. The extra $\U(1)_X$ is needed in order to get the right hypercharge assignment for all the SM fields, with $Y = T_{3R} + X$, and does not take part in the breaking.

The Nambu-Goldstone bosons can be written as (see \eqref{Sigma} for comparison)
\begin{equation}\label{GoldstoneSigma}
\Sigma(x) = \Sigma_0 \exp\Big(\frac{i T^{\hat a} \pi^{\hat a}(x)}{f}\Big),
\end{equation}
where $T^{\hat a}$ are the broken generators of $SO(5)$ and $\Sigma_0$ is a constant vector of unit modulus parametrizing the direction of breaking. In a suitable basis where the first 4 coordinates correspond to the $SO(4)_{\rm SM}\supset G_{\rm SM}$ generators and $\Sigma_0 = (\vec 0,1)$, \eqref{GoldstoneSigma} can be rewritten as
\begin{equation}\label{SigmaSO5decomposition}
\Sigma = \Big(\frac{\vec\pi}{\pi}\sin(\pi/f), \cos(\pi/f)\Big),
\end{equation}
where $\vec\pi = (\pi^1,\pi^2,\pi^3,\pi^4)$ and $\pi = \sqrt{\vec\pi^2}$. One sees that in a background where $\langle\vec\pi\rangle\neq 0$ there is a misalignment of an angle $\langle\pi\rangle/f$ between $SO(4)_{\rm SM}$ and the preserved $SO(4)$, so that $G_{\rm SM}$ is broken together with $SO(5)$.

Integrating out the strong dynamics, an effective Lagrangian for the vector fields is generated, in complete analogy with \eqref{vacpol},
\begin{equation}\label{effLagrangianSO5}
\L_{V, \rm eff} = -\frac{1}{2}\Pi_{0}^{\mu\nu}(q^2)A_{\mu}^{\hat a}A_{\nu}^{\hat a} - \frac{1}{2} \Pi_{1}^{\mu\nu}(q^2)A_{\mu}^{\tilde a}A_{\nu}^{\tilde a} - \frac{1}{2}\Pi_X^{\mu\nu}(q^2)X_{\mu}X_{\nu},
\end{equation}
where $\Pi_{\alpha}^{\mu\nu}(q^2) = \Pi_\alpha(q^2)(g^{\mu\nu} - q^{\mu} q^{\nu}/q^2)$, $A_{\mu}^{\hat a}$ and $A_{\mu}^{\tilde a}$ are the fields associated with the broken and unbroken generators of $SO(5)$, respectively, and $X_{\mu}$ are the $\U(1)_X$ vectors.
The $\Pi's$ at vanishing external momentum correspond to a mass terms for the vectors, the first derivatives determine the coupling constants. At all orders in $q^2$, \eqref{effLagrangianSO5} describes also a series of composite resonances excited by the strong dynamics, with masses corresponding to the poles of the form factors $\Pi(q^2)$.\footnote{In the large $N$ limit one can explicitly check that for each $\Pi(q^2)$ there is an infinite tower of narrow resonances with masses $m_{\rho_n}$.}

Only the $A_{\mu}^{\hat a}$ fields which are associated with the broken generators can get a mass term by spontaneous symmetry breaking. For the $\SU(2)_L\times \U(1)_Y$ gauge fields, \eqref{effLagrangianSO5} at $q^2 = 0$ can be identified with the leading term of the chiral Lagrangian for $\Sigma$ in a derivative expansion,
\begin{align}
\left. \L_{{V_{\rm EW}},\rm eff}\right |_{q^2 = 0} &\simeq \frac{f^2}{4}\tr\big[\Sigma^{\dag}\hat A_{\mu}\hat A^{\mu}\Sigma\big]\notag\\
&= \frac{f^2}{4}\sin^2\Big(\frac{\pi}{f}\Big)\tr\Big[\Big(W_{\mu}\!\cdot\!\hat{\mathcal{H}} + \hat{\mathcal{H}}\!\cdot\!\sigma_3\, YB_{\mu}\Big)^{\dag}\Big(W_{\mu}\!\cdot\!\hat{\mathcal{H}} + \hat{\mathcal{H}}\!\cdot\!\sigma_3\, YB_{\mu}\Big)\Big],
\end{align}
where $\hat A_{\mu}$ denotes collectively the part of the SM gauge fields corresponding to the broken $SO(5)$ generators, $\hat{\mathcal{H}} = \pi^aT^a/\pi$, and the last equality follows from the decomposition \eqref{SigmaSO5decomposition}. When $\pi$ gets a vev, in order to recover the correct form \eqref{chiral} it must be
\begin{align}
v &= f \sin\frac{\langle\pi\rangle}{f}, & \xi\equiv \frac{v^2}{f^2} = \sin^2\frac{\langle\pi\rangle}{f}.
\end{align}
Expanding the pion field around its vev, $\pi = \langle\pi\rangle + h$, one obtains
\begin{align}
f^2\sin^2\frac{\langle\pi\rangle + h}{f} &= f^2\Big[\sin^2\frac{\langle\pi\rangle}{f} + 2\sin\frac{\langle\pi\rangle}{f}\cos\frac{\langle\pi\rangle}{f}\Big(\frac{h}{f}\Big) + \Big(1 - 2\sin^2\frac{\langle\pi\rangle}{f}\Big)\Big(\frac{h}{f}\Big)^2 + \cdots\Big]\notag\\
&= v^2 + 2v\sqrt{1 - \xi}\,h + (1-2\xi)\, h^2 + \cdots,
\end{align}
which in turn gives also an explicit form for the Higgs couplings of \eqref{scalar},
\begin{align}
a &= \sqrt{1-\xi}, & b &= 1 - 2\xi.
\end{align}

For completeness, we report also the correction to the gauge couplings coming from the strong dynamics,
\begin{align}
\frac{1}{g^{2}} &= \frac{1}{g_0^2} + \Pi_0'(0), & \frac{1}{g'^2} &= \frac{1}{g'^2_0} + \Pi_0'(0) + \Pi_{X}'(0).
\end{align}

In a similar way one can derive the effective action for the top quark \cite{Agashe:2004rs,Contino:2006qr}
\begin{align}\label{effLagrangianfermions}
\L_{t, \rm eff} &= \bar q_L\,\slashed{p}\,\Pi_L(p^2) q_L + \bar t_R\,\slashed{p}\,\Pi_R(p^2) t_R + \Pi_m \bar q_L\hat{\mathcal{H}}\,t_R,
\end{align}
where
\begin{align}
\Pi_L &= \Pi_{0,L} + \Pi_{1,L}\cos(\pi/f),\label{topformfactorL}\\
\Pi_R &= \Pi_{0,R} + \Pi_{1,R}\cos(\pi/f),\label{topformfactorR}\\
\Pi_m &= \Pi_{1,m}\sin(\pi/f).\label{topformfactorM}
\end{align}
The poles of these form factors determine the masses of the fermionic resonances of the strong sector.
In the limit $p = 0$ one finds
\begin{equation}\label{topmassSO5}
m_t \simeq \frac{\xi\Pi_{1,m}}{\sqrt{(\Pi_{0,L} + \Pi_{1,L})(\Pi_{0,R} - \Pi_{1,R})}},
\end{equation}
while the anomalous coupling of the Higgs to fermions is
\begin{equation}
c = \sqrt{1-\xi}.
\end{equation}

\subsection{The Higgs potential and symmetry breaking}\label{CHM/PNGB/potential}

If the symmetry under $G$ were exact, the Higgs potential would vanish because of the shift symmetry \eqref{shift}. As a consequence, one would have no EWSB and a perfectly massless Higgs boson.
It is thus clear that the global symmetry $G$ has to be only approximate. Its explicit breaking, as observed above, comes from the fact that only a subgroup -- in our case just the electroweak group of the SM -- is gauged. The effective potential of the Higgs field (which will not respect the full symmetry under $G$) is thus generated through radiative corrections by the Standard Model fields, mainly vector bosons and the top quark, which are the ones to which it couples most strongly. This is the same mechanism that generates the mass difference between the pions in QCD.

There are in general two conditions that have to be satisfied by the potential:
\begin{itemize}
\item[$(i)$] the correct value of the Higgs vev has to be reproduced,
\item[$(ii)$] the Higgs boson mass has to be $m_h = 125$ GeV.
\end{itemize}
This is where the hierarchy problem appears in composite Higgs models: in a model with generic order one parameters, since the only energy scale in the theory is $f$, the EWSB scale that is generated will also be of the same order, up to some model-dependent factor. In order to have a significant splitting between the two scales, a certain amount of fine-tuning between different terms in the potential has to be introduced. In other words, if the groups $H$ and $G_{\rm SM}$ are not exactly aligned, radiative corrections tend to spoil completely the alignment, generating EWSB at the scale $f$ and a Higgs mass of the same order. The misalignment between the two groups has to be fine-tuned to the weak scale, and the parameter
$\xi = v^2/f^2$
is a good measure of the amount of tuning. The Higgs mass will then be proportional to this quantity.

In the minimal $SO(5)/SO(4)$ model one can easily derive the one-loop Coleman-Weinberg potential for the Nambu-Goldstone Higgs from \eqref{effLagrangianSO5} and \eqref{effLagrangianfermions}, in terms of the form factors. Neglecting hypercharge the vector contribution reads
\begin{equation}
V_{W}(h) = \frac{9}{2}\int\frac{d^4 q}{(2\pi)^4}\log\Big(1 + \frac{1}{4}\frac{\Pi_1(q^2)}{\Pi_0(q^2)}\sin^2\frac{h}{f}\Big).
\end{equation}
Although giving $h$ a mass, thus breaking the Goldstone symmetry, this form of the potential doesn't trigger EWSB since it has a minimum for $h = 0$. The reason is that the $W_{\mu}$ field alone, ignoring the small contribution from $g'$, respects the custodial symmetry and can't induce a $\SU(2)_L\times \SU(2)_R\to \SU(2)_V$ breaking.

The leading contribution is however the one from the top quark, which reads \cite{Agashe:2004rs,Contino:2006qr}
\begin{equation}
V_t(h) = -2 N_c\int\frac{d^4p}{(2\pi)^4}\Big[\log\Pi_L + \log\Big(p^2\Pi_L\Pi_R - \Pi_m^2\Big)\Big],
\end{equation}
where $N_c = 3$ is the number of colours.
Expanding the potential for small $\Pi$'s, after substituting the expressions (\ref{topformfactorL}), (\ref{topformfactorR}), yields the simple approximate form
\begin{equation}
V(h) \simeq \alpha\cos\frac{h}{f} - \beta\sin^2\frac{h}{f},
\end{equation}
where $\alpha$ and $\beta$ are functions of the form factors. Now, because of the two contrasting contributions, the potential has a nontrivial minimum at
\begin{equation}
\xi = \sin^2\frac{\langle h\rangle}{f} = 1 - \Big(\frac{\alpha}{2\beta}\Big)^2
\end{equation}
if $\alpha\leq 2\beta$. This shows explicitly the previous statement: a sizable splitting between the scales $v$ and $f$, i.e. a small $\xi$, requires a tuning of $\alpha \approx 2\beta$. In a natural theory with order one parameters, $\xi$ will also be of order one, and $f$ of the order of the weak scale.

The Higgs mass is
\begin{equation}
m_h^2 = \left. \frac{\partial^2 V(h)}{\partial h^2}\right|_{\langle h\rangle} = \frac{2\beta\xi}{f^2}.
\end{equation}
One can check that
\begin{equation}
\beta \simeq 2N_c\int \frac{d^4p}{(2\pi)^4}\frac{1}{-p^2}\frac{\Pi_{1,m}^2}{(\Pi_{0,L} + \Pi_{1,L})(\Pi_{0,R} - \Pi_{1,R})} \approx \frac{N_c}{8\pi^2} \frac{m_t^2}{\xi} m_{\psi}^2,
\end{equation}
where $m_\psi$ is the mass of the lightest resonance that cuts-off the integral, and where we used the expression \eqref{topmassSO5} for $m_t$. Finally, one finds
\begin{equation}\label{higgsmassSO5}
m_h^2 \simeq N_c\frac{m_t^2 m_{\psi}^2}{4\pi^2 f^2} \equiv N_c \frac{m_t^2 Y^2}{4\pi^2},
\end{equation}
where the paramter $Y$ is defined as $m_{\psi} = Y f$.

\section{Partial compositeness}\label{CHM/partialcompositeness}

For phenomenological purposes, it is often sufficient to consider only the lightest resonances of each type,\footnote{Sometimes one may want to include also the next-to-lightest resonances. These are required, for instance, to get a finite prediction for the Higgs potential \cite{Redi:2012ha}.} which are probably also the only ones that are potentially observable at the LHC. In this limit, an efficient way to describe the coupling of this new strong sector to the Standard Model fields is the one of {\it partial compositeness}\cite{Kaplan:1991dc,Contino:2006nn}.

In this picture, every SM field has at least one composite partner with the same $\SU(2)_L\times \U(1)_Y$ quantum numbers. Bilinear mass mixing terms between operators with the same quantum numbers are generated by the renormalization-group running. Due to the presence of these mixing terms, the physical mass eigenstates are a superposition of elementary and composite fields. This phenomenon is completely analogous to the {\it photon-rho mixing} in QCD.

The situation can be described as follows. Let us denote collectively by $A_{\mu}$ the elementary gauge fields, with coupling $g_{\rm el}$, and with $\rho_{\mu}$ the composite vector resonances. The Lagrangian for vector fields, including all the possible quadratic terms, becomes
\begin{equation}\label{partialV}
\L_V = -\frac{1}{2}\tr\left[\rho_{\mu\nu}\rho^{\mu\nu}\right] - \frac{1}{2}\tr\left[F_{\mu\nu}F^{\mu\nu}\right] + \frac{m_{\rho}^2}{2}\left(\rho_{\mu}^2 - 2 \tan\phi\, \rho_{\mu}A^{\mu} + \tan\phi^2 A_{\mu}^2\right),
\end{equation}
where the mixing angle $\phi$ is a small parameter.
The size of the $A_{\mu}^2$ term has been fixed by requiring that there is a massless eigenstate which corresponds to the physical gauge boson. Indeed, diagonalizing \eqref{partialV} yields the two eigenstates
\begin{align}
W_{\mu} &= \cos\phi\, A_{\mu} + \sin\phi\, \rho_{\mu}, & 
W_{\mu}^* &= -\sin\phi\, A_{\mu} + \cos\phi\, \rho_{\mu},
\end{align}
with masses respectively $m_W = 0$ and $m_* = m_{\rho} / \cos\phi\simeq m_{\rho}$.
The coupling between $W_{\mu}$ and the composite Higgs field $H$ is obtained by projection onto the strong sector, and is
\begin{equation}
g^2 = g_{\rho}^2\sin^2\phi,
\end{equation}
where $g$ is the SM weak coupling. With the additional condition that the gauge coupling in the field strength $W_{\mu\nu}$ be equal to $g$, we get
\begin{align}
\tan\phi &= \frac{g_{\rm el}}{g_{\rho}}, & g &= \frac{g_{\rm el}g_{\rho}}{\sqrt{g_{\rm el}^2 + g_{\rho}^2}},
\end{align}
and the smallness of the mixing angle $\phi$, imposed by electroweak precision measurements, implies $g_{\rho} \gg g_{\rm el}\simeq g$.

The same setup can be translated to the fermions, although with some complication due to the inclusion of flavour. The coupling of quarks and leptons to the Higgs - and therefore to the strongly interacting sector - suggests that also fermionic resonances are excited close to the compositeness scale. In the partial compositeness approach, these are described by a set of heavy, composite fermions $Q_i$ with the same $\SU(3)\times \SU(2)_L\times \U(1)_Y$ quantum numbers of the standard fermions, with which they mix through bilinear mass terms.
Since we want these fermions to have masses above the electroweak scale, we will assume that they are vector-like with Dirac masses $m_{Q_i}$. The composite Lagrangian \eqref{partialV} is then extended as follows
\begin{equation}\begin{aligned}\label{compositefermions}
{\L}_c &= \L_V + |D_{\mu}\mathcal{H}|^2 + \bar Q_i(i\slashed{D} - m_{Q_i})Q_i + Y^{ij} \bar Q_i \mathcal{H} Q_j + {\rm h.c.},
\end{aligned}\end{equation}
where $Y$ is some Yukawa coupling of the strong sector involving fermions with the appropriate quantum numbers, and we use the matrix notation $\mathcal{H} = (H^c, H)$ for the Higgs.
In models where the Higgs is a pseudo-Goldstone boson, if the mass of the fermions and the strong Yukawa interactions come from the same kind of Yukawa-like composite operator after breaking of the global symmetry, one can take the couplings $Y^{ij}$ to be proportional to the parameter $Y$ in \eqref{higgsmassSO5}, and approximately identify the heavy fermion masses with $m_{\psi} = Y f$, up to a model dependent factor of order one. The implications of this connection between the Higgs mass and flavour physics will be discussed in chapter~\ref{CHMbounds}.

The mixing Lagrangian for the fermions can be written as
\begin{equation}\label{fermionmixings}
\L_{{\rm mix}, f} = m_{Q_i} \bar Q_i\lambda_q^{ij} q_j,
\end{equation}
where $\lambda_q$ is some matrix of dimensionless parameters in flavour space and we denoted generically with $q_i$ the elementary fermions. Using a singular-value-decomposition
\begin{equation}\label{mixinglambda}
\lambda_q = U_q^{\dag}\cdot \lambda_q^{\rm diag} \cdot V_q,
\end{equation}
the diagonalization of the mass matrix for light fermions leads to the zero mass eigenstates
\begin{align}
\tilde q_i &= \cos\varphi_i q_i + \sin\varphi_i\, U_{ij}Q_j,
\end{align}
which correspond to the SM chiral fermions before EWSB. At the same time, in the $v=0$ limit, the remaining states have mass $m_{Q_i}$ or $m_{Q_i}\cos\varphi_i$, respectively if they mix or do not mix with the elementary fermions. The mixing angles are defined as $\tan\varphi_i = \lambda_i^{\rm diag}$.
Notice how the matrices $V_f$ on the right-hand side of \eqref{mixinglambda} are unphysical and can be eliminated rotating the elementary fields.

The SM Yukawas, which couple the physical quarks $\tilde q$ to the composite Higgs boson, and give them a mass after EWSB, are obtained again by projection onto the strong sector, and read
\begin{align}\label{SMyukawa}
y_u^{ij} &= \sum_{k,l}\sin\varphi_{q_i}U_{ik}^qY_U^{kl}(U_{lj}^u)^{\dag} \sin\varphi_{u_j}, & y_d^{ij} &= \sum_{k,l}\sin\varphi_{q_i}U_{ik}^q Y_D^{kl}(U_{lj}^d)^{\dag} \sin\varphi_{d_j},
\end{align}
where $Y_U$ and $Y_D$ are the strong Yukawa couplings for the up-type and down-type quark partners, respectively.

It is clear that the parameters $Y$ and $\lambda$ are subject to the constraint of reproducing the correct flavour structure at the electroweak scale, which - as thoroughly discussed in the previous chapters - is in general not a straightforward property. There are different approaches to this problem, which will be the topic of the next chapter.

\subsection{Effective Lagrangians for partial composite fermions}\label{CHM/partialcompositeness/fermions}

The set of composite fermion partners is not uniquely determined. Since the strong sector is invariant under the unbroken global group $H$, only complete representations of $H$ are allowed. In particular they have to come at least in complete representations of the custodial group $\SU(2)_L\times \SU(2)_R\simeq \SO(4)$.
Moreover, in theories where the Higgs is a PNGB, as a consequence of the larger global symmetry of the Lagrangian the heavy fermions have to form also complete representations of $G$, although mass splittings of order $f$ are allowed inside these representations as a consequence of the spontaneous breaking $G\to H$. There are however different choices that can be made concerning the specific representation of $H$ (or $G$).

The choice of the fermion representations has important implications for the electroweak precision constraints. Here we classify the fermions under representations of the custodial $\SU(2)_L\times \SU(2)_R\times \U(1)_X$, the extra $\U(1)_X$ being needed in order to get the right hypercharges for the quarks. We will consider the three simplest possible cases, which are also most frequently considered in the literature.
\subsubsection{The doublet model}
The elementary $\SU(2)_L$ quark doublets $q_L$ mix with composite vector-like $\SU(2)_L$ doublets $Q$, one per generation. The elementary quark singlets, $u_R$ and $d_R$, couple both to an $\SU(2)_R$ doublet $R = (U~D)^T$.
For concreteness, the part of the Lagrangian involving fermions reads\footnote{\label{Ytildefootnote}Note that we have omitted ``wrong-chirality'' Yukawa couplings like $\tilde Y^{ij}{\rm tr}[\bar Q_R^i\mathcal{H}R_L^j]$ for simplicity. They are not relevant for the tree-level electroweak and flavour constraints and do not add qualitatively new effects to the loop contributions to the $T$ parameter.}
\begin{align}
\L_s^\text{doublet} &=
-\bar Q^i m_{Q}^i Q^i
-\bar R^i m_{R}^i R^i
+ \left( Y^{ij} \text{tr}[ \bar Q^i_L \mathcal H R_R^j]  + \text{h.c} \right)
\,,
\label{doubletL}\\
\L_\text{mix}^\text{doublet} &=
m_{Q}^j\lambda_{L}^{ij}\bar q_L^i Q_{R}^j
+
m_{R}^i\lambda_{Ru}^{ij}\bar U_L^i u_{R}^j
+
m_{R}^i\lambda_{Rd}^{ij}\bar D_L^i d_{R}^j
\,.\label{doubletLmix}
\end{align}
where $\mathcal H=(i\sigma_2H^*,H)$. Everywhere $i,j$ are flavour indices. Note that the two doublets $Q$ and $R$ can be embedded in the spinorial representation $\four$ of $SO(5)$ of the minimal composite Higgs model MCHM$_4$ \cite{Agashe:2004rs}.

\subsubsection{The triplet model}
The elementary $\SU(2)_L$ quark doublets mix with a  composite $L = (\two,\two)_{2/3}$ of $\SU(2)_L\times \SU(2)_R\times \U(1)_X$, and the elementary quark singlets couple both to a composite triplet $R=(\one,\three)_{2/3}$. The model also contains a $(\three,\one)_{2/3}$ to preserve LR symmetry. The part of the Lagrangian involving fermions reads
\begin{align}
\L_s^\text{triplet} &=
-\text{tr}[ \bar L^i m_{L}^i L^i ]
-\text{tr}[ \bar R^i m_{R}^i R^i ]
-\text{tr}[ \bar R^{\prime\, i} m_{R}^iR^{\prime\, i}]\notag\\
&\quad+ Y^{ij} \text{tr}[ \bar L_L^i \mathcal H R_R^j] + Y^{ij}\text{tr}[\mathcal H\,  \bar L_L^i R_R^{\prime\, j}]  + \text{h.c}
\,,
\label{tripletL} \\
\L_\text{mix}^\text{triplet} &=
m_{L}^j\lambda_{L}^{ij}\bar q_L^i Q_{R}^j
+
m_{R}^i\lambda_{Ru}^{ij}\bar U_L^i u_{R}^j
+
m_{R}^i\lambda_{Rd}^{ij}\bar D_L^i d_{R}^j
\,.\label{tripletLmix}
\end{align}
where $Q$ is the $\SU(2)_L$ doublet with $T_{3R}=-\frac{1}{2}$ contained in $L$ and $U, D$ are the elements in the triplet $R$ with charge 2/3 and -1/3 respectively. Note that $Q$, $R$ and $R'$ together form a complete representation of $SO(5)$ since $\ten = \four \oplus \three \oplus \three$. The triplet model is thus equivalent to the minimal composite Higgs model MCHM$_{10}$ with fermions in the adjoint representation $\ten$ of $SO(5)$ \cite{Contino:2006qr}.

\subsubsection{The bidoublet model}
The elementary $\SU(2)_L$ quark doublets mix with a $L_U = (\two,\two)_{2/3}$ and a $L_D = (\two,\two)_{-1/3}$ of $\SU(2)_L\times \SU(2)_R\times \U(1)_X$, the former giving masses to up-type quarks, the latter to down-type quarks. The elementary up and down quark singlets couple to a $(\one,\one)_{2/3}$ and a $(\one,\one)_{-1/3}$ respectively.
\begin{align}
\L_s^\text{bidoublet} &=
-\text{tr}[ \bar L_U^i m_{Q_u}^i L_U^i ]
-\bar U^i m_{U}^i U^i
+ \left( Y_U^{ij} \text{tr}[ \bar L_U^i \mathcal H ]_L U_R^j  + \text{h.c} \right)
+ (U\to D)
\,,\label{bidoubletL}
\\
\L_\text{mix}^\text{bidoublet} &=
m_{Q_u}^j\lambda_{Lu}^{ij}\bar q_L^i Q_{Ru}^j
+
m_{U}^i\lambda_{Ru}^{ij}\bar U_L^i u_{R}^j
+ (U,u\to D,d)
\,,\label{bidoubletLmix}
\end{align}
where again $Q_u$ and $Q_d$ are the doublets in $L_U$ and $L_D$ which have the same gauge quantum numbers of the SM left-handed quark doublet. Note that here, as opposed to the previous models, there are two different left-handed mixing angles for up- and down-type quarks, $s_{Lu}$ and $s_{Ld}$. The composite fields $L_U$ and $U$, as well as $L_D$ and $D$, together form a complete representation of $SO(5)$ since $\five = \four \oplus \one$. The bidoublet model thus corresponds to the minimal composite Higgs model MCHM$_5$ with fermions in the fundamental representation $\five$ of $SO(5)$ \cite{Contino:2006qr}.

\bigskip

The field content in all three cases is summarized in table~\ref{tab:fields}. We avoid a more detailed discussion of the relation between the above simple effective Lagrangians and more basic models which include the Higgs particle as a pseudo-Goldstone boson. Here it suffices to say that the above Lagrangians are suitable to catch the main phenomenological properties of more fundamental models. For this to be the case, the truly basic assumption is that the lowest elements of towers of resonances, either of spin-$\frac{1}{2}$ or of spin 1, normally occurring in more complete models, are enough to describe the main phenomenological consequences, at least in as much as tree-level effects are considered.

\begin{table}
\renewcommand{\arraystretch}{1.15}
\centering
\begin{tabular}{lccccc}
\hline
model && $\SU(3)_c$ & $\SU(2)_L$ & $\SU(2)_R$ & $\U(1)_X$ \\
\hline
\multirow{2}{1.5cm}{doublet}
&$Q$ & $\three$ & $\two$ & $\one$ & ${1}/{6}$ \\
&$R$ & $\three$ & $\one$ & $\two$ & ${1}/{6}$ \\
\hline
\multirow{3}{1.5cm}{triplet}
&$L$ & $\three$ & $\two$ & $\two$ & ${2}/{3}$ \\
&$R$ & $\three$ & $\one$ & $\three$ & ${2}/{3}$ \\
&$R'$ & $\three$ & $\three$ & $\one$ & ${2}/{3}$ \\
\hline
\multirow{4}{1.5cm}{bidoublet}
&$L_U$ & $\three$ & $\two$ & $\two$ & ${2}/{3}$ \\
&$L_D$ & $\three$ & $\two$ & $\two$ & $-{1}/{3}$ \\
&$U$ & $\three$ & $\one$ & $\one$ & ${2}/{3}$ \\
&$D$ & $\three$ & $\one$ & $\one$ & $-{1}/{3}$ \\
\hline
\end{tabular}
\caption{Quantum numbers of the fermionic resonances in the three models considered. All composite fields come in vector-like pairs. The $X$ charge is related to the standard hypercharge as $Y=T_{3R}+X$.}
\label{tab:fields}
\end{table}

\chapter{Flavour, electroweak observables and a 125 GeV composite Higgs}\label{CHMbounds}

If the Higgs boson is a composite particle by a suitable new strong interaction, the problem of describing flavour in a way consistent with current experiments is non trivial.
In addition, it is well known that in theories with a strong coupling at the TeV scale there are in general relevant contributions to the electroweak precision observables. Both these aspects - flavour and EWPT - play an important role in setting bounds on the masses and the other parameters of the theory.

The discovery of a Higgs-like particle \cite{Chatrchyan:2012ufa,Aad:2012tfa} puts the additional constraint of its measured mass of about 125 GeV on the models where it is interpreted as a (natural) pseudo-Nambu-Goldstone boson.
Here we shall concentrate our attention on the compatibility of such interpretation of the newly found particle with constraints from flavour and electroweak precision tests. 

The common  features  emerging from the modelling of the strong dynamics responsible for the existence of the composite pseudo-Goldstone Higgs boson, as described in the previous chapter, are: 
\begin{enumerate}[$(i)$]
\item a breaking scale of the global symmetry group, $f$, somewhat larger than the EWSB scale $v \approx 246$~GeV;
\item a set of $\rho$-like vector resonances of typical mass $m_\rho = g_\rho f$;
\item a set of spin-$\frac{1}{2}$  resonances, vector-like under the Standard Model gauge group, of typical mass $m_\psi = Y f$;
\item bilinear mass-mixing terms between the composite and the elementary fermions, ultimately responsible for the masses of the elementary fermions themselves \cite{Kaplan:1991dc}.
\end{enumerate}

These same mass mixings are crucial in explicitly breaking the global symmetry of the strong dynamics, i.e. in triggering EWSB, with a resulting Higgs boson mass
\begin{equation}\label{mh}
m_h = C \frac{\sqrt{N_c}}{\pi} m_t Y,
\end{equation}
where $N_c=3$ is the number of colours, $m_t$ is the top mass and $C$ is a model dependent coefficient of ${\Ord}(1)$, barring  unnatural fine-tunings \cite{Contino:2006qr,Redi:2012ha,Pomarol:2012qf, Matsedonskyi:2012ym, Marzocca:2012zn, Panico:2012uw}. This very equation makes manifest that the measured mass of 125 GeV calls for a semi-perturbative coupling $Y$ of the fermion resonances and, in turn, for their relative lightness, if one wants to insist on a breaking scale $f$ not too distant from $v$ itself. For a reference value of $f = 500\text{--}700$ GeV, which in PNGB Higgs models is enough to bring all the Higgs signal strengths in agreement with the currently measured values \cite{Giardino:2013bma},  one expects fermion resonances with typical mass not exceeding about 1 TeV, of crucial importance for their direct searches at the LHC. These searches are currently sensitive to masses in the  500--700 GeV range, depending on the charge of the spin-$\frac{1}{2}$ resonance and on the decay channel~\cite{CMS:2012ab,Chatrchyan:2012vu,Chatrchyan:2012af,ATLAS:2012hpa}.  In this work we aim to investigate the compatibility of this feature with 
flavour and EWPT.

To address this question, keeping the discussion simple and possibly not too model dependent, we follow the {\it partial compositeness} approach introduced in section \ref{CHM/partialcompositeness}. The vector resonances transform in the adjoint representation of a global symmetry respected by the strong sector, which contains the SM gauge group. To protect the $T$ parameter from tree-level contributions, we take this symmetry to be $G_c=\SU(3)_c\times \SU(2)_L\times \SU(2)_R\times \U(1)_X$. We assume all vector resonances to have mass $m_\rho$ and coupling $g_\rho$. For simplicity we also assume the composite fermions to have all the same mass $m_{\psi}$.

Quark masses and mixings are generated after electroweak symmetry breaking from the composite-elementary mixing.
While the effective SM Yukawa couplings $\hat y_{u,d}$ must have the known hierarchical form, the Yukawa couplings in the strong sector, $Y_{U,D}$, could be structureless {\em anarchic} matrices (see e.g. \cite{Grossman:1999ra,Huber:2000ie,Gherghetta:2000qt,Agashe:2004cp,Blanke:2008zb,Bauer:2009cf,KerenZur:2012fr,Csaki:2008zd}).
However, to ameliorate flavour problems, one can also impose global flavour symmetries on the strong sector.

We consider a number of  different options for the transformation properties of the spin-$\frac{1}{2}$ resonances under the global symmetries of the strong dynamics, motivated by the need to be consistent with the constraints from the EWPT, as well as different options for the flavour structure/symmetries, motivated by  the many significant flavour bounds.
We analyze in succession the constraints from flavour and EWPT in the various cases, considering for each case the three fermion representations introduced in section \ref{CHM/partialcompositeness/fermions} and defined by the Lagrangians \eqref{doubletL}--\eqref{bidoubletLmix}, which are customary in the literature.

\section{General electroweak precision constraints}\label{CHMbounds/EWPT}

In this section we discuss electroweak precision constraints that hold independently of the flavour structure. Among the models considered, only \UtreLC\ is subject to additional electroweak constraints, to be discussed in section~\ref{CHMbounds/U3/EWPT}.

We recall the standard Yukawa couplings \eqref{SMyukawa} in matrix notation,
\begin{align}\label{SMYuk}
\hat y_u &\approx s_{Lu} \cdot U_{Lu} \cdot Y_U \cdot U_{Ru}^\dagger \cdot s_{Ru}
\end{align}
where
\begin{gather}
\lambda_{Lu}=\diag(\lambda_{Lu1},\lambda_{Lu2},\lambda_{Lu3}) \cdot U_{Lu}\,,
\\
\lambda_{Ru}=U_{Ru}^{\dag}\cdot\diag(\lambda_{Ru1},\lambda_{Ru2},\lambda_{Ru3})\,,
\\
s_{X}^{ii}={\lambda_{X i}}/{\sqrt{1+(\lambda_{X i})^2}},
~ X = L,R~,
\end{gather}
and similarly for $\hat y_d$. Here and in the following, the left-handed mixings are different for $u$ and $d$ quarks, $s_{Lu}\neq s_{Ld}$, only in the bidoublet model. For later convenience we define the parameters
\begin{align}
x_t &\equiv \frac{\lambda_{Lt}}{\lambda_{Rt}}, & z &\equiv \frac{\lambda_{Lt}}{\lambda_{Lb}}.
\end{align}

\subsection{Oblique corrections}\label{CHMbounds/EWPT/oblique}

As well known, the $S$ parameter receives a tree-level contribution, which for degenerate composite vectors reads \cite{Contino:2006nn}
\begin{equation}
S = \frac{8 \pi v^2}{m_\rho^2} ~,
\label{S-corr}
\end{equation}
independently of the choice of fermion representations. It is also well known that $S$ and $T$ both get at one loop model-independent ``infrared-log'' contributions \cite{Barbieri:2007bh}
\begin{equation}
\hat{S} =  \left(\frac{v}{f}\right)^2 \frac{g^2}{96\pi^2} \log{\frac{m_\rho}{m_h}}~,
\qquad
 \hat{T} = -   \left(\frac{v}{f}\right)^2 \frac{3 g^2 t^2_w}{32\pi^2} \log{\frac{m_\rho}{m_h}}~.
 \label{inf-logs}
\end{equation}
where $\hat{S} = \alpha_\text{em}/(4 s^2_w)S$ and $\hat{T}=\alpha_\text{em}T$.

Experimentally, a recent global electroweak fit after the discovery of the Higgs boson \cite{Baak:2012kk} finds $S - S_{\text{SM}}=0.03\pm0.10$ and $T- T_{\text{SM}}=0.05\pm0.12$. Requiring $2\sigma$ consistency with these results of the tree level correction to $S$, eq.~(\ref{S-corr}), which largely exceeds the infrared logarithmic contribution of (\ref{inf-logs}) and has the same sign, gives the bound
\begin{equation}
m_\rho>2.6\,\text{TeV} \,.
\label{eq:boundmrho}
\end{equation}

The one loop correction to the $T$ parameter instead strongly depends on the choice of the fermion representations. 
We present here simplified formulae valid in the three models \eqref{doubletL}, \eqref{tripletL} and \eqref{bidoubletL} for a common fermion resonance mass $m_\psi$ and developed to first nonvanishing order in $\lambda_{Lt}, \lambda_{Rt}$, as such only valid for small $s_{Lt}, s_{Rt}$. 

In the {doublet model} the leading contribution to $\hat{T}$, proportional to $\lambda_{Rt}^4$, reads
\begin{equation}
\hat{T} = \frac{71}{140} \frac{N_c}{16 \pi^2} \frac{m_t^2}{m_\psi^2} \frac{Y^2}{x_t^2}\,.
\label{eq:T-doublet}
\end{equation}

In the {bidoublet model} one obtains from a leading $\lambda_{Lt}^4$ term 
\begin{equation}
\hat{T} = - \frac{107}{420}\frac{N_c}{16 \pi^2} \frac{m_t^2}{m_\psi^2} x_t^2 Y_U^2\,.
\label{eq:T-bidoublet}
\end{equation}

In the {triplet model} the leading contributions are
\begin{equation}
\hat{T} = \Big(\log\frac{\Lambda^2}{m_\psi^2} - \frac{1}{2}\Big)\frac{N_c}{16 \pi^2} \frac{m_t^2}{m_\psi^2} \frac{Y^3}{y_t x_t}\,, \quad \text{and} \quad \hat{T} = \frac{197}{84}\frac{N_c}{16 \pi^2} \frac{m_t^2}{m_\psi^2} x_t^2 Y^2\,,
\label{eq:T-triplet}
\end{equation}
where the first comes from $\lambda_{Rt}^2$ and the second from $\lambda_{Lt}^4.$ Note the logarithmically divergent contribution to the $\lambda_{Rt}^2$ term that is related to the explicit breaking of the $\SU(2)_R$ symmetry in the elementary-composite fermion mixing and would have to be cured in a more complete model. More details on the derivation of formulae \eqref{eq:T-doublet}, \eqref{eq:T-bidoublet} and \eqref{eq:T-triplet} are given in appendix~\ref{app:T}.

Imposing the experimental bound at $2\sigma$, eqs.~(\ref{eq:T-doublet}), (\ref{eq:T-bidoublet}), (\ref{eq:T-triplet}) give rise to the bounds on the first line in table~\ref{tab:bounds-ew} (where we set $\log{(\Lambda/m_\psi)} = 1$). Here however there are two caveats. First, as mentioned, these equations are only valid for small mixing angles. Furthermore, for moderate values of $f$, a cancellation could take place between the fermionic contributions and the infrared logs of the bosonic contribution to $T$.
As we shall see, the bounds  from $S$ and $T$ are anyhow not the strongest ones that we will encounter: they are compatible with  $m_\psi\lesssim 1$ TeV for $Y = 1$ to $2$ and $g_\rho = 3$ to $5$. Note that here and in the following $m_{\psi}$ represents the mass of the composite fermions that mix with the elementary ones, whereas, as already noticed, the ``custodians'' have mass $m_{\psi}/\sqrt{1+(\lambda_X)^2}$.

\begin{table}[tbp]
\renewcommand{\arraystretch}{1.3}
\centering
\begin{tabular}{cccc}
\hline
Observable & \multicolumn{3}{c}{Bounds on $m_{\psi}$ [TeV]} \\
& doublet  & triplet & bidoublet \\
\hline
$T$ & $0.28 ~Y/x_t$ &$0.51 ~\sqrt{Y^3/x_t}$, \; $0.60  ~x_t Y$  & $0.25 ~x_tY_U$\\
$R_b$ ($g_{Zbb}^L$) & $5.6  ~\sqrt{x_tY}$ & & $6.5  ~Y_D \sqrt{x_t/Y_{U}}/z$ \\
$B\to X_s\gamma$ ($g_{Wtb}^R$) & $0.44  ~\sqrt{Y/x_t}$  & $0.44  ~\sqrt{Y/x_t}$ & 0.61\\
\hline
\end{tabular}
\caption{Lower bounds on the fermion resonance mass $m_\psi=Y f$ in TeV from electroweak precision observables. A blank space means no significant bound.}
\label{tab:bounds-ew}
\end{table}

\subsection{Modified $Z$ couplings}\label{CHMbounds/EWPT/Zcouplings}

In all three models for the electroweak structure, fields with different $\SU(2)_L$ quantum numbers mix after electroweak symmetry breaking, leading to modifications in $Z$ couplings which have been precisely measured at LEP. Independently of the flavour structure, an important constraint comes from the $Z$ partial width into $b$ quarks, which deviates by $2.5\sigma$ from its best-fit SM value \cite{Baak:2012kk}
\begin{align}
R_b^\text{exp} &= 0.21629(66)~,
&
R_b^\text{SM} &= 0.21474(3)~.
\label{eq:Rbexp}
\end{align}
Writing the left- and right-handed $Z$ couplings as
\begin{equation}
\frac{g}{c_w}\bar b \gamma^\mu \left[
(-\tfrac{1}{2}+\tfrac{1}{3}s_w^2+\delta g^L_{Zbb}) P_L
+(\tfrac{1}{3}s^2_w+\delta g^R_{Zbb})P_R
\right]bZ_\mu
\,,
\end{equation}
one gets
\begin{align}
\delta g_{Zbb}^L &=
\frac{v^2Y_{D}^2}{2m_{D}^2}\frac{xy_t}{Y_U} \,a+
\frac{g_\rho^2v^2}{4m_\rho^2}\frac{xy_t}{Y_U}\,b \,,
&
\delta g_{Zbb}^R &=
\frac{v^2Y_{D}^2}{2m_{D}^2}\frac{y_b^2 Y_U}{x_t y_t Y_D^2} \,c+
\frac{g_\rho^2v^2}{4m_\rho^2}\frac{y_b^2 Y_U}{x_t y_t Y_D^2}\,d \,,
\label{eq:Zbb1}
\end{align}
with the coefficients
\begin{center}
\begin{tabular}{c|ccc}
& doublet & triplet & bidoublet \\
\hline
$a$ & $1/2$   & $0$ & $1/(2 z^2)$\\
$b$ & $1/2$ & $0$ & $1/z^2$
\end{tabular}
\qquad
\begin{tabular}{c|ccc}
& doublet & triplet & bidoublet \\
\hline
$c$ & $-1/2$   & $-1/2$ & $0$\\
$d$ & $-1/2$ & $-1$ & $0$
\end{tabular}
\end{center}
The vanishing of some entries in (\ref{eq:Zbb1}) can be simply understood by the symmetry considerations of ref.~\cite{Agashe:2006at}. 
As manifest from their explicit expressions the contributions proportional to $a$ and $c$ come from mixings between elementary and composite fermions with different $\SU(2)\times \U(1)$ properties, whereas the contributions proportional to $b$ and $d$ come from $\rho$-$Z$ mixing. Taking $Y_U=Y_D=Y$, $m_D = Yf$ and $m_\rho = g_\rho f$, all these contributions scale however in the same way as $1/(f^2 Y)$.

It is important to note that $\delta g^L_{Zbb}$ is always positive or 0, while $\delta g^R_{Zbb}$ is always negative or 0, while the sign of the SM couplings is opposite. As a consequence, in all 3 models considered, the tension in eq.~(\ref{eq:Rbexp}) is {\em always increased}. Allowing the discrepancy to be at most $3\sigma$, we obtain the numerical bounds in the second row of table~\ref{tab:bounds-ew}. The bound on $m_\psi$ in the doublet model is highly significant since $x_t Y > 1$, whereas it is irrelevant in the triplet model and can be kept under control in the bidoublet model for large enough $z$ (but see below). In the triplet model, there is a bound from the modification of the right-handed coupling, which is however insignificant.

\subsection{Right-handed $W$ couplings}\label{CHMbounds/EWPT/Wcouplings}

Analogously to the modified $Z$ couplings, also the $W$ couplings are modified after EWSB. Most importantly, a right-handed coupling of the $W$ to quarks is generated. The most relevant experimental constraint on such coupling is the branching ratio of $B\to X_s\gamma$, because a right-handed $Wtb$ coupling lifts the helicity suppression present in this loop-induced decay in the SM \cite{Vignaroli:2012si}. Writing this coupling as
\begin{equation}
\frac{g}{\sqrt{2}}\delta g^R_{Wtb} 
(\bar t \gamma^\mu
P_R
b)W_\mu^+
\,,
\end{equation}
one gets
\begin{align}
\delta g_{Wtb}^R &=
\frac{v^2Y_UY_D}{2m_{Q}m_{U}}\frac{y_b}{x_t Y_U} \,a+
\frac{g_\rho^2v^2}{4m_\rho^2}\frac{y_b}{x_t Y_U}\,b \,,
\label{eq:Wtb1}
\end{align}
with the coefficients
\begin{center}
\begin{tabular}{c|ccc}
& doublet & triplet & bidoublet \\
\hline
$a$ & $1$ & $1$ & $-2x_t y_t/Y$\\
$b$ & $1$ & $1$ & $0$
\end{tabular}
\end{center}
The coefficients in the bidoublet model vanish at quadratic order in the elementary-composite mixings as a consequence of a discrete symmetry \cite{Agashe:2006at}. The nonzero value for $a$ in the table is due to the violation of that symmetry at quartic order \cite{Vignaroli:2012si}.
The contribution to the Wilson coefficient $C_{7,8}$, defined as in \cite{Altmannshofer:2012az}, reads
\begin{equation}
C_{7,8} = \frac{m_t}{m_b} \frac{\delta g^R_{Wtb}}{V_{tb}} A_{7,8}(m_t^2/m_W^2)
\end{equation}
where $A_7(m_t^2/m_W^2)\approx -0.80$ and $A_8(m_t^2/m_W^2)\approx -0.36$.

Since the $B\to X_s\gamma$ decay receives also UV contributions involving composite dynamics, we impose the conservative bound that the SM plus the infrared contributions above do not exceed the experimental branching ratio by more than $3\sigma$. In this way we find the bound in the last row of table~\ref{tab:bounds-ew}.

\section{The anarchic model} \label{CHMbounds/anarchy}

In the anarchic model, the $Y_{U,D}$ are anarchic matrices, with all entries of similar order, and the Yukawa hierarchies are generated by hierarchical mixings $\lambda$.
From a low energy effective theory point of view, the
requirement to reproduce the observed quark masses and mixings fixes the relative size of the mixing parameters up to -- a priori unknown -- functions of the elements in $Y_{U,D}$. We follow the common approach to replace functions of Yukawa couplings by appropriate powers of ``average'' Yukawas $Y_{U*,D*}$, keeping in mind that this introduces $\Ord(1)$ uncertainties in all observables. In this convention, assuming $\lambda_{X3}\gg\lambda_{X2}\gg\lambda_{X1}$, the quark Yukawas are given by
\begin{align}
y_{u}&=Y_{U*} s_{Lu 1}s_{Ru 1}~,&
y_{c}&=Y_{U*} s_{Lu 2}s_{Ru 2}~,&
y_{t}&=Y_{U*} s_{Lu 3}s_{Ru 3}~.
\label{eq:anarchy-yukawas}
\end{align}
and similarly for the $Q=-1/3$ quarks. In the doublet and triplet models, the entries of the CKM matrix are approximately given by
\begin{align}
V_{ij} &\sim \frac{s_{L i}}{s_{L j}} ~,
\label{eq:anarchy-ckm-doublet}
\end{align}
where $i<j$. Using eqs.~(\ref{eq:anarchy-yukawas}) and (\ref{eq:anarchy-ckm-doublet}), one can trade all but one of the $s_{L,R}$ for known quark masses and mixings. We choose the free parameter as
\begin{equation}
x_t \equiv s_{L3}/s_{Ru3}.
\label{eq:x-doublet}
\end{equation}
In the bidoublet model, instead of (\ref{eq:anarchy-ckm-doublet}) one has
 in general two different contributions to $V_{ij}$%
\begin{align}
V_{ij} &\sim
\frac{s_{Ld i}}{s_{Ld j}}\pm \frac{s_{Lu i}}{s_{Lu j}} ~.
\label{eq:anarchy-ckm-bidoublet}
\end{align}
Given the values of all quark masses and mixings, the hierarchy $\lambda_{X3}\gg\lambda_{X2}\gg\lambda_{X1}$ is only compatible with $s_{Lu i}/s_{Lu j}$ being at most comparable to $s_{Ld i}/s_{Ld j}$. In view of this, the two important parameters are
\begin{align}
x_t &\equiv s_{Lt}/s_{Rt} ~,
&
z &\equiv s_{Lt}/s_{Lb} ~.
\label{eq:x-bidoublet}
\end{align}
The requirement to reproduce the large top quark Yukawa ($m_t = \frac{y_t}{\sqrt{2}}v$)
\begin{equation}
y_t = s_{Lt} Y_{U*} s_{Rt},
\end{equation}
restricts $x_t$ to a limited range around one,\footnote{\text{In our numerical analysis, we will take $y_t=0.78$, which is the running $\MS$ coupling at 3~TeV.}}
\begin{equation}
\frac{y_t}{Y_{U*}} < x_t < \frac{Y_{U*}}{y_t} ~,
\end{equation}
while we take $z$ throughout to be greater than or equal to 1.

If the composite sector is flavour-anarchic, then all the possible flavour-violating operators of section~\ref{Flavour/EFT} (with the composite fermions replacing the SM quarks) are generated by the strong interaction with a suppression scale of at most $\Lambda\lesssim 4\pi f$. The presence of these operators would be by far excluded by flavour bounds if there was a sizable communication of their effects to the elementary sector. Nevertheless, the smallness of the mass mixings $\lambda$ makes the situation different.

Any flavour-changing light-quark bilinear is obtained by mixing on the corresponding composite-quark bilinear. The flavour effects are more suppressed for light quarks, which, having a little amount of compositeness, have smaller mixing angles $\varphi_i$. Since those are also the ones for which the flavour bounds are stronger, in many cases one gets results consistent with a compositeness scale low enough to be natural.
However, problems arise if one considers \CP\ violation in the neutral $K$ system, for which the experimental bounds are very strong (see section~\ref{sec:DF=2}). With a compositeness scale $f\lesssim{\rm TeV}$, even for an appropriate choice of the other paramteres, the coefficient of $\Q_4$ is one or two orders of magnitude above the experimental bound $\Im(C_4) < 2.6\times 10^{-11}$.

We now discuss constraints that are specific to the anarchic model, as defined above, and that hold in addition to the bounds described in the previous section. From now on we identify $Y_{U*}$ and $Y_{D*}$ with the parameter $Y$ of \eqref{mh}.

\subsection{Tree-level $\Delta F=2$ FCNCs}\label{CHMbounds/anarchy/DF2}

In the anarchic model exchanges of gauge resonances give rise to $\Delta F=2$ operators at tree level. Up to corrections of order $v^2/f^2$, the Wilson coefficients of the operators
\begin{align}
Q_V^{dLL} &= (\bar d^i_L\gamma^\mu d^j_L)(\bar d^i_L\gamma^\mu d^j_L) \,,
&
Q_V^{dRR} &= (\bar d^i_R\gamma^\mu d^j_R)(\bar d^i_R\gamma^\mu d^j_R) \,,
\\
Q_V^{dLR} &= (\bar d^i_L\gamma^\mu d^j_L)(\bar d^i_R\gamma^\mu d^j_R) \,,
&
Q_S^{dLR} &= (\bar d^i_R d^j_L)(\bar d^i_L d^j_R) \,,
\end{align}
can be written as
\begin{align}
C_X^{dAB} &= \frac{g_\rho^2}{m_\rho^2}
g_{Ad}^{ij}g_{Bd}^{ij}
c_X^{dAB},& A,B &= L,R,\quad X = V,S,
\label{eq:DF2}
\end{align}
and with the obvious replacements for up-type quarks, relevant for $D$-$\bar D$ mixing.

The couplings $g_{qA}^{ij}$ with $i\neq j$ contain two powers of elementary-composite mixings. In the doublet and triplet models, one can use eqs.~(\ref{eq:anarchy-yukawas})--(\ref{eq:x-doublet}) to write them as
\begin{align}
g_L^{ij}&\sim s_{Ldi}s_{Ldj}\sim \xi_{ij} \frac{x_t y_t}{Y} \,,
\\
g_{Ru}^{ij}& \sim s_{Rui}s_{Ruj}\sim \frac{y_{u^i}y_{u^j}}{Y y_t x_t \xi_{ij}} \,,
\\
g_{Rd}^{ij}& \sim s_{Rdi}s_{Rdj}\sim \frac{y_{d^i}y_{d^j}}{Y y_t x_t \xi_{ij}} \,,
\end{align}
where $\xi_{ij} = V_{tj}V_{ti}^*$.
In the bidoublet model,
one has
\begin{align}
g_{Ld}^{ij}\sim g_{Lu}^{ij}&\sim \xi_{ij} \frac{x_t y_t}{Y_U} \,,
&
g_{Rd}^{ij}&\sim z^2\frac{Y_U}{Y_D^2}\frac{y_{d^i}y_{d^j}}{y_t x_t \xi_{ij}} \,.
&
g_{Ru}^{ij}&\sim \frac{y_{u^i}y_{u^j}}{Y_Uy_t x_t \xi_{ij}} \,.
\end{align}

The coefficients $c^{AB}_X$ can be written as
\begin{align}
c_V^{dLL} &= -\frac{1}{6} -\frac{1}{2}\left[X(Q)^2+T_{3L}(Q)^2+T_{3R}(Q)^2\right]
\,,\label{cVdLL}\\
c_V^{dRR} &= -\frac{1}{6} -\frac{1}{2}\left[X(D)^2+T_{3L}(D)^2+T_{3R}(D)^2\right]
\,,\label{cVdRR}\\
c_V^{dLR} &= \frac{1}{6}-\left[X(Q)X(D)+T_{3L}(Q)T_{3L}(D)+T_{3R}(Q)T_{3R}(D)\right]
\,,\label{cVdLR}\\
c_S^{dLR} &= 1
\,,\label{cSdLR}
\end{align}
where the first terms come from heavy gluon exchange and the terms in brackets from neutral heavy gauge boson exchange. $Q$ refers to the charge $-1/3$ fermion mixing with $q$ and $D$ to the charge $-1/3$ fermion mixing with $d_R$. In the bidoublet model, we consider only the contribution from $Q_u$ for simplicity, which is enhanced if $z>1$.
The numerical coefficients relevant for the models discussed above are collected in table~\ref{tab:df2coeffs}.

\begin{table}[t]
\renewcommand{\arraystretch}{1.5}
\centering%
\begin{tabular}{cccc}
\hline
 & doublet & triplet & bidoublet\\
\hline
$c_V^{dLL}$ & $-\frac{11}{36}$ & $-\frac{23}{36}$ & $-\frac{23}{36}$\\
$c_V^{dRR}$ & $-\frac{11}{36}$ & $-\frac{8}{9}$ & $-\frac{2}{9}$\\
$c_V^{dLR}$ & $\frac{5}{36}$ & $-\frac{7}{9}$ & $\frac{7}{18}$\\
$c_S^{dLR}$ & 1 & 1 & 1\\
\hline
\end{tabular}
\caption{Coefficients relevant for $\Delta F = 2$ operators in anarchy and $\U(2)^3$.}
\label{tab:df2coeffs}
\end{table}

The experimental bounds on the real and imaginary parts of the Wilson coefficients have been given in \cite{Isidori:2011qw,Calibbi:2012at}.
Since the phases of the coefficients can be of order one and are uncorrelated, we derive the bounds assuming the phase to be maximal. We obtain the bounds in the first eight rows of table~\ref{tab:bounds-anarchy}.
As is well known, by far the strongest bound, shown in the first row, comes from the scalar left-right operator in the kaon system which is enhanced by RG evolution and a chiral factor. Note in particular the growth with $z$ of the bound in the bidoublet case, which counteracts the $1/z$ behaviour of the bound from $R_b$. But also the left-left vector operators in the kaon, $B_d$ and $B_s$ systems lead to bounds which are relevant in some regions of parameter space. The bounds from the $D$ system are subleading.

\begin{table}[tbp]
\renewcommand{\arraystretch}{1.3}
\centering
\begin{tabular}{clll}
\hline
Observable & \multicolumn{3}{c}{Bounds on $m_{\psi}$ [TeV]} \\
& doublet & triplet & bidoublet \\
\hline
$\epsilon_K$ $(Q_S^{LR})$ & $14 $  & $14 $& $14  ~z $ \\
$\epsilon_K$ $(Q_V^{LL})$ & $2.7 ~x_t$& $3.9  ~x_t$  & $3.9  ~x_t$ \\
$B_d$-$\bar B_d$ $(Q_S^{LR})$ & $0.7 $ & $0.7  $ & $0.7 $ \\
$B_d$-$\bar B_d$ $(Q_V^{LL})$ & $2.3 ~x_t$ & $3.4  ~x_t$ & $3.4  ~x_t$ \\
$B_s$-$\bar B_s$ $(Q_S^{LR})$ & $0.6 $ & $0.6  $ & $0.6 $ \\
$B_s$-$\bar B_s$ $(Q_V^{LL})$ & $2.3 ~x_t$ & $3.4  ~x_t$ & $3.4  ~x_t$ \\
$D$-$\bar D$ $(Q_S^{LR})$ & $0.5 $ & $0.5  $ & $0.5 $ \\
$D$-$\bar D$ $(Q_V^{LL})$ & $0.4 ~x_t$ & $0.6  ~x_t$ & $0.6  ~x_t$ \\
$K_L\to\mu\mu$ ($f$--$\psi$) & $0.56  ~\sqrt{Y/x_t}$  & $0.56  ~\sqrt{Y/x_t}$ &\\
$K_L\to\mu\mu$ ($Z$--$\rho$) & $0.39  ~\sqrt{Y/x_t}$  & $0.56  ~\sqrt{Y/x_t}$ &\\
\hline
\end{tabular}
\caption{Flavour bounds on the fermion resonance mass $m_\psi$ in TeV in the anarchic model.}
\label{tab:bounds-anarchy}
\end{table}

\subsection{Flavour-changing $Z$ couplings}\label{CHMbounds/anarchy/Zcouplings}

Similarly to the modified flavour-conserving $Z$ couplings discussed in section~\ref{CHMbounds/EWPT/Zcouplings}, also {\em  flavour-changing} $Z$ couplings are generated in the anarchic model. In the triplet and doublet models, as well as in the bidoublet model, since the down-type contributions to the CKM matrix are not smaller than the up-type contributions in \eqref{eq:anarchy-ckm-bidoublet}, one has
\begin{gather}
\delta g_{Zd^id^j}^L \sim \frac{s_{Ldi}s_{Ldj}}{s_{Lb}^2} ~\delta g_{Zbb}^L \sim \xi_{ij} ~ \delta g_{Zbb}^L~, 
\label{eq:ZbsL}
\\
\delta g_{Zd^id^j}^R \sim \frac{s_{Rdi}s_{Rdj}}{s_{Rb}^2}~ \delta g_{Zbb}^R \sim \frac{y_{d^i}y_{d^j}}{y_b^2 \xi_{ij}} ~ \delta g_{Zbb}^R~.
\label{eq:ZbsR}
\end{gather}

In the $b\to s$ case, a global analysis of inclusive and exclusive $b\to s\ell^+\ell^-$ decays \cite{Altmannshofer:2012az} finds $|\delta g_{Zbs}^{L,R}|\lesssim 8\times 10^{-5}$, while in the $s\to d$ case, one finds $|\delta g_{Zsd}^{L,R}|\lesssim 6\times 10^{-7}$ from the $K_L\to\mu^+\mu^-$ decay \cite{Buras:2011ph}.\footnote{The decay $K^+\to\pi^+\nu\bar\nu$ leads to a bound $|\delta g_{Zsd}^{L,R}|\lesssim 3 \times10^{-6}$ at 95\% C.L. and is thus currently weaker than $K_L\to\mu^+\mu^-$, even though it is theoretically much cleaner.} Using (\ref{eq:ZbsL}) one finds that the resulting constraints on the left-handed coupling are comparable for $b\to s$ and $s \to d$. Since they are about a factor of 3 weaker than the corresponding bound from $Z\to b\bar b$, we refrain from listing them in table~\ref{tab:bounds-anarchy}, but their presence shows that the strong bound from $R_b$ cannot simply be circumvented by a fortuitous cancellation.
In the case of the right-handed coupling, one finds that the constraint from $K_L\to\mu^+\mu^-$ is an order of magnitude stronger than the one from $b\to s\ell^+\ell^-$, and also much stronger than the bound on the right-handed coupling coming from $Z\to b\bar b$. The numerical bounds we obtain are shown in the last two rows of table~\ref{tab:bounds-anarchy} from the contributions with fermion or gauge boson mixing separately since, in constrast to $Z\to b\bar b$, the two terms are multiplied by different $\Ord(1)$ parameters in the flavour-violating case.

\subsection{Loop-induced chirality-breaking effects}\label{CHMbounds/anarchy/loop}

Every flavour changing effect discussed so far originates from tree-level chirality-conserving interactions of the vector bosons, either the elementary $W_\mu$ and $Z_\mu$ or the composite $\rho_\mu$. At loop level, chirality-breaking interactions occur as well, most notably with the photon and the gluon, which give rise  in general to significant  $\Delta F=1$ flavour-changing effects ($b\rightarrow s \gamma$, $\epsilon_K^\prime$, $\Delta A_{\CP}(D\rightarrow PP)$), as well as to electric dipole moments of the light quarks. In the weak mixing limit between the elementary and the composite fermions, explicit calculations of some of the  $\Delta F=1$ effects have been made in \cite{Agashe:2008uz,Vignaroli:2012si,Gedalia:2009ws}, obtaining bounds in the range $m_\psi > (0.5\text{--}1.5)Y$\,TeV. For large \CP-violating phases the generated EDMs for the light quarks can be estimated  consistent with the current limit on the neutron EDM only if $m_\psi > (3\text{--}5)Y$\,TeV, where the limit is obtained from the 
analysis of \cite{Barbieri:2012bh}. 

\subsection{Direct bounds on vector resonances}\label{CHMbounds/anarchy/jjresonances}

Direct production of vector resonances and subsequent decay to light quarks can lead to a peak in the invariant mass distribution of $pp\to jj$ events at the LHC. The production cross section of a gluon resonance in $pp$ collisions reads
\begin{align}
\sigma(pp\to G^*) &=  \frac{2\pi}{9s}
\left[
(|g_L^u|^2+|g_R^u|^2) \mathcal L_{u\bar u}(s,m_\rho^2)
+
(|g_L^d|^2+|g_R^d|^2) \mathcal L_{d\bar d}(s,m_\rho^2)
\right],
\label{eq:sigmapprho}
\end{align}
where
\begin{equation}\label{partonlumi}
\mathcal L_{q\bar q}(s,\tilde s)=
\int_{\hat s/s}^1 \frac{dx}{x}
f_q(x,\mu)
f_{\bar q}\!\left(\frac{\tilde s}{xs},\mu\right)
\end{equation}
is the parton-parton luminosity function at partonic (hadronic) center of mass energy $\sqrt{\tilde s}$ ($\sqrt{s}$)
and the couplings $g_{L,R}^{u,d}$ are defined as $\mathcal L \supset \bar u_L \gamma^\mu T^a g_L^u
u_L G^*_\mu$.
The total width reads
\begin{equation}
\Gamma(G^*\to q \bar q) = \sum_{q=u,d}\sum_{i=1}^3\frac{m_\rho}{48\pi}\left(|g^{q^i}_L|^2+|g^{q^i}_R|^2\right)
\end{equation}
while the branching ratio to dijets is simply the width without the top contribution divided by the total width.\footnote{Neglecting the top quark mass in the kinematics, which is a good approximation for multi-TeV resonances still allowed by the constraints}
The coupling of a first generation quark mass eigenstate to a heavy vector resonance receives contributions from fermion composite-elementary mixing as well as vector boson composite-elementary mixing. For example, the coupling of the up quark to the gluon resonance reads
\begin{equation}
\bar u \gamma^\mu T^a\left(
g_\rho s_{Lu}^2 P_L +g_\rho s_{Ru}^2 P_R + \frac{g_3^2}{g_\rho}
\right)u G^*_\mu ~.
\label{eq:gG}
\end{equation}
In the anarchic model, due to the small degree of compositeness of first generation quarks, the coupling of vector resonances to a first generation quark-antiquark pair is dominated by mixing with the SM gauge bosons and thus suppressed by $g_\text{el}^2/g_\rho$. For a 3~TeV gluon resonance at the LHC with $\sqrt{s}=8$~TeV we expect
\begin{equation}
\sigma(pp\to G^*) = \frac{2\pi}{9s}\frac{g_3^4}{g_\rho^2}
\left[\mathcal L_{u\bar u}(s,m_\rho^2) +\mathcal L_{d\bar d}(s,m_\rho^2)\right]
\approx
\frac{5 ~\text{fb}}{g_\rho^2}\,.
\end{equation}
The ATLAS collaboration has set an upper bound of 7~fb on the cross section times branching ratio to two jets times the acceptance \cite{ATLAS:2012joa}, and a similar bound has been obtained by CMS \cite{CMS:2012eza}. Given that the gluon resonance will decay dominantly to top quarks, we conclude that the bound is currently not relevant, even for small $g_\rho$.

A similar argument holds in the case of the dijet angular distribution, which can be used to constrain local four-quark operators mediated by vector resonances. Following the discussion in section~\ref{CHMbounds/U3/compositeness}, in the limit of vanishing elementary-composite mixings for the light generations we obtain the bound
\begin{equation}
m_\rho > \frac{4.5~\text{TeV}}{g_\rho}
\end{equation}
which, in combination with the bound on $m_\rho$ from the $S$ parameter, is irrelevant for $g_\rho\gtrsim1.5$.

\section{A flavour-symmetric composite sector: $\U(3)^3$}\label{CHMbounds/U3}

One way to describe flavour in partial composite models, introduced in \cite{Barbieri:2008zt},\footnote{But actually effectively implemented before in a 5-dimensional example in \cite{Cacciapaglia:2007fw}.} is to assume that a sufficiently large flavour symmetry is respected by the strong sector, and shared by some of the elementary fermions. In this way one tries to ameliorate the flavour problem of the anarchic model, at the price of giving up the potential dynamical explanation of the generation of flavour hierarchies.

Concretely, in the $\U(3)^3$ models \cite{Cacciapaglia:2007fw,Barbieri:2008zt,Redi:2011zi}
one assumes the strong sector -- i.e. the Yukawa couplings\footnote{We omit the Yukawa terms $\tilde{\lambda}_U \bar{Q}^u_R H U_L + \tilde{\lambda}_D \bar{Q}^d_R \tilde{H} D_L + \text{h.c.}$, since they do not affect our considerations.}
 of the heavy fermions as well as their mass terms in \eqref{doubletL}, \eqref{tripletL} or \eqref{bidoubletL} -- to be invariant under the diagonal group $\U(3)_{Q+U+D}$ or $\U(3)_{Q^u+U}\times \U(3)_{Q^d+D}$. Furthermore, to minimize possible new flavour effects, one assumes that this symmetry is extended to a subset of the elementary quarks. The composite-elementary mixings are the only sources of breaking of the flavour symmetry of the composite sector and of the $\U(3)_{q}\times \U(3)_{u}\times \U(3)_{d}$ flavour symmetry of the elementary sector. We consider two choices:
\begin{enumerate}
\item In {\it left-compositeness}, to be called {\UtreLC} in short, the left mixings are proportional to the identity, thus linking $q_L$ to $Q$ into $\U(3)_{Q+U+D+q}$ in the doublet and triplet models. In the bidoublet model, since $q_L$ couples to both $Q^u$ and $Q^d$, the symmetry group $\U(3)_{Q^u+U}\times \U(3)_{Q^d+D}$ is broken down to $\U(3)_{Q^u+Q^d+U+D+q}$.
The right mixings $\lambda_{Ru}$, $\lambda_{Rd}$ are the only source of $\U(3)^3$ breaking.

\item In {\it right-compositeness}, to be called {\UtreRC} in short, the right mixings link $u_R$ to $U$ and $d_R$ to $D$ into $\U(3)_{Q^u+U+u}\times \U(3)_{Q^d+D+d}$, while
the left mixings $\lambda_{Lu}$, $\lambda_{Ld}$ are the only source of $\U(3)^3$ breaking.

\end{enumerate}
In the two cases the flavour-violating $\lambda$'s transform under the $\U(3)^3$ symmetry as $(\three,\threebar,\one)$ and $(\three,\one,\threebar)$, playing the role of the spurions of MFV.

All the composite-elementary mixings are then fixed by the known quark masses and CKM angles, up to the parameters $x_t$ (and, in the bidoublet model, $z$), which are defined as in \eqref{eq:x-doublet},\eqref{eq:x-bidoublet}.
Compared to the anarchic case, one now expects the presence of resonances related to the global symmetry $\U(3)_{Q+U+D}$ or $\U(3)_{Q^u+U}\times \U(3)_{Q^d+D}$, which in the following will be called flavour gauge bosons\footnote{We will only allow flavour gauge bosons related to the $\SU(3)$ subgroups of the $\U(3)$ factors.} and assumed to have the same masses $m_\rho$ and $g_\rho$ as the gauge resonances. Note that left-compositeness can be meaningfully defined for any of the three cases for the fermion representations, whereas right-compositeness allows to describe flavour violations only in the bidoublet model.

The limited ability to calculate in a strongly interacting theory makes it seem difficult to go beyond EFT considerations. Due to the specific way of flavour breaking by mass mixings this is actually not the case: only some of the flavour breaking operators survive among the ones discussed in section~\ref{Flavour/symmetries/MFV}. As we will illustrate, there are substantial differences if compared with generic EFT theory expectations, with a neat distinction between Right- and Left-compositeness.
We are interested in the relevant flavour violating operators involving the elementary fields to all orders in the strong interaction dynamics, i.e. after integrating out the heavy composite degrees of freedom.

We now discuss the constraints specific to \Utre. In \UtreLC\ the sizable degree of compositeness of light left-handed quarks leads to additional contributions to electroweak precision observables; in \UtreRC\ FCNCs arise at the tree level. In both cases collider bounds on the compositeness of light quarks place important constraints. This analysis follows and extends the analysis in \cite{Redi:2011zi}.

\subsection{Electroweak precision constraints specific to $\U(3)^3$}\label{CHMbounds/U3/EWPT}

The bounds from $R_b$ as well as the $S$ and $T$ parameters discussed in section~\ref{CHMbounds/EWPT} are also valid in $\U(3)^3$, with one modification: in  \UtreLC\ the contributions to the $\hat T$ parameter proportional to $s_{Lt}^4$ have to be multiplied by 3 since all three generations of left-handed up-type quarks contribute. The corresponding
bounds remain nevertheless relatively mild.

In addition, an important constraint arises from the partial width of the $Z$ into hadrons normalized to the partial width into leptons, which was measured precisely at LEP
\begin{align}
R_h^\text{exp} &= 20.767(25)~,
&
R_h^\text{SM} &= 20.740(17)~,
\end{align}
showing a $1.1\sigma$ tension with the best-fit SM prediction \cite{Baak:2012kk}.

In \UtreLC\ the modified left-handed $Z$ couplings of up and down quarks are equal to the ones of the $t$ and $b$ respectively, while the same is true in \UtreRC\ for the right-handed modified couplings. Analogously to the discussion in section \ref{CHMbounds/EWPT/Zcouplings}, one can write the modified $Z$ coupling of the top as
\begin{equation}
\frac{g}{c_w}\bar t \gamma^\mu \left[
(\tfrac{1}{2}-\tfrac{2}{3}s_w^2+\delta g^L_{Ztt}) P_L
+(-\tfrac{2}{3}s^2_w+\delta g^R_{Ztt})P_R
\right]tZ_\mu
\,,
\end{equation}
and one has
\begin{align}
\delta g_{Ztt}^L &=
\frac{v^2Y_{U}^2}{2m_{U}^2}\frac{x_ty_t}{Y_U} \,a+
\frac{g_\rho^2v^2}{4m_\rho^2}\frac{x_ty_t}{Y_U}\,b \,,
&
\delta g_{Ztt}^R &=
\frac{v^2Y_{U}^2}{2m_{U}^2}\frac{y_t}{x_t Y_U} \,c+
\frac{g_\rho^2v^2}{4m_\rho^2}\frac{y_t}{x_t Y_U}\,d \,,
\label{eq:Zbb}
\end{align}
with
\begin{center}
\begin{tabular}{c|ccc}
& doublet & triplet & bidoublet \\
\hline
$a$ & $-1/2$   & $-1$ & $-1/2$\\
$b$ & $-1/2$ & $-1$ & $-1$
\end{tabular}
\qquad
\begin{tabular}{c|ccc}
& doublet & triplet & bidoublet \\
\hline
$c$ & $1/2$   & $0$ & $0$\\
$d$ & $1/2$ & $0$ & $0$
\end{tabular}
\end{center}
Since the right-handed $Z$ coupling to $b$ and $t$ receives no contribution in the bidoublet model, there is no additional bound from $R_h$ in \UtreRC.
In \UtreLC we find the numerical bounds shown in table~\ref{tab:bounds-U3LC}.

In \UtreLC\ an additional bound arises from violations of quark-lepton universality.
Writing the $W$ couplings as
\begin{equation}
\frac{g}{\sqrt{2}}(1+\delta g^L_{W}) \bar u\, V_{ui}\gamma^\mu 
P_L
d_i W_\mu^+
\,,
\end{equation}
we find
\begin{align}
\delta g_{W}^L &=
\frac{Y_{U}^2v^2}{2m_{U}^2}\frac{x_ty_t}{Y_U} \,a_u +
\frac{Y_{D}^2v^2}{2m_{D}^2}\frac{x_ty_t}{Y_U} \,a_d +
\frac{g_\rho^2v^2}{4m_\rho^2}\frac{x_ty_t}{Y_U}\,b \,,
\end{align}
with
\begin{center}
\begin{tabular}{c|ccc}
& doublet & triplet & bidoublet \\
\hline
$a_u$ & $-1/2$ & $-1/2$ & $-1/2$
\\
$a_d$ & $-1/2$ & $-1/2$ & $-1/(2z^2)$
\\
$b$ & $-1$ & $-1$ & $-1$
\end{tabular}
\end{center}
The usual experimental constraint on the strength of the $W\bar{u}d_i$ couplings, normalized to the leptonic ones, is expressed by $(1+\delta g^L_{W})^2 \sum_i|V_{ui}|^2-1=(-1\pm6)\times10^{-4}$, which, from the unitarity of the $V_{ij}$ matrix, becomes $2 \delta g^L_{W}= (-1\pm6)\times10^{-4}$.  By requiring it to be fulfilled within $2\sigma$, we find the numerical bounds in table~\ref{tab:bounds-U3LC}.

Finally we note that, in contrast to the anarchic case, there are no {\em flavour-changing} $Z$ couplings neither in \UtreLC\ nor in \UtreRC. In the former case this is a general property of chirality-conserving bilinears (see next section), while in the latter it is a consequence of the fact that only the down-type mixings $\lambda_{Ld}$ affect the $Z$ vertex, which thus becomes flavour-diagonal in the mass basis.

\subsection{Tree-level  $\Delta F=2$ FCNCs}\label{CHMbounds/U3/DF2}

In $\U(3)^3_\text{RC}$  the effective Yukawa couplings have the form 
\begin{align}
&\bar{q}_L \hat{s}_{Lu} Y_U s_{Ru} u_R, & &\bar{q}_L\hat{s}_{Ld} Y_D s_{Rd} d_R,
\label{Yuk_RC}
\end{align}
where $\hat{s}_{Lu}$ is a generic $3\times 3$ mixing matrix and $Y_U$, $s_{Ru}$ are both proportional to the unit matrix. In $\U(3)^3_\text{LC}$ the role of the mixings is reversed and the Yukawa couplings take the form  
\begin{align}
&\bar{q}_L{s}_{Lu} Y_U \hat {s}_{Ru} u_R, & &\bar{q}_L {s}_{Ld} Y_D \hat {s}_{Rd} d_R.
\label{Yuk_LC}
\end{align}
At the same time, before going to the physical basis, the relevant interactions with the composite resonances have the form in $\U(3)^3_\text{RC}$ 
\begin{equation}
\rho_\mu (\bar{q}_L \hat{s}_{Lu} \gamma_\mu \hat{s}_{Lu}^{\dag} q_L),
\label{int_RC}
\end{equation}
and in $\U(3)^3_\text{LC}$ 
\begin{equation}
\rho_\mu (\bar{q}_L {s}_{Lu} \gamma_\mu {s}_{Lu}^* q_L).
\label{int_LC}
\end{equation}
In $\U(3)^3_\text{RC}$ the physical bases for up and down quarks are reached by proper $3\times 3$ unitary transformations that diagonalize $\hat{s}_{Lu}$ and $\hat{s}_{Ld}$
\begin{equation}
U_L^u\hat{s}_{Lu} U_R^{u\dag}=  \hat{s}_{Lu}^{\rm diag}~~~~ U_L^d\hat{s}_{Ld} U_R^{d\dag}=  \hat{s}_{Ld}^{\rm diag},
\end{equation} 
 so that the CKM matrix is $V = U_L^u U_L^{d\dag}$. In the same physical basis the interaction  (\ref{int_RC}) in the down sector becomes
\begin{equation}
\rho_\mu (\bar{d}_L V^{\dag} \hat{s}_{Lu}^{\rm diag} \gamma_\mu (\hat{s}_{Lu}^{\rm diag})^* V d_L) \approx 
\rho_\mu s_{Lt}^2 \xi_{ij} (\bar{d}_{Li}  \gamma_\mu  d_{Lj}),~~~~~\xi_{ij} = V_{ti}^* V_{tj}.
\label{int_dRC}
\end{equation}
Note that the ratio of the third to  the second entry in  $\hat{s}_{Lu}^{\rm diag}$ equals $y_t/y_c$. On the other hand a similar procedure for $\U(3)^3_\text{LC}$ leaves (\ref{int_LC}) unaltered since ${s}_{Lu}$ is proportional to the identity matrix.

Equations \eqref{int_LC} and \eqref{int_dRC} make clear that, while in \UtreLC\ there are no tree-level FCNCs at all \cite{Redi:2011zi}, minimally flavour violating tree-level FCNCs are generated in \UtreRC\ \cite{Barbieri:2012uh,Redi:2012uj}. The Wilson coefficients of $\Delta F=2$ operators are given by (\ref{eq:DF2}), with the couplings 
\begin{align}
g_{Ld}^{ij} &= \xi_{ij}\frac{x_ty_t}{Y_U} \,,
&
g_{Rd}^{ij} &\approx 0 \,.
\label{U(3)RC-FC}
\end{align}
The coefficients $c_X^{qAB}$ are not the ones listed in table~\ref{tab:df2coeffs}, since in $\U(3)^3$ there is an additional contribution to \eqref{cVdLL}--\eqref{cSdLR} from flavour gauge bosons. However the only relevant $\Delta F = 2$ operator is $Q_V^{dLL}$ in \UtreRC, for which one obtains $c_V^{dLL} = -{29}/{36}$ instead of the value reported in the table.

One obtains the bounds shown in table~\ref{tab:bounds-U3RC}. The bound from $D$-$\bar D$ mixing turns out to be numerically irrelevant.

We stress that, in contrast to the anarchic case, there is no $\Ord(1)$ uncertainty in these bounds since the composite Yukawas are proportional to the identity. Furthermore, since the model is minimally flavour violating, there is no contribution to the meson mixing phases and the new physics effects in the $K$, $B_d$ and $B_s$ systems are prefectly correlated.

\begin{table}[tbp]
\renewcommand{\arraystretch}{1.3}
\centering
\begin{tabular}{cccc}
\hline
Observable & \multicolumn{3}{c}{Bounds on $m_{\psi}$ [TeV]} \\
&  doublet & triplet  & bidoublet\\
\hline
$R_h$ & $7.2 ~\sqrt{x_tY}$  &$6.0 ~\sqrt{x_tY}$  &$4.9 ~\sqrt{x_tY_U}$\\
$V_\text{CKM}$ & $7.4 ~\sqrt{x_tY}$ & $7.4 ~\sqrt{x_tY}$ & $6.0 ~\sqrt{x_tY_U}$ \\
$pp\to jj$ ang. dist. & $3.4  ~x_t$ & $4.2  ~x_t$ & $4.2  ~x_t$  \\
\hline
\end{tabular}
\caption{Lower bounds on the fermion resonance mass $m_\psi$ in TeV in \UtreLC.}
\label{tab:bounds-U3LC}
\end{table}

\begin{table}[tbp]
\renewcommand{\arraystretch}{1.3}
\centering
\begin{tabular}{cc}
\hline
Observable & Bounds on $m_{\psi}$ [TeV] \\
\hline
$\epsilon_K(Q^{LL}_V)$ & $3.7 ~x_t$ \\
$B_d$-$\bar B_d$ & $3.2 ~x_t$ \\
$B_s$-$\bar B_s$  & $3.6 ~x_t$\\
$pp\to jj$ ang. dist. & $3.0/x_t$ \\
\hline
\end{tabular}
\caption{Lower bounds on the fermion resonance mass $m_\psi$ in TeV in \UtreRC\ (bidoublet model).}
\label{tab:bounds-U3RC}
\end{table}

\subsection{Loop-induced chirality-breaking effects}\label{CHMbounds/U3/loop}

Flavour-changing chirality-breaking effects in $\U(3)^3$ occur only when elementary-composite mixings are included inside the loops. At least for moderate mixings, the bounds are of the form $m_\psi > (0.5\text{--}1.5) \sqrt{Y/x_t} $ TeV in the \UtreLC\ case, or $m_\psi > (0.5\text{--}1.5)\sqrt{Y x_t}$ TeV in the \UtreRC\ case. The stronger bounds from  quark EDMs, similar to the ones of the anarchic case,  disappear if the strong sector conserves \CP. This is automatically realized, in our effective Lagrangian description, if the ``wrong chirality'' Yukawas vanish or are aligned in phase with the $Y$'s. On the contrary, in the anarchic case this condition is in general not sufficient to avoid large EDMs.

\subsection{Compositeness constraints}\label{CHMbounds/U3/compositeness}

Since one chirality of first-generation quarks has a sizable degree of compositeness in the $\U(3)^3$ models, a significant constraint arises from the angular distribution of high-mass dijet events at LHC, which is modified by local four-quark operators obtained after integrating out the heavy vector resonances related to the global $\SU(3)_c\times \SU(2)_L\times \SU(2)_R\times \U(1)_X$ as well as the flavour symmetry in the strong sector, $\U(3)$ in the case of \UtreLC\ and $\U(3)\times \U(3)$ in the case of \UtreRC.

In general, there are ten four-quark operators relevant in the dijet angular distribution \cite{Domenech:2012ai}, which we list here for convenience
\begin{align}
\mathcal O^{(1)}_{uu}&=(\bar{u}_R\gamma^\mu u_R)(\bar{u}_R\gamma_\mu u_R)~,
&
\mathcal O^{(1)}_{dd}&=(\bar{d}_R\gamma^\mu d_R)(\bar{d}_R\gamma_\mu d_R)~,
\nonumber\\
\mathcal O^{(1)}_{ud}&=(\bar{u}_R\gamma^\mu u_R)(\bar{d}_R\gamma_\mu d_R)~,
&
\mathcal O^{(8)}_{ud}&=(\bar{u}_R\gamma^\mu T^A u_R)(\bar{d}_R\gamma_\mu T^A d_R)~,
\nonumber\\
\mathcal O^{(1)}_{qq}&=(\bar{q}_L\gamma^\mu q_L)(\bar{q}_L\gamma_\mu q_L)~,
&
\mathcal O^{(8)}_{qq}&=(\bar{q}_L\gamma^\mu T^A q_L)(\bar{q}_L\gamma_\mu T^A q_L)~,
\nonumber\\
\mathcal O^{(1)}_{qu} &= (\bar{q}_{L} \gamma^\mu q_{L}) (\bar{u}_{R} \gamma_{\mu}  u_R)~,
&
\mathcal O^{(8)}_{qu} &= (\bar{q}_{L} \gamma^\mu T^A q_{L}) (\bar{u}_{R} \gamma_{\mu}  T^A u_R)~,
\nonumber\\
\mathcal O^{(1)}_{qd} &= (\bar{q}_{L} \gamma^\mu q_{L}) (\bar{d}_{R} \gamma_{\mu}  d_R)~,
&
\mathcal O^{(8)}_{qd} &= (\bar{q}_{L} \gamma^\mu T^A q_{L}) (\bar{d}_{R} \gamma_{\mu}  T^A d_R)~.
\label{eq:jj-operators}
\end{align}
The coupling of a first generation quark mass eigenstate to a heavy vector resonance was given in \eqref{eq:gG}. Neglecting electroweak gauge couplings, one can then write the Wilson coefficients of the above operators in general as
\begin{equation}
C_{ab}^{(1,8)} = \frac{g_\rho^2}{m_\rho^2}\left[
s_{a}^2
s_{b}^2
c_{ab}^{(1,8)}
+
\left(
\frac{g_3^4}{g_\rho^4}
-(s_{a}^2+s_{b}^2)
\frac{g_3^2}{g_\rho^2}
\right)
d_{ab}^{(1,8)}
\right],
\end{equation}
where $(a,b)=(q,u,d)$ and $s_{u,d}^2\equiv s_{Ru,d}^2$, $s_q^2\equiv s_{L}^2$ (in the bidoublet model, for simplicity we will neglect terms with $s_{Ld}^2$ over terms with $s_{Lu}^2$).
The numerical coefficients $c_{ab}^{(1,8)}$ depend on the electroweak structure and on the flavour group and are collected in table~\ref{tab:dijet} together with the $d_{ab}^{(1,8)}$.

\begin{table}[tbp]
\renewcommand{\arraystretch}{1.5}
\centering
\begin{tabular}{lcccccccccc}
\hline
&$c_{uu}^{(1)}$ & $c_{dd}^{(1)}$ & $c_{ud}^{(1)}$ & $c_{ud}^{(8)}$ &
$c_{qq}^{(1)}$ & $c_{qq}^{(8)}$ & $c_{qu}^{(1)}$ & $c_{qu}^{(8)}$ & $c_{qd}^{(1)}$ & $c_{qd}^{(8)}$ \\
\hline
doublet $\U(3)^3_\text{LC}$ & $-\frac{17}{36}$ & $-\frac{17}{36}$ & $-\frac{1}{9}$ & $-1$ & $-\frac{5}{36}$ & $-1$ & $-\frac{13}{36}$ & $-1$ & $-\frac{13}{36}$ & $-1$ \\
triplet $\U(3)^3_\text{LC}$& $-\frac{5}{9}$ & $-\frac{19}{18}$ & $-\frac{7}{9}$ & $-1$ & $-\frac{17}{36}$ & $-1$ & $-\frac{7}{9}$ & $-1$ & $-\frac{7}{9}$ & $-1$ \\
bidoublet $\U(3)^3_\text{LC}$ &  $-\frac{5}{9}$ & $-\frac{7}{18}$ & $-\frac{1}{9}$ & $-1$ & $-\frac{17}{36}$ & $-1$ & $-\frac{7}{9}$ & $-1$ & $-\frac{1}{9}$ & $-1$ \\
bidoublet $\U(3)^3_\text{RC}$ & $-\frac{5}{9}$ & $-\frac{7}{18}$ & $\frac{2}{9}$ & $-1$ & $-\frac{17}{36}$ & $-1$ & $-\frac{7}{9}$ & $-1$ & $\frac{2}{9}$ & $-1$ \\
\hline
\hline
&$d_{uu}^{(1)}$ & $d_{dd}^{(1)}$ & $d_{ud}^{(1)}$ & $d_{ud}^{(8)}$ &
$d_{qq}^{(1)}$ & $d_{qq}^{(8)}$ & $d_{qu}^{(1)}$ & $d_{qu}^{(8)}$ & $d_{qd}^{(1)}$ & $d_{qd}^{(8)}$ \\
\hline
all models & $-\frac{1}{6}$ & $-\frac{1}{6}$ & $0$ & $-1$ & $0$ & $-\frac{1}{2}$ & $0$ & $-1$ & $0$ & $-1$\\
\hline
\end{tabular}
\caption{Coefficients $c_{ab}^{(1,8)}$ relevant for dijet bounds in the doublet, triplet and bidoublet models as well as the coefficients $d_{ab}^{(1,8)}$, which are independent of the flavour and electroweak groups.}
\label{tab:dijet}
\end{table}

The only relevant operators in \UtreLC\ are $\mathcal O_{qq}^{(1,8)}$. Their Wilson coefficients read
\begin{align}
C_{qq}^{(1)} = -\frac{a}{36}\frac{g_\rho^2}{m_\rho^2}\left(\frac{x_t y_t}{Y_U}\right)^2\,,
\qquad
C_{qq}^{(8)} = -\frac{g_\rho^2}{m_\rho^2}\left(\frac{x_t y_t}{Y_U}\right)^2\,,
\end{align}
where $a=5$ in the doublet model and $a=17$ in the triplet and bidoublet models.
Using the updated version of \cite{Domenech:2012ai},
we obtain the bound in the last row of table~\ref{tab:bounds-U3LC}.
In \UtreRC\ the operators with right-handed quarks are relevant, i.e. $\mathcal O_{uu,dd,ud}^{(1)}$ and $\mathcal O_{ud}^{(8)}$.
Numerically, we find the bound on $\mathcal O_{uu}^{(1)}$ to give the most significant constraint on the model parameters. Its Wilson coefficient reads
\begin{align}
C_{uu}^{(1)} = -\frac{5}{9}\frac{g_\rho^2}{m_\rho^2}\left(\frac{y_t}{x_t Y_U}\right)^2\,.
\end{align}
and the resulting numerical constraint is shown in the last row of table~\ref{tab:bounds-U3RC}.

\subsection{Direct bounds on vector resonances}\label{CHMbounds/U3/jjresonances}

As discussed in section~\ref{CHMbounds/anarchy/jjresonances}, direct bounds on $m_\rho$ are obtained from searches for peaks in the invariant mass of dijets at LHC. In $\U(3)^3$ the production cross sections can be larger than in the anarchic case due to the sizable degree of compositeness of first-generation quarks, which causes a large contribution to \eqref{eq:gG} coming from fermion mixings. Neglecting the contribution due to mixing of the vector resonances with the gauge bosons, the production cross section of a gluon resonance reads
\begin{align}
\sigma(pp\to G^*) &=  \frac{2\pi}{9s}g_\rho^2
\left[
s_{L,Ru}^4 \mathcal L_{u\bar u}(s,m_\rho^2)
+
s_{L,Rd}^4 \mathcal L_{d\bar d}(s,m_\rho^2)
\right],
\label{eq:sigmapprhoU3}
\end{align}
where the $\mathcal{L}$'s are the parton-parton luminosity functions \eqref{partonlumi}, and where the $L$ is valid in \UtreLC\ and the $R$ in \UtreRC.
In \UtreLC\ the branching ratio to two jets reads approximately
\begin{align}
\text{BR}(G^*\to jj) =
\frac{2 s_{Lu}^4+3 s_{Ld}^4+s_{Rb}^4}{3 s_{Lu}^4+s_{Rt}^4+3 s_{Ld}^4+s_{Rb}^4
}\,,
\end{align}
and is typically larger than in the anarchic case. Similarly, in \UtreRC\ one has
\begin{equation}
\text{BR}(G^*\to jj) =
\frac{
2 s_{Ru}^4
+
s_{Lb}^4
+
3 s_{Rd}^4
}{
s_{Lt}^4
+
3 s_{Ru}^4
+
s_{Lb}^4
+
3 s_{Rd}^4
}\,.
\end{equation}

To judge if the most recent experimental bounds by ATLAS \cite{ATLAS:2012joa} and CMS \cite{CMS:2012eza} have already started to probe the $\U(3)^3$ parameter space, we evaluate the cross section for maximal mixing, i.e. $x_t=Y/y_t$ in \UtreLC\ and $x_t=y_t/Y$ in \UtreRC, for a 3~TeV gluon resonance, i.e. only marginally heavier than allowed by the $S$ parameter (cf. table~\ref{tab:bounds-ew}). For \UtreLC\ we obtain 
\begin{align}
\sigma(pp\to G^*) \approx 13g_\rho^2 ~\text{fb}~,
\qquad
\text{BR}(G^*\to jj) \approx 58\% ~(83\%) \text{ for } Y=1 ~(4\pi)~,
\end{align}
and for \UtreRC
\begin{align}
\sigma(pp\to G^*) \approx 30g_\rho^2 ~\text{fb}~,
\qquad
\text{BR}(G^*\to jj) \approx 69\% ~(67\%) \text{ for } Y=1 ~(4\pi)~.
\end{align}
This is to be compared to the ATLAS bound $\sigma\times\text{BR}\times A<7~\text{fb}$, where $A$ is the acceptance. We conclude that, assuming an acceptance of the order of 60\% \cite{ATLAS:2012joa}, maximal mixing is on the border of exclusion in \UtreLC\ and already excluded in \UtreRC\ for a 3~TeV gluon resonance.
We note however that maximal mixing is already disfavoured by the indirect bounds discussed above.

\section{A flavour-symmetric composite sector: $\U(2)^3$}\label{CHMbounds/U2}

In $\U(2)^3$ models one considers the $\U(2)_q\times \U(2)_u\times \U(2)_d$ symmetry described in chapter~\ref{U2}, under which the first two generations of quarks transform as doublets and the third generation as singlets, broken in specific directions dictated by minimality \cite{Barbieri:2011ci,Barbieri:2012uh}. Compared to \Utre, one has a larger number of free parameters, but can break the flavour symmetry {\em weakly}, since the large top Yukawa is invariant under \Udue.
Let us define
\begin{align}
Q ^{u} &= \begin{pmatrix}\boldsymbol{Q^{u}}\\ Q_3^{u}\end{pmatrix}, & U &= \begin{pmatrix}\boldsymbol{U}\\T \end{pmatrix}, & q_L &= \begin{pmatrix}\ELqL\\ \EHqL\end{pmatrix}, & u_R &= \begin{pmatrix}\ELuR\\ \EHuR\end{pmatrix},
\end{align}
where the first two generation doublets are written in boldface, and the same for down-type quarks.

Analogously to the \Utre\ case, in the strong sector the flavour groups are $\U(2)_{Q+U+D}$ or $\U(2)_{Q^u+U}\times \U(2)_{Q^d+D}$, and again we expect the presence of flavour gauge bosons associated with the global symmetries of the strong sector.

We consider the two following situations:

\begin{enumerate}
\item In {\it left-compositeness}, to be called {\UdueLC}, the left mixings are diagonal with the first two entries equal to each other and the only sources of $\U(2)^3$ breaking reside in the right-handed mixings.
\item In {\it right-compositenss}, to be called {\UdueRC}, the right mixings are diagonal with the first two entries equal to each other and the only sources of $\U(2)^3$ breaking reside in the left-handed mixings.
\end{enumerate}
As before right-compositeness can be meaningfully defined only in the bidoublet model.
The mixing Lagrangians in the two cases are respectively\footnote{We write the Lagrangians for the bidoublet model. The doublet and triplet cases are analogous, with $Q^u$ and $Q^d$ replaced by a single $Q$.}
\begin{align}
\mathcal{L}_\text{mix}^{\U(2)^3_\text{LC}} &=
m_{U3}\lambda_{Lu3}\, \EHqLbar \CHquR +
m_{U2}\lambda_{Lu2} \,\ELqLbar\CLquR +
m_{U3}\lambda_{Ru3}\, \CHuLbar \EHuR
\notag\\
&+m_{U2}\,d_u\, (\CLuLbar\V)\EHuR +
m_{U2}\,\CLuLbar \Delta_u \ELuR +
\text{h.c.}
+ (u,U,t,T\to d,D,b,B)
\label{mixing2L}
\end{align}
and
\begin{align}
\mathcal{L}_\text{mix}^{\U(2)^3_\text{RC}} &=
m_{U3}\lambda_{Ru3}\, \CHuLbar\EHuR +
m_{U2}\lambda_{Ru2} \, \CLuLbar\ELuR +
m_{U3}\lambda_{L(u)3}\, \EHqLbar\CHquR
\notag\\
&+m_{U3}\,d_u\, (\ELqLbar\V)\CHquR +
m_{U2}\,\ELqLbar \Delta_u \CLquR +
\text{h.c.}
+ (u,U,t,T\to d,D,b,B).
\label{mixing2R}
\end{align}
The diagonal mixings in the first line of \eqref{mixing2L} and \eqref{mixing2R} break the symmetry of the strong sector down to $\U(2)_q\times \U(2)_u\times \U(2)_d$. This symmetry is in turn broken minimally by the spurions $\V$, $\Delta_u$ and $\Delta_d$ transforming as in \eqref{DeltaY}, \eqref{V}. Here we do not discuss the case of generic $\U(2)^3$ breaking \cite{Barbieri:2012bh}.

Using $\U(2)^3$ transformations of the quarks they can be put in the simple form
\begin{align}
\V &= \begin{pmatrix} 0 \\ \epsilon_L \end{pmatrix},&
\Delta_u &=
\begin{pmatrix}
c_u & s_u e^{i\alpha_u} \\
-s_u e^{-i\alpha_u} & c_u 
\end{pmatrix}
\begin{pmatrix}
\lambda_{Xu1} & 0 \\
0 & \lambda_{Xu2}
\end{pmatrix}, & (u\leftrightarrow d),
\end{align}
where $X=R,L$ in left- and right-compositeness, respectively.
The SM Yukawa couplings \eqref{SMYuk} can be written in terms of the spurions as
\begin{align}\label{SMYukLC}
\hat y_u &= \begin{pmatrix}a_u\, \Delta_u & b_t e^{i\phi_t}\V\\ 0 & y_t\end{pmatrix}, &
\hat y_d &= \begin{pmatrix}a_d\, \Delta_d & b_b e^{i\phi_b}\V\\ 0 & y_b\end{pmatrix},
\end{align}
where
\begin{align}
y_t &= Y_{U3}s_{Lu3}s_{Ru3},
\\
a_u &= Y_{U2}s_{Lu2},& b_t &= Y_{U2}s_{Lu2}\,d_u, & &\text{ in left-compositeness},\\ a_u &= Y_{U2}s_{Ru2},& b_t &= Y_{U3}s_{Ru3}\,d_u,& &\text{ in right-compositeness},
\end{align}
$s_{Xi} = \lambda_{Xi}/\sqrt{1 + (\lambda_{Xi})^2}$, and similarly for $a_d$, $b_b$ and $y_b$.
Here and in the following we consider all the parameters real, factoring out the phases everywhere as in \eqref{SMYukLC}. The $\hat y_{u,d}$ are diagonalized to a sufficient level of approximation by the pure unitary transformations of the left-handed quarks\footnote{Redefining the phases of the quarks one can move all the phases in $\hat y_u$ to $\hat y_d$, in which case one recovers the transformations $L_u$ and $U_d$ of \eqref{U2transformationsreal}. Here we choose a different convention for simplicity.}
\begin{align}\label{UL}
U_u &\simeq \begin{pmatrix}c_{u} & s_u e^{i\alpha_u} & -s_u s_t e^{i(\alpha_u + \phi_t)}\\
-s_u e^{-i\alpha_u} & c_u & -c_u s_t e^{i\phi_t}\\
0 & s_t e^{-i\phi_t} & 1
\end{pmatrix}, &
U_d &\simeq \begin{pmatrix}c_{d} & s_d e^{i\alpha_d} & -s_d s_b e^{i(\alpha_d + \phi_b)}\\
-s_d e^{-i\alpha_d} & c_d & -c_d s_b e^{i\phi_b}\\
0 & s_b e^{-i\phi_b} & 1
\end{pmatrix},\end{align}
where
\begin{align}
s_t &= Y_{U2}s_{Lu2}\frac{d_u\epsilon_L}{y_t}, & s_b &= Y_{D2}s_{Ld2}\frac{d_d\epsilon_L}{y_b},& &\text{in left-compositeness},
\label{eq:stbLC}
\\
s_t &= Y_{U3}s_{Ru3}\frac{d_u\epsilon_L}{y_t}, & s_b &= Y_{D3}s_{Rd3}\frac{d_d\epsilon_L}{y_b},& &\text{in right-compositeness}.
\end{align}

The CKM matrix is $V = U_uU_d^{\dag}$ and, after a suitable redefinition of quark phases, takes the form \eqref{CKM}.

In \UdueLC\ and \UdueRC\ the first and second generation elementary-composite mixings are expected to be significantly smaller than the third generation ones, so that the electroweak 
precision constraints and the collider phenomenology are virtually identical to the anarchic case and the most serious problems plaguing the \Utre\ models are absent. The most important difference concerns the flavour constraints.

\subsection{Tree-level $\Delta F = 2$ FCNCs}\label{CHMbounds/U2/DF2}

\begin{figure}[tb]
\centering
\includegraphics[width=\textwidth]{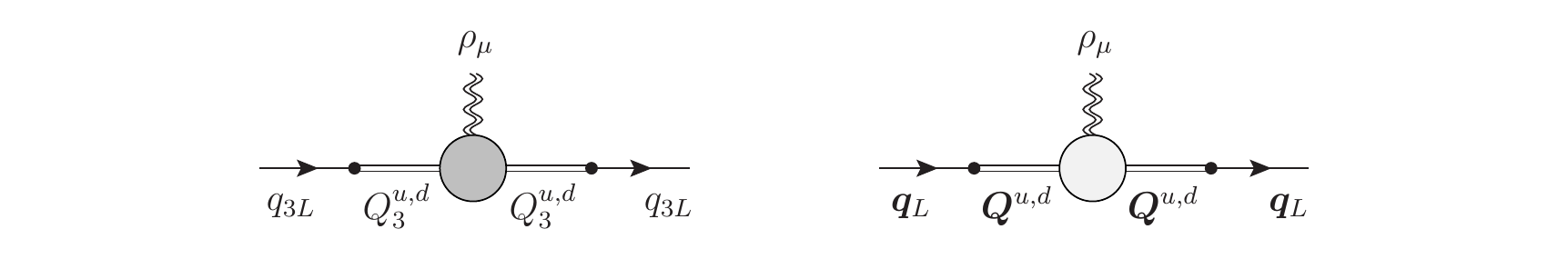}
\caption{Leading contribution to flavour conserving bilinears in \UdueLC. Flavour violation arises after rotating to the physical quark basis, because of the mismatch between the coefficients of the $\EHqLbar\gamma_\mu \EHqL$  and $\ELqLbar\gamma_\mu \ELqL$ currents.\label{B}}
\end{figure}

The Wilson coefficients of $\Delta F=2$ operators generated in \UdueLC\ and \UdueRC\ are again given by (\ref{eq:DF2}). Since in $\U(2)^3$ all the flavour effects are generated by mixing with third generation partners, which are not charged under any of the $\U(2)$ flavour groups, there is no relevant additional effect coming from flavour gauge bosons, and also the coefficients of table \ref{tab:df2coeffs} are valid.

Equations \eqref{int_RC}, \eqref{int_LC} remain formally true in $\U(2)^3$  as well, with the following qualifications.  $Y_U, s_{Ru},  s_{Lu}$ are no longer proportional to the identity but are still  diagonal with only the first  two entries equal to each other. At the same time minimal breaking of $\U(2)^3$ leads to a special form of the matrices $ \hat{s}_{Lu},  \hat{s}_{Ru}$ that allows to diagonalize approximately the Yukawa couplings  by pure left unitary transformations of the form \eqref{UL}.

In $\U(2)^3_\text{RC}$ these transformations lead to exactly the same equation as (\ref{int_dRC}), 
and the flavour-changing couplings are the same as in \UtreRC,
\begin{align}
g_{Ld}^{i3} &= \xi_{i3} \,\frac{x_ty_t}{Y_U} \,,
&
g_{Ld}^{12} &= \xi_{12} \, \frac{x_t y_t}{Y_U} \,,
&
g_{Rd}^{ij} &\approx 0 \,.
\end{align}
Thus, as in \UtreRC, there is no new phase in meson-antimeson mixing and the NP effects in the $K$, $B_d$ and $B_s$ systems are perfectly correlated. The resulting bounds are shown in table~\ref{tab:bounds-U2RC}.

\begin{table}[tbp]
\renewcommand{\arraystretch}{1.3}
\centering
\begin{tabular}{cc}
\hline
Observable & Bounds on $m_{\psi}$ [TeV] \\
\hline
$\epsilon_K(Q^{LL}_V)$ & $3.3 ~x_t$ \\
$B_d$-$\bar B_d$ & $2.8 ~x_t$ \\
$B_s$-$\bar B_s$  & $3.1 ~x_t$\\
\hline
\end{tabular}
\caption{Lower bounds on the fermion resonance mass $m_\psi$ in TeV in \UdueRC\ (bidoublet model).}
\label{tab:bounds-U2RC}
\end{table}

In the $\U(2)^3_\text{LC}$ case, instead, the analogous of equation (\ref{int_LC}) in the down sector becomes
\begin{equation}
\rho_\mu (\bar{d}_L U_d {s}_{Lu} \gamma_\mu {s}_{Lu}^* U_{d}^{\dag} d_L) \approx 
\rho_\mu s_{Lt}^2 \chi_{ij} (\bar{d}_{Li}  \gamma_\mu  d_{Lj}),~~~~~\chi_{ij} = U^d_{i3} U^{d*}_{j3}.
\label{int_dLC}
\end{equation}
Remember that, contrary to the $\U(3)^3_\text{RC}$ case, ${s}_{Lu}$, although still diagonal, is not proportional to the unit matrix. Hence a flavour violation survives (see figure~\ref{B}), with flavour-changing couplings that read
\begin{align}
g_{Ld}^{i3} &= \xi_{i3} \, r_b\frac{x_t y_t}{Y_U} \,,
&
g_{Ld}^{12} &= \xi_{12} \, |r_b|^2 \frac{x_t y_t}{Y_U} \,,
&
g_{Rd}^{ij} &\approx 0 \,,
\label{FCU2LC}
\end{align}
where 
\begin{equation}
\label{eq:rb}
r_b \equiv \frac{s_b}{s} e^{i(\chi - \phi_b)}
\end{equation}
is a free complex parameter, and $s_b e^{i\phi_b} - s_t e^{i\phi_t}\equiv s e^{i\chi}$.
As a consequence there is a new, universal phase in $B_d$ and $B_s$ mixing, while the $K$-$\bar K$ amplitude is aligned in phase with the SM. We find the bounds in table~\ref{tab:bounds-U2LC}.

Note that in \UdueLC, for $Y_{U2}\sim Y_{D2}\sim \Ord(1)$ and $d_u,d_d \lesssim \Ord(1)$, (\ref{eq:stbLC}) leads to two possibilities:
\begin{enumerate}
 \item $s_t\ll s_b$, i.e. $|r_b|\approx1$;
 \item $s_t\sim s_b\sim|V_{cb}|$, which allows $|r_b|$ to deviate from 1 but requires at the same time $s_{Lu2}\epsilon_L\sim |V_{cb}|$.
\end{enumerate}
In the first case one would have $m_{\psi} \gtrsim 1\text{--}1.5$ TeV from the flavour bounds of table \ref{tab:bounds-U2LC}, while in the second case, if the parameter $|r_b|$ is somewhat less than 1, these bounds can be in agreement with experiment even for light fermion resonances. One can obtain a minimal value of $m_\psi \simeq 0.6$ TeV for $|r_b| \sim 0.25$ and $Y \sim 1$. However, to avoid a too large $\U(2)^3$-breaking -- i.e. a large $\epsilon_L$ -- the mixing angle of the first generations quarks $s_{Lu2}$ cannot be too small. This in turn has to be confronted with the lower bounds on $m_\psi$ from $R_h$, $V_{\text{CKM}}$ and the dijet angular distribution shown in table \ref{tab:bounds-U2LC-sL2}: to make them consistent with $m_\psi \simeq 0.6$ TeV, it must 
be $\epsilon_L \gtrsim 0.3$. Note anyhow that we are not treating $\epsilon_L$ as an expansion parameter.

In the bidoublet model, in addition to \eqref{int_dLC} there are also the terms coming from the mixing with the $\bar Q^d\gamma_{\mu}Q^d$ current, which are suppressed as $1/z^2$.
In the up-quark sector with right-compositeness only this suppressed contribution from $Q^d$ gives rise to flavour violation, while in left-compositeness the analog of \eqref{int_dLC} holds, with $U_d$ replaced by $U_u$.
Note that the contribution to the $\Delta C=2$ operator is proportional to $|1-r_b|^2$, so it cannot be reduced simultaneously with the contribution in the down-quark sector. However, it turns out that it is numerically insignificant. Since furthermore the contribution is real -- a general prediction of the \Udue\ symmetry for $1\leftrightarrow 2$ transitions -- the expected improvement of the bound on \CP\ violation in $D$-$\bar D$ mixing will have no impact.

Flavour violation from chirality-conserving right-handed quark bilinears is suppressed, a general property of the Minimal $\U(2)^3$ framework \cite{Barbieri:2011ci,Barbieri:2012uh}.

\begin{table}[tbp]
\renewcommand{\arraystretch}{1.3}
\centering
\begin{tabular}{cccc}
\hline
Observable & \multicolumn{3}{c}{Bounds on $m_{\psi}$ [TeV]} \\
& doublet & triplet& bidoublet  \\
\hline
$\epsilon_K(Q^{LL}_V)$ & $2.3 ~x_t|r_b|^2$ & $3.3  ~x_t|r_b|^2$ & $3.3  ~x_t|r_b|^2$ \\
$B_d$-$\bar B_d$ & $2.3 ~x_t|r_b|$ & $3.4  ~x_t|r_b|$ & $3.4  ~x_t|r_b|$ \\
$B_s$-$\bar B_s$ & $2.3 ~x_t|r_b|$ & $3.4  ~x_t|r_b|$ & $3.4  ~x_t|r_b|$ \\
$K_L\to\mu\mu$ & $3.8 ~\sqrt{x_tY}|r_b|$ && $3.8  ~Y_D |r_b| \sqrt{x_t/Y_{U}}/z$ \\
$b\to s\ell\ell$ & $3.5 ~\sqrt{x_tY|r_b|}$ && $3.5  ~Y_D \sqrt{x_t|r_b|/Y_{U}}/z$ \\
\hline
\end{tabular}
\caption{Lower bounds on the fermion resonance mass $m_\psi$ in TeV in \UdueLC.}
\label{tab:bounds-U2LC}
\end{table}

\begin{table}[tbp]
\renewcommand{\arraystretch}{1.3}
\centering
\begin{tabular}{cccc}
\hline
Observable & \multicolumn{3}{c}{Bounds on $m_{\psi}$ [TeV]} \\
&  doublet & triplet  & bidoublet\\
\hline
$R_h$ & $7.2 ~s_{L2}Y_2$  &$6.8 ~s_{L2}Y_2$  &$5.6 ~s_{Lu2}Y_{U2}$\\
$V_\text{CKM}$ & $8.4 ~s_{L2}Y_2$  &$6.8 ~s_{L2}Y_2$  &$6.8 ~s_{Lu2}Y_{U2}$\\
$pp\to jj$ ang. dist. & $4.3  ~s_{L2}^2Y_{2}$ & $5.3  ~s_{L2}^2Y_{2}$ & $5.3  ~s_{Lu2}^2Y_{U2}$  \\
\hline
\end{tabular}
\caption{Lower bounds on the fermion resonance mass $m_\psi$ in TeV in \UdueLC\ from left-handed 1st and 2nd generation quarks mixed with the composite resonances by an angle $s_{Lu2}$.}
\label{tab:bounds-U2LC-sL2}
\end{table}

\subsection{Loop-induced chirality-breaking effects}\label{CHMbounds/U2/loop}

Flavour-changing chirality-braking effects are produced at one loop in the strong interactions. One expects in general  flavour-changing chirality-breaking effects in $\U(2)^3$ with bounds on the fermion resonances similar to the one of the anarchic case,  $m_\psi > (0.5\text{--}1.5)Y$\,TeV.
With \CP\ conservation in the strong sector, however, the contributions to the quark EDMs would arise only at higher orders in the $\U(2)^3$ breaking terms, so that they would not be significant for the current limit on the neutron EDM.

\begin{figure}[tb]
\centering
\subfigure[Right-handed compositeness \label{uno}]{
\includegraphics[width=\textwidth]{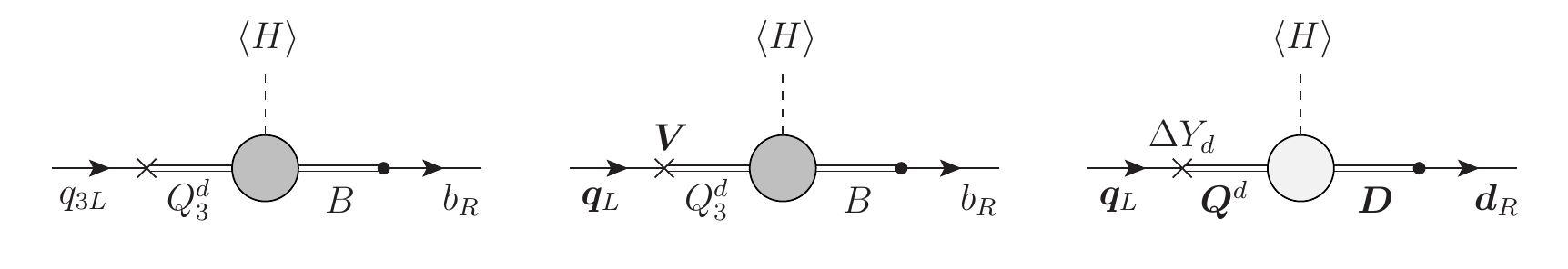}
}
\subfigure[Left-handed compositeness \label{due}]{
\includegraphics[width=\textwidth]{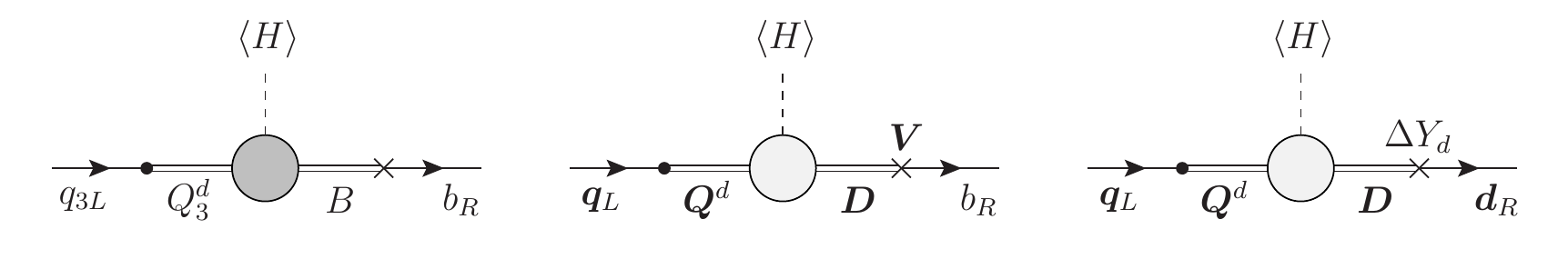}
}
\caption{Leading contributions to chirality breaking bilinears (mass terms) from insertions of composite fermions, in the two cases of right- and left-handed compositeness. Crosses denote the mixings that break flavour symmetry, while dots denote the diagonal mixings. The strong dynamics in the first two diagrams in $(a)$ is the same, as it is in the last two diagrams in $(b)$. Magnetic dipole terms have the same structure.\label{A}}
\end{figure}

In figure~\ref{A} we show the leading terms contributing to the effective Yukawa couplings of the down quarks to the Higgs boson. In the case of left-compositeness the strong Yukawa couplings of the Higgs doublet to the heavy quarks, although flavour conserving, are different for the third generation with respect to the first two. Let us now consider the analogous diagrams with a photon or gluon line also attached to the strong Yukawa-coupling vertex, i.e. an operator contributing, when the Higgs boson gets a vev, to the dipole moments of the heavy quarks. Here too the dipole moment of the third-generation composite quarks is in general different from the one of the first two generations, but, what is more important, the two dipole moments will not be in the same ratio as the corresponding Yukawa couplings. In turn this produces a misalignment in flavour space of the mass versus the magnetic moment operator, i.e. a residual flavour breaking interaction with the size expected in the generic EFT approach, except for the fact that the presence of a new phase requires \CP\ violation in the strong sector.

In the case of right-compositeness, on the contrary, the ratio of the coefficients of the $\EHqLbar\EHdR$ and $\EHqLbar\sigma_{\mu\nu}\EHdR$ bilinears coincides with the ratio of the coefficients of $\ELqLbar\EHdR$ and $\ELqLbar\sigma_{\mu\nu}\EHdR$, because the same heavy quarks mediate the two effects. For this reason, when going to the physical quark basis, the dipole terms are diagonalized along with the mass terms, and there are no flavour-changing effects at leading order.

\subsection{Flavour-changing $Z$ couplings}\label{CHMbounds/U2/Zcouplings}

In \UdueRC\ flavour-changing $Z$ couplings are absent at tree level. In \UdueLC\ the left-handed couplings do arise, while the right-handed couplings are strongly suppressed. Similarly to the anarchic case, one can write them as
\begin{gather}
\delta g_{Zbd^i}^L \sim  \xi_{i3} ~ r_b ~ \delta g_{Zbb}^L~,
\qquad
\delta g_{Zsd}^L \sim  \xi_{12} ~ |r_b|^2 ~ \delta g_{Zbb}^L~.
\label{eq:ZbsL-U2}
\end{gather}
One obtains the bounds in the last two lines of table~\ref{tab:bounds-U2LC}, which are weaker than the analogous bounds from $R_b$ unless $|r_b|>1$.
An important difference with respect to the anarchic case is the absence of sizable flavour-changing {\em right-handed} $Z$ couplings, which can be probed e.g. in certain angular observables in $B\to K^*\mu^+\mu^-$ decays \cite{Altmannshofer:2008dz}.

\section{Summary and comparison of bounds}\label{CHMbounds/summary}

We have analyzed the bounds coming from flavour observables, electroweak precision test, and the constraint coming from the Higgs mass in a number of different models of a composite Higgs, investigating different options for the generation of the flavour pattern of the SM quarks. Here we summarize the most relevant bounds for each model, trying to delineate a picture of what are the lowest allowed values for fermion resonance masses compatible with all the constraints, and of which are the observables most sensitive to new physics in each case.

\bigskip

In {\it Anarchy}, if the bound coming from the $Q_S^{LR}$ contribution to $\epsilon_K$ is taken at face value, the fermion resonances should be far too heavy to be consistent with a naturally light Higgs boson and certainly unobservable, either directly or indirectly. Note in particular the growth of this bound with $z$ in the bidoublet model.

In view of the fact that this bound carries an $\Ord(1)$ uncertainty, one might however speculate on what happens if this constraint is ignored.
As visible from table~\ref{tab:bounds-anarchy}, with the exception of the first line, all the strongest bounds  on $m_\psi$ in the bidoublet or in the triplet models can be reduced down to about 1 TeV by taking $x_t = \frac{1}{3}$ to $\frac{1}{4}$. This however correspondingly requires $Y = 3$ to $4$ (and maximal right-handed mixing) which pushes up the bounds from $K_L\rightarrow \mu^+ \mu^-$ and is not consistent with $m_\psi = Yf$ and $f \gtrsim 0.5$~TeV. The loop-induced chirality-breaking effects on $\epsilon^\prime$ and $\Delta A_{\CP}$ in $D\rightarrow P P$ decays would also come into play. Altogether, even neglecting the bound from $\epsilon_K(Q_S^{LR})$, fermion resonances below about 1.5 TeV seem hard to conceive in anarchic models.

\bigskip

In $\U(3)^3$ models with Minimal Flavour Violation, as apparent from tables~\ref{tab:bounds-U3LC} and \ref{tab:bounds-U3RC}, a fermion resonance at about 1 TeV is disfavoured.
In \UtreLC\ the crucial constrains come from the EWPT due to the large mixing of the first generations quarks in their left component. Note that $x_t Y$ cannot go below $y_t\sim1$. In  \UtreRC\ there is a clash between the tree-level $\Delta F=2$ FCNC effects, which decrease with $x_t$, and the bound from the $pp\to jj$  angular distributions due to the composite nature of the light quarks in their right component, which goes like $1/x_t$.
We stress again that these conclusions are more robust than in the anarchic case, since there is no uncertainty related to the composite Yukawas, which are flavour universal in the \Utre\ case.

\bigskip

Two important differences distinguish the $\U(2)^3$ case from the $\U(3)^3$ one:
\begin{itemize}
\item[$(i)$] both for the bidoublet (at  large enough $z$) and for the triplet models, the  bounds from the EWPT or from compositeness become irrelevant;
\item[$(ii)$] a single complex parameter correlates the various observables, $r_b$ in the  \UdueLC\  case.
\end{itemize}
As apparent from table~{\ref{tab:bounds-U2LC}, values of $x_t$ and $r_b$ somewhat smaller than one can reduce the bounds on the fermion resonance mass at or even below the 1 TeV level. This is also formally possible in \UdueRC, where $r_b=1$, but requires $x_t \lesssim 0.3$, i.e. $Y \gtrsim 3$, not consistent with $m_\psi = Y f$ and $f \gtrsim 0.5$~TeV.

\begin{table}[tbp]
\renewcommand{\arraystretch}{1.3}
\centering
\begin{tabular}{cccc}
\hline
& doublet & triplet & bidoublet \\
\hline
\CircledA & $4.9$ & $1.7$ & $1.2\, *$ \\
\UtreLC & $6.5$ & $6.5$ & $5.3$ \\
\UtreRC &-&-& $3.3$ \\
\UdueLC & $4.9$ & $0.6$ & $0.6$ \\
\UdueRC &-&-& $1.1\, *$ \\
\hline
\end{tabular}
\caption{Minimal fermion resonance mass $m_\psi$ in TeV  compatible with all the bounds (except for the $Q_S^{LR}$ contribution to $\epsilon_K$ in the anarchic model), fixing $\Ord(1)$ parameters in anarchy to 1 and assuming the parameter $|r_b|$ in \UdueLC\ to be $\sim 0.2$. The bounds with a $*$ are obtained for a  value of  $Y \approx 2.5$, that minimizes the flavour and EWPT constraints consistently with $m_\psi = Y f$ and $f \gtrsim 0.5$ TeV.}
\label{tab:mmin}
\end{table}

\bigskip

Table~\ref{tab:bounds-ew} tries to summarize all in one go the content of the more detailed tables \ref{tab:bounds-ew} to \ref{tab:bounds-U2RC} discussed throughout the paper, taking into account all constraints from flavour and EWPT. For any given case, this table estimates a lowest possible value for the mass of the composite fermions that mix with the elementary ones and which are heavier than the ``custodians'' by a factor of $\sqrt{1 + (\lambda_X)^2}$. In the case of {\it anarchy} we are neglecting the constraint coming from $\epsilon_K$ (first line of table~\ref{tab:bounds-anarchy}, particularly problematic for the bidoublet model, maybe accidentally suppressed) and the various 
$\Ord(1)$ factors that plague most of the other flavour observables in table~\ref{tab:bounds-anarchy}.
In every case we also neglect the constraint coming from  one-loop chirality-breaking operators, relevant to direct \CP\ violation both in the $K$ and in the $D$ systems, as well as to the quark electric dipole moments. This is a subject that  deserves further detailed study.
Note that the bounds with a $*$ (bidoublet model with anarchic or  \UdueRC\  flavour structure) are obtained for a value of $Y \approx 2.5$, that minimizes the flavour and EWPT constraints consistently with $m_\psi = Y f$ and $f~\gtrsim~0.5$~TeV.

We also note that measurements of Higgs boson properties, which have not been considered here, amount to lower bounds on the decay constant $f$ in the case of PNGB Higgs models, and are currently probing values of $500\text{--}700$ GeV. Once these bounds improve, tables~\ref{tab:bounds-ew}~to~\ref{tab:bounds-U2RC} allow a straightforward qualitative understanding of their impact on flavour and electroweak observables. Since our predictions are based on a simple partial compositeness Lagrangian, they are in fact independent of the details of the Higgs sector and can even be applied to other theories, like 4D duals of Randall-Sundrum models.

\definecolor{green}{cmyk}{0.5,0,1,0.2}
\definecolor{lightgray}{rgb}{0.7,0.7,0.7}
\newcommand{\si}{{\color{green}\footnotesize$\bigstar$}}
\newcommand{\no}{{\color{lightgray}$\circ$}}

\begin{table}
\renewcommand{\arraystretch}{1.3}
\centering
\begin{tabular}{cccccc}
\hline
& ~\CircledA~ & \UtreLC & \UtreRC & \UdueLC & \UdueRC \\
\hline
$\epsilon_K$, $\Delta M_{d,s}$ & \si & \no & \si & \si & \si \\
$\Delta M_{s}/\Delta M_{d}$ & \si & \no & \no & \no & \no \\
$\phi_{d,s}$ & \si & \no & \no & \si & \no \\
$\phi_s-\phi_d$ & \si & \no & \no & \no & \no \\
$C_{10}$ & \si & \no & \no & \si & \no \\
$C_{10}'$ & \si & \no & \no & \no & \no \\
\hline
$pp\to jj$ & \no & \si & \si & \no & \no \\
$pp\to q'q'$ & \si & \no & \no & \si & \si \\
\hline
\end{tabular}
\caption{Observables where NP effects could show up with realistic experimental and/or lattice improvements in the most favourable cases. The observables are, from top to to bottom: the direct \CP\ violating parameter in $K$-$\bar K$ mixing and the $B_d$ and $B_s$ mass differences (as well as their ratio), the mixing phases $\phi_d,\phi_s$ in the $B_d$ and $B_s$ systems (as well as their difference), the Wilson coefficient of the axial vector semi-leptonic operator relevant for $b\to s\ell^+\ell^-$ transitions $C_{10}$ and its chirality-flipped counterpart $C_{10}'$, the angular distribution of dijet events at LHC as discussed above and the direct production of fermion resonances at LHC.}
\label{tab:dna}
\end{table}

The general message that emerges from table~\ref{tab:mmin}, taken at face value,  is pretty clear. To accommodate top partners at or below 1 TeV is often not  possible and requires a judicious choice of the underlying model: an approximate $\U(2)^3$ flavour symmetry appears favorite, if not necessary. There are two simple reasons for the emergence of $\U(2)^3$:
\begin{itemize}
\item[$(i)$] in common with $\U(3)^3$, the suppression of flavour changing effects in four-fermion operators with both left- and right-handed currents, present in the anarchic case;
\item[$(ii)$] contrary to $\U(3)^3$ but as in anarchy, the disentanglement of the properties (their degree of compositeness) of the first and third generation of quarks.
\end{itemize}

The source of the constraint that plays the dominant role in the various cases is diverse. Sometimes more than one observable gives comparable constraints. This is reflected in table~\ref{tab:dna}, which summarizes 
where possible new physics
effects could show up
(for some observables with more experimental
data, for others if  lattice parameters and/or other theoretical inputs are improved). We keep in this table every possible case even though  some of them, according to table~\ref{tab:mmin},   would have to live with a fine tuned Higgs boson squared mass and, as such, appear less motivated.

\part{Supersymmetric Higgs bosons}

\chapter{Supersymmetry}\label{SUSY}

If physics at the Fermi scale turns out to be weakly interacting and the Higgs boson is not part of a composite sector, then the naturalness problem has to be solved by means of some symmetry, accurately cancelling the radiative corrections of the Higgs mass as it happens for fermions and vector bosons. In this chapter we show that supersymmetry (SUSY) - the maximal extension of the Poincar\'e group of space-time transformations - does this job, and we review very briefly the main properties of supersymmetric models. We discuss how the necessity of breaking supersymmetry reintroduces a fine-tuning problem, which once more requires the SM to be modified close to the electroweak scale.

The remarkable property of supersymmetry of solving the hierarchy problem without introducing a threshold of strong dynamics, thus remaining perturbative in principle up to very high energies, justifies the great attention devoted to the subject in the literature \cite{Fayet:1976cr,Wess:1974tw,Wess:1973kz,Iliopoulos:1974zv,Girardello:1981wz,Dimopoulos:1981zb,Barbieri:1982eh,Chamseddine:1982jx,Hall:1983iz,Nilles:1983ge,Dine:1981rt,AlvarezGaume:1981wy,AlvarezGaume:1983gj,Giudice:1998bp}. Here we will not try to cover in detail the broad plethora of topics related to supersymmetry, which range from collider studies to models of dark matter or gauge coupling unification, not to mention its relevance in string theory. We will rather concentrate on the aspects which are related to the phenomenology of the recently discovered Higgs boson.

As for composite Higgs models, one of the main consequences of supersymmetry at the weak scale is the presence of a number of new particles coupled to the SM ones, with masses which lie in the range explorable by high-energy collision experiments at the LHC. The lack of signals so far from the direct production of supersymmetric particles calls for a careful analysis of the parameter space of SUSY models, and perhaps for a reconsideration of the strategy to search for supersymmetry. 
Focussing on naturalness, which to us still looks to be the best motivated guideline, special attention is paid to the s-particles that have the largest influence on the quadratic terms in the Higgs potential, namely the two stops, the left-handed sbottom, the higgsinos and, indirectly through the stops, the gluino \cite{Dimopoulos:1995mi,Cohen:1996vb,Barbieri:2009ev,Papucci:2011wy}.

In the first part of the chapter, where we introduce the basic concepts of supersymmetry, we follow the approach of ref.\cite{Terning:2006bq}, to which we refer for further details. In section~\ref{SUSY/MSSM} we introduce the minimal extension of the Standard Model which is able to incorporate supersymmetry, the well-studied MSSM, focussing in particular on the Higgs sector of that model. We discuss in particular the issue of accommodating a 125 GeV Higgs boson in this context, which is in manifest tension with the naturalness requirement of light top partners and small radiative corrections.

A general prediction of supersymmetric extensions of the SM is the presence of an extended Higgs sector with more than one scalar degree of freedom. The mixing of the observed Higgs boson with the extra states has a relevant impact on the signal strengths measured at the LHC, which in turn can be used to set bounds on the parameters space of the Higgs system. Here we discuss the problem in the context of the MSSM, while we dedicate the next chapter to the more appealing case of the NMSSM.

\section{Basics of supersymmetry}\label{SUSY/Basics}

Consider the following Lagrangian
\begin{align}\label{example}
\L &= \L_H + \L_\psi + |\partial_{\mu}\phi_L|^2 + |\partial_\mu\phi_R|^2 - m_L^2|\phi_L|^2 - m_R^2|\phi_R|^2\notag\\
& - \frac{y}{\sqrt{2}}h\bar \psi_L \psi_R - \frac{\lambda}{2}h^2(|\phi_L|^2 + |\phi_R|^2) - h(\mu_L|\phi_L|^2 + \mu_R|\phi_R|^2),
\end{align}
which contains the SM Higgs and a fermion $\psi$ of mass $m_\psi$, together with their Yukawa interaction, and we introduced two new complex scalar fields $\phi_L$ and $\phi_R$ which interact with the Higgs.
If we compute the radiative correction to the Higgs mass in this model, using a momentum cut-off for simplicity, we find
\begin{align}\label{higgsmassexample}
\delta m_h^2 &= \frac{1}{16 \pi^2}\Big(2\lambda\Lambda^2 - (\lambda m_L^2 + \mu_L^2)\log\frac{\Lambda^2 + m_L^2}{m_L^2} - (\lambda m_R^2 + \mu_R^2)\log\frac{\Lambda^2 + m_R^2}{m_R^2}\\
&\quad - 2 y^2 \Lambda^2 + 6 y^2 m_\psi^2 \log\frac{\Lambda^2 + m_\psi^2}{m_\psi^2}\Big).
\end{align}
We see that if $\lambda = y^2$, then the two quadratic ``divergences'' associated with the scalar and fermion loops cancel out exactly. If in addition $m_L^2 = m_R^2 = m_\psi^2$ and $\mu_L^2 = \mu_R^2 = 2\lambda m_\psi^2$, then also the logarithmic terms cancel, and the Higgs mass is not renormalized. As we will now see, the Lagrangian \eqref{example} enjoys a very special symmetry if these relations are satisfied. This property of protecting scalar masses from large radiative corrections makes such a model, with a scalar associated to every fermion degree of freedom, a perfect candidate for natural physics beyond the Standard Model.

\subsection{The supersymmetric algebra}\label{SUSY/Basics/Algebra}

Take a very simple Lagrangian which contains only the kinetic terms of a complex scalar and a free chiral fermion, plus the quadratic action for a non-propagating field $\mathcal{F}$,
\begin{equation}\label{LSUSYfree}
S = \int d^4 x\,( \L_s + \L_f + \L_{\rm aux}) = \int d^4 x\,\Big(|\partial_{\mu}\phi|^2 + i\psi^{\dag}\bar\sigma_{\mu}\partial_{\mu}\psi + |\mathcal{F}|^2\Big),
\end{equation}
using the standard notation
\begin{align}
\sigma^{\mu}_{\alpha\dot\alpha} &= (1, \sigma^i), & \bar\sigma^{\mu}_{\alpha\dot\alpha} &= (1, -\sigma^i),
\end{align}
and consider the following transformation rules
\begin{align}
\phi &\to \phi + \epsilon^{\alpha}\psi_{\alpha},\label{scalartrans}\\
\psi_{\alpha} &\to \psi_{\alpha} -i (\sigma^{\nu}\epsilon^{\dag})_{\alpha}\partial_{\nu}\phi + \epsilon_{\alpha}\mathcal{F},\label{fermiontrans}\\
\mathcal{F}&\to \mathcal{F} -i \epsilon^{\dag}\bar\sigma^{\mu}\partial_{\mu}\psi,\label{auxtrans}
\end{align}
where $\epsilon^\alpha$ is a small parameter that has to be a fermion.
Computing the variation of \eqref{LSUSYfree}
\begin{equation}
\delta\L_s + \delta\L_f + \delta\L_{\rm aux} = \partial_{\mu}\big(\epsilon\sigma^{\mu}\bar\sigma^{\nu}\psi\partial_{\nu}\phi^* - \epsilon\psi\partial^{\mu}\phi^* + \epsilon^{\dag}\psi^{\dag}\partial^{\mu}\phi + i\psi^{\dag}\bar\sigma^{\mu}\epsilon \mathcal{F}\big),
\end{equation}
one explicitly verifies that, since this is a total derivative, the action is invariant, $\delta S = 0$.

The commutator of two infinitesimal transformations turns out to be
\begin{align}\label{SUSYcommutator}
(\delta_{\epsilon_1}\delta_{\epsilon_2} - \delta_{\epsilon_2}\delta_{\epsilon_1})X = -i(\epsilon_1\sigma^{\mu}\epsilon_2^{\dag} - \epsilon_2\sigma^{\mu}\epsilon_1^{\dag})\partial_{\mu}X,
\end{align}
for any of the fields $X = (\phi,\phi^*,\psi,\psi^{\dag},\mathcal{F},\mathcal{F}^*)$. Note that this last property holds only because of the presence of the auxiliary field $\mathcal{F}$. Without that field an additional term proportional to the equations of motion $\bar\sigma^{\mu}\partial_{\mu}\psi$ would be present, and \eqref{SUSYcommutator} would be true only for on-shell fields.

If one writes the infinitesimal transformations \eqref{scalartrans}, \eqref{fermiontrans}, \eqref{auxtrans} as $\exp(i\epsilon^{\alpha} Q_{\alpha})$ acting on the different fields, then one has, in terms of the generators $Q_{\alpha}$,
\begin{equation}\begin{aligned}\label{SUSYalgebra}
\{ Q_{\alpha}, Q_{\dot\beta}^{\dag}\} &= 2\sigma^{\mu}_{\alpha\dot\beta} P_{\mu},\\
\{Q_{\alpha}, Q_{\beta}\} &= \{Q_{\dot\alpha}^{\dag}, Q_{\dot\beta}^{\dag}\} = 0,
\end{aligned}\end{equation}
where the anti-commutators arise from the commutators like $[ \epsilon_1 Q, \epsilon_2 Q ]$ because of the fermionic nature of the $\epsilon$'s. The {\it supercharges} $Q_{\alpha}$, $Q_{\dot\alpha}^{\dag}$, thus form a closed algebra together with the generator of space-time traslations $P_{\mu}$, called the {\it supersymmetric algebra}. This is the only possible extension of the Poincar\'e group which is not the direct product with an internal symmetry group. Such an extension, which would be forbidden by the Coleman-Mandula theorem, is possible at all only because the Poincar\'e algebra has been extended to a $Z_2$-graded Lie algebra which contains also fermionic anti-commuting generators.

A consequence of the anticommutation relations \eqref{SUSYalgebra} is that the Hamiltonian operator is always positive definite, and its expectation value on the supersymmetric vacuum must vanish. We have indeed
\begin{equation}
\langle 0| H |0\rangle = \langle 0|P^0|0\rangle = \frac{1}{2}\langle 0|\sum_{\alpha} Q_{\alpha}Q_{\dot\alpha}^{\dag} + Q_{\dot\alpha}^{\dag}Q_{\alpha}|0\rangle = 0,
\end{equation}
since $Q_{\alpha}|0\rangle = 0$ for every $\alpha$ if $|0\rangle$ preserves SUSY.

The multiplet of fields $(\phi,\psi,\mathcal{F})$ and their hermitian conjugates form a complete representation of the SUSY algebra. Since the supercharges are spinorial quantities, when they are applied to a state $|\psi\rangle$ they change its spin by $\pm 1/2$, and any supersymmetry transformation thus relates bosons with fermions. In the rest frame of a particle of mass $m$, \eqref{SUSYalgebra} reduces to a Clifford algebra
\begin{align}
\{Q_{\alpha},Q^{\dag}_{\dot\beta}\} &= 2m\delta_{\alpha\dot\beta}, & \{Q_{\alpha},Q_{\beta}\} &= \{Q^{\dag}_{\dot\alpha},Q^{\dag}_{\dot\beta}\} = 0,
\end{align}
which now allows to use $Q_{\alpha}$, $Q^{\dag}_{\dot\alpha}$ as raising and lowering operators in order to construct the complete multiplets of states\footnote{In the superspace formalism each multiplet forms the components of a single {\it superfield}.}
\begin{align}
&|\Omega_s\rangle, & &Q^{\dag}_1|\Omega_s\rangle, & &Q^{\dag}_2|\Omega_s\rangle, & &Q^{\dag}_1 Q^{\dag}_2|\Omega_s\rangle,
\end{align}
starting from a ``vacuum'' state $|\Omega_s\rangle$ of spin $s$ which is annihilated by all the $Q$'s:
\begin{itemize}
\item The massive {\it chiral multiplet} is obtained starting from a spin-0 state $|\Omega_0\rangle$, and contains a Majorana fermion and two real scalars (one complex scalar);
\item The massive {\it vector multiplet} is obtained starting from a spin-$\frac{1}{2}$ state $|\Omega_{1/2}\rangle$, and contains a massive vector, two Majorana fermions and a real scalar.
\end{itemize}
In a similar fashion one constructs the multiplets for massless fields:
\begin{itemize}
\item The massless chiral multiplet contains a Weyl fermion and a complex scalar;
\item The massless vector multiplet contains a gauge vector boson and a Weyl fermion.
\end{itemize}
We will not consider other multiplets which contain states with spin greater than 1.

\subsection{Supersymmetric Lagrangians}\label{SUSY/Basics/Lagrangian}

Using the transformations \eqref{scalartrans}, \eqref{fermiontrans}, \eqref{auxtrans} one can go beyond the free theory \eqref{LSUSYfree} and derive the most general structure that a supersymmetric Lagrangian can have \cite{Wess:1974tw,Wess:1973kz,Iliopoulos:1974zv}. The most general renormalizable interaction for a set of chiral multiplets $\phi_i$, $\psi_i$, $\mathcal{F}_i$ is
\begin{equation}\label{LSUSYint}
\L_{\rm int} = -\frac{1}{2} W_{ij}(\phi,\phi^*)\psi_i\psi_j + W_i\mathcal{F}_i - V(\phi,\phi^*) + \text{ h.c.}
\end{equation}
First of all, notice that a scalar potential $V(\phi,\phi^*)$ is forbidden at this stage, since
\begin{equation}
\delta V = \frac{\partial V}{\partial\phi_i}\delta\phi_i + \frac{\partial V}{\partial \phi_i^*}\delta\phi_i^* = \frac{\partial V}{\partial \phi_i}\epsilon\psi_i + \frac{\partial V}{\partial\phi_i^*}\epsilon^{\dag}\psi_i^{\dag},
\end{equation}
and this cannot be canceled by the variation of any of the other terms in $\L_{\rm int}$.

One can check that $\L_{\rm int}$ is SUSY-invariant, up to total derivatives, if and only if
\begin{align}
W_{ij} &= \frac{\partial^2 W}{\partial\phi_i\partial\phi_j}, & W_i &= \frac{\partial W}{\partial \phi_i},
\end{align}
where $W(\phi)$ is a holomorphic function of the scalar fields, i.e. $\partial W/\partial\phi_j^* = 0$, called the {\it superpotential}.\footnote{If one allows generic non-renormalizable interactions, additional terms with a particular, non-holomorphic complex structure can be present in what is called the {\it K\"ahler potential}} By power-counting one sees from \eqref{LSUSYint} that $W_{ij}$ can be at most linear in the fields, in order to have a renormalizable Lagrangian, and thus the superpotential has the form
\begin{equation}
W(\phi_i) = \frac{1}{2}m_{ij}\phi_i\phi_j + \frac{1}{6}y_{ijk}\phi_i\phi_j\phi_k + J_i\phi_i.
\end{equation}

A scalar potential is now generated when integrating out the non-propagating auxiliary fields and replacing them with their on-shell values $\mathcal{F}_i = -W_i$. From the quadratic term $\mathcal{F}_i\mathcal{F}_i^*$ we get
\begin{equation}\label{SUSYpotential}
V(\phi,\phi^*) = \sum_i\left|\frac{\partial W}{\partial\phi_i}\right|^2.
\end{equation}
Note that if there is a linear term in the superpotential, then $V$ contains a constant and supersymmetry is broken.

The full Lagrangian for a generic set of interacting chiral superfields, the {\it Wess-Zumino} Lagrangian, reads
\begin{align}
\L_{\rm WZ} &= |\partial_{\mu}\phi_j|^2 + i\psi^{\dag}_j\bar\sigma^{\mu}\partial_{\mu}\psi_j - \frac{1}{2}m_{jk}\psi_j\psi_k - \frac{1}{2}y_{jkl}\phi_j\psi_k\psi_l + \text{h.c.} - V(\phi,\phi^*),\label{WZ}\\
V(\phi,\phi^*) &= \frac{1}{2}m_{jl}m_{lk}\phi_j^*\phi_k + \frac{1}{2}m_{jm}y_{klm}^*\phi_j\phi_k^*\phi_l^* + \frac{1}{8}y_{jkn}y_{lmn}^*\phi_j\phi_k\phi_l^*\phi_m^* + \text{h.c.}\label{WZpotential}
\end{align}
Notice that the scalar mass coincides with the fermion mass, and the relations between the scalar cubic and quartic couplings and the Yukawa coupling, required in \eqref{example} to cancel the divergences, hold. Equation \eqref{example} is in fact just a particular case of the general Wess-Zumino Lagrangian, and the absence of radiative corrections in the Higgs mass in \eqref{higgsmassexample} is the consequence of a more general {\it non-renormalization} theorem of supersymmetry which states that the superpotential $W$ is not renormalized at any order in perturbation theory.

In a similar way to what discussed up to here, a supersymmetric Lagrangian for gauge fields can be derived, starting from the massless vector multiplet. As in the previous case, auxiliary fields $\mathcal{D}^a$ have to be introduced along with the vectors $A_\mu^a$ and the fermions $\lambda^a$, in order to close the algebra of SUSY transformations also for off-shell fields. The pure super-Yang-Mills Lagrangian reads
\begin{equation}\label{SYM}
\L_{\rm SYM} = -\frac{1}{4}F_{\mu\nu}^a F_{\mu\nu}^a + i\lambda^{\dag\,a}\bar\sigma^{\mu}D_{\mu}\lambda^a + \frac{1}{2}\mathcal{D}^a \mathcal{D}^a.
\end{equation}
Including also chiral matter multiplets, the complete gauge Lagrangian becomes
\begin{equation}\label{SYMWZ}
\L = \L_{\rm SYM} + \L_{\rm WZ} - \sqrt{2}g\Big((\phi_i^* T^a\psi_i)\lambda^a + \text{h.c.}\Big) + g(\phi_i^* T^a \phi_i)\mathcal{D}^a.
\end{equation}
The scalar potential gets now a contribution from both the chiral and the vector auxiliary fields, and reads
\begin{equation}
V(\phi,\phi^*) = \left. \mathcal{F}_i^* \mathcal{F}_i + \frac{1}{2}\mathcal{D}^a \mathcal{D}^a\right|_{\rm on-shell} = \left|\frac{\partial W}{\partial\phi_i}\right|^2 + \frac{1}{2}g^2(\phi_i^* T^a \phi_i)^2.
\end{equation}

\subsection{Soft SUSY-breaking}\label{SUSY/Basics/Soft}

Supersymmetry requires the existence of an equal number of bosonic and fermionic states with the same quantum numbers and degenerate masses, related by a SUSY transformation within the same supermultiplet. The elementary particles observed in experiments do not have this property, and therefore SUSY cannot be an exact symmetry of Nature.

If nevertheless one wants to preserve the peculiar property of protecting the scalar masses against high-energy radiative corrections, one has to break supersymmetry in some special way. More specifically, one does not want to spoil the UV structure of the theory that determines the form of the divergences in the radiative corrections \eqref{higgsmassexample}.

The cancellation of quadratic divergences depends only on the adimensional parameters like the scalar quartic couplings, the gauge couplings and the Yukawas, which are thus fixed to their supersymmetric values. At the cost of introducing some logarithmic dependence on the high-energy cut-off -- which is nevertheless not a problem for naturalness arguments -- one can add generic soft SUSY-breaking terms to the Lagrangian, i.e. operators of dimension smaller than four which are not relevant in the UV \cite{Girardello:1981wz,Dimopoulos:1981zb}. Loop effects generated by such soft terms will not generate quadratic divergences in the high scale $\Lambda$. However, as can also be seen in \eqref{higgsmassexample}, logarithms and finite terms, if they do not cancel out exactly, are still proportional to the squared mass of the particle which generates them. The soft SUSY-breaking scale $\Lambda_{\rm soft}$ can thus not be too far away from the weak scale in order not to generate a {\it little hierarchy problem}. This is in perfect agreement with the discussion in section \ref{SM/hierarchy}, since $\Lambda_{\rm soft}$ is the scale at which the SM is modified.

A general set of soft SUSY-breaking terms is described by\footnote{A mass term for the fermions can be reabsorbed in the superpotential. The non-holomorphic trilinear coupling $c_{ijk}\phi_i^*\phi_j\phi_k$ may introduce quadratic divergences if a gauge-singlet is present, and a linear term $J_i\phi_i$ is allowed only for gauge singlets.}
\begin{equation}
\L_{\rm soft} = -\frac{1}{2}m_{\lambda}\lambda^a\lambda^a - \tilde m_{ij}^2\phi_i\phi_j^* - \frac{1}{2}b_{ij}\phi_i\phi_j - \frac{1}{6}a_{ijk}\phi_i\phi_j\phi_k  + {\rm h.c.}.
\end{equation}

There are many different possibilities for the dynamical mechanism which generates such breaking terms. Generically one needs a spontaneous breaking of supersymmetry. This should happen in a different sector with respect to the SM one, with the breaking then being communicated to the supersymmetric SM by some mediator through loop processes. The most common scenarios of this kind are {\it gauge-mediation} \cite{Dine:1981rt,AlvarezGaume:1981wy,Giudice:1998bp,Meade:2008wd}, where SUSY-breaking is transmitted to the SM through ``messenger'' fields with gauge interactions, and {\it gravity-mediation} \cite{Barbieri:1982eh,Chamseddine:1982jx,Hall:1983iz,AlvarezGaume:1983gj}, where the SUSY-breaking terms are generated at the Planck scale by gravitational interactions, and are suppressed by some power of $\mpl$. We will not treat these problems here, and take an effective field theory approach, considering generic soft Lagrangians with the size of the parameters constrained by experiment.

\section{The Minimal Supersymmetric Standard Model}\label{SUSY/MSSM}

We have already seen that the particles of the Standard Model do not fit into complete representations of the superalgebra. While the problem of mass splittings could be overcome through the introduction of suitable soft terms, the different quantum numbers of bosons and fermions force us to include new particles in order to complete each multiplet which contains a SM field.\footnote{The only boson/fermion pair with the same $\SU(3)\times \SU(2)\times \U(1)$ quantum numbers are the Higgs doublet and the left-handed leptons. If the Higgs field were the sparticle associated with one of the leptons, its vev would break lepton number.}

The Minimal Supersymmetric Standard Model (MSSM) is the model with the minimal number of new fields added to the SM in order to have a supersymmetric theory. This corresponds to the following situation:
\begin{itemize}
\item[$(i)$] a set of complex scalar fields, collectively denoted as {\it sfermions} (squarks or sleptons), one associated to each chiral fermion,
\item[$(ii)$] a set of chiral fermions, collectively denoted as {\it gauginos} or {\it higgsinos}, one for every gauge or Higgs boson respectively.
\end{itemize}
The particle content of the MSSM is summarized in table~\ref{tab:MSSM}.
\begin{table}
\renewcommand{\arraystretch}{1.3}
\centering%
\begin{tabular}{cccccc}
Superfield & Bosons & Fermions & $\SU(3)_c$ & $\SU(2)_L$ & $\U(1)_Y$\\
\hline
$Q_i$ & $\tilde q_i$ & $q_{Li}$ & $\three$ & $\two$ & 1/6\\
$U_i$ & $\tilde u_i$ & $u_{Ri}$ & $\three$ & $\one$ & 2/3\\
$D_i$ & $\tilde d_i$ & $d_{Ri}$ & $\three$ & $\one$ & -1/3\\
\hline
$L_i$ & $\tilde \ell_i$ & $\ell_{Li}$ & $\one$ & $\two$ & -1/2\\
$E_i$ & $\tilde e_i$ & $e_{Ri}$ & $\one$ & $\one$ & -1\\
\hline
$H_u$ & $H_u$ & $\tilde H_u$ & $\one$ & $\two$ & 1/2\\
$H_d$ & $H_d$ & $\tilde H_d$ & $\one$ & $\two$ & -1/2\\
\hline
$W$ & $W_{\mu}^a$ & $\tilde W^a$ & $\one$ & $\three$ & 0\\
$B$ & $B_{\mu}$ & $\tilde B$ & $\one$ & $\one$ & 0\\
$G$ & $G_{\mu}^a$ & $\tilde G^a$ & $\eight$ & $\one$ & 0
\end{tabular}
\caption{Field content of the MSSM.\label{tab:MSSM}}
\end{table}

A peculiar property of SUSY is the need of the presence of two Higgs doublets, $H_u$ and $H_d$, each of them coupled either to the up-type fermions or the down-type fermions, respectively. This fact is easily understood in terms of the holomorphy of the superpotential: gauge invariance requires fields of opposite hypercharge to couple to the two different types of fermions, but a term $H^c \bar q_L u_R$ which contains the hermitian conjugate of $H$ is not allowed, so a new field has to be introduced. At the same time, the second Higgs field is also needed to cancel the new $\U(1)$ anomalies of the higgsinos.

The gauge vector multiplets $W$, $B$, $G$ are described by a super-Yang-Mills Lagrangian as in \eqref{SYM}, while the quark, lepton and Higgs superfields are chiral multiplets described by a Wess-Zumino Lagrangian \eqref{WZ}, extended to include gauge invariance. Soft SUSY breaking terms for all the fields are present in order to reproduce the correct, non-supersymmetric values, of all the dimensionful parameters.

Baryon number ($B$) and lepton number ($L$) violating terms in the superpotential, such as e.g. $QLD$, are set to zero by imposing invariance under the matter parity
\begin{equation}
R = (-1)^{3(B-L) + 2s},
\end{equation}
where $s$ is the spin. This discrete symmetry has very important consequences for the phenomenology of SUSY models, since it requires the superpartners to be produced always in couples, starting from an interaction of SM particles, and it stabilizes the lightest supersymmetric particle (LSP), $\chi_{\rm LSP}$, which then becomes an excellent dark matter candidate. Versions of the MSSM without $R$-parity, which have also been studied, have a very different phenomenology \cite{Barbier:2004ez}.

\subsection{The Higgs sector of the MSSM}\label{SUSY/MSSM/Higgs}

Let us denote the components of the two Higgs doublets as
\begin{align}
H_u &= \begin{pmatrix}H_u^0\\ H_u^+\end{pmatrix}, & H_d &= \begin{pmatrix}H_d^-\\ H_d^0\end{pmatrix}.
\end{align}
The superpotential for the $H_u$ and $H_d$ superfields reads
\begin{equation}
W_H = y_u^{ij} H_u Q_i U_j - y_d^{ij}H_d Q_i D_j - y_e^{ij} H_d L_i E_j + \mu H_u H_d.
\end{equation}
The last term is allowed by all the symmetries and gives rise to mass terms, as well as cubic and quartic interactions for the Higgses through $\mathcal{F}$-terms.
The full scalar Higgs potential, including $\mathcal{F}$-terms, $\mathcal{D}$-terms and soft breaking terms, reads
\begin{align}
V(H_u,H_d) &= (|\mu|^2 + m_{H_u}^2)(|H_u^0|^2 + |H_u^+|^2) + (|\mu|^2 + m_{H_d}^2)(|H_d^0|^2 + |H_d^-|^2)\notag\\
&\quad + B\hspace{-0.04cm}\mu (H_u^+ H_d^- - H_u^0 H_d^0) + {\rm h.c.} + \frac{g^2}{2}|H_u^+ H_d^{0*} + H_u^0 H_d^{-*}|^2\notag\\
&\quad + \frac{g^2 + g^{\prime 2}}{8}(|H_u^0|^2 + |H_u^+|^2 - |H_d^0|^2 - |H_d^-|^2)^2.\label{scalarpotSUSY}
\end{align}
All the parameters have to be chosen in a suitable way in order to reproduce EWSB in a correct way. This is not automatic as in the SM, because of the presence of more parameters and more fields.

First of all, a vev for the charged fields $H_u^+$ and $H_d^-$ should be excluded, because this would break electrodynamics. Notice that one of the two Higgs vevs can always be put in the form
\begin{equation}
\langle H_u\rangle = \frac{1}{\sqrt{2}}\begin{pmatrix}v_u\\ 0\end{pmatrix},\qquad v_u\in\mathbb{R}
\end{equation}
through $\SU(2)_L\times \U(1)_Y$ transformations, as one typically does in the SM. The minimum condition then reduces to
\begin{equation}
\left. \frac{\partial V}{\partial H_u^+}\right|_{\langle H_u^+\rangle = 0} = \Big(B\mu + \frac{g^2}{2} H_u^{0*}H_d^{0*}\Big)H_d^- = 0,
\end{equation}
which is generally satisfied for $\langle H_u^+\rangle = \langle H_d^-\rangle = 0$. It is also straightforward to check that \CP\ is not broken by the Higgs vev, since the only term of the potential which contains the eventual phase $\varphi_{H_d}$ of $\langle H_d^0\rangle$ is $(H_u^0 H_d^0$ + h.c.) $\propto -\cos\varphi_{H_d}$, and so the phase vanishes at the minimum.

Next, by minimizing the scalar potential w.r.t. $H_u^0$, $H_d^0$ one finds the following expressions
\begin{align}
|\mu|^2 + m_{H_u}^2 = B\mu\cot\beta + \frac{m_Z^2}{2}\cos 2\beta,\label{susyvev1}\\
|\mu|^2 + m_{H_d}^2 = B\mu\tan\beta - \frac{m_Z^2}{2}\cos 2\beta\label{susyvev2},
\end{align}
where
\begin{align}
\tan\beta&\equiv\frac{v_u}{v_d}, & v^2 = v_u^2 + v_d^2,
\end{align}
and $v_u$, $v_d$ are the vevs of $H_u$ and $H_d$, respectively. The two relations \eqref{susyvev1}, \eqref{susyvev2}, which determine the weak scale $m_Z$ in terms of $\mu$ and the soft parameters $m_{H_u}$, $m_{H_d}$, $B$ reintroduce a fine-tuning problem. If the soft breaking scale is too large, different terms have to cancel out to a high precision in order to give the correct value of the electroweak scale. Moreover, the $\mu$ term is a supersymmetric quantity which comes from the superpotential, and might in principle be related to a higher scale than the smaller soft masses. There are many possible interpretations of this smallness of $\mu$ \cite{Kim:1983dt,Giudice:1988yz}.

It is convenient to rotate the Higgs fields to the basis
\begin{equation}
\begin{pmatrix}H\\h\end{pmatrix} = \begin{pmatrix}s_{\beta} & -c_{\beta}\\c_{\beta} & s_{\beta}\end{pmatrix}\begin{pmatrix}H_d\\ H_u\end{pmatrix},
\end{equation}
where $s_\beta = \sin\beta$, $c_\beta = \cos\beta$, and only the $h$ doublet acquires the vev $v$. The charged and \CP-odd components of $h$ coincide now with the would-be Goldstone bosons $\chi$ and $\eta$ which are eaten-up by the longitudinal polarizations of $W$ and $Z$ after EWSB. On the contrary, the charged and \CP-odd components of $H$ are physical fields $H^\pm$, $A^0$. Their masses can be read off the potential \eqref{scalarpotSUSY},
\begin{align}
m_A^2 &= \frac{B\mu}{s_\beta c_\beta},\label{mAMSSM}\\
m_{H^\pm} &= m_A^2 + m_W^2.\label{mHpMSSM}
\end{align}
Finally, the mass matrix for the two \CP-even neutral states reads, in the basis $(H_u,H_d)$,
\begin{equation}\label{treemassmatrixMSSM}
M^2 = \begin{pmatrix} m_A^2 c_\beta^2 + m_Z^2 s_\beta^2 & -(m_A^2 + m_Z^2)s_\beta c_\beta\\
-(m_A^2 + m_Z^2)s_\beta c_\beta & m_A^2 s_\beta^2 + m_Z^2 c_\beta^2\end{pmatrix}.
\end{equation}
Diagonalizing $M^2$ one finds the masses of the physical eigenstates $h_1$ and $h_3$\footnote{We choose this notation for later convenience in the context of the NMSSM where three states are present. Note in particular the difference from the standard MSSM notation, where the \CP-even mass eigenstates are $h, H$.}
\begin{equation}\label{physicalmhMSSM}
m_{h_{1,3}} = \frac{1}{2}\Big(m_A^2 + m_Z^2 \pm\sqrt{(m_A^2 + m_Z^2)^2 - 4 m_A^2 m_Z^2\cos^2 2\beta}\Big),
\end{equation}
as well as the mixing angle $\alpha$ between $H_u$ and $H_d$
\begin{equation}
\tan 2\alpha = \frac{m_A^2 + m_Z^2}{m_A^2 - m_Z^2}\tan 2\beta.
\end{equation}
Equation \eqref{physicalmhMSSM} in the limit of large $m_A$ reduces to an upper bound to the tree-level lightest Higgs mass
\begin{equation}\label{upperboundmhMSSM}
m_h^2 < m_Z^2\cos^2 2\beta < m_Z^2,
\end{equation}
which is by far too light to explain the observed value of 125 GeV.

\section{Supersymmetry and naturalness}

Since SUSY is broken above the weak scale, radiative corrections affect all the parameters of the MSSM. One the one hand these corrections are suitable to fit the predicted value of the Higgs mass with the experimental data, a task which would be not possible at  tree-level because of the bound \eqref{upperboundmhMSSM}. On the other hand, quadratic divergences proportional to the masses of the sparticles may constitute a problem for naturalness if they are too high. As we will see, there is a significative tension between these two effects in the MSSM. One finds a better situation in the NMSSM, which will be the subject of the next chapter.

\subsection{Radiative corrections to the Higgs mass}\label{SUSY/MSSM/deltaHiggs}

The relations \eqref{mAMSSM}, \eqref{mHpMSSM}, \eqref{physicalmhMSSM}, all coming from the potential \eqref{scalarpotSUSY}, are valid only at tree-level. Loop contributions, in particular the ones coming from third generation quarks and squarks, are nevertheless relevant.

The corrections to the neutral mass matrix \eqref{treemassmatrixMSSM} can be parametrized as
\begin{equation}\label{massmatrixMSSM}
M^2 = \begin{pmatrix}
M^2_{11} + \delta^2_{11} & M^2_{12} + \delta^2_{12}\\
M^2_{12} + \delta^2_{12} & M^2_{22} + \delta^2_{22}
\end{pmatrix}.
\end{equation}
The dominant contribution to $\delta_{22}$ arising from the large tree-level coupling to the top-stop system, including also two-loop leading-log QCD effects, reads \cite{Carena:1995bx}
\begin{align}\label{deltamhMSSM}
\delta_{22}^2 &=
\frac{3 m_t^4}{4\pi^2 v^2\sin^2\beta}\Big[\log\frac{M_S^2}{M_t^2} + \frac{\tilde X_t}{2}
+ \frac{1}{16\pi^2}\Big(\frac{3 m_t^2}{v^2} - 32\pi\alpha_s(M_t)\Big)\Big(\tilde X_t \log\frac{M_S^2}{M_t^2} + \log^2\frac{M_S^2}{M_t^2}\Big)\Big].
\end{align}
Here
\begin{align}
\tilde X_t &= \frac{2\tilde A_t^2}{M_S^2}\Big(1 - \frac{A_t^2}{12 M_S^2}\Big), & \tilde A_t &= A_t - \mu\cot\beta = \frac{m_{\tilde t_1}^2 - m_{\tilde t_1}^2}{2 m_t}\sin 2\theta_t,
\end{align}
where the SUSY-breaking scale can be taken as a mean value of the two stop masses $\tilde t_1, \tilde t_2$, $M_S^2\approx \langle m_{\tilde t}^2\rangle$, $\theta_t$ is the mixing angle of the stop mass matrix
\begin{equation}
\tilde M^2_t \simeq \begin{pmatrix}
\tilde m_{Q_3}^2 + m_t^2 & m_t \tilde A_t\\
m_t \tilde A_t & \tilde m_{U_3}^2 + m_t^2 
\end{pmatrix} +{\Ord}(g^2v^2),
\end{equation}
$M_t$ is the pole top mass, and $m_t$ is the running top mass evaluated at the pole. Diagonalizing \eqref{massmatrixMSSM} one gets for the physical Higgs masses
\begin{equation}\label{mhhMSSM}
m_{h_{1,3}} = \frac{1}{2}\Big(m_A^2 + m_Z^2 + \delta_{22}^2 \pm\sqrt{(m_A^2 + m_Z^2 + \delta_{22}^2)^2 - 4 m_A^2 m_Z^2\cos^2 2\beta - 4 M_{22}^2\delta_{22}^2}\Big).
\end{equation}
In the limit of large $m_A$ this reduces to a modified upper bound to the lightest Higgs mass
\begin{equation}\label{upperboundmhdeltat}
m_{h}^2 < m_Z^2\cos^2 2\beta + \delta_{22}^2\sin^2\beta \equiv m_Z^2\cos^2 2\beta + \Delta_t^2\equiv m_{hh}^2.
\end{equation}
Leading corrections to the $\mathcal{D}$-terms generate also an additional contribution to $M^2$, which cannot be neglected for average and large $\tan\beta$, and translates into a negative shift of $\Delta_t$ which can be as large as $\sim$ 5 GeV \cite{Carena:1995bx}.

The corrections $\delta_{12}^2$ and $\delta_{11}^2$ to the other matrix elements in \eqref{massmatrixMSSM} are suppressed by the first and second power of $\tilde A_t \mu / M_S^2$. The leading terms of these corrections are reported in \cite{Djouadi:2005gj}. They can safely be neglected with respect to $\delta_{22}^2$ as long as $\mu A_t$ is not bigger than the average soft breaking mass, as required by naturalness. Subleading contributions from the bottom quark and squarks are also present, and become relevant for large $\tan\beta$ where the bottom Yukawa coupling is enhanced.

The leading correction to formula \eqref{mHpMSSM} for the charged Higgs mass instead arises at order $\alpha_w m_t^2$ and reads
\begin{equation}\label{corrmHp}
m_{H^\pm}^2 = m_A^2 + m_W^2 + \frac{3 G_{\mu} m_W^2}{4\sqrt{2}\pi^2}\Big(\frac{m_t^2}{\sin^2\beta} + \frac{m_b^2}{\cos^2\beta}\Big)\log\frac{M_S^2}{M_t^2}.
\end{equation}
In the decoupling limit of large $m_A$, both $m_{h_3}$ and $m_{H^\pm}$ coincide approximately with $m_A$.

\subsection{A natural supersymmetric spectrum}

\begin{figure}[t]
\centering%
\includegraphics[width=0.49\textwidth]{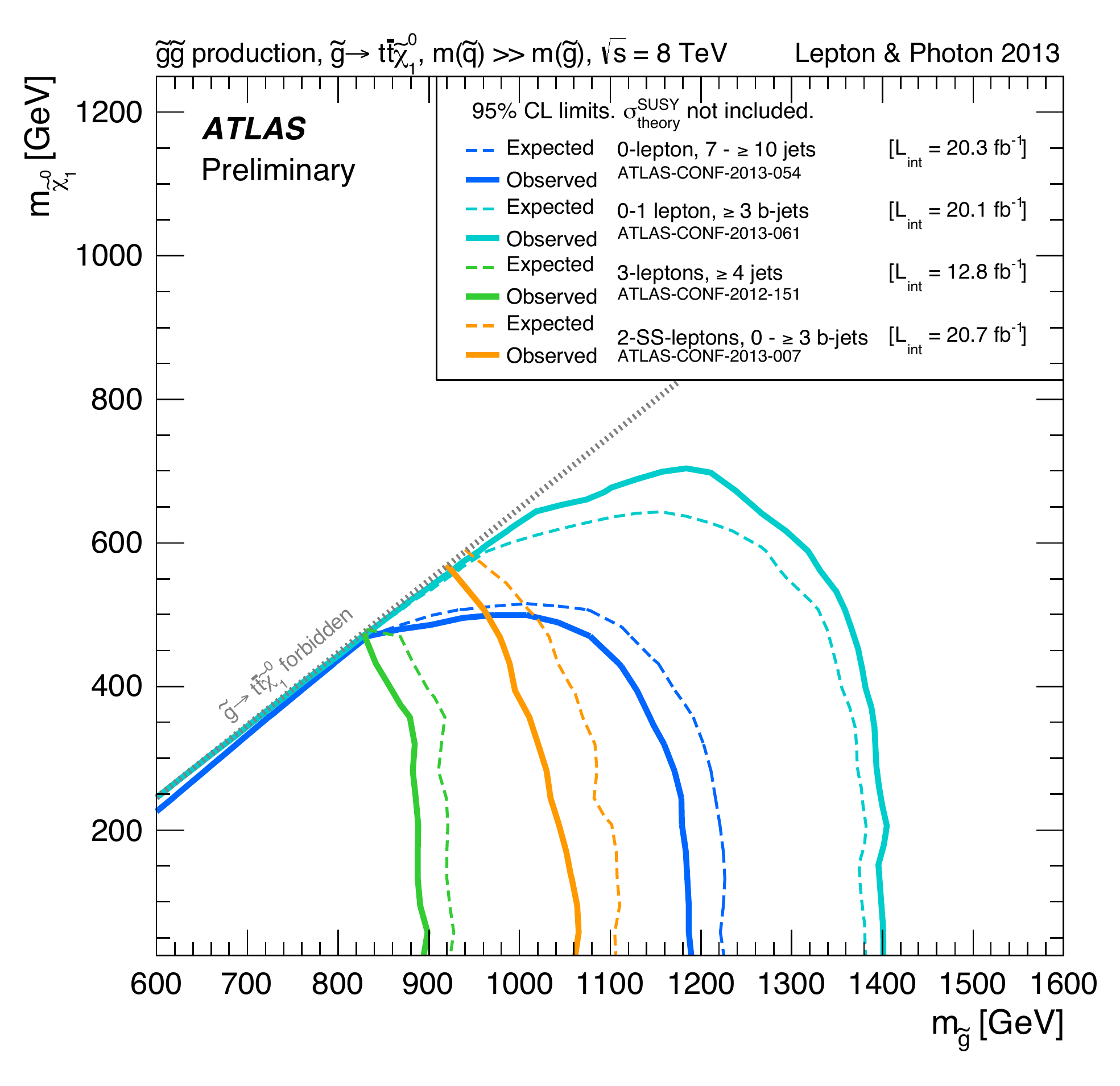}\hfill
\raisebox{.15cm}{\includegraphics[width=0.5\textwidth]{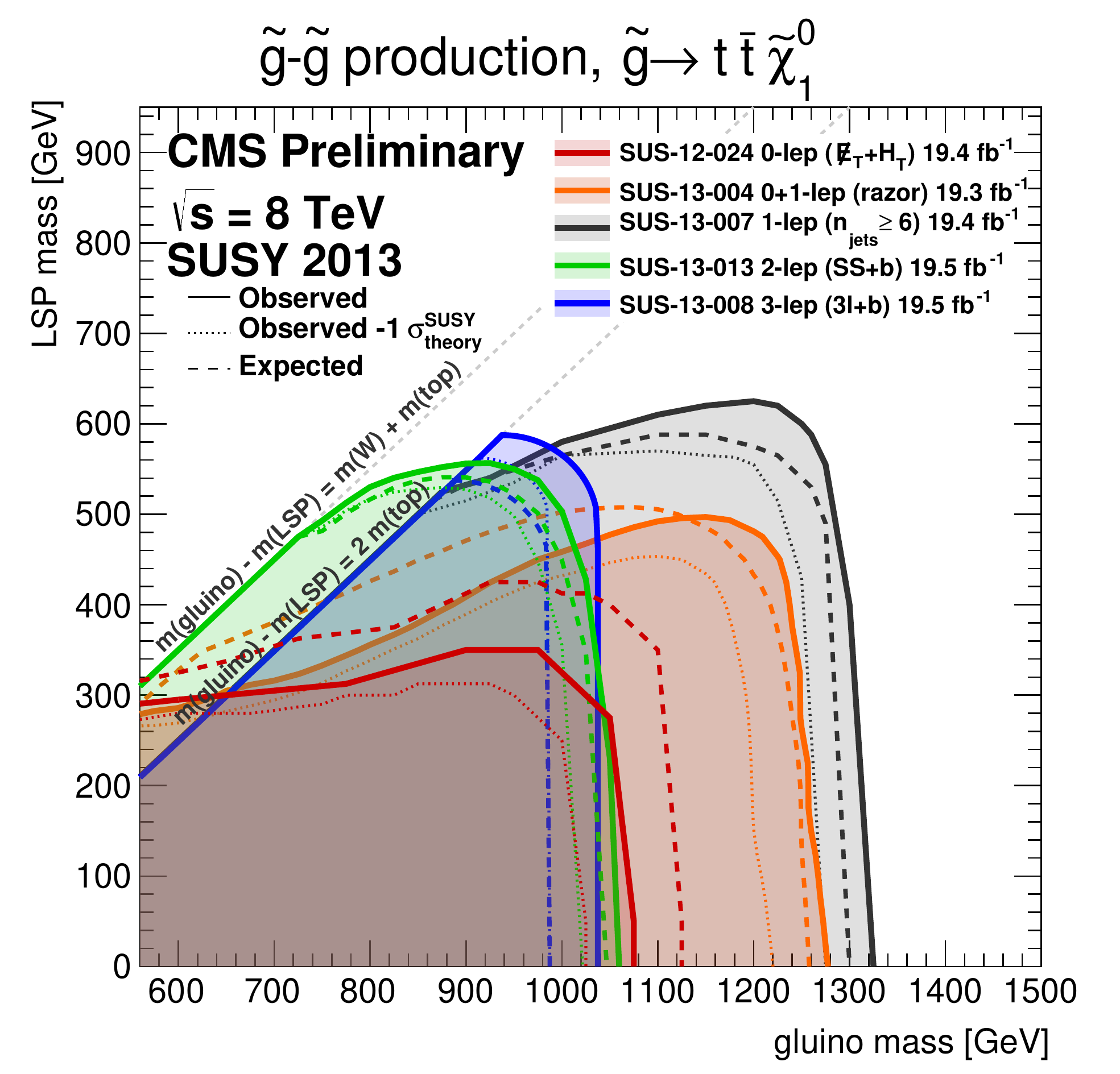}}\hfill
\caption{95\% C.L. exclusions for $\tilde g\to t \bar t \tilde \chi_1^0$ from the ATLAS \cite{ATLAS:2013zze,ATLAS:2013zzf,ATLAS:2012epx,ATLAS:2013tma} (left) and CMS \cite{Chatrchyan:2013wxa,CMS:wwa,CMS:zzb,CMS:zzc,CMS:zzd} (right) experiments in the plane $(m_{\tilde g}, m_{\tilde \chi^0_1})$, where $\tilde g$ is the gluino and $\tilde \chi_1^0$ is the lightest neutralino.\label{gluino}}
\end{figure}

Naturalness requires the mass of the third generation superpartners to be close to the electroweak scale. Although these masses enter only logarithmically in the radiatively corrected expressions \eqref{deltamhMSSM} for the physical Higgs masses, the parameters of the potential are additively renormalized and get quadratic corrections $\propto \tilde m_{\rm soft}^2$. In particular for $m_{H_u}$, which gets the largest contribution from the top-stop sector, one has \cite{Gherghetta:2012gb}
\begin{equation}\label{tuningmHu}
\delta m_{H_u}^2 = -\frac{3 y_t^2}{16\pi^2}\big(\tilde m_{Q_3}^2 + \tilde m_{U_3}^2 + A_t^2\big)\log\frac{\Lambda^2}{M_{S}^2}
\end{equation}
between the average scale $M_S$ of soft masses and the high scale $\Lambda$ at which these soft masses are generated, which can be considered as the ``input'' scale for RGE running. This in turn reflects into a fine-tuning of the weak scale, since $m_{H_u}$ enters the definition of $v$ through \eqref{susyvev1}. One can define the fine-tuning in $v$ as a sum of the individual tunings with respect to different parameters
\begin{equation}
\frac{d \log v^2}{d \log \tilde m_i} =\frac{\tilde m_i^2}{v^2}\frac{d m_{H_u}^2}{d \tilde m_i^2}\frac{d v^2}{d m_{H_u}^2} + \cdots,
\end{equation}
where the dots indicate subleading contributions from the other parameters $m_{H_d}$, $\mu$, and $\tilde m_i = (\tilde m_{Q_3}, \tilde m_{U_3}, A_t)$. The dependence of $v$ on $m_{H_u}$ can be read off eq. \eqref{susyvev1},
\begin{equation}
\frac{d v^2}{d m_{H_u}^2} = \frac{2 v^2}{m_Z^2 \cos^2 2\beta} \simeq - 2\frac{v^2}{m_Z^2} +{\Ord}\Big(\frac{1}{\tan^2\beta}\Big),
\end{equation}
so that using \eqref{tuningmHu} one finally gets, in the limit of large $\tan\beta$,
\begin{equation}
\Delta = \sum_i \left| \frac{d\log v^2}{d\log \tilde m_i}\right| \simeq \frac{3 y_t^2}{8\pi^2}\frac{\tilde m_{Q_3}^2 + \tilde m_{U_3}^2 + A_t^2}{m_Z^2}\log\frac{\Lambda^2}{M_S^2}.
\end{equation}
As expected, this quantity depends logarithmically on the high SUSY-breaking scale and quadratically on the soft squark masses. One gets an overall fine-tuning of about 1\% for a lightest stop mass $m_{\tilde t_1}$ of 1 TeV, generic $A_t\approx m_t$ and maximal mixing.

\begin{figure}[t]
\centering%
\includegraphics[width=.63\textwidth]{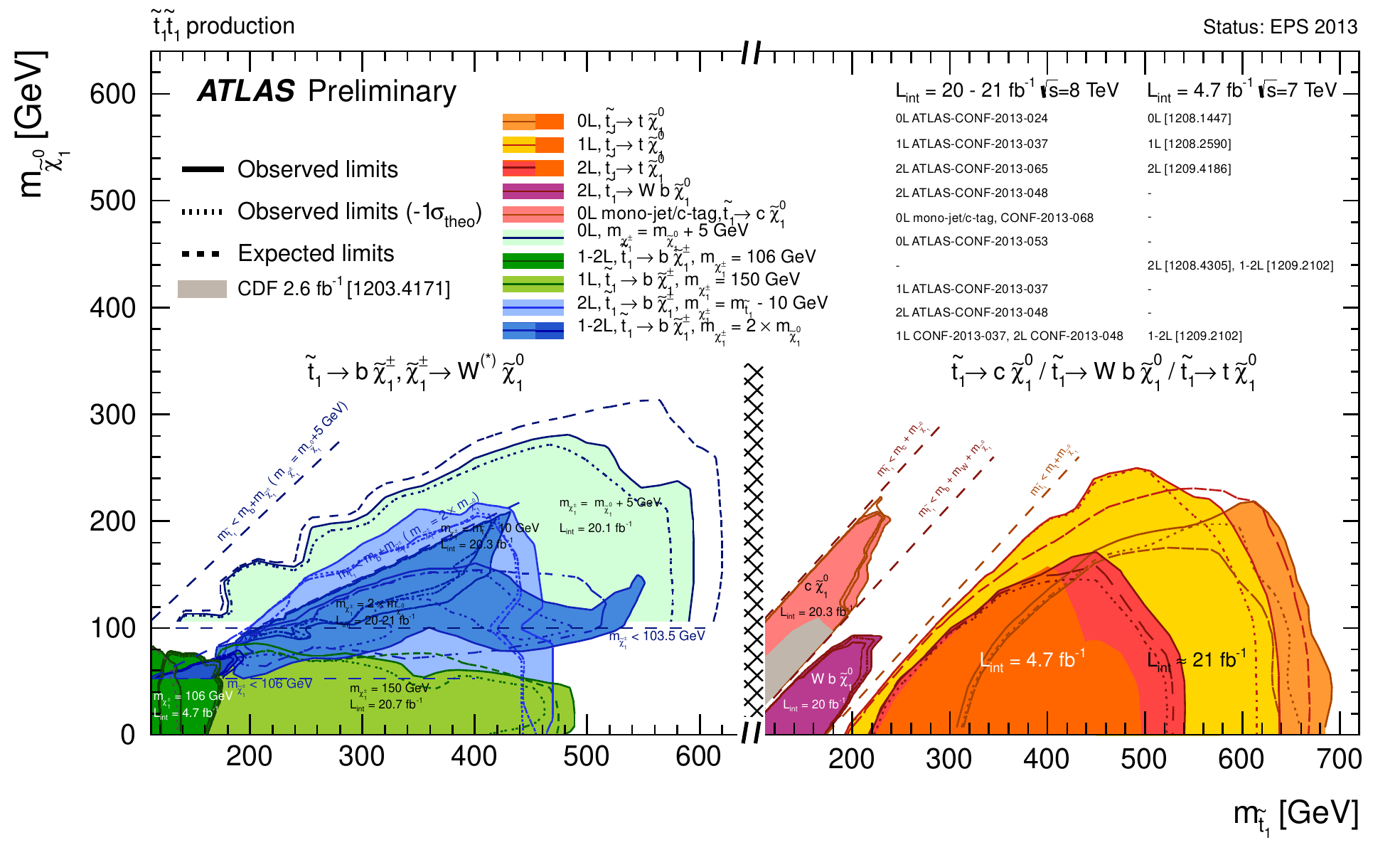}
\raisebox{2.2mm}{\includegraphics[width=.36\textwidth]{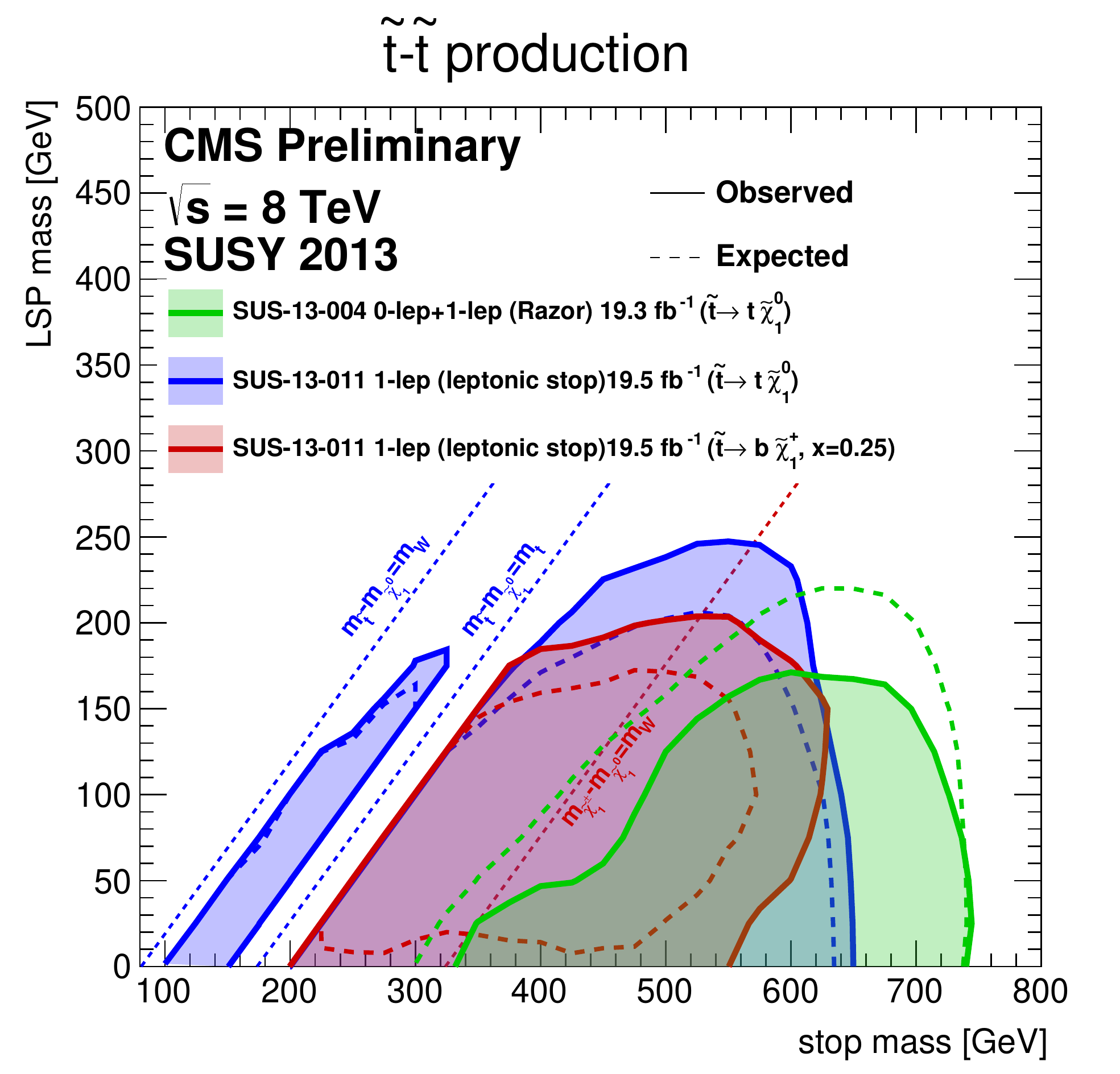}}
\caption{95\% C.L. exclusions from ATLAS \cite{ATLAS:2013cma,ATLAS:2013pla,ATLAS:2013zza,ATLAS:2013zzb,ATLAS:2013zzc,ATLAS:2013zzd,Aad:2012ywa,Aad:2012xqa,Aad:2012uu,Aad:2012tx,Aad:2012yr,ATLAS:2013zzd} (left) and CMS \cite{CMS:zzb,Chatrchyan:2013xna} (right) on $\tilde t_1\to t\chi_1^\pm$ and $\tilde t_1\to t\chi_1^0$, in the plane $(m_{\tilde t_1}, m_{\tilde \chi_1^0})$, where $\tilde t_1$ is the lightest stop and $\tilde\chi_1^\pm, \tilde \chi_1^0$ are the chargino and the neutralino.\label{stop}}
\end{figure}

The contribution from squarks other than the stop are irrelevant here, since they are suppressed by the much smaller Yukawa couplings of the light generations. For a given amount of fine-tuning, the naturalness bounds on their soft masses are therefore much weaker, and the first two generation squarks and the right-handed sbottom can easily lie in the multi-TeV range without affecting the fine-tuning of the weak scale.\footnote{Note that this splitting between the third and the first two generations of squarks is also an intriguing feature from the point of view of the $\U(2)^3$ flavour symmetry discussed in chapter~\ref{U2}, whose implications in the context of supersymmetry have been studied in \cite{Barbieri:2011ci}.}

\begin{figure}[t]
\begin{center}
\includegraphics[width=.48\textwidth]{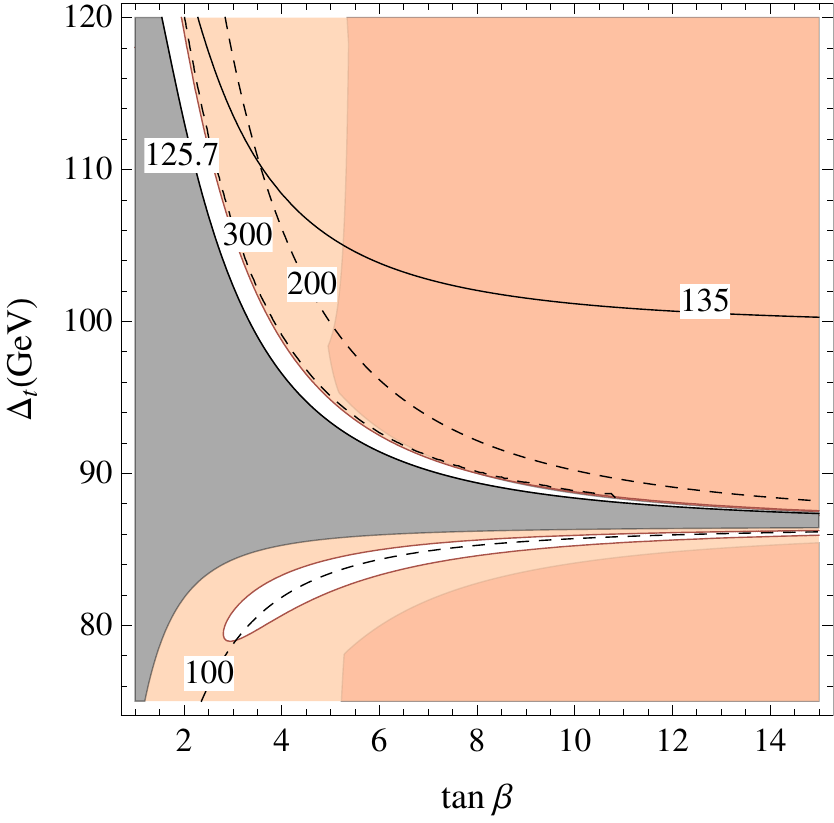}\hfill
\raisebox{.1cm}{\includegraphics[width=.48\textwidth]{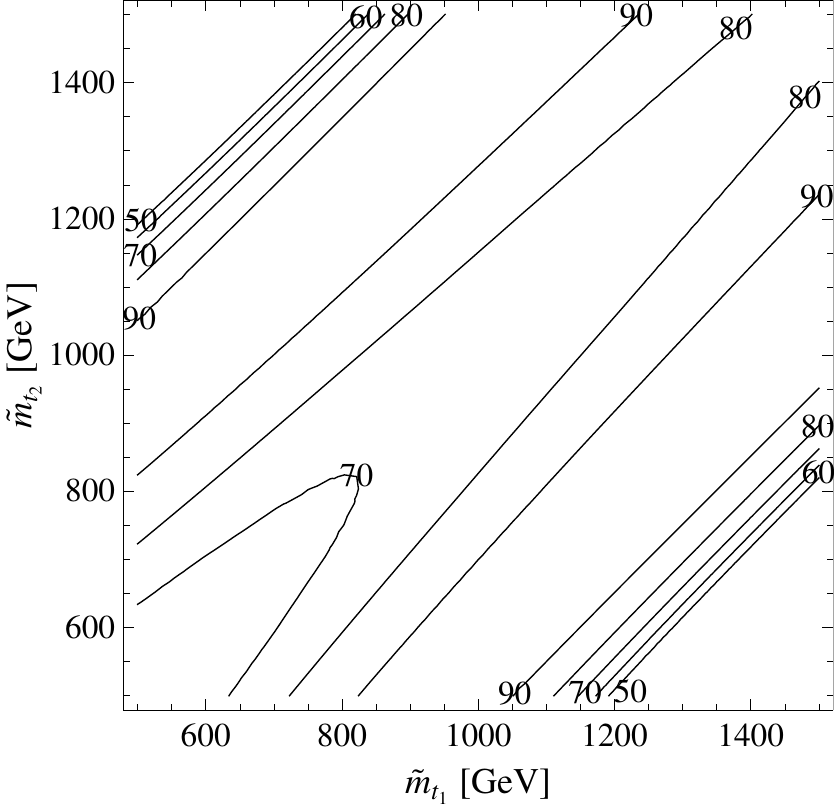}}
\caption{\label{fig:MSSM-1}\small MSSM. Left: Isolines of $m_{hh}$ (solid) and $m_{H^\pm}$ (dashed), the gray region is unphysical because of $m^2_A<0$.  Light colored regions are excluded at 95$\%$ C.L. by the Higgs fit, the red regions are excluded by CMS direct searches for $A,H\to \tau^+\tau^-$. Right: Isolines of $\Delta_t$ as a function of the two physical stop masses $m_{\tilde t_1}$, $m_{\tilde t_2}$, including the D-term corrections and for $\tan\beta = 5$.}
\end{center}
\end{figure}

Naturalness arguments thus suggest a mass spectrum of supersymmetric particles with the following properties:
\begin{itemize}
\item third generation top superpartners, namely the left- and right-handed stops and the left-handed sbottom, below about 1 TeV in order to have a fine-tuning of less than about a percent;
\item not-too-heavy gluinos, which renormalize the stop masses through QCD loops and enter the Higgs mass at two loops;
\item light higgsinos, which have tree-level masses determined by the $\mu$ parameter (in the range of a few hundreds of GeV);
\item first two generation squarks not constrained by naturalness, which can be as heavy as several TeV.
\end{itemize}
The relevant s-particles to look for, from the point of view of natural supersymmetry, are thus the third generation squarks and the gluinos. In figures~\ref{gluino} and \ref{stop} we report the most recent exclusion plots from the direct searches at the CMS and ATLAS experiments at the LHC on the masses of these particles \cite{ATLAS:2013zze,ATLAS:2013zzf,ATLAS:2012epx,ATLAS:2013tma,Chatrchyan:2013wxa,CMS:wwa,CMS:zzb,CMS:zzc,CMS:zzd,ATLAS:2013cma,ATLAS:2013pla,ATLAS:2013zza,ATLAS:2013zzb,ATLAS:2013zzc,ATLAS:2013zzd,Aad:2012ywa,Aad:2012xqa,Aad:2012uu,Aad:2012tx,Aad:2012yr,ATLAS:2013zzd,Chatrchyan:2013xna}. From figure~\ref{gluino} we see that a gluino mass below 1 TeV is safely excluded by both experiments, while searches in some particular decay channel reach up to about 1.4 TeV. In figure~\ref{stop} the ATLAS (left) and CMS (right) exclusions for a stop decaying into top and neutralino and into bottom and chargino are shown. These searches are presently sensitive up to 700 GeV for the lightest stop mass in the worst case of a light neutralino, but there are still allowed regions of the parameter space with a somewhat heavier $\tilde\chi^0_1$ where the stop can be as light as about 300--400 GeV. Although somewhat problematic, these limits are still too weak to put the whole naturalness picture into serious trouble.

In the MSSM, however, an additional problem has to be considered.
From \eqref{upperboundmhdeltat} one can extract the value of $\Delta_t$ required to reproduce a Higgs mass of 125.7 GeV, assuming for the moment that the particle observed at the LHC coincides with the lightest Higgs state, $h_{\rm LHC} = h_1$. In the limiting case where $\cos 2\beta\to 1$ one has $\Delta_t^2 = m_h^2 - m_Z^2\simeq (86.5\, {\rm GeV})^2$. The situation as a function of $\tan\beta$ is shown in figure~\ref{fig:MSSM-1} left, where the grey area is unphysical. The impact of such high values of $\Delta_t$ on the stop masses can be read off figure \ref{fig:MSSM-1} right, where we show the isolines of $\Delta_t$ in the plane $(m_{\tilde t_1}, m_{\tilde t_2})$. One can see that in order to reproduce the measured value of $m_{h_1}$ at least one of the stops has to lie above a TeV. The area below the grey region in figure~\ref{fig:MSSM-1} left corresponds to a lightest Higgs mass below 125 GeV, where the observed state is identified with the heavier of the two Higgs bosons. Important constraints on both configurations arise from the measurement of the couplings of $h_{\rm LHC}$, as we will now discuss.

\section{Higgs couplings in the MSSM}\label{SUSY/HiggsMSSM}

The couplings of the physical states in the MSSM are modified with respect to those of a standard Higgs boson because of the mixing of $h$ with $H$. The couplings of $h_1$ read
\begin{align}\label{h1couplingsMSSM}
\frac{g_{h_1tt}}{g^{\text{SM}}_{htt}} &= R_{\tilde t}\Big(\cos\delta +\frac{\sin\delta}{\tan\beta}\Big), & \frac{g_{h_1bb}}{g^{\text{SM}}_{hbb}} &= \cos\delta -\sin\delta \tan\beta , & \frac{g_{h_1VV}}{g^{\text{SM}}_{hVV}} &=   \cos\delta,
\end{align}
in terms of the mixing angle $\delta = \alpha - \beta + \pi/2$ between $h$ and $H$. The factor $R_{\tilde t}$ includes the modification of the $h_1 t t$ coupling due to stop-loop effects. Here we neglect such an effect and set $R_{\tilde t}$ to 1.

\begin{figure}[t]
\begin{center}
\includegraphics[width=0.48\textwidth]{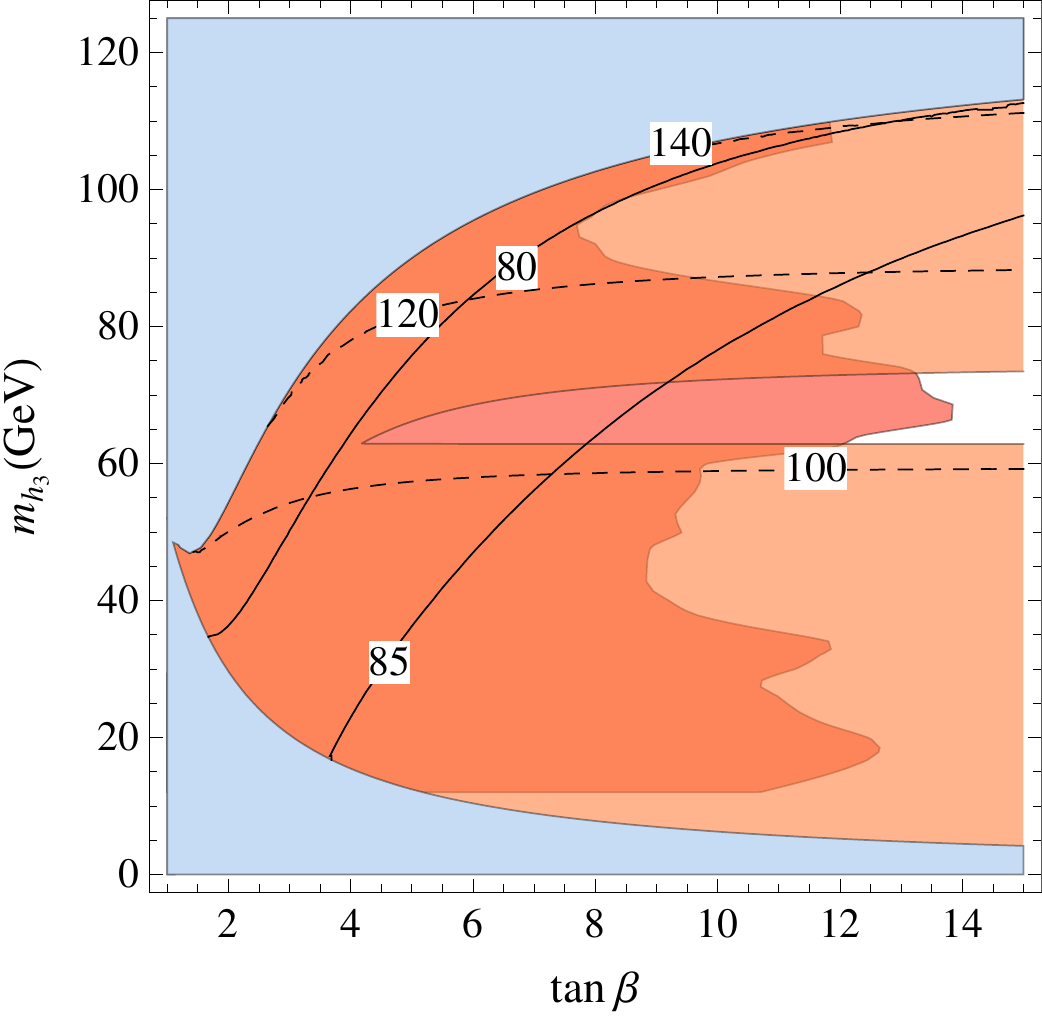}\hfill
\includegraphics[width=0.48\textwidth]{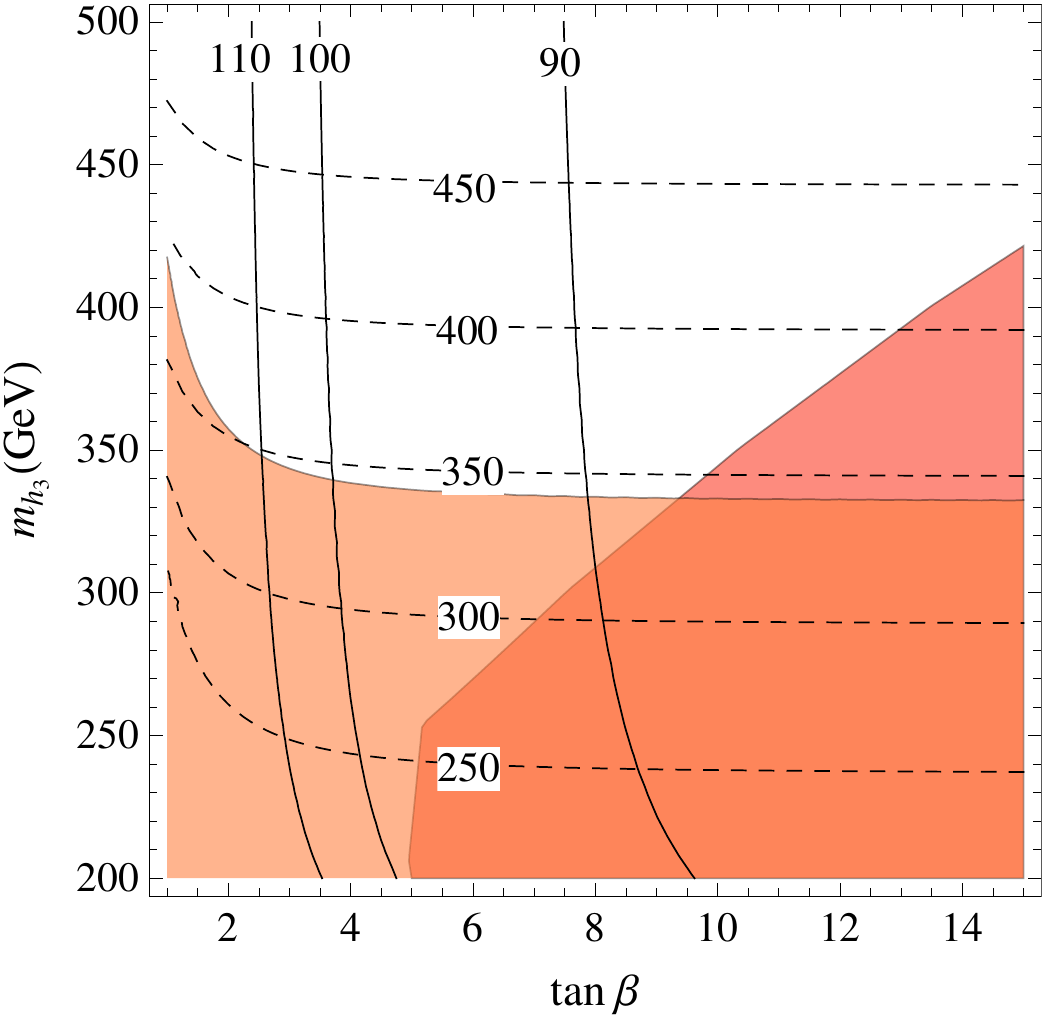}
\caption{\label{fig7} MSSM. Isolines of $\Delta_t$ (solid) and $m_{H^\pm}$ (dashed) at $(\mu A_t)/\langle m_{\tilde{t}}^2\rangle \ll 1$. The orange regions are excluded at 95\% C.L. by the fit of the couplings of $h_{\rm LHC}$, the blue region is unphysical. Left: $h_{\rm LHC}>h_3$, red region is excluded by LEP direct searches for $h_3 \to b\bar b$. Right: $h_{\rm LHC}<h_3$, red region is excluded by CMS direct searches for $A, H\to \tau^+\tau^-$ \cite{CMS:gya}.}
\end{center}
\end{figure}

The modified couplings \eqref{h1couplingsMSSM} affect both the production cross-sections of $h_{1,3}$ as well as their branching ratios.
One can then perform a fit of all ATLAS \cite{ATLAS:2013rma,ATLAS:2013qma,ATLAS:2013pma,ATLAS:2013oma,ATLAS:2013nma,ATLAS:2013mma,ATLAS:2013wla,ATLAS:moriond1,ATLAS:moriond2,ATLAS:moriond3}, CMS \cite{CMS:ril,CMS:xwa,CMS:bxa,CMS:utj,Chatrchyan:2013vaa,CMS:zwa,CMS:gya,CMS:moriond1,CMS:moriond2,CMS:moriond3,CMS:moriond4} and TeVatron \cite{tevatron:2013} data collected so far on the various signal strengths of $h_{\text{LHC}}$ in order to set an upper bound on the mixing angle $\delta$. In figure \ref{fig:MSSM-1} the result of such a fit is shown in the plane $(\tan\beta, \Delta_t)$, together with the isolines of $m_{hh}$, as defined in \eqref{mhhMSSM}, and of the charged Higgs mass $m_{H^\pm}$. Notice how the fit alone restricts the allowed region to two tiny slices of the parameters space, respectively for $m_{hh} < m_{h_{\rm LHC}}$ and $m_{hh} > m_{h_{\rm LHC}}$. The same situation is reflected in figure~\ref{fig:MSSM-1} in terms of the physical parameters $\tan\beta$ and $m_{h_3}$, for the two cases. As already discussed, it is clear from the figures that a large value of $\Delta_t$ is needed to make the MSSM consistent with a 125 GeV Higgs boson. Note that in the plane $(\tan{\beta}, m_{h_3})$ the isolines of  $\Delta_t$ are increasingly  large at lower $\tan{\beta}$: a sign of increasing fine tuning.

The projection of the measurements of the signal strengths of 
$h_{\text{LHC}}$ at the LHC with 14 TeV of center-of-mass energy, assuming central values equal to one, is expected to scrutinize most of the parameter space. We have checked that this is indeed the case with the indirect sensitivity to $m_{h_3}$ in the right panel of figure~\ref{fig7}, which will be excluded up to about 1 TeV, as well as with the closure of the white region in the left side of the same figure. A more detailed discussion of the projected sensitivities to the signal strengths at LHC14 is carried out in section~\ref{NMSSM/Sdecoupled}.

Notice that a similar exclusion from the Higgs fit will hold also for the \CP-odd and charged Higgs bosons, whose masses are fixed in terms of the one of $h_3$ in the MSSM.

A warning should be kept in mind, however, relevant to the case $h_3 < h_{\text{LHC}}$:  the one loop corrections to the mass matrix controlled by $(\mu A_t)/\langle m_{\tilde{t}}^2\rangle$ modify the left side of figure \ref{fig7} for $(\mu A_t)/\langle m_{\tilde{t}}^2\rangle \gtrsim 1$, changing in particular the currently and projected allowed regions.

\begin{figure}[t]
\begin{center}
\includegraphics[width=.48\textwidth]{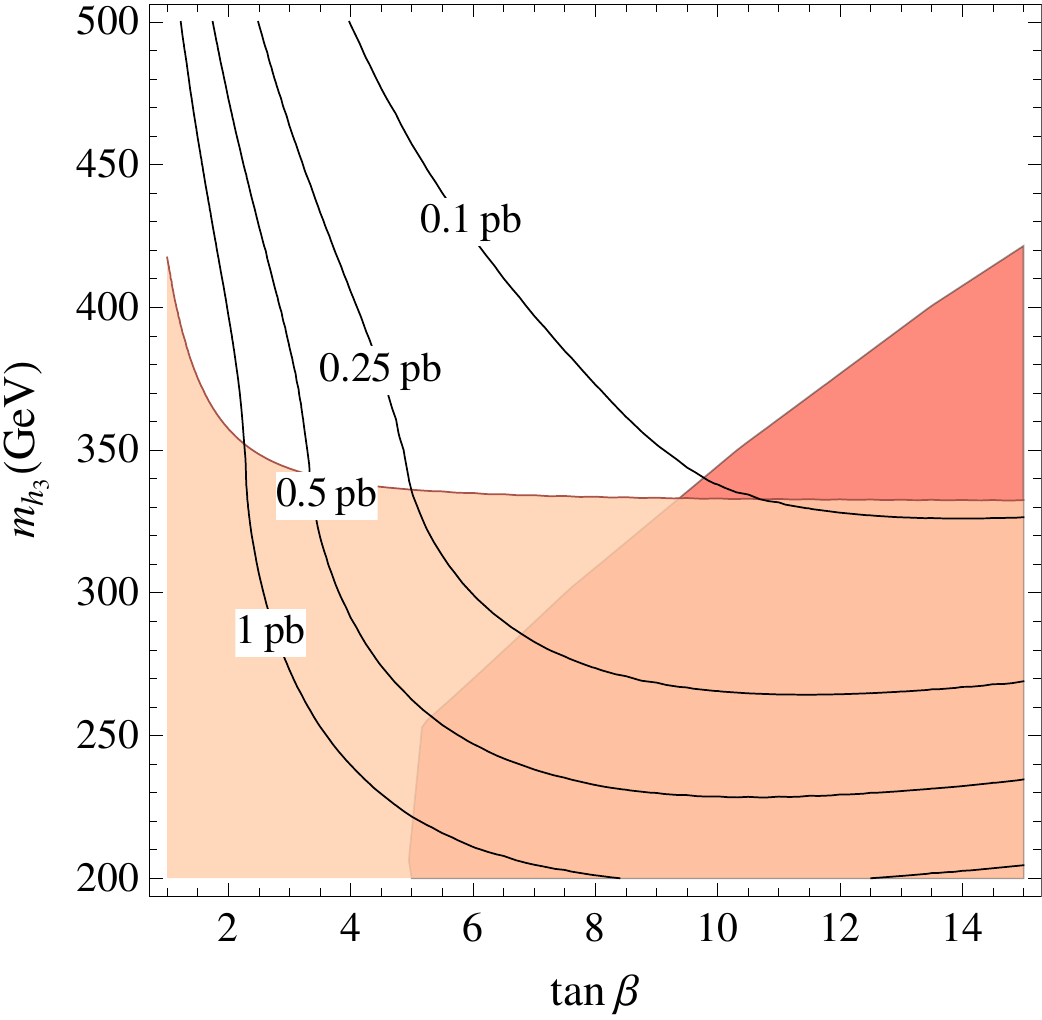}\hfill
\includegraphics[width=.48\textwidth]{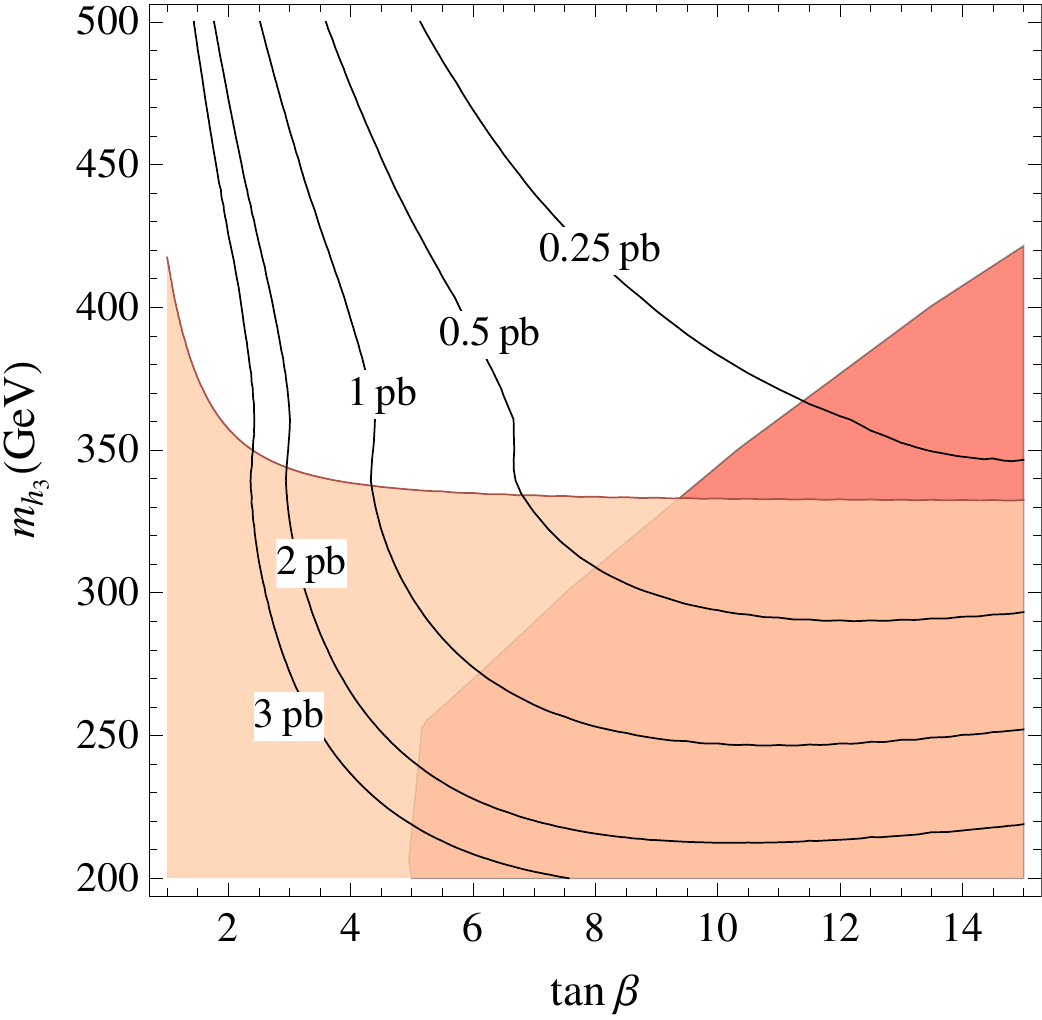}
\caption{\label{fig:MSSM-Xsec}\small MSSM. Isolines of gluon fusion production cross section $\sigma(gg\to h_3)$. Light colored region is excluded at 95$\%$ C.L., the red region is excluded by CMS direct searches for $A,H\to \tau^+\tau^-$. Left: LHC8. Right: LHC14.}
\end{center}
\end{figure}

The couplings of  $h_3$ are
\begin{align}
\frac{g_{h_3tt}}{g^{\text{SM}}_{htt}} &=\sin\delta-\frac{\cos\delta}{\tan\beta}, & \frac{g_{h_3bb}}{g^{\text{SM}}_{hbb}}&=\sin\delta+\tan\beta\cos\delta, & \frac{g_{h_3VV}}{g^{\text{SM}}_{hVV}}&= \sin\delta.
\label{h3couplingsMSSM}
\end{align}
These couplings allow to compute the gluon-fusion production cross section of $h_3$ by means of \cite{Cheung:2013bn}
\begin{equation}
\sigma(gg\rightarrow h_3) = \sigma^{\text{SM}}\left(gg\rightarrow H(m_{h_3})\right)
\Big|\mathcal{A}_t \frac{g_{h_3tt}}{g^{\text{SM}}_{htt}} + \mathcal{A}_b \frac{g_{h_3bb}}{g^{\text{SM}}_{hbb}}\Big|^2,
\end{equation}
where
\begin{equation}
\mathcal{A}_{t,b} =  \frac{F_{\frac{1}{2}}(\tau_{t,b})}{F_{\frac{1}{2}}(\tau_t) + F_{\frac{1}{2}}(\tau_b)},   ~~~~~~\tau_i = 4 \frac{m_i^2}{m_{h_3}^2},
\end{equation}
and $F_{\frac{1}{2}}(\tau)$ is a one-loop function that can be found e.g. in \cite{Azatov:2012qz,Carmi:2012in}. We used the values of $\sigma^{\text{SM}}$ at NNLL precision provided in \cite{Dittmaier:2011ti}, and the running masses $m_{t,b}$ at NLO precision. We checked the validity of this choice by performing the same computation both with the use of masses at LO precision and K-factors \cite{Anastasiou:2009kn}, and with the program {\sc Higlu} \cite{Spira:1995rr,Spira:1995mt}, finding in both cases an excellent agreement. Our results are also in very good agreement with the ones recently presented in \cite{Arbey:2013jla} and \cite{Djouadi:2013vqa}.
We show in figures~\ref{fig:MSSM-Xsec}--\ref{fig:MSSM-BRf} the results for the gluon-fusion production cross sections and the widths of $h_3$ for the MSSM in the $m_{h_3} > m_{h_{\rm LHC}}$ case, in the $(\tan\beta, m_{h_3})$ plane of figure~\ref{fig7}.

\begin{figure}[t]
\begin{center}
\includegraphics[width=.48\textwidth]{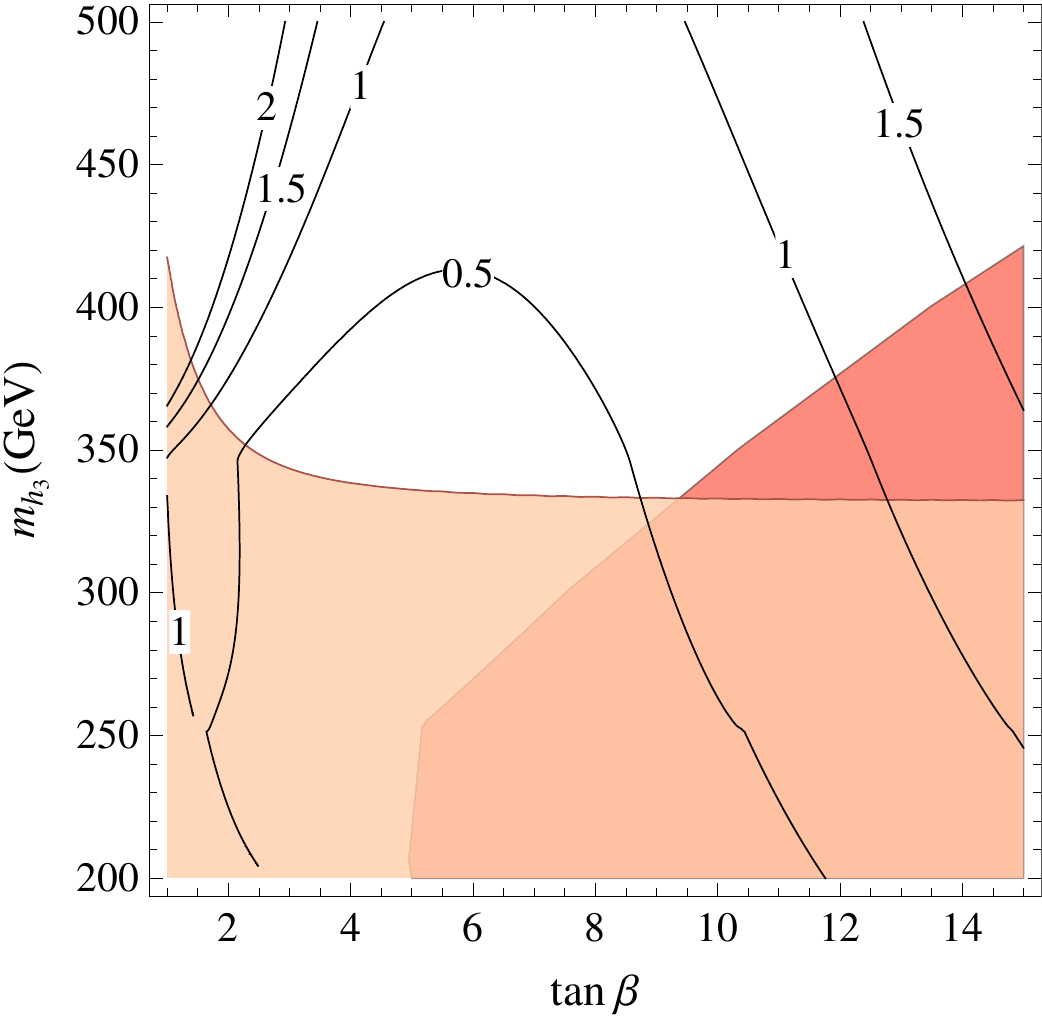}\hfill
\includegraphics[width=.48\textwidth]{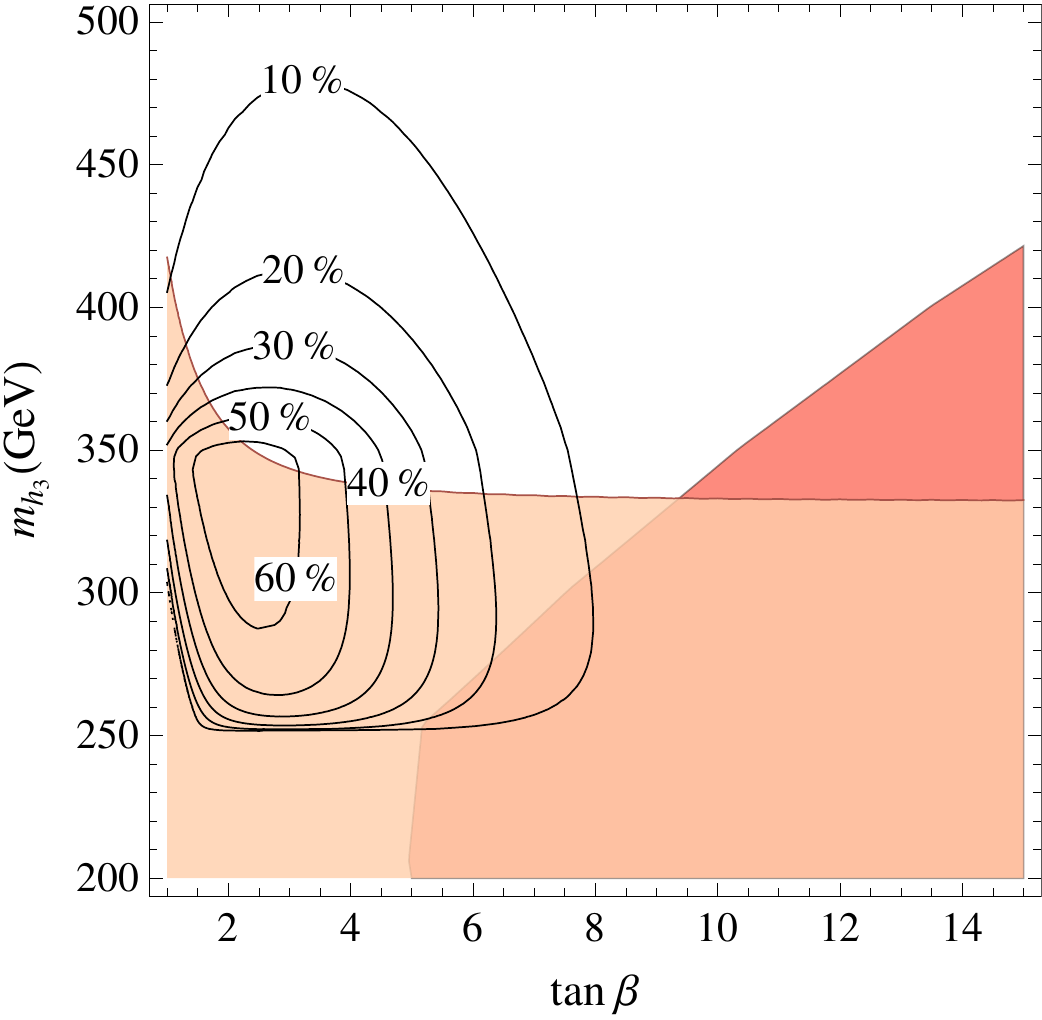}
\caption{\label{fig:MSSM-BRs}\small MSSM. Left: Isolines of the total width $\Gamma_{h_3}$ (GeV). Right: Isolines of BR$(h_3\to h h)$. The light colored region is excluded at 95\% C.L., the red region is excluded by CMS direct searches.}
\end{center}
\end{figure}
\begin{figure}[t!]
\begin{center}
\includegraphics[width=.48\textwidth]{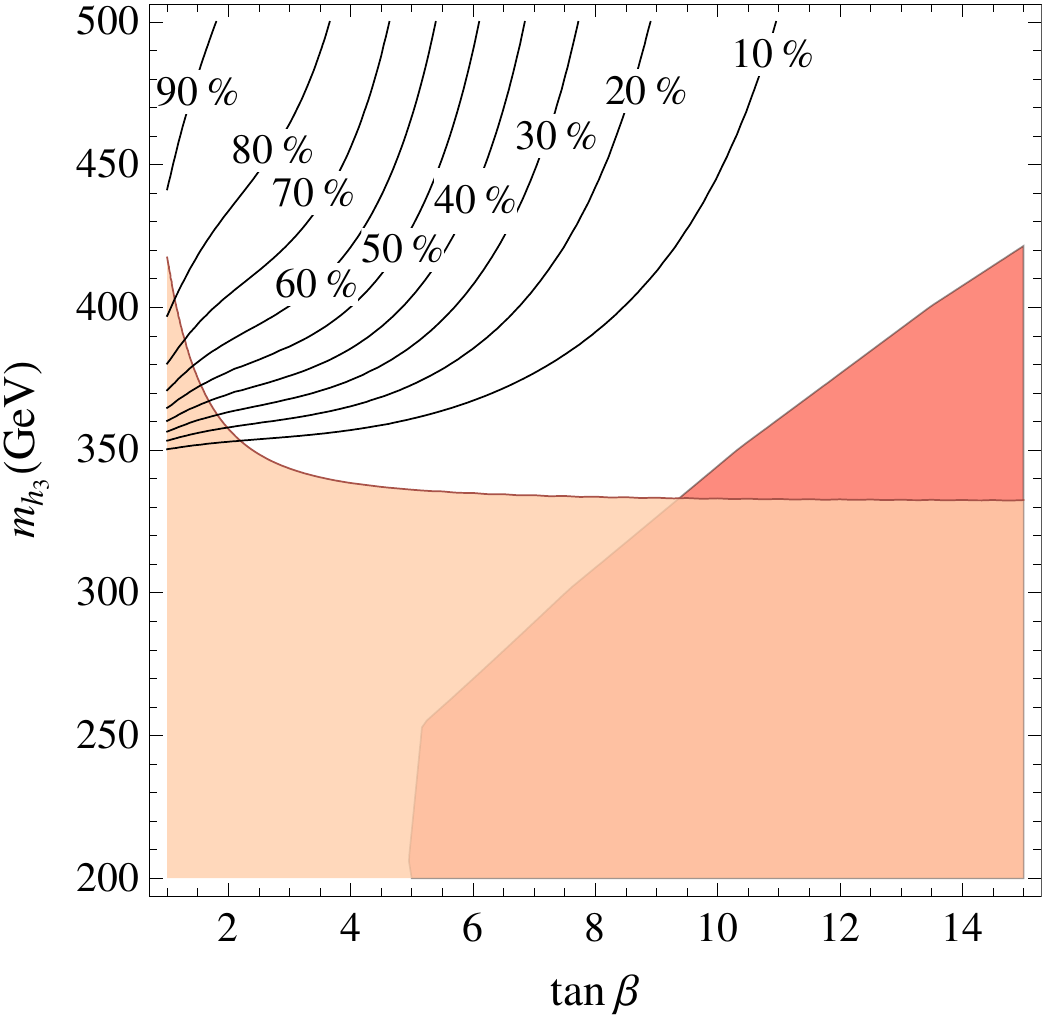}\hfill
\includegraphics[width=.48\textwidth]{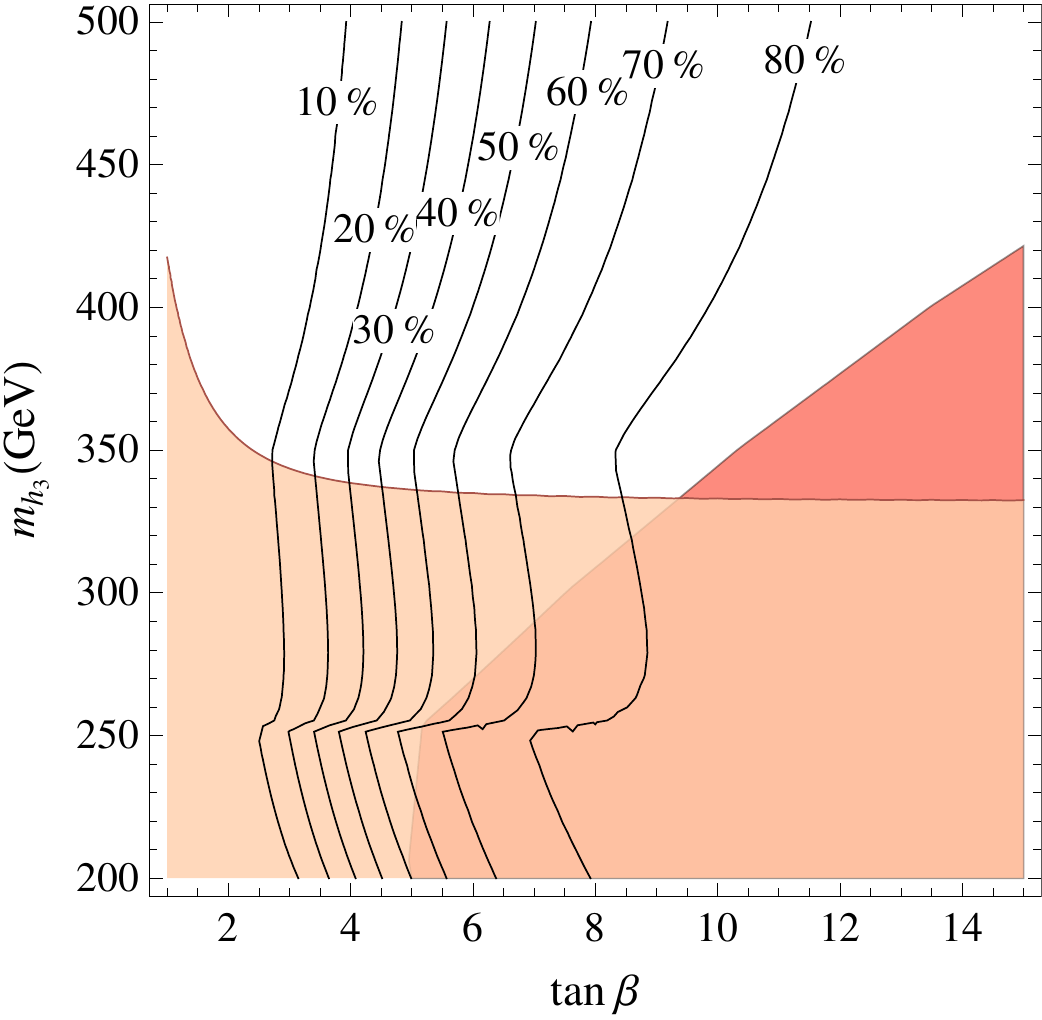}
\caption{\label{fig:MSSM-BRf}\small MSSM. Left: isolines of BR$(h_3\to t\bar t)$. Right: Isolines of BR$(h_3\to b \bar{b})$. The light colored region is excluded at 95$\%$ C.L., the red region is excluded by CMS direct searches.}
\end{center}
\end{figure}

In the same plane $\sigma (gg\rightarrow A)$ is also determined, since the pseudoscalar mass $m_A$ is related to $m_{h_3}$ through \eqref{mHpMSSM}, which allows to delimit the currently excluded region by the direct searches for $A, h_3\rightarrow \tau^+ \tau^-$. Such a region is known to be significant, especially for growing $\tan{\beta}$. In figures~\ref{fig:MSSM-1}--\ref{fig:MSSM-BRf} we draw the region excluded by such search, as inferred from \cite{CMS:gya}. In the left panel of figure~\ref{fig7}, instead, we draw with the same red colour the region excluded by the LEP searches for $h\to b\bar b$, which are relevant for $m_{h} < 115$ GeV.

\chapter{The Higgs sector of a most natural NMSSM}\label{NMSSM}

The mounting evidence for a Higgs boson $h_{\text{LHC}}$ at 125 GeV with Standard-Model-like properties \cite{Chatrchyan:2012ufa,Aad:2012tfa,CMS:ril,CMS:xwa,CMS:bxa,CMS:utj,Chatrchyan:2013vaa,CMS:zwa,CMS:gya,ATLAS:2013rma,ATLAS:2013qma,ATLAS:2013pma,ATLAS:2013oma,ATLAS:2013nma,ATLAS:2013mma,ATLAS:2013wla,tevatron:2013} has a significant impact on the issue of naturalness in the context of supersymmetry.
While the identification of the new resonance with the lightest Higgs boson of a supersymmetric model is a well-motivated possibility, we have already seen that the measured mass is in serious tension with maintaining naturalness in the MSSM, requiring either large stop masses, above a TeV, in the case of negligible mixing, or a large trilinear $A_t$ term. At the heart of this problem is the fact that the quartic terms of the Higgs potential in the MSSM are controlled by the weak gauge couplings. 

The situation is different in the Next-to-Minimal Supersymmetric Standard Model (NMSSM), where a singlet superfield $S$ couples to the Higgs superfields, $H_u$ and $H_d$, via the Yukawa-like coupling $\lambda S H_u H_d$ \cite{Fayet:1974pd} (see \cite{Ellwanger:2009dp} for a review).  
On the one hand, the inclusion of this coupling allows to get a 126 GeV Higgs boson mass with both the stop masses well below a TeV. On the other hand, values of $\lambda \gtrsim 1$ suppress the sensitivity of the Higgs vacuum expectation value with respect to changes in the soft supersymmetry-breaking masses, thus still keeping the fine tuning at a moderate level even for stop masses up to 1 TeV \cite{Barbieri:2006bg,Hall:2011aa,Agashe:2012zq}.%
\footnote{A recent analysis \cite{Gherghetta:2012gb} finds that the fine tuning in the NMSSM with a scale-invariant superpotential can be above $5\%$ for stop masses up to $1.2$ TeV and gluino masses up to 3 TeV for $\lambda \approx 1$, moderate $\tan{\beta}$ and the messenger scale at 20 TeV.}
We shall comment below on how a coupling $\lambda \gtrsim 1$ can be made compatible with gauge coupling unification.

In general terms, to see whether the newly found resonance at 126 GeV is part of an extended Higgs system is a primary task of the current and future experimental studies. Given the above motivations, this appears to be especially true  for the extra Higgs states of the NMSSM, which might be the lightest new particles of a suitable supersymmetric model, except perhaps for the lightest supersymmetric particle (LSP).
A particularly important question is how the measurements of the couplings of $h_{\text{LHC}}$, current and foreseen, bear on this issue, especially in comparison with the potential of the direct searches of new Higgs states, and in a very similar fashion to what discussed in section~\ref{SUSY/MSSM} in the case of the MSSM.\footnote{For recent studies see e.g. refs. \cite{Gupta:2012fy,DAgnolo:2012mj}.}

Not the least difficulty that one encounters in attacking these problems is the number of parameters that enter the Higgs system of the NMSSM, especially if one does not want to stick to a particular version of it but rather wishes to consider the general case.
Here we aim at an analytic understanding of the properties of the Higgs system of the general NMSSM, trying to keep under control as much as possible the complications due to the proliferation of model parameters and avoid the use of benchmark points.

The content of this chapter is the following. In section~\ref{NMSSM/Lagrangian} we define the model and we establish some relations between the physical parameters of the \CP-even Higgs system valid in the general NMSSM. In sections \ref{NMSSM/Sdecoupled} and \ref{NMSSM/Hdecoupled} we consider two limiting cases in which one of the \CP-even scalars is decoupled, determining in each situation the sensitivity of the measurements of the couplings of $h_{\text{LHC}}$, current and foreseen, as well as the production cross sections and the branching ratios (BR) for the new intermediate scalar. In section~\ref{NMSSM/general} we consider some peculiarities of the more general case where all the three states mix with each other.
To conclude, in section~\ref{NMSSM/unification} we illustrate a possible simple and generic extension of the NMSSM that can make it compatible with standard gauge unification even for a coupling $\lambda \gtrsim 1$.

\section{The Next-to-Minimal Supersymmetric Standard Model}\label{NMSSM/Lagrangian}

The Next-to-Minimal Supersymmetric Standard Model differs from the MSSM only in the Higgs sector. In addition to the two doublets $H_u$ and $H_d$ one adds a third chiral superfield $S$ to the model, singlet under all the gauge groups of the SM. The singlet $S$ interacts at tree-level only with itself and with the other two Higgses through a cubic coupling $\lambda S H_u H_d$ \cite{Fayet:1974pd}. The most general superpotential of the NMSSM then reads
\begin{equation}
W_{\rm NMSSM} = W_{\rm MSSM} + \frac{m_S^2}{2} S^2 + \frac{\kappa}{3} S^3 + \lambda H_u H_d S.
\end{equation}
A scalar potential for $S$ is generated as usual through $\mathcal{F}$-terms. $\mathcal{D}$-terms are not involved there since the singlet has no gauge interactions. Including also the soft SUSY-breaking terms, which are of course present for all the fields, the scalar potential for the Higgs sector reads
\begin{align}
V(H_u, H_d, S) &= \left |\mu + \lambda S\right |^2\left(|H_u^+|^2 + |H_u^0|^2 + |H_d^-|^2 + |H_d^0|^2\right) + V_{\mathcal{D}}(H_u, H_d) \notag\\
&+ \left |m_S^2 S + \kappa S^2 +  \lambda (H_u^+ H_d^- - H_u^0 H_d^0)\right |^2 + V_{\rm soft}(H_u, H_d, S),
\end{align}
where $V_{\mathcal{D}}$ are the same $\mathcal{D}$-terms of \eqref{scalarpotSUSY}.

The cubic interaction of the singlet with $H_u$ and $H_d$ generates an effective $\mu$-term when $S$ takes a vev $v_S$. Particular versions of the NMSSM, where $\mu$ is forbidden at tree-level by some symmetry, provide thus a solution to the $\mu$-problem. Two relevant examples of this kind are the scale-invariant NMSSM, where all the dimensionful couplings do vanish, and the Peccei-Quinn-invariant NMSSM, where the superpotential satisfies a global Peccei-Quinn symmetry \cite{Peccei:1977ur,Peccei:1977hh} which forbids both $m_S$ and $\kappa$, in addition to the $\mu$-term.

As we now see, the most important consequence of the $\lambda$ coupling is to add a new contribution to the tree-level masses of the Higgs bosons, thus reducing the value of the radiative correction.

\subsection{Physical parameters of the \CP-even Higgs system}\label{NMSSM/Lagrangian/parameters}

Assuming a negligibly small violation of \CP\ in the Higgs sector, we take as a starting point the form of the squared mass matrix of the neutral \CP-even Higgs system in the general NMSSM: 
\begin{equation}\label{scalar_mass_matrix}
{\M}^2=\left(
\begin{array}{ccc}
m_Z^2 \cos^2\beta+m_A^2 \sin^2\beta & \left(v^2 \lambda ^2-m_A^2-m_Z^2\right) \cos\beta \sin\beta &  v M_1  \\
 \left(v^2 \lambda ^2-m_A^2-m_Z^2\right) \cos\beta \sin\beta & m_A^2 \cos^2\beta+m_Z^2 \sin^2\beta +\delta_t^2 &  v M_2  \\
  v  M_1 &  v M_2 & M_3^2
\end{array}
\right)
\end{equation}
in the basis $\mathcal{H} = (H_d^0, H_u^0, S)^T$. In this equation 

\begin{equation}
\label{mHcharged}
m_A^2 = m_{H^{\pm}}^2 - m_W^2 +\frac{\lambda^2 v^2}{2},
\end{equation}
where $m_{H^{\pm}}$ is the physical mass of the single charged Higgs boson, $v \simeq 174$ GeV, and
\begin{equation}
\delta_t^2 = \Delta_t^2(m_{\tilde t_1}, m_{\tilde t_2}, \theta_{\tilde t})/ \sin^2\beta
\label{delta-t}
\end{equation}
is the well-known effect of the top-stop loop corrections to the quartic coupling of $H_u$ \eqref{deltamhMSSM}, with $m_{\tilde t_{1,2}}$ and $\theta_{\tilde t}$ physical stop masses and mixing.
We neglect the analogous correction \eqref{corrmHp} to equation \eqref{mHcharged} \cite{Djouadi:2005gj}, which lowers $m_{H^{\pm}}$ by less than 3 GeV for stop masses below 1 TeV. More importantly we have also not included in \eqref{scalar_mass_matrix} the one loop corrections to the $12$ and $11$ entries, respectively proportional to the first and second power of $(\mu A_t)/\langle m_{\tilde{t}}^2\rangle$, to which we shall return.
We leave unspecified the other parameters in eq. (\ref{scalar_mass_matrix}), $M_1, M_2, M_3$, which are not directly related to physical masses and  depend on the particular NMSSM under consideration. Note that here, unlike in the MSSM case, the quantity $m_A$ in \eqref{scalar_mass_matrix} and \eqref{mHcharged} does not correspond to the physical mass of a \CP-odd scalar, but has to be considered just as a parameter of the Lagrangian.

The vector of the three physical mass eigenstates  $\mathcal{H}_{\rm ph}$ is related to the original scalar fields by
\begin{equation}
\mathcal{H} = R^{12}_{\alpha} R^{23}_{\gamma} R^{13}_{\sigma} \mathcal{H}_{\rm ph} \equiv R \mathcal{H}_{\rm ph},
\label{rotation_matrix}
\end{equation}
where $R^{ij}_\theta$ is the rotation matrix in the $i j$ sector by the angle $\theta = \alpha,\gamma,\sigma$. For later purposes it is useful to define also the basis $(H,h,s)^T = R_{\beta-\pi/2}^{12}\mathcal{H}$ where the field $h = c_\beta H_u + s_\beta H_d$ takes all the vev.

Defining $ \mathcal{H}_{\rm ph} = (h_3, h_1, h_2)^T$, we have
\begin{equation}
R^T {\M}^2 R = \diag(m_{h_3}^2, m_{h_1}^2, m_{h_2}^2).
\label{diag_matrix}
\end{equation}
We identify $h_1$ with the state found at the LHC, so that $m_{h_1} = 125.7$ GeV.
From (\ref{rotation_matrix}) $h_1$ is related to the original fields by
\begin{equation}
h_1 = c_{\gamma} (-s_{\alpha} H_d + c_\alpha H_u) + s_{\gamma} S,
\end{equation}
where $s_\theta = \sin{\theta}, c_\theta = \cos{\theta}$. Similar relations, also involving the angle $\sigma$, hold for $h_2$ and $h_3$.

These angles determine  the couplings of $h_1 = h_{\text{LHC}}$  to the fermions or to vector boson pairs, $VV = WW, ZZ$, normalized to the corresponding couplings of the SM Higgs boson. Defining  $\delta = \alpha - \beta +\pi/2$, they are given by (see also \cite{Cheung:2013bn,Choi:2012he})
\begin{equation}
\frac{g_{h_1tt}}{g^{\text{SM}}_{htt}}= c_\gamma(c_\delta +\frac{s_\delta}{\tan\beta}),~~\frac{g_{h_1bb}}{g^{\text{SM}}_{hbb}}= c_\gamma(c_\delta -s_\delta \tan\beta ),~~\frac{g_{h_1VV}}{g^{\text{SM}}_{hVV}}=  c_\gamma c_\delta.
\label{h1couplings}
\end{equation}
\begin{figure}
\begin{center}
\includegraphics[width=.48\textwidth]{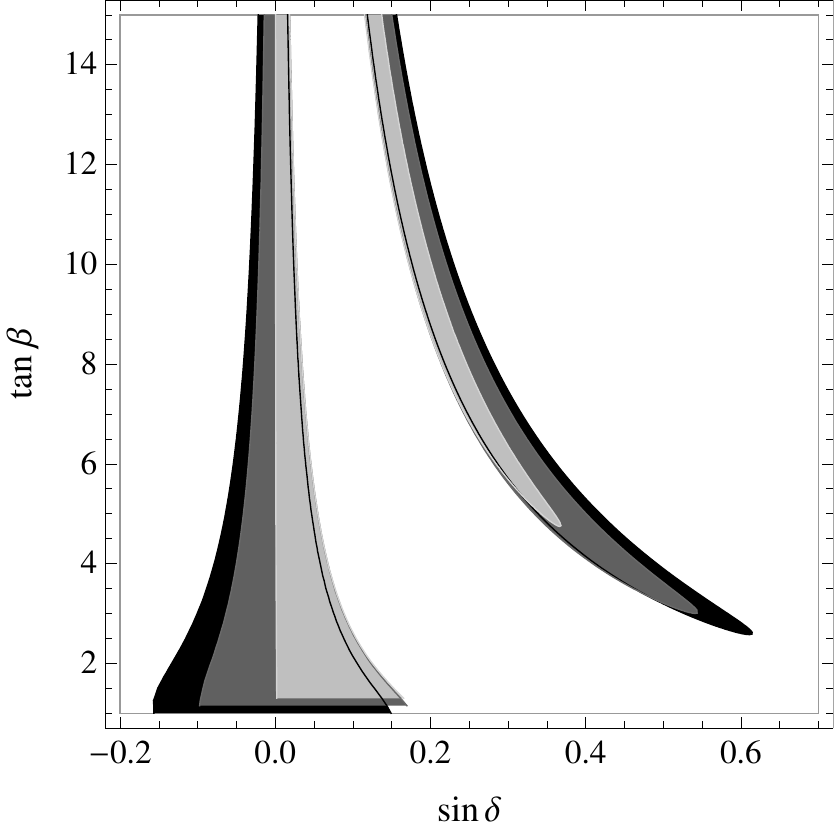}\hfill
\includegraphics[width=.48\textwidth]{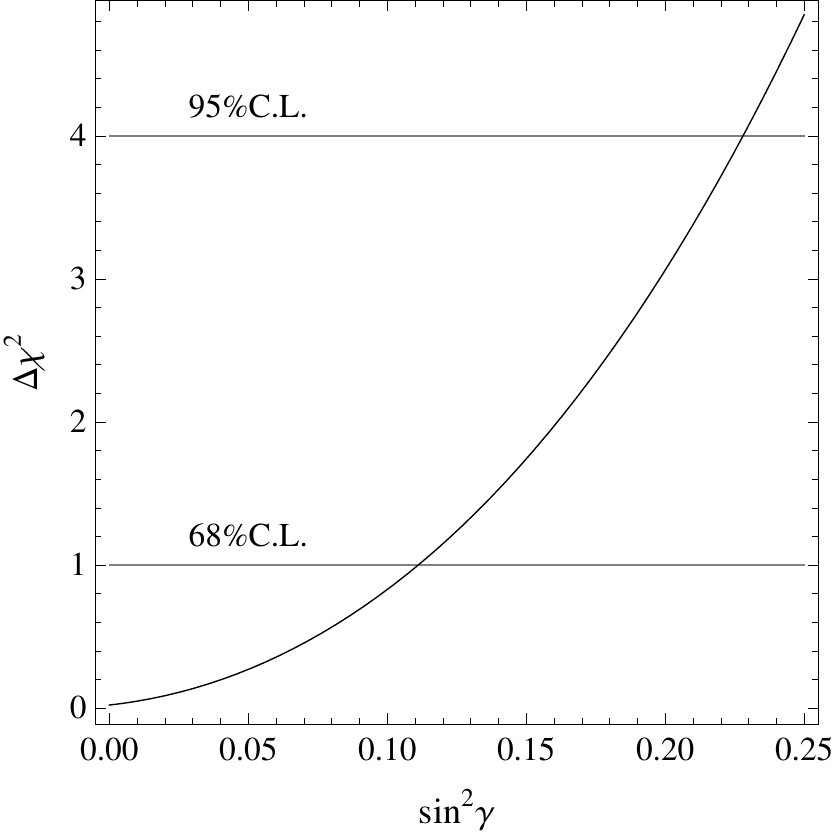}
\caption{\label{fig:FIT}\small Fit of the measured signal strengths of $h_1 = h_{\text{LHC}}$. Left: 3-parameter fit of $\tan \beta$, $s_{\delta}$ and $s_{\gamma}^{2}$. The allowed regions at 95\% C.L. are given for $s_{\gamma}^{2} = 0$ (black), $0.15$ (dark grey), and $0.3$ (light grey). The regions overlap in part, but their borders are also shown. Right: Fit of $s_{\gamma}^{2}$ in the case of $\delta = 0$.
}
\end{center}
\end{figure}
The fit of all ATLAS \cite{ATLAS:2013rma,ATLAS:2013qma,ATLAS:2013pma,ATLAS:2013oma,ATLAS:2013nma,ATLAS:2013mma,ATLAS:2013wla,ATLAS:moriond1,ATLAS:moriond2,ATLAS:moriond3}, CMS \cite{CMS:ril,CMS:xwa,CMS:bxa,CMS:utj,Chatrchyan:2013vaa,CMS:zwa,CMS:gya,CMS:moriond1,CMS:moriond2,CMS:moriond3,CMS:moriond4} and TeVatron \cite{tevatron:2013} data collected so far on the various signal strengths of $h_{\text{LHC}}$ gives the bounds on $\delta$ for different fixed values of $\gamma$ shown in the left of figure~\ref{fig:FIT} and the bound on $\gamma$ for $\delta= 0$ shown in the right of figure~\ref{fig:FIT}. To make this fit, we adapt the code provided by the authors of \cite{Giardino:2013bma}. As stated below, we do not include in this fit any supersymmetric loop effects. Note that in the region of $s_\delta$ close to zero, a larger $s_\gamma^2$ forces $\delta$ to take a larger central value.

The matrix equation (\ref{diag_matrix}) restricted to the $1 2$ sector gives three relations between the mixing angles $\alpha, \gamma, \sigma$ and the physical masses $m_{h_1,h_2,h_3}, m_{H^{\pm}}$ for any given value of $\lambda, \tan{\beta}$ and $\Delta_t$. In terms of the $2\times 2$ submatrix $M^2$ in the $1 2$ sector of ${\M}^2$, eq. (\ref{scalar_mass_matrix}),
 these relations can be made explicit as
\begin{align}
  s_\gamma^{2} &=  \frac{ \det M^{2} + m_{h_1}^{2} (m_{h_1}^{2} - \tr M^{2})}{(m_{h_1}^{2} - m_{h_2}^{2}) 
  (m_{h_1}^{2} - m_{h_3}^{2}) }, 
  \label{eq:sin:gamma:general}
  \\
  s_\sigma^{2} &= \frac{m_{h_2}^{2} - m_{h_1}^{2}}{m_{h_2}^{2} - m_{h_3}^{2}} \; \frac{ \det M^{2} + m_{h_3}^{2} (m_{h_3}^{2} - \tr M^{2}) }
  { \det M^{2} - m_{h_2}^{2} m_{h_3}^{2} + m_{h_1}^{2} (m_{h_2}^{2} + m_{h_3}^{2} - \tr M^{2}) },
  \label{eq:sin:sigma:general}
  \end{align}
  \begin{align}
     s_{2\delta}&=
  \Big[ 
    2 s_\sigma c_\sigma s_\gamma \left(m_{h_3}^2-m_{h_2}^2\right) \left(2 \tilde M^2_{11}-m_{h_1}^2c_\gamma^2 -m_{h_2}^2(s_\gamma^2+s_\sigma^2c_\gamma^2) - m_{h_3}^2(c_\sigma^2+s_\gamma^2 s_\sigma^2)\right) \notag
    \\ 
    & +2 \tilde M^2_{12} \left(m_{h_3}^2 \left(c_\sigma^2-s_\gamma^2 s_\sigma^2\right)+m_{h_2}^2 \left(s_\sigma^2-s_\gamma^2  c_\sigma^2\right)-m_{h_1}^2 c_\gamma^2 \right)
  \Big]
  \notag \\
  & \times \Big[ \left(m_{h_3}^2-m_{h_2}^2 s_\gamma^2- m_{h_1}^2 c_\gamma^2\right)^2
  +\left(m_{h_2}^2-m_{h_3}^2\right)^2 c_\gamma^4 s_\sigma^4 \notag\\
   &+2 \left(m_{h_2}^2-m_{h_3}^2\right) \left(m_{h_3}^2+m_{h_2}^2 s_\gamma^2-m_{h_1}^2 \left(1+s_\gamma^2\right)\right) c_\gamma^2 s_\sigma^2
  \Big]^{-1}, \label{eq:sin:2alpha:general} \end{align}
where $s_\theta = \sin\theta$, $c_\theta = \cos\theta$ and $\tilde M^2= R_{\beta-\pi/2}M^2 R_{\beta-\pi/2}^T$ is the mass matrix in the basis $(H,h)$.
These expressions for the mixing angles do not involve the unknown parameters $M_1, M_2, M_3$, which depend on the specific NMSSM. Their values in particular cases may limit the range of the physical parameters $m_{h_1,h_2,h_3}, m_{H^{\pm}}$ and $\delta, \gamma, \sigma$ but cannot affect equations (\ref{eq:sin:gamma:general}), (\ref{eq:sin:sigma:general}) and (\ref{eq:sin:2alpha:general}). To our knowledge, analytical expressions for the mixing angles in the general NMSSM have not been presented before.

To simplify the analysis we consider two limiting cases:
\begin{enumerate}
\item $H$ decoupled: In eq. (\ref{scalar_mass_matrix}) $m_A^2 \gg v M_1, v M_2$ or $m_{h_3} \gg m_{h_1,h_2}$ and $\sigma, \delta \rightarrow 0$,

\item Singlet decoupled: In eq. (\ref{scalar_mass_matrix}) $M_3^2 \gg v M_1, v M_2$ or $m_{h_2} \gg m_{h_1,h_3}$ and $\sigma, \gamma \rightarrow 0$,
\end{enumerate}
but we use (\ref{eq:sin:gamma:general})--(\ref{eq:sin:2alpha:general}) to control the size of the deviations from the limiting cases when the heavier mass is lowered.

When considering the couplings of the \CP-even scalars to SM particles, relevant to their production and decays, we shall not include any supersymmetric loop effect other than the one that gives rise to (\ref{delta-t}). This is motivated by the kind of spectrum outlined in the previous section, with all s-particles at their ``naturalness limit'', and provides in any event a useful well-defined reference case.  We also do not include any invisible decay of the  \CP-even scalars, e.g. into a pair of neutralinos. To correct for this is straightforward with all branching ratios and signal rates that will have to be multiplied by a factor $\Gamma_{\rm vis}/(\Gamma_{\rm vis} + \Gamma_{\rm inv})$. Finally we do not consider here the two neutral \CP-odd scalars, since in the general NMSSM both their masses and their composition in terms of the original fields depend upon extra parameters not related to the masses and the mixings of the \CP-even states nor to the mass of the charged Higgs.

\section{Singlet decoupled}\label{NMSSM/Sdecoupled}

\begin{figure}[t]
\begin{center}
\includegraphics[width=0.55\textwidth]{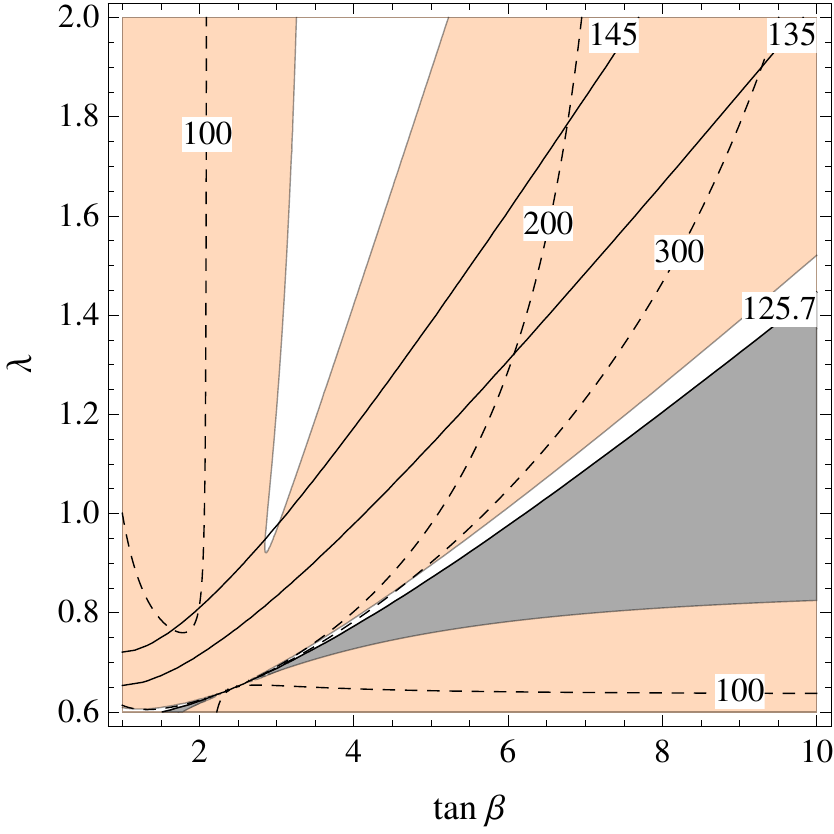}
\caption{\label{fig:mSdecoupled-1}\small Singlet decoupled. Isolines of $m_{hh}$ \eqref{mhh} (solid) and of $m_{H^\pm}$ (dashed). The grey region is unphysical due to $m_{H^{\pm}}^{\,2} < 0$,
the colored regions are excluded at 95$\%$ C.L.}
\end{center}
\end{figure}

Let us consider the limit $M_3^2 \gg v M_1, v M_2$ in (\ref{scalar_mass_matrix}), which corresponds to $m_{h_2} \gg m_{h_1}, m_{h_3}$ and $\sigma, \gamma \rightarrow 0$, where the singlet decouples from the other two neutral Higgs states.
In this case the three relations that have led to (\ref{eq:sin:gamma:general})--(\ref{eq:sin:2alpha:general}) become
\begin{align}\label{Sdecalpha}
\sin 2\alpha &= \sin 2\beta \; \frac{\lambda^2 v^2-m_Z^2-m_A^2|_{m_{h_1}}}{m_A^2|_{m_{h_1}} +m_Z^2 +\delta_t^2 -2m_{h_1}^2},\\
m_{h_3}^2&= m_A^2|_{m_{h_1}}+m_Z^2 +\delta_t^2 -m_{h_1}^2,\label{Sdecm3}
\end{align}
where
\begin{equation}\label{mA_mh}
m_A^2\big|_{m_{h_1}}=\frac{\frac{\lambda^2v^2}{2}\big(\frac{\lambda^2v^2}{2}-m_Z^2\big)\sin^2 2\beta-m_{h_1}^2(m_{h_1}^2-m_Z^2-\delta_t^2)-m_Z^2\delta_t^2 \cos^2\beta}{m_{hh}^2-m_{h_1}^2},
\end{equation}
and
\begin{equation}\label{mhh}
m_{hh}^2 = m_Z^2 c_{2\beta}^2 + \frac{\lambda^2 v^2}{2} s_{2\beta}^2 + \Delta_t^2
\end{equation}
is the second diagonal entry in the square mass matrix $\tilde M$ of the reduced basis $(H,h)$.
Identifying
$h_1$ with the resonance found at the LHC, since $m_{h_1}$ is known, this determines the masses of $h_3$ and $H^{\pm}$ for any given value of $\lambda$ and $\tan{\beta}$.

From our point of view, one of the main motivations for considering the NMSSM is the possibility to account for the mass of $h_{\rm LHC}$ with not too big values of the stop masses. For this reason we take $\Delta_t = 75$ GeV, which can be obtained e.g. for an average stop mass of about 700 GeV and maximal mixing $\theta_t\simeq 45^\circ$. In turn, as will be momentarily seen, the consistency of \eqref{Sdecalpha}--\eqref{mA_mh} requires not too small values of the coupling $\lambda$. It turns out in fact that for any value of $\Delta_t\lesssim 85$ GeV the dependence on $\Delta_t$ itself can be neglected, so that $m_{h_3}$, $m_{H^\pm}$ and $\delta$ are determined by $\tan\beta$ and $\lambda$ only. For the same reason it is legitimate to neglect the one-loop corrections to the 11 and 12 entries of the mass matrix \eqref{scalar_mass_matrix} as long as $\mu A_t/\langle m_{\tilde t}^2\rangle\lesssim 1$.

Fixing $\Delta_t$, the knowledge of $\delta$ in every point of the $(\lambda,\tan\beta)$ plane allows to draw the currently excluded regions from the measurements of the signal strengths of $h_{\rm LHC}$, as determined by a two-parameter fit of $\tan\beta$, $\sin\delta$, and shown in figure~\ref{fig:mSdecoupled-1}. This fit results in an allowed region which is virtually the same as the one with $\gamma = 0$ in the left side of figure~\ref{fig:FIT}.
Negative searches at LHC of $h_3 \rightarrow \bar{\tau} \tau$ may also exclude a further portion of the 
parameter space for $m_{h_3} > m_{h_1}$. When drawing  the currently excluded regions, we are not considering the possible decays of $h_{\text{LHC}}$ and/or of $h_3$ into invisible particles, such as dark matter, or into any final state undetected because of backgrounds, like e.g. a pair of light pseudo-scalars. The existence of such decays, however,  would not alter in any significant way the excluded regions from the measurements of the signal strengths of $h_{\text{LHC}}$, because the inclusion in the fit of the LHC  data of an invisible branching ratio of $h_{\text{LHC}}$, $\mathrm{BR}_{\rm inv}$, leaves essentially unchanged the allowed range for $\delta$ at different $\tan{\beta}$ values, provided $\mathrm{BR}_{\rm inv} \lesssim 0.2$.

\begin{figure}[t!]
\begin{center}
\includegraphics[width=0.48\textwidth]{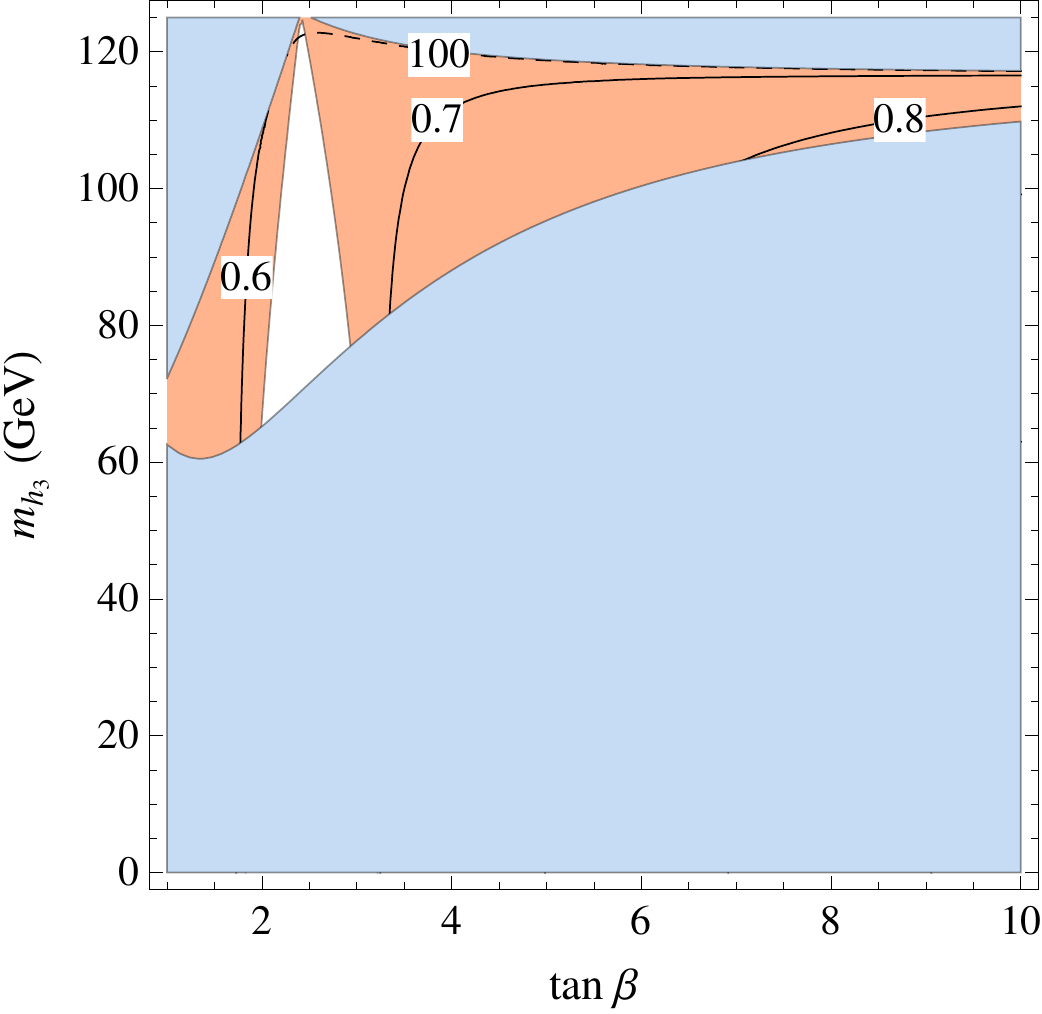}\hfill
\includegraphics[width=0.48\textwidth]{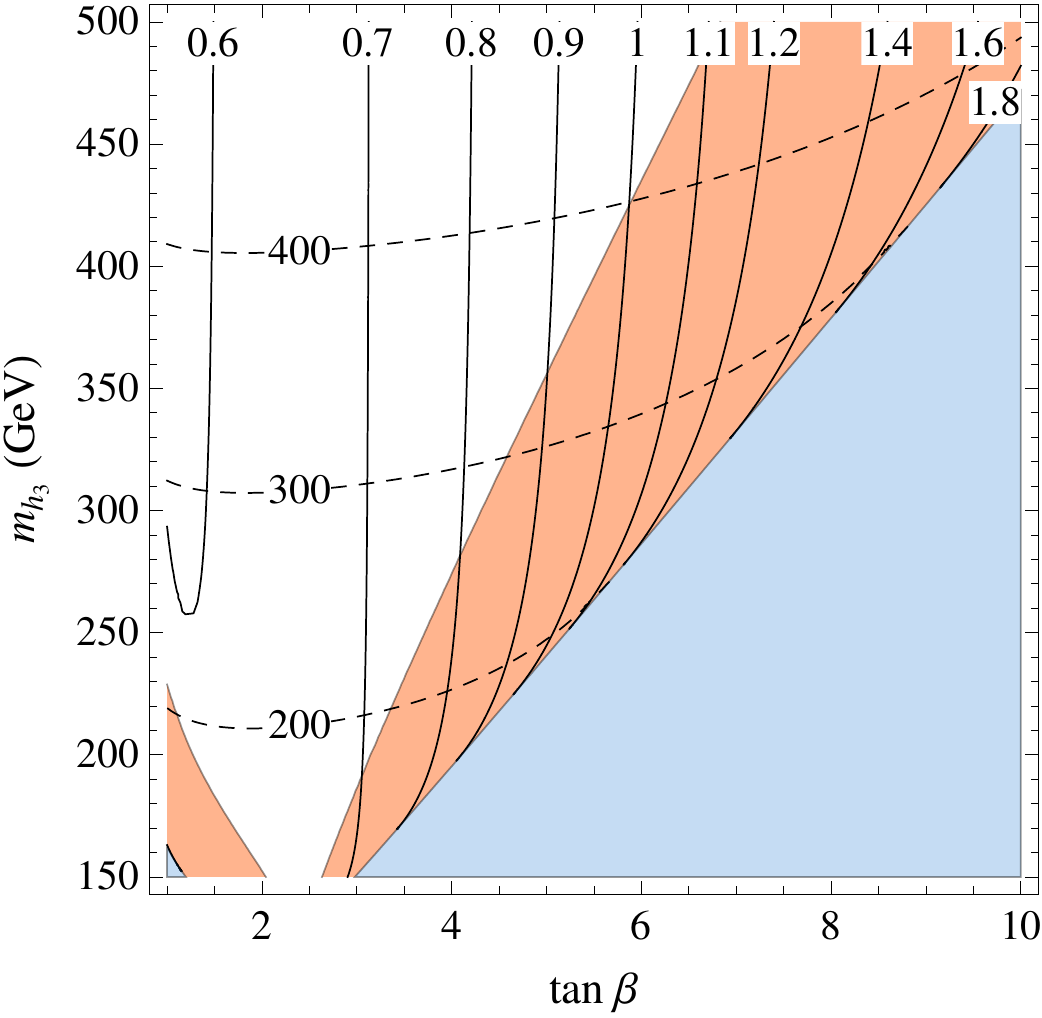}
\caption{\label{fig1} Singlet decoupled. Isolines of $\lambda$ (solid) and $m_{H^\pm}$ (dashed). Left: $h_{\rm LHC}>h_3$. Right: $h_{\rm LHC}<h_3$. The orange region is excluded at 95\% C.L. by the experimental data for the signal strengths of $h_1 = h_{\rm LHC}$. The blue region is unphysical.}
\end{center}
\end{figure}

In figure~\ref{fig1} we reflect the allowed regions in the plane $(\tan\beta, m_{h_3})$ and show the isolines of $\lambda$ and $m_{H^\pm}$, both for $m_{h_1} < m_{h_3}$ and  for $m_{h_3} < m_{h_1}$. In the first case, when inverting $\lambda$ as a function of $\tan\beta, m_{h_3}$, there are two possible solutions. If the left panel of figure~\ref{fig1} we show only the one which corresponds to the narrow allowed region with $m_{hh}$ close to 126 GeV. The other allowed region in figure \ref{fig:mSdecoupled-1}, corresponding to the other solution for $\lambda$ when translated to the ($\tan\beta, m_{h_3}$) plane, always implies a charged Higgs mass $m_{H^{\pm}}$ below 150 GeV, which is disfavored by indirect constraints \cite{Misiak:2006zs}. Note that this region, corresponding to the allowed region with large $\delta$ in figure~\ref{fig:FIT}, is mainly allowed because of the large error in the measurement of the $b\bar b$ coupling of $h_{\text{LHC}}$. Reducing this error down to about 30\% around $g_{h_1bb}/g_{hbb}^{\rm SM}\simeq 1$ would exclude the region.

Note, as anticipated, that $\lambda$ is bound to be above about 0.6. To go to lower values of $\lambda$ would require considering $\Delta_t\gtrsim 85$ GeV, i.e. heavier stops. On the other hand in this singlet-decoupled case lowering $\lambda$ and raising $\Delta_t$ makes the NMSSM more close to the MSSM.

The couplings of  $h_3$ are the same as in the MSSM and are given by \eqref{h3couplingsMSSM}.
They allow to compute the gluon-fusion production cross section of $h_3$ using the same procedure of the MSSM case that we described in section~\ref{SUSY/HiggsMSSM}. We performed the computation starting from the NNLL SM cross section using both NLO running masses \cite{Dittmaier:2011ti}}, and LO masses plus K-factors \cite{Anastasiou:2009kn}, performing a further check with the program {\sc Higlu} \cite{Spira:1995rr,Spira:1995mt}, finding an excellent agreement between all the procedures. The results for the cross section are shown in figure~\ref{fig:mSdecoupled-Xsec} for $m_{h_3} > m_{h_1}$.

\begin{figure}[t]
\begin{center}
\includegraphics[width=.48\textwidth]{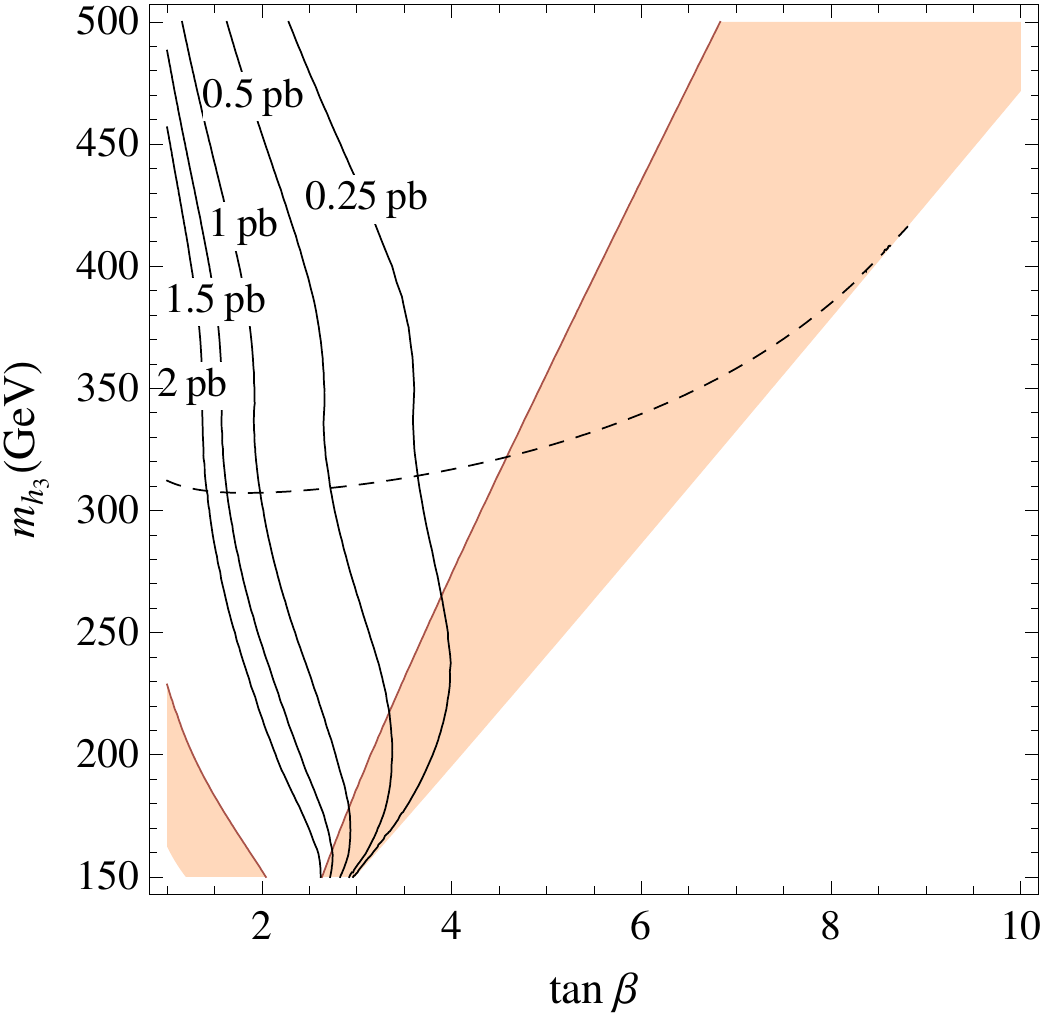}\hfill
\includegraphics[width=.48\textwidth]{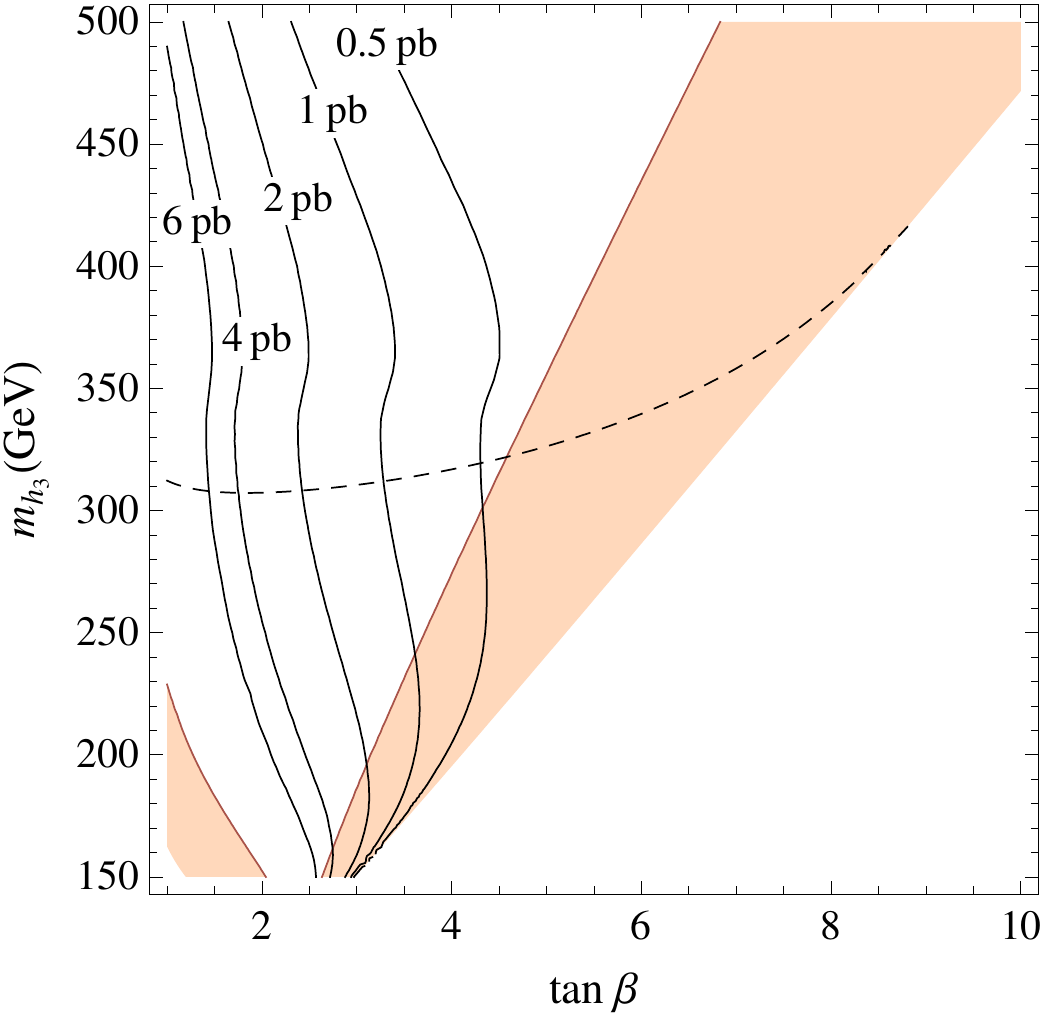}
\caption{\label{fig:mSdecoupled-Xsec}\small Singlet decoupled. Isolines of gluon fusion production cross section $\sigma(gg\to h_3)$. The colored regions are excluded at 95$\%$ C.L., and the dashed line shows $m_{H^\pm}=300$ GeV. Left: LHC8. Right: LHC14.}
\end{center}
\end{figure}

\begin{figure}[t!]
\begin{center}
\includegraphics[width=.48\textwidth]{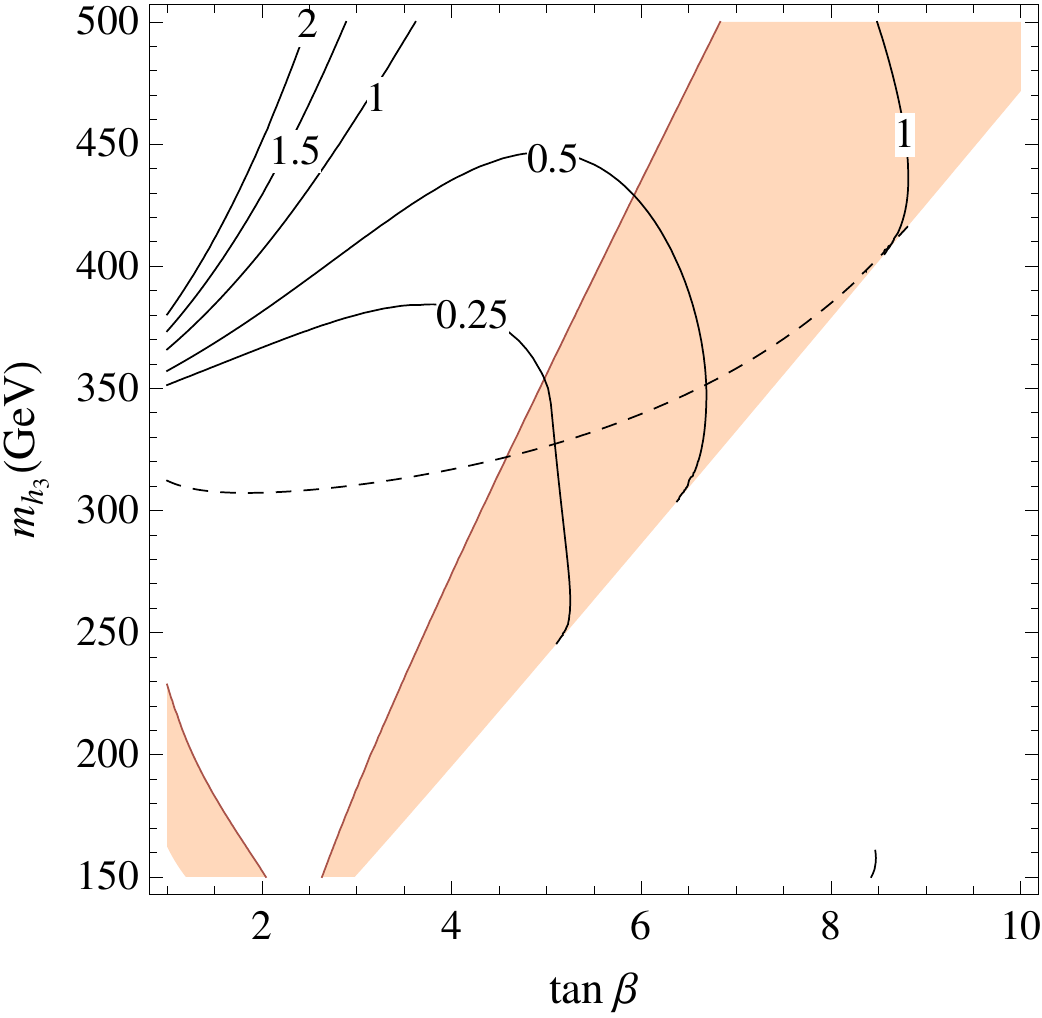}\hfill
\includegraphics[width=.48\textwidth]{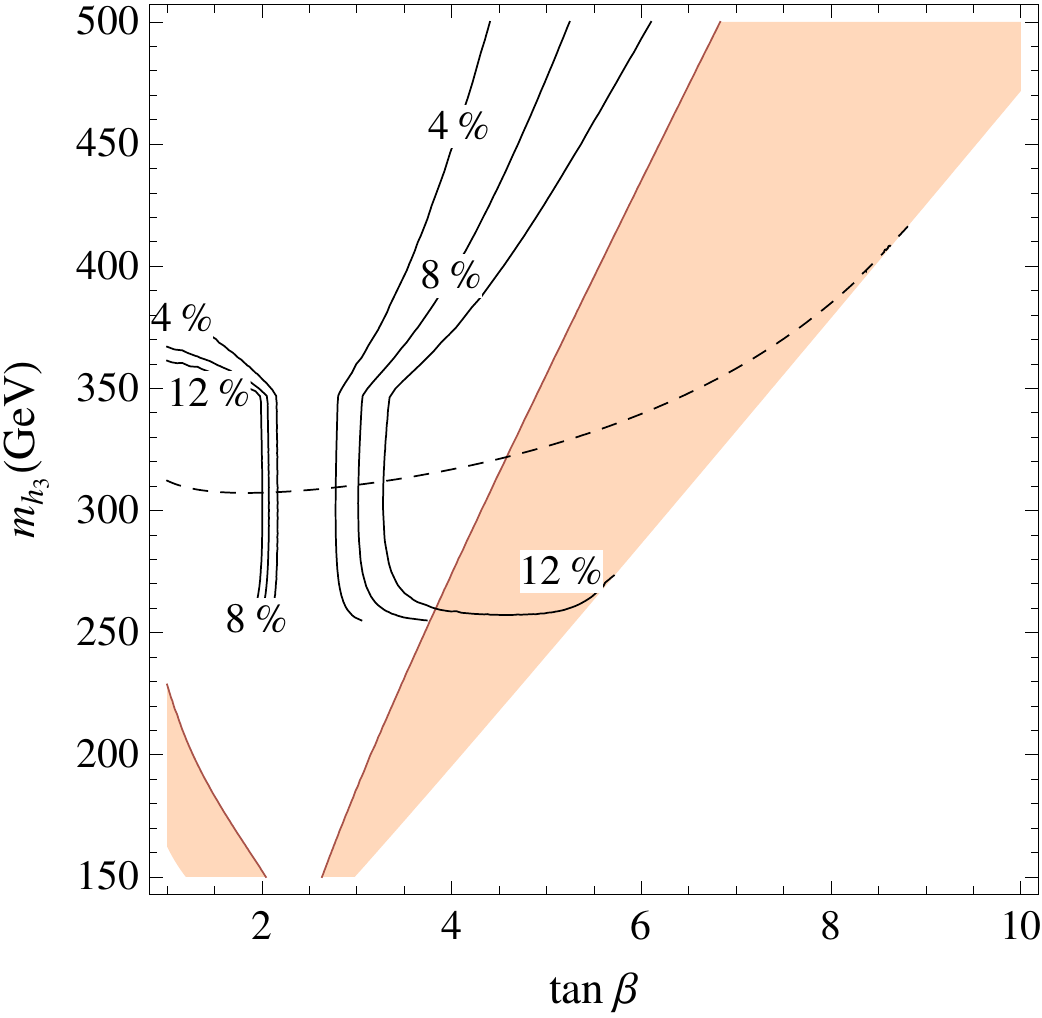}
\caption{\label{fig:mSdecoupled-BRs}\small Singlet decoupled. Left: isolines of the total width $\Gamma_{h_3}$(GeV). Right: isolines of BR$(h_3\!\to \!h h)$. The colored regions are excluded at 95$\%$ C.L., and the dashed line shows $m_{H^\pm}=300$ GeV.}
\vspace{2.5cm}
\includegraphics[width=.48\textwidth]{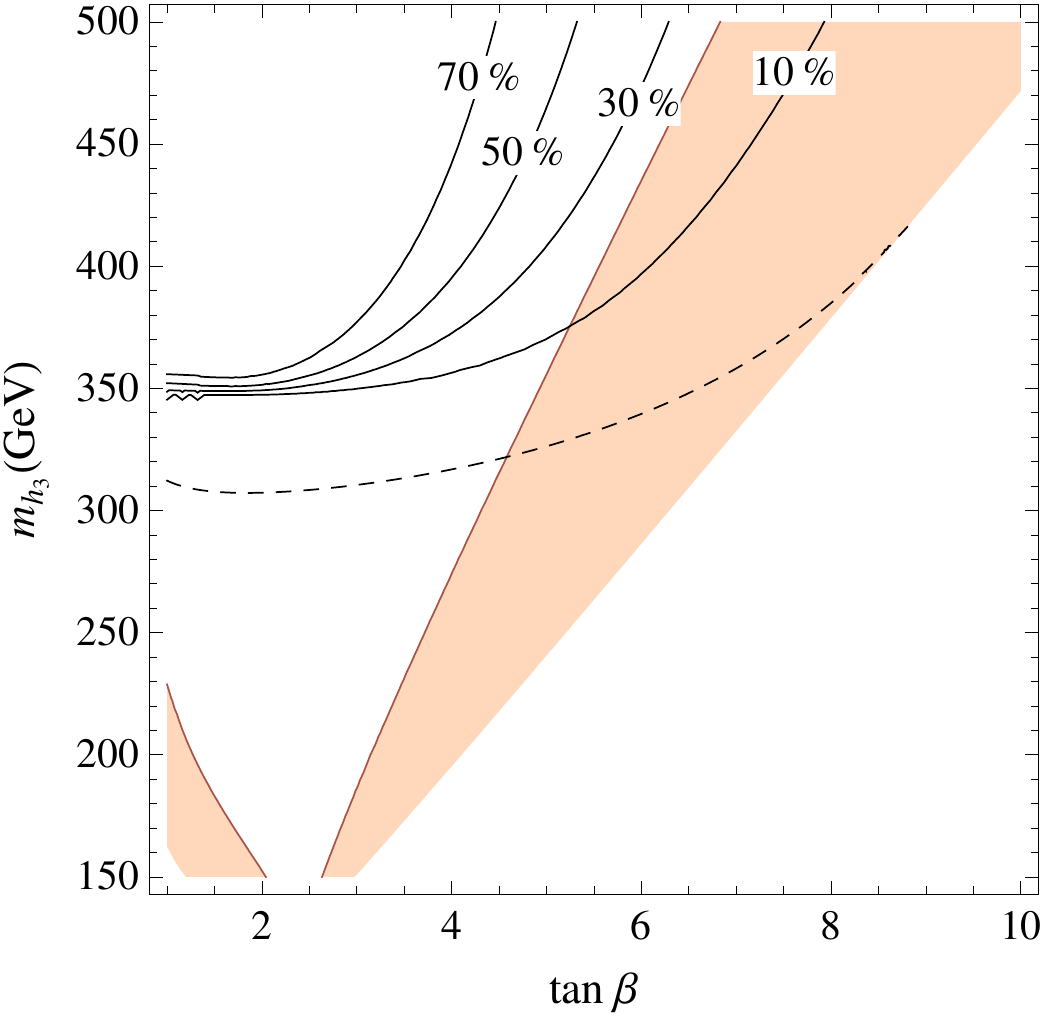}\hfill
\includegraphics[width=.48\textwidth]{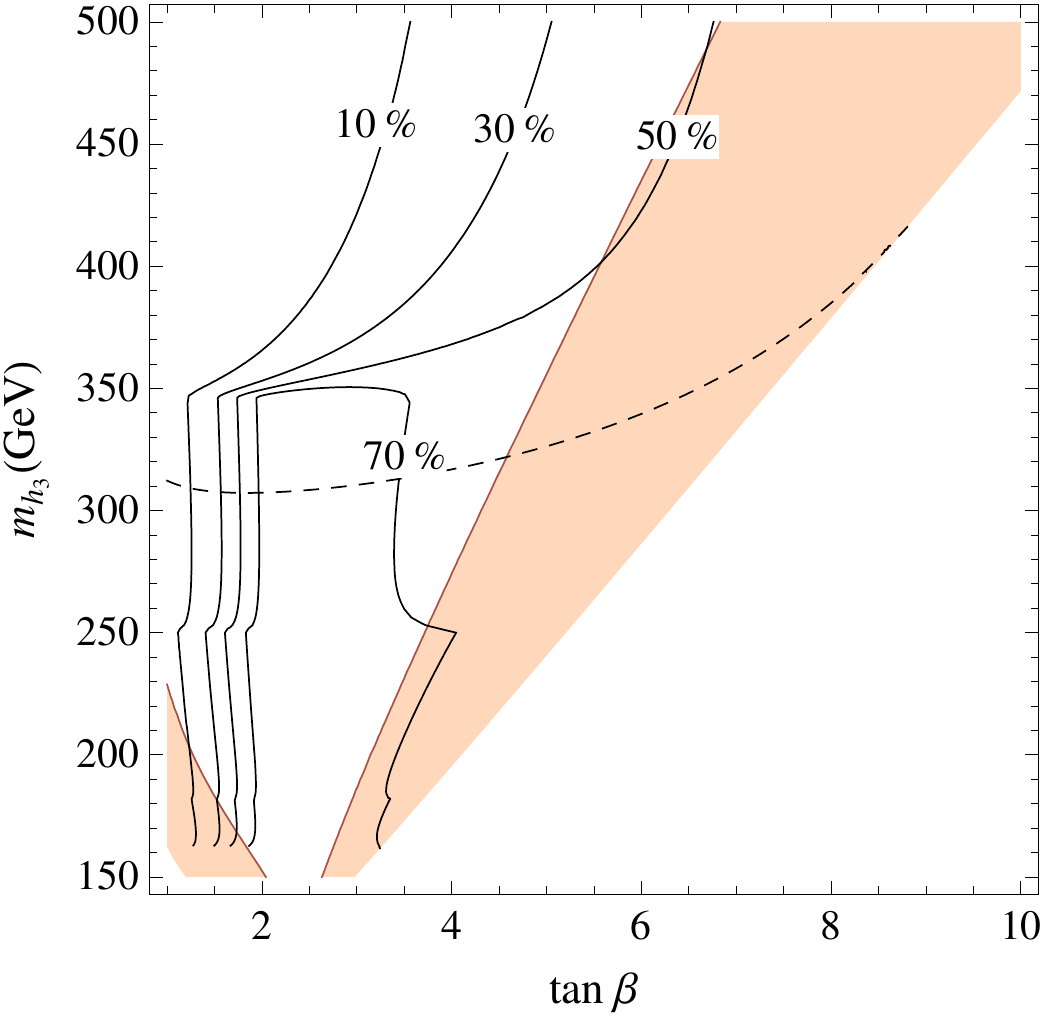}
\caption{\label{fig:mSdecoupled-BRf}\small Singlet decoupled. Left: Isolines of BR$(h_3\to t\bar t)$. Right: Isolines of BR$(h_3\to b \bar{b})$. The colored regions are excluded at 95$\%$ C.L., and the dashed line shows $m_{H^\pm}=300$ GeV.}
\end{center}
\end{figure}

The coupling of $h_3$ to the lighter state $\displaystyle\frac{g_{h_3 h_1^2}}{2} h_3 h_1^2$ and the triple Higgs coupling $\displaystyle\frac{g_{h_1^3}}{6}h_1^3$ are given by
\begin{align}
g_{h_3h_1^2}&=\frac{1}{2 v} \left[\Big(m_Z^2 + \frac{\lambda^2 v^2}{2}\Big) \sin \delta + 3 \Big(m_Z^2 - \frac{\lambda^2 v^2}{2}\Big) \sin(4 \beta + 3 \delta)\right] \nonumber \\
 &- \frac{3 \Delta_t^2}{v} \frac{\cos(\beta + \delta) \sin^2(\beta+\delta)}{\sin^3\beta},\\
\frac{g_{h_1^3}}{g^{\text{SM}}_{h_1^3}} &= \frac{\big(m_Z^2 + \frac{\lambda^2 v^2}{2}\big) \cos \delta + \big(m_Z^2 - \frac{\lambda^2v^2}{2}\big) \cos(4 \beta + 3 \delta)}{2 m_{h_1}^2} + \frac{\Delta_t^2}{m_{h_1}^2} \frac{\sin^3(\beta+\delta)}{\sin^3\beta}.
\end{align}
Figures~\ref{fig:mSdecoupled-BRs} and \ref{fig:mSdecoupled-BRf} show the most relevant widths of $h_3$.

\begin{figure}[t]
\begin{center}
\includegraphics[width=0.48\textwidth]{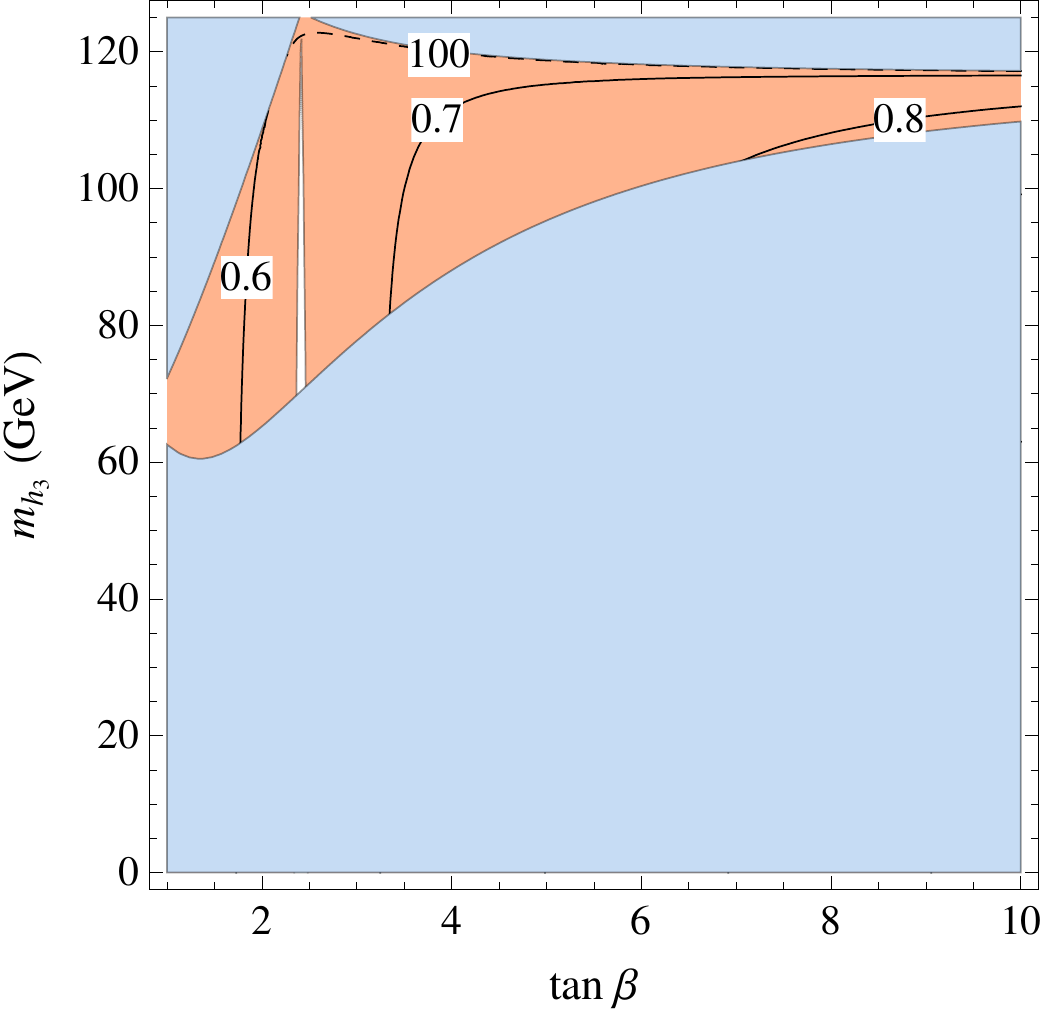}\hfill
\includegraphics[width=0.48\textwidth]{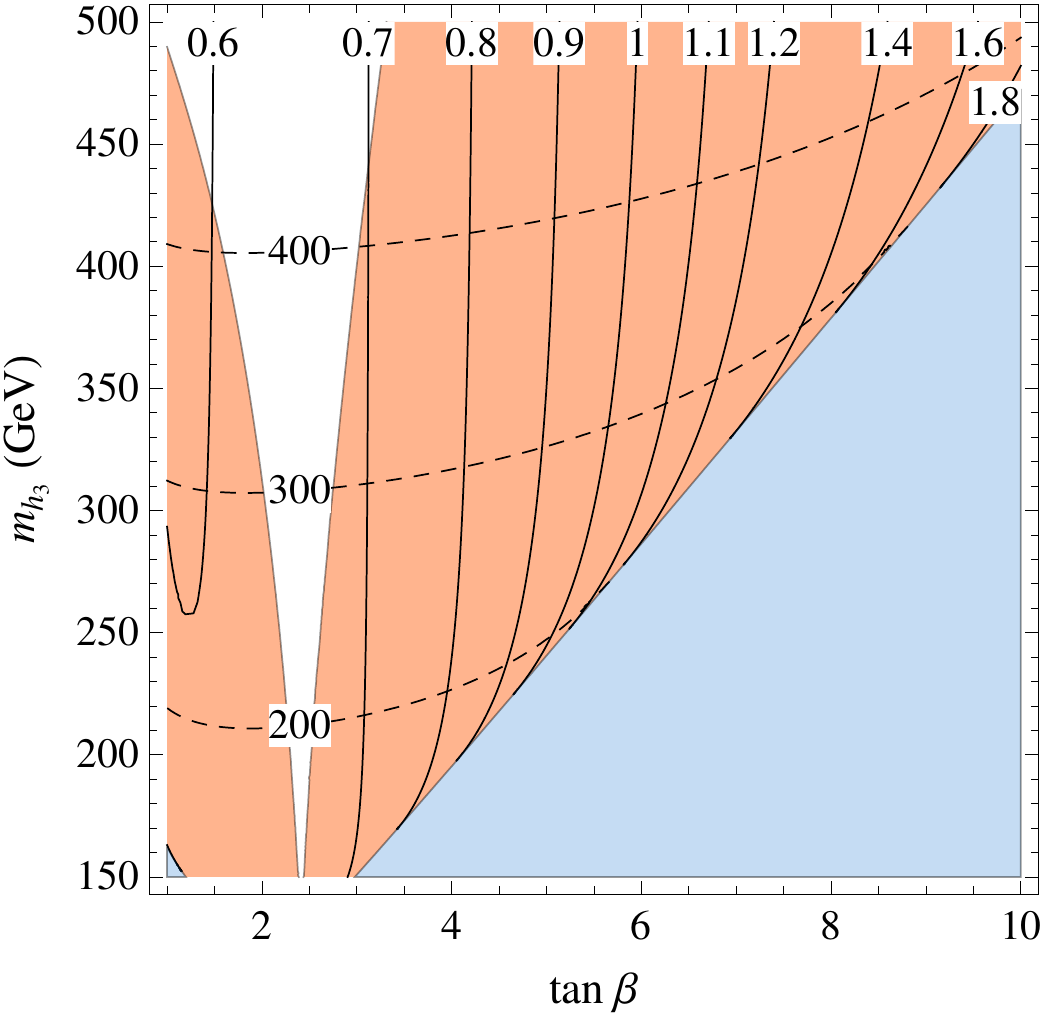}
\caption{\label{fig2} Singlet decoupled. Isolines of $\lambda$ (solid) and $m_{H^\pm}$ (dashed). Left: $h_{\rm LHC}>h_3$. Right: $h_{\rm LHC}<h_3$. The orange region would be excluded at 95\% C.L. by the experimental data for the signal strengths of $h_1 = h_{\rm LHC}$ with SM central values and projected errors at the LHC14 as discussed in the text. The blue region is unphysical.}
\end{center}
\end{figure}

The significant constraint set on figure~\ref{fig1} by the  current measurements of the signal strengths of $h_{\text{LHC}}$ suggests that an improvement of such measurements, as foreseen in the coming stage of LHC, could lead to an effective exploration of most of the relevant parameter space. 
To quantify this we have considered the impact on the fit of the measurements of the signal strengths of $h_{\text{LHC}}$ with the projected errors at LHC14 with $300~\mathrm{fb}^{-1}$ by ATLAS\cite{ATLAS-collaboration:2012iza} and CMS\cite{CMS:note}, shown in table \ref{tab1}.
The result  is shown in figure~\ref{fig2}, again both for $m_{h_3} < m_{h_1}$ and for $m_{h_1} < m_{h_3}$, assuming SM central values for the signal strengths.

\begin{table}[b]
\begin{center}
\begin{tabular}{ccc}
& ATLAS & CMS \\
\hline
$h \to \gamma \gamma$ & 0.16 & 0.15 \\
$h \to Z Z$ & 0.15 & 0.11 \\
$h \to W W$ & 0.30 & 0.14 \\
$V h \to V b \bar{b}$ & -- & 0.17 \\
$h \to \tau \tau$ & 0.24 & 0.11 \\
$h \to \mu \mu$ & 0.52 & -- \\
\end{tabular}
\caption{\label{tab1}Projected uncertainties of the measurements of the signal strengths of $h_{\text{LHC}}$, normalized to the SM, at the 14 TeV LHC with $300~\mathrm{fb}^{-1}$.}
\end{center}
\end{table}
Needless to say, the direct search of the extra \CP-even states will be essential either in presence of a possible indirect evidence from the signal strengths or to fully cover the parameter space for $m_{h_3} > m_{h_1}$. To this end, under the stated assumptions, all production cross sections and branching ratios for the $h_3$ state are determined in every point of the $(\tan{\beta}, m_{h_3})$ plane.

One may ask if the electro-weak precision tests (EWPT) set some further constraint on the parameter space explored so far. In the singlet-decoupled case the mixing between the two doublets can in principle lead to important effects, which are however limited by the constraint on the mixing angle $\alpha$ or the closeness to zero of $\delta = \alpha - \beta +\pi/2$ already demanded by the measurements of the signal strengths of $h_{\text{LHC}}$.\footnote{Notice that in the fully mixed situation there may be relevant regions of the parameter space still allowed by the fit with a largish $\delta$ (see e.g. figure~\ref{fig:mSdecoupled-1}).  This could further constrain the small allowed regions, but the precise contributions to the EWPT depend on the value of the masses of the \CP-odd scalars, which in the generic NMSSM are controlled by further parameters.}
Since in the $\delta = 0$ limit every extra effect on $\hat{S}$ and $\hat{T}$  vanishes, this explains why the EWPT do not impose further constraints on the parameter space that we have considered.

\section{Doublet decoupled}\label{NMSSM/Hdecoupled}

\begin{figure}
\begin{center}
\includegraphics[width=0.48\textwidth]{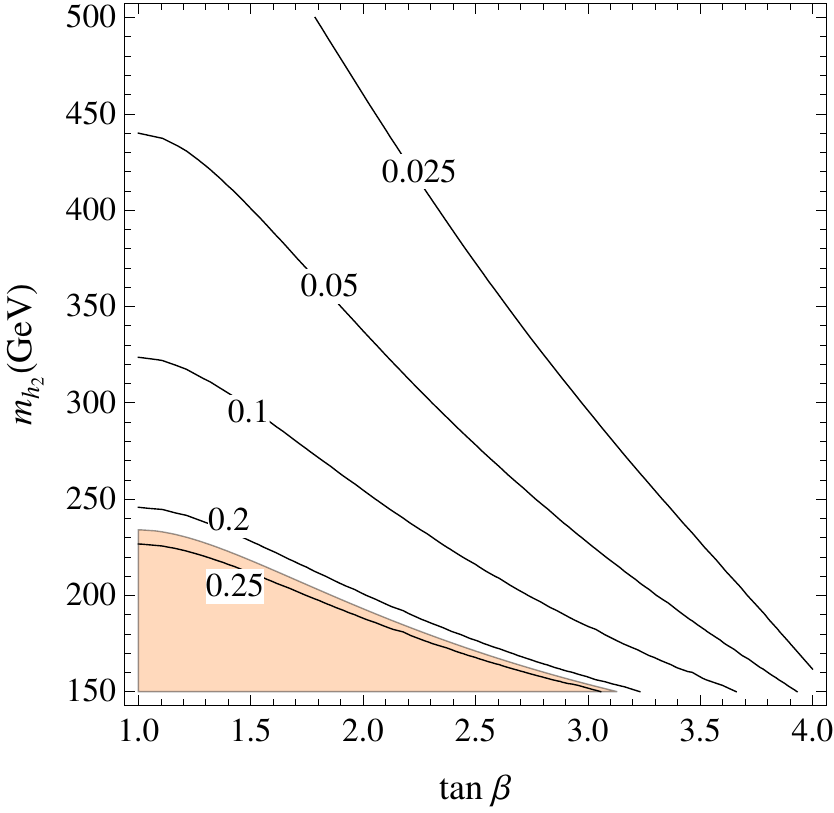}\hfill
\includegraphics[width=0.48\textwidth]{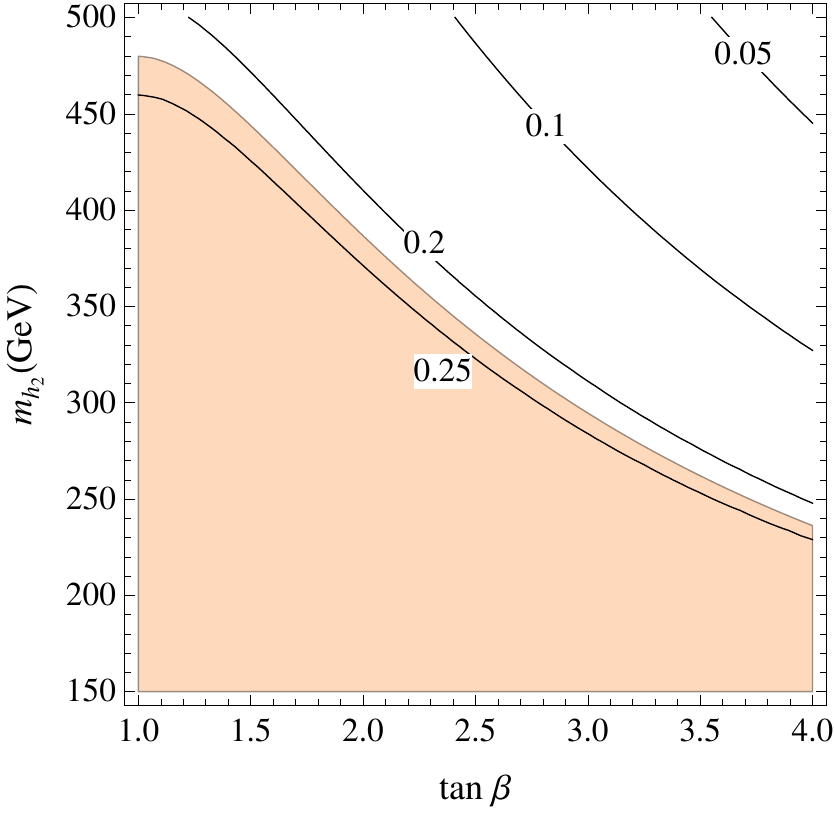}
\caption{\label{fig:mAdecoupled-1}\small $H$ decoupled. Isolines of $\sin^2\gamma$. Left: $\lambda=0.8$. Right: $\lambda=1.4$. The colored region is excluded at 95\% C.L. by the experimental data for the signal strengths of $h_1 = h_{\rm LHC}$.}
\end{center}\end{figure}

\begin{figure}
\begin{center}
\subfigure[\label{fig:mAdecoupled-Xsec8a}8 TeV, $\lambda=0.8$]{\includegraphics[width=.48\textwidth]{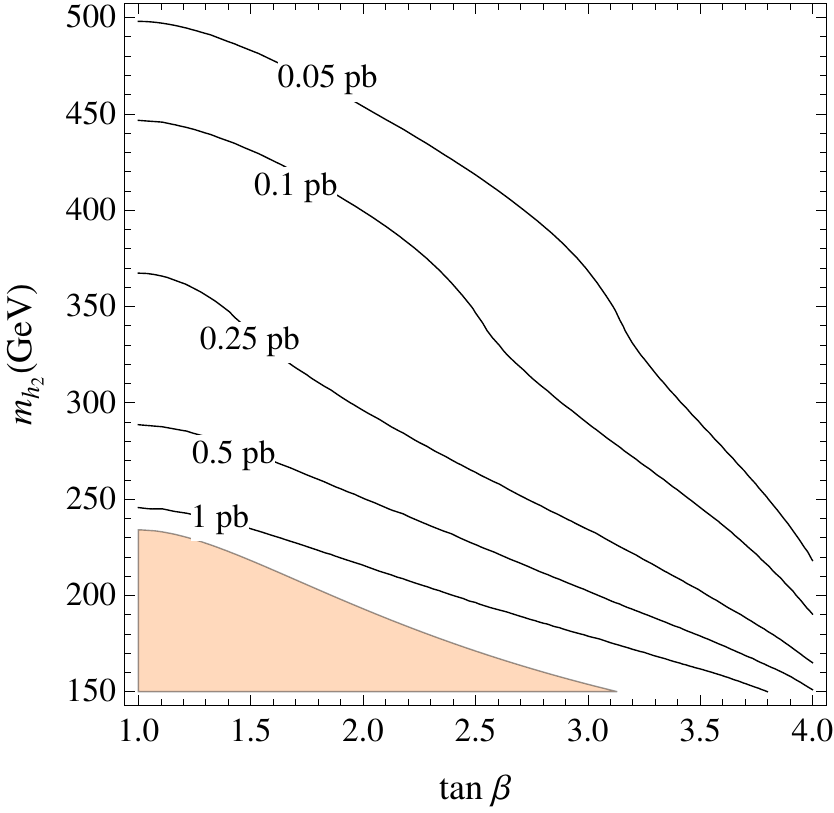}}\hfill
\subfigure[\label{fig:mAdecoupled-Xsec8b}8 TeV, $\lambda=1.4$]{\includegraphics[width=.48\textwidth]{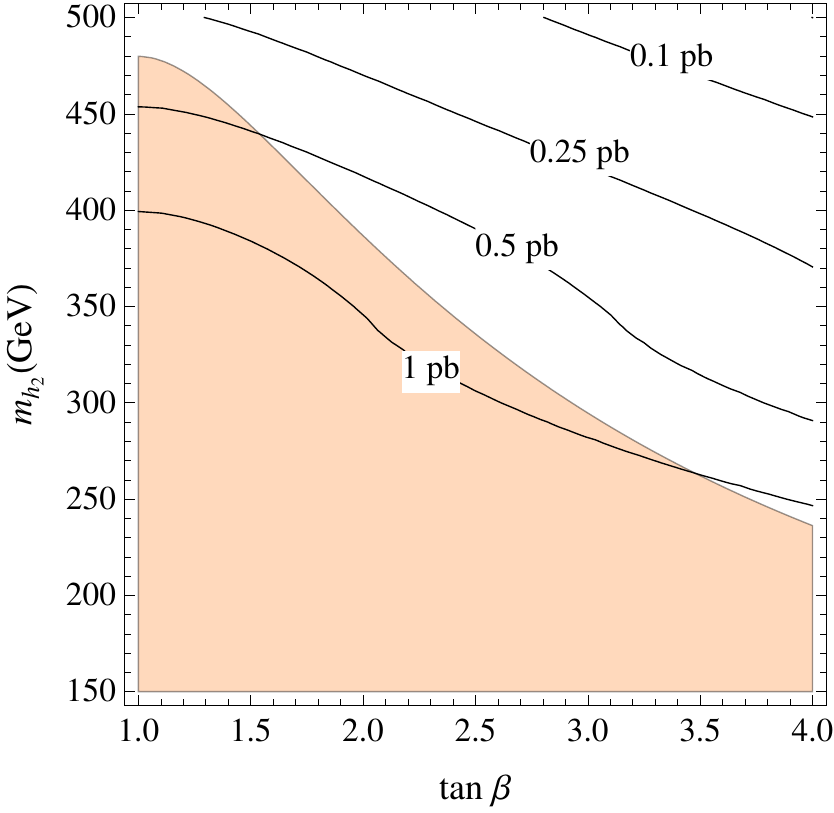}}\vspace{0.8cm}\\
\subfigure[\label{fig:mAdecoupled-Xsec14c}14 TeV, $\lambda=0.8$]{\includegraphics[width=.48\textwidth]{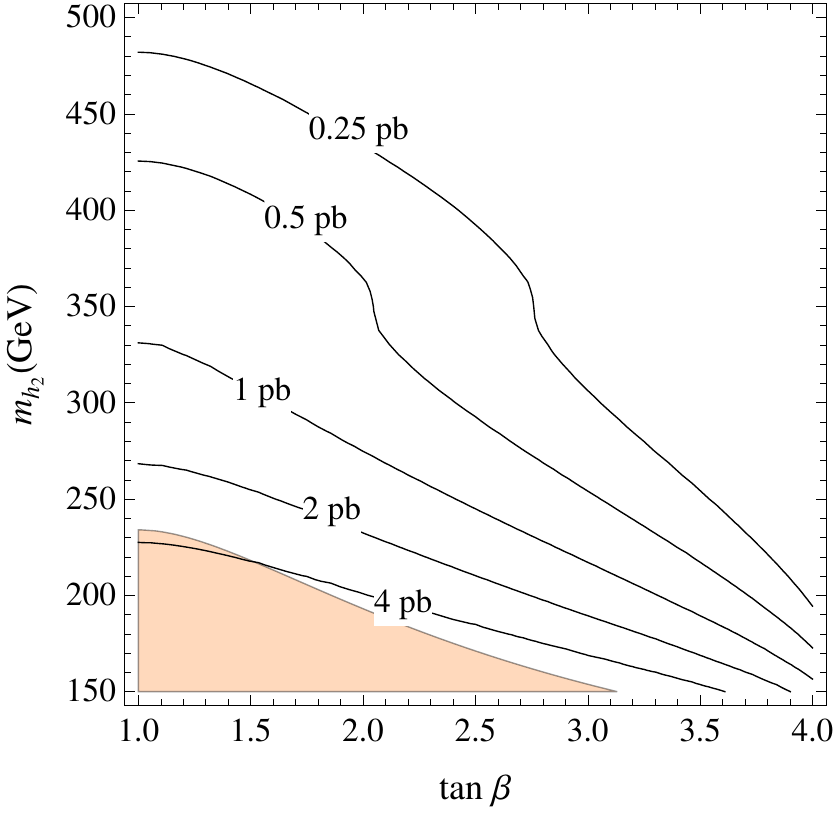}}\hfill
\subfigure[\label{fig:mAdecoupled-Xsec14d}14 TeV, $\lambda=1.4$]{\includegraphics[width=.48\textwidth]{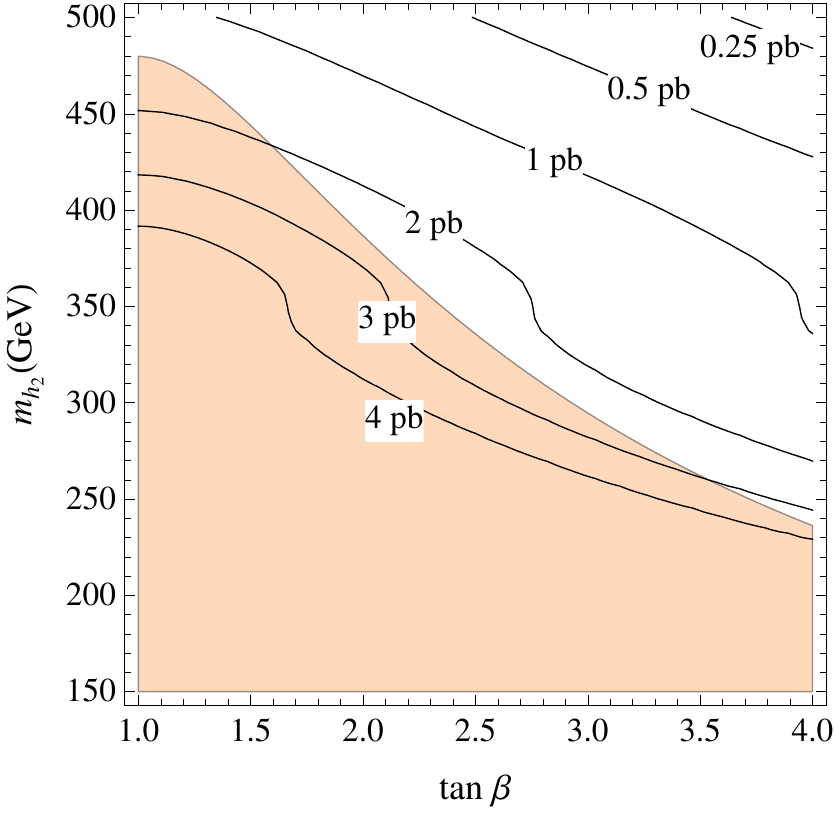}}
\caption{\label{fig:mAdecoupled-Xsec}\small $H$ decoupled. Isolines of gluon fusion cross section $\sigma(gg\to h_2)$ at LHC8 and LHC14, for the values $\lambda=0.8$ and $\lambda=1.4$. The colored region is excluded at 95$\%$ C.L.}
\end{center}
\end{figure}

As we are going to see, the situation changes significantly when considering the case where the singlet $S$ mixes with the doublet with SM couplings.
To study this limiting case, it is best to go in the basis $(H, h, s)$ with $H=s_\beta H_d - c_\beta H_u$ and $h=c_\beta H_d + s_\beta H_u$, and let $H\approx h_3$ decouple, so that $\sigma, \delta \rightarrow 0$. For the remaining nonvanishing angle $\gamma$ one has 
\begin{equation}
\sin^2{\gamma}= \frac{m_{hh}^2-m_{h_1}^2}{m_{h_2}^2-m_{h_1}^2}.
\label{sin2gamma}
\end{equation}
Note that in this case there is only a single relation between the mixing angle and the mass of the extra \CP-even state $m_{h_2}$ involving only $\lambda$, $\tan\beta$ and $\Delta_t$. For this reason one has to fix two parameters in order to produce plots similar to the ones of the previous section.

Due to the singlet nature of $S$, it is straightforward to see that the couplings of $h_1 = h_{\text{LHC}}$ and $h_2$ to fermions or to vector boson pairs, $VV = WW, ZZ$, normalized to the same couplings of the SM Higgs boson, are given by
\begin{equation}
\frac{g_{h_1ff}}{g^{\text{SM}}_{hff}} = \frac{g_{h_1VV}}{g^{\text{SM}}_{hVV}}= c_\gamma, ~~~~~
\frac{g_{h_2ff}}{g^{\text{SM}}_{hff}}= \frac{g_{h_2VV}}{g^{\text{SM}}_{hVV}}= - s_\gamma.
\end{equation}
As a consequence for $m_{h_2} > m_{h_1}/2$ none of the branching ratios of $h_1 = h_{\rm LHC}$ and $h_2$ gets modified with respect to the ones of a SM Higgs boson with the corresponding mass, whereas their production cross sections, or the various signal strengths, are reduced by a common factor $c_\gamma^2$ or $s_\gamma^2$ respectively for $h_1$ and $h_2$.
The fit of all experimental data collected so far gives the bound $s_\gamma^2 < 0.22$ at 95\% C.L., as shown in the right side of figure~\ref{fig:FIT}.

\begin{figure}[t!]
\begin{center}
\includegraphics[width=.48\textwidth]{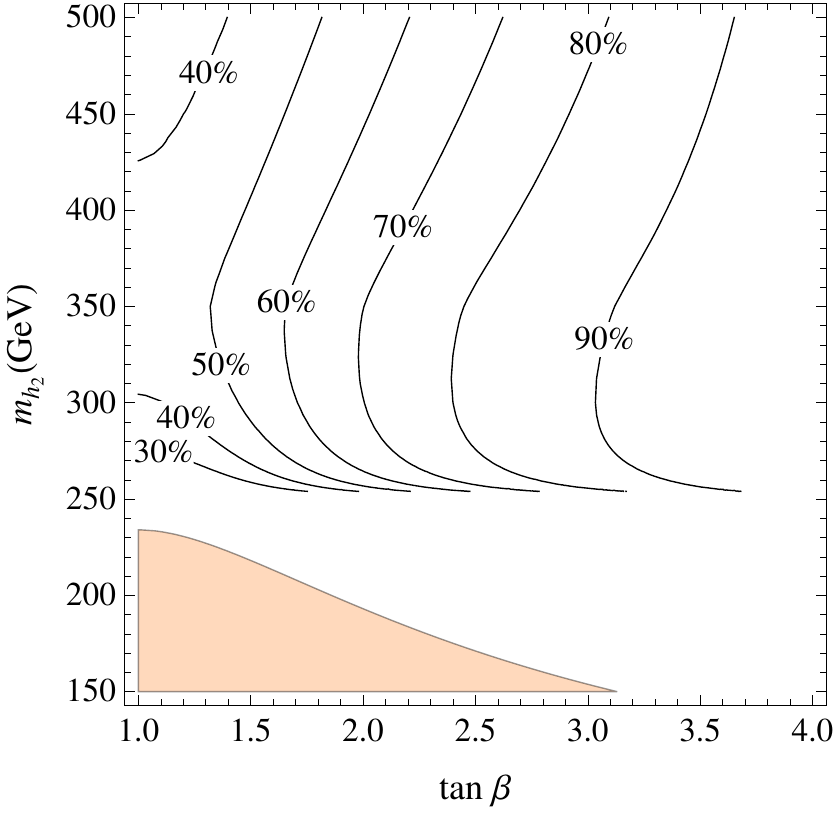}\hfill
\includegraphics[width=.48\textwidth]{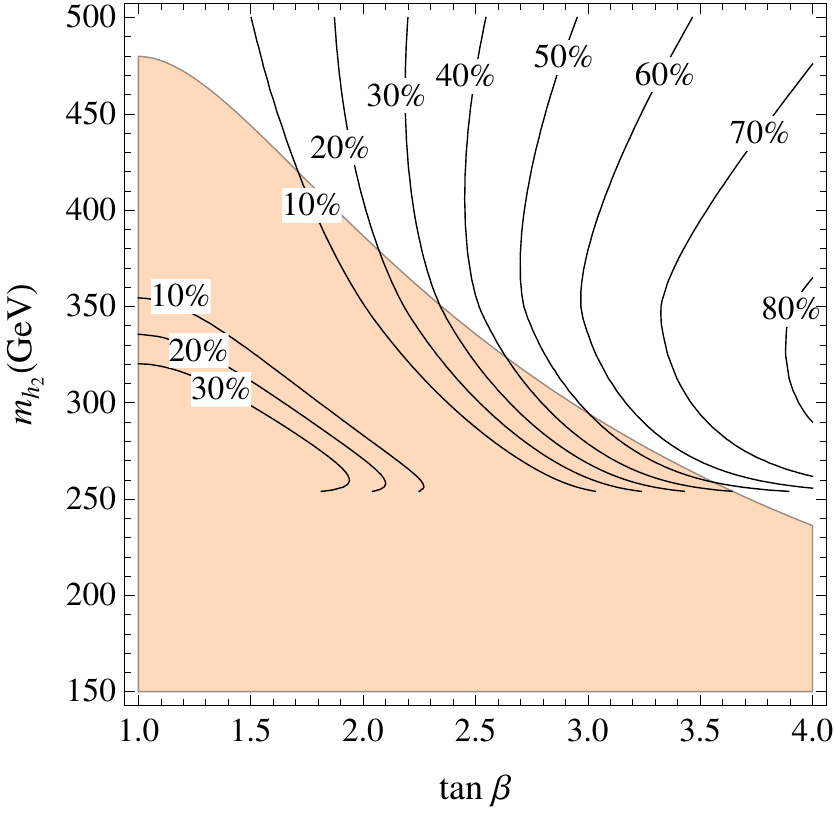}
\caption{\label{fig:mAdecoupled-hh}\small $H$ decoupled. Isolines of BR$(h_2\to h h )$. Left: $\lambda=0.8$ and $v_S=2v$. Right: $\lambda=1.4$ and $v_S=v$. The colored region is excluded at 95$\%$ C.L.}
\end{center}
\end{figure}

\begin{figure}
\begin{center}
\includegraphics[width=.48\textwidth]{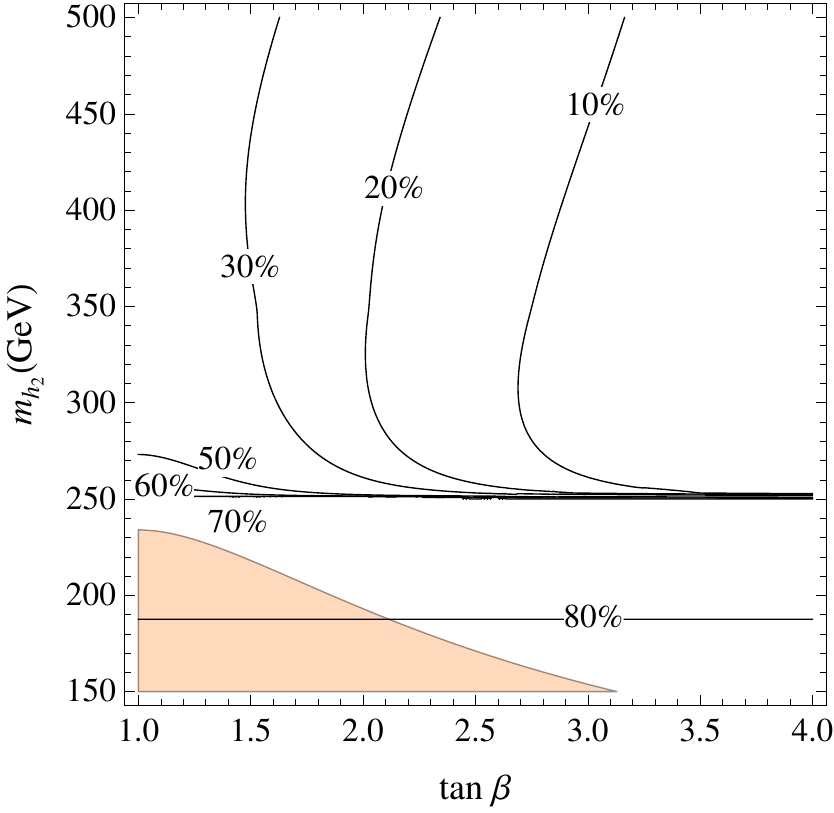}\hfill
\includegraphics[width=.48\textwidth]{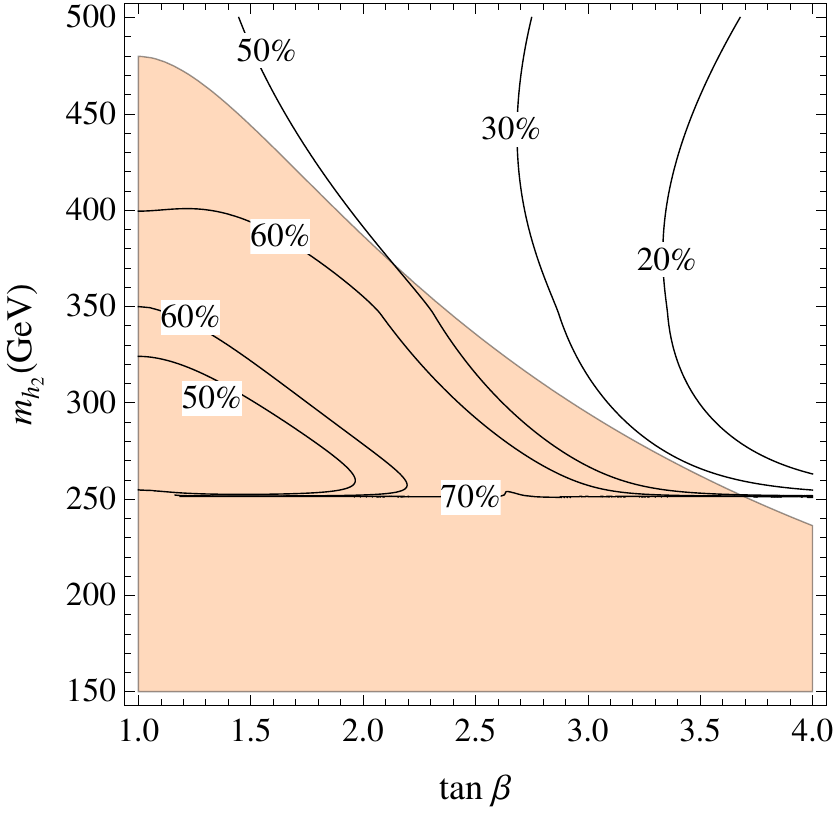}
\caption{\label{fig:mAdecoupled-WW}\small $H$ decoupled. Isolines of BR$(h_2\to W^+W^- )$. Left: $\lambda=0.8$ and $v_S=2v$. Right: $\lambda=1.4$ and $v_S=v$. The colored region is excluded at 95$\%$ C.L.}
\end{center}
\end{figure}

Let us first consider the case in which $m_{h_1} < m_{h_2}$.
Upon use of (\ref{sin2gamma}) the impact of the fit to the $h_{\rm LHC}$ couplings on the parameter space is shown in figure~\ref{fig:mAdecoupled-1} for $\lambda = 0.8$ and 1.4, together with the isolines of different values of $s_\gamma^2$ that might be probed by future improvements in the measurements of the $h_1$ signal strengths. Larger values of $\lambda$ already exclude a significant portion of the parameter space, at least for moderate $\tan{\beta}$, as preferred by naturalness. Here we are taking a fixed value of $\Delta_t = 75$~GeV as in the singlet-decoupled case. As already stated, as long as one stays at $\Delta_t \lesssim 85$ GeV, in a range of moderate fine tuning, and $\lambda \gtrsim 0.8$, our results do not depend significantly on $\Delta_t$.

In the same $(\tan{\beta}, m_{h_2})$ plane of figure~\ref{fig:mAdecoupled-1} and for the same values of $\lambda$, figure~\ref{fig:mAdecoupled-Xsec} shows the gluon-fusion production cross sections of $h_2$ at LHC for 8 or 14 TeV center-of-mass energies, where we rescaled by $s_\gamma^2$ the (next-to-next-to-leading logarithmic) NNLL ones provided in \cite{Dittmaier:2011ti}. All other $h_2$ production cross sections, relative to the gluon-fusion one, scale as in the SM with $m_{h_{\text{SM}}} = m_{h_2}$.

\begin{figure}[t!]
\begin{center}
\includegraphics[width=.48\textwidth]{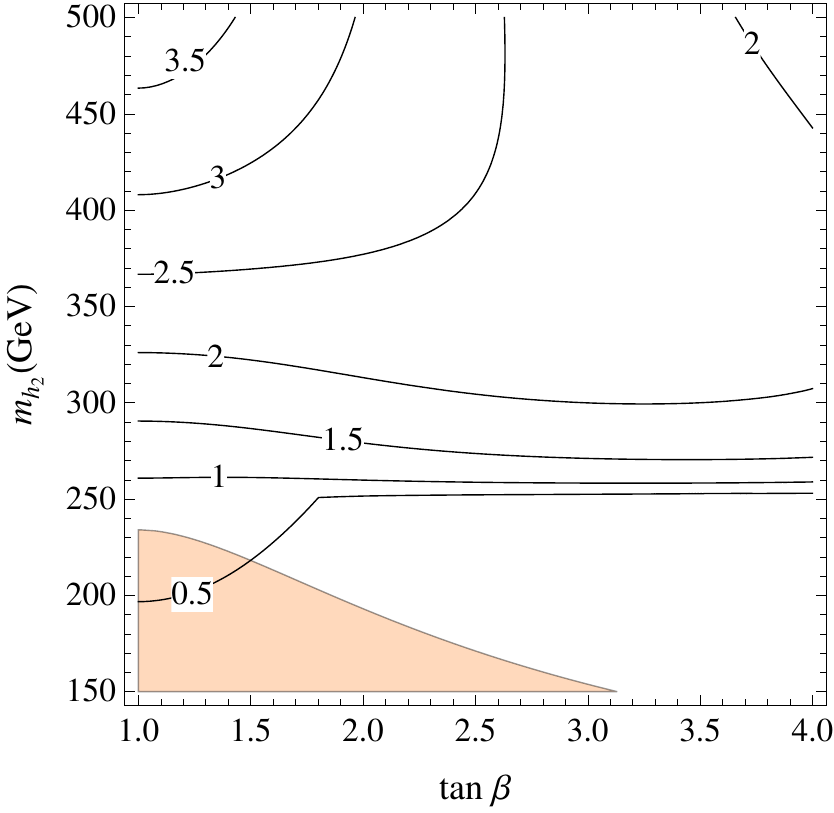}\hfill
\includegraphics[width=.48\textwidth]{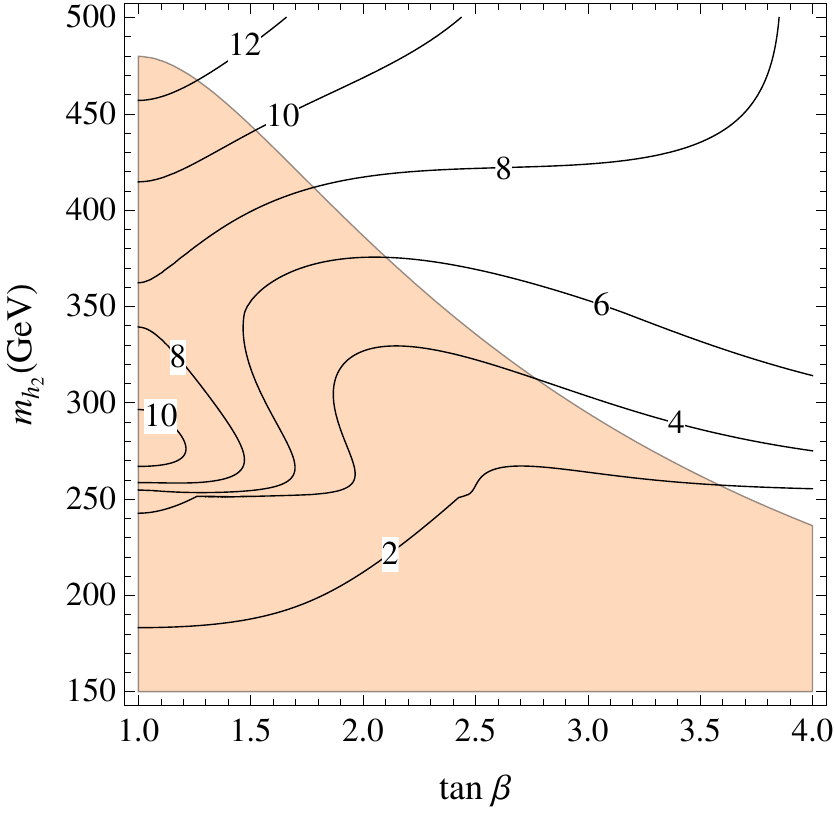}
\caption{\label{fig:mAdecoupled-width}\small $H$ decoupled. Isolines of the total width $\Gamma_{h_2}(\text{GeV})$. Left: $\lambda=0.8$ and $v_S=2v$. Right: $\lambda=1.4$ and $v_S=v$. The colored region is excluded at 95$\%$ C.L.}
\end{center}
\end{figure}

\begin{figure}
\begin{center}
\includegraphics[width=.48\textwidth]{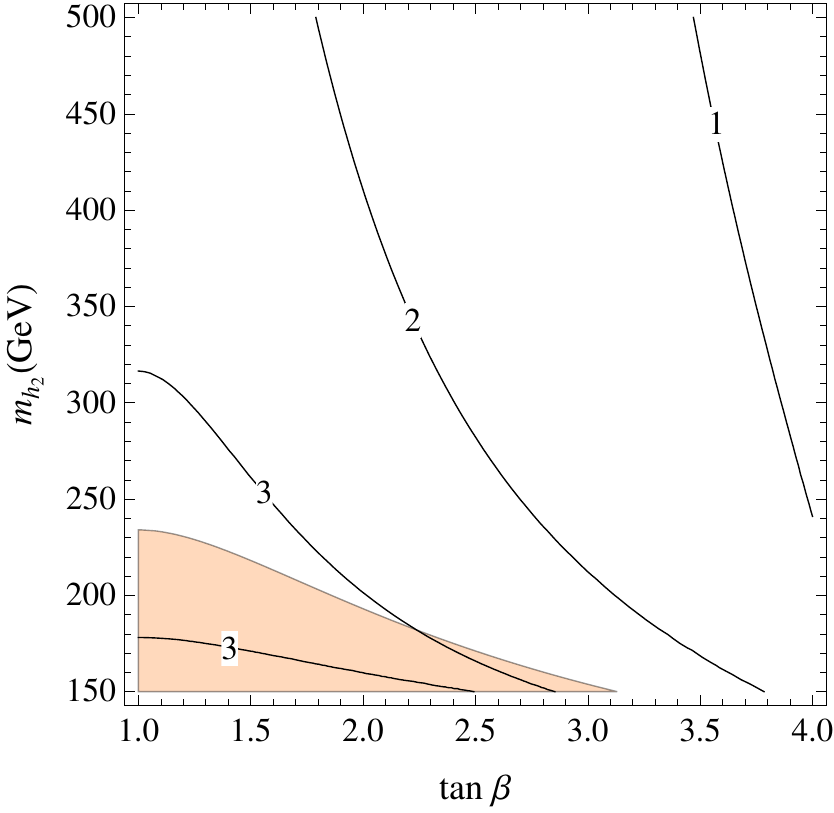}\hfill
\includegraphics[width=.48\textwidth]{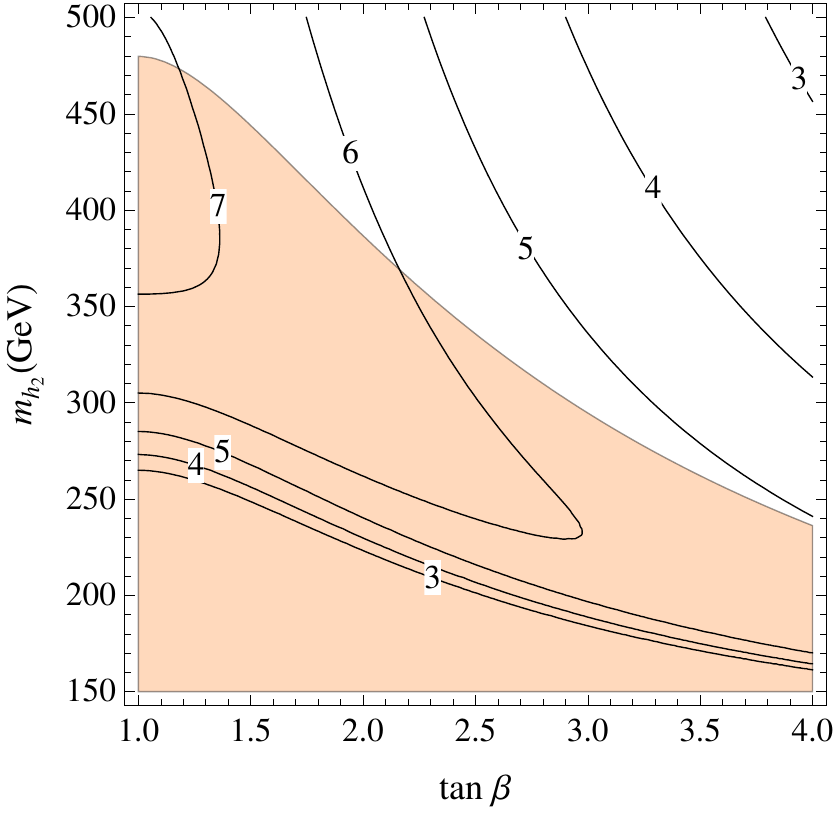}
\caption{\label{fig:mAdecoupled-cubic}\small $H$ decoupled. Isolines of $g_{hhh}/g_{hhh}^{\text{SM}}$. Left: $\lambda=0.8$ and $v_S=2v$. Right: $\lambda=1.4$ and $v_S=v$. The colored region is excluded at 95$\%$ C.L.}
\end{center}
\end{figure}

To determine the decay properties of $h_2$ it is crucial to know its coupling $(g_{h_2h_1^2}/2) h_2h_1^2$ to the lighter state. In the general NMSSM and in the large-$m_H$ limit considered in this section, the leading $\lambda^2$-term contribution to this coupling, as well as the one to the cubic $h_1$ coupling $(g_{h_1^3}/6) h_1^3$, are given by
\begin{align}
g_{h_2h_1^2}&=\frac{\lambda ^2 v}{16} \left(4 \frac{v_S}{v} \cos\gamma +12 \frac{v_S}{v} \cos3 \gamma-7  \sin\gamma +12  \cos4 \beta \cos^2\gamma \sin\gamma +9  \sin3 \gamma \right)\notag\\
&- \frac{3}{v}\Delta_t^2 \cos^2\gamma\sin\gamma,\\
\frac{g_{h_1^3}}{g^{\text{SM}}_{h_1^3}}&=\frac{\lambda ^2 v^2}{16 m_{h_1}^2} \cos\gamma \left(10 - 4  \cos4 \beta \cos^2\gamma - 6 \cos2 \gamma + 8 \frac{v_S}{v} \sin2 \gamma\right) + \dfrac{\Delta_t^2}{m_{h_1}^2} \cos^3\gamma,
\end{align}
where $v_S$ is the vev of the singlet.
Figures~\ref{fig:mAdecoupled-width} and \ref{fig:mAdecoupled-hh} show the total width of $h_2$ and its branching ratio into a pair of light states for some choices of $v_S$.
The other most significant decay mode of $h_2$ is into a $W$ pair, with a branching ratio given in figure~\ref{fig:mAdecoupled-WW}. Figure~\ref{fig:mAdecoupled-cubic} shows the triple $h_1$ coupling normalized to the SM one.

These results depend on the value taken by $v_S$. In particular we note that the Higgs fit still allows the triple Higgs coupling to get a relative enhancement of a factor of a few (with a negative or positive sign) with respect to the Standard Model one, thus yielding potentially large effects in Higgs pair production cross sections\cite{Baglio:2012np}.

\begin{figure}[t]
\begin{center}
\includegraphics[width=0.48\textwidth]{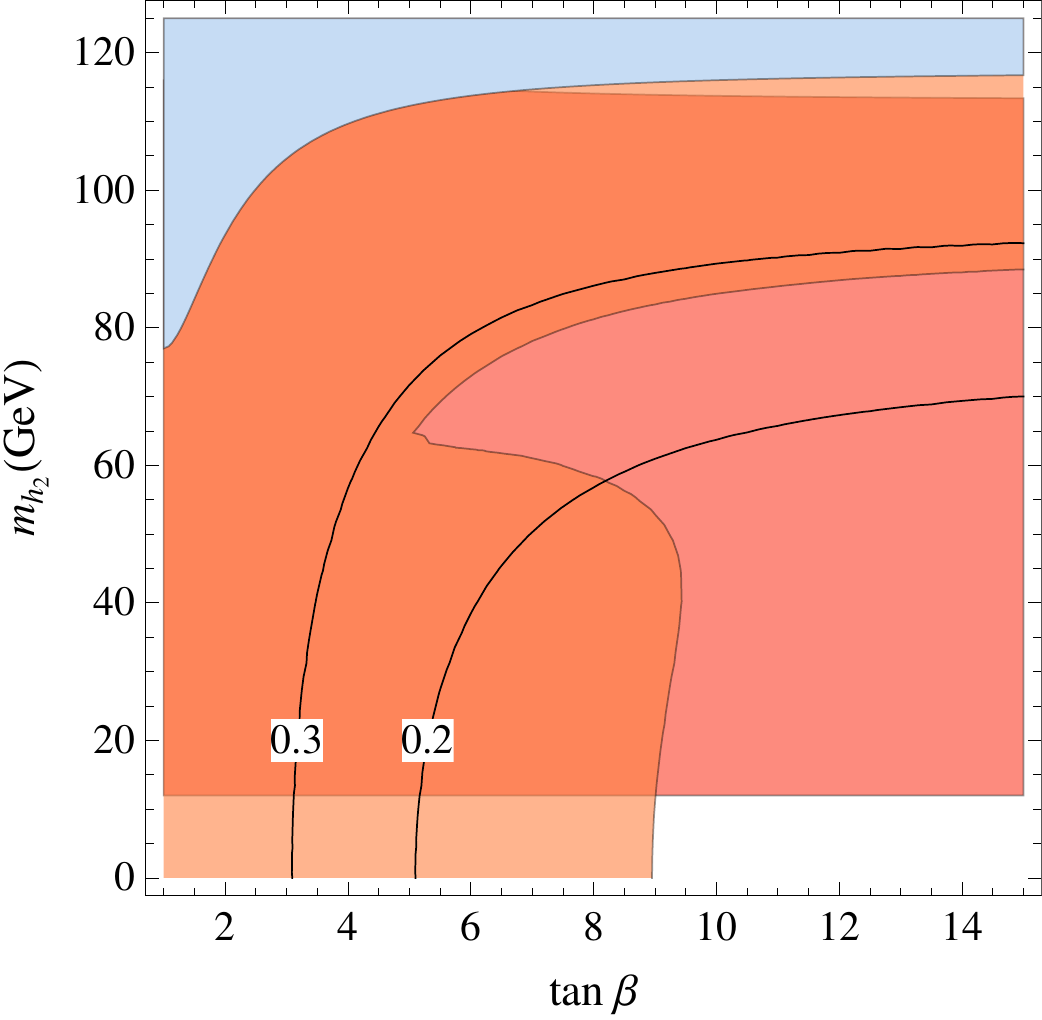}\hfill
\includegraphics[width=0.48\textwidth]{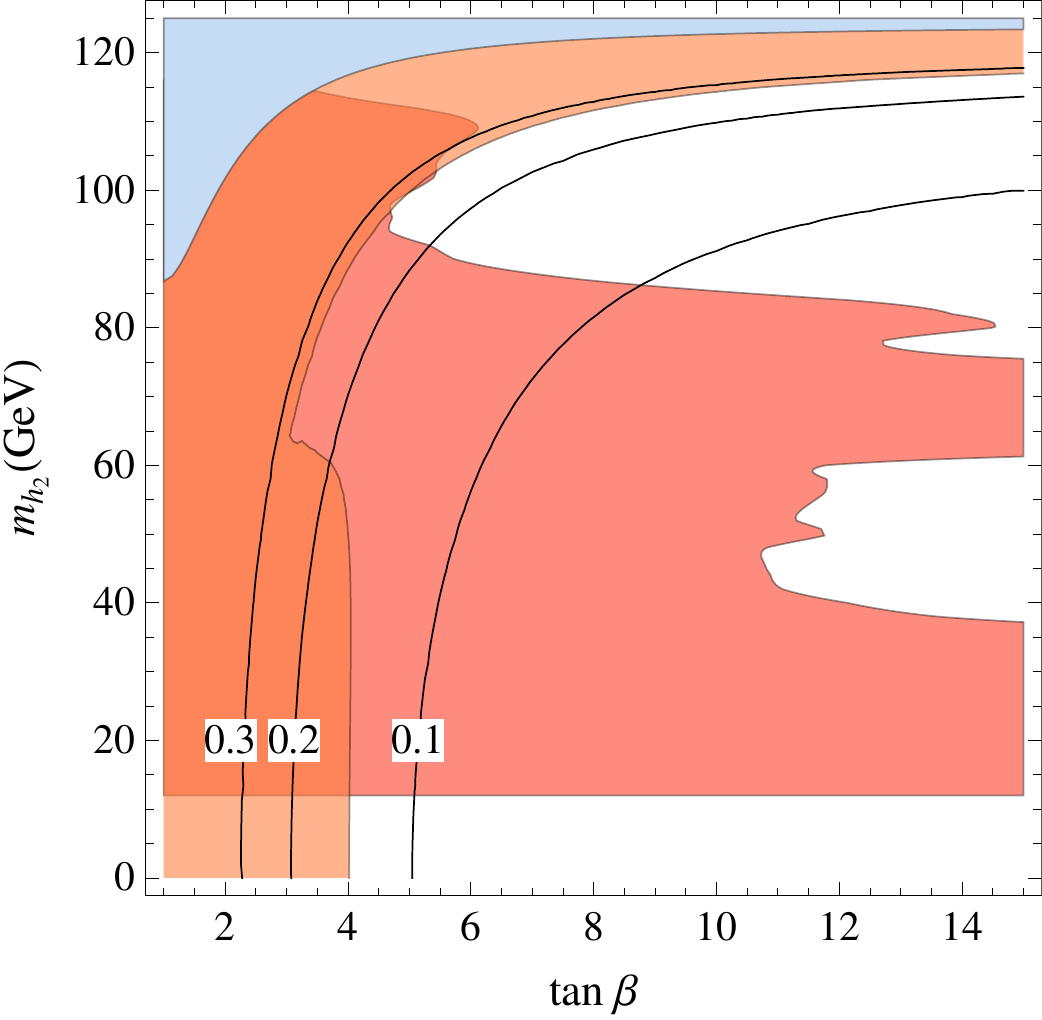}
\caption{\label{fig3} $H$-decoupled. Isolines of $s^2_\gamma$. $\lambda=0.1$ and $v_S=v$. Left: $\Delta_t=75$ GeV. Right: $\Delta_t=85$ GeV. Orange and blue regions as in figure~\ref{fig1}. The red region is excluded by LEP direct searches for $h_2\to b\bar b$.}
\end{center}
\end{figure}%
\begin{figure}[t]
\begin{center}
\includegraphics[width=0.48\textwidth]{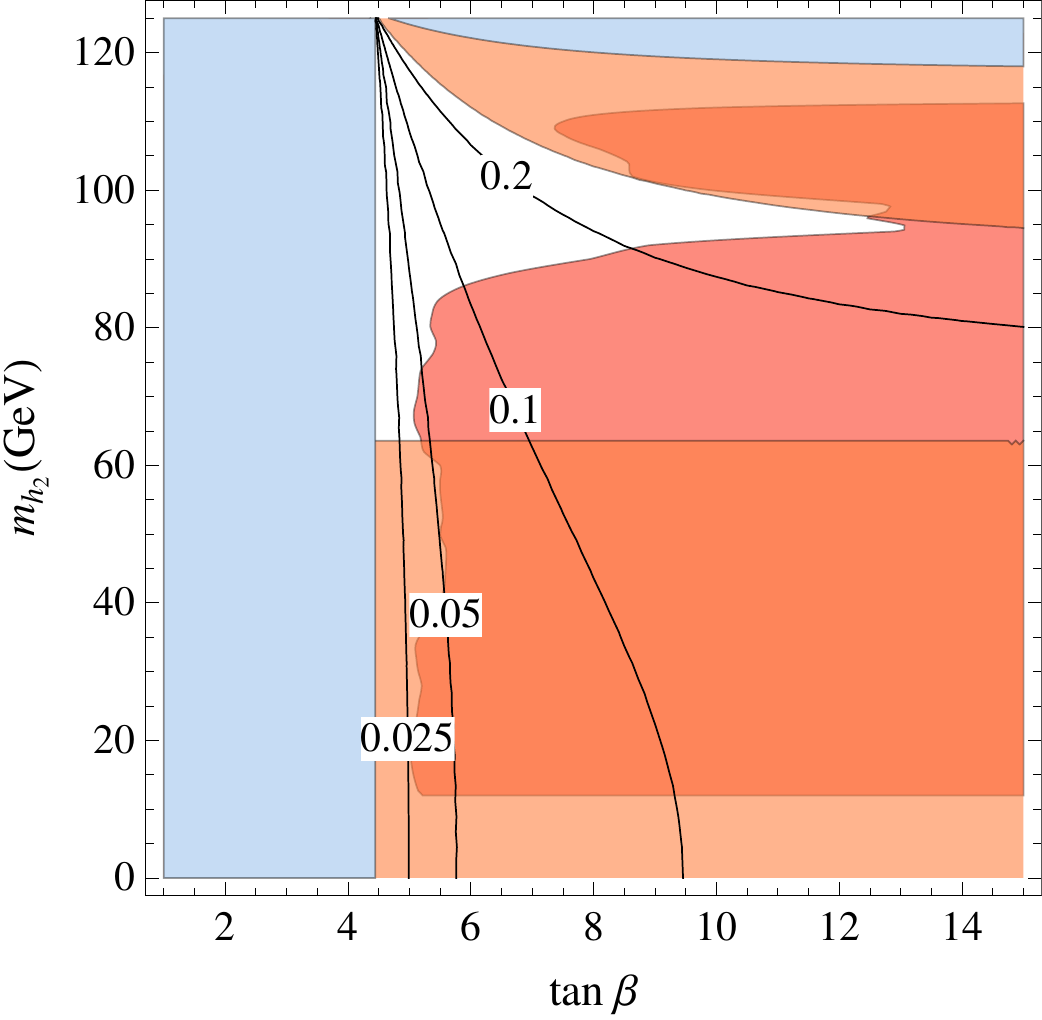}\hfill
\includegraphics[width=0.48\textwidth]{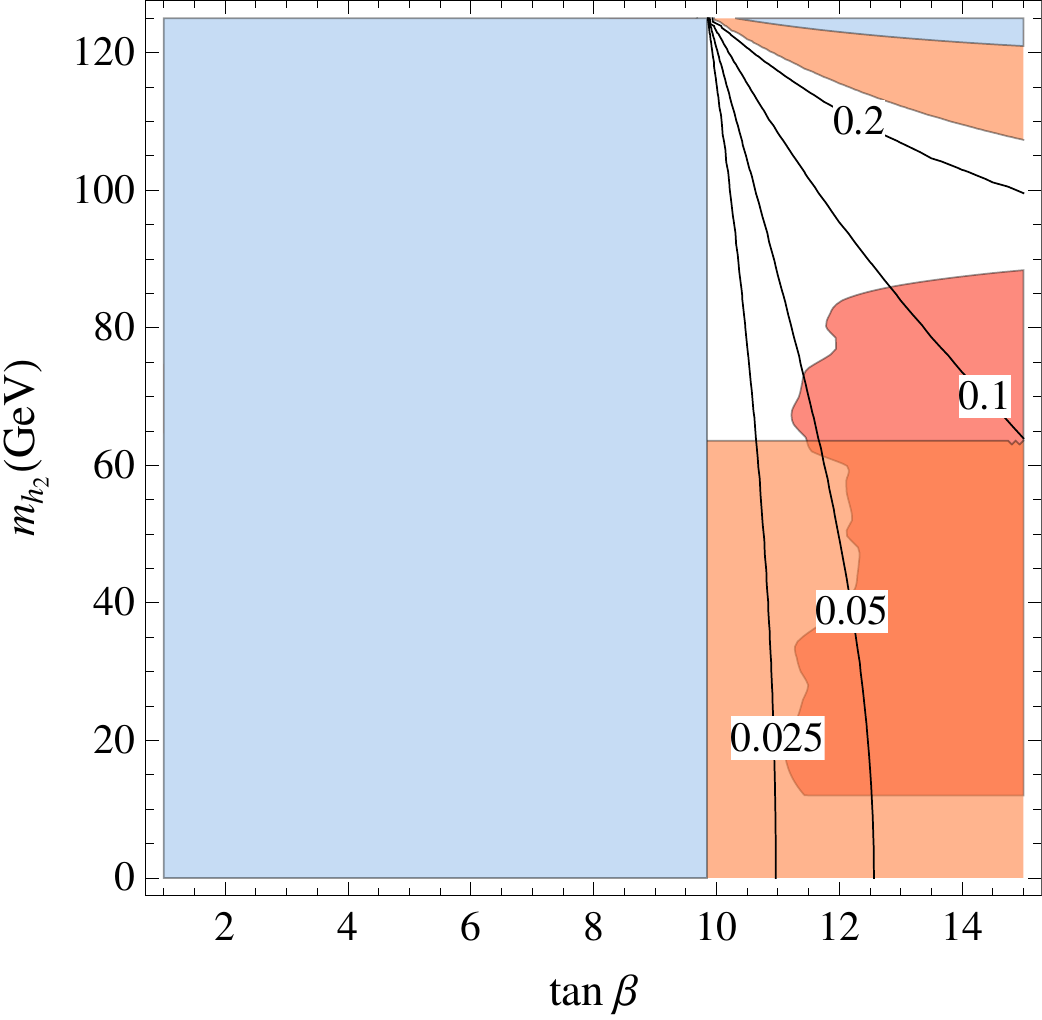}
\caption{\label{fig4} $H$-decoupled. Isolines of $s^2_\gamma$. $\Delta_t=75$ GeV and $v_S=v$. Left: $\lambda=0.8$. Right: $\lambda=1.4$. Orange and blue regions as in figure~\ref{fig1}. The red region is excluded by LEP direct searches for $h_2\to b\bar b$.}
\end{center}
\end{figure}
When $m_{h_2} < m_{h_1}$ we consider both the low and the large $\lambda$ cases.
The low $\lambda$ case $(\lambda = 0.1)$ is shown in figure~\ref{fig3} for two values of $\Delta_t$ together with the isolines of $s^2_{\gamma}$.
As in the previous sections we do not include any invisible decay mode except for $h_{\text{LHC}}\rightarrow h_2 h_2$ when kinematically allowed.\footnote{To include $h_{\text{LHC}}\rightarrow h_2 h_2$ we rely on the triple Higgs couplings as computed by retaining only the $\lambda^2$-contributions.
This is a defendable approximation for $\lambda$ close to unity, where $h_{\text{LHC}}\rightarrow h_2 h_2$ is important. In the low $\lambda$ case the $\lambda^2$-approximation can only be taken as indicative, but there $h_{\text{LHC}}\rightarrow h_2 h_2$ is less important.} Here an invisible branching ratio  of $h_{\text{LHC}}$, $\mathrm{BR}_{\rm inv}$, would strengthen the bound on the mixing angle to $s^2_\gamma < (0.22 - 0.78 \mathrm{BR}_{\rm inv})$.

For $\lambda$  close to unity we take again $\Delta_t = 75$ GeV, but any choice lower than this would not change the conclusions. The currently allowed region is shown in figure~\ref{fig4} for two values of $\lambda$. Note that, for large $\lambda$, no solution is possible at low enough $\tan{\beta}$, since, before mixing, $m^2_{hh}$ in eq.~\eqref{mhh} has to be below the mass squared of $h_{\text{LHC}}$.

\begin{figure}[t]
\begin{center}
\includegraphics[width=0.48\textwidth]{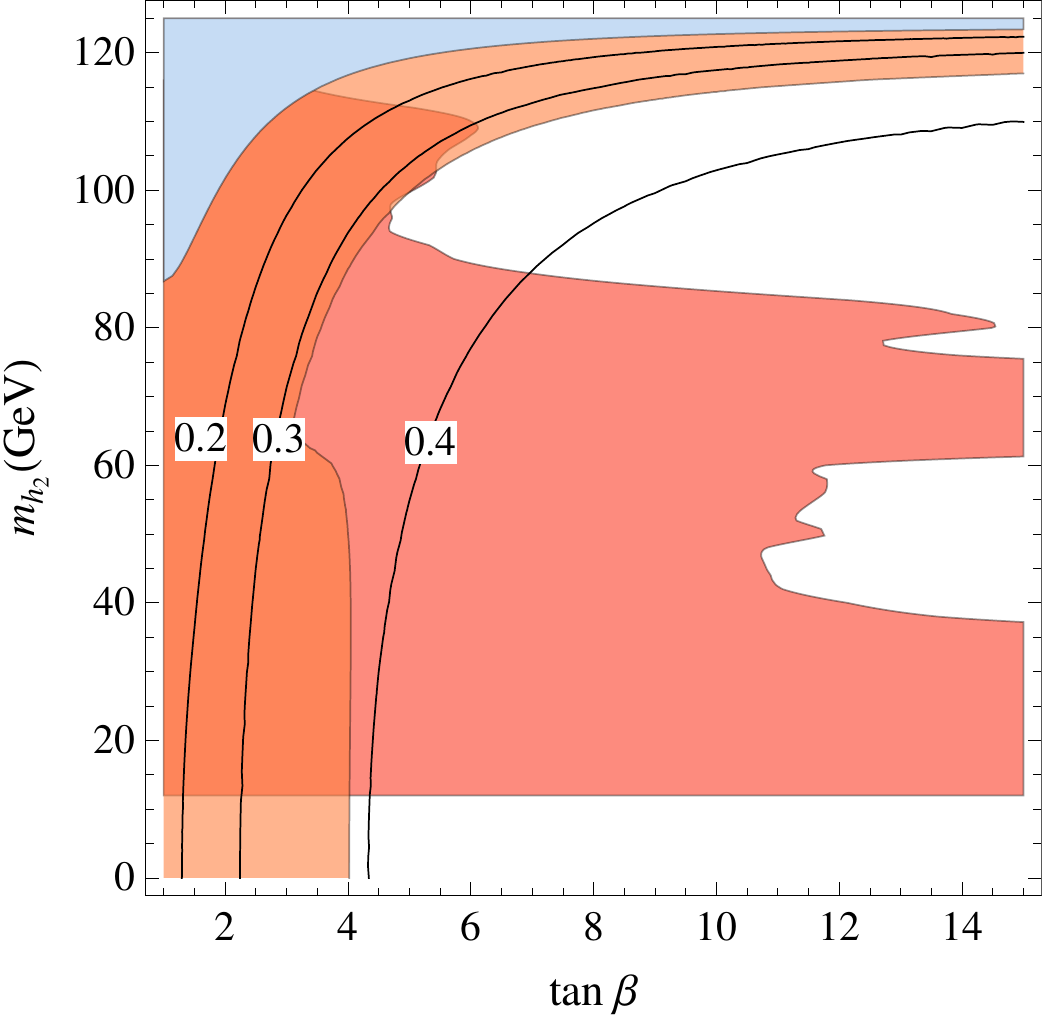}\hfill
\includegraphics[width=0.48\textwidth]{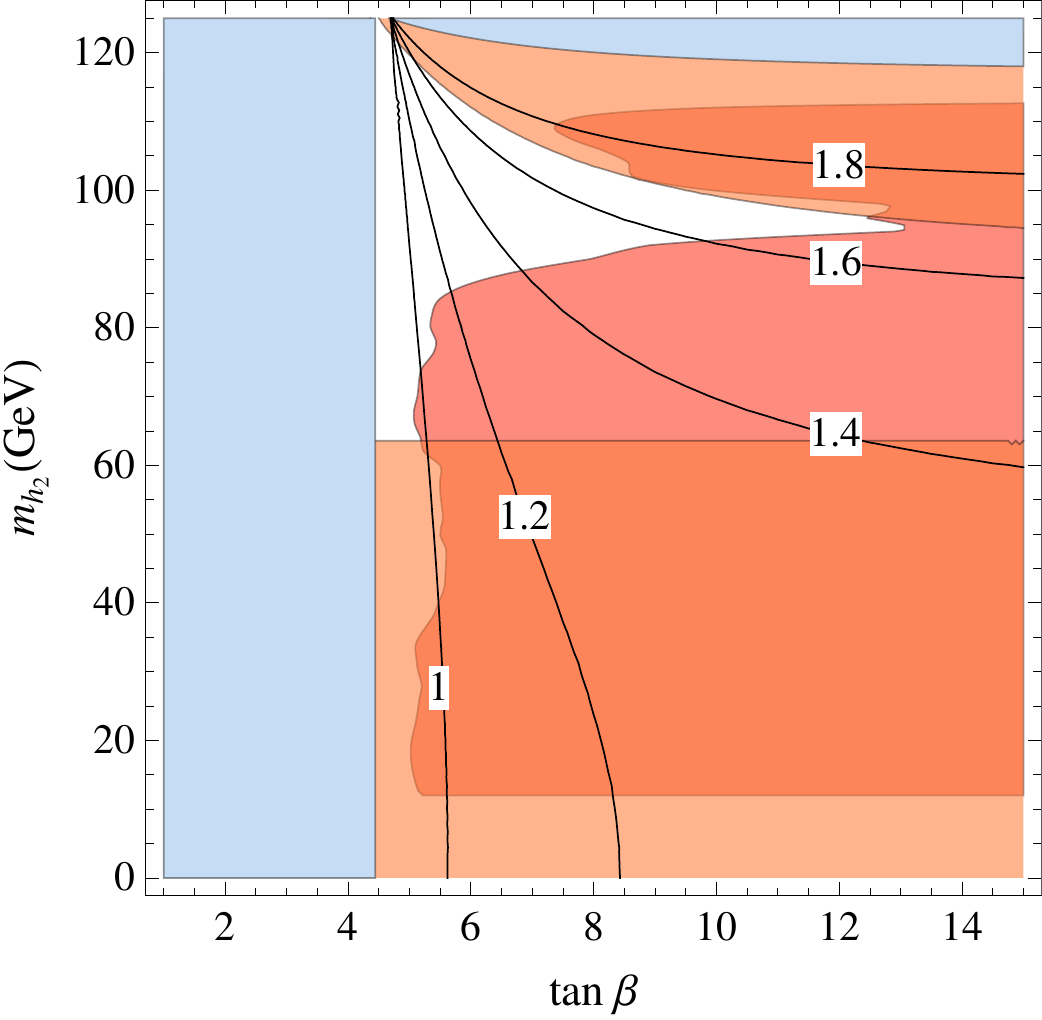}
\caption{\label{fig5} $H$-decoupled. Isolines of $g_{h^3}/g_{h^3}\big|_{\rm SM}$. Left: $\lambda=0.1$, $\Delta_t=85$ GeV and $v_S=v$. Right: $\lambda=0.8$, $\Delta_t=75$ GeV and $v_S=v$. Orange and blue regions as in figure~\ref{fig1}. The red region is excluded by LEP direct searches for $h_2\to b\bar b$.}
\end{center}
\end{figure}

How will it be possible to explore the regions of parameter space currently still allowed in this $m_{h_2} < m_{h_1} (\ll m_{h_3})$ case in view of the reduced couplings of the lighter state? Unlike in the singlet-decoupled case, the improvement in the measurements of the signal strengths of $h_{\text{LHC}}$ is not going to play a major role. Based on the projected sensitivity of table \ref{tab1}, the bound on the mixing angle will be reduced to $s^2_\gamma < 0.15$ at $95\%$ C.L. A significant  deviation from the case of the SM can occur in the cubic $h_{\text{LHC}}$-coupling $g_{h^3_1}$, like in the $m_{h_1} < m_{h_2}$ case, as shown in figure~\ref{fig5}.
The LHC14 in the high-luminosity regime is expected to get enough sensitivity to be able to see such deviations \cite{Dolan:2012rv,ATLAS-collaboration:2012iza,Goertz:2013kp}. 
At that point, on the other hand,  the searches for directly produced s-partners should have already given some clear indications on the relevance of the entire picture.

Finally, we have directly checked that EWPT do not put any further constraint on the parameter space also in the $H$-decoupled case. The reason is different with respect to the singlet-decoupled case. Here the reduced couplings of $h_{\text{LHC}}$ to the weak bosons lead to well known asymptotic formulae for the corrections to the $\hat{S}$ and $\hat{T}$ parameters \cite{Barbieri:2007bh}
\begin{align}
\Delta \hat S &=  + \frac{\alpha}{48\pi s_w^2}s^2_\gamma \log \frac{m_{h_2}^2}{m_{h_{\text{LHC}}}^2}, & \Delta \hat T &=      - \frac{3\alpha}{16\pi c_w^2}s^2_\gamma \log \frac{m_{h_2}^2}{m_{h_{\text{LHC}}}^2}
\end{align}
valid for $m_{h_2}$ sufficiently heavier that $h_{\text{LHC}}$. The correlation of $s^2_\gamma$ with $m_{h_2}$ given in eq. \eqref{sin2gamma} leads therefore to a rapid decoupling of these effects. 
The one loop effect on $\hat{S}$ and $\hat{T}$ becomes also vanishingly small as 
 $m_{h_2}$ and  $h_{\text{LHC}}$ get close to each other, since in the degenerate limit any mixing can be redefined away and only the standard doublet contributes as in the SM.

\section{Three-state mixing}\label{NMSSM/general}

\begin{figure}
\begin{center}
\includegraphics[width=.48\textwidth]{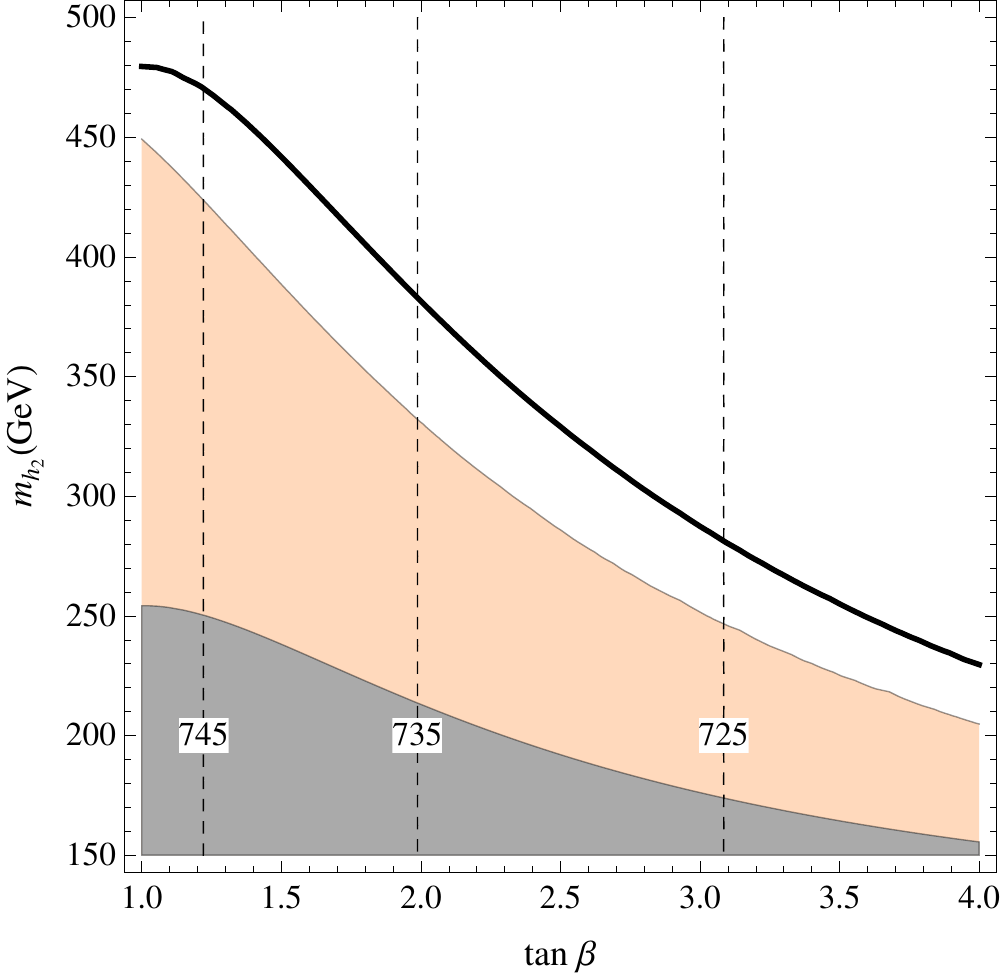}\hfill
\includegraphics[width=.48\textwidth]{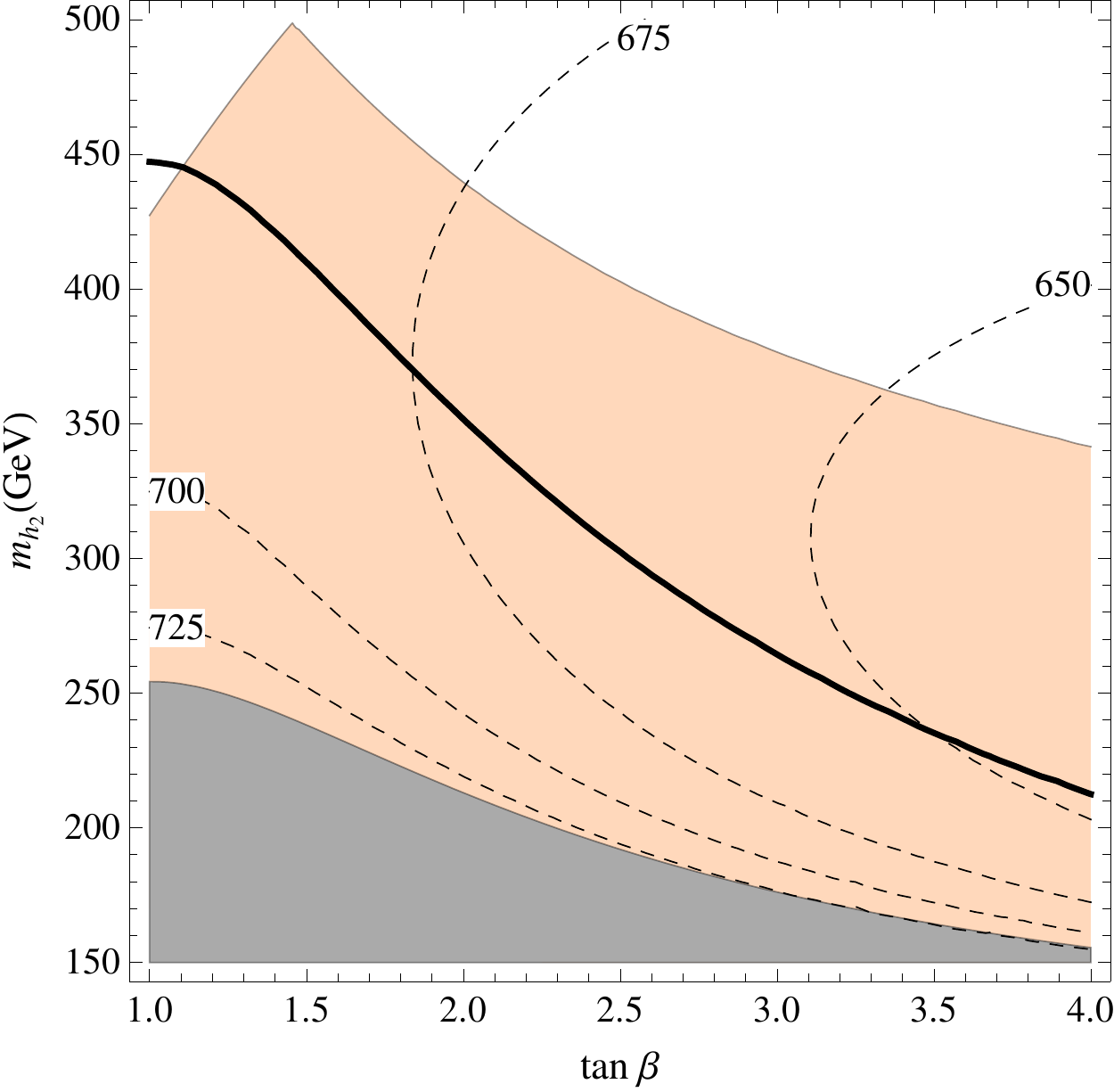}
\caption{\label{fig:mA:quasidecoupled}\small $H$ ``almost decoupled'' with $\lambda = 1.4$ and $m_{h_3} = 750$ GeV. The dashed isolines are for $m_{H^{\pm}}$. Left: $\sin^2\sigma=0$. Right: $\sin^2\sigma=0.25$. The colored region is excluded at $95\%$ C.L. In the grey area there is no solution for $\delta$. The thick line shows the na\"ive exclusion limit from $s_{\gamma}^{2}$ only.}
\end{center}
\end{figure}

In the general case where none of the states is completely decoupled the three angles $\delta$, $\gamma$ and $\sigma$ can all be different from zero, and the three masses $m_{h_2}$, $m_{h_3}$ and $m_{H^\pm}$ are all virtually independent. This complicates the analysis with respect to the simpler limiting cases of the previous sections, although several consistency relations among them are present.
In figure~\ref{fig:mA:quasidecoupled} we show the excluded regions in the plane $(\tan\beta,m_{h_2})$ for $m_{h_3} = 750$ GeV and $\lambda = 1.4$, setting $s^2_{\sigma}$ to two different values in order to fix $m_{H^\pm}$. When $s^2_{\sigma} = 0$ one recovers the previous $H$ decoupled case in the limit $m_{h_3}\to\infty$. With respect to this case, both $\gamma$ and $\delta$ are free parameters in the fit to the couplings of $h_{\rm LHC}$, and as a consequence the bounds are milder than what is expected from using only $\gamma$. If $s^2_{\sigma} \neq 0$, $h_2$ and $h_3$ are not decoupled, and their masses can not be split very consistently with all the other constraints. This is reflected in a broader excluded region for low $m_{h_2}$ in the right side of figure~\ref{fig:mA:quasidecoupled}, where we take $s^2_{\sigma} = 0.25$.

\subsection{The diphoton signal}\label{NMSSM/general/diphoton}
\begin{figure}[t]
\begin{center}
\includegraphics[width=0.48\textwidth]{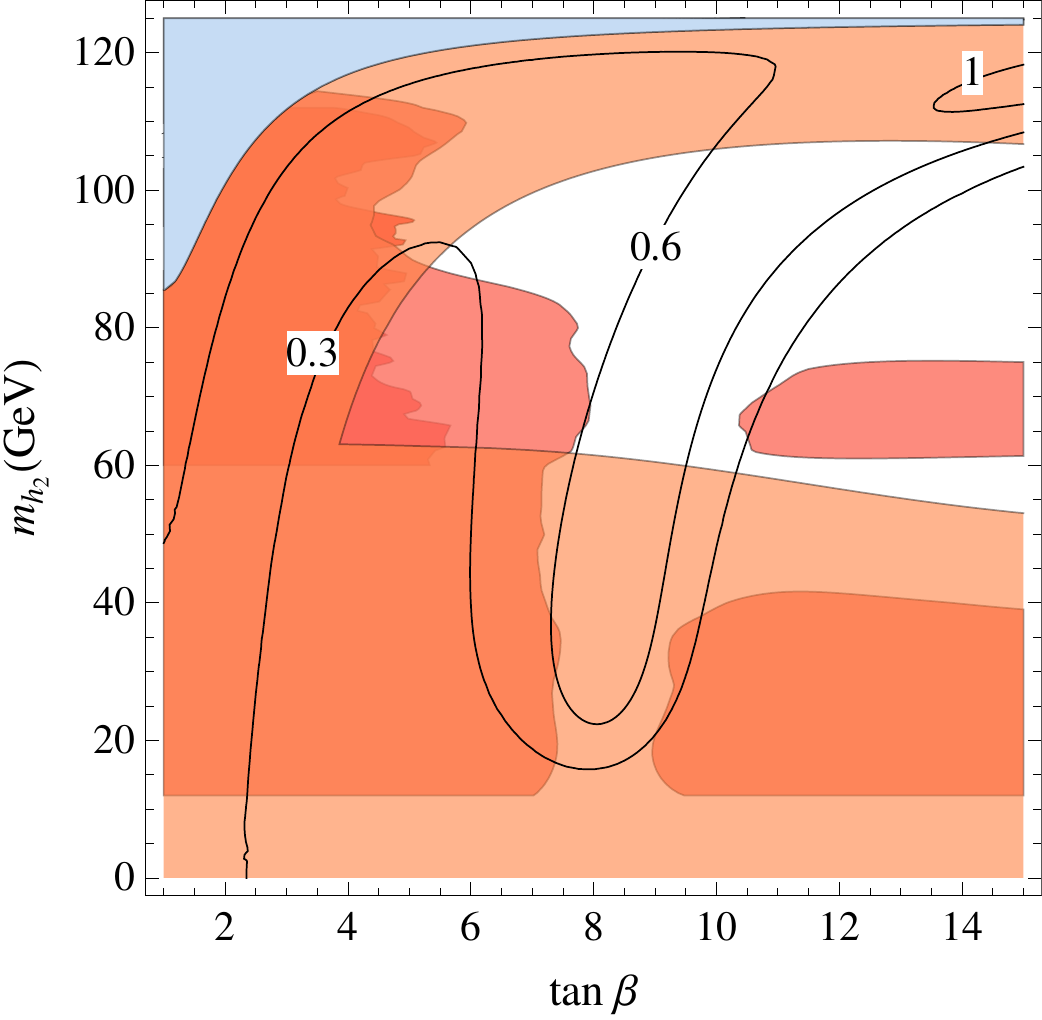}\hfill
\includegraphics[width=0.48\textwidth]{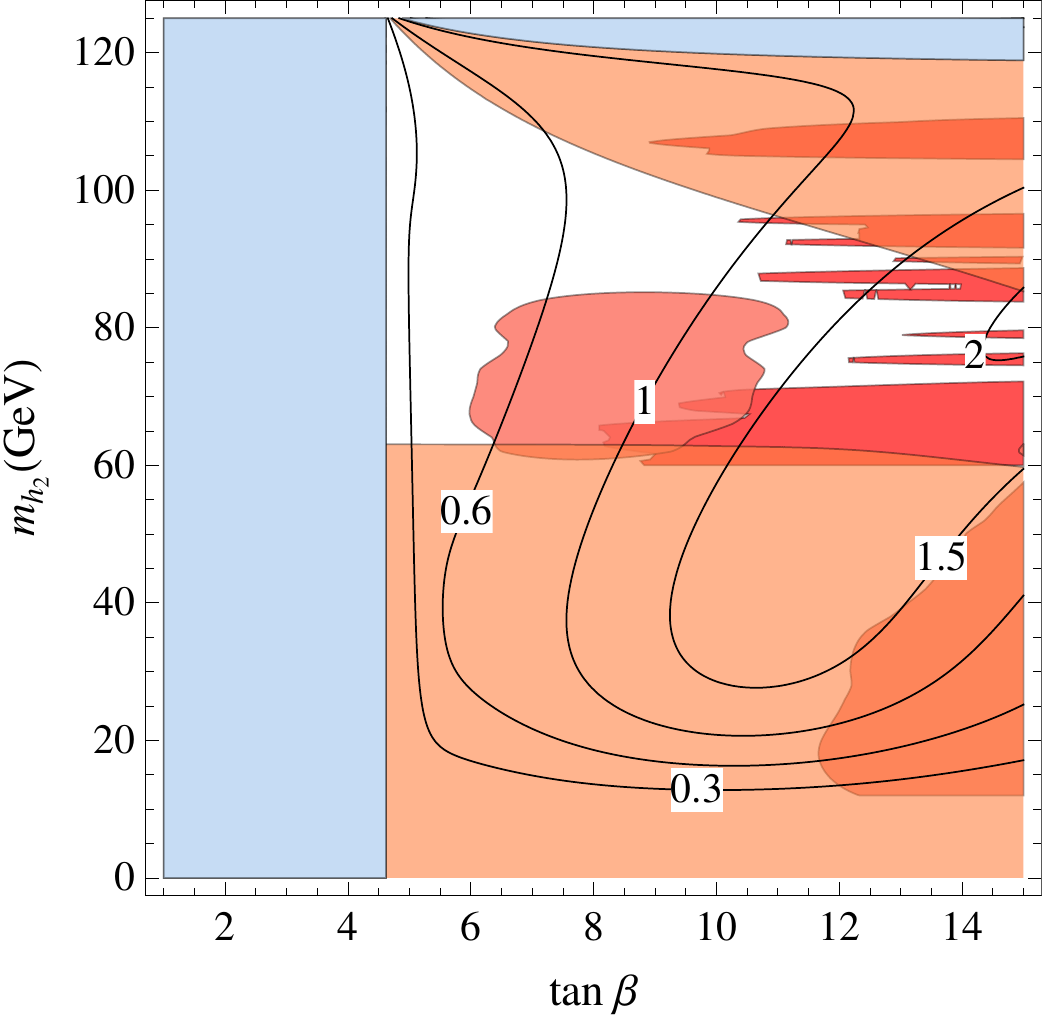}
\caption{\label{fig6} Fully mixed situation. Isolines of the ratio between the signal strength of $h_2 \to \gamma \gamma$ and the one for a SM Higgs with mass $m_{h_2}$. We take  $m_{h_3}=500$ GeV, $s^2_\sigma=0.001$ and $v_s = v$. Left: $\lambda=0.1$, $\Delta_t=85$ GeV. Right: $\lambda=0.8$, $\Delta_t=75$ GeV. Orange and blue regions as in figure~\ref{fig1}. The red and dark red regions are excluded by LEP direct searches for $h_2\to b\bar b$ and $h_2 \to$ hadrons respectively.}
\end{center}
\end{figure}
The phenomenological exploration of the situation considered in the previous section could be significantly influenced if the third state were not fully decoupled. 
As a relevant example we still consider the case of a state $h_2$ lighter than $h_{\text{LHC}}$, lowering $m_{h_3}$ to 500 GeV, to see if it could have an enhanced signal strength into $\gamma \gamma$.
Using \eqref{eq:sin:gamma:general}--\eqref{eq:sin:2alpha:general}, for fixed values of $\sigma$, $\lambda$ and $\Delta_t$, the two remaining angles $\alpha$ (or $\delta = \alpha - \beta + \pi/2$) and $\gamma$ are determined in any point of the $(\tan{\beta}, m_{h_2})$ plane and so are all the branching ratios of $h_2$ and of $h_{\text{LHC}}$. More precisely $\delta$ is fixed up to the sign of $s_\sigma c_\sigma s_\gamma$ (see first line of eq.~\eqref{eq:sin:2alpha:general}). Notice that this is the only physical sign that enters the observables we are considering.

The corresponding situation is represented in figure~\ref{fig6}, for two choices of $\lambda$ and $\Delta_t$. The sign of $s_\sigma c_\sigma s_\gamma$ has been taken negative in order to suppress BR$(h_2\to b \bar b)$. This constrains $s_\sigma^2$ to be very small in order to leave a region still not excluded by the signal strengths of $h_{\text{LHC}}$, with $\delta$ small and negative. To get a signal strength for $h_2\rightarrow \gamma \gamma$ close to the SM one for the corresponding mass is possible for a small enough value of $s^2_\gamma$, while the dependence on $m_{h_3}$ is weak for values of $m_{h_3}$ greater than 500 GeV. Note that the suppression of the coupling of $h_2$ to $b$-quarks makes it necessary to consider the negative LEP searches for $h_2 \to$ hadrons \cite{Searches:2001aa}, which have been performed down to $m_{h_2} = 60$~GeV.

Looking at the similar problem when $h_2$ is heavier than $h_{\text{LHC}}$, we find it harder to get a signal strength close to the SM one, although this might be possible for a rather special choice of the parameters.\footnote{An increasing significance of the excess found by CMS \cite{CMS:zza} at 136 GeV would motivate such special choice.} Our purpose here is more to show that in the fully mixed situation the role of the measured signal strengths of $h_{\text{LHC}}$, either current or foreseen, plays a crucial role.

\section{The NMSSM at $\lambda\gtrsim 1$ and gauge coupling unification}\label{NMSSM/unification}

As said in the Introduction we are particularly interested in the NMSSM at $\lambda$ close to one and moderate $\tan{\beta}$ to limit the fine tuning. At least in the $H$ decoupled case we have seen in section~\ref{NMSSM/Hdecoupled} that this is consistent with current data. On the other hand a well known objection to the NMSSM at $\lambda \gtrsim 1$ is its compatibility with  gauge coupling unification. Requiring $\lambda$ to stay semi-perturbative up to the grand unified theory (GUT) scale bounds $\lambda$ at the weak scale at  about $0.7$ \cite{Espinosa:1991gr}. This value is in fact influenced by the presence of vector-like matter in full $\SU(5)$ multiplets that slows down the running of $\lambda$  by increasing  the gauge couplings at high energies. However, even adding three vector-like five-plets at 1 TeV, in which case $\alpha_G$ still remains perturbative, does not allow $\lambda$ at the weak scale to go above $0.8$ \cite{Masip:1998jc,Barbieri:2007tu}.

There are several ways \cite{Harnik:2003rs,Chang:2004db,Birkedal:2004zx,Delgado:2005fq,Gherghetta:2011wc,Craig:2011ev,Csaki:2011xn,Hardy:2012ef} in which $\lambda$ could go to 1--1.5  without spoiling unification nor affecting  the  consequences at the weak scale of the NMSSM Lagrangian, as treated above. One further possibility also makes use of two vector-like five-plets as follows. For ease of exposition let us call them  $F_{u,d} + \bar{F}_{u,d}$, where $F_u$ is a $5$ and $F_d$ a $\bar{5}$, thus containing one $\SU(2)$ doublet each, $h_u$ and $h_d$, with the same quantum numbers of the standard $H_u, H_d$ used so far. Correspondingly $\bar{F}_{u,d}$ contain two doublets that we call $\bar{h}_{u,d}$. Needless to say all these are superfields. Let us further assume that the superpotential is such that:
\begin{itemize}
\item The five-plets interact with a singlet $S$ and pick up $\SU(5)$-invariant masses consistently with a Peccei-Quinn symmetry;
\item The standard doublets $H_u, H_d$ mix by mass terms with $h_u$ and $h_d$, still maintaining the Peccei-Quinn symmetry, and do not interact directly with $S$.
\end{itemize}
The corresponding superpotential is 
\begin{equation}
W = \lambda_S S F_u F_d + M_u F_u \bar{F}_u + M_d F_d \bar{F}_d + m_u H_u \bar{h}_u + m_d H_d \bar{h}_d + \lambda_t H_u Q t,
\label{f}
\end{equation}
where we have also made explicit the  Yukawa coupling of the top to $H_u$.
Below these masses, all taken to be comparable, one is left with three massless supermultiplets after eliminating the mixings through diagonalization of the quadratic terms: 
\begin{equation}
S,~~~ \hat{H}_u = c_u H_u + s_u h_u,~~~\hat{H}_d = c_d H_d + s_d h_d,
\end{equation}
which interact through the superpotential
\begin{equation}
\hat{W}= \hat{\lambda} S \hat{H}_u \hat{H}_d + \hat{\lambda}_t \hat{H}_u Q t, ~~~\hat{\lambda} = \lambda_S s_u s_d,~~~~\hat{\lambda}_t=\lambda_t c_u .
\end{equation}
This superpotential, completed by Peccei-Quinn symmetry breaking terms at the Fermi scale, defines the effective NMSSM as discussed so far. To cure the growth of $\hat{\lambda}$ at increasing energies, the masses in (\ref{f}) will have to be crossed while $\hat{\lambda}$ is still semi-perturbative. For $\hat{\lambda} =1$--$1.5$ these masses are above 1000 TeV.

 At greater energies the running of the gauge couplings is affected, compared to the standard supersymmetric case,  by the supermultiplet $H_u$ with a top Yukawa coupling increased by a factor $1/c_u$ and by the  degenerate complete $\SU(5)$ multiplets $F_{u,d} + \bar{F}_{u,d}$. To avoid a Landau pole before the GUT scale in the top Yukawa coupling,  $c_u$ has to be bigger than about $1/\sqrt{2}$. As to the effect of $F_{u,d} + \bar{F}_{u,d}$, they do not alter the relative one loop running of the gauge couplings but might give rise to an exceeding growth of all of them before $M_{\text{GUT}}$ due to the presence of the coupling $\lambda_S$, which at some point will get strong. To avoid this, a change of regime in the $\SU(5)$-symmetric sector will have to intervene to keep under control the anomalous dimensions of the $F_{u,d}, \bar{F}_{u,d}$ superfields.

\section{Summary and comparison of bounds}

Given the current experimental informations, the Higgs sector of the NMSSM appears to allow a minimally fine-tuned description of electro-weak symmetry breaking, at least in the context of supersymmetric extensions of the SM. Motivated by this fact we have have analyzed the different signal strengths of the newly found scalar resonance, together with the current bounds from direct searches, in order to outline a possible overall strategy to search for signs of the \CP-even states.

To have a simple characterization of the properties of the extended Higgs system we have focussed on relations between physical parameters. We have also not included any radiative effects of superpartners  other than the top-stop loop corrections to the quartic Higgs couplings, \eqref{delta-t}, which mostly affect the properties of the lightest \CP-even state that we identify with $h_{\rm LHC}$. While providing at least a useful reference case, we think that this is motivated by the consideration of superpartner masses at their ``naturalness limit''. Furthermore we have considered separately two limiting cases of the NMSSM, in which one \CP-even scalar is decoupled: either the singlet component or the doublet $H=-s_\beta H_d + c_\beta H_u$. This allows to work in general with four effective parameters: $\lambda, \tan{\beta}$, $\Delta_t$ in (\ref{delta-t}) and the mass of the intermediate \CP-even scalar, $h_2$ or $h_3$ in the two cases. The mass of $h_1=h_{\rm LHC}$ sets one relation among these parameters. A 
further relation exists in the singlet-decoupled case, of which the MSSM is a limit ($\lambda=0$).
We have not been sticking to a particular NMSSM, which might imply specific constraints on the physical parameters that we consider, but we have assumed to live in the case of negligible \CP\ violation in the Higgs sector. In each of the limiting cases we have considered both the situation where $h_{\rm LHC}$ is the lightest scalar, and the possibility -- still allowed by the data in some specific case -- that there is a lighter \CP-even state with reduced couplings.

Within this framework, even though the signal strengths of $h_{\rm LHC}$ are close to those expected in the SM, they still allow for a new further state nearby, unlike in the case of the MSSM, where a \CP-even scalar heavier than $h_{\mathrm{LHC}}$ and below about 300 GeV is unlikely \cite{DAgnolo:2012mj}. This is true in both the limiting cases that we have considered, as visible in figures~\ref{fig:mSdecoupled-1} and \ref{fig:mAdecoupled-1}, to be contrasted with figure~\ref{fig:MSSM-1}. On the other hand, the same figures show that the measured signal strengths of $h_{\rm LHC}$ do limit the possible values of $\lambda$, at least for moderate values of $\tan{\beta}$, which is the region mostly motivated by naturalness: $\lambda \approx 1$ is still largely allowed in the $H$-decoupled case, whereas it is borderline in the singlet decoupled situation. As commented upon in section~\ref{NMSSM/unification}, we think that $\lambda\gtrsim 1$ can be compatible with gauge coupling unification.

Not surprisingly,  a clear difference emerges between the singlet-decoupled and the $H$-decoupled cases: the influence on the signal strengths of  $h_{\text{LHC}}$ of the mixing with a doublet or with a singlet makes the relative effects visible at different levels. A quantitative estimate of the sensitivity of the foreseen measurements at LHC14 with 300 $\text{fb}^{-1}$ makes it likely that the singlet-decoupled case will be thoroughly explored, while the singlet-mixing effects could remain hidden.
We also found that, in the MSSM with $(\mu A_t)/\langle m_{\tilde{t}}^2\rangle \lesssim 1$, the absence of deviations in the $h_{\text{LHC}}$ signal strengths would push the mass of the other Higgs bosons up to a TeV.

Needless to say, in any case the direct searches will be essential with a variety of possibilities discussed in the literature.\footnote{A first attempt at studying heavier Higgs decays in the NMSSM with $\lambda > 1$ was made in \cite{Cavicchia:2007dp}.} Most importantly from this point of view, the new states behave quite differently in the two cases, especially in what concerns their decay properties. The state $h_2$ of the $H$-decoupled case has a large BR into a pair of $h_1$, whenever it is allowed by phase space, with $VV$ as subdominant decay (figures~\ref{fig:mAdecoupled-hh},\ref{fig:mAdecoupled-WW}). With the production cross sections shown in figure~\ref{fig:mAdecoupled-Xsec} its direct search at LHC8 or LHC14 may be challenging, although perhaps not impossible \cite{Caterina,Gouzevitch:2013qca,Chatrchyan:2013yoa}. It is also interesting that, in the $H$-decoupled case, large deviations from the SM value are possible in the triple Higgs coupling $g^3_{h_{\rm LHC}}$, contrary to the $S$-decoupled and MSSM cases.
On the other hand the reduced value of $\lambda$ allowed in the singlet decoupled case makes the $b\bar{b}$ channel, and so the $\tau\bar{\tau}$, most important, below the $t\bar{t}$ threshold (figures~\ref{fig:mSdecoupled-BRs} and \ref{fig:mSdecoupled-BRf}).  This makes the state $h_3$ relatively more similar to the \CP-even $H$ state of the MSSM (figures~\ref{fig:MSSM-BRs} and \ref{fig:MSSM-BRf}), 
which is being 
actively searched. 
An important point in this comparison is that in the  NMSSM there are two \CP-odd states and that the lightest of the two may be quite different from the state $A$ of the MSSM unless in a special $S$-decoupled case.

Finally, in case of a positive signal pointing to a second Higgs boson, direct or indirect, it may be important to try to interpret it in a fully mixed scheme, involving all the three \CP-even states. To this end the analytic relations of the mixing angles to the physical masses given in equations~\eqref{eq:sin:gamma:general}--\eqref{eq:sin:2alpha:general} offer a useful tool, as illustrated in the examples of a $\gamma\gamma$ signal of figure~\ref{fig6}.

\part{A critical Higgs boson}

\chapter{Renormalization of the Standard Model at two loops}\label{2loops}

After having considered in the previous chapters the best motivated scenarios of physics at the Fermi scale, as suggested by the principle of naturalness, in this last part of the work we are concerned the possibility that an unnatural Standard Model remains a valid description of Nature up to very high scales, and ask ourselves what are the consequences of the recent discovery of the Higgs for this scenario. Although suffering a severe fine-tuning problem when viewed in an effective field theory framework, there are good reasons to explore carefully the implications of such a picture.

As we have already discussed in detail, the measured value of the Higgs boson mass $M_h=\Mhexperr$~\cite{Giardino:2013bma,CMS:ril,CMS:xwa,ATLAS:2013oma,ATLAS:2013nma} is a bit high for
supersymmetry and a bit low for composite models, making theoretical
interpretations not quite straightforward.
On the other hand,
it lies well within the parameter window in which the SM can
be extrapolated all the way up to the Planck mass $\mpl $, with no
problem of consistency other than remaining in the dark about
naturalness. Remarkably, in the context of the SM the measured value
of $M_h$ is special because it corresponds to a near-critical
situation in which the Higgs vacuum does not reside in the
configuration of minimal energy, but in a metastable state close to a
phase transition~\cite{Degrassi:2012ry} (for earlier considerations see~\cite{Krive:1976sg,Krasnikov:1978pu,Maiani:1977cg,Politzer:1978ic,Hung:1979dn,Cabibbo:1979ay,Linde:1979ny,Lindner:1985uk,Lindner:1988ww,Sher:1988mj,Arnold:1989cb,Arnold:1991cv,Sher:1993mf,Altarelli:1994rb,Casas:1994qy,Espinosa:1995se,Casas:1996aq,Schrempp:1996fb,Hambye:1996wb,Isidori:2001bm,Espinosa:2007qp,Ellis:2009tp};
for related studies see~\cite{Holthausen:2011aa,EliasMiro:2011aa,Chen:2012faa,Lebedev:2012zw,EliasMiro:2012ay,Rodejohann:2012px,Bezrukov:2012sa,Datta:2012db,Alekhin:2012py,Chakrabortty:2012np,Anchordoqui:2012fq,Masina:2012tz,Chun:2012jw,Chung:2012vg,Chao:2012mx,Lebedev:2012sy,Nielsen:2012pu,Kobakhidze:2013tn,Tang:2013bz,Klinkhamer:2013sos,He:2013tla,Chun:2013soa,Jegerlehner:2013cta}).

The extrapolation of the SM up to high energies implies the computation of the renormalization-group running of all the relevant parameters from the EW scale up to the Planck scale. Moreover, when matching the running parameters to the physical measured quantities, threshold corrections at the weak scale -- i.e. constant, renormalization scheme-dependent radiative corrections -- have also to be taken into account. The extreme closeness of the Higgs mass to a critical point separating two phases with very different physical properties motivates an accurate study which includes higher-order loop effects.

In this chapter we calculate the $\MS$ quartic Higgs coupling $\lambda(\mub)$ and the
top Yukawa coupling $y_t(\mub)$ at NNLO precision (two loops) in terms
of physical observables: the pole masses of the Higgs ($M_h$), of the
top ($M_t$), of the $Z$ ($M_Z$), of the $W$ ($M_W$), the
$\MS$ strong coupling $\alpha_3(M_Z)$, and the Fermi
constant $G_\mu$.
We improve on the study in ref.~\cite{Degrassi:2012ry}
where two-loop threshold corrections to $\lambda(\mub)$ had been computed
in the limit of vanishing weak gauge couplings, and two-loop electroweak threshold
corrections to $y_t(\mub)$ had been neglected.
As a byproduct of our
two-loop calculation of  $\lambda(\mub)$ we also obtain the 
$\MS$ quadratic Higgs coupling $m_h^2(\mub)$ at the
 NNLO level.

Recently, many authors have contributed towards the completion of the
calculation of the renormalization-group evolution
($\beta$-functions and thresholds) of the sizeable SM couplings at
NNLO precision. We summarize the present status of these calculations
in tables~\ref{tab:statusRGE}, \ref{tab:status}. Our new calculation of threshold
corrections, together with the results collected in
table~\ref{tab:status}, will allow us to refine the determination of the
critical value of $M_h$ that ensures absolute vacuum stability within
the Standard Model up to the Planck scale.

\begin{table}[t]
\begin{center}
\begin{tabular}{lllll}
\hline
& LO & NLO & NNLO & NNNLO\\ 
 & 1 loop & 2 loop & 3 loop & 4 loop\\   
 \hline\\[-5mm]
 $g_{3}$ & full~\hbox {\cite{Gross:1973id,Politzer:1973fx}} & $\Ord(\alpha_3^2)$~\hbox{\cite{Caswell:1974gg,Jones:1974mm}} & $\Ord(\alpha_3^3)$~\hbox{\cite{Tarasov:1980au,Larin:1993tp}} & $\Ord(\alpha_3^4)$~\hbox{\cite{vanRitbergen:1997va,Czakon:2004bu}}\\
  &  & $\Ord(\alpha_3\alpha_{1,2})$~\hbox{\cite{Jones:1981we}} & $\Ord(\alpha_3^2\alpha_t)$~\hbox{\cite{Steinhauser:1998cm}} & \\
 &  & full~\hbox{\cite{Machacek:1983tz}} & full~\hbox{\cite{Mihaila:2012fm,Mihaila:2012pz}} & \\[2mm]
$g_{1,2}$ & full~\hbox {\cite{Gross:1973id,Politzer:1973fx}} & full~\hbox{\cite{Machacek:1983tz}} & full~\hbox{\cite{Mihaila:2012fm,Mihaila:2012pz}} & --- \\[2mm]
$y_t$ & full~\hbox{\cite{Cheng:1973nv}} & $\Ord(\alpha_t^2,\alpha_3\alpha_t)$~\hbox{\cite{Fischler:1982du}} & full~\hbox{\cite{Chetyrkin:2012rz,Bednyakov:2012en}} & --- \\
 &  &  full~\hbox{\cite{Machacek:1983fi}}& & \\[2mm]
$\lambda, m_h^2$ & full~\hbox{\cite{Cheng:1973nv}} & full~\hbox{\cite{Machacek:1984zw,Luo:2002ey}} &full~\hbox{\cite{Chetyrkin:2013wya,Bednyakov:2013eba}} & --- \\[1mm]
\hline 
\end{tabular}
\caption{Present status of higher-order computations of the renormalization-group equations ($\beta$-functions) of the SM parameters. Here and in the following $g_2$ and $g_Y$ stand for the $\SU(2)_L$ and $\U(1)_Y$ gauge couplings, while $g_3$ is the strong gauge coupling..\label{tab:statusRGE}}
\end{center}
\end{table}
\begin{table}[t]
\begin{center}
\begin{tabular}{lllll}
\hline
 &LO & NLO & NNLO & NNNLO\\
 & 0 loop & 1 loop & 2 loop & 3 loop\\   
 \hline
$g_{2}$ &$ 2M_W/V$
& full~\hbox{\cite{Sirlin:1980nh,Marciano:1980pb}} & full~\hbox{\cite{Buttazzo:2013uya}}    & --- 
\\[2mm] 
$g_{Y}$ &$ 2\sqrt{M_Z^2-M_W^2}/V$
& full~\hbox{\cite{Sirlin:1980nh,Marciano:1980pb}} & full~\hbox{\cite{Buttazzo:2013uya}}   & --- 
\\[2mm] 
$y_t$ &$\sqrt{2} M_t/V$ &   $\Ord(\alpha_3)$~\hbox{\cite{Tarrach:1980up}}& $\Ord(\alpha_3^2,\alpha_3\alpha_{1,2})$~\hbox{\cite{Bezrukov:2012sa}}   & $\Ord(\alpha_3^3)$~\hbox{\cite{Chetyrkin:1999ys,Chetyrkin:1999qi,Melnikov:2000qh}}  \\
 & &   $\Ord(\alpha)$~\hbox{\cite{Hempfling:1994ar}} &   full~\cite{Buttazzo:2013uya} &   \\[2mm]
$\lambda$ & $M_h^2/2V^2$& full~\hbox{\cite{Sirlin:1985ux}} & for $g_{1,2}=0$~\hbox{\cite{Degrassi:2012ry}}
 & --- \\
 & &  & full~\cite{Buttazzo:2013uya}
 &  \\[2mm]
$m_h^2$ & $M_h^2$& full~\hbox{\cite{Sirlin:1985ux}} &  full~\cite{Buttazzo:2013uya} & --- \\
 \hline
\end{tabular}
\caption{\label{tab:status} Present status of higher-order computations of weak-scale thresholds.
With the results of this chapter the calculation of the SM parameters at NNLO precision is complete. Here we have defined $V \equiv (\sqrt{2} G_\mu)^{-1/2}$ and $g_1=\sqrt{5/3} g_Y$.
}
\end{center}
\end{table}%

\section{Computing the  $\MS$ parameters with two-loop accuracy}\label{Vacuumstability/MSparameters}
\label{sec:strategy}

In this section we first outline the general strategy followed to determine the 
 $\MS$ parameters in terms of physical observables at the
two-loop level. Then, in  sections~\ref{sec2.1} and \ref{sec2.2}, 
we will discuss the results for the quartic and quadratic Higgs couplings, respectively, while
section~\ref{sec2.3} is dedicated to the calculation of two-loop threshold 
corrections to the top Yukawa coupling, and the corrections to the electroweak gauge couplings are presented in section~\ref{sec2.4}.

\subsection{Gauge invariance}
First of all, {all $\MS$ parameters have gauge-invariant
  renormalization group equations~\cite{Caswell:1974cj} (see also \cite{Muta:1987mz}) and are gauge invariant},
as we now prove.\footnote{Gauge invariance of fermion pole masses has
  been proved in
  refs.~\cite{Atkinson:1979ut,Breckenridge:1994gs,Kronfeld:1998di} and here we
  generalize their proof.}  Let us consider a generic $\MS$ coupling
$\theta$ measuring the strength of a gauge invariant term in the
Lagrangian and a generic gauge fixing parametrized by $\xi$ (for
example the $R_\xi$ gauges). The definition of
$\theta$ in terms of the bare coupling $\theta_0$ is 
\begin{equation}
\mub^{d-4}\theta_0=\sum_{k=0}^{\infty}\frac{c_k(\theta,\xi)}{(d-4)^k}, 
\label{lambda-definition}
\end{equation}
where the $c_k$ are defined to be the residues at the divergence $d=4$. 
The important point is that $c_0 = \theta$, with no dependence on $\xi$.
Since $\theta_0$ is gauge independent, we have
\begin{equation}
0= \mub^{d-4}\frac{d \theta_0}{d\xi}=\frac{d \theta}{d\xi}
+\sum_{k=1}^{\infty}\frac{1}{(d-4)^k}\frac{dc_k(\theta,\xi)}{d\xi}.
\end{equation}
Since this equation is valid for any $d$, and $\theta$ has no poles at $d = 4$ by definition, we obtain
$d\theta/d\xi=0$, namely that is $\theta$ is gauge invariant (as well as all the residues $c_k$).\footnote{Notice
that this proof does not apply to the Higgs vev $v$, because it is
not the coefficient of a gauge-invariant term in the Lagrangian.}

\subsection{Renormalization}

To determine the $\MS$ parameters in terms of physical 
observables two strategies can be envisaged:
\begin{itemize}
\item[$(i)$] Perform an  
$\MS$ renormalization to obtain directly the
  $\MS$ quantity of interest
 in terms of  $\MS$ parameters. Then  express the
$\MS$ parameters in terms of the physical ones via appropriately 
derived two-loop relations. 
\item[$(ii)$] Use a renormalization scheme in which the
renormalized parameters   are directly expressed in terms of
 physical observables (we call this scheme generically on-shell
(OS) and label quantities in this scheme with an \os). Then  relate the 
parameters  as expressed in the OS 
scheme to their $\MS$ counterparts we are looking for. 
\end{itemize}
This last step can be easily done using the relation  
\begin{equation}
\theta_0 = \theta_\os - \delta \theta_\os = 
\theta(\mub)-{\delta  \theta}_{\MS}
\label{eq:g1} 
\end{equation}
or 
\begin{equation}
\theta(\mub) = \theta_\os - \delta \theta_\os + 
{\delta  \theta}_{ \MS}\, ,
\label{eq:g2} 
\end{equation}
where $\theta_0$ is the bare parameter, $\theta (\mub)$ ($\theta_\os$) is
the renormalized $ \MS$ (OS) version and 
$\delta  \theta_{\MS }$ ($\delta  \theta_{\os }$) the
corresponding counterterm. By definition $\delta  \theta_{\MS }$
subtracts only the terms proportional to powers of 
$1/\epsilon$ and $\gamma -\ln (4 \pi)$ in dimensional regularization, 
with $d= 4- 2\, \epsilon$ being the space-time dimension. Concerning  the structure of the $1/\epsilon$ poles
in the  OS and $ \MS$ counterterms, one  notices that it should 
be   identical once the poles in the OS counterterms are expressed in terms of 
$\MS$ quantities. Then, after this
operation is performed, the desired $\theta(\mub)$ is obtained from
\begin{equation}
\theta(\mub) = \theta_\os - \left. \delta \theta_\os \right|_{\rm fin} +
\Delta_\theta,
\label{eq:g3} 
\end{equation}
where the subscript `fin' denotes the finite part of the quantity involved 
and $\Delta$ is the two-loop finite contribution that is obtained
when the OS parameters entering the  1/$\epsilon$ pole in the OS counterterm 
are  expressed in terms of $\MS$ quantities -- the finite
contribution coming from the ${\Ord}(\epsilon)$ part of the shifts.

\begin{table}[t]
\begin{center}
$$\begin{array}{rclll}
\hline
M_W &=& 80.384\pm0.014\,\GeV & \hbox{Pole mass of the $W$ boson} & \hbox{\cite{Group:2012gb,Alcaraz:2006mx}}\\
M_Z &=& 91.1876\pm0.0021\,\GeV & \hbox{Pole mass of the $Z$ boson} & \hbox{\cite{Beringer:1900zz}}\\
M_h &=& \Mhexperr & \hbox{Pole mass of the Higgs boson}\!\!\!\! & \hbox{\cite{Giardino:2013bma,CMS:ril,CMS:xwa,ATLAS:2013oma,ATLAS:2013nma,ATLAS:2013mma}}\\
M_t &=& 173.10\pm0.59\pm0.3\,\GeV & \hbox{Pole mass of the top quark} & \hbox{\cite{Lancaster:2011wr,CMS:2012fya,CMS:2012awa,ATLAS:2012coa,ATLAS:2013zzh,ATLAS:2013zzi}}\\
(\sqrt{2} G_\mu)^{-1/2}\!\!\! &=&246.21971\pm0.00006\GeV\!\!& \hbox{Fermi constant for $\mu$ decay} & \hbox{\cite{Tishchenko:2012ie}}\\
\alpha_3(M_Z) &=& 0.1184\pm 0.0007& \hbox{$\MS$ gauge $\SU(3)_c$ coupling} & \hbox {\cite{Bethke:2012jm}}\\
\hline
\end{array}$$
\caption{\label{tab:values} Input values of the SM observables used to fix the SM fundamental parameters
$\lambda, m, y_t, g_2, g_Y$.  The pole top mass $M_t$ is a na\"ive average of
TeVatron, CMS, ATLAS measurements, all extracted from difficult Monte Carlo modellings of top decay and production
in hadronic collisions.  Furthermore, $M_t$
is also affected by a non-perturbative theoretical uncertainty of order $\Lambda_{\rm QCD}$,
that we quantify as $\pm0.3\GeV$.}
\end{center}
\end{table}%

In the following we adopt strategy $(ii)$. 
The quantities of interests
$\theta = (m^2, \lambda, v, y_t, g_2, g_1)$ -- i.e. the quadratic and
quartic couplings in the Higgs potential, the vacuum expectation value,
the top Yukawa coupling, the $\SU(2)_L$ and $\U(1)_Y$ gauge couplings
$g_2$ and $g_Y$
(with $g_1 = \sqrt{5/3}\, g_Y$ being the hypercharge coupling rewritten in 
$\SU(5)$ normalization) -- are directly 
determined in terms of the pole masses\footnote{We will indicate throughout this chapter physical measurable quantities, such as pole masses, with capital letters, and any bare quantity with a subscript $0$.}
 of the Higgs ($M_h$),
of the top ($M_t$), of the $Z$ ($M_Z$), of the $W$ ($M_W$),  the Fermi
constant $G_\mu$  and
the $\MS$ strong coupling $\alpha_3(M_Z)$.
Their input values are listed in table~\ref{tab:values}.
Then, using eq.~(\ref{eq:g3}), the $\MS$ quantities are obtained. We notice
that the weak-scale values for the 
$\MS$ gauge couplings at the scale $\mub$ are given in terms
of $G_\mu,\, M_W$ and $M_Z$ 
and not in terms of the fine structure constant and the weak mixing angle
at the $M_Z$ scale as usually done.

The classical Higgs doublet $H_0$ is defined by eq.~\eqref{Higgsfield}
in terms of the physical Higgs field $h$, and of the neutral
and charged would-be Goldstone bosons $\eta$ and $\chi$.  The
renormalization of the Higgs potential \eqref{higgspot}, was
discussed at the one-loop level in~\cite{Sirlin:1985ux} and extended at the
two-loop level in~\cite{Degrassi:2012ry}. We refer to these papers for
details. We recall that in ref.~\cite{Degrassi:2012ry} the renormalized vacuum
is identified with the minimum of the radiatively corrected
potential\footnote{This condition is enforced choosing the tadpole
  counterterm to cancel completely the tadpole graphs.}  and it is
defined through $G_\mu$. Writing the relation between the Fermi
constant and the bare vacuum as 
\begin{equation}
\frac{G_\mu}{\sqrt2}=
\frac{1}{2v_0^2}(1+ \Delta r_\ss0)\, ,
\label{Deltar0} 
\end{equation}
one gets
\begin{equation}
v_\os^2= \frac1{\sqrt2 G_\mu}, \qquad 
\delta v^2_\os = -\frac{\Delta r_\ss0}{\sqrt2 G_\mu} . 
\label{deltav2}
\end{equation}
The quadratic and quartic couplings in the Higgs potential in the on-shell renormalization scheme are defined through
$M_h$ via
\begin{equation}
\lambda_{\os}=\frac{G_\mu}{\sqrt2}M_h^2 , \qquad m_\os^2 = M_h^2 .
\end{equation}
Writing the counterterm for the quartic Higgs coupling as
\begin{equation}
\delta \lambda_\os = \delta^{(1)} \lambda_\os + \delta^{(2)} \lambda_\os\ ,
\end{equation}
where the superscript indicates the loop order, one finds
\bea 
\delta^{(1)} \lambda_\os & = & \frac{G_\mu}{\sqrt2} M_h^2 
\left\{\Delta r_\ss0^{(1)}+ 
\frac{1}{M_h^2} \left[\frac{T^{(1)}}{v_\os}+ \delta^{(1)} M_h^2 
\right]\right\}\, ,
\label{eq:lh1}\\
\delta^{(2)} \lambda_\os &=& \frac{G_\mu}{\sqrt2} M_h^2 \left\{\Delta r_\ss0^{(2)}+ 
\frac{1}{M_h^2} \left[\frac{T^{(2)}}{v_\os}
+\delta^{(2)} M_h^2 \right]+ \right. \nonumber \\ 
&& \left.-  \Delta r_\ss0^{(1)} 
\left(\Delta r_\ss0^{(1)}+ \frac{1}{M_h^2}\left[\frac{3\, T^{(1)}}{2\,v_\os}
+\delta^{(1)} M_h^2  \right]\right)\right\} \, .
\label{eq:lh2}
\eea 
In (\ref{eq:lh1}), (\ref{eq:lh2}) $i T$ represents the sum of the tadpole 
diagrams with external leg extracted, and  $ \delta M_a^2$ labels
the mass counterterm for the particle $a$. 

Similarly, one finds for the counterterm of the quadratic Higgs coupling in the
potential
\bea 
\delta^{(1)} m_\os^2 & = & 
 3\frac{T^{(1)}}{v_\os}+ \delta^{(1)} M_h^2  \, ,
\label{eq:mh1}\\
\delta^{(2)} m_\os^2 &=& 
3 \frac{T^{(2)}}{v_\os}
+ \delta^{(2)} M_h^2  - \frac{3\, T^{(1)}}{2\,v_\os}\Delta r_\ss0^{(1)}\, .
\label{eq:mh2}
\eea 
The top Yukawa and gauge  couplings are fixed using $M_t$, $M_W$ and $M_Z$ via
\begin{equation}
M_t = \frac{y_{t_{\os}}}{\sqrt2} v_\os, \qquad M_W^2 = \frac{g^2_{2_{\os}}}4 v^2_\os, \qquad
 M_Z^2 = \frac{g^2_{2_{\os}} +g^2_{Y_{\os}}}4 v_\os^2    ,
\end{equation}
or 
\begin{equation}
y_{t_{\os}} = 2 \left( \frac{G_\mu}{\sqrt2} M_t^2 \right)^{1/2}   , ~~
g_{2_{\os}} = 2 \left( \sqrt2\,G_\mu \right)^{1/2}   M_W, ~~
g_{Y_{\os}} = 2 \left( \sqrt2\,G_\mu \right)^{1/2}  \sqrt{M_Z^2 -M^2_W}~.
\label{eq:g5}
\end{equation}
The corresponding counterterms are found to be
\bea 
\delta^{(1)} y_{t_{\os}} &=& 2 \left( \frac{G_\mu}{\sqrt2} M_t^2 \right)^{1/2} 
        \left( \frac{\delta^{(1)} M_t}{ M_t} +    \frac{\Delta r_\ss0^{(1)}}2
 \right) ,
\label{eq:yt1} \\
\delta^{(2)} y_{t_{\os}} &=& 2 \left( \frac{G_\mu}{\sqrt2}  M_t^2 \right)^{1/2}  
        \left( \frac{\delta^{(2)} M_t}{  M_t } +  \frac{\Delta r_\ss0^{(2)}}2 -
 \frac{\Delta r_\ss0^{(1)}}2 \left[ \frac{\delta^{(1)} M_t}{ M_t} + 
   \frac{3\,\Delta r_\ss0^{(1)}}4 \right] \right)  ,
\label{eq:yt2}
\eea
for the top Yukawa coupling,
\bea
\delta^{(1)} g_{2_{\os}} & = &  \left( \sqrt2\,G_\mu \right)^{1/2} M_W \left(
\frac{\delta^{(1)} M_W^2}{M_W^2} +   \Delta r_\ss0^{(1)} \right),
\label{eq:G2} \\
\delta^{(2)} g_{2_{\os}} & = &  \left( \sqrt2\,G_\mu \right)^{1/2} M_W \left(
\frac{\delta^{(2)} M_W^2}{M_W^2} +   \Delta r_\ss0^{(2)} \right. \nonumber\\
&&-\left.\frac{\Delta r_\ss0^{(1)}}{2}\left[\frac{\delta^{(1)} M_W^2}{M_W^2} +   \frac{3\Delta r_\ss0^{(1)}}{2}\right]+\frac{1}{4}\left(\frac{\delta^{(1)} M_W^2}{M_W^2}\right)^{2}\right),
\label{eq:G22} 
\eea
for the $\SU(2)_L$ gauge coupling, and
\bea
\delta^{(1)} g_{Y_{\os}} &=&  \left( \sqrt2\,G_\mu \right)^{1/2}  
\sqrt{M_Z^2 -M^2_W}~ \left( 
\frac{\delta^{(1)} M_Z^2 - \delta^{(1)} M_W^2}{M_Z^2 -M^2_W} +  \Delta r_\ss0^{(1)}
\right),
\label{eq:G1}\\
\delta^{(2)} g_{Y_{\os}} &=&  \left( \sqrt2\,G_\mu \right)^{1/2}  
\sqrt{M_Z^2 -M^2_W}~ \left[ 
\frac{\delta^{(2)} M_Z^2 - \delta^{(2)} M_W^2}{M_Z^2 -M^2_W} +  \Delta r_\ss0^{(2)}\right. \nonumber\\
&&\left.-\frac{\Delta r_\ss0^{(1)}}{2}\!\left(\frac{\delta^{(1)} M_Z^2 - \delta^{(1)} M_W^2}{M_Z^2 -M^2_W} +  \frac{3\Delta r_\ss0^{(1)}}{2}\right)\!+\frac{1}{4}\left(\frac{\delta^{(1)} M_Z^2 - \delta^{(1)} M_W^2}{M_Z^2 -M^2_W} \right)^{\!\!2}
\right]~~~~ 
\label{eq:G12}
\eea
for the hypercharge gauge coupling.

\section{Two-loop correction to the Higgs quartic coupling} 
\label{sec2.1}

The $\MS$ Higgs quartic coupling is given by
\begin{equation}
\lambda(\mub) = 
\frac{G_\mu}{\sqrt2}M_h^2  + \lambda^{(1)}(\mub) +  \lambda^{(2)}(\mub),
\label{eq:lambda}
\end{equation}
with
\bea
 \lambda^{(1)}(\mub) &=& - \left. \delta^{(1)} \lambda_\os \right|_{\rm fin} 
\label{eq:lmu1}\, ,\\
 \lambda^{(2)}(\mub) &=& -\left. \delta^{(2)} \lambda_\os \right|_{\rm fin} 
+ \Delta_{\lambda}~. 
\label{eq:lmu2}
\eea
The one-loop contribution in (\ref{eq:lambda}), $\lambda^{(1)}$,
is given by the finite part of equation (\ref{eq:lh1}). 
Concerning the two-loop part, $\lambda^{(2)} (\mub)$, the  QCD corrections 
were presented in refs.~\cite{Bezrukov:2012sa,Degrassi:2012ry},
and the  two-loop electroweak (EW) part, $\lambda_{\rm EW}^{(2)}(\mub)$,
 was computed in 
ref.~\cite{Degrassi:2012ry} in the so-called gauge-less limit of the SM, in which the 
electroweak gauge interactions are switched off.
The main advantage of this limit results in a simplified evaluation
of  $ \Delta r_\ss0^{(2)}$. The computation of the two-loop
EW part in the full SM requires instead the
complete evaluation of this quantity and we outline here the 
derivation of $\lambda^{(2)}_{\rm EW}(\mub)$ starting from the term 
$ \Delta r_\ss0^{(2)}$ in $\delta^{(2)} \lambda_\os$.
 
\subsection{Two-loop correction to the muon decay constant}

We recall that the Fermi constant 
is defined in terms of the muon lifetime $\tau_\mu$ as computed
in the 4-fermion $V-A$ Fermi  theory supplemented by QED interactions. 
We extract $G_\mu$ from  $\tau_\mu$ via
\begin{equation}
\label{eq:taumu}
\frac{1}{\tau_\mu} = \frac{G_\mu^2 m_\mu^5}{192\pi^3} F(\frac{m_e^2}{m_\mu^2}) 
(1 + \Delta q)(1+\frac{3m_\mu^2}{5M_W^2}) \ ,
\end{equation}
where $F(\rho)=1-8\rho+8\rho^3-\rho^4-12\rho^2\ln\rho=0.9981295$ 
(for $\rho=m_e^2/m_\mu^2$) is the phase space factor and
$\Delta q=  \Delta q^{(1)} + \Delta q^{(2)}=(-4.234+0.036)\times10^{-3}$
are the QED corrections computed at one~\cite{Kinoshita:1958ru} and two 
loops~\cite{vanRitbergen:1999fi}.
From the measurement $\tau_\mu = (2196980.3\pm2.2)\,{\rm ps}$~\cite{Tishchenko:2012ie} 
we find
$G_\mu = 1.1663781(6)~10^{-5}/\GeV^2$.
This is $1\sigma$ lower than the value quoted in~\cite{Tishchenko:2012ie} because we do not 
follow the convention of including
in the definition of $G_\mu$ itself the last term of\eq{taumu}, which is the
contribution from dimension-8 SM operators. 

\bigskip

The computation
of $ \Delta r_\ss0$ requires  the subtraction of the QED corrections
by matching the result in the SM with that in the Fermi theory. However,
it is well known that the Fermi theory is renormalizable to all order in the
electromagnetic interaction but to lowest order in $G_\mu$ due to a Ward
identity that becomes manifest if the 4-fermion interaction is rewritten 
via a Fierz transformation in the  ``charge retention order''. 
As a consequence, in the
limit of neglecting the fermion masses,  $ \Delta r_\ss0$ as computed in the
Fermi theory vanishes and we are just left with the calculation in the SM.\footnote{
We explicitly verified that $ \Delta r_\ss0$ vanishes when computed in the Fermi
theory.}

Starting from equation (\ref{Deltar0}) we write $\Delta r_\ss0$ as a sum of
different terms:
\begin{equation}
\Delta r_\ss0= \Vcal_W- \frac{\A_{WW}}{M^2_{W,0}} + 
2\, v_0^2\, {\B}_W + \E + \M\, , 
\label{Deltar-expansion}
\end{equation}
where $M_{W,0}$ is the bare W mass;  $\A_{WW}=\A_{WW}(0)$ is the $W$
self-energy at zero momentum;
$\Vcal_W$ is the vertex contribution; ${\B}_W$ is the box
contribution; $\E$ is the term due to the renormalization of the
external legs; $\M$ is the mixed contribution due to product of
different objects among $\Vcal_W$, $\A_{WW}$, $\B_W$ and
$\E$ (see below for an explicit expression at two-loops). All
quantities in (\ref{Deltar-expansion}) are computed at zero
external momenta. We point out that in the right-hand side of (\ref{Deltar-expansion}) no tadpole
contribution is included because  of our choice of identifying the
renormalized vacuum with the minimum of the radiatively corrected potential. 
As a consequence $\Delta r_\ss0$ is a gauge-dependent quantity.

From eq.~(\ref{Deltar-expansion}) the one-loop  term is  
given by
\begin{equation} 
\Delta r_\ss0^{(1)}=\Vcal_W^{(1)} - \frac{\A_{WW}^{(1)}}{M^2_{W}}  + 
\frac{\sqrt{2}}{G_\mu} \, {\B}_W^{(1)} + {\E}^{(1)}  \ ,
\label{dr1}
\end{equation}
where  we have used that ${\M}^{(1)}=0$, while at two-loops
\bea 
\Delta r_\ss0^{(2)} &=& \Vcal_W^{(2)}- \frac{\A_{WW}^{(2)}}{M^2_{W}} 
+ \sqrt{2}\, \frac{{\B}^{(2)}_W}{G_\mu} + {\E}^{(2)}+{\M}^{(2)}+\,
 \nonumber \\
&& -\delta^{(1)}M^2_W \,\frac{\A_{WW}^{(1)}}{M^4_{W}} + \frac{\sqrt{2}}{G_\mu}
\, {\B}_W^{(1)}\left(\Vcal_W^{(1)}- \frac{\A_{WW}^{(1)}}{M^2_{W}} 
+ \sqrt{2}\, \frac{{\B}^{(1)}_W}{G_\mu} + {\E}^{(1)}\right).
\label{dr2}
\eea
 Here
\begin{equation}
\delta^{(1)}M^2_W =  \Re \Pi_{WW}(M_W^2)
\end{equation}
with $ \Pi_{WW}(M_W^2)$  the $W$ boson  self-energy  evaluated at external 
momentum  equal to $M_W$, and 
\begin{equation}
{\M}^{(2)} =\frac{\sqrt 2}{G_\mu}{\E}^{(1)}\,{\B}^{(1)}_W
+ \sum_{i< j}{\E}^{(1)}_i{\E}^{(1)}_j+{\E}^{(1)} \Vcal^{(1)}-
\left({\E}^{(1)}+\Vcal^{(1)}\right)\frac{\A_{WW}^{(1)}}{M_W^2}~.
\label{mixedSM}
\end{equation}
The indices $i,j$ in (\ref{mixedSM}) label
the different species in the muon decay: $\mu$, $e$, $\nu_\mu$ and $\nu_e$
with the sum that runs over $i< j$ because the terms with $i=j$ are included in 
$\E^{(2)}$.

We recall that $\Delta r_\ss0$ is an infrared (IR) safe quantity but not 
ultraviolet finite. However,  the $\E$ and $\B_W$ terms
in (\ref{dr1}) and (\ref{dr2}) contain IR-divergent contributions from 
photon diagrams. To separate the UV-divergent terms from the IR ones we 
regulated the latter giving a small  mass to the photon. We then explicitly 
verified the cancellation of all IR divergent contributions.

\subsection{Higgs two-point function and tadpole}

The other proper two-loop contributions  to $\lambda^{(2)}_{\rm EW} (\mub)$
are the two-loop tadpole diagrams and the two-loop Higgs boson 
mass counterterm. The Higgs mass counterterm, not taking into account 
negligible width effects, is given by  
\begin{equation}
\delta M_h^2 = \Re \Pi_{hh}(M_h^2)
\label{eq:mass}
\end{equation}
with $ \Pi_{hh}(M_h^2)$  the Higgs  self-energy  evaluated at external momentum 
equal to $M_h$. The Higgs mass counterterm as defined in eq.~(\ref{eq:mass})
is a gauge-dependent quantity. Yet, 
as proved at the beginning of section~\ref{sec:strategy}, $\lambda(\mub)$ is
a gauge-invariant object.

The diagrams contributing to $\delta^{(2)} \lambda_\os$
were generated using the {\sc Mathematica} package {\sc FeynArts}~\cite{Hahn:2000kx}.
The reduction of the two-loop diagrams to scalar integrals was done using 
the code {\sc Tarcer} \cite{Mertig:1998vk}  that uses the 
Tarasov's algorithm \cite{Tarasov:1997kx} and it is now part of the {\sc
 FeynCalc}~\cite{Mertig:1990an} package. In order to extract the 
$V_W$ and ${\B}_W$ terms in $\Delta r_\ss0$ from the relevant
diagrams we used the  projector presented in ref.~\cite{Awramik:2002vu}.
The two-loop self-energy diagrams at external momenta different from zero 
were reduced to the set of loop-integral basis functions introduced
in ref.~\cite{Martin:2003qz}. The evaluation of the basis functions was 
done numerically using the code {\sc Tsil}~\cite{Martin:2005qm}.

The two loop correction to $\lambda$ is the sum of a QCD term and of an electroweak (EW) term.
The QCD correction $\lambda_{\rm QCD}^{(2)}(\mub)$ is reported as an approximated formula in eq.~(47) of~\cite{Degrassi:2012ry}. 
For simplicity here we present it also in a numerical form:
\begin{equation}
\lambda_{\rm QCD}^{(2)}(\mub=M_{t}) = 
\frac{g_3^{2}}{(4\pi)^4}\bigg[
- 23.89 +0.12 \left(\frac{M_h}{\rm GeV} - 125\right) - 0.64
\left(\frac{M_t}{\rm GeV} - 173\right)\bigg].
\end{equation}
The result for  $\lambda_{\rm EW}^{(2)}(\mub)$  is  too long to be
displayed explicitly. 
Here we present it in a numerical form valid around the measured values of $M_{h}$ and $M_{t}$. Using the inputs in table \ref{tab:values} we find
\begin{equation}
\lambda_{\rm EW}^{(2)}(\mub=M_{t}) = 
\frac{1}{(4\pi)^4}\bigg[
- 9.45 -0.11 \left(\frac{M_h}{\rm GeV} - 125\right) - 0.21
\left(\frac{M_t}{\rm GeV} - 173\right)\bigg].
\label{eq:lh21}
\end{equation}
This numerical expression is accidentally very close to 
the gaugeless limit of the SM presented 
in eq.~(2.45) of~\cite{Degrassi:2012ry}.
Furthermore, as a check of our result, we verified that in the (physically irrelevant) limit $M_h=0$, 
it agrees with an independent computation of $\lambda^{(2)}$  performed
using the known results for the two-loop effective potential in the Landau gauge.

\section{Two-loop correction to the Higgs mass term} 
\label{sec2.2}

The  result for the mass term in the Higgs 
potential can be easily obtained from that on $\lambda(\mub)$. We write
\begin{equation}
m^2(\mub) = M_h^2 + \delta^{(1)} m^{2}(\mub) + \delta^{(2)} m^{2}(\mub),
\label{eq:m2}
\end{equation}
with
\bea
\delta^{(1)} m^{2}(\mub) &=&  - \left. \delta^{(1)} m^2_\os \right|_{\rm fin},\\
\delta^{(2)} m^{2}(\mub) & = & - \left. \delta^{(2)} m^2_\os \right|_{\rm fin} 
+ \Delta_{m^2}~. 
\label{eq:m21}
\eea
The one-loop contribution $\delta^{(1)} m^{2}(\mub)$
is given by the finite part of eq.~(\ref{eq:mh1}). 
The two-loop corrections
$\delta^{(2)} m^{2} (\mub)$ can be divided into a QCD contribution plus an EW contribution.

The QCD contribution,  $\delta^{(2)}_{\rm QCD} m^{2} (\mub)$,
can be obtained evaluating the relevant diagrams via a Taylor series in
$x_{ht} \equiv M_h^2/M_t^2$ up to fourth order
\bea
\delta^{(2)}_{\rm QCD} m^{2}  (\mub) &=& \frac{G_\mu M_t^4}{\sqrt{2}(4 \pi)^4}\, N_c\, C_F
g_3^2 \bigg[  -96 +
(41 -12 \ln^2\frac{M_t^2}{\mub^2} +12 \ln^2\frac{M_t^2}{\mub^2} ) x_{ht}+
\nonumber\\
& & \qquad\qquad\qquad\qquad
+  \frac{122 }{135} x^2_{ht} +
 \frac{1223 }{3150} x_{ht}^3 +  \frac{43123 }{661500} x_{ht}^4\bigg],
\eea
where $N_c$ and $C_F$ are colour factors ($N_c=3,\, C_F= 4/3$), such that it is numerically approximated as
\begin{equation}
\delta^{(2)}_{\rm QCD} m^{2}  (\mub=M_t)  = \frac{g_{3}^{2}M_{h}^{2}}{(4\pi)^4}\bigg[
-140.50 +2.89 \left(\frac{M_h}{\rm GeV} - 125\right) -3.71
\left(\frac{M_t}{\rm GeV} - 173\right)\bigg].
\end{equation}
The two-loop EW part,  $\delta^{(2)}_{\rm EW} m^{2} (\mub)$,
can be obtained as a byproduct of the calculation of 
$\lambda_{\rm EW}^{(2)}(\mub)$. Also in this case the result is too long to
be displayed and we present an interpolating formula. Using the inputs in 
table~\ref{tab:values} we find
\begin{equation}
\delta_{\rm EW}^{(2)} m^2 (\mub=M_{t})= \frac{M_{h}^{2}}{(4\pi)^4}\bigg[
-149.47 +2.53 \left(\frac{M_h}{\rm GeV} - 125\right) - 4.69
\left(\frac{M_t}{\rm GeV} - 173\right)\bigg].
\label{eq:m22}
\end{equation}

\section{Two loop correction to the  top Yukawa coupling}
\label{sec2.3}

The $\MS$ top Yukawa coupling is given by
\begin{equation}
y_t(\mub) = 
2 \left(\frac{G_\mu}{\sqrt2}  M_t^2\right)^{1/2} + y_t^{(1)}(\mub)
+ y_t^{(2)}(\mub),
\label{eq:topYuk}
\end{equation}
with 
\bea
 y_t^{(1)}(\mub) &=&   - \left. \delta^{(1)} y_{t{_\os}} \right|_{\rm fin} ,\\
 y_t^{(2)}(\mub) &=& -  \left. \delta^{(2)} y_{t_{\os}} \right|_{\rm fin} + 
\Delta_{y_t}~.
\eea
According to (\ref{eq:yt1}), (\ref{eq:yt2}) the corrections to the tree-level
value of $y_t$ are given in terms of $ \Delta r_\ss0$ and the top mass
counterterm. Regarding the latter, a general discussion  on the mass 
counterterm for unstable  fermions in parity-nonconserving theories is 
presented in  ref.~\cite{Kniehl:2008cj}. Writing the fermion self-energy as
\bea
\Sigma(p) & =& \Sigma_1 (p) + \Sigma_2(p) \gamma_5,\\
\Sigma_{1,2} (p) &=& \slashed{p} B_{1,2} (p^2) + m_0 A_{1,2} (p^2) ,
\label{KS}
\eea 
the fermion propagator is given by
\begin{equation}
i S(p) = \frac{i}{\slashed{p} - m_0 - \Sigma(p)} = 
\frac{i}{\slashed{p} - m_0 - \Sigma_{\rm eff} (p)} 
\bigg[1 - \frac{\Sigma_2 (p)}{\slashed{p} -\Sigma_1(p) + m_0 [1 + 2 A_1(p^2)]} \gamma_5
\bigg],
\label{eq:prop}
\end{equation}
where $m_0$ is the bare fermion mass and
\begin{equation}
\Sigma_{\rm eff} (p) = \Sigma_1 (p) +
\frac{ \Sigma_2 (p) \left[ \Sigma_2 (p) -2 m_0 A_2 (p^2)\right]}{\slashed{p}
          -\Sigma_1(p) + m_0[1 + 2 A_1(p^2)]}~.
\end{equation}
Identifying the position $\slashed{p} = \tilde M$ of the complex pole in eq.~(\ref{eq:prop})
by
\begin{equation}
\tilde M = m_0 + \Sigma_{\rm eff}(\tilde M)
\end{equation}
and parametrizing $\tilde M  = M -i \Gamma/2$ with $M$ the pole mass of
the unstable fermion and $\Gamma$ its width,
the mass counterterm for the unstable fermion is found to be
\begin{equation}
\delta M   =  \Re \Sigma_{\rm eff} (\tilde M)~.
\end{equation}
Specializing the above discussion to the top we find, including up to
two-loop contributions,
\begin{equation}
\delta M_t = \Re \left[\Sigma_1 (\tilde{M}_t) +
\frac{ \Sigma_2 (M_t) \left[ \Sigma_2 (M_t) -2 M_t A_2 (M_t^2)\right]}{2 M_t}
\right],
\label{eq:Mt}
\end{equation}
with $\tilde{M}_t = M_t -i \Gamma_t/2$.
The mass counterterm defined in
eq.~(\ref{eq:Mt})  is expressed in terms of the self-energy diagrams only,
without including the tadpole  contribution. While this definition follows 
from our choice of identifying the
renormalized vacuum with the minimum of the radiatively corrected potential, it
gives rise to a $\delta M_t$ that is gauge-dependent and, as a consequence,
in this framework, the $\MS$ top mass, $M_t(\mub)$, is a 
gauge-dependent  quantity. However, a $\MS$ mass is not a 
physical quantity nor a Lagrangian parameter
and  therefore the requirement of gauge-invariance is not 
mandatory. A  gauge-invariant definition of $M_t(\mub)$ can be obtained by 
including the tadpole contribution in the mass 
counterterm~\cite{Hempfling:1994ar}.
However, with this choice the relation between the pole and $\MS$
masses of top quark acquires a very large electroweak 
correction~\cite{Jegerlehner:2012kn}.
The top Yukawa coupling computed in this paper is a parameter of the Lagrangian,
and thereby does not suffer of these problems.

Concerning the two-loop contributions in eq.~(\ref{eq:topYuk}), we have computed
the QCD corrections to the one-loop term and the two-loop EW contribution.

These contributions are too long to
be displayed explicitly, and we report them as interpolating formul\ae.
Using the inputs in  table \ref{tab:values} we find
\bea\nonumber
y^{(2)}_{t}(\mub=M_{t}) &=& \frac{1}{(4\pi)^4}\bigg[
5.22 -0.01 \left(\frac{M_h}{\rm GeV} - 125\right) + 0.15 
\left(\frac{M_t}{\rm GeV} - 173\right)\bigg]+\\
 &&+ \frac{g_{3}^{2}}{(4\pi)^4}\bigg[
-7.53 +0.09\left(\frac{M_h}{\rm GeV} - 125\right) - 0.23 
\left(\frac{M_t}{\rm GeV} - 173\right)\bigg]+\nonumber  \\
 &&+ \frac{g_{3}^{4}}{(4\pi)^4}\bigg[
-145.08 - 0.84 
\left(\frac{M_t}{\rm GeV} - 173\right)\bigg],
\eea
where the last term is the well known pure QCD contribution; the second term is the mixed QCD/EW contribution that agrees with~\cite{Bezrukov:2012sa};
 the first term is the pure EW contribution computed in this paper for the first time.
 
\section{Two-loop correction to the weak gauge couplings} 
\label{sec2.4}
The $g_{2}$ and $g_{Y}$ gauge couplings are given by 

\begin{align}
g_{2}(\mub) &= 2 (\sqrt{2}G_{\mu})^{1/2} M_{W}+ g_{2}^{(1)}(\mub) + g_{2}^{(2)}(\mub),\nonumber \\
g_{Y}(\mub) &= 2 (\sqrt{2}G_{\mu})^{1/2} \sqrt{M_{Z}^{2}-M_{W}^{2}}+ g_{Y}^{(1)}(\mub) + g_{Y}^{(2)}(\mub),
\label{eq:g12}
\end{align}
with
\begin{align}
g_2^{(1)}(\mub) &=   - \left. \delta^{(1)} g_{2{_\os}} \right|_{\rm fin}\ , &
g_2^{(2)}(\mub) = -  \left. \delta^{(2)} g_{2_{\os}} \right|_{\rm fin} + \Delta_{g_2}\ , \nonumber \\
g_Y^{(1)}(\mub) &=   - \left. \delta^{(1)} g_{Y{_\os}} \right|_{\rm fin}\ , &
g_Y^{(2)}(\mub) = -  \left. \delta^{(2)} g_{Y_{\os}} \right|_{\rm fin} + \Delta_{g_y}\ .
\label{eq:deltag12}
\end{align}
The one-loop contributions in (\ref{eq:deltag12}) are given by the finite parts of (\ref{eq:G2})--(\ref{eq:G1}). 
Also in this case the results of the two-loop corrections are too long to be displayed and we present them with interpolating formulas. Using the inputs in 
table~\ref{tab:values} we find
\begin{align}
g^{(2)}_{2} (\mub=M_{t}) &= \frac{1}{(4\pi)^4}\bigg[
5.83 +0.01 \left(\frac{M_h}{\rm GeV} - 125\right) +0.01
\left(\frac{M_t}{\rm GeV} - 173\right)\bigg]+ \nonumber\\
&+\frac{g_{3}^{2}}{(4\pi)^4}\bigg[
7.98 +0.01
\left(\frac{M_t}{\rm GeV} - 173\right)\bigg].
\label{eq:g2MS}
\end{align}
and
\begin{align}\nonumber
g^{(2)}_{Y} (\mub=M_{t})&= \frac{1}{(4\pi)^4}\bigg[
-12.58 - 0.01 \left(\frac{M_h}{\rm GeV} - 125\right) - 0.11
\left(\frac{M_t}{\rm GeV} - 173\right)\bigg]+ \\
&+\frac{g_{3}^{2}}{(4\pi)^4}\bigg[
-23.12 -0.14
\left(\frac{M_t}{\rm GeV} - 173\right)\bigg]
\label{eq:g2MS}
\end{align}

\section{SM couplings at the electroweak scale}\label{sec:inputs}

In this section we give practical results for the SM  parameters 
$\theta=\{\lambda, m^2, y_t, g_2, g_Y\}$
computed in terms of the observables $M_h,M_t,M_W,M_Z, G_\mu$ and $\alpha_3(M_Z)$, whose measured values are listed in table~\ref{tab:values}.
Each $\MS$ parameter $\theta$ is expanded in loops as
\begin{equation}
\theta = \theta^{(0)}+\theta^{(1)}+\theta^{(2)}+\cdots\label{eq:exp}
\end{equation}
where
\begin{enumerate}
\item the tree-level values $\theta^{(0)}$ are listed in table~\ref{tab:status};
\item the one-loop corrections $\theta^{(1)}$ are analytically given in appendix~\ref{app:1loop};
\item the two-loop corrections $\theta^{(2)}$ are computed in section~\ref{sec:strategy}.
\end{enumerate}
After combining these corrections, we give in the following the numerical values
for the SM parameters renormalized at the top pole mass $M_t$ in the
$\MS$ scheme.
Table~\ref{tab:123} reports their central values extrapolated at the renormalization scale $\mub=M_t$.

\begin{table}[t]
\renewcommand{\arraystretch}{1.0}
 \begin{center}
\begin{tabular}{cccccc}
 $\mub=M_t$ & $\lambda$  & $y_t$ & $g_2$ & $g_Y$ & $m_h/\GeV$ \\ \hline
 LO & 0.13023 & 0.99425 & 0.65294 & 0.34972 & 125.66 \\
 NLO & 0.12879 & 0.94953 & 0.64755 & 0.35937 & 132.80 \\
 NNLO & 0.12711 & 0.93849 &  0.64822&  0.35761& 132.03 \\
\end{tabular}
\caption{Values of the  fundamental SM parameters computed at tree level, one loop, two loops
in the $\MS$ scheme and renormalized at $\mub=M_t$ for the central values of the measurements listed in table~\ref{tab:values}.
\label{tab:123}}
\end{center}
\end{table}
 
 \subsection{The Higgs quartic coupling}

For the Higgs quartic coupling, defined by writing the SM potential as in \eqref{higgspot},
we find
\begin{equation}
\lambda(\mub=M_t) = 0.12711+0.00206\left( \frac{M_h}{\GeV}-\Mhexp \right)-0.00004 \left( \frac{M_t}{\GeV}-\Mtexp \right)\pm
{0.00030}_{\rm th}~.
\end{equation} 
The dependence on $M_t$ is  small because $\lambda$ is renormalized at $M_t$ itself.
Here and below the theoretical uncertainty is estimated from the dependence on $\bar\mu$ 
(varied around $M_t$ by one order of magnitude) of the higher-order unknown 3 loop corrections.
Such dependence is extracted from the known SM RGE at 3 loops (as summarized in appendix~\ref{app:SM-RGE}).

\subsection{The Higgs mass term}

For the mass term of the Higgs doublet in the SM Lagrangian (normalized such that $m_h=M_h$ at tree level)  we find 
 \begin{equation}
 \frac{m_h(\mub=M_t)}{\GeV}=132.03+0.94 \left( \frac{M_h}{\GeV}-\Mhexp \right)+0.17\left( \frac{M_t}{\GeV}-\Mtexp \right) \pm 0.15_{\rm th}.
 \end{equation}
 
\subsection{The top Yukawa coupling}

For the top Yukawa coupling we get
\begin{equation}
y_t(\mub=M_t) = 0.93697 +0.00550 \left( \frac{M_t}{\GeV}-\Mtexp \right)
 -0.00042 \asdiff \pm{0.00050}_{\rm th}~. 
\label{eq:ht_ew}
\end{equation}
The central value differs from the NNLO value in table~\ref{tab:123} because we include here also the NNNLO (3 loop)
pure QCD effect~\cite{Chetyrkin:1999ys,Chetyrkin:1999qi,Melnikov:2000qh}. The theoretical uncertainty is estimated accordingly,
and does not take into account the non-perturbative theoretical uncertainty of order $\Lambda_{\rm QCD}$ in the definition of $M_t$.

\subsection{The weak gauge couplings}

For the weak gauge couplings $g_2$ and $g_Y$ computed at NNLO accuracy in terms of $M_W$ and $M_Z$ we find
\begin{align}
g_2(\bar\mu=M_t) &= 0.64822 +0.00004 \left(\! \frac{M_t}{\GeV}-\Mtexp \right)+ 0.00011 \MWdiff
, \\
g_Y(\bar\mu=M_t) &= 0.35761 +0.00011 \left(\! \frac{M_t}{\GeV}-\Mtexp \right)- 0.00021 \MWdiff,
\end{align}
where the adopted value for $M_W$ and its experimental error are reported in table~\ref{tab:values}.

\subsection{The strong gauge coupling}

Table~\ref{tab:values} contains the value of $\alpha_3(M_Z)$, as extracted from the global fit of~\cite{Bethke:2012jm} in the effective SM with 5 flavours.
Including RG running from $M_Z$ to $M_t$
at 4 loops in QCD and at 2 loops in the electroweak gauge interactions,
and 3 loop QCD matching at $M_t$ to the full SM with 6 flavours, we get
\begin{equation}
g_3(\mub=M_t)  =   
1.1666 
+0.00314\asdiff -0.00046 \left( \frac{M_t}{\GeV}-\Mtexp \right).
\end{equation}

The SM parameters can be renormalized to any other desired energy by solving the SM renormalization group equations summarized in appendix~\ref{app:SM-RGE}.
For completeness, we  include in the one- and two-loop
RG equations the contributions of the small bottom and tau Yukawa
couplings, as computed from the $\MS$ $b$-quark mass,
$m_b(m_b)=4.2\GeV$, and from $M_\tau=1.777\GeV$.  Within the $\MS$
scheme $\beta$ functions are gauge-independent~\cite{Caswell:1974cj}; similarly
the $\MS$ parameters are gauge independent too.

\chapter{Stability of the electroweak vacuum}\label{Vacuumstability}

A very puzzling and intriguing outcome of the Higgs discovery has
been the finding that $M_h$ lies very close to the boundary between
stability and metastability regions.
Near-criticality of the SM vacuum may be the most
important message we have learnt so far from experimental data on the
Higgs boson, if no new physics is found at the Fermi scale. In that case, near-criticality gives us a unique opportunity to obtain
information about physics taking place at energy scales well beyond
the reach of any collider experiment.

This fact is the main
motivation for a refined NNLO calculation of the SM Higgs potential
at large field values and of the critical value of the Higgs mass $M_h$ required for absolute vacuum stability. Indeed, the special Higgs mass found by ATLAS
and CMS is so close to criticality that any statement about stability
or metastability of the EW vacuum requires a careful analysis of
theoretical and experimental errors.

First, in section~\ref{sec:Planck}, we extrapolate the SM couplings up to the Planck scale using the results of the previous chapter. We then analyze in section~\ref{phasediagram} the phase diagram of the EW vacuum in terms of different SM parameters, both as evaluated at the weak scale and at the Planck scale, showing that in many cases the various masses and couplings lie very close to the boundary between stability and instability.

The discovered proximity to
criticality also naturally stimulates many theoretical speculations on
its possible hidden significance or on special matching conditions at
very high energy scales. In the last part of the chapter we will explore different possible
implications of our improved computation of the Higgs quartic coupling
extrapolated to very high scales.

\section{Extrapolation of the SM up to the Planck scale}
\label{sec:Planck}

\begin{figure}[t]
\centering%
\includegraphics[width=0.63\textwidth]{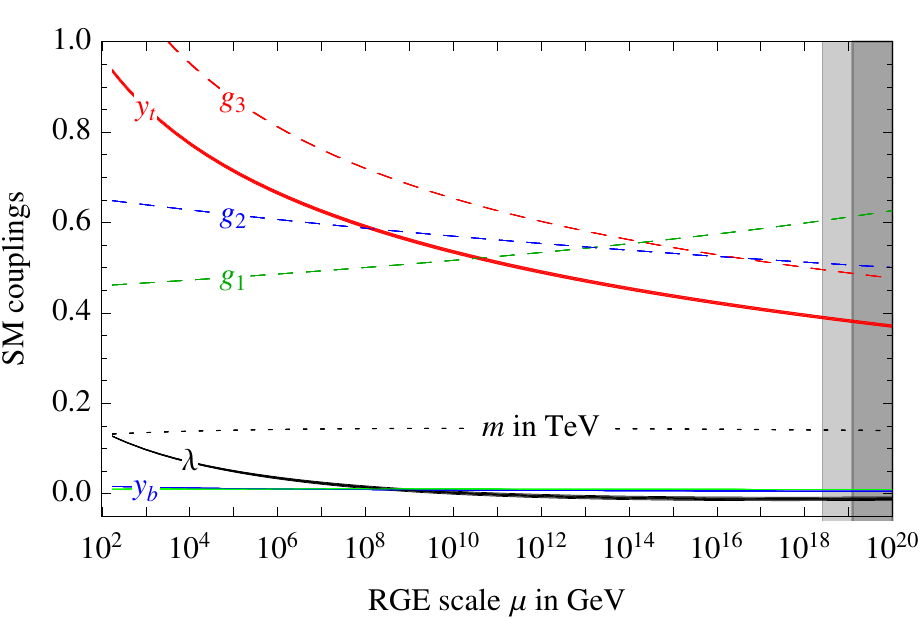}
\caption{
Renormalization of the SM  gauge couplings 
$g_1=\sqrt{5/3}g_Y, g_2, ~g_3$, of the top, bottom and $\tau$ couplings ($y_t$, $y_b$, $y_\tau$),
 of the Higgs quartic coupling $\lambda$ and of the Higgs mass parameter $m$.
All parameters are defined in the  $\MS$ scheme.
We include two-loop thresholds at the weak scale and three-loop RG equations.
The thickness indicates the $\pm1\sigma$ uncertainties in $M_t,M_h,\alpha_3$.}
\label{fig:run1} 
\end{figure}

\subsection{SM couplings at the Planck scale}\label{sec:scc}

The first issue we want to address concerns the size of the SM
coupling constants. When we try to extract information from the values
of the coupling constants, it is reasonable to analyse their values
not at the weak scale, but at some high-energy scale where we believe
the SM matches onto some extended theory.  So, using our NNLO results, we extrapolate the SM couplings
from their weak-scale values (as determined in section~\ref{sec:inputs}) to higher
energies.  

The evolution of the SM couplings up to a large cut-off scale is shown in figure~\ref{fig:run1}. 
 At the Planck mass, we find the following values of the SM parameters:
\begin{align}
g_1(\mpl )&= 0.6133+0.0003 \Mtdiff -0.0006 \MWdiff \\
g_2(\mpl )&= 0.5057 \\
g_3(\mpl)&= 0.4873+ 0.0002 \asdiff \\
y_t(\mpl )&= 0.3813+ 0.0051\Mtdiff - 0.0021\asdiff  \\
\label{eq:lammp}
\lambda (\mpl ) &=  -0.0113- 0.0065\Mtdiff +\\   \nonumber
&+0.0018\asdiff +0.0029\Mhdiff \\
m(\mpl) &= 140.3\GeV +1.6\GeV\Mhdiff+
\\&-0.25\GeV\Mtdiff+0.05\GeV\asdiff\nonumber
\end{align}
All Yukawa couplings, other than the one of the top quark,
  are very small. This is the well-known flavour problem of the SM,
  which we discussed in chapter~\ref{Flavour}.

The three gauge couplings and the top Yukawa coupling remain perturbative and 
  are fairly weak at high energy, becoming roughly equal in the
  vicinity of the Planck mass. The near equality of the gauge
  couplings may be viewed as an indicator of an underlying grand
  unification even within the simple SM, once we allow for threshold
  corrections of the order of 10\% around a scale of about
  $10^{16}$~GeV (of course disregarding the acute naturalness problem, as we are doing throughout this chapter). It is amusing to note
  that the ordering of the coupling constants at low energy is
  completely overturned at high energy. The (properly normalized)
  hypercharge coupling $g_1$ becomes the largest coupling in the SM
  already at scales of about $10^{14}$~GeV, and the weak coupling
  $g_2$ overcomes the strong coupling at about $10^{16}$~GeV. The top
  Yukawa becomes smaller than any of the gauge couplings at scales
  larger than about $10^{10}$~GeV. 
  
\begin{figure}[t]
\centering%
\includegraphics[width=0.45\textwidth]{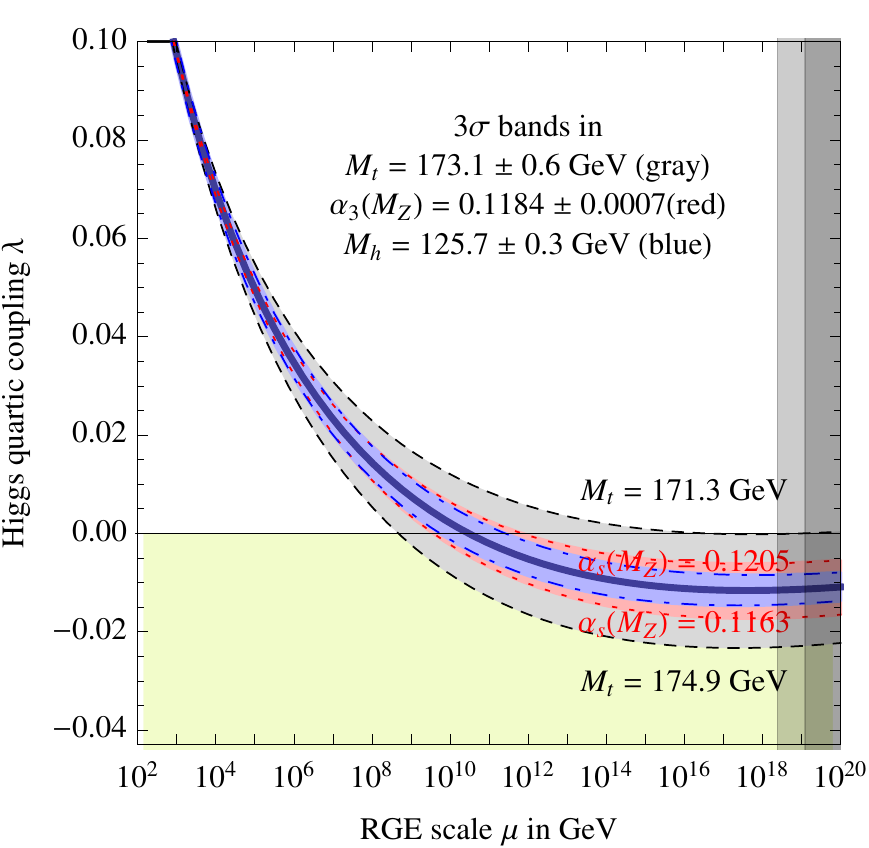}\qquad
\includegraphics[width=0.46\textwidth]{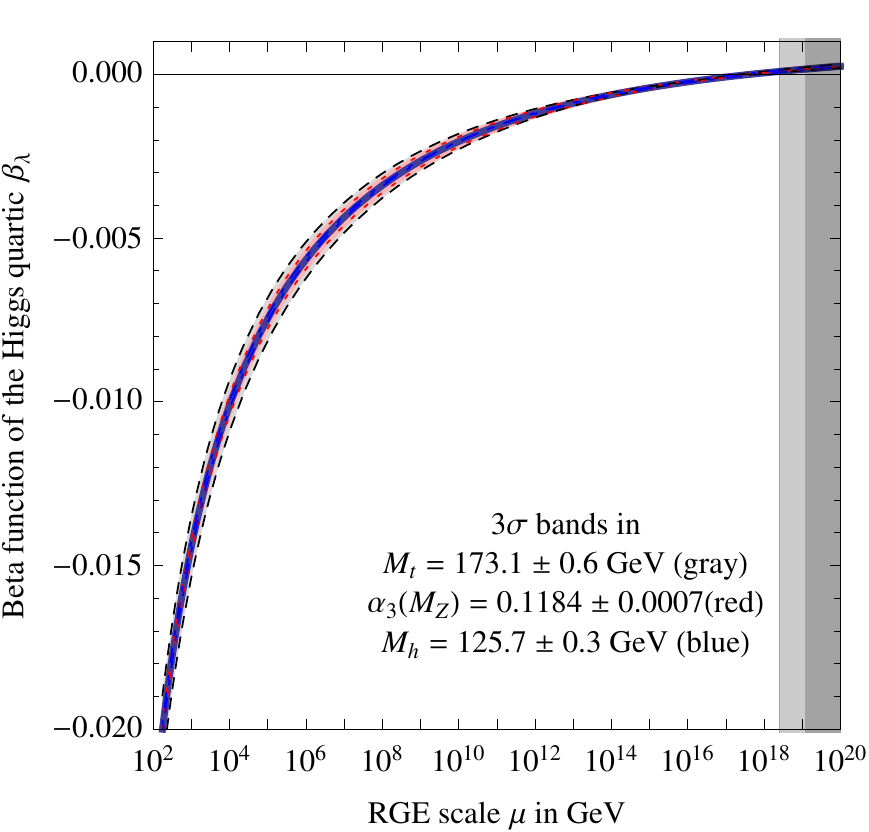}
\caption{RG evolution of $\lambda$ (left) and of $\beta_\lambda$ (right) varying 
$M_t$ (grey), $\alpha_3(M_Z)$ (red) and $M_h$ (blue) by $\pm 3\sigma$. The grey shadings cover values of the RG scale above the Planck mass 
$\mpl \approx 1.2\times 10^{19}\GeV$,
and above the reduced Planck mass\, $\xbar{M}_{\rm Pl} = \mpl /\sqrt{8\pi}$.\label{rattazzi}}
\end{figure}
\begin{figure}[t]
\centering%
\includegraphics[width=0.45\textwidth]{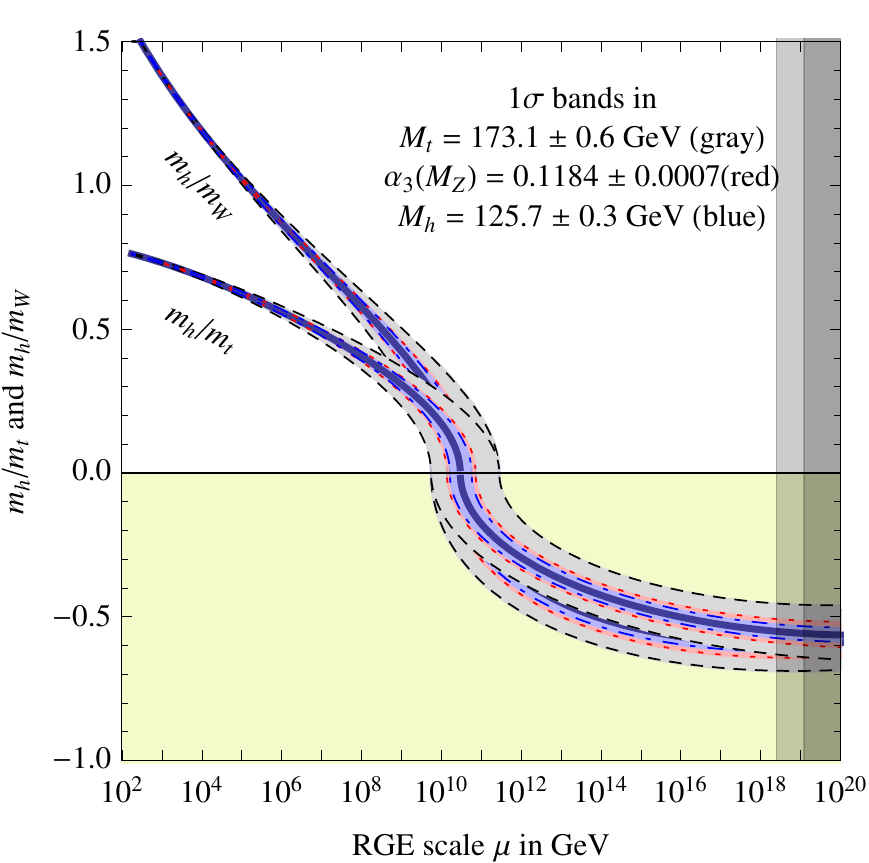}\qquad
\includegraphics[width=0.45\textwidth]{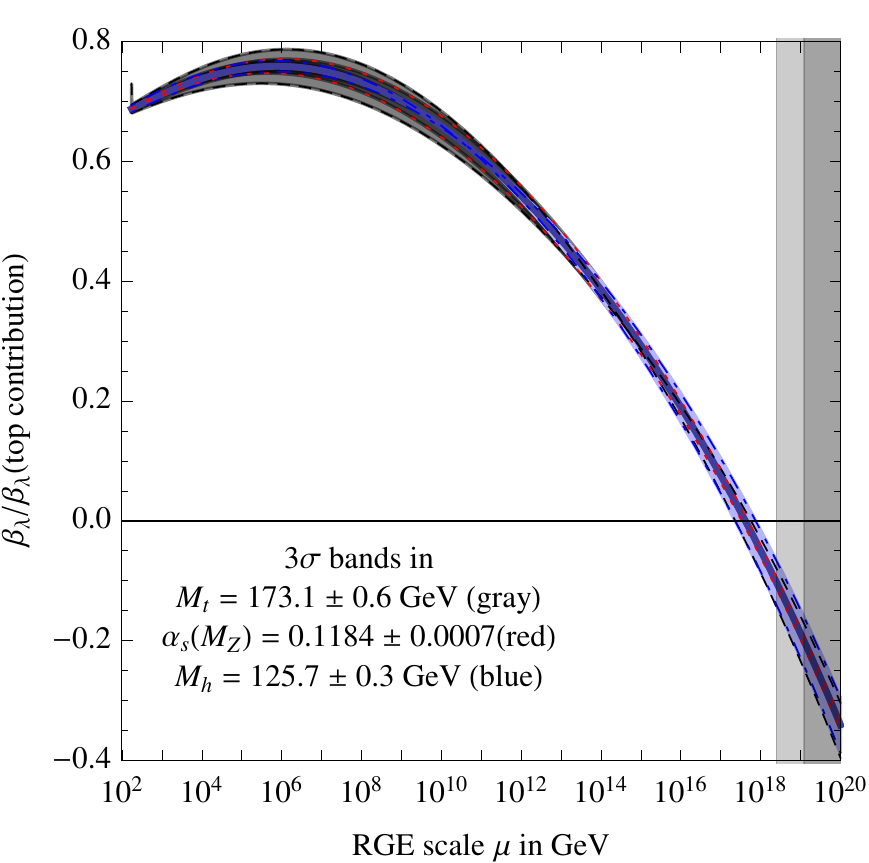}
\caption{Same as figure~\ref{rattazzi}, with more ``physical" normalizations. Left: the Higgs quartic coupling is 
compared with the top Yukawa and weak gauge coupling through the ratios ${\rm sign}(\lambda)\sqrt{4|\lambda|}/y_t$ and ${\rm sign}(\lambda)\sqrt{8|\lambda|}/g_2$, which correspond to the ratios of running masses $m_h/m_t$ and $m_h/m_W$, respectively. Right: the Higgs quartic $\beta$-function is shown in units of its top-quark contribution, $\beta_\lambda^{\rm top} =-3y_t^4/8\pi^2$.
\label{fig:rattazzi}}
\end{figure}

The Higgs quartic coupling remains weak in the entire energy domain
below $\mpl $. It decreases with energy crossing $\lambda =0$
at a scale of about $10^{10}$~GeV, see the left panel of figure~\ref{rattazzi}. Indeed, $\lambda$ is the only SM coupling that is
allowed to change sign during the RG evolution because it is not
multiplicatively renormalized. For all other SM couplings, the $\beta$
functions are proportional to their respective couplings and crossing
zero is not possible. This corresponds to the fact that $\lambda =0$
is not a point of enhanced symmetry.

In the left panel of figure~\ref{fig:rattazzi} we compare the size of $\lambda$
with the top Yukawa coupling $y_t$ and the gauge coupling $g_2$,
choosing a normalization such that each coupling is equal to the
corresponding particle mass, up to the same proportionality
constant. In other words, we are plotting the ratios 
\begin{equation}
\hbox{${\rm
  sign}(\lambda)\times\sqrt{4|\lambda|}/y_t\qquad$ and $\qquad{\rm
  sign}(\lambda)\times\sqrt{8|\lambda|}/g_2$} \ ,
\end{equation}
equal to the ratios of running
masses $m_h/m_t$ and $m_h/m_W$, respectively.  
Except for the region
in which $\lambda$ vanishes, the Higgs quartic coupling looks fairly
``normal" with respect to the other SM couplings. Nonetheless, the RG
effect reduces significantly the overall size of $\lambda$ in its
evolution from low to high energy. 
Although the central values of Higgs and top masses do not favour a
scenario with vanishing Higgs self coupling at the Planck scale
$\mpl $ -- a possibility originally proposed in
ref.~\cite{Froggatt:1995rt,Froggatt:2001pa} and discussed more recently in
ref.~\cite{Burgess:2001tj,Isidori:2007vm,Bezrukov:2009db,Shaposhnikov:2009pv,Degrassi:2012ry}
-- the smallness of $\lambda$ around $\mpl $ offers reasons for speculation, 
as we will discuss later. 

Another important feature of the RG evolution of $\lambda$ is the
slowing down of the running at high energy. As shown in
the right panel of figure~\ref{rattazzi}, the corresponding Higgs quartic
$\beta$-function vanishes at a scale of about
$10^{17}$--$10^{18}$~GeV. In order to quantify the degree of
cancellation in the $\beta$-function, we plot in figure~\ref{fig:rattazzi}
(right) $\beta_\lambda$ in units of its pure top
contribution. The vanishing of $\beta_\lambda$ looks more like an
accidental cancellation between various large contributions, rather
than an asymptotic approach to zero. Given that the $\beta$-functions
of the other SM couplings are all different than zero, it is not evident to find valid symmetry or dynamical reasons for the vanishing of
$\beta_\lambda$ alone near $\mpl $. However, the smallness of
$\beta_\lambda$ (and $\lambda$) at high energy implies that tiny
variations of the input values of the couplings at $\mpl $ lead to
wide fluctuations of the instability scale, thus justifying our
refined calculation.

\section{The SM phase diagram}\label{phasediagram}

\subsection{Derivation of the stability bound}

In order to compute the stability bound on the Higgs
mass one has to study the full effective potential and identify the
critical Higgs field above which the potential becomes smaller than the value at
the EW vacuum. We will refer to such critical energy as the
instability scale $\Lambda_I$.

A first estimate of the instability scale
can be obtained by approximating the effective potential with its
RG-improved tree level expression
\begin{equation}
V_{\rm eff}^{\rm tree}(h) \approx \frac{\lambda(\mub = h)}{4}h^4.
\end{equation}
In this approximation the stability condition simply corresponds to the requirement that $\lambda(\Lambda) \geq 0$ for every $\Lambda < \mpl$.
This analysis shows that the instability scale, corresponding approximately to the scale at which the running quartic coupling turns negative, occurs at energies much bigger than
the EW scale. Thus, for our purposes, the approximation of neglecting $v$ with respect to the value of the field $h$ is amply justified. Under this assumption, as pointed out in \cite{Casas:1994qy}, one can always define an effective coupling $\lambda_{\rm eff}$ such that the full effective potential in the relevant region $h\gg v$ becomes
\begin{equation}
V_{\rm eff} (h)=\lambda_{\rm eff}(h)\frac{h^4}{4}. \label{eff-potential-high-h}
\end{equation}
The effective quartic coupling takes the form
\begin{equation}\label{lambdaeffcorr}
\lambda_{\rm eff}(h) =e^{4\Gamma(h)} \Big[ \lambda(\bar\mu=h) + \lambda_{\rm eff}^{(1)}(\bar\mu=h) + \lambda_{\rm eff}^{(2)}(\bar\mu=h)\Big],
\end{equation}
where
\begin{align}\label{Gammaeffpot}
\Gamma(h) &= \int_{M_t}^h \gamma(\mub)\, d\log(\mub),
\end{align}
$\gamma(\mub)$ is the Higgs anomalous dimension, and the quantities $\lambda^{(1)}_{\rm eff}, \lambda^{(2)}_{\rm eff}$ can be extracted from the effective potential at two loops~\cite{Ford:1992pn} and are explicitly given in appendix~\ref{eff-potential-app}.

Notice that the stability bound is scheme and gauge independent.
While intermediate steps of the computation
(threshold corrections, higher-order RG equations, and the effective potential)
are scheme-dependent, the values of the effective potential at its local minima are scheme-independent physical observables,
and thus the stability condition has the same property.

We find that the instability scale, defined as the scale at which $\lambda_{\rm eff}$ vanishes, is
\begin{equation}
\log_{10}\frac{\Lambda_I}{\GeV} = 11.3 + 1.0\Mhdiff-1.2\Mtdiff + 0.4 \asdiff .
\label{eqlambdai}
\end{equation}
The scale $\Lambda_0$ at which the $\MS$ running coupling $\lambda$ vanishes
is a scheme-dependent quantity and is  slightly smaller than the scale $\Lambda_I$. We find
$\Lambda_0 \approx 0.15 \Lambda_I$, with the same dependence on the SM parameters as in \eqg{eqlambdai}.

\subsection{The SM phase diagram in terms of Higgs and top masses}

\begin{figure}[t]
\centering%
\includegraphics[width=0.45\textwidth]{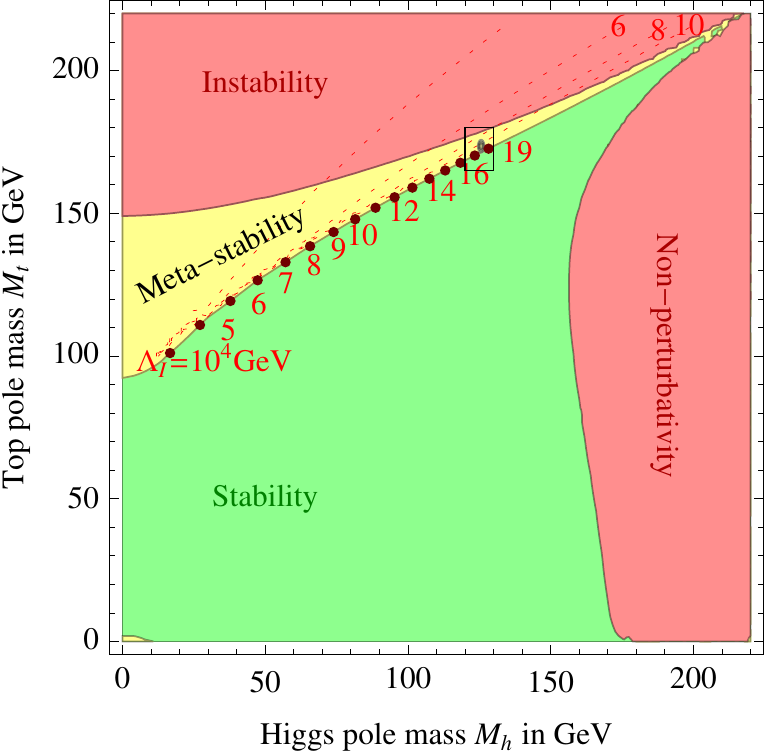}\qquad
\includegraphics[width=0.46\textwidth]{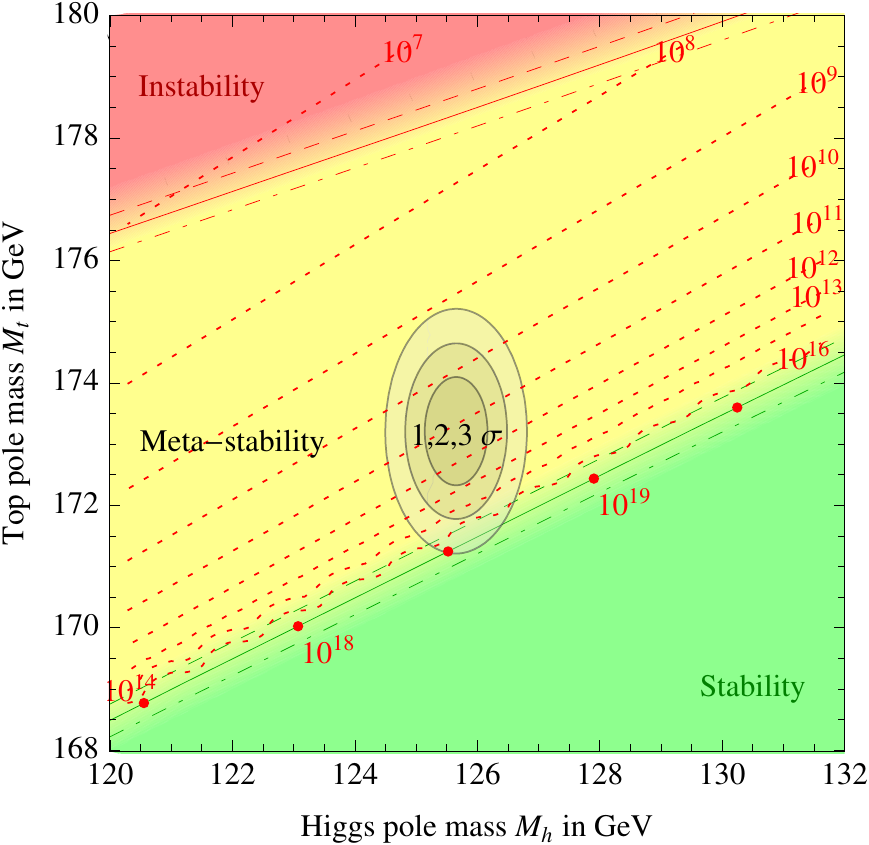}
\caption{Left: SM phase diagram in terms of Higgs and top pole masses.
The plane is divided into regions of absolute stability, meta-stability, 
instability of the SM vacuum, and non-perturbativity of the Higgs quartic 
coupling. The top Yukawa coupling becomes non-perturbative for 
$M_t>230\, \GeV$. The dotted contour-lines show the instability scale 
$\Lambda_I$ in $\GeV$ assuming $\alpha_3(M_Z)=0.1184$.
Right: Zoom in the region of the preferred experimental range of $M_h$ 
and $M_t$ (the grey areas denote the allowed region at 1, 2, and 3$\sigma$).
The three  boundary lines correspond to 1$\sigma$ variations of 
$\alpha_3(M_Z)=0.1184\pm 0.0007$, and the grading of the 
colours indicates the size of the theoretical error.  
\label{fig:regions}}
\end{figure}

The two most important parameters that determine the various EW phases
of the SM are the Higgs and top-quark masses. In figure~\ref{fig:regions} we
update the phase diagram given in ref.~\cite{Degrassi:2012ry} with our
improved calculation of the evolution of the Higgs quartic
coupling. The regions of stability, metastability, and instability of
the EW vacuum are shown both for a broad range of $M_h$ and $M_t$, and
after zooming into the region corresponding to the measured
values. The uncertainty from $\alpha_3$ and from theoretical errors
are indicated by the dashed lines and the colour shading along the
borders. Also shown are contour lines of the instability scale
$\Lambda_I$.

As previously noticed in ref.~\cite{Degrassi:2012ry}, the measured values of
$M_h$ and $M_t$ appear to be rather special, in the sense that they
place the SM vacuum in a near-critical condition, at the border
between stability and metastability.  In the neighbourhood of the
measured values of $M_h$ and $M_t$, the stability condition is well
approximated by 
\begin{equation}
M_h > 129.1 \GeV+ 2.0 (M_t-\Mtexp\GeV) -0.5\GeV\asdiff \pm 0.3 \GeV
\ .
\label{eq:stability}
\end{equation}
The quoted uncertainty comes only from higher order perturbative corrections.
Other non-perturbative uncertainties associated with the relation between the measured value of the top mass
and the actual definition of the top pole mass used here (presumably of the order of $\Lambda_{\rm QCD}$) are buried inside the parameter $M_t$ in \eqg{eq:stability}. For this reason we include a theoretical error in the top pole mass and take
$M_t = (\Mtexp \pm \Mterr_{\rm exp} \pm 0.3_{\rm th})\GeV$. Combining in quadrature
theoretical uncertainties with experimental errors, we find for the stability condition
\begin{equation} 
M_h > (129.1 \pm 1.5)\GeV
\end{equation} 
From this result we conclude that vacuum stability of the SM up to the
Planck scale is excluded at $2.2\sigma$ (98.6\% C.L.~one-sided). 
Since the main source of uncertainty in \eqg{eq:stability}
comes from $M_t$, any refinement in the measurement of the top mass is
of great importance for the question of EW vacuum stability.

Since the experimental error on the Higgs mass is already fairly small
and will be further reduced by future LHC analyses, it is becoming
more appropriate to express the stability condition in terms of the pole
top mass. We can express the stability condition of \eqg{eq:stability} as 
\begin{equation}
M_t < (171.53\pm 0.15\pm 0.23_{\alpha_3} \pm 0.15_{M_h})\GeV=
(171.53\pm 0.42)\GeV .
\end{equation}
In the latter equation we combined in quadrature the theoretical uncertainty
with the experimental uncertainties on $M_h$ and $\alpha_3$.

\subsection{The SM phase diagram in terms of Planck-scale couplings}

\begin{figure}[t]
\centering%
\includegraphics[width=0.45\textwidth]{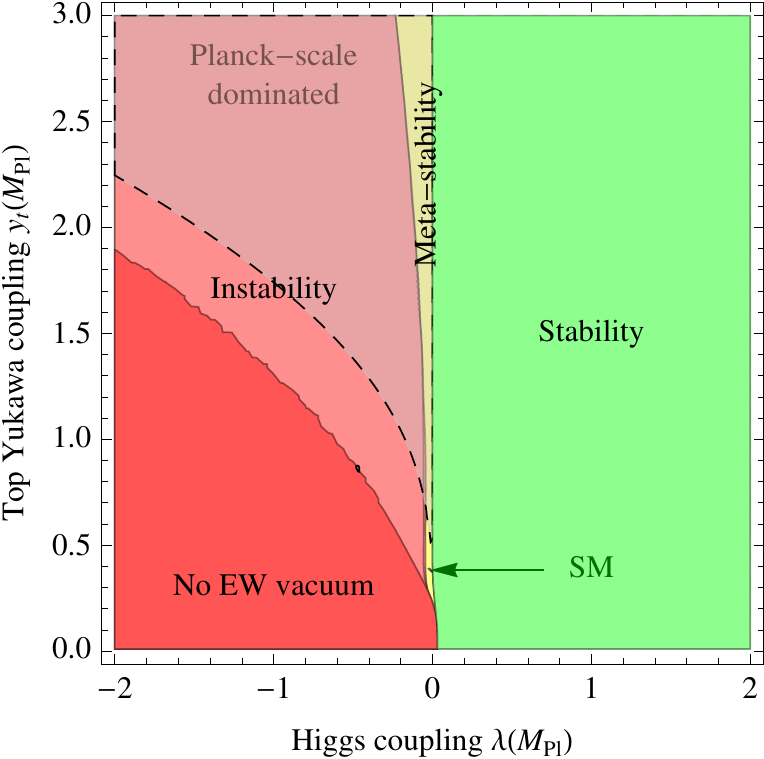}\qquad\includegraphics[width=0.45\textwidth]{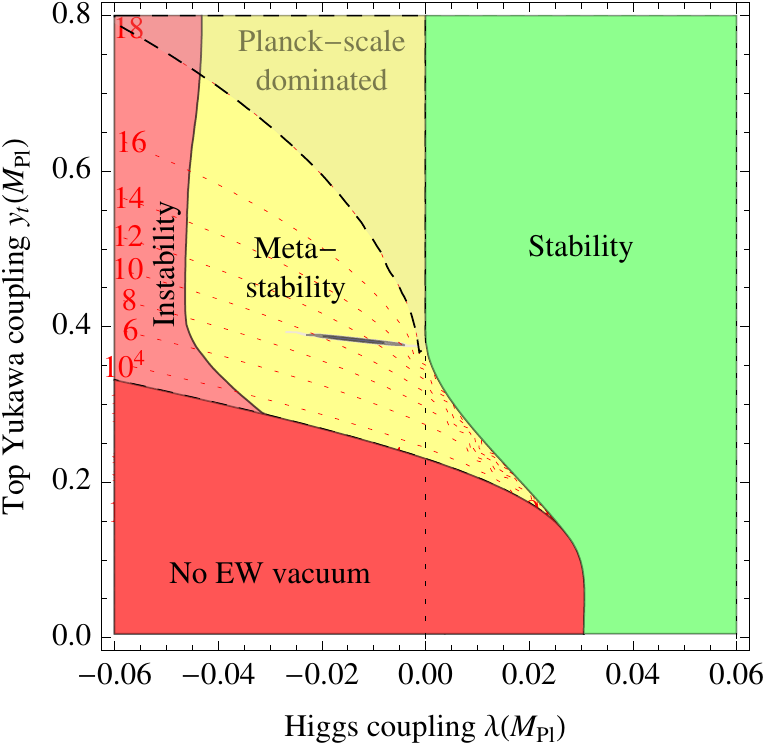}
\caption{Left: SM phase diagram in terms of quartic Higgs coupling $\lambda$ and top Yukawa coupling $y_t$ renormalized at the Planck scale. The region where the instability scale $\Lambda_I$ is larger than $10^{18}\, \GeV$
is indicated as `Planck-scale dominated'.
Right: Zoom around the experimentally measured values of the couplings, which correspond to the thin ellipse roughly at the centre of the panel.
The dotted lines show contours of $\Lambda_I$ in $\GeV$.  
\label{fig:regionsPlanck}}
\end{figure}

The discovery of the SM near-criticality has led to many theoretical
speculations~\cite{Feldstein:2006ce,Degrassi:2012ry,Holthausen:2011aa,EliasMiro:2011aa,Chen:2012faa,Lebedev:2012zw,EliasMiro:2012ay,Rodejohann:2012px,Bezrukov:2012sa,Datta:2012db,Alekhin:2012py,Chakrabortty:2012np,Anchordoqui:2012fq,Masina:2012tz,Chun:2012jw,Chung:2012vg,Chao:2012mx,Lebedev:2012sy,Nielsen:2012pu,Kobakhidze:2013tn,Tang:2013bz,Klinkhamer:2013sos,He:2013tla,Chun:2013soa,Jegerlehner:2013cta,Bezrukov:2009db,Shaposhnikov:2009pv}. In
order to address such speculations and to investigate if the measured
value of $M_h$ is really special in the SM, it is more appropriate to
study the phase diagram in terms of the Higgs quartic and the top
Yukawa coupling evaluated at some high-energy scale, rather than at
the weak scale. This is because of our theoretical bias that the SM is
eventually embedded into a new framework at short distances, possibly
as short as the Planck length. Therefore, it is more likely that
information about the underlying theory is directly encoded in the
high-energy coupling constants. For this reason in
figure~\ref{fig:regionsPlanck} we recast the phase diagram of
figure~\ref{fig:regions} in terms of $\lambda (\mpl )$ and $y_t (\mpl )$. The
diagram is shown in a broad range of couplings allowed by
perturbativity, and also after zooming into the interesting
region. The new area denoted as `no EW vacuum' corresponds to a
situation in which $\lambda$ is negative at the weak scale, and
therefore the usual Higgs vacuum does not exist. In the region denoted
as `Planck-scale dominated' the instability scale $\Lambda_I$ is
larger than $10^{18}\, \GeV$. In this situation we expect that both
the Higgs potential and the tunnelling rate receive large
gravitational corrections and any assessment about vacuum stability
becomes unreliable.

From figure~\ref{fig:regionsPlanck} we can infer more than just criticality
of the Higgs quartic coupling. Indeed, this figure shows that the
measured values of the Higgs and top masses lie in the region
corresponding not only to the lowest possible values of $\lambda (\mpl
)$ allowed by (meta)stability, but also to the smallest possible value
of $y_t(\mpl )$, once $\lambda (\mpl )$ has been selected.

Indeed, for small Higgs quartic ($\lambda (\mpl )<0.02$),
there is a non-vanishing minimum value of $y_t(\mpl )$ required to
avoid instability. 
This special feature is related to the approximate vanishing of $\beta_\lambda$ around the Planck mass.
Indeed, for
fixed $\lambda (\mpl )$, the top Yukawa coupling has the effect to
stabilize the potential, as we evolve from high to low
energies. Without a sizeable contribution from $y_t$, the gauge
couplings tend to push $\lambda$ towards more smaller (and eventually
negative) values, leading to an instability. Therefore, whenever
$\lambda (\mpl )$ is small or negative, a non-zero $y_t(\mpl )$ is
necessary to compensate the destabilizing effect of gauge
couplings.

It is a remarkable coincidence that the measured values of the Higgs
and top masses correspond rather precisely to the simultaneous minima
of both $\lambda (\mpl )$ and $y_t(\mpl )$. In other words, it is
curious that not only do we live in the narrow vertical yellow stripe
of figure~\ref{fig:regionsPlanck} -- the minimum of $\lambda (\mpl )$ --
but also near the bottom of the funnel -- the minimum of $y_t (\mpl
)$. Near-criticality holds for both the Higgs quartic and the top
Yukawa coupling.

From the left panel of figure~\ref{fig:regionsPlanck} it is evident
that, even when we consider the situation in terms of high-energy
couplings, our universe appears to live under very special
conditions. The interesting theoretical question is to understand if
the apparent peculiarity of $\lambda (\mpl )$ and $y_t (\mpl )$ carry
any important information about phenomena well beyond the reach of any
collider experiment. Of course this result could be just an accidental
coincidence, because in reality the SM potential is significantly
modified by new physics at low or intermediate scales. Indeed, the
Higgs naturalness problem corroborates this possibility. However, both
the reputed violation of naturalness in the cosmological constant and
the present lack of new physics at the LHC cast doubts on the validity
of the naturalness criterion for the Higgs boson. Of course, even
without a natural EW sector, there are good reasons to believe in the
existence of new degrees of freedom at intermediate energies.
Neutrino masses, dark matter, axion, inflation, baryon asymmetry
provide good motivations for the existence of new dynamics below the
Planck mass. However, for each of these problems we can imagine
solutions that either involve physics well above the instability scale
or do not significantly modify the shape of the Higgs potential. As a
typical example, take the see-saw mechanism. As shown in
ref.~\cite{EliasMiro:2011aa}, for neutrino masses smaller than 0.1~eV
(as suggested by neutrino-oscillation data without mass degeneracies),
either neutrino Yukawa couplings are too small to modify the running
of $\lambda$ or the right-handed neutrino masses are larger than the
instability scale. In other words, a see-saw neutrino does not modify
our conclusions about stability of the EW vacuum.  Couplings of
weak-scale dark matter to the Higgs boson are constrained to be small
by WIMP direct searches (although dark-matter particles with weak
interactions would modify the running of the weak gauge couplings,
making the Higgs potential more stable).

Thus, it is not inconceivable that the special values of $\lambda
(\mpl )$ and $y_t (\mpl )$ carry a significance and it is worth to
investigate their consequences.

Finally, we notice that extrapolating SM parameters above the Planck scale ignoring gravity
(this is a questionable assumption) the hypercharge couplings hits a Landau pole at about $10^{42}\GeV$.
Demanding perturbativity up to such scale (rather than up to the Planck scale),
the  bounds on the top and Higgs masses become stronger by about 10 and 20 GeV respectively,
and their measured values still lie in the region that can be extrapolated up to such high scale.

\subsection{The SM phase diagram in terms of gauge couplings}
\label{sec:gaugescan}

\begin{figure}[t]
\centering%
\includegraphics[width=0.45\textwidth]{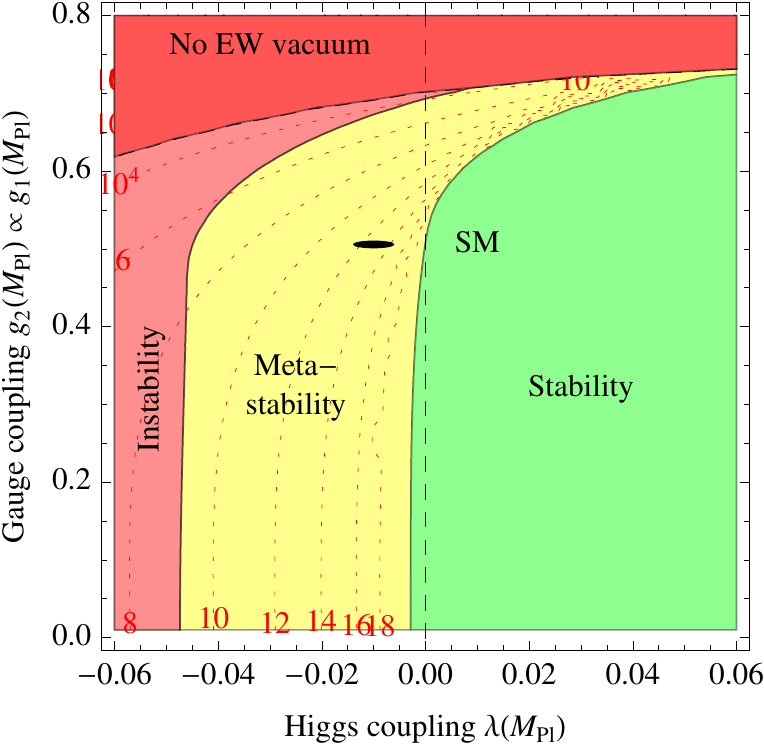}\qquad
\includegraphics[width=0.45\textwidth]{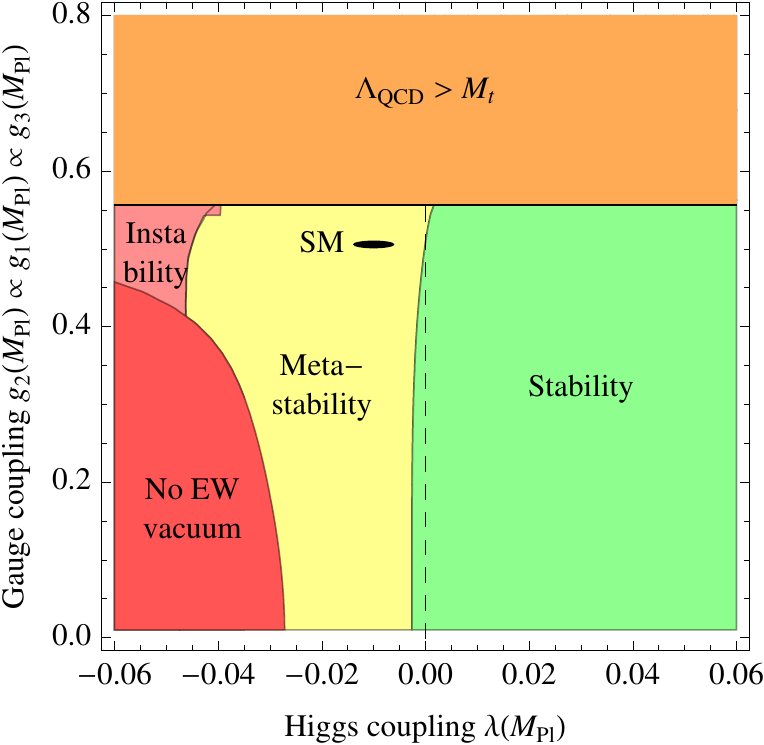}
\caption{SM phase diagram in terms of the Higgs quartic coupling $\lambda (\mpl )$ and of the gauge coupling $g_2(\mpl )$.
{Left}: A common rescaling factor is applied to the electro-weak gauge couplings $g_1$ and $g_2$, while $g_3$ is kept constant.
{Right}: A common rescaling factor is applied to all SM gauge couplings $g_1,g_2,g_3$, such that a $10\%$ increase in the
strong gauge coupling at the Planck scale makes $\Lambda_{\rm QCD}$ larger than the weak scale.
The measured values of the couplings correspond to the small ellipse marked 
as `SM'.
\label{fig:regionsPlanckg}}
\end{figure}

So far we have been studying the phase diagram in terms of Higgs and
top masses or couplings, keeping the other SM parameters fixed. This
is reasonable, since the EW vacuum is mostly influenced by the Higgs
and top quark. It is however interesting to study how a variation in the other couplings affects our results.

We start by considering the scanning of weak couplings defined at some
high-energy scale, which we identify with $\mpl $. The impact of the
gauge couplings $g_1$ and $g_2$ can be understood from the leading
terms of the RG equation for the Higgs quartic coupling
\begin{equation}
(4\pi)^2\, \frac{d\lambda}{d\ln\mub^2} =-3y_t^4+6y_t^2\lambda +12\lambda^2 +\frac{9}{16} \left( g_2^4 +\frac{2}{5} g_2^2g_1^2+\frac{3}{25}g_1^4\right) -\frac{9}{2}\lambda \left( g_2^2+\frac{g_1^2}{5}\right) +\cdots.
\end{equation}
For small $\lambda(\mpl )$, the weak gauge couplings have the effect
of reducing even further the Higgs quartic coupling in its evolution
towards lower energies, thus contributing to destabilize the
potential. For large $\lambda(\mpl )$, instead, they tend to make $\lambda$
grow at lower energy.

We quantify the situation by plotting in figure~\ref{fig:regionsPlanckg} the SM phase diagram in terms of $\lambda(\mpl )$ and $g_2(\mpl)$. 
In the left panel, for simplicity, we vary the hypercharge coupling $g_1(\mpl )$
by keeping fixed the ratio $g_1(\mpl )/g_2(\mpl )=1.22$ as in the SM,
while $y_t(\mpl )$ and $g_3(\mpl )$ are held to their SM values.  As
in previous cases, also the phase diagram in terms of weak gauge
couplings shows the peculiar characteristic of the SM parameters to
live close to the phase boundary.\footnote{Note that the figure is zoomed
around the region of the physical values, so that the proximity to the
boundary is not emphasized.}
The weak gauge
couplings in the SM lie near the maximum possible values that do not
lead to a premature decay of the EW vacuum. Were $g_2$ and $g_1$
50\% larger than their actual values, we wouldn't be here speculating on
the peculiarity of the Higgs mass.

Next, we discuss the impact of the strong gauge coupling
constant.  In the right panel of figure~\ref{fig:regionsPlanckg} we show again the phase
diagram in the plane $\lambda (\mpl )$, $g_2(\mpl )$, this time obtained by
varying all three gauge couplings by a common rescaling factor. The
top Yukawa coupling $y_t(\mpl )$ is held fixed at its SM value and so,
as the other couplings scan, the top mass does not correspond to the
measured value.

The coupling $g_3$ affects $\beta_\lambda$ only at two loops, but it
has a more important role in the RG evolution of the top Yukawa
coupling, whose leading terms are given by
\begin{equation}
(4\pi)^2\, \frac{dy_t^2}{d\ln\mub^2} = y_t^2 \left( \frac92 y_t^2-8g_3^2-\frac94 g_2^2-\frac{17}{20} g_1^2\right)+\cdots.
\end{equation}
When the value of $g_3$ is reduced at fixed $y_t(\mpl )$, the
low-energy top Yukawa coupling becomes smaller. This reduces the
stabilizing effect of the top for a given $\lambda (\mpl )$ and
explains the appearance in figure~\ref{fig:regionsPlanckg} (right) at
small gauge couplings of a region where $\lambda$ is
negative at the weak scale and there is no EW vacuum.

On the other hand, when $g_3$ is increased, the value of $\Lambda_{\rm QCD}$ 
grows rapidly. Whenever
\begin{equation}
\alpha_3 (\mpl ) > \frac{6\pi}{21 \ln (\mpl / M_t)},
\end{equation}
which corresponds to $g_3 (\mpl) > 0.54$, the value of $\Lambda_{\rm QCD}$ 
becomes larger than $M_t$, preventing a perturbative
extrapolation from the Planck to the weak scale. As shown in
figure~\ref{fig:regionsPlanckg} (right), this region is reached as soon as
the SM gauge couplings are increased by only 11\%. Once again, the SM
gauge couplings live near the top of the range allowed by simple
extrapolations of the minimal theory.

\begin{figure}[t]
\centering%
\includegraphics[width=0.45\textwidth]{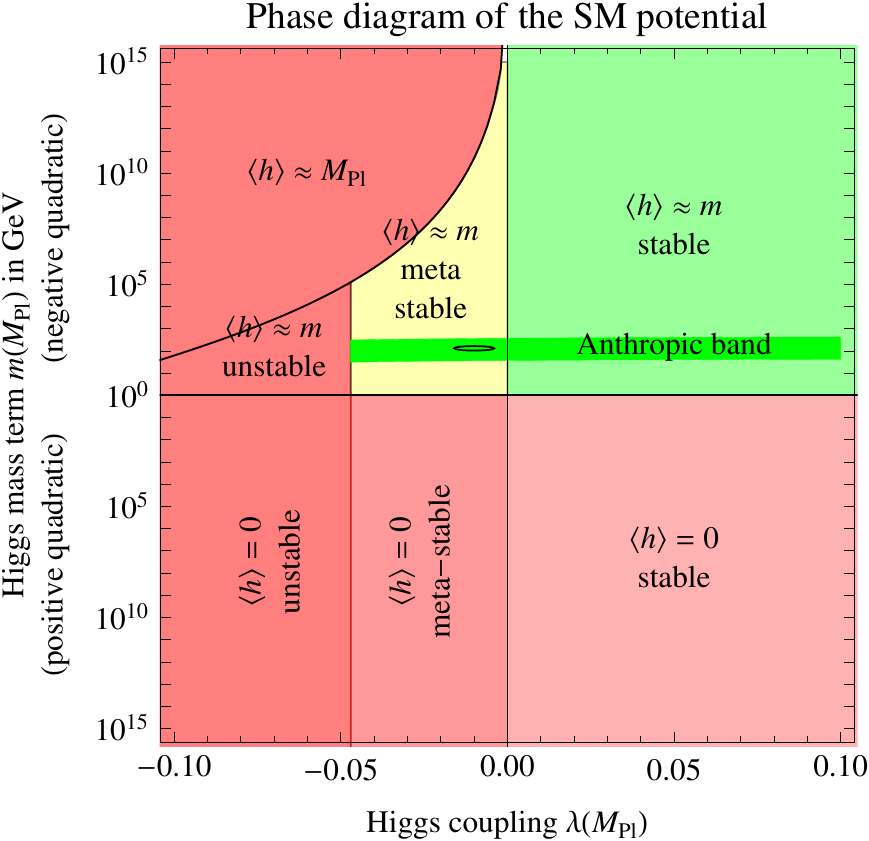}
\caption{Phase diagram of the SM in terms of the parameters of the Higgs potential evaluated at the Planck scale. In the metastability region, there is an upper bound on $m$ from the requirement of a Higgs vacuum at a finite field value. 
The green region is simple thanks to the fact that $\beta(\lambda)=0$ at $\mpl $.
On the vertical axis we plot $|m(\mpl )|$, in the case of negative (above) and positive (below) Higgs quadratic term.
\label{fig:VSMphases}}
\end{figure}

\subsection{The SM phase diagram in terms of Higgs potential parameters}

The Higgs mass parameter $m_h$ in the Higgs potential is the origin of
the well-known naturalness problem. Here we show that the simple
requirement of the existence of a non-trivial EW vacuum sets an upper
bound on $m_h$, which is completely independent of any naturalness
argument.

Let us start by considering the tree-level Higgs potential in \eqg{higgspot}.  For
$m_h^2>0$ and $\lambda>0$, the potential has the usual
non-trivial vacuum at $\langle h \rangle
=v=m_h/\sqrt{2\lambda}$. However, since $v$ is proportional to $m_h$ and
$\lambda$ is negative above the instability scale $\Lambda_I$, the
Higgs vacuum at finite field value no longer exists when $m_h^2$ is too
large. The upper bound on $m_h^2$ can be estimated by considering the
minimisation condition of the potential, including only the
logarithmic running of $\lambda$, but neglecting the evolution of $m_h$
(which is a good approximation, as shown in figure~\ref{fig:run1}):
\begin{equation}
\left[ 2\lambda (v) +\frac{\beta_\lambda (v)}{2}\right]v^2 =m_h^2.
\label{eq:minbeta}
\end{equation}
For values of $v$ in the neighbourhood of $\Lambda_I$, we can
approximate\footnote{In this analysis, we can safely neglect the non-logarithmic corrections to the effective potential and so we do not distinguish between $\lambda$ and $\lambda_{\rm eff}$.} $\lambda(v) \approx \beta_\lambda (\Lambda_I ) \ln
v/\Lambda_I$ and $ \beta_\lambda (v) \approx \beta_\lambda (\Lambda_I
)$. Then we see that \eqg{eq:minbeta} has a solution only if
\begin{equation}
m_h^2 <
-\beta_\lambda(\Lambda_I)\, e^{-3/2} \Lambda_I^2.
\label{limstab}
\end{equation}
Note that $\beta_\lambda(\Lambda_I)$ is negative in the SM.

Figure~\ref{fig:VSMphases} shows the SM phase diagram in terms of the
parameters $\lambda(\mpl )$ and $m_h(\mpl )$. 
The sign of each one of these parameters corresponds to different phases of the theory,
such that  $\lambda(\mpl )=m_h(\mpl )=0$ is a tri-critical point.

The region denoted by
`$\langle h\rangle \approx \mpl $' corresponds to the case in which \eqg{limstab} is
not satisfied and there is no SM-like vacuum, while the Higgs field
slides to large values. In the region of practical interest, the upper
limit on $m_h$ is rather far from its actual physical value $m_h=M_h$,
although it is much stronger than $\mpl $, the ultimate ultraviolet
cutoff of the SM. A much more stringent bound on $m_h$ can be derived
from anthropic considerations~\cite{Agrawal:1997gf} and the
corresponding band in parameter space is shown in
figure~\ref{fig:VSMphases}.

\begin{figure}[t]
$$\includegraphics[width=0.45\textwidth]{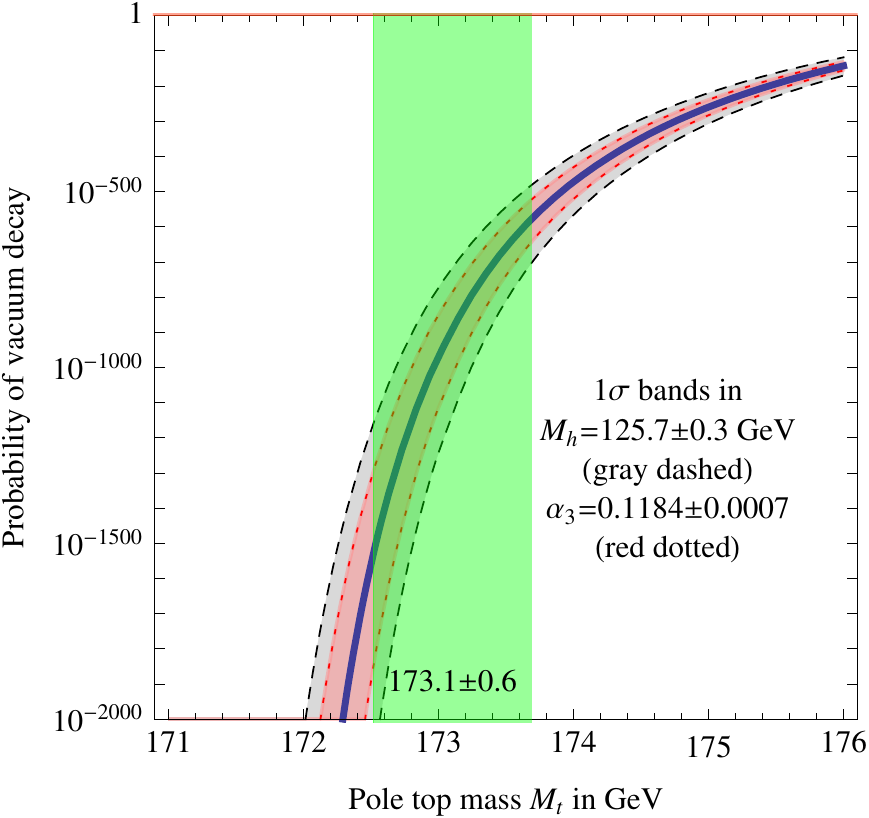}\qquad\includegraphics[width=0.45\textwidth]{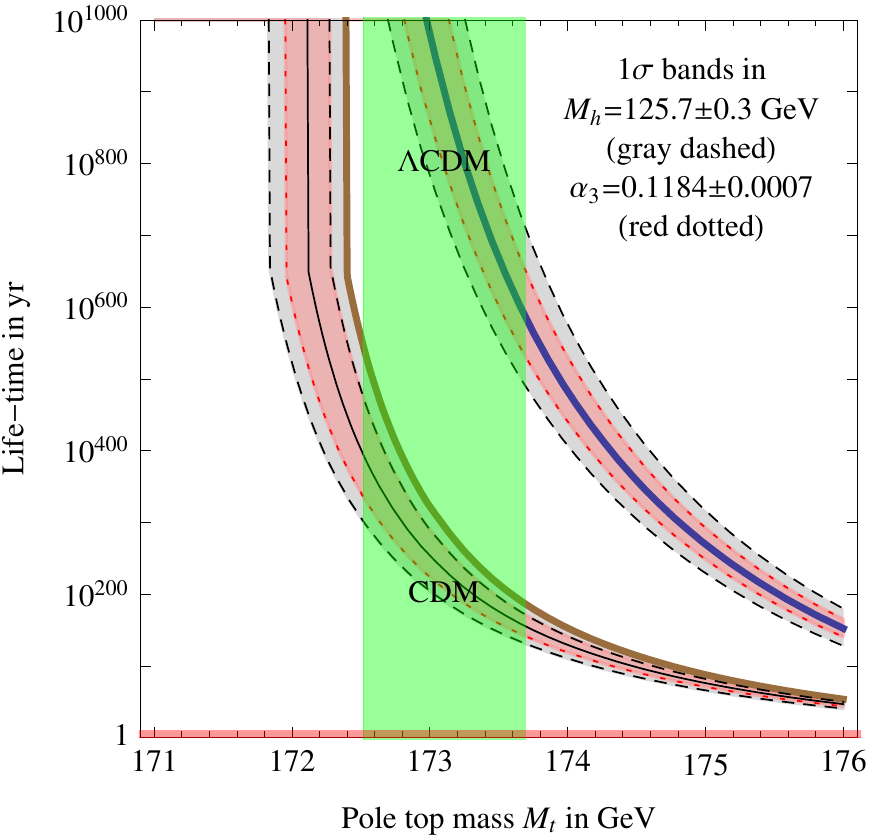}$$
\caption{{Left:} The probability that electroweak vacuum decay happened in our past light-cone,
taking into account the expansion of the universe. {Right}:
The life-time of the electroweak vacuum, with two different assumptions for future cosmology: universes dominated by the
cosmological constant ($\Lambda$CDM) or by dark matter (CDM).
\label{fig:lifetime}}
\end{figure}

\section{Lifetime of the SM vacuum}

The measured values of $M_h$ and $M_t$ indicate that the SM Higgs
vacuum is not the true vacuum of the theory and that our universe is
potentially unstable. The rate of quantum tunnelling out of the EW
vacuum is given by the probability $d\wp/dV\, dt$ of nucleating a
bubble of true vacuum within a space volume $dV$ and time interval
$dt$~\cite{Kobzarev:1974cp,Coleman:1977py,Callan:1977pt}
\begin{equation}
d\wp =dt\,dV ~ \Lambda_B^4\, e^{-S(\Lambda_B)} \, ,
\label{eq:probd}
\end{equation}
where $S(\Lambda_B)$ is the action of the bounce of size 
$R=\Lambda_B^{-1}$, given by
\begin{equation}
S(\Lambda_B)=\frac{8\pi^2}{3|\lambda(\Lambda_B)|}.
\end{equation}

At the classical level, the Higgs theory with only quartic coupling is
scale-invariant and the size of the bounce $\Lambda_B^{-1}$ is
arbitrary. The RG flow breaks scale invariance and the tree level
action gets replaced by the one-loop action, as calculated in
ref.~\cite{Isidori:2001bm}.  
Then, $\Lambda_B$ is determined as the scale at
which $\Lambda_B^4 e^{-S(\Lambda_B)}$ is maximized.  In practice this
roughly amounts to minimizing $\lambda(\Lambda_B)$, which corresponds
to the condition $\beta_\lambda (\Lambda_B)=0$.  As long as
$\Lambda_B\ll \mpl $, gravitational effects are irrelevant, since
corrections to the action in minimal Einstein gravity are given by
$\delta S_G = 256\pi^3\Lambda_B^2/45|\lambda|\mpl
^2$~\cite{Isidori:2007vm}. The effect of gravitational corrections is
to slow down the tunnelling rate~\cite{Coleman:1980aw}.  Whenever
$\Lambda_B >\mpl $, one can only obtain a lower bound on the
tunnelling probability by setting $\lambda (\Lambda_B)=\lambda (\mpl)$. 
For the physical values of $M_h$ and $M_t$, our results are fairly
insensitive to Planckian dynamics.

The total probability $\wp$ for vacuum decay to have occurred during the 
history of the universe can be computed by integrating eq.\eq{probd} over the
space-time volume of our past light-cone,
\begin{equation} 
\int dt \,dV = \int_0^{t_0}  dt~\int_{|x|<a(\eta_0-\eta )}d^3x = \frac{4\pi}{3} \int_0^{\eta_0} d\eta\, a^4 (\eta_0-\eta)^3 \approx \frac{0.15}{H_0^4}.
\label{eq:tunrate}
\end{equation}
Here $a$ is the scale factor, $\eta$ is conformal time ($d\eta /dt =1/a$), $\eta_0\approx 3.4/H_0$ is the present conformal time
and $H_0\approx 67.4\,{\rm km/sec~Mpc}$ is the present Hubble rate.
Equation~(\ref{eq:tunrate}) roughly amounts to saying that the `radius' of the universe is given by $cT_U$, where $T_U\approx 0.96/H_0$ is the present age.
The present value of the vacuum-decay probability $\wp$ is 
\begin{equation}
\wp_0 = 0.15~ \frac{\Lambda_B^4}{H_0^4}\, e^{-S(\Lambda_B)} \, ,
\label{eq:probdoggi}
\end{equation}
and is dominated by late times: this makes our result more robust,
since it is independent of the early cosmological history. In the left panel of figure~\ref{fig:lifetime} we plot, as a function of the top mass, the
probability $\wp_0$ that the EW vacuum had decayed during the past
history of the universe. We find that the probability is spectacularly
small, as a consequence of the proximity of the SM parameters to the
boundary with the region of absolute stability.

The lifetime of the present EW vacuum $\tau_{\rm EW}$ depends on the
future cosmological history. If dark energy shuts off and the future
universe is matter dominated, the space-time volume of the past
light-cone at time $t_0$ is given by
\begin{equation}
\int dt \,dV = \frac{4\pi}{3} \int_0^{\eta_0} d\eta\, a^4 (\eta_0-\eta)^3=\frac{16\, \pi}{1485\, H_0^4}.
\end{equation}
Here $H_0$ is the Hubble parameter at time $t_0$, and we have
performed the integral using the relations
$a^{1/2}=H_0\eta/2=(3H_0t/2)^{1/3}$ and $t_0=2/(3H_0)$, valid in a
matter-dominated flat universe. The lifetime $\tau_{\rm EW}$ is given by the time at which $\wp=1$. In the case of matter domination,
\begin{equation}
\tau_{\rm EW}^{\rm CDM} =\left( \frac{55}{3\pi}\right)^{1/4} \frac{e^{S(\Lambda_B)/4}} {\Lambda_B} \approx \frac{T_U} {\wp_0^{1/4}}
\end{equation}
where $\wp_0$ is given in \eqg{eq:probdoggi} and shown in figure~\ref{fig:lifetime} left.

If instead the universe keeps being accelerated by the cosmological
constant, entering into a de Sitter phase with Hubble constant $H =
H_0 \sqrt{\Omega_\Lambda}$, at a time $t_0$ in the far future the
volume of the past light-cone will be
\begin{equation} 
\int dt~dV = \frac{4\pi}{3} \int_0^{\eta_0} d\eta\, a^4 (\eta_0-\eta)^3= \frac{4\pi }{3H^4} \left[ Ht_0 -\frac{11}{6}+{\Ord}(e^{-Ht_0})\right] .
\end{equation}
Here we have used the relations $a=(1-H\eta)^{-1}=e^{Ht}$, valid in a 
vacuum-energy dominated universe. The lifetime 
for the case of vacuum energy domination is equal to
\begin{equation}
\tau_{\rm EW}^{\Lambda\rm CDM} = \frac{3H^3e^{S(\Lambda_B)}} {4\pi \Lambda_B^4} \approx \frac{0.02~T_U}{ \wp_0} 
\end{equation}
 
The lifetime of the present EW vacuum is plotted in
figure~\ref{fig:lifetime} in both cases of matter or vacuum-energy
domination. As shown, the SM vacuum is likely to survive for times
that are enormously longer than any significant astrophysical age
(e.g. the sun will exhaust its fuel in about five billion
years).

\section{Interpretations of the high-energy SM couplings}\label{interpretations}

In this last section we attempt some possible explanation for the special values of the measured SM parameters. The first possible interpretation is the result of new dynamics
occurring at some high-energy scale, the others find their most
natural implementations in a {\it multiverse} of vacua such as the one predicted by string theory.

\subsection{Matching conditions}\label{match}

The special value of the Higgs quartic coupling could be the result of
a matching condition with some high-energy theory in the vicinity of
$\mpl $. There are many examples of theories able to drive
$\lambda(\mpl )$ to zero. An approximate Goldstone or shift symmetry at a high scale \cite{Hebecker:2012qp,Redi:2012ad}, for instance, can provide a vanishing potential for the Higgs forbidding any non-derivative interaction, as explained in section~\ref{CHM/PNGB}. An appealing explanation comes also from supersymmetry, where radiatively-stable flat directions provide a valid
justification of vanishing quartic couplings for scalar particles that
have other kinds of interactions at zero momentum. In this way,
supersymmetry convincingly evades the problem, encountered by
Goldstone bosons, of explaining why $\lambda \approx 0$ is compatible
with sizeable gauge and Yukawa couplings of the Higgs boson. The
scheme can be automatically realized in $\mathcal{N}=2$ supersymmetry \cite{Fox:2002bu,Benakli:2012cy}, while a
dynamical vacuum alignment with $\tan\beta \approx 1$ is required in
the case of $\mathcal{N}=1$ supersymmetry \cite{Hall:2009nd,Giudice:2011cg,Cabrera:2011bi,Arbey:2011ab,Ibanez:2013gf,Hebecker:2013lha}. Further explanations proposed in the literature include an infrared
fixed-point of some transplanckian physics~\cite{Shaposhnikov:2009pv} or a power-law running in a quasi-conformal theory.

Present data suggest
that an exact zero of $\lambda$ is reached at scales of about
$10^{10}~\text{--}~10^{12}$~GeV, see eq.~(\ref{eqlambdai}), well below the Planck mass. It is not
difficult to imagine theories that give $\lambda(\mpl )$ in agreement
with \eqg{eq:lammp} as a result of a vanishing matching condition at $M_{\rm Pl}$ 
modified by threshold corrections.

Note also that the smallness of the Higgs quartic $\beta$-function at high energy is the key ingredient that allows for the possibility of extending the SM up to a matching scale much larger than $\Lambda_I$. If $\lambda$ ran fast above $\Lambda_I$, it would rapidly trigger vacuum instability and the region of metastability would be limited to SM cut-off scales only slightly larger than $\Lambda_I$. This is another peculiarity of the measured values of $M_h$ and $M_t$.  

\subsection{Statistical interpretation}
 
Statistical properties of a multiverse in which the coupling constants fluctuate offer alternatives to
dynamical determinations of $\lambda(\mpl )$ from matching conditions
with new theories.

In order to describe this situation, we introduce a set of scalar
fields $\Phi_i$ ($i=1,\dots ,N$), each having $p$ different vacuum
configurations, which model some Planckian dynamics. The total number of possible vacua is $p^N$, which is
huge for large $N$. If this multiverse of vacua is a viable candidate
to host somewhere a region where the cosmological constant is
$10^{120}$ times smaller than $\mpl ^4$, as it is in our universe, then it is reasonable that
$p^N$ should be at least $10^{120}$. So we envisage a situation in
which $N$ is at least $\Ord(10^2)$, which is not inconceivable in a string-theory framework.  

To describe the variation of the SM couplings we assume 
that the SM fields are coupled to the fields $\Phi_i$ in the most general way,
\begin{equation}
\L=-\frac{Z_G(\Phi)}{4} F_{\mu \nu} F^{\mu \nu} +Z_H(\Phi) \left| D_\mu H\right|^2 +\left( iZ_\psi (\Phi) \bar \psi 
{D\!\!\!\!/\,}\psi +Y_{ab}(\Phi)\bar\psi_a  H \psi_b +{\rm h.c.}\right) -\Lambda(\Phi) |H|^4.
\end{equation}
Here $F_{\mu \nu}$ and $\psi$ collectively denote the SM gauge and fermion fields, and $H$ is the Higgs doublet. The physical SM coupling constants are given by
\begin{equation}
g=Z_G^{-1/2},~~~y_t=Z_{t_L}^{-1/2}Y_tZ_{H}^{-1/2}Z_{t_R}^{-1/2},~~~\lambda = Z_H^{-2}\Lambda ~,
\label{coupmulti}
\end{equation}
where the functions $Z_{G,\psi,H}$, $Y$, and $\Lambda$ are evaluated
at a vacuum of the fields $\Phi_i$. Since the fields $\Phi_i$ have $p^N$
vacua, the SM couplings effectively scan in this multiverse. The
coupling constants in \eqref{coupmulti} are evaluated at the high-energy
scale, here identified with $\mpl $, where the new dynamics is
integrated out.

For simplicity, we consider the toy example proposed in
ref.~\cite{ArkaniHamed:2005yv}, in which each field $\Phi_i$ has two
vacua ($p=2$) called $\Phi_i^{(+)}$ and $\Phi_i^{(-)}$. We also assume
that each of the functions $Z_{G,\psi,H}$, $Y$, $\Lambda$ (let us call
them collectively $Z$, to simplify notation) can be split as a sum of
the contributions of the different fields,\footnote{This is a consistent hypothesis, as long as the fields $\Phi$ are
mutually weakly-interacting. In this case, any mixed interaction is
generated only by small loop effects and can be ignored.}
\begin{equation}
Z(\Phi_1,\dots,\Phi_n)=\sum_{i=1}^NZ_i(\Phi_i),~~~~~~~~Z=\{Z_{G}, Z_{\psi},Z_{H}, Y,\Lambda \}.
\end{equation}
Under this
hypothesis, the $2^N$ values of $Z$ corresponding to the vacua of
$\Phi$ can written as
\begin{equation}
Z =\sum_{i=1}^N\left( Z_i^{(S)}+\eta_iZ_i^{(D)}\right), ~~~Z_i^{(S)}=\frac{Z_i(\Phi_i^{(+)})+Z_i(\Phi_i^{(-)})}{2},~~~Z_i^{(D)}=\frac{Z_i(\Phi_i^{(+)})-Z_i(\Phi_i^{(-)})}{2},
\end{equation}
where $\eta_i=\pm 1$. Each of the $2^N$ vacua (and each of the $2^N$ values of 
$Z$) is then labeled by the vector $\eta =(\eta_i,\dots,\eta_N)$.

The normalized probability distribution of $Z$ within the multiverse of vacua 
is given by
\begin{equation}
\rho(Z)=2^{-N} \sum_\eta \delta \left( Z-N\bar Z-\sum_{i=1}^N \eta_iZ_i^{(D)}\right),
\label{sumeta}
\qquad
\bar Z \equiv \frac{1}{N}\sum_{i=1}^NZ_i^{(S)}.
\end{equation}
Using the central limit theorem, the discrete sum over the $2^N$
configurations of $\eta$ in \eqref{sumeta} can be approximated for large
$N$ with a Gaussian distribution~\cite{ArkaniHamed:2005yv}
\begin{equation}
\rho(Z)=\frac{1}{\sqrt{2\pi N\Delta^2}}\exp\left[ -\frac{(Z-N\bar Z)^2}{2N\Delta^2}\right] ,\qquad
\Delta^2 =\frac{1}{N}\sum_{i=1}^N {Z_i^{(D)}}^2 .
\end{equation}
This shows that $Z$ peaks around $N\bar Z$ with an
approximately flat distribution in the range $|Z-N\bar Z|<\sqrt{N}\Delta$.

For generic couplings, we expect that $\bar Z$ and $\Delta$ are
quantities of order unity, and thus $Z$ is ${\Ord}(N)$ with a
relative uncertainty of order $1/\sqrt{N}$. Plugging this result
(which is valid for $Z=Z_{G,\psi,H}$, $Y$, $\Lambda$) into
\eqref{coupmulti}, we find
\begin{equation}
g, y_t \sim \frac{1}{\sqrt{N}},\qquad \lambda \sim \frac{1}{N}.
\label{coupnn}
\end{equation}
For $N\sim 100$, we obtain that gauge and top-Yukawa couplings are
predicted to be ${\Ord}(10^{-1})$ at around $\mpl $, while the Higgs
quartic coupling is ${\Ord}(10^{-2})$, in good qualitative agreement
with experimental data. Indeed,  adopting a `physical' normalization of couplings as in figure~\ref{fig:rattazzi} left, the
SM predicts $g_{1,2,3} (\mpl )/\sqrt{2}\approx y_t (\mpl )\approx
\sqrt{4|\lambda (\mpl )|}\approx 0.3$.   

The different behaviour with $N$ in \eqref{coupnn} arises because
$\lambda$ is a quartic coupling, while $g$ and $y_t$ are cubic
couplings.  Note that this framework suggests a hierarchy between $g$,
$y_t$ on one side, and $\lambda$ on the other side, but does not
predict that $\lambda$ should vanish at $\mpl $, again as indicated by
data. Actually, since $\lambda$ scans by a relative amount ${\Ord}(1/\sqrt{N})$, a vanishing value of $\lambda(\mpl )$ turns out to
be fairly improbable in this setup.

\subsection{Criticality as an attractor}

We can also envisage a different situation within the multiverse hypothesis, motivated by
the observation that the measured value of both $M_h$ and $v$ (or equivalently the parameters $m_h$ and $\lambda$ of the potential) looks special, in the
sense that they correspond to near-critical parameters separating two
phases.
Indeed, as remarked in ref.~\cite{Giudice:2006sn}, also Higgs
naturalness can be viewed as a problem of near-criticality between  
the broken and unbroken EW phases. 
This leads to the speculation that within the multiverse
critical points are attractors and the
probability density in the space of parameters is peaked around the boundaries
between different phases.
In this picture generic universes are likely to live
near critical lines, and the Higgs
parameters found in our universe are not at all special, but correspond to the most likely occurrence among all the possible values which they can assume.

There are many natural phenomena in which near-criticality emerges as
an attractor~\cite{Bak:244579}. A typical example is given by the slope angle of
sand dunes. While one could expect to find in a beach sand dunes with
any possible slope angles, in practice the vast majority of dunes have
a slope angle roughly equal to the so-called ``angle of repose". The
angle of repose is the steepest angle of descent, which is achieved
when the material forming the pile is at a critical condition on the
verge of sliding. The angle of repose depends on size and shape of the
material granularity, and for sand is usually about 30--35
degrees. The typicality of finding sand dunes with slope angles near
the critical value is simply understood in terms of the forces that
shape dunes. Wind builds up the dune moving sand up to the top;
gravity makes the pile collapse under its own weight when the dune is
too steep. As a result, near-criticality is the most likely condition,
as a compromise between two competing effects.

Something similar could happen with the Higgs parameters. Suppose that the vacuum structure of the fields $\Phi$ participating
in Planckian dynamics prefers low values of the Higgs quartic coupling, so that its probability distribution is not uniform, but is a
monotonically decreasing function of $\lambda$.
Once $\lambda$ becomes smaller than
the critical value, the Higgs potential develops an instability at
large field values. If tunnelling is sufficiently fast, the Higgs
field slides towards Planckian scales.
Such large Higgs-field configurations
will in general affect the scalar potential of the fields $\Phi$,
which will readjust into a different vacuum structure. The new vacua
will give a different probability distribution for the Higgs quartic
coupling and it is imaginable that now larger values of
$\lambda$ are preferred. In summary: universes in the stable or
metastable phases will experience pressure towards small $\lambda$;
universes in the unstable phase will experience pressure towards large
$\lambda$. As a result, the most probable universes lie around the
critical line separating the two phases.

In an alternative scenario, when the Higgs field is destabilized the large negative value of the potential may form a bubble of AdS space with a negative cosmological constant of order
$-\mpl ^4$ in its interior.\footnote{Recall, however, that in our universe the cosmological constant is 120 orders of magnitude smaller than the na\"ive estimate $M_{\rm Pl}^4$.} Such regions of space would rapidly
contract and finally disappear. Therefore, the cosmological evolution
removes regions that correspond to unstable EW vacua, leaving again the vast
majority of universes crowded around the critical boundary.

Finally, the proximity of
our universe to an inhospitable phase, as shown in
figure~\ref{fig:regionsPlanck}, could be viewed as an indication that the anthropic principle
is at work, in a way similar to the case of the
cosmological constant~\cite{Weinberg:1987dv}. One can assume, as
before, that the probability distribution function of $\lambda(\mpl )$
in the multiverse is skewed towards the lowest possible values, making
it more likely for our universe to live in the leftmost region of
figure~\ref{fig:regionsPlanck}. The anthropic boundary of EW instability
limits the allowed parameter space, giving a justification of why our
universe is `living dangerously', with conditions for stability barely
satisfied.

\section{Partial summary on vacuum stability}

The measurement of the Higgs mass $M_h$ has determined the last
unknown parameter of the SM, fixing the Higgs quartic coupling $\lambda$. In order to investigate whether this result contains
any useful information about physics at shorter distances, we extrapolated $\lambda$ to high energy in search for
clues, assuming that no new physics degrees of freedom apart from the standard model ones are relevant. Just as high-energy extrapolations of the gauge coupling
constants gave us hints about a possible grand unification of
fundamental forces, so the extrapolation of $\lambda$ has revealed an
unexpected feature of the SM. The intriguing result is that, assuming the validity of
the SM up to very high energy scales, the measured value of $M_h$ is
near-critical, in the sense that it places the EW vacuum right at the
border between absolute stability and metastability. Because of the
present experimental uncertainties on the SM parameters (mostly the
top quark mass), we cannot conclusively establish the fate of the EW
vacuum, although metastability is now preferred at 98.6\% C.L.

The special coincidence found in the value of $M_h$ warrants a refined
calculation of the high-energy extrapolation of $\lambda$. In chapter~\ref{2loops}
we extracted the fundamental SM parameters $\lambda$ (quartic Higgs coupling), $m_h$ (Higgs mass term), $y_t$
(top quark Yukawa coupling) 
from the precisely measured values of the Higgs, top, $W$ and $Z$ masses and from
the Fermi constant at full NNLO, by performing dedicated two-loop computations.
All couplings have been extrapolated to large energies using the RGE equations, now known at NNLO (three loops).
We could then compute the effective potential known with two-loop accuracy.

A second objective was to investigate the significance
of the measured value of $M_h$, in view of its high-energy
extrapolation. A first observation is that $\lambda$, together with
all other SM coupling constants, remains perturbative in the entire
energy domain between the Fermi and the Planck scales.
The most important result concerns the stability of the Higgs
potential. The critical condition for stability is defined as the
vanishing of the effective coupling $\lambda_{\rm eff}$ of 
\eqref{eff-potential-high-h} at some energy scale $\Lambda_I$. We find
$\Lambda_I= 10^{10}$--$10^{12}$~GeV, suggesting that the instability
is reached well below the Planck mass. The presence of an instability
at an intermediate scale could be interpreted as a sign of a
new-physics threshold around $\Lambda_I$.
It is suggestive that
neutrino masses, axion, and inflation give independent indications for
new dynamics at roughly similar energy scales. The hypothetical new
physics could be responsible for a matching condition $\lambda \approx
0$ at a scale near $\Lambda_I$.

Another peculiarity found in the extrapolation of $\lambda$ is its
slow running at high energy. This is due to a combination of two
factors: the reduction of all SM couplings at high energy and an
accidental zero of $\beta_\lambda$ at a scale of about
$10^{17}$--$10^{18}$~GeV. It is the slow running of $\lambda$ at high
energy that saves the EW vacuum from premature collapse, in a
situation where $\Lambda_I\ll \mpl$. Were $\beta_\lambda$ large and
negative above $\Lambda_I$, we could not live with an instability
scale much smaller than the cutoff scale, without being confronted
with early vacuum decay. Unfortunately, for the moment we have no way
to tell whether this special condition allowing for a prolonged vacuum
lifetime is just a numerical coincidence or an important feature of
the SM.

At any rate, the smallness of $\beta_\lambda$ at high energy makes it
possible to assume that there is no new-physics threshold around
$\Lambda_I$ and that the SM continues to be valid up to the
quantum-gravity scale, since the tunnelling probability remains
small. In this context, the value of $\lambda (\mpl )$ may be regarded
as `normal' for a SM coupling. Indeed, as discussed in
section~\ref{sec:scc}, the ratios $\sqrt{4|\lambda|}/y_t$ and
$\sqrt{8|\lambda|}/g_2$ (which, at low energy, correspond to $M_h/M_t$
and $M_h/M_W$, respectively) are of order unity both at the Fermi and
Planck scales. The vanishing of $\lambda$ at an intermediate scale
could then be purely accidental. After all, the Higgs quartic is the
only SM coupling that can cross zero during its RG evolution, since
$\lambda =0$ is not a point of enhanced symmetry.

In our view, the most interesting aspect of the measured value of
$M_h$ is its near-criticality. We have thoroughly
studied the condition of near-criticality in terms of the SM
parameters at a high scale, which we identified with the Planck
mass. This procedure is more appropriate than a study in terms of
physical particle masses, since it is more likely that special
features are exhibited by high-energy parameters, just like in the
case of gauge coupling unification.
We have found that near-criticality is manifest also when we explore
the phase diagram as a function of high-energy SM couplings. Moreover,
we found evidence for multiple near-critical conditions: the
measured SM parameters roughly correspond to the minimum values of
both the Higgs quartic coupling $\lambda (\mpl )$ 
and the top Yukawa
coupling $y_t (\mpl )$ (at fixed gauge couplings) that allow for the existence of a sufficiently
long-lived EW vacuum. Furthermore, at fixed top Yukawa coupling, the maximum possible values of the gauge
couplings $g(\mpl )$ are preferred.

Finally, in section~\ref{interpretations} we have illustrated a set of possible scenarios which could explain the values of the SM couplings at high energies. It is interesting that explanations can be found in the context of the multiverse without relying on anthropic arguments: near-criticality can be achieved by cosmological selection and/or by probability distributions following from some unknown Planck-scale dynamics.

\addtocontents{toc}{\linespread{2}\selectfont}

\chapter{Summary and conclusions}

After about two years of operation of the LHC and the remarkable discovery of a Higgs-like particle of 125 GeV mass, the view of a natural Fermi scale is still under scrutiny, with three different lines of investigation: the more precise measurements of the properties of the same Higgs-like boson, the direct searches of new particles that are expected to accompany the Higgs boson, and several precision measurements in flavour physics.
While no clear hint of physics beyond the Standard Model has appeared up to now in either of these searches, the negative results allow to put limits on the masses and couplings of the expected new particles, which in turn translate into valuable informations about the nature of the physical phenomena taking place at the Fermi scale.

Taking naturalness as a guiding principle, the lightness of the Higgs boson calls for the existence of new physics related to EWSB in the TeV range, not too far above the electroweak scale. In most motivated extensions of the SM this implies also the appearance of new flavour physics phenomena at similar energies. For this to be the case, however, some physical mechanism must be operative to avoid the presence of unseen large deviations from the CKM picture of flavour and \CP\ violation: if present at all they must be kept under some control. This is a problem that one has to face in any realistic model of BSM physics at the weak scale.

The approximate $\U(2)^3$ symmetry exhibited by the quark sector of the SM Lagrangian can provide such a mechanism of suppression of the unwanted flavour and \CP\ effects.
In the first part of this thesis we have explored this possibility in an effective field theory approach, where all the BSM degrees of freedom are encoded in the Wilson coefficients of a set of effective operators of dimension greater than four. We have considered in particular the hypothesis that the breaking of the $\U(2)^3$ symmetry is governed only by the parameters responsible for the light quark masses and mixings; we have then generalized this assumption to the generic case where all the possible breaking terms are included.
 
In both cases, the main message that emerges from these general EFT considerations is that the wealth of current flavour physics data is still broadly compatible with new phenomena hiding at an interesting scale of about 3 TeV $\approx 4\pi v$. This is clearly a relevant energy range for BSM theories of EWSB, either perturbative or strongly interacting. We find this a particularly significant fact from many different points of view:  most of all it gives  reasons to think that new physics searches in the flavour sector may be about to explore an interesting realm of phenomena.

There are many observables where deviations from the SM could show up in view of near-future experimental progress; among these particular attention is due to \CP\ violation in the mixing of the $B_s$ system and in rare $B$ and $K$ decays, as described in chapter~\ref{U2}. 

Suppose that a significant deviation from the SM emerged in the experiments to come, which could be accounted for in the effective framework described above. How could one tell that $\U(2)^3$ is the relevant approximate symmetry, without uncovering by direct production the underlying dynamics (supersymmetry, a new strong interaction or whatever)? The best way would be to study the correlation between $B$ decays with $s$ and $d$ quarks in the final state, which would have to be the same as in the SM.
Such correlation is in fact also expected in MFV. However, the only way to have effects in MFV similar to the ones discussed in Minimal $\U(2)^3$ requires the presence of two Higgs doublets, one coupled to the up quarks and one to the down quarks, with large values of the usual $\tan{\beta}$ parameter \cite{DAmbrosio:2002ex,Kagan:2009bn}. This, in turn, would have other characteristic effects  not necessarily expected in Minimal $\U(2)^3$. 
In the case of small $\tan\beta$ or with one Higgs doublet only, distinguishing MFV from $\U(2)^3$ would be straightforward by means of the additional effects like \CP\ violation in $B_s$ mixing or non-universal contributions to $B\to K\nu\bar\nu$ vs.\ $K\to\pi\nu\bar\nu$ decays.

The relation of these considerations with natural models of electroweak symmetry-breaking is manifest: both in supersymmetry and composite models it is non trivial how specific realizations cope with tests of flavour and \CP\ violation. In particular if the Higgs boson is a composite object,
solving the flavour problem is of primary importance for any realistic model.
A remarkable fact is that the $\U(2)^3$ symmetry naturally provides for a distinction between the third and the first two generations, a feature that is required both in supersymmetry and in composite Higgs models for several different reasons.
More in general, the interplay of measurements in flavour, electroweak and Higgs physics may help answering the relevant question of whether the mechanism ultimately responsible for EWSB is weakly or strongly interacting.

Finally, a coherent picture of flavour physics at the weak scale may also shed light on another fundamental, and perhaps not unrelated question in particle physics: what is the origin of the three generations of matter and of their peculiar pattern of masses and mixings?

\bigskip

One way to implement a natural Fermi scale is to make the Higgs particle, one or more of them, a pseudo-Goldstone boson of a new strong interaction in the few TeV range. 
A meaningful question is then if and how a Higgs boson of 125 GeV mass fits into this picture, which requires spin-$\frac{1}{2}$ resonances, partners of the top, with a semi-perturbative coupling to the strong sector and a mass not exceeding about 1 TeV.

Not the least difficulty in addressing this question is the variety of  possible specific implementations of the pseudo-Goldstone-boson-Higgs picture, especially with regard to the different representations of the spin-$\frac{1}{2}$ resonances and the various ways to describe flavour. A further problem is represented by the limited calculability of key observables in potentially complete models, due to their strongly interacting nature. 
To circumvent these difficulties, we have adopted some simple partial-compositeness Lagrangians and assumed that they catch the basic phenomenological properties of the theories under consideration.

Remarkably, some of the strongest constraints on models of partial compositeness come from flavour physics, enforced by the relative smallness of the coupling of fermionic resonances in the composite sector required to reproduce the measured value of the Higgs mass. This is especially true for the case of a strong dynamics with anarchic flavour structure, where the suppression of dangerous flavour effects relies entirely on the smallness of the elementary-composite mixings for the light generations of quarks. The introduction of a large flavour symmetry presents some additional difficulties with respect to the effective field theory analysis, mainly because of the new relations among parameters that are introduced. Minimal flavour violation based on $\U(3)^3$ requires a large degree of compositeness of (at least one chirality of) the light quarks, which is strongly constrained by collider studies. The requirement of both a nontrivial flavour structure in the strong sector, and the disentanglement of the properties of the first and third generation of quarks naturally leads to $\U(2)^3$.

We have seen in chapter~\ref{CHMbounds} how the $\U(2)^3$ symmetry and its breaking can be implemented in a generic composite Higgs model, and we have studied its peculiar consequences for flavour and \CP\ violation, comparing it to the cases of anarchy and MFV. The general message that emerges from our analysis
is pretty clear: top partners with a mass equal or less than about 1 TeV, as required by naturalness, are often not compatible with all the constraints, and one has to choose accurately the underlying model in order to accommodate them. An approximate $\U(2)^3$ symmetry appears favorite, if not necessary.

The attempt to include many different possibilities, though motivated, is also a limit of our analysis. A next step might consist in selecting a few emerging cases to analyze them in more detail, perhaps going beyond the partial-compositeness  effective description. For this we think that table~\ref{tab:mmin} offers a useful criterion.
It is in any event important and a priori non trivial that some models with a suitable structure emerge that look  capable of accommodating a 125 GeV Higgs boson without too much fine tuning, i.e. with top partners in an interesting mass range for discovery at the LHC.

\bigskip

Is the newly found Higgs boson alone or is it a member of a family of scalar particles? Now that we know that one Higgs boson exists this is a compelling question, both {\it per se} and in many motivated different scenarios. While also non-minimal composite Higgs models can include additional scalar states, a context where their presence is not just a possibility, but they are required for consistency, is the one of supersymmetric extensions of the SM. In particular, naturalness requires at least another doublet of Higgs particles to lie close to the Fermi scale.
It is in fact not inconceivable that part of such an extended Higgs system be the lightest fragments of the entire supersymmetric particle spectrum, with the possible exception of the LSP.

It is thus of particular interest to know which impact the measurements -- current and foreseen -- of the different signal strengths of the newly found resonance, as well as of its mass, have on this problem and how they compare with the potential of the direct searches of extra states.
It turns out that the properties of the Higgs boson discovered at the LHC play, once again, a dominant role in determining the features of a natural supersymmetric model at the Fermi scale.

In the MSSM a well-known fact is that a 125 GeV Higgs mass can be obtained only with a large radiative contribution from the top-stop sector, which in turn implies a sizeable amount of fine-tuning. In addition, always in the MSSM, a second doublet is excluded below about 300 GeV
from the determination of the various Higgs decay rates, reaching at present a 30\% precision in many channels, which would be influenced by the mixing with a light scalar state.

In the NMSSM the introduction of an additional scalar singlet coupled to the doublets of the MSSM alleviates many fine-tuning problems: it allows for lighter stops by virtue of the larger tree-level contribution to the Higgs mass, and at the same time it reduces the tuning of the weak scale $v$. 
A difficulty in a thorough study of the Higgs sector of the NMSSM is nevertheless represented by the large number of parameters, many of which related to the singlet potential and thus at present not measurable. In chapter~\ref{NMSSM} we focussed on the \CP-even sector, assuming a negligibly small \CP\ violation in the Higgs system. We tried to formulate our analysis as much as possible in terms of observable physical quantities, and without sticking to a particular model, in view of forthcoming searches to be performed at the LHC in a systematic way. In addition, we have made a number of simplifying assumptions on the parameter space, motivated by naturalness requirements. In our view the advantages of having an overall coherent analytic picture justify the introduction of these assumptions.

There are a few interesting simplified configurations where one of the \CP-even states is decoupled which allow a relatively simple description, at the same time catching the main phenomenological features of the extended Higgs sector. The main message that emerges from the analysis of these cases is that a new further state nearby is allowed by present data, unlike in the case of the MSSM. A variety of possible phenomenologies is expected for these new states, depending on their masses and mixings. 

We kept a particular attention on the region of parameter space at low $\tan\beta$ and $\lambda\approx 1$, which is most relevant for naturalness. Such large values of the singlet-Higgs coupling imply the presence of a threshold of strong dynamics well below the GUT scale. This is nevertheless compatible with gauge-coupling unification, as we demonstrated in a simple toy-model with partially composite Higgs fields.

It will be interesting to follow the progression of the searches of the Higgs system, directly or indirectly through the more precise measurements of the properties of the state already found at the LHC. It is also clear that the combined improvement in both these directions -- even if pretty challenging in some cases -- will allow to extensively explore the parameter space of the NMSSM in the next few years.
We believe that the framework outlined in chapter~\ref{NMSSM} should allow one to systematize these searches in a clear way. We also think that they should be pursued actively and independently from the searches of the superpartners.

\bigskip

In spite of the absence of any signal of the physical phenomena associated with a natural Fermi scale, the LHC has already provided relevant new informations about the possible nature of weak-scale physics -- be it strongly or weakly interacting -- first through the discovery of the Higgs boson and then through the more detailed determination of its properties. But is naturalness really a good guiding principle to explore the physics of EWSB? The first run of the LHC has also shown that Nature at the weak scale is necessarily fine-tuned to some amount. Recall also that, while the LHC had the certainty of a discovery (the Higgs boson or whatever else) guaranteed by the need for a consistent theory, the appearance of any further new phenomenon is not assured by any theorem.
What we have tried to show in the different parts of this thesis is that there are several possibilities where a natural Fermi scale with a mild amount of fine-tuning is compatible with the various experimental constraints, in particular with partners of the top quark -- either composite or supersymmetric -- and extra Higgs bosons within the range of forthcoming experimental searches. We think, therefore, that naturalness still remains the most motivated scenario for physics at the electroweak scale, as it already was in many other contexts of particle physics.

The second run of the LHC at a center-of-mass energy of 13 or 14 TeV will in any case be crucial for the answer to that question. If no departure from the SM will be found there, a shift of paradigm will very likely be required in order to understand the physical phenomena above the electroweak scale. A worrisome aspect of such an eventuality would be the impossibility to ignore very-short-distance physics, not accessible to direct experimental investigation. On the other hand, an intriguing side of this fact is that Fermi-scale physics may then provide some insight on unknown Planck-scale phenomena.

\bigskip

An interesting possibility in this direction, which has been studied in chapter~\ref{Vacuumstability}, comes from the impressive proximity of various SM couplings to their critical value associated with the stability of the electroweak vacuum. If the SM is assumed to hold without modifications up to very high energies, then the effective Higgs potential develops an instability at a scale of about $10^{11}$ GeV; its vacuum state $v\simeq 246$ GeV is metastable, with a decay-time much longer than the age of the universe. The Higgs quartic and top Yukawa couplings are critical parameters separating two phases: a tiny variation with respect to their measured values would stabilize the potential or lead to a fast decay of the vacuum. The LHC, measuring the last free parameter of the SM, has thus already provided valuable informations about short-distance physics even in the worst case where there is no BSM physics at the Fermi scale.

Provided it is not just a fortuitous coincidence, an
explanation of near-criticality almost necessarily requires the
existence of an underlying statistical system.
A multiverse of vacua where the coupling constants scan through the parameter-space and are statistically determined can provide such kind of a system.
Near-criticality may emerge there from
an appropriate probabilistic pressure in the space of coupling
constants, together with the anthropic requirement that selects
universes in which the life-friendly EW vacuum is sufficiently
long-lived. The principle of {\it living dangerously} populates universes
close to the boundary of a hospitable phase, just as it is conjectured
to happen in the case of the cosmological constant.

Near-criticality could also find an explanation in
the multiverse without any reference to anthropic reasoning. In
Nature there exist statistical systems in which criticality is an
attractor point of their dynamical evolution. If a similar
phenomenon took place in the multiverse, then the majority of
universes would populate regions close to phase transitions. Such an explanation of near-criticality of the Higgs mass
could also provide a link to the naturalness problem, since the
smallness of the mass parameter $m$ in the Higgs potential is
near-critical with respect to the EW symmetry-breaking phase
transition. It is indeed a remarkable experimental fact that both
$\lambda$ and $m$, the two parameters of the Higgs potential, happen
to lie very close to boundaries between different phases of the
SM. So, according to this interpretation, 
near-criticality of the Higgs
parameters would be a fairly generic property
and our universe would be unexceptional.

\bigskip

We have tried to summarize in this thesis a variety of very different scenarios of physics at the Fermi scale, both strongly and weakly interacting, natural and not. All of these scenarios are well motivated, and, if they should be verified with the forthcoming experimental progress, they would represent a major breakthrough in our understanding of Nature at very small scales. The diversity of the various presented topics makes our discussion inevitably partial and incomplete. We have thus focussed on a few situations which to us look most relevant from the point of view of naturalness and of the impact of near-future experimental improvements. The last part of the work constitutes perhaps an exception to this general attitude. Flavour and Higgs physics and their interplay are the main guidelines for all the considered subjects.

Whether or not physics at and beyond the electroweak scale can be ascribed to one of the frameworks considered here will be clear as soon as the LHC will provide new data at a higher collision energy. In any case, the progress of the next decade will give us a first clear picture of the phenomena taking place at the Fermi scale and may provide an answer to several fundamental questions of particle physics.

\addtocontents{toc}{\linespread{1}\selectfont}

\chapter*{Acknowledgements}
\addcontentsline{toc}{chapter}{Acknowledgements}
I am deeply grateful to Riccardo Barbieri for his patient and careful supervision during this thesis work and the many collaborations of the last two years, and for having taught me the basic concepts of doing scientific research, along with a lot of physics.

\noindent{I am indebted to Christophe Grojean, who made it possible for me to stay a long period at CERN under his supervision, profiting from a very stimulating environment.}

\noindent{I would like to thank Filippo Sala, David M. Straub, Andrea Tesi, Giuseppe Degrassi, Pier P. Giardino, Gian F. Giudice, Kristjan Kannike, Alberto Salvio, Alessandro Strumia for the pleasant and fruitful collaborations that led to the results presented in this work, and Damiano Anselmi for the collaboration at the beginning of my PhD.}

\noindent{Thanks also for the many interesting discussions about physics -- but not only -- to Paolo Campli, Caterina Vernieri, Enrico Bertuzzo, Marco Farina, Enirco Trincherini, Oleksii Matsedonskyi, David Pirtskhalava and all my officemates and colleagues at Scuola Normale; to my colleagues and friends from Pisa Diego Redigolo, Andrea Pallottini, Marco Tinivella, Giuseppe Vitagliano, Luca Morescalchi, Michele Mancarella; to Riccardo Torre, Andrea Thamm, Ennio Salvioni, Ferdinando Giordano, Raffaele T. D'Agnolo, Paolo Lodone, Christine Hartmann, David Marzocca and to all the people that I have met in Geneva and at the many conferences all over the world.


\noindent{Thanks to Francesco Fidecaro, Giancarlo Cella, Mauro Dell'Orso, Gabriele Vajente for letting me rediscover the beauty of classical mechanics, showing me how to teach it.}

\noindent{I am very grateful to my family
as well as to all the teachers from whom I learned a lot about physics, mathematics and science in general, in particular Carla Fr\"ohlich, Giuseppe Buttazzo, Pasquale Maiano, Mihail Mintchev, Pietro Menotti.}

\noindent{Thanks to CERN, the Galileo Galilei Institute in Florence, the Institute of High-Energy Physics in Beijing, the Technical University of Munich for hospitality, and, it goes without saying, to Scuola Normale Superiore.}

\noindent{Finally, a big `thank you' to all the people that have been close during these years, to my family and to my friends, a page not being enough to mention them all.}

\noindent{This work has been partly supported by a Marie Curie Early ITN Fellowship of the European Community's Seventh Framework Programme under contract PITN-GA-2008-237920-UNILHC, and by MIUR under contracts 2008XM9HLM, 2010YJ2NYW-010.}

\cleardoublepage

\phantomsection

\appendix

\addtocontents{toc}{\linespread{2}\selectfont}

\chapter{Loop functions for flavour amplitudes}\label{app:loops}

\addtocontents{toc}{\linespread{1}\selectfont}

The loop functions appearing in equations \eqref{DS2}--\eqref{btogammag} are, in the 't Hooft-Feynman gauge for the gauge bosons \cite{Buras:1998raa},
\begin{align}
B_0(x) &= \frac{1}{4}\left(\frac{x}{1 - x} + \frac{x\log x}{(x-1)^2}\right),\\
C_0(x) &= \frac{x}{8}\left(\frac{x-6}{x-1} + \frac{3x + 2}{(x-1)^2}\log x\right),\\
D_0(x) &= -\frac{4}{9}\log x + \frac{25x^2 - 19x^3}{36(x-1)^3} + \frac{x^2(5x^2 - 2x - 6)}{18(x-1)^4}\log x,\\
E_0(x) &= -\frac{2}{3}\log x + \frac{x^2(15 - 16 x + 4x^2)}{6(1-x)^4}\log x + \frac{x(18-11x-x^2)}{12(1-x)^3},\\
D_0'(x) &= -\frac{8x^3 + 5x^2 - 7x}{12(1-x)^3} + \frac{x^2(2-3x)}{2(1-x)^4}\log x,\\
E_0'(x) &= -\frac{x(x^2 - 5x - 2)}{4(1-x)^3} + \frac{3}{2}\frac{x^2}{(1-x)^4}\log x,\\
S_0(x, x) &= \frac{4x - 11x^2 + x^3}{4(1-x)^2} - \frac{3x^2\log x}{2(1-x)^3},\\
S_0(x, y) &= y\left(\log\frac{x}{y} - \frac{3x}{4(1-x)} + \frac{3x^2\log x}{4(1-x)^2}\right) +{\Ord}(y^2).
\end{align}

\chapter{Quark bilinears in Generic $\U(2)^3$}\label{app:bilinears}

In the basis where the spurions take the form \eqref{genericV}, \eqref{genericY} the chirality-conserving bilinears of \eqref{ccL}--\eqref{ccd} can be written in the form
\begin{align}\label{ccmatrix}
X_{Lu}^{\rm kin} &= A_{uL}\mathbbm{1} + B_{uL} L_{23}^u\I_{32}^L (L_{23}^u)^{T}, & X_{Lu}^{\rm int,\alpha} &= A_{uL}^{\alpha}\mathbbm{1} + B_{uL}^{\alpha}U_{23}^{u,\alpha}\I_{32}^L (U_{23}^{u,\alpha})^{\dag},\\
X_{Ru}^{\rm kin} &= A_{uR}\mathbbm{1} + B_{uR} (R_{23}^u)^T\I_{32}^{Ru} R_{23}^u, & X_{Ru}^{\rm int,\alpha} &= A_{uR}^{\alpha}\mathbbm{1} + B_{uR}^{\alpha}(U_{23}^{u,\alpha})^{\dag}\I_{32}^{Ru} U_{23}^{u,\alpha},
\end{align}
plus analogous expressions for the down sector, where $\I_{32}^{I} = \diag(0, \Ord(\epsilon^2_{I}), 1)$, the $A$'s and $B$'s are real functions of the parameters of \eqref{ccL}--\eqref{ccd}. The chirality-breaking bilinears of \eqref{cb_u},\eqref{cb_d} are
\begin{align}
Y_u &= \lambda_t (L_{23}^u\I_3 R_{23}^u + L_{12}^u\Delta \tilde Y^{\rm diag}_u V_{12}^u), & M_u^{\beta} &= \lambda_t(a_u^{\beta}{U}_{23}^{u,\beta}\I_3{V}_{23}^{u,\beta} + d_u^{\beta} L_{12}^u\Delta \tilde Y^{\rm diag}_u V_{12}^u),\\
Y_d &= \lambda_b (U_{23}^d\I_3 R_{23}^d + U_{12}^d\Delta \tilde Y^{\rm diag}_d V_{12}^d), & M_d^{\beta} &= \lambda_b(a_d^{\beta}{U}_{23}^{d,\beta}\I_3{V}_{23}^{d,\beta} + d_d^{\beta} U_{12}^d\Delta \tilde Y^{\rm diag}_d V_{12}^d),
\end{align}
where $\I_3 = \diag(0,0,1)$, $\Delta \tilde Y_{u,d}^{\rm diag} = \diag(y_{u,d}, y_{c,s},0)$, and $y_{u,d,c,s}$ are the diagonal entries of $\Delta Y_{u,d}^{\rm diag}$. Everywhere $U_{ij}$ ($V_{ij}$) stand for unitary left (right) matrices in the $(i,j)$ sector, while $L_{ij}$ ($R_{ij}$) indicate orthogonal left (right) matrices. In particular $U_{12}^d = \Phi_L L_{12}^d$ and $V_{12}^{u,d} = \Phi_R^{u,d} R_{12}^{u,d}$.

Going to the physical mass basis through the transformations \eqref{U2transformationsreal}, \eqref{U2transformationsrealR} the chirality-conserving operators become
\begin{align}
X_{dL}^{{\rm int},\alpha}&\mapsto A_{dL}^{\alpha}\mathbbm{1} + B_{dL}^{\alpha}(U_{12}^d)^{\dag} U_{23}^{d,\alpha}\I_{32}^L(U_{23}^{d,\alpha})^{\dag}U_{12}^d,\\
X_{dR}^{{\rm int},\alpha}&\mapsto A^{\alpha}_{dR}\mathbbm{1} + B^{\alpha}_{dR}(V_{12}^d)^{\dag} V_{23}^{d,\alpha}\I_{32}^{Rd}(V_{23}^{d,\alpha})^{\dag}V_{12}^d,
\end{align}
and the $\sigma_{\mu\nu}$-terms are
\begin{equation}
M_d^{\beta}\mapsto \lambda_b\big(a_d^{\beta}(U_{12}^d)^{\dag} U_{23}^{\beta}\I_3 V_{23}^{\beta} V_{12}^d + c_d^{\beta} \Delta\tilde Y_d^{\rm diag}\big),
\end{equation}
plus analogous expressions for the up sector.

Here we give the results obtained after rotation to the mass basis for all possible quark bilinears, factorizing out explicitly all the phases, CKM matrix elements and quark masses. We define $\xi_{ij} = V_{ti}^*V_{tj}$, and $\zeta_{ij} = V_{ib}V_{jb}^*$. In chirality-breaking bilinears $\alpha = \gamma(g)$ for (chromo)electric dipole operators, while $\alpha = {\rm cb}$ for other generic interaction bilinears. All the parameters are real.

\section{Up quark sector}

Chirality-conserving LL and RR currents:
\begin{align}
X_{12}^{uL} &= c_D\zeta_{uc}, & X_{13}^{uL} &= c_t e^{i\phi_t}\zeta_{ut}, & X_{23}^{uL} &= c_t e^{i\phi_t}\zeta_{ct},\\
X_{12}^{uR} &= \tilde c_D e^{i(\phi_1^u - \phi_2^u)}\zeta_{uc}\frac{s^u_R}{s^u_L}\Big(\frac{\epsilon^u_R}{\epsilon_L}\Big)^2, & X_{13}^{uR} &= \tilde c_t e^{i(\tilde\phi_t + \phi_1^u)}\zeta_{ut}\frac{s^u_R}{s^u_L}\frac{\epsilon^u_R}{\epsilon_L}, & X_{23}^{uR} &= \tilde c_t e^{i(\tilde\phi_t + \phi_2^u)}\zeta_{ct}\frac{\epsilon^u_R}{\epsilon_L}.\label{firstOcoefficient}
\end{align}

\noindent Flavour-conserving dipole operators:
\begin{align}
M_{11}^u &= c_D^{\rm \alpha}e^{i(\phi_D^{\rm \alpha} - \phi_1^u)}\zeta_{uu}\frac{s^u_R}{s^u_L}\frac{\epsilon^u_R}{\epsilon_L}, & M_{22}^u &= c_D^{\rm \alpha}e^{i(\phi_D^{\rm \alpha} - \phi_2^u)}\zeta_{cc}\frac{\epsilon^u_R}{\epsilon_L}, & M_{33}^u &= a_t e^{i\alpha_t}.
\end{align}

\noindent Flavour-changing, chirality-breaking operators:
\begin{align}
M_{12}^u &= c_D^{\rm \alpha}e^{i(\phi_D^{\rm \alpha} - \phi_2^u)}\zeta_{uc}\frac{\epsilon^u_R}{\epsilon_L}, & M_{21}^u  &= c_D^{\rm \alpha}e^{i(\phi_D^{\rm \alpha} - \phi_1^u)}\zeta_{uc}^*\frac{s^u_R}{s^u_L}\frac{\epsilon^u_R}{\epsilon_L},\\
M_{13}^u &= c_t^{\rm \alpha}e^{i\phi_t^{\rm \alpha}}\zeta_{ut}, & M_{31}^u &= \tilde c_t^{\rm \alpha}e^{i(\tilde\phi_t^{\rm \alpha} - \phi_1^u)}\zeta_{ut}^*\frac{s^u_R}{s^u_L}\frac{\epsilon^u_R}{\epsilon_L},\\
M_{23}^u &= c_t^{\rm \alpha}e^{i\phi_t^{\rm \alpha}}\zeta_{ct}, & M_{32}^u &= \tilde c_t^{\rm \alpha}e^{i(\tilde\phi_t^{\rm \alpha} - \phi_2^u)}\zeta_{ct}^*\frac{\epsilon^u_R}{\epsilon_L}.
\end{align}

\section{Down quark sector}

Chirality-conserving LL and RR currents:
\begin{align}
X_{12}^{dL} &= c_K\xi_{ds}, & X_{13}^{dL} &= c_B e^{i\phi_B}\xi_{db}, & X_{23}^{dL} &= c_B e^{i\phi_B}\xi_{sb},\\
X_{12}^{dR} &= \tilde c_K e^{i(\phi_1^d - \phi_2^d)}\xi_{ds}\frac{s^d_R}{s^d_L}\Big(\frac{\epsilon^d_R}{\epsilon_L}\Big)^2, & X_{13}^{dR} &= \tilde c_B e^{i(\tilde\phi_B + \phi_1^d)}\xi_{db}\frac{s^d_R}{s^d_L}\frac{\epsilon^d_R}{\epsilon_L}, & X_{23}^{dR} &= \tilde c_B e^{i(\tilde\phi_B + \phi_2^d)}\xi_{sb}\frac{\epsilon^d_R}{\epsilon_L}.
\end{align}

\noindent Flavour-conserving dipole operators:
\begin{align}
M_{11}^d &= \lambda_b\,c_K^{\rm \alpha}e^{i(\phi_K^{\rm \alpha} - \phi_1^d)}\xi_{dd}\frac{s^d_R}{s^d_L}\frac{\epsilon^d_R}{\epsilon_L}, & M_{22}^d &= \lambda_b\,c_K^{\rm \alpha}e^{i(\phi_K^{\rm \alpha} - \phi_2^d)}\xi_{ss}\frac{\epsilon^d_R}{\epsilon_L}, & M_{33}^d &= \lambda_b\,a_b e^{i\alpha_b}.
\end{align}

\noindent Flavour-changing, chirality-breaking operators:
\begin{align}
M_{12}^d &= \lambda_b\,c_K^{\rm \alpha}e^{i(\phi_K^{\rm \alpha} - \phi_2^d)}\xi_{ds}\frac{\epsilon^d_R}{\epsilon_L}, & M_{21}^d  &= \lambda_b\,c_K^{\rm \alpha}e^{i(\phi_K^{\rm \alpha} - \phi_1^d)}\xi_{ds}^*\frac{s^d_R}{s^d_L}\frac{\epsilon^d_R}{\epsilon_L},\\
M_{13}^d &= \lambda_b\,c_B^{\rm \alpha}e^{i\phi_B^{\rm \alpha}}\xi_{db}, & M_{31}^d &= \lambda_b\,\tilde c_B^{\rm \alpha}e^{i(\tilde\phi_B^{\rm \alpha} - \phi_1^d)}\xi_{db}^*\frac{s^d_R}{s^d_L}\frac{\epsilon^d_R}{\epsilon_L},\\
M_{23}^d &= \lambda_b\,c_B^{\rm \alpha}e^{i\phi_B^{\rm \alpha}}\xi_{sb}, & M_{32}^d &= \lambda_b\,\tilde c_B^{\rm \alpha}e^{i(\tilde\phi_B^{\rm \alpha} - \phi_2^d)}\xi_{sb}^*\frac{\epsilon^d_R}{\epsilon_L}.\label{lastOcoefficient}
\end{align}

\chapter{Contributions to $\hat T$ from partial compositeness}\label{app:T}

We compute the one-loop contributions to $\Delta \rho = \hat T$ arising from top partner loops in models with partial compositeness. We do the calculation in the elementary-composite basis for fermions and vector bosons.

Consider a composite sector which has a custodial $\SU(2)_L\times \SU(2)_R$ symmetry, and the corresponding $\rho_L$ and $\rho_R$ vector resonances. After electroweak symmetry breaking, the resonances mix with the elementary $W$ bosons in the following way:
\begin{align}\label{mixings}
\raisebox{-.5cm}{\includegraphics[width=0.2\textwidth]{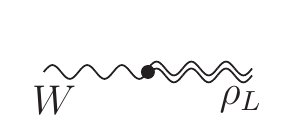}}\!\!\!\!\! &= i\frac{g}{g_{\rho}}m_{\rho}^2\equiv -im_{\rho}^2\Delta_L,&
\raisebox{-.5cm}{\includegraphics[width=0.2\textwidth]{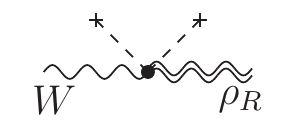}}\!\!\!\!\! &= i\frac{g}{g_{\rho}}\frac{v^2 g_{\rho}^2}{4}\equiv -im_{\rho}^2\Delta_R.
\end{align}

Since the composite sector is symmetric under the custodial $\SU(2)_L\times \SU(2)_R$, elementary fermion insertions are needed in the loop in order to have a non-vanishing contribution. Moreover at least four Higgs insertions are needed overall to generate the operator ${\O}_H = (H^{\dag}D_{\mu}H)^2$.
The contributions from the following two-point functions contribute to $\hat T$ after elementary-composite mixings on the external legs:
\begin{itemize}
\item $\hat T^{LL}\propto \Delta_L^2$ from $W$-$\rho_L$ mixing on both external legs of $\langle\rho_L\rho_L\rangle$: four Higgs insertions inside the fermion loop are needed;
\item $\hat T^{WL}\propto \Delta_L$ from $W$-$\rho_L$ mixing on the $\rho_L$ leg of $\langle W\rho_L\rangle$ and $\langle\rho_L W\rangle$: four Higgs insertions inside the fermion loop;
\item $\hat T^{RR}\propto \Delta_R^2$ from $W$-$\rho_R$ mixing on both external legs of $\langle\rho_R\rho_R\rangle$: all the Higgs insertions are in the $L$-$R$ mixings;
\item $\hat T^{LR}\propto \Delta_L\Delta_R$ from $W$-$\rho_L$ and $W$-$\rho_R$ mixing on the external legs of $\langle\rho_L\rho_R\rangle$ and $\langle\rho_R\rho_L\rangle$: two Higgs insertions are in the $L$-$R$ mixing and two are in the fermion loop;
\item $\hat T^{WR}\propto \Delta_R$ from $W$-$\rho_R$ mixing on the $\rho_R$ leg of $\langle W\rho_R\rangle$ and $\langle\rho_R W\rangle$: two Higgs insertions are in the $L$-$R$ mixing and two are in the fermion loop.
\end{itemize}

Here we give the leading terms in $v/f$ and in the elementary/composite quark mixings in the one-doublet (1D), two-bidoublets (2B) and triplet (T) models. We consider only third-generation quarks, since all the other mixings are negligible.

\section{Doublet model}

In the limit $\tilde Y\to 0$ one gets\footnote{i.e. neglecting the composite Yukawa couplings $\tilde Y$ with non-standard chirality, see footnote at p. \pageref{Ytildefootnote}.}
\begin{align}
\hat T^{LL}_{1D} &= \frac{13}{105}\frac{N_c}{(4 \pi)^2}\frac{m_t^2 Y^2}{M_f^2 x_t^2}, & \hat T^{WL}_{1D} &= 0,\\
\hat T^{LR}_{1D} &= \frac{13}{60}\frac{N_c}{(4 \pi)^2}\frac{m_t^2 Y^2}{M_f^2 x_t^2} & \hat T^{WR}_{1D} &= 0,\\
\hat T^{RR}_{1D} &= \frac{1}{6}\frac{N_c}{(4 \pi)^2}\frac{m_t^2 Y^2}{M_f^2 x_t^2},
\end{align}
and the total contribution is thus
\begin{equation}
\hat T_{1D} = \frac{71}{140}\frac{N_c}{(4 \pi)^2}\frac{m_t^2 Y^2}{M_f^2 x_t^2}.
\end{equation}
Including also the $\tilde Y$ terms one has
\begin{equation}
\hat T_{1D} = \frac{N_c}{(4 \pi)^2}\frac{m_t^2 Y^2}{M_f^2 x_t^2}\left(\frac{1}{210}(26 + 6 X + 9 X^2 - X^3 + 5 X^4) + R^2(13 + X + 8 X^2) + \frac{R^4}{6}\right),
\end{equation}
where $X\equiv \tilde Y/Y$ and $R\equiv M_f / (Y f)$.

\section{Bidoublet model}

In the limit $\tilde Y\to 0$ one gets
\begin{align}
\hat T^{LL}_{2B} &= -\frac{4}{105}\frac{N_c}{(4 \pi)^2}\frac{m_t^2 x_t^2 Y^2}{M_f^2}, & \hat T^{WL}_{2B} &= \frac{1}{15}\frac{N_c}{(4 \pi)^2}\frac{m_t^2 x_t^2 Y^2}{M_f^2},\\
\hat T^{LR}_{2B} &= -\frac{1}{5}\frac{N_c}{(4 \pi)^2}\frac{m_t^2 x_t^2 Y^2}{M_f^2}, & \hat T^{WR}_{2B} &= -\frac{5}{12}\frac{N_c}{(4 \pi)^2}\frac{m_t^2 x_t^2 Y^2}{M_f^2},\\
\hat T^{RR}_{2B} &= \frac{1}{3}\frac{N_c}{(4 \pi)^2}\frac{m_t^2 x_t^2 Y^2}{M_f^2},
\end{align}
and the total contribution is
\begin{equation}
\hat T_{2B} = -\frac{107}{420}\frac{N_c}{(4 \pi)^2}\frac{m_t^2 x_t^2 Y^2}{M_f^2}.
\end{equation}
Including also the $\tilde Y$ terms one has
\begin{equation}
\hat T_{2B} = \frac{N_c}{(4 \pi)^2}\frac{m_t^2 Y^2 x_t^2}{M_f^2}\left(\frac{12 + 272 X + 485 X^2 - 302 X^3 + 82 X^4}{420} - R^2\frac{37 + 98 X - 24 X^2}{60} + \frac{R^4}{3}\right),
\end{equation}
where $X\equiv \tilde Y/Y$ and $R\equiv M_f / (Y f)$.

\section{Triplet model}

The leading contributions are proportional to $\lambda_R^2$. In the limit of vanishing $\tilde Y$, and regularizing with a momentum cut-off, one gets:
\begin{align}
\hat T^{LL}_{T,R} &= -\frac{1}{2}\frac{N_c}{(4 \pi)^2}\frac{m_t^2 Y^3}{M_f^2 y_t x_t}, & \hat T^{WL}_{T,R} &= 0,\\
\hat T^{LR}_{T,R} &= 0, & \hat T^{WR}_{T,R} &= 0,\\
\hat T^{RR}_{T,R} &= \frac{N_c}{(4 \pi)^2}\frac{m_t^2 Y^3}{M_f^2 y_t x_t}\log\Big(\frac{\Lambda^2}{M_f^2} + 1\Big),
\end{align}
and the total $\lambda_R^2$ contribution is
\begin{equation}
\hat T_{T,R} = \left(\log\Big(\frac{\Lambda^2}{M_f^2} + 1\Big) - \frac{1}{2}\right)\frac{N_c}{(4 \pi)^2}\frac{m_t^2 Y^3}{M_f^2 y_t x_t}.
\end{equation}
Including also the $\tilde Y$ terms one has
\begin{equation}\begin{aligned}
\hat T_{T,R} = \frac{N_c}{(4 \pi)^2}\frac{m_t^2 Y^3}{M_f^2 y_t x_t}&\left(-\frac{15 - 2 X^2 - 26 X^2 + 24 X^3 - 17 X^4}{480}\right.\\
&\left.\quad+ \frac{R^2}{48}X(5X - 1) + R^4\log\Big(\frac{\Lambda^2}{M_f^2} + 1\Big)\right).
\end{aligned}\end{equation}

There are many diagrams that give contributions to $\hat T$ proportional to $\lambda_L^4$. In the limit $\tilde Y\to 0$ one gets
\begin{align}
\hat T^{LL}_{T,L} &= \frac{3}{7}\frac{N_c}{(4 \pi)^2}\frac{m_t^2 Y^2 x_t^2}{M_f^2}, & \hat T^{WL}_{T,L} &= \frac{8}{15}\frac{N_c}{(4 \pi)^2}\frac{m_t^2 x_t^2 Y^2}{M_f^2},\\
\hat T^{LR}_{T,L} &= \frac{19}{30}\frac{N_c}{(4 \pi)^2}\frac{m_t^2 x_t^2 Y^2}{M_f^2}, & \hat T^{WR}_{T,L} &= \frac{5}{12}\frac{N_c}{(4 \pi)^2}\frac{m_t^2 x_t^2 Y^2}{M_f^2},\\
\hat T^{RR}_{T,L} &= \frac{1}{3}\frac{N_c}{(4 \pi)^2}\frac{m_t^2 x_t^2 Y^2}{M_f^2},
\end{align}
and the total $\lambda_L^4$ contribution is
\begin{equation}
\hat T_{T,L} = \frac{197}{84}\frac{N_c}{(4 \pi)^2}\frac{m_t^2 x_t^2 Y^2}{M_f^2}.
\end{equation}
Including also the $\tilde Y$ terms one has
\begin{equation}\begin{aligned}
\hat T_{T,L} = \frac{N_c}{(4 \pi)^2}\frac{m_t^2 Y^2 x_t^2}{M_f^2}&\left(\frac{404 + 1168 X + 737 X^2 - 78 X^3 - 2 X^4}{420}\right.\\
&\left.\quad + R^2\frac{63 + 101 X - 22 X^2}{60} + \frac{R^4}{3}\right).
\end{aligned}\end{equation}

\chapter{Weak scale thresholds at one loop}\label{app:1loop}

We summarize here the one-loop corrections $\theta^{(1)}$
to the various SM parameters 
\begin{equation} 
\theta=\{\lambda, m, y_t, g_2, g_Y\}= 
\theta^{(0)}+\theta^{(1)}+\theta^{(2)}+\cdots.
\end{equation}
We perform one-loop computations in a generic $\xi$ gauge,
confirming that $\theta^{(1)}$ is gauge-independent, as it should.
Our expressions for $\theta^{(1)}$ are equivalent to the well known
expressions in the literature.  We write $\theta^{(1)}$ in terms of the
finite parts of the the Passarino-Veltman functions 
\begin{align}
A_0 (M) &=
M^2(1-\ln\frac{M^2}{\mub^2}), &  B_0(p;M_1,M_2) &= -\int_0^1
\ln\frac{xM_1^2+(1-x) M_2^2-x(1-x)p^2}{\mub^2}dx\ .  
\end{align} 
The dependence of $\theta^{(1)}$ on the renormalization scale $\mub$ reproduces
the well known one-loop RGE equations for $\theta$.
Below we report the expressions valid in the limit $M_b=M_\tau=0$; the negligible effect of
light fermions masses is included in our full code.

\section{The quartic Higgs coupling}

The one-loop result is obtained from equation \eqref{eq:lh1}:
\begin{align}
\delta^{(1)}\lambda &=
\frac{1}{(4\pi)^2V^4}\Re\bigg[  3M_t^2(M_h^2-4M_t^2) B_0(M_h;M_t,M_t)+3M_h^2 A_0(M_t)\notag\\
&+\frac{1}{4}\left(M_h^4-4 M_h^2 M_Z^2+12 M_Z^4\right)B_0(M_h;M_Z,M_Z) +\frac{M_h^2(7M_W^2-4 M_Z^2)}{2(M_Z^2-M_W^2)}  A_0(M_Z)    \notag\\
& +\frac{1}{2}(M_h^4-4M_h^2M_W^2+12M_W^4)B_0(M_h;M_W,M_W)-\frac{3M_h^2 M_W^2}{2(M_h^2-M_W^2)} A_0(M_h)\notag\\
&+\frac{M_h^2}{2}\left(\frac{3 M_h^2}{M_h^2-M_W^2} -\frac{3 M_W^2}{M_Z^2 - M_W^2} -11\right) A_0(M_W) +\frac{9}{4} M_h^4 B_0(M_h;M_h,M_h)\notag\\
&+\frac{1}{4}(M_h^4 +M_h^2(M_Z^2+2M_W^2-6 M_t^2)-8(M_Z^4+2M_W^4))
 \bigg] .
\label{d1Mh}
\end{align}
Each one of the terms in \eqref{eq:lh1} is gauge dependent, e.g.\
the one-loop correction to muon decay is
\begin{align}
{\left. \Delta r^{(1)}_0 \right|_{\rm fin}}&=\frac{1}{(4\pi V)^2}  
\bigg[3M_t^2 -M_W^2-\frac{M_Z^2}{2}-\frac{M_h^2}{2}+\frac{3M_W^2 A_0(M_h)}{M_h^2-M_W^2}-6A_0(M_t) \notag\\
&+\frac{6M_W^2-3M_Z^2}{M_W^2-M_Z^2}A_0(M_Z) + \bigg(9-\frac{3M_h^2}{M_h^2-M_W^2}-\frac{3M_W^2}{M_W^2-M_Z^2}\bigg)A_0(M_W)\notag\\
&+ 2A_0(\sqrt{\xi} M_W)+ A_0(\sqrt{\xi} M_Z) \bigg]
\end{align}
and the gauge dependence cancels out in the sum $\lambda^{(1)}(\mub)$.

\section{The Higgs mass term}

The correction is obtained from eq.~\eqref{eq:mh1}:
\begin{align}\label{Higgsmassthreshold1loop}
\delta^{(1)} m^{2}  &=
\frac{1}{(4\pi)^2 V^2}\Re \bigg[  6M_t^2(M_h^2-4M_t^2)B_0(M_h;M_t,M_t)+ 24 M_t^2 A_0(M_t)\nonumber\\
& +(M_h^4-4M_h^2M_W^2+12M_W^4)B_0(M_h;M_W,M_W)-2(M_h^2+ 6M_W^2) A_0(M_W) \nonumber  \\
&+\frac{1}{2}\left(M_h^4-4 M_h^2 M_Z^2+12 M_Z^4\right)B_0(M_h;M_Z,M_Z) -(M_h^2+ 6M_Z^2) A_0(M_Z)  \nonumber  \\
&+\frac{9}{2} M_h^4 B_0(M_h;M_h,M_h)-3M_h^2 A_0(M_h) \bigg] \ .
\end{align}

\section{The top Yukawa coupling}

The gauge-invariant one-loop correction to the top Yukawa coupling is obtained 
from~\eqref{eq:yt1}
\begin{align} \nonumber
\delta^{(1)} y_t  &=\frac{M_t}{\sqrt{2}V^3 (4\pi)^2}
\Re\bigg[
- \left(M_h^2-4 M_t^2\right) B_0\left(M_t;M_h,M_t\right) 
\\ &+  \nonumber
\frac{M_t^2 (80 M_W^2 M_Z^2\!-64 M_W^4\!-\!7
   M_Z^4)+40 M_W^2 M_Z^4-32 M_W^4 M_Z^2\!-\!17 M_Z^6 }{9  M_t^2
   M_Z^2}  B_0\left(M_t;M_t,M_Z\right)
  \\ &+ 
   \frac{\left(M_t^2 M_W^2+M_t^4-2 M_W^4\right)}{ M_t^2} B_0\left(M_t;0,M_W\right) +
    \left(\frac{3 M_W^2}{M_W^2-M_h^2}+1\right) A_0\left(M_h\right)\nonumber
   \\ &+
  \left(\frac{3 M_h^2}{M_h^2-M_W^2}+\frac{2
   M_W^2}{M_t^2}+\frac{3 M_W^2}{M_W^2-M_Z^2}-10\right) A_0\left(M_W\right)\nonumber
   \\ &+\nonumber
   \frac{\left(36 M_t^2 M_Z^2-56 M_W^2 M_Z^2+64 M_W^4-17
   M_Z^4\right)}{9  M_t^2 M_Z^2} A_0\left(M_t\right)
   \\ &+\nonumber
   \left( \frac{3 M_W^2}{M_Z^2-M_W^2} + \frac{32 M_W^4 - 40 M_W^2 M_Z^2 +17 M_Z^4}{9 M_t^2 M_Z^2} -3\right) A_0\left(M_Z\right) 
   \\ &+
    \frac{M_h^2}{2} - 3 M_t^2 - 9 M_W^2 +\frac{7 M_Z^2}{18}+\frac{64 M_W^2}{9 M_Z^2} \bigg]
- \frac{M_t g_3^2}{\sqrt{2}V(4\pi)^2}  \left(\frac{8
   A_0\left(M_t\right)}{M_t^2}+\frac{8}{3}\right) .
   \end{align} 
   
\section{The weak gauge couplings}

The one-loop correction to the $\SU(2)_L$ gauge coupling is obtained from
eq.~\eqref{eq:G2}: 
\begin{align}
\delta^{(1)} g_2  &= \frac{2M_W}{(4\pi)^2V^3} \Re\bigg[
\left(\frac{M_h^4}{6 M_W^2}-\frac{2 M_h^2}{3}+2 M_W^2\right) B_0\left(M_W;M_h,M_W\right)
\nonumber
   \\ &\nonumber
   +\left(-\frac{M_t^4}{M_W^2}-M_t^2+2
   M_W^2\right) B_0\left(M_W;0,M_t\right)
      \\ &\nonumber
      +\frac{1}{6} \left(-\frac{48 M_W^4}{M_Z^2}+\frac{M_Z^4}{M_W^2}-68 M_W^2+16
   M_Z^2\right) B_0\left(M_W;M_W,M_Z\right)
         \\ &\nonumber
        + \frac{1}{6}\left(M_h^2
   \left(\frac{9}{M_h^2-M_W^2}+\frac{1}{M_W^2}\right)+\frac{M_Z^2}{M_W^2}+M_W^2
   \left(\frac{9}{M_W^2-M_Z^2}+\frac{48}{M_Z^2}\right)-27\right)  A_0\left(M_W\right) 
      \\ &\nonumber
      + \left( 2- \frac{M_h^2 \left( M_h^2 + 8 M_W^2\right)}{6 M_W^2 \left(M_h^2 - M_W^2 \right)}\right) A_0\left(M_h\right) + \left(\frac{M_t^2}{M_W^2}+1\right) A_0\left(M_t\right)
      \\ &\nonumber
      +\frac{1}{6} \left(\frac{24 M_W^2}{M_Z^2} - \frac{M_Z^2}{M_W^2} + \frac{9 M_W^2}{M_Z^2 - M_W^2} - 17\right) A_0\left(M_Z\right) 
      \\ &
      +\frac{1}{36} \left(-3 M_h^2+18 M_t^2+\frac{288 M_W^4}{M_Z^2}-374 M_W^2-3 M_Z^2\right)\bigg] \ .
\end{align}
The one-loop correction to the $\U(1)_Y$ gauge coupling is obtained 
from eq.~\eqref{eq:G1}: 
 \begin{align} \nonumber
 \delta^{(1)} g_Y &= \frac{2 \sqrt{M_Z^2-M_W^2}}{(4\pi)^2 V^3}\Re\bigg[
   \left(\frac{88}{9} - \frac{124 M_W^2}{9 M_Z^2} + \frac{M_h^2+34 M_W^2}{6 (M_Z^2-M_W^2)} \right) A_0\left(M_Z\right)
 \\ &\nonumber+
    \frac{M_h^2-4 M_W^2}{2 (M_h^2-   M_W^2)}   A_0\left(M_h\right)  +
   \left(-\frac{7}{9} - \frac{M_t^2}{M_Z^2-M_W^2} +\frac{64 M_W^2}{9 M_Z^2} \right) A_0\left(M_t\right)
    \\ &\nonumber+
    \frac{M_h^4+ 2M_W^2 (M_W^2-15M_Z^2) +3M_H^2 (2 M_W^2 + 7 M_Z^2) }{6
   \left(M_h^2-M_W^2\right) \left(M_W^2-M_Z^2\right)}A_0\left(M_W\right)
    \\ &\nonumber 
   -\frac{M_t^4+M_W^2 M_t^2-2 M_W^4   }{M_W^2-M_Z^2}B_0\left(M_W;0,M_t\right)
   -\frac{M_h^4-4 M_Z^2 M_h^2+12 M_Z^4 }{6 (M_W^2-M_Z^2)} B_0\left(M_Z;M_h,M_Z\right)
    \\& \nonumber +
    \frac{M_h^4-4 M_W^2 M_h^2+12 M_W^4 }{6(   M_W^2- M_Z^2)}B_0\left(M_W;M_h,M_W\right)
    \\ &\nonumber+
    \frac{M_Z^6-48 M_W^6-68 M_Z^2 M_W^4+16 M_Z^4 M_W^2 }{6 M_Z^2
   \left(M_W^2-M_Z^2\right)}B_0\left(M_W;M_W,M_Z\right)
    \\ &\nonumber +
    \frac{1}{9}\left(7 M_t^2 - 23 M_W^2 +17 M_Z^2 -\frac{64 M_t^2 M_W^2}{M_Z^2} - \frac{9 M_W^2 (M_t^2-M_W^2)}{M_Z^2-M_W^2}\right) B_0\left(M_Z;M_t,M_t\right)
    \\ &\nonumber+
    \frac{M_Z^6 -48 M_W^6-68 M_Z^2 M_W^4+16 M_Z^4 M_W^2  }{6M_Z^2 \left(M_Z^2-M_W^2 \right)}B_0\left(M_Z;M_W,M_W\right) \\ & +
 \frac{1}{36} \bigg(\frac{576 M_W^4}{M_Z^2}-242 M_W^2-3 M_h^2 + 257
   M_Z^2 + \frac{36 M_W^2}{M_Z^2-M_W^2}+ M_t^2 \Big(82-\frac{256 M_W^2}{M_Z^2}\Big)\bigg)
   \bigg].
   \end{align}

\chapter{SM RGE equations up to three loops}\label{app:SM-RGE}

We list here the known results for the
renormalization group equations  up to three-loop order
for the SM couplings
$g_1,g_2,g_3,y_t$ and $\lambda$ in the $\MS$ scheme.
We write numerically those three-loop coefficients that involve the $\zeta_3$ constant.
Stopping for simplicity at two loops,
we also write RGE equations for the smaller bottom and tau Yukawa couplings and their contributions
to the RGE of the large couplings.
Our numerical code includes full RGE at 3 loops.

\section{Gauge couplings}

RGE for the hypercharge gauge coupling in GUT normalization ($g_1^2 = 5 g_Y^2/3$):
\begin{eqnarray}
\frac{dg_1^2}{d\ln\bar\mu^2} &=& \frac{g_1^4}{(4\pi)^2} \bigg[\frac{41 }{10}\bigg]
+ \frac{g_1^4}{(4\pi)^4} \bigg[ \frac{44 g_3^2}{5}+\frac{27 g_2^2}{10}+\frac{199 g_1^2}{50}-\frac{17 y_t^2}{10}
-\frac{y_b^2}{2}-\frac{3 y_{\tau }^2}{2} \bigg]  
\nonumber \\ && 
+ \frac{g_1^4}{(4\pi)^6} \bigg[ y_t^2\left( \frac{189 y_t^2}{16}-\frac{29 g_3^2}{5}-\frac{471 g_2^2}{32}
-\frac{2827 g_1^2}{800}\right) + \lambda \left( -\frac{36 \lambda}{5} +\frac{9 g_2^2}{5}+ \frac{27 g_1^2}{25}\right) 
\nonumber \\ &&
+\frac{297 g_3^4}{5}+\frac{789 g_2^4}{64}-\frac{388613 g_1^4}{24000}-\frac{3 g_3^2 g_2^2}{5} 
-\frac{137 g_3^2g_1^2}{75}+\frac{123 g_2^2g_1^2}{160} \bigg].
     \end{eqnarray}
RGE for the $\SU(2)_L$ gauge coupling:
\begin{eqnarray}
\frac{dg_2^2}{d\ln\bar\mu^2} &=& \frac{g_2^4}{(4\pi)^2} \bigg[-\frac{19}{6}\bigg] + 
\frac{g_2^4}{(4\pi)^4} \bigg[ 12 g_3^2+\frac{35 g_2^2}{6}+\frac{9 g_1^2}{10}-\frac{3 y_t^2}{2}
-\frac{3y_b^2}{2}-\frac{y_{\tau }^2}{2} \bigg]  
\nonumber \\ && 
+ \frac{g_2^4}{(4\pi)^6} \bigg[ y_t^2\left( \frac{147 y_t^2}{16}-7 g_3^2-\frac{729 g_2^2}{32}
-\frac{593 g_1^2}{160}\right) + \lambda \left( -3 \lambda +\frac{3 g_2^2}{2}+ \frac{3 g_1^2}{10} \right) 
\nonumber \\ &&
+81 g_3^4+\frac{324953 g_2^4}{1728}-\frac{5597 g_1^4}{1600}+39 g_3^2 g_2^2 
-\frac{g_3^2g_1^2}{5}+\frac{873 g_2^2g_1^2}{160} \bigg].
    \end{eqnarray}
RGE for the strong gauge coupling, including also pure QCD terms at 4 loops:
\begin{align}
\frac{dg_3^2}{d\ln\bar\mu^2} &= -7\frac{g_3^4}{(4\pi)^2}
+ \frac{g_3^4}{(4\pi)^4} \bigg[ -26 g_3^2+\frac{9 g_2^2}{2}+\frac{11 g_1^2}{10}-2 y_t^2-2 y_b^2 \bigg]  
\nonumber \\ &
+ \frac{g_3^4}{(4\pi)^6} \bigg[ y_t^2\left( 15 y_t^2-40 g_3^2-\frac{93 g_2^2}{8}
-\frac{101 g_1^2}{40}\right)
+\frac{65 g_3^4}{2}+\frac{109 g_2^4}{8}-\frac{523 g_1^4}{120}\nonumber\\
&+21 g_3^2 g_2^2 
+\frac{77g_3^2g_1^2}{15}-\frac{3 g_2^2g_1^2}{40} \bigg]  - 2472.28\frac{g_3^{10}}{(4\pi)^8} \ .    
\end{align}

\section{Higgs quartic coupling}

RGE for the Higgs quartic coupling:
 \begin{align}
\frac{d\lambda}{d\ln\bar\mu^2} &= \frac{1}{(4\pi)^2} \bigg[\lambda  \left(12 \lambda +6 y_t^2+6 y_b^2+2 y_{\tau }^2-\frac{9 g_2^2}{2}-\frac{9 g_1^2}{10}\right) -3 y_t^4-3 y_b^4-y_{\tau }^4\nonumber\\
&+\frac{9 g_2^4}{16}+\frac{27 g_1^4}{400}+\frac{9 g_2^2 g_1^2}{40} \bigg]  
+ \frac{1}{(4\pi)^4} \bigg[\lambda ^2 \left(-156\lambda -72 y_t^2-72 y_b^2 -24 y_{\tau }^2+54 g_2^2+\frac{54 g_1^2}{5}\right)\nonumber \\
&+\lambda y_t^2 \left( -\frac{3 y_t^2}{2}-21 y_b^2+40 g_3^2
+\frac{45 g_2^2}{4}+\frac{17 g_1^2}{4}\right)
+\lambda y_b^2 \left( -\frac{3 y_b^2}{2}+40 g_3^2+\frac{45 g_2^2}{4}+\frac{5 g_1^2}{4}\right)\nonumber \\
&+\lambda y_\tau^2 \left( -\frac{y_\tau^2}{2}+\frac{15 g_2^2}{4}+\frac{15 g_1^2}{4}\right)
+ \lambda \left( -\frac{73 g_2^4}{16}+ \frac{1887 g_1^4}{400}+\frac{117 g_2^2 g_1^2}{40} \right)\nonumber \\
&+ y_t^4 \left( 15 y_t^2-3y_b^2-16g_3^2-\frac{4 g_1^2}{5}\right)
+y_t^2 \left( -\frac{9 g_2^2}{8}-\frac{171 g_1^4}{200}+\frac{63 g_2^2g_1^2}{20}\right)\nonumber \\
&+y_b^4 \left( -3 y_t^2+15y_b^2-16g_3^2+\frac{2 g_1^2}{5}\right)
+y_b^2 \left( -\frac{9 g_2^2}{8}+\frac{9 g_1^4}{40}+\frac{27 g_2^2g_1^2}{20}\right) +y_\tau^4 \left( 5 y_\tau^2-\frac{6 g_1^2}{5}\right)\nonumber \\
&+y_\tau^2 \left( -\frac{3 g_2^4}{8}-\frac{9 g_1^4}{8}+\frac{33 g_2^2g_1^2}{20}\right) 
+\frac{305 g_2^6}{32} -\frac{3411 g_1^6}{4000} -\frac{289 g_2^4 g_1^2}{160} -\frac{1677 g_2^2 g_1^4}{800} 
\bigg]
 \nonumber \\ &
 + \frac{1}{(4\pi)^6} \bigg[\lambda ^3 (6011.35 \lambda +873 y_t^2-387.452 g_2^2-77.490 g_1^2)   
+\lambda ^2 y_t^2(1768.26 y_t^2+160.77 g_3^2\nonumber \\
&-359.539 g_2^2-63.869 g_1^2)   
+\lambda ^2 ( -790.28 g_2^4-185.532 g_1^4 -316.64 g_2^2 g_1^2)\nonumber \\
&+\lambda  y_t^4 ( -223.382 y_t^2 -662.866 g_3^2-5.470 g_2^2-21.015 g_1^2)\nonumber \\
&+\lambda  y_t^2 (356.968 g_3^4-319.664 g_2^4-74.8599 g_1^4 +15.1443 g_3^2 g_2^2+17.454 g_3^2 g_1^2+5.615 g_2^2 g_1^2)\nonumber \\
&+\lambda  g_2^4 (-57.144 g_3^2+865.483 g_2^2 +79.638  g_1^2) +\lambda  g_1^4 (-8.381 g_3^2+61.753 g_2^2+28.168 g_1^2)\nonumber \\
&+  y_t^6 (-243.149 y_t^2+250.494 g_3^2+74.138 g_2^2+33.930 g_1^2)
\nonumber \\ &
+  y_t^4 (-50.201 g_3^4+15.884 g_2^4+15.948 g_1^4+13.349 g_3^2 g_2^2+17.570 g_3^2 g_1^2-70.356 g_2^2 g_1^2)
\nonumber \\ &
+  y_t^2g_3^2(16.464 g_2^4 +1.016  g_1^4+11.386 g_2^2  g_1^2)
+  y_t^2g_2^4( 62.500 g_2^2+13.041  g_1^2)
\nonumber \\ &
+  y_t^2g_1^4(10.627 g_2^2 +11.117 g_1^2)
+  g_3^2( 7.536 g_2^6+0.663  g_1^6+1.507 g_2^4 g_1^2+1.105 g_2^2  g_1^4)
\nonumber \\ &
-114.091 g_2^8-1.508 g_1^8-37.889 g_2^6 g_1^2+6.500 g_2^4 g_1^4-1.543 g_2^2 g_1^6\bigg].
 \end{align}

\section{Higgs mass term}

RGE for the Higgs mass term:
\begin{align}
\frac{dm_h^2}{d\ln\bar\mu^2}&= \frac{m_h^2}{(4\pi)^2} \bigg[ 6 \lambda +3  y_t^2 +3 y_b^2 + y_ {\tau }^2
-\frac{9 g_ 2^2}{4} -\frac{9 g_ 1^2 }{20} \bigg]
\nonumber \\ &
 + \frac{m_h^2}{(4\pi)^4} \bigg[\lambda  \left(-30 \lambda -36 y_t^2 -36 y_b^2 -12  y_ {\tau }^2
 +36 g_ 2^2+\frac{36 g_ 1^2}{5} \right)
\nonumber \\ &
 +y_t^2 \left(-\frac{27 y_t^2}{4} -\frac{21 y_b^2}{2}  +20 g_ 3^2
+\frac{45 g_ 2^2}{8}  +\frac{17 g_ 1^2}{8}  \right)
 +y_b^2 \left(-\frac{27 y_b^2}{4}  +20 g_ 3^2
 +\frac{45 g_ 2^2}{8}  +\frac{5 g_ 1^2}{8}  \right)
\nonumber \\ &
+y_\tau^2 \left(-\frac{9 y_\tau^2}{4}  +\frac{15 g_ 2^2}{8}  +\frac{15 g_ 1^2}{8}  \right) 
-\frac{145}{32} g_ 2^4 +\frac{1671}{800} g_ 1^4+\frac{9 g_ 2^2 g_ 1^2}{16} \bigg]
\nonumber \\ &
 + \frac{m_h^2}{(4\pi)^6} \bigg[\lambda^2  \bigg(1026 \lambda +\frac{297 y_t^2}{2}  
 -192.822 g_ 2^2-38.564 g_ 1^2 \bigg)
 +\lambda y_t^2 (347.394  y_t^2\nonumber \\
 &+80.385 g_ 3^2 -318.591 g_ 2^2-59.699 g_ 1^2)
 -\lambda ( 64.5145 g_ 2^4 + 65.8056 g_1^4 + 37.8231 g_2^2 g_1^2)\nonumber \\
& + y_t^4 ( 154.405  y_t^2 - 209.24 g_3^2 - 3.82928 g_2^2 - 7.50769 g_1^2)\nonumber \\
& + y_t^2 ( 178.484 g_3^4 - 102.627 g_2^4 - 27.721 g_1^4 + 7.572 g_3^2 g_2^2
+ 8.727 g_3^2 g_1^2 + 11.470 g_2^2 g_1^2)\nonumber \\
& -g_2^4 ( 28.572  g_3^2 - 301.724 g_2^2 - 9.931 g_1^2)
 -g_1^4 ( 4.191 g_3^2 - 9.778 g_2^2 - 8.378 g_1^2) \bigg].
     \end{align}

\section{Yukawa couplings}

RGE for the top Yukawa coupling:
\begin{align}
\frac{dy_t^2}{d\ln\bar\mu^2}&= \frac{y_t^2}{(4\pi)^2} \bigg[ \frac{9 y_t^2}{2}+\frac{3 y_b^2}{2}+y_{\tau }^2
-8 g_3^2-\frac{9 g_2^2}{4}-\frac{17 g_1^2}{20}\bigg]\nonumber\\
&+\frac{y_t^2}{(4\pi)^4} \bigg[ y_t^2 \left( -12 y_t^2-\frac{11 y_b^2}{4}-\frac{9 y_\tau^2}{4}-12 \lambda 
+36 g_3^2+\frac{225 g_2^2}{16}+\frac{393 g_1^2}{80}\right)
\nonumber \\ &
+y_b^2 \left( -\frac{y_b^2}{4}+\frac{5 y_{\tau }^2}{4} 
+4 g_3^2+\frac{99 g_2^2}{16}  +\frac{7 g_1^2}{80} \right)
+y_\tau^2 \left( -\frac{9 y_\tau^2}{4}+\frac{15}{8} g_2^2+\frac{15}{8} g_1^2\right)
\nonumber \\ &
+6\lambda^2 -108 g_3^4 -\frac{23 g_2^4}{4}+ \frac{1187 g_1^4}{600}+9 g_3^2 g_2^2+ 
   \frac{19}{15} g_3^2 g_1^2-\frac{9}{20} g_2^2 g_1^2 \bigg]
\nonumber \\ &
+\frac{y_t^2}{(4\pi)^6} \bigg[ 
y_t^4 \left( 58.6028 y_t^2 +198 \lambda -157 g_3^2 -\frac{1593 g_2^2}{16}-\frac{2437 g_1^2}{80}\right)\\
&+y_t^2 \left( 363.764 g_3^4 +16.990 g_2^4-24.422 g_1^4+48.370 g_3^2 g_2^2+18.074 g_3^2 g_1^2
+34.829 g_2^2 g_1^2 \right)\nonumber\\
&+\lambda y_t^2 \left( \frac{15 \lambda }{4}+16 g_3^2 -\frac{135 g_2^2}{2}-\frac{127 g_1^2}{10}\right)
+\lambda \left( -\frac{171 g_2^4}{16}-\frac{1089 g_1^4}{400}+\frac{117 g_2^2 g_1^2}{40} \right)
 \nonumber \\ &
+\lambda^2 \left( -36 \lambda +45 g_2^2 +9 g_1^2\right) -619.35 g_3^6+169.829 g_2^6+16.099 g_1^6+73.654 g_3^4 g_2^2\nonumber\\
&-15.096 g_3^4 g_1^2
-21.072 g_3^2 g_2^4
-22.319 g_3^2 g_1^4- \frac{321}{20}g_3^2 g_2^2 g_1^2-4.743 g_2^4 g_1^2
 -4.442 g_2^2 g_1^4\nonumber
\Big].
 \end{align}
RGE for the bottom Yukawa coupling  (up to two loops):
\begin{align}
\frac{dy_b^2}{d\ln\bar\mu^2}&= \frac{y_b^2}{(4\pi)^2} \bigg[ \frac{3 y_t^2}{2}+\frac{9 y_b^2}{2}+y_{\tau }^2
-8 g_3^2-\frac{9 g_2^2}{4}-\frac{g_1^2}{4}\bigg]
\nonumber \\ &
+\frac{y_b^2}{(4\pi)^4} \bigg[ y_t^2 \left( -\frac{y_t^2}{4}-\frac{11 y_b^2}{4}+\frac{5 y_\tau^2}{4} 
+4 g_3^2+\frac{99 g_2^2}{16}+\frac{91 g_1^2}{80}\right)
\nonumber \\ &
+y_b^2 \left( -12y_b^2-\frac{9 y_{\tau }^2}{4}-12 \lambda 
+36 g_3^2+\frac{225 g_2^2}{16}  +\frac{237 g_1^2}{80} \right)
+y_\tau^2 \left( -\frac{9 y_\tau^2}{4}+\frac{15}{8} g_2^2+\frac{15}{8} g_1^2\right)
\nonumber \\ &
+6\lambda^2 -108 g_3^4 -\frac{23 g_2^4}{4}- \frac{127 g_1^4}{600}+9 g_3^2 g_2^2+ 
   \frac{31}{15} g_3^2 g_1^2-\frac{27}{20} g_2^2 g_1^2 \bigg].
 \end{align}
RGE for the tau Yukawa coupling  (up to two loops):
\begin{align}
\frac{dy_\tau^2}{d\ln\bar\mu^2}&= \frac{y_\tau^2}{(4\pi)^2} \bigg[ 3 y_t^2+3 y_b^2+\frac{ 5 y_{\tau }^2}{2}
-\frac{9 g_2^2}{4}-\frac{9 g_1^2}{4}\bigg]+
\frac{y_\tau^2}{(4\pi)^4} \bigg[6\lambda^2  -\frac{23 g_2^4}{4}+ \frac{1371 g_1^4}{200}
   +\frac{27}{20} g_2^2 g_1^2
\nonumber \\ 
&+y_t^2 \left(20 g_3^2+\frac{45 g_2^2}{8}+\frac{17 g_1^2}{8} -\frac{27y_t^2}{4}+\frac{3 y_b^2}{2}-\frac{27 y_\tau^2}{4} \right)
\\ &
+y_b^2 \left(20 g_3^2+\frac{45 g_2^2}{8}  +\frac{5 g_1^2}{8}  -\frac{27 y_b^2}{4}-\frac{27 y_{\tau }^2}{4}\right)
+y_\tau^2 \left(\frac{165}{16} g_2^2+\frac{537}{80} g_1^2 -3 y_\tau^2-12\lambda\right) \bigg]. \nonumber
 \end{align}

\bigskip

\chapter{SM effective potential at two loops}\label{eff-potential-app}

The Higgs effective potential including one-loop and two-loop corrections in Landau gauge
for $h\gg v$ is given by \eqref{eff-potential-high-h},
where \cite{Casas:1994qy,Degrassi:2012ry} 
\begin{equation}
\lambda_{\rm eff}(h) =e^{4\Gamma(h)} \bigg[ \lambda(\bar\mu=h) + \lambda_{\rm eff}^{(1)}(\bar\mu=h) + \lambda_{\rm eff}^{(2)}(\bar\mu=h)\bigg].
\end{equation}
All running couplings are evaluated at $\bar\mu=h$. Here, $\Gamma(h)\equiv \int_{M_t}^h \gamma(\bar\mu)d\ln\bar\mu$, 
where the Higgs field anomalous dimension is
\begin{align}
\gamma &= \frac{1}{(4\pi)^2} \bigg[ \frac94 g_2^2 + \frac{9}{20} g_1^2 -3 y_t^2-3y_b^2-y_\tau^2 \bigg] 
\nonumber \\&+
\frac{1}{(4\pi)^4}\bigg[ 
y_t^2 \left(-\frac{3 y_b^2}{2}-\frac{17 g_1^2}{8}-\frac{45 g_2^2}{8}-20   g_3^2+\frac{27 y_t^2}{4}\right)- y_{\tau }^2
   \left(\frac{15 g_1^2}{8}+\frac{15
   g_2^2}{8}-\frac{9 y_{\tau}^2}{4}\right)
      \nonumber\\& +
  y_b^2        \left(-\frac{5 g_1^2}{8}-\frac{45 g_2^2}{8}-20 g_3^2+\frac{27 y_b^2}{4}\right)
  -\frac{1293 g_1^4}{800}-\frac{27}{80} g_2^2
   g_1^2+\frac{271 g_2^4}{32}-6 \lambda ^2\bigg] \nonumber \\ 
&+
\frac{1}{(4\pi)^6}\bigg[ 
-9 g_1^2 \lambda ^2-45 g_2^2 \lambda ^2+1.07 g_1^4 \lambda +3.57 g_2^2
   g_1^2 \lambda +
   8.92 g_2^4 \lambda
      \nonumber\\& +
      14.99 g_1^4 y_t^2+14.13 g_1^2 y_t^4-13.21
   g_2^2 g_1^2 y_t^2-8.73 g_3^2 g_1^2 y_t^2+40.11 g_2^2 y_t^4+
      79.05 g_3^2
   y_t^4
         \nonumber\\& +
         23.40 g_2^4 y_t^2-178.48 g_3^4 y_t^2-7.57 g_2^2 g_3^2 y_t^2-5.26
   g_1^6+1.93 g_2^2 g_1^4+
      4.19 g_3^2 g_1^4
            \nonumber\\& +
            1.81 g_2^4 g_1^2-158.51 g_2^6+28.57
   g_2^4 g_3^2+36 \lambda ^3+\frac{135}{2} \lambda ^2 y_t^2-45.00 \lambda 
   y_t^4-60.13 y_t^6
\bigg].
 \end{align}
The one-loop correction is
\begin{align}
\lambda_{\rm eff}^{(1)} &=
\frac{1}{(4\pi)^2}\bigg[  \frac{3g_2^4}{8} ( \log \frac{g_2^2}{4} - \frac{5}{6}+2\Gamma)
+\frac{3}{16}(g_2^2+g_Y^2)^2 (\log\frac{g_2^2+g_Y^2}{4} - \frac56  + 2\Gamma)\nonumber \\
&
-3 y_t^4 (\log\frac{y_t^2}{2}-\frac32+2\Gamma)+
3\lambda^2(4 \log\lambda-6  + 3\log 3+8\Gamma)\bigg]\ . 
\end{align}
The two-loop correction is 
\begin{align}
\lambda_{\rm eff}^{(2)} &=\frac{1}{(4\pi)^4}\bigg[ 
          8g_3^2y_t^4 \left(3r_t^2-8  r_t+9\right) +\frac{1}{2} y_t^6 \left(-6 r_t r_W-3 r_t^2+48 r_t-6 r_{tW}-69-\pi ^2\right)
                      \nonumber\\&+
   \frac{3y_t^2 g_2^4}{16}    \left(8 r_W+4 r_Z-3  r_t^2-6 r_t r_Z-12 r_t+12 r_{tW}+15+2 \pi ^2\right)
   \nonumber\\&+
   \frac{y_t^2g_Y^4}{48}
   \left(27 r_t^2-54 r_t  r_Z-68r_t -28 r_Z+189\right)+
   \frac{y_t^2 g_2^2 g_Y^2 }{8} (9 r_t^2-18 r_t r_Z+4r_t+44 r_Z-57)
      \nonumber\\&+
      \frac{g_2^6}{192}  (
    36 r_t r_Z\!+\!54 r_t^2\!-\!414 r_W r_Z\!+\!69 r_W^2\!+\!1264 r_W\!+\!156 r_Z^2\!+\!632 r_Z-\!144r_{tW}-2067\!+\!90 \pi ^2)  
      \nonumber\\&+
     \frac{ g_2^4 g_Y^2}{192} (12 r_t r_Z-6 r_t^2-6 r_W (53 r_Z+50)+213 r_W^2+4 r_Z
   (57 r_Z-91)+817+46 \pi ^2)
      \nonumber\\&+
     \frac{ g_2^2 g_Y^4}{576} (132 r_t r_Z-66 r_t^2+306 r_W r_Z-153 r_W^2-36 r_W+924 r_Z^2-4080 r_Z+4359+218 \pi ^2)\nonumber
      \\&+
      \frac{g_Y^6}{576} (6
   r_Z (34 r_t+3 r_W-470)-102 r_t^2-9 r_W^2+708 r_Z^2+2883+206 \pi ^2)
\nonumber       \\&+
     \frac{   y_t^4 }{6}
\left(4 g_Y^2 (3 r_t^2-8r_t+9)-9 g_2^2 \left(r_t-r_W+1\right)\right)+
   \frac{3}{4} \left(g_2^6-3 g_2^4 y_t^2+4 y_t^6\right)\mathrm{Li}_2\frac{g_2^2}{2y_t^2}
         \nonumber\\&+
      \frac{y_t^2}{48}  \xi(\frac{g_2^2+g_Y^2}{2 y_t^2} )
         \left(9 g_2^4
-6 g_2^2 g_Y^2+17 g_Y^4+2y_t^2 \big(7 g_Y^2-73 g_2^2+\frac{64 g_2^4}{g_Y^2+g_2^2}\big) \right)
   \nonumber\\&+
    \frac{g_2^2}{64} \xi(\frac{g_2^2+g_Y^2}{g_2^2}) \left(18 g_2^2 g_Y^2+g_Y^4-51 g_2^4-\frac{48
   g_2^6}{g_Y^2+g_2^2}\right)
   \bigg]\ . 
\label{large-h-effeP}\end{align}
Here we have given $\lambda_{\rm eff}^{(2)}$ in the approximation $\lambda =0$, which is well justified around the instability region. The full expression of $\lambda_{\rm eff}^{(2)}$ can be found in ref.~\cite{Degrassi:2012ry}.
Moreover, we have defined
\begin{equation}
\xi(z)\equiv \sqrt{z^2-4z}\left[2\log^2\Big(\frac{z-\sqrt{z^2-4z}}{2z}\Big)-\log^2 z-4{\rm Li}_2\Big(\frac{z-\sqrt{z^2-4z}}{2z}\Big)+\frac{\pi^2}{3}\right],
\end{equation}
where  ${\rm Li}_2
$ is the dilogarithm function, and
\begin{equation}
r_W = \log \frac{g_2^2}{4}+2\Gamma \ ,
\qquad r_Z = \log\frac{g_2^2+g_Y^2}{4}+2\Gamma\ , \qquad
r_t = \log\frac{y_t^2}{2}+2\Gamma \ ,
\end{equation}
\begin{equation}
r_{tW} = (r_t-r_W)\left[ \log \left(\frac{y_t^2}{2}-\frac{g_2^2}{4}\right)+2\Gamma \right] \ .
\end{equation}

\addcontentsline{toc}{chapter}{Bibliography}

\bibliographystyle{style}
\renewcommand{\bibname}{B\lowercase{ibliography}}
\bibliography{tesi}

\begin{thebibliography}{100}
\providecommand{\url}[1]{\texttt{#1}}
\providecommand{\urlprefix}{URL }
\providecommand{\bibAnnoteFile}[1]{%
  \IfFileExists{#1}{\begin{quotation}\noindent\textsc{Key:} #1\\
  \textsc{Annotation:}\ \input{#1}\end{quotation}}{}}
\providecommand{\bibAnnote}[2]{%
  \begin{quotation}\noindent\textsc{Key:} #1\\
  \textsc{Annotation:}\ #2\end{quotation}}
\providecommand{\eprint}[2][]{\url{#2}}

\bibitem{Barbieri:2012uh}
R.~Barbieri, D.~Buttazzo, F.~Sala and D.~M. Straub,
  \href{http://dx.doi.org/10.1007/JHEP07(2012)181}{\emph{JHEP} \textbf{1207}
  (2012) 181}, \href{http://arxiv.org/abs/1203.4218}{{\tt arXiv:1203.4218
  [hep-ph]}}.
\bibAnnoteFile{Barbieri:2012uh}

\bibitem{Barbieri:2012bh}
R.~Barbieri, D.~Buttazzo, F.~Sala and D.~M. Straub,
  \href{http://dx.doi.org/10.1007/JHEP10(2012)040}{\emph{JHEP} \textbf{1210}
  (2012) 040}, \href{http://arxiv.org/abs/1206.1327}{{\tt arXiv:1206.1327
  [hep-ph]}}.
\bibAnnoteFile{Barbieri:2012bh}

\bibitem{Barbieri:2012tu}
R.~Barbieri, D.~Buttazzo, F.~Sala, D.~M. Straub and A.~Tesi,
  \href{http://dx.doi.org/10.1007/JHEP05(2013)069}{\emph{JHEP} \textbf{1305}
  (2013) 069}, \href{http://arxiv.org/abs/1211.5085}{{\tt arXiv:1211.5085
  [hep-ph]}}.
\bibAnnoteFile{Barbieri:2012tu}

\bibitem{Barbieri:2013hxa}
R.~Barbieri, D.~Buttazzo, K.~Kannike, F.~Sala and A.~Tesi,
  \href{http://dx.doi.org/10.1103/PhysRevD.87.115018}{\emph{Phys. Rev.}
  \textbf{D 87} (2013) 115018}, \href{http://arxiv.org/abs/1304.3670}{{\tt
  arXiv:1304.3670 [hep-ph]}}.
\bibAnnoteFile{Barbieri:2013hxa}

\bibitem{Barbieri:2013nka}
R.~Barbieri, D.~Buttazzo, K.~Kannike, F.~Sala and A.~Tesi,
  \href{http://dx.doi.org/10.1103/PhysRevD.88.055011}{\emph{Phys. Rev.}
  \textbf{D 88} (2013) 055011}, \href{http://arxiv.org/abs/1307.4937}{{\tt
  arXiv:1307.4937 [hep-ph]}}.
\bibAnnoteFile{Barbieri:2013nka}

\bibitem{Buttazzo:2013uya}
D.~Buttazzo, G.~Degrassi, P.~P. Giardino, G.~F. Giudice, F.~Sala \emph{et~al.},
  \href{http://dx.doi.org/10.1007/JHEP12(2013)089}{\emph{JHEP} \textbf{1312}
  (2013) 089}, \href{http://arxiv.org/abs/1307.3536}{{\tt arXiv:1307.3536
  [hep-ph]}}.
\bibAnnoteFile{Buttazzo:2013uya}

\bibitem{Glashow:1961tr}
S.~Glashow, \href{http://dx.doi.org/10.1016/0029-5582(61)90469-2}{\emph{Nucl.
  Phys.} \textbf{22} (1961) 579}.
\bibAnnoteFile{Glashow:1961tr}

\bibitem{Weinberg:1967tq}
S.~Weinberg, \href{http://dx.doi.org/10.1103/PhysRevLett.19.1264}{\emph{Phys.
  Rev. Lett.} \textbf{19} (1967) 1264}.
\bibAnnoteFile{Weinberg:1967tq}

\bibitem{Salam:1968rm}
A.~Salam, \emph{Conf. Proc.} \textbf{C680519} (1968) 367.
\bibAnnoteFile{Salam:1968rm}

\bibitem{ALEPH:2005ab}
ALEPH, DELPHI, L3, OPAL and SLD Collaborations, LEP Electroweak Working Group,
  SLD Electroweak and Heavy Flavour Groups,
  \href{http://dx.doi.org/10.1016/j.physrep.2005.12.006}{\emph{Phys. Rept.}
  \textbf{427} (2006) 257}, \href{http://arxiv.org/abs/hep-ex/0509008}{{\tt
  arXiv:hep-ex/0509008}}.
\bibAnnoteFile{ALEPH:2005ab}

\bibitem{Chatrchyan:2012ufa}
CMS Collaboration,
  \href{http://dx.doi.org/10.1016/j.physletb.2012.08.021}{\emph{Phys. Lett.}
  \textbf{B 716} (2012) 30}, \href{http://arxiv.org/abs/1207.7235}{{\tt
  arXiv:1207.7235 [hep-ex]}}.
\bibAnnoteFile{Chatrchyan:2012ufa}

\bibitem{Aad:2012tfa}
ATLAS Collaboration,
  \href{http://dx.doi.org/10.1016/j.physletb.2012.08.020}{\emph{Phys. Lett.}
  \textbf{B 716} (2012) 1}, \href{http://arxiv.org/abs/1207.7214}{{\tt
  arXiv:1207.7214 [hep-ex]}}.
\bibAnnoteFile{Aad:2012tfa}

\bibitem{Giardino:2013bma}
P.~P. Giardino, K.~Kannike, I.~Masina, M.~Raidal and A.~Strumia,
  \href{http://arxiv.org/abs/1303.3570}{{\tt arXiv:1303.3570 [hep-ph]}}.
\bibAnnoteFile{Giardino:2013bma}

\bibitem{CMS:ril}
CMS Collaboration,
  \href{http://cds.cern.ch/record/1530524}{CMS-PAS-HIG-13-001}.
\bibAnnoteFile{CMS:ril}

\bibitem{CMS:xwa}
CMS Collaboration,
  \href{http://cds.cern.ch/record/1523767}{CMS-PAS-HIG-13-002}.
\bibAnnoteFile{CMS:xwa}

\bibitem{CMS:bxa}
CMS Collaboration,
  \href{http://cds.cern.ch/record/1523673}{CMS-PAS-HIG-13-003}.
\bibAnnoteFile{CMS:bxa}

\bibitem{CMS:utj}
CMS Collaboration,
  \href{http://cds.cern.ch/record/1528271}{CMS-PAS-HIG-13-004}.
\bibAnnoteFile{CMS:utj}

\bibitem{Chatrchyan:2013vaa}
CMS Collaboration, \href{http://arxiv.org/abs/1307.5515}{{\tt arXiv:1307.5515
  [hep-ex]}}.
\bibAnnoteFile{Chatrchyan:2013vaa}

\bibitem{CMS:zwa}
CMS Collaboration,
  \href{http://cds.cern.ch/record/1523681}{CMS-PAS-HIG-13-009}.
\bibAnnoteFile{CMS:zwa}

\bibitem{CMS:gya}
CMS Collaboration,
  \href{http://cds.cern.ch/record/1493521}{CMS-PAS-HIG-12-050}.
\bibAnnoteFile{CMS:gya}

\bibitem{ATLAS:2013rma}
ATLAS Collaboration,
  \href{http://cds.cern.ch/record/1523683}{ATLAS-CONF-2013-009}.
\bibAnnoteFile{ATLAS:2013rma}

\bibitem{ATLAS:2013qma}
ATLAS Collaboration,
  \href{http://cds.cern.ch/record/1523695}{ATLAS-CONF-2013-010}.
\bibAnnoteFile{ATLAS:2013qma}

\bibitem{ATLAS:2013pma}
ATLAS Collaboration,
  \href{http://cds.cern.ch/record/1523696}{ATLAS-CONF-2013-011}.
\bibAnnoteFile{ATLAS:2013pma}

\bibitem{ATLAS:2013oma}
ATLAS Collaboration,
  \href{http://cds.cern.ch/record/1523698}{ATLAS-CONF-2013-012}.
\bibAnnoteFile{ATLAS:2013oma}

\bibitem{ATLAS:2013nma}
ATLAS Collaboration,
  \href{http://cds.cern.ch/record/1523699}{ATLAS-CONF-2013-013}.
\bibAnnoteFile{ATLAS:2013nma}

\bibitem{ATLAS:2013mma}
ATLAS Collaboration,
  \href{http://cds.cern.ch/record/1523727}{ATLAS-CONF-2013-014}.
\bibAnnoteFile{ATLAS:2013mma}

\bibitem{ATLAS:2013wla}
ATLAS Collaboration,
  \href{http://cds.cern.ch/record/1527126}{ATLAS-CONF-2013-030}.
\bibAnnoteFile{ATLAS:2013wla}

\bibitem{Wilson:1970ag}
K.~G. Wilson, \href{http://dx.doi.org/10.1103/PhysRevD.3.1818}{\emph{Phys.
  Rev.} \textbf{D 3} (1971) 1818}.
\bibAnnoteFile{Wilson:1970ag}

\bibitem{Gildener:1976ai}
E.~Gildener, \href{http://dx.doi.org/10.1103/PhysRevD.14.1667}{\emph{Phys.
  Rev.} \textbf{D 14} (1976) 1667}.
\bibAnnoteFile{Gildener:1976ai}

\bibitem{Gildener:1976ih}
E.~Gildener and S.~Weinberg,
  \href{http://dx.doi.org/10.1103/PhysRevD.13.3333}{\emph{Phys. Rev.} \textbf{D
  13} (1976) 3333}.
\bibAnnoteFile{Gildener:1976ih}

\bibitem{tHooft:1979bh}
G.~'t~Hooft, \emph{NATO Adv. Study Inst. Ser. B Phys.} \textbf{59} (1980) 135.
\bibAnnoteFile{tHooft:1979bh}

\bibitem{Maiani:1979cx}
L.~Maiani, \emph{Conf. Proc.} \textbf{C7909031} (1979) 1.
\bibAnnoteFile{Maiani:1979cx}

\bibitem{Kaplan:1983fs}
D.~B. Kaplan and H.~Georgi,
  \href{http://dx.doi.org/10.1016/0370-2693(84)91177-8}{\emph{Phys. Lett.}
  \textbf{B 136} (1984) 183}.
\bibAnnoteFile{Kaplan:1983fs}

\bibitem{Georgi:1984af}
H.~Georgi and D.~B. Kaplan,
  \href{http://dx.doi.org/10.1016/0370-2693(84)90341-1}{\emph{Phys. Lett.}
  \textbf{B 145} (1984) 216}.
\bibAnnoteFile{Georgi:1984af}

\bibitem{Contino:2003ve}
R.~Contino, Y.~Nomura and A.~Pomarol,
  \href{http://dx.doi.org/10.1016/j.nuclphysb.2003.08.027}{\emph{Nucl. Phys.}
  \textbf{B 671} (2003) 148}, \href{http://arxiv.org/abs/hep-ph/0306259}{{\tt
  arXiv:hep-ph/0306259}}.
\bibAnnoteFile{Contino:2003ve}

\bibitem{Agashe:2004rs}
K.~Agashe, R.~Contino and A.~Pomarol,
  \href{http://dx.doi.org/10.1016/j.nuclphysb.2005.04.035}{\emph{Nucl. Phys.}
  \textbf{B 719} (2005) 165}, \href{http://arxiv.org/abs/hep-ph/0412089}{{\tt
  arXiv:hep-ph/0412089}}.
\bibAnnoteFile{Agashe:2004rs}

\bibitem{Giudice:2007fh}
G.~Giudice, C.~Grojean, A.~Pomarol and R.~Rattazzi,
  \href{http://dx.doi.org/10.1088/1126-6708/2007/06/045}{\emph{JHEP}
  \textbf{0706} (2007) 045}, \href{http://arxiv.org/abs/hep-ph/0703164}{{\tt
  arXiv:hep-ph/0703164}}.
\bibAnnoteFile{Giudice:2007fh}

\bibitem{Contino:2006qr}
R.~Contino, L.~{Da Rold} and A.~Pomarol,
  \href{http://dx.doi.org/10.1103/PhysRevD.75.055014}{\emph{Phys. Rev.}
  \textbf{D 75} (2007) 055014}, \href{http://arxiv.org/abs/hep-ph/0612048}{{\tt
  arXiv:hep-ph/0612048}}.
\bibAnnoteFile{Contino:2006qr}

\bibitem{Djouadi:2013vqa}
A.~Djouadi and J.~Quevillon, \href{http://arxiv.org/abs/1304.1787}{{\tt
  arXiv:1304.1787 [hep-ph]}}.
\bibAnnoteFile{Djouadi:2013vqa}

\bibitem{Charles:2011va}
CKM fitter group,
  \href{http://dx.doi.org/10.1103/PhysRevD.84.033005}{\emph{Phys. Rev.}
  \textbf{D 84} (2011) 033005}, \href{http://arxiv.org/abs/1106.4041}{{\tt
  arXiv:1106.4041 [hep-ph]}}. Updates on
  \href{http://ckmfitter.in2p3.fr/}{http://ckmfitter.in2p3.fr/}.
\bibAnnoteFile{Charles:2011va}

\bibitem{Derkach:2013da}
UTfit Collaboration, \href{http://arxiv.org/abs/1301.3300}{{\tt arXiv:1301.3300
  [hep-ph]}}. Updates on
  \href{http://www.utfit.org/UTfit/}{http://www.utfit.org/}.
\bibAnnoteFile{Derkach:2013da}

\bibitem{Isidori:2010kg}
G.~Isidori, Y.~Nir and G.~Perez,
  \href{http://dx.doi.org/10.1146/annurev.nucl.012809.104534}{\emph{Ann. Rev.
  Nucl. Part. Sci.} \textbf{60} (2010) 355},
  \href{http://arxiv.org/abs/1002.0900}{{\tt arXiv:1002.0900 [hep-ph]}}.
\bibAnnoteFile{Isidori:2010kg}

\bibitem{Chivukula:1987py}
R.~Chivukula and H.~Georgi,
  \href{http://dx.doi.org/10.1016/0370-2693(87)90713-1}{\emph{Phys. Lett.}
  \textbf{B 188} (1987) 99}.
\bibAnnoteFile{Chivukula:1987py}

\bibitem{Hall:1990ac}
L.~Hall and L.~Randall,
  \href{http://dx.doi.org/10.1103/PhysRevLett.65.2939}{\emph{Phys. Rev. Lett.}
  \textbf{65} (1990) 2939}.
\bibAnnoteFile{Hall:1990ac}

\bibitem{DAmbrosio:2002ex}
G.~D'Ambrosio, G.~F. Giudice, G.~Isidori and A.~Strumia,
  \href{http://dx.doi.org/10.1016/S0550-3213(02)00836-2}{\emph{Nucl. Phys.}
  \textbf{B 645} (2002) 155}, \href{http://arxiv.org/abs/hep-ph/0207036}{{\tt
  arXiv:hep-ph/0207036}}.
\bibAnnoteFile{DAmbrosio:2002ex}

\bibitem{Kaplan:1991dc}
D.~B. Kaplan,
  \href{http://dx.doi.org/10.1016/S0550-3213(05)80021-5}{\emph{Nucl. Phys.}
  \textbf{B 365} (1991) 259}. Revised version.
\bibAnnoteFile{Kaplan:1991dc}

\bibitem{Contino:2006nn}
R.~Contino, T.~Kramer, M.~Son and R.~Sundrum,
  \href{http://dx.doi.org/10.1088/1126-6708/2007/05/074}{\emph{JHEP}
  \textbf{0705} (2007) 074}, \href{http://arxiv.org/abs/hep-ph/0612180}{{\tt
  arXiv:hep-ph/0612180}}.
\bibAnnoteFile{Contino:2006nn}

\bibitem{Degrassi:2012ry}
G.~Degrassi, S.~Di~Vita, J.~Elias-Miro, J.~R. Espinosa, G.~F. Giudice
  \emph{et~al.}, \href{http://dx.doi.org/10.1007/JHEP08(2012)098}{\emph{JHEP}
  \textbf{1208} (2012) 098}, \href{http://arxiv.org/abs/1205.6497}{{\tt
  arXiv:1205.6497 [hep-ph]}}.
\bibAnnoteFile{Degrassi:2012ry}

\bibitem{Barbieri:2007gi}
R.~Barbieri, ``{Ten Lectures on the ElectroWeak Interactions}'' (Edizioni della
  Normale, 2007), \href{http://arxiv.org/abs/0706.0684}{{\tt arXiv:0706.0684
  [hep-ph]}}.
\bibAnnoteFile{Barbieri:2007gi}

\bibitem{Higgs:1964ia}
P.~W. Higgs, \href{http://dx.doi.org/10.1016/0031-9163(64)91136-9}{\emph{Phys.
  Lett.} \textbf{12} (1964) 132}.
\bibAnnoteFile{Higgs:1964ia}

\bibitem{Englert:1964et}
F.~Englert and R.~Brout,
  \href{http://dx.doi.org/10.1103/PhysRevLett.13.321}{\emph{Phys. Rev. Lett.}
  \textbf{13} (1964) 321}.
\bibAnnoteFile{Englert:1964et}

\bibitem{Guralnik:1964eu}
G.~Guralnik, C.~Hagen and T.~Kibble,
  \href{http://dx.doi.org/10.1103/PhysRevLett.13.585}{\emph{Phys. Rev. Lett.}
  \textbf{13} (1964) 585}.
\bibAnnoteFile{Guralnik:1964eu}

\bibitem{Baak:2012kk}
M.~Baak, M.~Goebel, J.~Haller, A.~Hoecker, D.~Kennedy \emph{et~al.},
  \href{http://dx.doi.org/10.1140/epjc/s10052-012-2205-9}{\emph{Eur. Phys. J.}
  \textbf{C 72} (2012) 2205}, \href{http://arxiv.org/abs/1209.2716}{{\tt
  arXiv:1209.2716 [hep-ph]}}.
\bibAnnoteFile{Baak:2012kk}

\bibitem{Farina:2013mla}
M.~Farina, D.~Pappadopulo and A.~Strumia,
  \href{http://dx.doi.org/10.1007/JHEP08(2013)022}{\emph{JHEP} \textbf{1308}
  (2013) 022}, \href{http://arxiv.org/abs/1303.7244}{{\tt arXiv:1303.7244
  [hep-ph]}}.
\bibAnnoteFile{Farina:2013mla}

\bibitem{Cabibbo:1963yz}
N.~Cabibbo, \href{http://dx.doi.org/10.1103/PhysRevLett.10.531}{\emph{Phys.
  Rev. Lett.} \textbf{10} (1963) 531}.
\bibAnnoteFile{Cabibbo:1963yz}

\bibitem{Kobayashi:1973fv}
M.~Kobayashi and T.~Maskawa,
  \href{http://dx.doi.org/10.1143/PTP.49.652}{\emph{Prog. Theor. Phys.}
  \textbf{49} (1973) 652}.
\bibAnnoteFile{Kobayashi:1973fv}

\bibitem{Buras:1998raa}
A.~J. Buras, \href{http://arxiv.org/abs/hep-ph/9806471}{{\tt
  arXiv:hep-ph/9806471}}.
\bibAnnoteFile{Buras:1998raa}

\bibitem{Lunghi:2008aa}
E.~Lunghi and A.~Soni,
  \href{http://dx.doi.org/10.1016/j.physletb.2008.07.015}{\emph{Phys. Lett.}
  \textbf{B 666} (2008) 162}, \href{http://arxiv.org/abs/0803.4340}{{\tt
  arXiv:0803.4340 [hep-ph]}}.
\bibAnnoteFile{Lunghi:2008aa}

\bibitem{Buras:2008nn}
A.~J. Buras and D.~Guadagnoli,
  \href{http://dx.doi.org/10.1103/PhysRevD.78.033005}{\emph{Phys. Rev.}
  \textbf{D 78} (2008) 033005}, \href{http://arxiv.org/abs/0805.3887}{{\tt
  arXiv:0805.3887 [hep-ph]}}.
\bibAnnoteFile{Buras:2008nn}

\bibitem{Altmannshofer:2009ne}
W.~Altmannshofer, A.~J. Buras, S.~Gori, P.~Paradisi and D.~M. Straub,
  \href{http://dx.doi.org/10.1016/j.nuclphysb.2009.12.019}{\emph{Nucl. Phys.}
  \textbf{B 830} (2010) 17}, \href{http://arxiv.org/abs/0909.1333}{{\tt
  arXiv:0909.1333 [hep-ph]}}.
\bibAnnoteFile{Altmannshofer:2009ne}

\bibitem{Lunghi:2010gv}
E.~Lunghi and A.~Soni,
  \href{http://dx.doi.org/10.1016/j.physletb.2011.02.016}{\emph{Phys. Lett.}
  \textbf{B 697} (2011) 323}, \href{http://arxiv.org/abs/1010.6069}{{\tt
  arXiv:1010.6069 [hep-ph]}}.
\bibAnnoteFile{Lunghi:2010gv}

\bibitem{Bevan:2010gi}
UTfit Collaboration, \emph{PoS} \textbf{ICHEP2010} (2010) 270,
  \href{http://arxiv.org/abs/1010.5089}{{\tt arXiv:1010.5089 [hep-ph]}}.
\bibAnnoteFile{Bevan:2010gi}

\bibitem{Brod:2011ty}
J.~Brod and M.~Gorbahn,
  \href{http://dx.doi.org/10.1103/PhysRevLett.108.121801}{\emph{Phys. Rev.
  Lett.} \textbf{108} (2012) 121801},
  \href{http://arxiv.org/abs/1108.2036}{{\tt arXiv:1108.2036 [hep-ph]}}.
\bibAnnoteFile{Brod:2011ty}

\bibitem{Bosch:1999wr}
S.~Bosch, A.~Buras, M.~Gorbahn, S.~Jager, M.~Jamin \emph{et~al.},
  \href{http://dx.doi.org/10.1016/S0550-3213(99)00694-X}{\emph{Nucl. Phys.}
  \textbf{B 565} (2000) 3}, \href{http://arxiv.org/abs/hep-ph/9904408}{{\tt
  arXiv:hep-ph/9904408}}.
\bibAnnoteFile{Bosch:1999wr}

\bibitem{Buchalla:1995vs}
G.~Buchalla, A.~J. Buras and M.~E. Lautenbacher,
  \href{http://dx.doi.org/10.1103/RevModPhys.68.1125}{\emph{Rev. Mod. Phys.}
  \textbf{68} (1996) 1125}, \href{http://arxiv.org/abs/hep-ph/9512380}{{\tt
  arXiv:hep-ph/9512380}}.
\bibAnnoteFile{Buchalla:1995vs}

\bibitem{Mertens:2011ts}
P.~Mertens and C.~Smith, \emph{JHEP} \textbf{1108} (2011) 069,
  \href{http://arxiv.org/abs/1103.5992}{{\tt arXiv:1103.5992 [hep-ph]}}.
\bibAnnoteFile{Mertens:2011ts}

\bibitem{Pospelov:2000bw}
M.~Pospelov and A.~Ritz,
  \href{http://dx.doi.org/10.1103/PhysRevD.63.073015}{\emph{Phys. Rev.}
  \textbf{D 63} (2001) 073015}, \href{http://arxiv.org/abs/hep-ph/0010037}{{\tt
  arXiv:hep-ph/0010037}}.
\bibAnnoteFile{Pospelov:2000bw}

\bibitem{Baker:2006ts}
C.~Baker, D.~Doyle, P.~Geltenbort, K.~Green, M.~van~der Grinten \emph{et~al.},
  \href{http://dx.doi.org/10.1103/PhysRevLett.97.131801}{\emph{Phys. Rev.
  Lett.} \textbf{97} (2006) 131801},
  \href{http://arxiv.org/abs/hep-ex/0602020}{{\tt arXiv:hep-ex/0602020}}.
\bibAnnoteFile{Baker:2006ts}

\bibitem{Feldmann:2008ja}
T.~Feldmann and T.~Mannel,
  \href{http://dx.doi.org/10.1103/PhysRevLett.100.171601}{\emph{Phys. Rev.
  Lett.} \textbf{100} (2008) 171601},
  \href{http://arxiv.org/abs/0801.1802}{{\tt arXiv:0801.1802 [hep-ph]}}.
\bibAnnoteFile{Feldmann:2008ja}

\bibitem{Kagan:2009bn}
A.~L. Kagan, G.~Perez, T.~Volansky and J.~Zupan,
  \href{http://dx.doi.org/10.1103/PhysRevD.80.076002}{\emph{Phys. Rev.}
  \textbf{D 80} (2009) 076002}, \href{http://arxiv.org/abs/0903.1794}{{\tt
  arXiv:0903.1794 [hep-ph]}}.
\bibAnnoteFile{Kagan:2009bn}

\bibitem{Colangelo:2008qp}
G.~Colangelo, E.~Nikolidakis and C.~Smith,
  \href{http://dx.doi.org/10.1140/epjc/s10052-008-0796-y}{\emph{Eur. Phys. J.}
  \textbf{C 59} (2009) 75}, \href{http://arxiv.org/abs/0807.0801}{{\tt
  arXiv:0807.0801 [hep-ph]}}.
\bibAnnoteFile{Colangelo:2008qp}

\bibitem{Barbieri:2011ci}
R.~Barbieri, G.~Isidori, J.~Jones-Perez, P.~Lodone and D.~M. Straub,
  \href{http://dx.doi.org/10.1140/epjc/s10052-011-1725-z}{\emph{Eur. Phys. J.}
  \textbf{C 71} (2011) 1725}, \href{http://arxiv.org/abs/1105.2296}{{\tt
  arXiv:1105.2296 [hep-ph]}}.
\bibAnnoteFile{Barbieri:2011ci}

\bibitem{Barbieri:2011fc}
R.~Barbieri, P.~Campli, G.~Isidori, F.~Sala and D.~M. Straub,
  \href{http://dx.doi.org/10.1140/epjc/s10052-011-1812-1}{\emph{Eur. Phys. J.}
  \textbf{C 71} (2011) 1812}, \href{http://arxiv.org/abs/1108.5125}{{\tt
  arXiv:1108.5125 [hep-ph]}}.
\bibAnnoteFile{Barbieri:2011fc}

\bibitem{Hardy:2008gy}
J.~Hardy and I.~Towner,
  \href{http://dx.doi.org/10.1103/PhysRevC.79.055502}{\emph{Phys. Rev.}
  \textbf{C 79} (2009) 055502}, \href{http://arxiv.org/abs/0812.1202}{{\tt
  arXiv:0812.1202 [nucl-ex]}}.
\bibAnnoteFile{Hardy:2008gy}

\bibitem{Laiho:2009eu}
J.~Laiho, E.~Lunghi and R.~S. Van~de Water,
  \href{http://dx.doi.org/10.1103/PhysRevD.81.034503}{\emph{Phys. Rev.}
  \textbf{D 81} (2010) 034503}, \href{http://arxiv.org/abs/0910.2928}{{\tt
  arXiv:0910.2928 [hep-ph]}}.
\bibAnnoteFile{Laiho:2009eu}

\bibitem{Antonelli:2010yf}
M.~Antonelli, V.~Cirigliano, G.~Isidori, F.~Mescia, M.~Moulson \emph{et~al.},
  \href{http://dx.doi.org/10.1140/epjc/s10052-010-1406-3}{\emph{Eur. Phys. J.}
  \textbf{C 69} (2010) 399}, \href{http://arxiv.org/abs/1005.2323}{{\tt
  arXiv:1005.2323 [hep-ph]}}.
\bibAnnoteFile{Antonelli:2010yf}

\bibitem{Nakamura:2010zzi}
Particle Data Group,
  \href{http://dx.doi.org/10.1088/0954-3899/37/7A/075021}{\emph{J. Phys.}
  \textbf{G 37} (2010) 075021}.
\bibAnnoteFile{Nakamura:2010zzi}

\bibitem{Buras:2010pza}
A.~J. Buras, D.~Guadagnoli and G.~Isidori,
  \href{http://dx.doi.org/10.1016/j.physletb.2010.04.017}{\emph{Phys. Lett.}
  \textbf{B 688} (2010) 309}, \href{http://arxiv.org/abs/1002.3612}{{\tt
  arXiv:1002.3612 [hep-ph]}}.
\bibAnnoteFile{Buras:2010pza}

\bibitem{Kowalewski:2011zz}
BaBar Collaboration, \emph{PoS} \textbf{BEAUTY2011} (2011) 030.
\bibAnnoteFile{Kowalewski:2011zz}

\bibitem{Lunghi:2011xy}
E.~Lunghi and A.~Soni, \href{http://arxiv.org/abs/1104.2117}{{\tt
  arXiv:1104.2117 [hep-ph]}}.
\bibAnnoteFile{Lunghi:2011xy}

\bibitem{Buras:1990fn}
A.~J. Buras, M.~Jamin and P.~H. Weisz,
  \href{http://dx.doi.org/10.1016/0550-3213(90)90373-L}{\emph{Nucl. Phys.}
  \textbf{B 347} (1990) 491}.
\bibAnnoteFile{Buras:1990fn}

\bibitem{Asner:2010qj}
Heavy Flavor Averaging Group, \href{http://arxiv.org/abs/1010.1589}{{\tt
  arXiv:1010.1589 [hep-ex]}}.
\bibAnnoteFile{Asner:2010qj}

\bibitem{Brod:2010mj}
J.~Brod and M.~Gorbahn,
  \href{http://dx.doi.org/10.1103/PhysRevD.82.094026}{\emph{Phys. Rev.}
  \textbf{D 82} (2010) 094026}, \href{http://arxiv.org/abs/1007.0684}{{\tt
  arXiv:1007.0684 [hep-ph]}}.
\bibAnnoteFile{Brod:2010mj}

\bibitem{Abulencia:2006ze}
CDF Collaboration,
  \href{http://dx.doi.org/10.1103/PhysRevLett.97.242003}{\emph{Phys. Rev.
  Lett.} \textbf{97} (2006) 242003},
  \href{http://arxiv.org/abs/hep-ex/0609040}{{\tt arXiv:hep-ex/0609040}}.
\bibAnnoteFile{Abulencia:2006ze}

\bibitem{LHCb-TALK-2012-029}
LHCb Collaboration,
  \href{http://cds.cern.ch/record/1429149}{LHCb-TALK-2012-029}.
\bibAnnoteFile{LHCb-TALK-2012-029}

\bibitem{Charles:2013aka}
J.~Charles, S.~Descotes-Genon, Z.~Ligeti, S.~Monteil, M.~Papucci \emph{et~al.},
  \href{http://arxiv.org/abs/1309.2293}{{\tt arXiv:1309.2293 [hep-ph]}}.
\bibAnnoteFile{Charles:2013aka}

\bibitem{Altmannshofer:2011gn}
W.~Altmannshofer, P.~Paradisi and D.~M. Straub,
  \href{http://dx.doi.org/10.1007/JHEP04(2012)008}{\emph{JHEP} \textbf{1204}
  (2012) 008}, \href{http://arxiv.org/abs/1111.1257}{{\tt arXiv:1111.1257
  [hep-ph]}}.
\bibAnnoteFile{Altmannshofer:2011gn}

\bibitem{Altmannshofer:2013foa}
W.~Altmannshofer and D.~M. Straub, \href{http://arxiv.org/abs/1308.1501}{{\tt
  arXiv:1308.1501 [hep-ph]}}.
\bibAnnoteFile{Altmannshofer:2013foa}

\bibitem{Gedalia:2009kh}
O.~Gedalia, Y.~Grossman, Y.~Nir and G.~Perez,
  \href{http://dx.doi.org/10.1103/PhysRevD.80.055024}{\emph{Phys. Rev.}
  \textbf{D 80} (2009) 055024}, \href{http://arxiv.org/abs/0906.1879}{{\tt
  arXiv:0906.1879 [hep-ph]}}.
\bibAnnoteFile{Gedalia:2009kh}

\bibitem{Falk:2001hx}
A.~F. Falk, Y.~Grossman, Z.~Ligeti and A.~A. Petrov,
  \href{http://dx.doi.org/10.1103/PhysRevD.65.054034}{\emph{Phys. Rev.}
  \textbf{D 65} (2002) 054034}, \href{http://arxiv.org/abs/hep-ph/0110317}{{\tt
  arXiv:hep-ph/0110317}}.
\bibAnnoteFile{Falk:2001hx}

\bibitem{Falk:2004wg}
A.~F. Falk, Y.~Grossman, Z.~Ligeti, Y.~Nir and A.~A. Petrov,
  \href{http://dx.doi.org/10.1103/PhysRevD.69.114021}{\emph{Phys. Rev.}
  \textbf{D 69} (2004) 114021}, \href{http://arxiv.org/abs/hep-ph/0402204}{{\tt
  arXiv:hep-ph/0402204}}.
\bibAnnoteFile{Falk:2004wg}

\bibitem{Amhis:2012bh}
Heavy Flavor Averaging Group, \href{http://arxiv.org/abs/1207.1158}{{\tt
  arXiv:1207.1158 [hep-ex]}}. Updates on
  \href{http://www.slac.stanford.edu/xorg/hfag/}{http://www.slac.stanford.edu/%
xorg/hfag/}.
\bibAnnoteFile{Amhis:2012bh}

\bibitem{Isidori:2011qw}
G.~Isidori, J.~F. Kamenik, Z.~Ligeti and G.~Perez,
  \href{http://dx.doi.org/10.1016/j.physletb.2012.03.046}{\emph{Phys. Lett.}
  \textbf{B 711} (2012) 46}, \href{http://arxiv.org/abs/1111.4987}{{\tt
  arXiv:1111.4987 [hep-ph]}}.
\bibAnnoteFile{Isidori:2011qw}

\bibitem{Aaij:2011in}
LHCb Collaboration, \href{http://dx.doi.org/10.1103/PhysRevLett.108.129903,
  10.1103/PhysRevLett.108.111602}{\emph{Phys. Rev. Lett.} \textbf{108} (2012)
  111602}, \href{http://arxiv.org/abs/1112.0938}{{\tt arXiv:1112.0938
  [hep-ex]}}.
\bibAnnoteFile{Aaij:2011in}

\bibitem{CDF-Note-10784}
CDF Collaboration,
  \href{http://www-cdf.fnal.gov/physics/new/bottom/120216.blessed-CPVcharm10fb%
/cdf10784.pdf}{CDF Public Note 10784}.
\bibAnnoteFile{CDF-Note-10784}

\bibitem{Giudice:2012qq}
G.~F. Giudice, G.~Isidori and P.~Paradisi,
  \href{http://dx.doi.org/10.1007/JHEP04(2012)060}{\emph{JHEP} \textbf{1204}
  (2012) 060}, \href{http://arxiv.org/abs/1201.6204}{{\tt arXiv:1201.6204
  [hep-ph]}}.
\bibAnnoteFile{Giudice:2012qq}

\bibitem{Carvalho:2007yi}
ATLAS Collaboration,
  \href{http://dx.doi.org/10.1140/epjc/s10052-007-0434-0}{\emph{Eur. Phys. J.}
  \textbf{C 52} (2007) 999}, \href{http://arxiv.org/abs/0712.1127}{{\tt
  arXiv:0712.1127 [hep-ex]}}.
\bibAnnoteFile{Carvalho:2007yi}

\bibitem{Fox:2007in}
P.~J. Fox, Z.~Ligeti, M.~Papucci, G.~Perez and M.~D. Schwartz,
  \href{http://dx.doi.org/10.1103/PhysRevD.78.054008}{\emph{Phys. Rev.}
  \textbf{D 78} (2008) 054008}, \href{http://arxiv.org/abs/0704.1482}{{\tt
  arXiv:0704.1482 [hep-ph]}}.
\bibAnnoteFile{Fox:2007in}

\bibitem{Kamenik:2011dk}
J.~F. Kamenik, M.~Papucci and A.~Weiler,
  \href{http://dx.doi.org/10.1103/PhysRevD.85.071501}{\emph{Phys. Rev.}
  \textbf{D 85} (2012) 071501}, \href{http://arxiv.org/abs/1107.3143}{{\tt
  arXiv:1107.3143 [hep-ph]}}.
\bibAnnoteFile{Kamenik:2011dk}

\bibitem{CorderoCid:2007uc}
A.~Cordero-Cid, J.~Hernandez, G.~Tavares-Velasco and J.~Toscano,
  \href{http://dx.doi.org/10.1088/0954-3899/35/2/025004}{\emph{J. Phys.}
  \textbf{G 35} (2008) 025004}, \href{http://arxiv.org/abs/0712.0154}{{\tt
  arXiv:0712.0154 [hep-ph]}}.
\bibAnnoteFile{CorderoCid:2007uc}

\bibitem{Brod:2012ud}
J.~Brod, Y.~Grossman, A.~L. Kagan and J.~Zupan,
  \href{http://dx.doi.org/10.1007/JHEP10(2012)161}{\emph{JHEP} \textbf{1210}
  (2012) 161}, \href{http://arxiv.org/abs/1203.6659}{{\tt arXiv:1203.6659
  [hep-ph]}}.
\bibAnnoteFile{Brod:2012ud}

\bibitem{Isidori:2012yx}
G.~Isidori and J.~F. Kamenik,
  \href{http://dx.doi.org/10.1103/PhysRevLett.109.171801}{\emph{Phys. Rev.
  Lett.} \textbf{109} (2012) 171801},
  \href{http://arxiv.org/abs/1205.3164}{{\tt arXiv:1205.3164 [hep-ph]}}.
\bibAnnoteFile{Isidori:2012yx}

\bibitem{Aushev:2010bq}
T.~Aushev, W.~Bartel, A.~Bondar, J.~Brodzicka, T.~Browder \emph{et~al.},
  \href{http://arxiv.org/abs/1002.5012}{{\tt arXiv:1002.5012 [hep-ex]}}.
\bibAnnoteFile{Aushev:2010bq}

\bibitem{Merk:2011zz}
LHCb Collaboration, \emph{PoS} \textbf{BEAUTY2011} (2011) 039.
\bibAnnoteFile{Merk:2011zz}

\bibitem{Adam:2011ch}
MEG collaboration,
  \href{http://dx.doi.org/10.1103/PhysRevLett.107.171801}{\emph{Phys. Rev.
  Lett.} \textbf{107} (2011) 171801},
  \href{http://arxiv.org/abs/1107.5547}{{\tt arXiv:1107.5547 [hep-ex]}}.
\bibAnnoteFile{Adam:2011ch}

\bibitem{Bellgardt:1987du}
SINDRUM Collaboration,
  \href{http://dx.doi.org/10.1016/0550-3213(88)90462-2}{\emph{Nucl. Phys.}
  \textbf{B 299} (1988) 1}.
\bibAnnoteFile{Bellgardt:1987du}

\bibitem{Wintz:1998rp}
P.~Wintz, \emph{Conf. Proc.} \textbf{C980420} (1998) 534.
\bibAnnoteFile{Wintz:1998rp}

\bibitem{Aubert:2009ag}
BABAR Collaboration,
  \href{http://dx.doi.org/10.1103/PhysRevLett.104.021802}{\emph{Phys. Rev.
  Lett.} \textbf{104} (2010) 021802},
  \href{http://arxiv.org/abs/0908.2381}{{\tt arXiv:0908.2381 [hep-ex]}}.
\bibAnnoteFile{Aubert:2009ag}

\bibitem{Hayasaka:2010np}
K.~Hayasaka, K.~Inami, Y.~Miyazaki, K.~Arinstein, V.~Aulchenko \emph{et~al.},
  \href{http://dx.doi.org/10.1016/j.physletb.2010.03.037}{\emph{Phys. Lett.}
  \textbf{B 687} (2010) 139}, \href{http://arxiv.org/abs/1001.3221}{{\tt
  arXiv:1001.3221 [hep-ex]}}.
\bibAnnoteFile{Hayasaka:2010np}

\bibitem{Hisano:1995cp}
J.~Hisano, T.~Moroi, K.~Tobe and M.~Yamaguchi,
  \href{http://dx.doi.org/10.1103/PhysRevD.53.2442}{\emph{Phys. Rev.} \textbf{D
  53} (1996) 2442}, \href{http://arxiv.org/abs/hep-ph/9510309}{{\tt
  arXiv:hep-ph/9510309}}.
\bibAnnoteFile{Hisano:1995cp}

\bibitem{Arganda:2005ji}
E.~Arganda and M.~J. Herrero,
  \href{http://dx.doi.org/10.1103/PhysRevD.73.055003}{\emph{Phys. Rev.}
  \textbf{D 73} (2006) 055003}, \href{http://arxiv.org/abs/hep-ph/0510405}{{\tt
  arXiv:hep-ph/0510405}}.
\bibAnnoteFile{Arganda:2005ji}

\bibitem{Barbieri:1995tw}
R.~Barbieri, L.~J. Hall and A.~Strumia,
  \href{http://dx.doi.org/10.1016/0550-3213(95)00208-A}{\emph{Nucl. Phys.}
  \textbf{B 445} (1995) 219}, \href{http://arxiv.org/abs/hep-ph/9501334}{{\tt
  arXiv:hep-ph/9501334}}.
\bibAnnoteFile{Barbieri:1995tw}

\bibitem{Contino:2010rs}
R.~Contino, \href{http://arxiv.org/abs/1005.4269}{{\tt arXiv:1005.4269
  [hep-ph]}}.
\bibAnnoteFile{Contino:2010rs}

\bibitem{Barbieri:2009tx}
R.~Barbieri, A.~Carcamo~Hernandez, G.~Corcella, R.~Torre and E.~Trincherini,
  \href{http://dx.doi.org/10.1007/JHEP03(2010)068}{\emph{JHEP} \textbf{1003}
  (2010) 068}, \href{http://arxiv.org/abs/0911.1942}{{\tt arXiv:0911.1942
  [hep-ph]}}.
\bibAnnoteFile{Barbieri:2009tx}

\bibitem{CMS:2012ab}
CMS Collaboration,
  \href{http://dx.doi.org/10.1016/j.physletb.2012.07.059}{\emph{Phys. Lett.}
  \textbf{B 716} (2012) 103}, \href{http://arxiv.org/abs/1203.5410}{{\tt
  arXiv:1203.5410 [hep-ex]}}.
\bibAnnoteFile{CMS:2012ab}

\bibitem{Chatrchyan:2012vu}
CMS Collaboration,
  \href{http://dx.doi.org/10.1016/j.physletb.2012.10.038}{\emph{Phys. Lett.}
  \textbf{B 718} (2012) 307}, \href{http://arxiv.org/abs/1209.0471}{{\tt
  arXiv:1209.0471 [hep-ex]}}.
\bibAnnoteFile{Chatrchyan:2012vu}

\bibitem{Chatrchyan:2012af}
CMS Collaboration, \href{http://dx.doi.org/10.1007/JHEP01(2013)154}{\emph{JHEP}
  \textbf{1301} (2013) 154}, \href{http://arxiv.org/abs/1210.7471}{{\tt
  arXiv:1210.7471 [hep-ex]}}.
\bibAnnoteFile{Chatrchyan:2012af}

\bibitem{ATLAS:2012hpa}
ATLAS Collaboration,
  \href{http://cds.cern.ch/record/1478217}{ATLAS-CONF-2012-130}.
\bibAnnoteFile{ATLAS:2012hpa}

\bibitem{Chatrchyan:2013qha}
CMS Collaboration, \href{http://arxiv.org/abs/1302.4794}{{\tt arXiv:1302.4794
  [hep-ex]}}.
\bibAnnoteFile{Chatrchyan:2013qha}

\bibitem{Chatrchyan:2013lga}
CMS Collaboration,
  \href{http://dx.doi.org/10.1103/PhysRevD.87.072005}{\emph{Phys. Rev.}
  \textbf{D 87} (2013) 072005}, \href{http://arxiv.org/abs/1302.2812}{{\tt
  arXiv:1302.2812 [hep-ex]}}.
\bibAnnoteFile{Chatrchyan:2013lga}

\bibitem{ATLAS:2012qjz}
ATLAS Collaboration,
  \href{http://cds.cern.ch/record/1493487}{ATLAS-CONF-2012-148}.
\bibAnnoteFile{ATLAS:2012qjz}

\bibitem{ATLAS:2013zzj}
ATLAS Collaboration,
  \href{http://cds.cern.ch/record/1562930}{ATLAS-CONF-2013-074}.
\bibAnnoteFile{ATLAS:2013zzj}

\bibitem{Nambu:1960tm}
Y.~Nambu, \href{http://dx.doi.org/10.1103/PhysRev.117.648}{\emph{Phys. Rev.}
  \textbf{117} (1960) 648}.
\bibAnnoteFile{Nambu:1960tm}

\bibitem{Goldstone:1961eq}
J.~Goldstone, \href{http://dx.doi.org/10.1007/BF02812722}{\emph{Nuovo Cim.}
  \textbf{19} (1961) 154}.
\bibAnnoteFile{Goldstone:1961eq}

\bibitem{Redi:2012ha}
M.~Redi and A.~Tesi,
  \href{http://dx.doi.org/10.1007/JHEP10(2012)166}{\emph{JHEP} \textbf{1210}
  (2012) 166}, \href{http://arxiv.org/abs/1205.0232}{{\tt arXiv:1205.0232
  [hep-ph]}}.
\bibAnnoteFile{Redi:2012ha}

\bibitem{Pomarol:2012qf}
A.~Pomarol and F.~Riva,
  \href{http://dx.doi.org/10.1007/JHEP08(2012)135}{\emph{JHEP} \textbf{1208}
  (2012) 135}, \href{http://arxiv.org/abs/1205.6434}{{\tt arXiv:1205.6434
  [hep-ph]}}.
\bibAnnoteFile{Pomarol:2012qf}

\bibitem{Matsedonskyi:2012ym}
O.~Matsedonskyi, G.~Panico and A.~Wulzer,
  \href{http://dx.doi.org/10.1007/JHEP01(2013)164}{\emph{JHEP} \textbf{1301}
  (2013) 164}, \href{http://arxiv.org/abs/1204.6333}{{\tt arXiv:1204.6333
  [hep-ph]}}.
\bibAnnoteFile{Matsedonskyi:2012ym}

\bibitem{Marzocca:2012zn}
D.~Marzocca, M.~Serone and J.~Shu,
  \href{http://dx.doi.org/10.1007/JHEP08(2012)013}{\emph{JHEP} \textbf{1208}
  (2012) 013}, \href{http://arxiv.org/abs/1205.0770}{{\tt arXiv:1205.0770
  [hep-ph]}}.
\bibAnnoteFile{Marzocca:2012zn}

\bibitem{Panico:2012uw}
G.~Panico, M.~Redi, A.~Tesi and A.~Wulzer,
  \href{http://dx.doi.org/10.1007/JHEP03(2013)051}{\emph{JHEP} \textbf{1303}
  (2013) 051}, \href{http://arxiv.org/abs/1210.7114}{{\tt arXiv:1210.7114
  [hep-ph]}}.
\bibAnnoteFile{Panico:2012uw}

\bibitem{Grossman:1999ra}
Y.~Grossman and M.~Neubert,
  \href{http://dx.doi.org/10.1016/S0370-2693(00)00054-X}{\emph{Phys. Lett.}
  \textbf{B 474} (2000) 361}, \href{http://arxiv.org/abs/hep-ph/9912408}{{\tt
  arXiv:hep-ph/9912408}}.
\bibAnnoteFile{Grossman:1999ra}

\bibitem{Huber:2000ie}
S.~J. Huber and Q.~Shafi,
  \href{http://dx.doi.org/10.1016/S0370-2693(00)01399-X}{\emph{Phys. Lett.}
  \textbf{B 498} (2001) 256}, \href{http://arxiv.org/abs/hep-ph/0010195}{{\tt
  arXiv:hep-ph/0010195}}.
\bibAnnoteFile{Huber:2000ie}

\bibitem{Gherghetta:2000qt}
T.~Gherghetta and A.~Pomarol,
  \href{http://dx.doi.org/10.1016/S0550-3213(00)00392-8}{\emph{Nucl. Phys.}
  \textbf{B 586} (2000) 141}, \href{http://arxiv.org/abs/hep-ph/0003129}{{\tt
  arXiv:hep-ph/0003129}}.
\bibAnnoteFile{Gherghetta:2000qt}

\bibitem{Agashe:2004cp}
K.~Agashe, G.~Perez and A.~Soni,
  \href{http://dx.doi.org/10.1103/PhysRevD.71.016002}{\emph{Phys. Rev.}
  \textbf{D 71} (2005) 016002}, \href{http://arxiv.org/abs/hep-ph/0408134}{{\tt
  arXiv:hep-ph/0408134}}.
\bibAnnoteFile{Agashe:2004cp}

\bibitem{Blanke:2008zb}
M.~Blanke, A.~J. Buras, B.~Duling, S.~Gori and A.~Weiler,
  \href{http://dx.doi.org/10.1088/1126-6708/2009/03/001}{\emph{JHEP}
  \textbf{0903} (2009) 001}, \href{http://arxiv.org/abs/0809.1073}{{\tt
  arXiv:0809.1073 [hep-ph]}}.
\bibAnnoteFile{Blanke:2008zb}

\bibitem{Bauer:2009cf}
M.~Bauer, S.~Casagrande, U.~Haisch and M.~Neubert,
  \href{http://dx.doi.org/10.1007/JHEP09(2010)017}{\emph{JHEP} \textbf{1009}
  (2010) 017}, \href{http://arxiv.org/abs/0912.1625}{{\tt arXiv:0912.1625
  [hep-ph]}}.
\bibAnnoteFile{Bauer:2009cf}

\bibitem{KerenZur:2012fr}
B.~Keren-Zur, P.~Lodone, M.~Nardecchia, D.~Pappadopulo, R.~Rattazzi
  \emph{et~al.},
  \href{http://dx.doi.org/10.1016/j.nuclphysb.2012.10.012}{\emph{Nucl. Phys.}
  \textbf{B 867} (2013) 429}, \href{http://arxiv.org/abs/1205.5803}{{\tt
  arXiv:1205.5803 [hep-ph]}}.
\bibAnnoteFile{KerenZur:2012fr}

\bibitem{Csaki:2008zd}
C.~Csaki, A.~Falkowski and A.~Weiler,
  \href{http://dx.doi.org/10.1088/1126-6708/2008/09/008}{\emph{JHEP}
  \textbf{0809} (2008) 008}, \href{http://arxiv.org/abs/0804.1954}{{\tt
  arXiv:0804.1954 [hep-ph]}}.
\bibAnnoteFile{Csaki:2008zd}

\bibitem{Barbieri:2007bh}
R.~Barbieri, B.~Bellazzini, V.~S. Rychkov and A.~Varagnolo,
  \href{http://dx.doi.org/10.1103/PhysRevD.76.115008}{\emph{Phys. Rev.}
  \textbf{D 76} (2007) 115008}, \href{http://arxiv.org/abs/0706.0432}{{\tt
  arXiv:0706.0432 [hep-ph]}}.
\bibAnnoteFile{Barbieri:2007bh}

\bibitem{Agashe:2006at}
K.~Agashe, R.~Contino, L.~{Da Rold} and A.~Pomarol,
  \href{http://dx.doi.org/10.1016/j.physletb.2006.08.005}{\emph{Phys. Lett.}
  \textbf{B 641} (2006) 62}, \href{http://arxiv.org/abs/hep-ph/0605341}{{\tt
  arXiv:hep-ph/0605341}}.
\bibAnnoteFile{Agashe:2006at}

\bibitem{Vignaroli:2012si}
N.~Vignaroli, \href{http://dx.doi.org/10.1103/PhysRevD.86.115011}{\emph{Phys.
  Rev.} \textbf{D 86} (2012) 115011},
  \href{http://arxiv.org/abs/1204.0478}{{\tt arXiv:1204.0478 [hep-ph]}}.
\bibAnnoteFile{Vignaroli:2012si}

\bibitem{Altmannshofer:2012az}
W.~Altmannshofer and D.~M. Straub,
  \href{http://dx.doi.org/10.1007/JHEP08(2012)121}{\emph{JHEP} \textbf{1208}
  (2012) 121}, \href{http://arxiv.org/abs/1206.0273}{{\tt arXiv:1206.0273
  [hep-ph]}}.
\bibAnnoteFile{Altmannshofer:2012az}

\bibitem{Calibbi:2012at}
L.~Calibbi, Z.~Lalak, S.~Pokorski and R.~Ziegler,
  \href{http://dx.doi.org/10.1007/JHEP07(2012)004}{\emph{JHEP} \textbf{1207}
  (2012) 004}, \href{http://arxiv.org/abs/1204.1275}{{\tt arXiv:1204.1275
  [hep-ph]}}.
\bibAnnoteFile{Calibbi:2012at}

\bibitem{Buras:2011ph}
A.~J. Buras, C.~Grojean, S.~Pokorski and R.~Ziegler,
  \href{http://dx.doi.org/10.1007/JHEP08(2011)028}{\emph{JHEP} \textbf{1108}
  (2011) 028}, \href{http://arxiv.org/abs/1105.3725}{{\tt arXiv:1105.3725
  [hep-ph]}}.
\bibAnnoteFile{Buras:2011ph}

\bibitem{Agashe:2008uz}
K.~Agashe, A.~Azatov and L.~Zhu,
  \href{http://dx.doi.org/10.1103/PhysRevD.79.056006}{\emph{Phys. Rev.}
  \textbf{D 79} (2009) 056006}, \href{http://arxiv.org/abs/0810.1016}{{\tt
  arXiv:0810.1016 [hep-ph]}}.
\bibAnnoteFile{Agashe:2008uz}

\bibitem{Gedalia:2009ws}
O.~Gedalia, G.~Isidori and G.~Perez,
  \href{http://dx.doi.org/10.1016/j.physletb.2009.10.097}{\emph{Phys. Lett.}
  \textbf{B 682} (2009) 200}, \href{http://arxiv.org/abs/0905.3264}{{\tt
  arXiv:0905.3264 [hep-ph]}}.
\bibAnnoteFile{Gedalia:2009ws}

\bibitem{ATLAS:2012joa}
ATLAS Collaboration,
  \href{http://cds.cern.ch/record/1460400}{ATLAS-CONF-2012-088}.
\bibAnnoteFile{ATLAS:2012joa}

\bibitem{CMS:2012eza}
CMS Collaboration,
  \href{http://cds.cern.ch/record/1462265}{CMS-PAS-EXO-12-016}.
\bibAnnoteFile{CMS:2012eza}

\bibitem{Barbieri:2008zt}
R.~Barbieri, G.~Isidori and D.~Pappadopulo,
  \href{http://dx.doi.org/10.1088/1126-6708/2009/02/029}{\emph{JHEP}
  \textbf{0902} (2009) 029}, \href{http://arxiv.org/abs/0811.2888}{{\tt
  arXiv:0811.2888 [hep-ph]}}.
\bibAnnoteFile{Barbieri:2008zt}

\bibitem{Cacciapaglia:2007fw}
G.~Cacciapaglia, C.~Csaki, J.~Galloway, G.~Marandella, J.~Terning
  \emph{et~al.},
  \href{http://dx.doi.org/10.1088/1126-6708/2008/04/006}{\emph{JHEP}
  \textbf{0804} (2008) 006}, \href{http://arxiv.org/abs/0709.1714}{{\tt
  arXiv:0709.1714 [hep-ph]}}.
\bibAnnoteFile{Cacciapaglia:2007fw}

\bibitem{Redi:2011zi}
M.~Redi and A.~Weiler,
  \href{http://dx.doi.org/10.1007/JHEP11(2011)108}{\emph{JHEP} \textbf{1111}
  (2011) 108}, \href{http://arxiv.org/abs/1106.6357}{{\tt arXiv:1106.6357
  [hep-ph]}}.
\bibAnnoteFile{Redi:2011zi}

\bibitem{Redi:2012uj}
M.~Redi, \href{http://dx.doi.org/10.1140/epjc/s10052-012-2030-1}{\emph{Eur.
  Phys. J.} \textbf{C 72} (2012) 2030},
  \href{http://arxiv.org/abs/1203.4220}{{\tt arXiv:1203.4220 [hep-ph]}}.
\bibAnnoteFile{Redi:2012uj}

\bibitem{Domenech:2012ai}
O.~Domenech, A.~Pomarol and J.~Serra,
  \href{http://dx.doi.org/10.1103/PhysRevD.85.074030}{\emph{Phys. Rev.}
  \textbf{D 85} (2012) 074030}, \href{http://arxiv.org/abs/1201.6510}{{\tt
  arXiv:1201.6510 [hep-ph]}}.
\bibAnnoteFile{Domenech:2012ai}

\bibitem{Altmannshofer:2008dz}
W.~Altmannshofer, P.~Ball, A.~Bharucha, A.~J. Buras, D.~M. Straub
  \emph{et~al.},
  \href{http://dx.doi.org/10.1088/1126-6708/2009/01/019}{\emph{JHEP}
  \textbf{0901} (2009) 019}, \href{http://arxiv.org/abs/0811.1214}{{\tt
  arXiv:0811.1214 [hep-ph]}}.
\bibAnnoteFile{Altmannshofer:2008dz}

\bibitem{Fayet:1976cr}
P.~Fayet and S.~Ferrara,
  \href{http://dx.doi.org/10.1016/0370-1573(77)90066-7}{\emph{Phys. Rept.}
  \textbf{32} (1977) 249}.
\bibAnnoteFile{Fayet:1976cr}

\bibitem{Wess:1974tw}
J.~Wess and B.~Zumino,
  \href{http://dx.doi.org/10.1016/0550-3213(74)90355-1}{\emph{Nucl. Phys.}
  \textbf{B 70} (1974) 39}.
\bibAnnoteFile{Wess:1974tw}

\bibitem{Wess:1973kz}
J.~Wess and B.~Zumino,
  \href{http://dx.doi.org/10.1016/0370-2693(74)90578-4}{\emph{Phys. Lett.}
  \textbf{B 49} (1974) 52}.
\bibAnnoteFile{Wess:1973kz}

\bibitem{Iliopoulos:1974zv}
J.~Iliopoulos and B.~Zumino,
  \href{http://dx.doi.org/10.1016/0550-3213(74)90388-5}{\emph{Nucl. Phys.}
  \textbf{B 76} (1974) 310}.
\bibAnnoteFile{Iliopoulos:1974zv}

\bibitem{Girardello:1981wz}
L.~Girardello and M.~T. Grisaru,
  \href{http://dx.doi.org/10.1016/0550-3213(82)90512-0}{\emph{Nucl. Phys.}
  \textbf{B 194} (1982) 65}.
\bibAnnoteFile{Girardello:1981wz}

\bibitem{Dimopoulos:1981zb}
S.~Dimopoulos and H.~Georgi,
  \href{http://dx.doi.org/10.1016/0550-3213(81)90522-8}{\emph{Nucl. Phys.}
  \textbf{B 193} (1981) 150}.
\bibAnnoteFile{Dimopoulos:1981zb}

\bibitem{Barbieri:1982eh}
R.~Barbieri, S.~Ferrara and C.~A. Savoy,
  \href{http://dx.doi.org/10.1016/0370-2693(82)90685-2}{\emph{Phys. Lett.}
  \textbf{B 119} (1982) 343}.
\bibAnnoteFile{Barbieri:1982eh}

\bibitem{Chamseddine:1982jx}
A.~H. Chamseddine, R.~L. Arnowitt and P.~Nath,
  \href{http://dx.doi.org/10.1103/PhysRevLett.49.970}{\emph{Phys. Rev. Lett.}
  \textbf{49} (1982) 970}.
\bibAnnoteFile{Chamseddine:1982jx}

\bibitem{Hall:1983iz}
L.~J. Hall, J.~D. Lykken and S.~Weinberg,
  \href{http://dx.doi.org/10.1103/PhysRevD.27.2359}{\emph{Phys. Rev.} \textbf{D
  27} (1983) 2359}.
\bibAnnoteFile{Hall:1983iz}

\bibitem{Nilles:1983ge}
H.~P. Nilles, \href{http://dx.doi.org/10.1016/0370-1573(84)90008-5}{\emph{Phys.
  Rept.} \textbf{110} (1984) 1}.
\bibAnnoteFile{Nilles:1983ge}

\bibitem{Dine:1981rt}
M.~Dine, W.~Fischler and M.~Srednicki,
  \href{http://dx.doi.org/10.1016/0370-2693(81)90590-6}{\emph{Phys. Lett.}
  \textbf{B 104} (1981) 199}.
\bibAnnoteFile{Dine:1981rt}

\bibitem{AlvarezGaume:1981wy}
L.~Alvarez-Gaume, M.~Claudson and M.~B. Wise,
  \href{http://dx.doi.org/10.1016/0550-3213(82)90138-9}{\emph{Nucl. Phys.}
  \textbf{B 207} (1982) 96}.
\bibAnnoteFile{AlvarezGaume:1981wy}

\bibitem{AlvarezGaume:1983gj}
L.~Alvarez-Gaume, J.~Polchinski and M.~B. Wise,
  \href{http://dx.doi.org/10.1016/0550-3213(83)90591-6}{\emph{Nucl. Phys.}
  \textbf{B 221} (1983) 495}.
\bibAnnoteFile{AlvarezGaume:1983gj}

\bibitem{Giudice:1998bp}
G.~Giudice and R.~Rattazzi,
  \href{http://dx.doi.org/10.1016/S0370-1573(99)00042-3}{\emph{Phys. Rept.}
  \textbf{322} (1999) 419}, \href{http://arxiv.org/abs/hep-ph/9801271}{{\tt
  arXiv:hep-ph/9801271}}.
\bibAnnoteFile{Giudice:1998bp}

\bibitem{Dimopoulos:1995mi}
S.~Dimopoulos and G.~Giudice,
  \href{http://dx.doi.org/10.1016/0370-2693(95)00961-J}{\emph{Phys. Lett.}
  \textbf{B 357} (1995) 573}, \href{http://arxiv.org/abs/hep-ph/9507282}{{\tt
  arXiv:hep-ph/9507282}}.
\bibAnnoteFile{Dimopoulos:1995mi}

\bibitem{Cohen:1996vb}
A.~G. Cohen, D.~Kaplan and A.~Nelson,
  \href{http://dx.doi.org/10.1016/S0370-2693(96)01183-5}{\emph{Phys. Lett.}
  \textbf{B 388} (1996) 588}, \href{http://arxiv.org/abs/hep-ph/9607394}{{\tt
  arXiv:hep-ph/9607394}}.
\bibAnnoteFile{Cohen:1996vb}

\bibitem{Barbieri:2009ev}
R.~Barbieri and D.~Pappadopulo,
  \href{http://dx.doi.org/10.1088/1126-6708/2009/10/061}{\emph{JHEP}
  \textbf{0910} (2009) 061}, \href{http://arxiv.org/abs/0906.4546}{{\tt
  arXiv:0906.4546 [hep-ph]}}.
\bibAnnoteFile{Barbieri:2009ev}

\bibitem{Papucci:2011wy}
M.~Papucci, J.~T. Ruderman and A.~Weiler,
  \href{http://dx.doi.org/10.1007/JHEP09(2012)035}{\emph{JHEP} \textbf{1209}
  (2012) 035}, \href{http://arxiv.org/abs/1110.6926}{{\tt arXiv:1110.6926
  [hep-ph]}}.
\bibAnnoteFile{Papucci:2011wy}

\bibitem{Terning:2006bq}
J.~Terning, ``{Modern supersymmetry: Dynamics and duality}'',
  \emph{International Series of Monographs on Physics}, vol. 132 (Oxford
  Science, 2006).
\bibAnnoteFile{Terning:2006bq}

\bibitem{Meade:2008wd}
P.~Meade, N.~Seiberg and D.~Shih,
  \href{http://dx.doi.org/10.1143/PTPS.177.143}{\emph{Prog. Theor. Phys.
  Suppl.} \textbf{177} (2009) 143}, \href{http://arxiv.org/abs/0801.3278}{{\tt
  arXiv:0801.3278 [hep-ph]}}.
\bibAnnoteFile{Meade:2008wd}

\bibitem{Barbier:2004ez}
R.~Barbier, C.~Berat, M.~Besancon, M.~Chemtob, A.~Deandrea \emph{et~al.},
  \href{http://dx.doi.org/10.1016/j.physrep.2005.08.006}{\emph{Phys. Rept.}
  \textbf{420} (2005) 1}, \href{http://arxiv.org/abs/hep-ph/0406039}{{\tt
  arXiv:hep-ph/0406039}}.
\bibAnnoteFile{Barbier:2004ez}

\bibitem{Kim:1983dt}
J.~E. Kim and H.~P. Nilles,
  \href{http://dx.doi.org/10.1016/0370-2693(84)91890-2}{\emph{Phys. Lett.}
  \textbf{B 138} (1984) 150}.
\bibAnnoteFile{Kim:1983dt}

\bibitem{Giudice:1988yz}
G.~Giudice and A.~Masiero,
  \href{http://dx.doi.org/10.1016/0370-2693(88)91613-9}{\emph{Phys. Lett.}
  \textbf{B 206} (1988) 480}.
\bibAnnoteFile{Giudice:1988yz}

\bibitem{Carena:1995bx}
M.~S. Carena, J.~Espinosa, M.~Quiros and C.~Wagner,
  \href{http://dx.doi.org/10.1016/0370-2693(95)00694-G}{\emph{Phys. Lett.}
  \textbf{B 355} (1995) 209}, \href{http://arxiv.org/abs/hep-ph/9504316}{{\tt
  arXiv:hep-ph/9504316}}.
\bibAnnoteFile{Carena:1995bx}

\bibitem{Djouadi:2005gj}
A.~Djouadi, \href{http://dx.doi.org/10.1016/j.physrep.2007.10.005}{\emph{Phys.
  Rept.} \textbf{459} (2008) 1},
  \href{http://arxiv.org/abs/hep-ph/0503173}{{\tt arXiv:hep-ph/0503173}}.
\bibAnnoteFile{Djouadi:2005gj}

\bibitem{ATLAS:2013zze}
ATLAS Collaboration,
  \href{http://cds.cern.ch/record/1547571}{ATLAS-CONF-2013-054}.
\bibAnnoteFile{ATLAS:2013zze}

\bibitem{ATLAS:2013zzf}
ATLAS Collaboration,
  \href{http://cds.cern.ch/record/1557778}{ATLAS-CONF-2013-061}.
\bibAnnoteFile{ATLAS:2013zzf}

\bibitem{ATLAS:2012epx}
ATLAS Collaboration,
  \href{http://cds.cern.ch/record/1493490}{ATLAS-CONF-2012-151}.
\bibAnnoteFile{ATLAS:2012epx}

\bibitem{ATLAS:2013tma}
ATLAS Collaboration,
  \href{http://cds.cern.ch/record/1522430}{ATLAS-CONF-2013-007}.
\bibAnnoteFile{ATLAS:2013tma}

\bibitem{Chatrchyan:2013wxa}
CMS Collaboration,
  \href{http://dx.doi.org/10.1016/j.physletb.2013.06.058}{\emph{Phys. Lett.}
  \textbf{B 725} (2013) 243}, \href{http://arxiv.org/abs/1305.2390}{{\tt
  arXiv:1305.2390 [hep-ex]}}.
\bibAnnoteFile{Chatrchyan:2013wxa}

\bibitem{CMS:wwa}
CMS Collaboration,
  \href{http://cds.cern.ch/record/1523786}{CMS-PAS-SUS-13-007}.
\bibAnnoteFile{CMS:wwa}

\bibitem{CMS:zzb}
CMS Collaboration,
  \href{http://cds.cern.ch/record/1596446}{CMS-PAS-SUS-13-004}.
\bibAnnoteFile{CMS:zzb}

\bibitem{CMS:zzc}
CMS Collaboration,
  \href{http://cds.cern.ch/record/1563301}{CMS-PAS-SUS-13-013}.
\bibAnnoteFile{CMS:zzc}

\bibitem{CMS:zzd}
CMS Collaboration,
  \href{http://cds.cern.ch/record/1547560}{CMS-PAS-SUS-13-008}.
\bibAnnoteFile{CMS:zzd}

\bibitem{Gherghetta:2012gb}
T.~Gherghetta, B.~von Harling, A.~D. Medina and M.~A. Schmidt,
  \href{http://dx.doi.org/10.1007/JHEP02(2013)032}{\emph{JHEP} \textbf{1302}
  (2013) 032}, \href{http://arxiv.org/abs/1212.5243}{{\tt arXiv:1212.5243
  [hep-ph]}}.
\bibAnnoteFile{Gherghetta:2012gb}

\bibitem{ATLAS:2013cma}
ATLAS Collaboration,
  \href{http://cds.cern.ch/record/1525880}{ATLAS-CONF-2013-024}.
\bibAnnoteFile{ATLAS:2013cma}

\bibitem{ATLAS:2013pla}
ATLAS Collaboration,
  \href{http://cds.cern.ch/record/1532431}{ATLAS-CONF-2013-037}.
\bibAnnoteFile{ATLAS:2013pla}

\bibitem{ATLAS:2013zza}
ATLAS Collaboration,
  \href{http://cds.cern.ch/record/1562840}{ATLAS-CONF-2013-065}.
\bibAnnoteFile{ATLAS:2013zza}

\bibitem{ATLAS:2013zzb}
ATLAS Collaboration,
  \href{http://cds.cern.ch/record/1562880}{ATLAS-CONF-2013-068}.
\bibAnnoteFile{ATLAS:2013zzb}

\bibitem{ATLAS:2013zzc}
ATLAS Collaboration,
  \href{http://cds.cern.ch/record/1547564?ln=en}{ATLAS-CONF-2013-048}.
\bibAnnoteFile{ATLAS:2013zzc}

\bibitem{ATLAS:2013zzd}
ATLAS Collaboration,
  \href{http://cds.cern.ch/record/1547570}{ATLAS-CONF-2013-053}.
\bibAnnoteFile{ATLAS:2013zzd}

\bibitem{Aad:2012ywa}
ATLAS Collaboration,
  \href{http://dx.doi.org/10.1103/PhysRevLett.109.211802}{\emph{Phys. Rev.
  Lett.} \textbf{109} (2012) 211802},
  \href{http://arxiv.org/abs/1208.1447}{{\tt arXiv:1208.1447 [hep-ex]}}.
\bibAnnoteFile{Aad:2012ywa}

\bibitem{Aad:2012xqa}
ATLAS Collaboration,
  \href{http://dx.doi.org/10.1103/PhysRevLett.109.211803}{\emph{Phys. Rev.
  Lett.} \textbf{109} (2012) 211803},
  \href{http://arxiv.org/abs/1208.2590}{{\tt arXiv:1208.2590 [hep-ex]}}.
\bibAnnoteFile{Aad:2012xqa}

\bibitem{Aad:2012uu}
ATLAS Collaboration,
  \href{http://dx.doi.org/10.1007/JHEP11(2012)094}{\emph{JHEP} \textbf{1211}
  (2012) 094}, \href{http://arxiv.org/abs/1209.4186}{{\tt arXiv:1209.4186
  [hep-ex]}}.
\bibAnnoteFile{Aad:2012uu}

\bibitem{Aad:2012tx}
ATLAS Collaboration,
  \href{http://dx.doi.org/10.1140/epjc/s10052-012-2237-1}{\emph{Eur. Phys. J.}
  \textbf{C 72} (2012) 2237}, \href{http://arxiv.org/abs/1208.4305}{{\tt
  arXiv:1208.4305 [hep-ex]}}.
\bibAnnoteFile{Aad:2012tx}

\bibitem{Aad:2012yr}
ATLAS Collaboration,
  \href{http://dx.doi.org/10.1016/j.physletb.2013.01.049}{\emph{Phys. Lett.}
  \textbf{B 720} (2013) 13}, \href{http://arxiv.org/abs/1209.2102}{{\tt
  arXiv:1209.2102 [hep-ex]}}.
\bibAnnoteFile{Aad:2012yr}

\bibitem{Chatrchyan:2013xna}
CMS Collaboration, \href{http://arxiv.org/abs/1308.1586}{{\tt arXiv:1308.1586
  [hep-ex]}}.
\bibAnnoteFile{Chatrchyan:2013xna}

\bibitem{ATLAS:moriond1}
F.~Hubaut (ATLAS Collaboration), talk at the Moriond 2013 EW session.
\bibAnnoteFile{ATLAS:moriond1}

\bibitem{ATLAS:moriond2}
E.~Mountricha (ATLAS Collaboration), talk at the Moriond 2013 QCD session.
\bibAnnoteFile{ATLAS:moriond2}

\bibitem{ATLAS:moriond3}
V.~Martin (ATLAS Collaboration), talk at the Moriond 2013 EW session.
\bibAnnoteFile{ATLAS:moriond3}

\bibitem{CMS:moriond1}
G.~Gomez-Ceballos (CMS Collaboration), talk at the Moriond 2013 EW session.
\bibAnnoteFile{CMS:moriond1}

\bibitem{CMS:moriond2}
M.~Shen (CMS Collaboration), talk at the Moriond 2013 EW session.
\bibAnnoteFile{CMS:moriond2}

\bibitem{CMS:moriond3}
B.~Mansoulie (CMS Collaboration), talk at the Moriond 2013 EW session.
\bibAnnoteFile{CMS:moriond3}

\bibitem{CMS:moriond4}
V.~Dutta (CMS Collaboration), talk at the Moriond 2013 EW session.
\bibAnnoteFile{CMS:moriond4}

\bibitem{tevatron:2013}
L.~\v{Z}ivkovi\'{c} (CDF and D0 Collaborations), talk at the Moriond 2013 EW
  session.
\bibAnnoteFile{tevatron:2013}

\bibitem{Cheung:2013bn}
C.~Cheung, S.~D. McDermott and K.~M. Zurek,
  \href{http://dx.doi.org/10.1007/JHEP04(2013)074}{\emph{JHEP} \textbf{1304}
  (2013) 074}, \href{http://arxiv.org/abs/1302.0314}{{\tt arXiv:1302.0314
  [hep-ph]}}.
\bibAnnoteFile{Cheung:2013bn}

\bibitem{Azatov:2012qz}
A.~Azatov and J.~Galloway,
  \href{http://dx.doi.org/10.1142/S0217751X13300044}{\emph{Int. J. Mod. Phys.}
  \textbf{A 28} (2013) 1330004}, \href{http://arxiv.org/abs/1212.1380}{{\tt
  arXiv:1212.1380 [hep-ph]}}.
\bibAnnoteFile{Azatov:2012qz}

\bibitem{Carmi:2012in}
D.~Carmi, A.~Falkowski, E.~Kuflik, T.~Volansky and J.~Zupan,
  \href{http://dx.doi.org/10.1007/JHEP10(2012)196}{\emph{JHEP} \textbf{1210}
  (2012) 196}, \href{http://arxiv.org/abs/1207.1718}{{\tt arXiv:1207.1718
  [hep-ph]}}.
\bibAnnoteFile{Carmi:2012in}

\bibitem{Dittmaier:2011ti}
LHC Higgs Cross Section Working Group,
  \href{http://arxiv.org/abs/1101.0593}{{\tt arXiv:1101.0593 [hep-ph]}}.
  Updates on
  \href{https://twiki.cern.ch/twiki/bin/view/LHCPhysics/CrossSections}{https:/%
/twiki.cern.ch/twiki/bin/view/LHCPhysics/CrossSections}.
\bibAnnoteFile{Dittmaier:2011ti}

\bibitem{Anastasiou:2009kn}
C.~Anastasiou, S.~Bucherer and Z.~Kunszt,
  \href{http://dx.doi.org/10.1088/1126-6708/2009/10/068}{\emph{JHEP}
  \textbf{0910} (2009) 068}, \href{http://arxiv.org/abs/0907.2362}{{\tt
  arXiv:0907.2362 [hep-ph]}}.
\bibAnnoteFile{Anastasiou:2009kn}

\bibitem{Spira:1995rr}
M.~Spira, A.~Djouadi, D.~Graudenz and P.~Zerwas,
  \href{http://dx.doi.org/10.1016/0550-3213(95)00379-7}{\emph{Nucl. Phys.}
  \textbf{B 453} (1995) 17}, \href{http://arxiv.org/abs/hep-ph/9504378}{{\tt
  arXiv:hep-ph/9504378}}.
\bibAnnoteFile{Spira:1995rr}

\bibitem{Spira:1995mt}
M.~Spira, \href{http://arxiv.org/abs/hep-ph/9510347}{{\tt
  arXiv:hep-ph/9510347}}.
\bibAnnoteFile{Spira:1995mt}

\bibitem{Arbey:2013jla}
A.~Arbey, M.~Battaglia and F.~Mahmoudi,
  \href{http://dx.doi.org/10.1103/PhysRevD.88.015007}{\emph{Phys. Rev.}
  \textbf{D 88} (2013) 015007}, \href{http://arxiv.org/abs/1303.7450}{{\tt
  arXiv:1303.7450 [hep-ph]}}.
\bibAnnoteFile{Arbey:2013jla}

\bibitem{Fayet:1974pd}
P.~Fayet, \href{http://dx.doi.org/10.1016/0550-3213(75)90636-7}{\emph{Nucl.
  Phys.} \textbf{B 90} (1975) 104}.
\bibAnnoteFile{Fayet:1974pd}

\bibitem{Ellwanger:2009dp}
U.~Ellwanger, C.~Hugonie and A.~M. Teixeira,
  \href{http://dx.doi.org/10.1016/j.physrep.2010.07.001}{\emph{Phys. Rept.}
  \textbf{496} (2010) 1}, \href{http://arxiv.org/abs/0910.1785}{{\tt
  arXiv:0910.1785 [hep-ph]}}.
\bibAnnoteFile{Ellwanger:2009dp}

\bibitem{Barbieri:2006bg}
R.~Barbieri, L.~J. Hall, Y.~Nomura and V.~S. Rychkov,
  \href{http://dx.doi.org/10.1103/PhysRevD.75.035007}{\emph{Phys. Rev.}
  \textbf{D 75} (2007) 035007}, \href{http://arxiv.org/abs/hep-ph/0607332}{{\tt
  arXiv:hep-ph/0607332}}.
\bibAnnoteFile{Barbieri:2006bg}

\bibitem{Hall:2011aa}
L.~J. Hall, D.~Pinner and J.~T. Ruderman,
  \href{http://dx.doi.org/10.1007/JHEP04(2012)131}{\emph{JHEP} \textbf{1204}
  (2012) 131}, \href{http://arxiv.org/abs/1112.2703}{{\tt arXiv:1112.2703
  [hep-ph]}}.
\bibAnnoteFile{Hall:2011aa}

\bibitem{Agashe:2012zq}
K.~Agashe, Y.~Cui and R.~Franceschini,
  \href{http://dx.doi.org/10.1007/JHEP02(2013)031}{\emph{JHEP} \textbf{1302}
  (2013) 031}, \href{http://arxiv.org/abs/1209.2115}{{\tt arXiv:1209.2115
  [hep-ph]}}.
\bibAnnoteFile{Agashe:2012zq}

\bibitem{Gupta:2012fy}
R.~S. Gupta, M.~Montull and F.~Riva,
  \href{http://dx.doi.org/10.1007/JHEP04(2013)132}{\emph{JHEP} \textbf{1304}
  (2013) 132}, \href{http://arxiv.org/abs/1212.5240}{{\tt arXiv:1212.5240
  [hep-ph]}}.
\bibAnnoteFile{Gupta:2012fy}

\bibitem{DAgnolo:2012mj}
R.~T. D'Agnolo, E.~Kuflik and M.~Zanetti,
  \href{http://dx.doi.org/10.1007/JHEP03(2013)043}{\emph{JHEP} \textbf{1303}
  (2013) 043}, \href{http://arxiv.org/abs/1212.1165}{{\tt arXiv:1212.1165
  [hep-ph]}}.
\bibAnnoteFile{DAgnolo:2012mj}

\bibitem{Peccei:1977ur}
R.~Peccei and H.~R. Quinn,
  \href{http://dx.doi.org/10.1103/PhysRevD.16.1791}{\emph{Phys. Rev.} \textbf{D
  16} (1977) 1791}.
\bibAnnoteFile{Peccei:1977ur}

\bibitem{Peccei:1977hh}
R.~Peccei and H.~R. Quinn,
  \href{http://dx.doi.org/10.1103/PhysRevLett.38.1440}{\emph{Phys. Rev. Lett.}
  \textbf{38} (1977) 1440}.
\bibAnnoteFile{Peccei:1977hh}

\bibitem{Choi:2012he}
K.~Choi, S.~H. Im, K.~S. Jeong and M.~Yamaguchi,
  \href{http://dx.doi.org/10.1007/JHEP02(2013)090}{\emph{JHEP} \textbf{1302}
  (2013) 090}, \href{http://arxiv.org/abs/1211.0875}{{\tt arXiv:1211.0875
  [hep-ph]}}.
\bibAnnoteFile{Choi:2012he}

\bibitem{Misiak:2006zs}
M.~Misiak, H.~Asatrian, K.~Bieri, M.~Czakon, A.~Czarnecki \emph{et~al.},
  \href{http://dx.doi.org/10.1103/PhysRevLett.98.022002}{\emph{Phys. Rev.
  Lett.} \textbf{98} (2007) 022002},
  \href{http://arxiv.org/abs/hep-ph/0609232}{{\tt arXiv:hep-ph/0609232}}.
\bibAnnoteFile{Misiak:2006zs}

\bibitem{ATLAS-collaboration:2012iza}
ATLAS Collaboration,
  \href{http://cds.cern.ch/record/1484890}{ATL-PHYS-PUB-2012-004}.
\bibAnnoteFile{ATLAS-collaboration:2012iza}

\bibitem{CMS:note}
CMS Collaboration, \href{http://cds.cern.ch/record/1494600}{CMS-NOTE-2012-006}.
\bibAnnoteFile{CMS:note}

\bibitem{Baglio:2012np}
J.~Baglio, A.~Djouadi, R.~Gr{\"o}ber, M.~M{\"u}hlleitner, J.~Quevillon
  \emph{et~al.}, \href{http://dx.doi.org/10.1007/JHEP04(2013)151}{\emph{JHEP}
  \textbf{1304} (2013) 151}, \href{http://arxiv.org/abs/1212.5581}{{\tt
  arXiv:1212.5581 [hep-ph]}}.
\bibAnnoteFile{Baglio:2012np}

\bibitem{Dolan:2012rv}
M.~J. Dolan, C.~Englert and M.~Spannowsky,
  \href{http://dx.doi.org/10.1007/JHEP10(2012)112}{\emph{JHEP} \textbf{1210}
  (2012) 112}, \href{http://arxiv.org/abs/1206.5001}{{\tt arXiv:1206.5001
  [hep-ph]}}.
\bibAnnoteFile{Dolan:2012rv}

\bibitem{Goertz:2013kp}
F.~Goertz, A.~Papaefstathiou, L.~L. Yang and J.~Zurita,
  \href{http://dx.doi.org/10.1007/JHEP06(2013)016}{\emph{JHEP} \textbf{1306}
  (2013) 016}, \href{http://arxiv.org/abs/1301.3492}{{\tt arXiv:1301.3492
  [hep-ph]}}.
\bibAnnoteFile{Goertz:2013kp}

\bibitem{Searches:2001aa}
LEP Higgs Working Group for Higgs boson searches,
  \href{http://arxiv.org/abs/hep-ex/0107034}{{\tt arXiv:hep-ex/0107034}}.
\bibAnnoteFile{Searches:2001aa}

\bibitem{CMS:zza}
CMS Collaboration,
  \href{http://cds.cern.ch/record/1558930}{CMS-PAS-HIG-13-016}.
\bibAnnoteFile{CMS:zza}

\bibitem{Espinosa:1991gr}
J.~Espinosa and M.~Quiros,
  \href{http://dx.doi.org/10.1016/0370-2693(92)91846-2}{\emph{Phys. Lett.}
  \textbf{B 279} (1992) 92}.
\bibAnnoteFile{Espinosa:1991gr}

\bibitem{Masip:1998jc}
M.~Masip, R.~Munoz-Tapia and A.~Pomarol,
  \href{http://dx.doi.org/10.1103/PhysRevD.57.5340}{\emph{Phys. Rev.} \textbf{D
  57} (1998) R5340}, \href{http://arxiv.org/abs/hep-ph/9801437}{{\tt
  arXiv:hep-ph/9801437}}.
\bibAnnoteFile{Masip:1998jc}

\bibitem{Barbieri:2007tu}
R.~Barbieri, L.~J. Hall, A.~Y. Papaioannou, D.~Pappadopulo and V.~S. Rychkov,
  \href{http://dx.doi.org/10.1088/1126-6708/2008/03/005}{\emph{JHEP}
  \textbf{0803} (2008) 005}, \href{http://arxiv.org/abs/0712.2903}{{\tt
  arXiv:0712.2903 [hep-ph]}}.
\bibAnnoteFile{Barbieri:2007tu}

\bibitem{Harnik:2003rs}
R.~Harnik, G.~D. Kribs, D.~T. Larson and H.~Murayama,
  \href{http://dx.doi.org/10.1103/PhysRevD.70.015002}{\emph{Phys. Rev.}
  \textbf{D 70} (2004) 015002}, \href{http://arxiv.org/abs/hep-ph/0311349}{{\tt
  arXiv:hep-ph/0311349}}.
\bibAnnoteFile{Harnik:2003rs}

\bibitem{Chang:2004db}
S.~Chang, C.~Kilic and R.~Mahbubani,
  \href{http://dx.doi.org/10.1103/PhysRevD.71.015003}{\emph{Phys. Rev.}
  \textbf{D 71} (2005) 015003}, \href{http://arxiv.org/abs/hep-ph/0405267}{{\tt
  arXiv:hep-ph/0405267}}.
\bibAnnoteFile{Chang:2004db}

\bibitem{Birkedal:2004zx}
A.~Birkedal, Z.~Chacko and Y.~Nomura,
  \href{http://dx.doi.org/10.1103/PhysRevD.71.015006}{\emph{Phys. Rev.}
  \textbf{D 71} (2005) 015006}, \href{http://arxiv.org/abs/hep-ph/0408329}{{\tt
  arXiv:hep-ph/0408329}}.
\bibAnnoteFile{Birkedal:2004zx}

\bibitem{Delgado:2005fq}
A.~Delgado and T.~M. Tait,
  \href{http://dx.doi.org/10.1088/1126-6708/2005/07/023}{\emph{JHEP}
  \textbf{0507} (2005) 023}, \href{http://arxiv.org/abs/hep-ph/0504224}{{\tt
  arXiv:hep-ph/0504224}}.
\bibAnnoteFile{Delgado:2005fq}

\bibitem{Gherghetta:2011wc}
T.~Gherghetta, B.~von Harling and N.~Setzer,
  \href{http://dx.doi.org/10.1007/JHEP07(2011)011}{\emph{JHEP} \textbf{1107}
  (2011) 011}, \href{http://arxiv.org/abs/1104.3171}{{\tt arXiv:1104.3171
  [hep-ph]}}.
\bibAnnoteFile{Gherghetta:2011wc}

\bibitem{Craig:2011ev}
N.~Craig, D.~Stolarski and J.~Thaler,
  \href{http://dx.doi.org/10.1007/JHEP11(2011)145}{\emph{JHEP} \textbf{1111}
  (2011) 145}, \href{http://arxiv.org/abs/1106.2164}{{\tt arXiv:1106.2164
  [hep-ph]}}.
\bibAnnoteFile{Craig:2011ev}

\bibitem{Csaki:2011xn}
C.~Csaki, Y.~Shirman and J.~Terning,
  \href{http://dx.doi.org/10.1103/PhysRevD.84.095011}{\emph{Phys. Rev.}
  \textbf{D 84} (2011) 095011}, \href{http://arxiv.org/abs/1106.3074}{{\tt
  arXiv:1106.3074 [hep-ph]}}.
\bibAnnoteFile{Csaki:2011xn}

\bibitem{Hardy:2012ef}
E.~Hardy, J.~March-Russell and J.~Unwin,
  \href{http://dx.doi.org/10.1007/JHEP10(2012)072}{\emph{JHEP} \textbf{1210}
  (2012) 072}, \href{http://arxiv.org/abs/1207.1435}{{\tt arXiv:1207.1435
  [hep-ph]}}.
\bibAnnoteFile{Hardy:2012ef}

\bibitem{Cavicchia:2007dp}
L.~Cavicchia, R.~Franceschini and V.~S. Rychkov,
  \href{http://dx.doi.org/10.1103/PhysRevD.77.055006}{\emph{Phys. Rev.}
  \textbf{D 77} (2008) 055006}, \href{http://arxiv.org/abs/0710.5750}{{\tt
  arXiv:0710.5750 [hep-ph]}}.
\bibAnnoteFile{Cavicchia:2007dp}

\bibitem{Caterina}
C.~Vernieri, ``{Search for resonances in the 4b final state at LHC with the CMS
  experiment}'', Ph.D. thesis, Scuola Normale Superiore, Pisa (2013). In
  preparation.
\bibAnnoteFile{Caterina}

\bibitem{Gouzevitch:2013qca}
M.~Gouzevitch, A.~Oliveira, J.~Rojo, R.~Rosenfeld, G.~P. Salam \emph{et~al.},
  \href{http://dx.doi.org/10.1007/JHEP07(2013)148}{\emph{JHEP} \textbf{1307}
  (2013) 148}, \href{http://arxiv.org/abs/1303.6636}{{\tt arXiv:1303.6636
  [hep-ph]}}.
\bibAnnoteFile{Gouzevitch:2013qca}

\bibitem{Chatrchyan:2013yoa}
CMS Collaboration,
  \href{http://dx.doi.org/10.1140/epjc/s10052-013-2469-8}{\emph{Eur. Phys. J.}
  \textbf{C 73} (2013) 2469}, \href{http://arxiv.org/abs/1304.0213}{{\tt
  arXiv:1304.0213 [hep-ex]}}.
\bibAnnoteFile{Chatrchyan:2013yoa}

\bibitem{Krive:1976sg}
I.~Krive and A.~D. Linde,
  \href{http://dx.doi.org/10.1016/0550-3213(76)90573-3}{\emph{Nucl. Phys.}
  \textbf{B 117} (1976) 265}.
\bibAnnoteFile{Krive:1976sg}

\bibitem{Krasnikov:1978pu}
N.~Krasnikov, \emph{Yad. Fiz.} \textbf{28} (1978) 549.
\bibAnnoteFile{Krasnikov:1978pu}

\bibitem{Maiani:1977cg}
L.~Maiani, G.~Parisi and R.~Petronzio,
  \href{http://dx.doi.org/10.1016/0550-3213(78)90018-4}{\emph{Nucl. Phys.}
  \textbf{B 136} (1978) 115}.
\bibAnnoteFile{Maiani:1977cg}

\bibitem{Politzer:1978ic}
H.~D. Politzer and S.~Wolfram,
  \href{http://dx.doi.org/10.1016/0370-2693(79)90746-9}{\emph{Phys. Lett.}
  \textbf{B 82} (1979) 242}. Erratum-ibid.
  \href{http://dx.doi.org/10.1016/0370-2693(79)91144-4}{{\bf B83} (1979) 421}.
\bibAnnoteFile{Politzer:1978ic}

\bibitem{Hung:1979dn}
P.~Q. Hung, \href{http://dx.doi.org/10.1103/PhysRevLett.42.873}{\emph{Phys.
  Rev. Lett.} \textbf{42} (1979) 873}.
\bibAnnoteFile{Hung:1979dn}

\bibitem{Cabibbo:1979ay}
N.~Cabibbo, L.~Maiani, G.~Parisi and R.~Petronzio,
  \href{http://dx.doi.org/10.1016/0550-3213(79)90167-6}{\emph{Nucl. Phys.}
  \textbf{B 158} (1979) 295}.
\bibAnnoteFile{Cabibbo:1979ay}

\bibitem{Linde:1979ny}
A.~D. Linde, \href{http://dx.doi.org/10.1016/0370-2693(80)90318-4}{\emph{Phys.
  Lett.} \textbf{B 92} (1980) 119}.
\bibAnnoteFile{Linde:1979ny}

\bibitem{Lindner:1985uk}
M.~Lindner, \href{http://dx.doi.org/10.1007/BF01479540}{\emph{Z. Phys.}
  \textbf{C 31} (1986) 295}.
\bibAnnoteFile{Lindner:1985uk}

\bibitem{Lindner:1988ww}
M.~Lindner, M.~Sher and H.~W. Zaglauer,
  \href{http://dx.doi.org/10.1016/0370-2693(89)90540-6}{\emph{Phys. Lett.}
  \textbf{B 228} (1989) 139}.
\bibAnnoteFile{Lindner:1988ww}

\bibitem{Sher:1988mj}
M.~Sher, \href{http://dx.doi.org/10.1016/0370-1573(89)90061-6}{\emph{Phys.
  Rept.} \textbf{179} (1989) 273}.
\bibAnnoteFile{Sher:1988mj}

\bibitem{Arnold:1989cb}
P.~B. Arnold, \href{http://dx.doi.org/10.1103/PhysRevD.40.613}{\emph{Phys.
  Rev.} \textbf{D 40} (1989) 613}.
\bibAnnoteFile{Arnold:1989cb}

\bibitem{Arnold:1991cv}
P.~B. Arnold and S.~Vokos,
  \href{http://dx.doi.org/10.1103/PhysRevD.44.3620}{\emph{Phys. Rev.} \textbf{D
  44} (1991) 3620}.
\bibAnnoteFile{Arnold:1991cv}

\bibitem{Sher:1993mf}
M.~Sher, \href{http://dx.doi.org/10.1016/0370-2693(93)91586-C}{\emph{Phys.
  Lett.} \textbf{B 317} (1993) 159},
  \href{http://arxiv.org/abs/hep-ph/9307342}{{\tt arXiv:hep-ph/9307342}}.
  Addendum-ibid. \href{http://dx.doi.org/10.1016/0370-2693(94)91078-2}{{\bf
  B331} (1994) 448}.
\bibAnnoteFile{Sher:1993mf}

\bibitem{Altarelli:1994rb}
G.~Altarelli and G.~Isidori,
  \href{http://dx.doi.org/10.1016/0370-2693(94)91458-3}{\emph{Phys. Lett.}
  \textbf{B 337} (1994) 141}.
\bibAnnoteFile{Altarelli:1994rb}

\bibitem{Casas:1994qy}
J.~Casas, J.~Espinosa and M.~Quiros,
  \href{http://dx.doi.org/10.1016/0370-2693(94)01404-Z}{\emph{Phys. Lett.}
  \textbf{B 342} (1995) 171}, \href{http://arxiv.org/abs/hep-ph/9409458}{{\tt
  arXiv:hep-ph/9409458}}.
\bibAnnoteFile{Casas:1994qy}

\bibitem{Espinosa:1995se}
J.~Espinosa and M.~Quiros,
  \href{http://dx.doi.org/10.1016/0370-2693(95)00572-3}{\emph{Phys. Lett.}
  \textbf{B 353} (1995) 257}, \href{http://arxiv.org/abs/hep-ph/9504241}{{\tt
  arXiv:hep-ph/9504241}}.
\bibAnnoteFile{Espinosa:1995se}

\bibitem{Casas:1996aq}
J.~Casas, J.~Espinosa and M.~Quiros,
  \href{http://dx.doi.org/10.1016/0370-2693(96)00682-X}{\emph{Phys. Lett.}
  \textbf{B 382} (1996) 374}, \href{http://arxiv.org/abs/hep-ph/9603227}{{\tt
  arXiv:hep-ph/9603227}}.
\bibAnnoteFile{Casas:1996aq}

\bibitem{Schrempp:1996fb}
B.~Schrempp and M.~Wimmer,
  \href{http://dx.doi.org/10.1016/0146-6410(96)00059-2}{\emph{Prog. Part. Nucl.
  Phys.} \textbf{37} (1996) 1}, \href{http://arxiv.org/abs/hep-ph/9606386}{{\tt
  arXiv:hep-ph/9606386}}.
\bibAnnoteFile{Schrempp:1996fb}

\bibitem{Hambye:1996wb}
T.~Hambye and K.~Riesselmann,
  \href{http://dx.doi.org/10.1103/PhysRevD.55.7255}{\emph{Phys. Rev.} \textbf{D
  55} (1997) 7255}, \href{http://arxiv.org/abs/hep-ph/9610272}{{\tt
  arXiv:hep-ph/9610272}}.
\bibAnnoteFile{Hambye:1996wb}

\bibitem{Isidori:2001bm}
G.~Isidori, G.~Ridolfi and A.~Strumia,
  \href{http://dx.doi.org/10.1016/S0550-3213(01)00302-9}{\emph{Nucl. Phys.}
  \textbf{B 609} (2001) 387}, \href{http://arxiv.org/abs/hep-ph/0104016}{{\tt
  arXiv:hep-ph/0104016}}.
\bibAnnoteFile{Isidori:2001bm}

\bibitem{Espinosa:2007qp}
J.~Espinosa, G.~Giudice and A.~Riotto,
  \href{http://dx.doi.org/10.1088/1475-7516/2008/05/002}{\emph{JCAP}
  \textbf{0805} (2008) 002}, \href{http://arxiv.org/abs/0710.2484}{{\tt
  arXiv:0710.2484 [hep-ph]}}.
\bibAnnoteFile{Espinosa:2007qp}

\bibitem{Ellis:2009tp}
J.~Ellis, J.~Espinosa, G.~Giudice, A.~Hoecker and A.~Riotto,
  \href{http://dx.doi.org/10.1016/j.physletb.2009.07.054}{\emph{Phys. Lett.}
  \textbf{B 679} (2009) 369}, \href{http://arxiv.org/abs/0906.0954}{{\tt
  arXiv:0906.0954 [hep-ph]}}.
\bibAnnoteFile{Ellis:2009tp}

\bibitem{Holthausen:2011aa}
M.~Holthausen, K.~S. Lim and M.~Lindner,
  \href{http://dx.doi.org/10.1007/JHEP02(2012)037}{\emph{JHEP} \textbf{1202}
  (2012) 037}, \href{http://arxiv.org/abs/1112.2415}{{\tt arXiv:1112.2415
  [hep-ph]}}.
\bibAnnoteFile{Holthausen:2011aa}

\bibitem{EliasMiro:2011aa}
J.~Elias-Miro, J.~R. Espinosa, G.~F. Giudice, G.~Isidori, A.~Riotto
  \emph{et~al.},
  \href{http://dx.doi.org/10.1016/j.physletb.2012.02.013}{\emph{Phys. Lett.}
  \textbf{B 709} (2012) 222}, \href{http://arxiv.org/abs/1112.3022}{{\tt
  arXiv:1112.3022 [hep-ph]}}.
\bibAnnoteFile{EliasMiro:2011aa}

\bibitem{Chen:2012faa}
C.-S. Chen and Y.~Tang,
  \href{http://dx.doi.org/10.1007/JHEP04(2012)019}{\emph{JHEP} \textbf{1204}
  (2012) 019}, \href{http://arxiv.org/abs/1202.5717}{{\tt arXiv:1202.5717
  [hep-ph]}}.
\bibAnnoteFile{Chen:2012faa}

\bibitem{Lebedev:2012zw}
O.~Lebedev, \href{http://dx.doi.org/10.1140/epjc/s10052-012-2058-2}{\emph{Eur.
  Phys. J.} \textbf{C 72} (2012) 2058},
  \href{http://arxiv.org/abs/1203.0156}{{\tt arXiv:1203.0156 [hep-ph]}}.
\bibAnnoteFile{Lebedev:2012zw}

\bibitem{EliasMiro:2012ay}
J.~Elias-Miro, J.~R. Espinosa, G.~F. Giudice, H.~M. Lee and A.~Strumia,
  \href{http://dx.doi.org/10.1007/JHEP06(2012)031}{\emph{JHEP} \textbf{1206}
  (2012) 031}, \href{http://arxiv.org/abs/1203.0237}{{\tt arXiv:1203.0237
  [hep-ph]}}.
\bibAnnoteFile{EliasMiro:2012ay}

\bibitem{Rodejohann:2012px}
W.~Rodejohann and H.~Zhang,
  \href{http://dx.doi.org/10.1007/JHEP06(2012)022}{\emph{JHEP} \textbf{1206}
  (2012) 022}, \href{http://arxiv.org/abs/1203.3825}{{\tt arXiv:1203.3825
  [hep-ph]}}.
\bibAnnoteFile{Rodejohann:2012px}

\bibitem{Bezrukov:2012sa}
F.~Bezrukov, M.~Y. Kalmykov, B.~A. Kniehl and M.~Shaposhnikov,
  \href{http://dx.doi.org/10.1007/JHEP10(2012)140}{\emph{JHEP} \textbf{1210}
  (2012) 140}, \href{http://arxiv.org/abs/1205.2893}{{\tt arXiv:1205.2893
  [hep-ph]}}.
\bibAnnoteFile{Bezrukov:2012sa}

\bibitem{Datta:2012db}
A.~Datta and S.~Raychaudhuri,
  \href{http://dx.doi.org/10.1103/PhysRevD.87.035018}{\emph{Phys. Rev.}
  \textbf{D 87} (2013), 3 035018}, \href{http://arxiv.org/abs/1207.0476}{{\tt
  arXiv:1207.0476 [hep-ph]}}.
\bibAnnoteFile{Datta:2012db}

\bibitem{Alekhin:2012py}
S.~Alekhin, A.~Djouadi and S.~Moch,
  \href{http://dx.doi.org/10.1016/j.physletb.2012.08.024}{\emph{Phys. Lett.}
  \textbf{B 716} (2012) 214}, \href{http://arxiv.org/abs/1207.0980}{{\tt
  arXiv:1207.0980 [hep-ph]}}.
\bibAnnoteFile{Alekhin:2012py}

\bibitem{Chakrabortty:2012np}
J.~Chakrabortty, M.~Das and S.~Mohanty,
  \href{http://dx.doi.org/10.1142/S0217732313500326}{\emph{Mod. Phys. Lett.}
  \textbf{A 28} (2013) 1350032}, \href{http://arxiv.org/abs/1207.2027}{{\tt
  arXiv:1207.2027 [hep-ph]}}.
\bibAnnoteFile{Chakrabortty:2012np}

\bibitem{Anchordoqui:2012fq}
L.~A. Anchordoqui, I.~Antoniadis, H.~Goldberg, X.~Huang, D.~Lust \emph{et~al.},
  \href{http://dx.doi.org/10.1007/JHEP02(2013)074}{\emph{JHEP} \textbf{1302}
  (2013) 074}, \href{http://arxiv.org/abs/1208.2821}{{\tt arXiv:1208.2821
  [hep-ph]}}.
\bibAnnoteFile{Anchordoqui:2012fq}

\bibitem{Masina:2012tz}
I.~Masina, \href{http://dx.doi.org/10.1103/PhysRevD.87.053001}{\emph{Phys.
  Rev.} \textbf{D 87} (2013) 053001},
  \href{http://arxiv.org/abs/1209.0393}{{\tt arXiv:1209.0393 [hep-ph]}}.
\bibAnnoteFile{Masina:2012tz}

\bibitem{Chun:2012jw}
E.~J. Chun, H.~M. Lee and P.~Sharma,
  \href{http://dx.doi.org/10.1007/JHEP11(2012)106}{\emph{JHEP} \textbf{1211}
  (2012) 106}, \href{http://arxiv.org/abs/1209.1303}{{\tt arXiv:1209.1303
  [hep-ph]}}.
\bibAnnoteFile{Chun:2012jw}

\bibitem{Chung:2012vg}
D.~J. Chung, A.~J. Long and L.-T. Wang,
  \href{http://dx.doi.org/10.1103/PhysRevD.87.023509}{\emph{Phys. Rev.}
  \textbf{D 87} (2013) 023509}, \href{http://arxiv.org/abs/1209.1819}{{\tt
  arXiv:1209.1819 [hep-ph]}}.
\bibAnnoteFile{Chung:2012vg}

\bibitem{Chao:2012mx}
W.~Chao, M.~Gonderinger and M.~J. Ramsey-Musolf,
  \href{http://dx.doi.org/10.1103/PhysRevD.86.113017}{\emph{Phys. Rev.}
  \textbf{D 86} (2012) 113017}, \href{http://arxiv.org/abs/1210.0491}{{\tt
  arXiv:1210.0491 [hep-ph]}}.
\bibAnnoteFile{Chao:2012mx}

\bibitem{Lebedev:2012sy}
O.~Lebedev and A.~Westphal,
  \href{http://dx.doi.org/10.1016/j.physletb.2012.12.069}{\emph{Phys. Lett.}
  \textbf{B 719} (2013) 415}, \href{http://arxiv.org/abs/1210.6987}{{\tt
  arXiv:1210.6987 [hep-ph]}}.
\bibAnnoteFile{Lebedev:2012sy}

\bibitem{Nielsen:2012pu}
H.~B. Nielsen, \href{http://arxiv.org/abs/1212.5716}{{\tt arXiv:1212.5716
  [hep-ph]}}.
\bibAnnoteFile{Nielsen:2012pu}

\bibitem{Kobakhidze:2013tn}
A.~Kobakhidze and A.~Spencer-Smith,
  \href{http://dx.doi.org/10.1016/j.physletb.2013.04.013}{\emph{Phys. Lett.}
  \textbf{B 722} (2013) 130}, \href{http://arxiv.org/abs/1301.2846}{{\tt
  arXiv:1301.2846 [hep-ph]}}.
\bibAnnoteFile{Kobakhidze:2013tn}

\bibitem{Tang:2013bz}
Y.~Tang, \href{http://dx.doi.org/10.1142/S0217732313300024}{\emph{Mod. Phys.
  Lett.} \textbf{A 28} (2013) 1330002},
  \href{http://arxiv.org/abs/1301.5812}{{\tt arXiv:1301.5812 [hep-ph]}}.
\bibAnnoteFile{Tang:2013bz}

\bibitem{Klinkhamer:2013sos}
F.~Klinkhamer, \href{http://dx.doi.org/10.1134/S002136401306009X}{\emph{JETP
  Lett.} \textbf{97} (2013) 297}, \href{http://arxiv.org/abs/1302.1496}{{\tt
  arXiv:1302.1496 [hep-ph]}}.
\bibAnnoteFile{Klinkhamer:2013sos}

\bibitem{He:2013tla}
X.-G. He, H.~Phoon, Y.~Tang and G.~Valencia,
  \href{http://dx.doi.org/10.1007/JHEP05(2013)026}{\emph{JHEP} \textbf{1305}
  (2013) 026}, \href{http://arxiv.org/abs/1303.4848}{{\tt arXiv:1303.4848
  [hep-ph]}}.
\bibAnnoteFile{He:2013tla}

\bibitem{Chun:2013soa}
E.~J. Chun, S.~Jung and H.~M. Lee,
  \href{http://dx.doi.org/10.1016/j.physletb.2013.06.055}{\emph{Phys. Lett.}
  \textbf{B 725} (2013) 158}, \href{http://arxiv.org/abs/1304.5815}{{\tt
  arXiv:1304.5815 [hep-ph]}}.
\bibAnnoteFile{Chun:2013soa}

\bibitem{Jegerlehner:2013cta}
F.~Jegerlehner, \href{http://arxiv.org/abs/1304.7813}{{\tt arXiv:1304.7813
  [hep-ph]}}.
\bibAnnoteFile{Jegerlehner:2013cta}

\bibitem{Gross:1973id}
D.~Gross and F.~Wilczek,
  \href{http://dx.doi.org/10.1103/PhysRevLett.30.1343}{\emph{Phys. Rev. Lett.}
  \textbf{30} (1973) 1343}.
\bibAnnoteFile{Gross:1973id}

\bibitem{Politzer:1973fx}
H.~D. Politzer,
  \href{http://dx.doi.org/10.1103/PhysRevLett.30.1346}{\emph{Phys. Rev. Lett.}
  \textbf{30} (1973) 1346}.
\bibAnnoteFile{Politzer:1973fx}

\bibitem{Caswell:1974gg}
W.~E. Caswell, \href{http://dx.doi.org/10.1103/PhysRevLett.33.244}{\emph{Phys.
  Rev. Lett.} \textbf{33} (1974) 244}.
\bibAnnoteFile{Caswell:1974gg}

\bibitem{Jones:1974mm}
D.~Jones, \href{http://dx.doi.org/10.1016/0550-3213(74)90093-5}{\emph{Nucl.
  Phys.} \textbf{B 75} (1974) 531}.
\bibAnnoteFile{Jones:1974mm}

\bibitem{Tarasov:1980au}
O.~Tarasov, A.~Vladimirov and A.~Y. Zharkov,
  \href{http://dx.doi.org/10.1016/0370-2693(80)90358-5}{\emph{Phys. Lett.}
  \textbf{B 93} (1980) 429}.
\bibAnnoteFile{Tarasov:1980au}

\bibitem{Larin:1993tp}
S.~Larin and J.~Vermaseren,
  \href{http://dx.doi.org/10.1016/0370-2693(93)91441-O}{\emph{Phys. Lett.}
  \textbf{B 303} (1993) 334}, \href{http://arxiv.org/abs/hep-ph/9302208}{{\tt
  arXiv:hep-ph/9302208}}.
\bibAnnoteFile{Larin:1993tp}

\bibitem{vanRitbergen:1997va}
T.~van Ritbergen, J.~Vermaseren and S.~Larin,
  \href{http://dx.doi.org/10.1016/S0370-2693(97)00370-5}{\emph{Phys. Lett.}
  \textbf{B 400} (1997) 379}, \href{http://arxiv.org/abs/hep-ph/9701390}{{\tt
  arXiv:hep-ph/9701390}}.
\bibAnnoteFile{vanRitbergen:1997va}

\bibitem{Czakon:2004bu}
M.~Czakon, \href{http://dx.doi.org/10.1016/j.nuclphysb.2005.01.012}{\emph{Nucl.
  Phys.} \textbf{B 710} (2005) 485},
  \href{http://arxiv.org/abs/hep-ph/0411261}{{\tt arXiv:hep-ph/0411261}}.
\bibAnnoteFile{Czakon:2004bu}

\bibitem{Jones:1981we}
D.~Jones, \href{http://dx.doi.org/10.1103/PhysRevD.25.581}{\emph{Phys. Rev.}
  \textbf{D 25} (1982) 581}.
\bibAnnoteFile{Jones:1981we}

\bibitem{Steinhauser:1998cm}
M.~Steinhauser, \href{http://dx.doi.org/10.1103/PhysRevD.59.054005}{\emph{Phys.
  Rev.} \textbf{D 59} (1999) 054005},
  \href{http://arxiv.org/abs/hep-ph/9809507}{{\tt arXiv:hep-ph/9809507}}.
\bibAnnoteFile{Steinhauser:1998cm}

\bibitem{Machacek:1983tz}
M.~E. Machacek and M.~T. Vaughn,
  \href{http://dx.doi.org/10.1016/0550-3213(83)90610-7}{\emph{Nucl. Phys.}
  \textbf{B 222} (1983) 83}.
\bibAnnoteFile{Machacek:1983tz}

\bibitem{Mihaila:2012fm}
L.~N. Mihaila, J.~Salomon and M.~Steinhauser,
  \href{http://dx.doi.org/10.1103/PhysRevLett.108.151602}{\emph{Phys. Rev.
  Lett.} \textbf{108} (2012) 151602},
  \href{http://arxiv.org/abs/1201.5868}{{\tt arXiv:1201.5868 [hep-ph]}}.
\bibAnnoteFile{Mihaila:2012fm}

\bibitem{Mihaila:2012pz}
L.~N. Mihaila, J.~Salomon and M.~Steinhauser,
  \href{http://dx.doi.org/10.1103/PhysRevD.86.096008}{\emph{Phys. Rev.}
  \textbf{D 86} (2012) 096008}, \href{http://arxiv.org/abs/1208.3357}{{\tt
  arXiv:1208.3357 [hep-ph]}}.
\bibAnnoteFile{Mihaila:2012pz}

\bibitem{Cheng:1973nv}
T.~Cheng, E.~Eichten and L.-F. Li,
  \href{http://dx.doi.org/10.1103/PhysRevD.9.2259}{\emph{Phys. Rev.} \textbf{D
  9} (1974) 2259}.
\bibAnnoteFile{Cheng:1973nv}

\bibitem{Fischler:1982du}
M.~Fischler and J.~Oliensis,
  \href{http://dx.doi.org/10.1016/0370-2693(82)90695-5}{\emph{Phys. Lett.}
  \textbf{B 119} (1982) 385}.
\bibAnnoteFile{Fischler:1982du}

\bibitem{Chetyrkin:2012rz}
K.~Chetyrkin and M.~Zoller,
  \href{http://dx.doi.org/10.1007/JHEP06(2012)033}{\emph{JHEP} \textbf{1206}
  (2012) 033}, \href{http://arxiv.org/abs/1205.2892}{{\tt arXiv:1205.2892
  [hep-ph]}}.
\bibAnnoteFile{Chetyrkin:2012rz}

\bibitem{Bednyakov:2012en}
A.~Bednyakov, A.~Pikelner and V.~Velizhanin,
  \href{http://dx.doi.org/10.1016/j.physletb.2013.04.038}{\emph{Phys. Lett.}
  \textbf{B 722} (2013) 336}, \href{http://arxiv.org/abs/1212.6829}{{\tt
  arXiv:1212.6829}}.
\bibAnnoteFile{Bednyakov:2012en}

\bibitem{Machacek:1983fi}
M.~E. Machacek and M.~T. Vaughn,
  \href{http://dx.doi.org/10.1016/0550-3213(84)90533-9}{\emph{Nucl. Phys.}
  \textbf{B 236} (1984) 221}.
\bibAnnoteFile{Machacek:1983fi}

\bibitem{Machacek:1984zw}
M.~E. Machacek and M.~T. Vaughn,
  \href{http://dx.doi.org/10.1016/0550-3213(85)90040-9}{\emph{Nucl. Phys.}
  \textbf{B 249} (1985) 70}.
\bibAnnoteFile{Machacek:1984zw}

\bibitem{Luo:2002ey}
M.-x. Luo and Y.~Xiao,
  \href{http://dx.doi.org/10.1103/PhysRevLett.90.011601}{\emph{Phys. Rev.
  Lett.} \textbf{90} (2003) 011601},
  \href{http://arxiv.org/abs/hep-ph/0207271}{{\tt arXiv:hep-ph/0207271}}.
\bibAnnoteFile{Luo:2002ey}

\bibitem{Chetyrkin:2013wya}
K.~Chetyrkin and M.~Zoller,
  \href{http://dx.doi.org/10.1007/JHEP04(2013)091}{\emph{JHEP} \textbf{1304}
  (2013) 091}, \href{http://arxiv.org/abs/1303.2890}{{\tt arXiv:1303.2890
  [hep-ph]}}.
\bibAnnoteFile{Chetyrkin:2013wya}

\bibitem{Bednyakov:2013eba}
A.~Bednyakov, A.~Pikelner and V.~Velizhanin,
  \href{http://dx.doi.org/10.1016/j.nuclphysb.2013.07.015}{\emph{Nucl. Phys.}
  \textbf{B 875} (2013) 552}, \href{http://arxiv.org/abs/1303.4364}{{\tt
  arXiv:1303.4364}}.
\bibAnnoteFile{Bednyakov:2013eba}

\bibitem{Sirlin:1980nh}
A.~Sirlin, \href{http://dx.doi.org/10.1103/PhysRevD.22.971}{\emph{Phys. Rev.}
  \textbf{D 22} (1980) 971}.
\bibAnnoteFile{Sirlin:1980nh}

\bibitem{Marciano:1980pb}
W.~Marciano and A.~Sirlin,
  \href{http://dx.doi.org/10.1103/PhysRevD.31.213}{\emph{Phys. Rev.} \textbf{D
  22} (1980) 2695}. Erratum-ibid.
  \href{http://dx.doi.org/10.1103/PhysRevD.22.2695}{{\bf D31} (1985) 213}.
\bibAnnoteFile{Marciano:1980pb}

\bibitem{Tarrach:1980up}
R.~Tarrach, \href{http://dx.doi.org/10.1016/0550-3213(81)90140-1}{\emph{Nucl.
  Phys.} \textbf{B 183} (1981) 384}.
\bibAnnoteFile{Tarrach:1980up}

\bibitem{Chetyrkin:1999ys}
K.~Chetyrkin and M.~Steinhauser,
  \href{http://dx.doi.org/10.1103/PhysRevLett.83.4001}{\emph{Phys. Rev. Lett.}
  \textbf{83} (1999) 4001}, \href{http://arxiv.org/abs/hep-ph/9907509}{{\tt
  arXiv:hep-ph/9907509}}.
\bibAnnoteFile{Chetyrkin:1999ys}

\bibitem{Chetyrkin:1999qi}
K.~Chetyrkin and M.~Steinhauser,
  \href{http://dx.doi.org/10.1016/S0550-3213(99)00784-1}{\emph{Nucl. Phys.}
  \textbf{B 573} (2000) 617}, \href{http://arxiv.org/abs/hep-ph/9911434}{{\tt
  arXiv:hep-ph/9911434}}.
\bibAnnoteFile{Chetyrkin:1999qi}

\bibitem{Melnikov:2000qh}
K.~Melnikov and T.~v. Ritbergen,
  \href{http://dx.doi.org/10.1016/S0370-2693(00)00507-4}{\emph{Phys. Lett.}
  \textbf{B 482} (2000) 99}, \href{http://arxiv.org/abs/hep-ph/9912391}{{\tt
  arXiv:hep-ph/9912391}}.
\bibAnnoteFile{Melnikov:2000qh}

\bibitem{Hempfling:1994ar}
R.~Hempfling and B.~A. Kniehl,
  \href{http://dx.doi.org/10.1103/PhysRevD.51.1386}{\emph{Phys. Rev.} \textbf{D
  51} (1995) 1386}, \href{http://arxiv.org/abs/hep-ph/9408313}{{\tt
  arXiv:hep-ph/9408313}}.
\bibAnnoteFile{Hempfling:1994ar}

\bibitem{Sirlin:1985ux}
A.~Sirlin and R.~Zucchini,
  \href{http://dx.doi.org/10.1016/0550-3213(86)90096-9}{\emph{Nucl. Phys.}
  \textbf{B 266} (1986) 389}.
\bibAnnoteFile{Sirlin:1985ux}

\bibitem{Caswell:1974cj}
W.~E. Caswell and F.~Wilczek,
  \href{http://dx.doi.org/10.1016/0370-2693(74)90437-7}{\emph{Phys. Lett.}
  \textbf{B 49} (1974) 291}.
\bibAnnoteFile{Caswell:1974cj}

\bibitem{Muta:1987mz}
T.~Muta, ``{Foundations of quantum chromodynamics: An Introduction to
  perturbative methods in gauge theories}'', \emph{World Sci. Lect. Notes
  Phys.}, vol.~5 (World Scientific, 1987).
\bibAnnoteFile{Muta:1987mz}

\bibitem{Atkinson:1979ut}
D.~Atkinson and M.~Fry,
  \href{http://dx.doi.org/10.1016/0550-3213(79)90033-6}{\emph{Nucl. Phys.}
  \textbf{B 156} (1979) 301}.
\bibAnnoteFile{Atkinson:1979ut}

\bibitem{Breckenridge:1994gs}
J.~Breckenridge, M.~Lavelle and T.~G. Steele,
  \href{http://dx.doi.org/10.1007/BF01571316}{\emph{Z. Phys.} \textbf{C 65}
  (1995) 155}, \href{http://arxiv.org/abs/hep-th/9407028}{{\tt
  arXiv:hep-th/9407028}}.
\bibAnnoteFile{Breckenridge:1994gs}

\bibitem{Kronfeld:1998di}
A.~S. Kronfeld, \href{http://dx.doi.org/10.1103/PhysRevD.58.051501}{\emph{Phys.
  Rev.} \textbf{D 58} (1998) 051501},
  \href{http://arxiv.org/abs/hep-ph/9805215}{{\tt arXiv:hep-ph/9805215}}.
\bibAnnoteFile{Kronfeld:1998di}

\bibitem{Group:2012gb}
CDF and D0 Collaborations, Tevatron Electroweak Working Group,
  \href{http://arxiv.org/abs/1204.0042}{{\tt arXiv:1204.0042 [hep-ex]}}.
\bibAnnoteFile{Group:2012gb}

\bibitem{Alcaraz:2006mx}
ALEPH, DELPHI, L3 and OPAL Collaborations, LEP Electroweak Working Group,
  \href{http://arxiv.org/abs/hep-ex/0612034}{{\tt arXiv:hep-ex/0612034}}.
\bibAnnoteFile{Alcaraz:2006mx}

\bibitem{Beringer:1900zz}
Particle Data Group,
  \href{http://dx.doi.org/10.1103/PhysRevD.86.010001}{\emph{Phys. Rev.}
  \textbf{D 86} (2012) 010001}.
\bibAnnoteFile{Beringer:1900zz}

\bibitem{Lancaster:2011wr}
CDF and D0 Collaborations, Tevatron Electroweak Working Group,
  \href{http://arxiv.org/abs/1107.5255}{{\tt arXiv:1107.5255 [hep-ex]}}.
\bibAnnoteFile{Lancaster:2011wr}

\bibitem{CMS:2012fya}
CMS Collaboration,
  \href{http://cds.cern.ch/record/1478194}{CMS-PAS-TOP-11-018}.
\bibAnnoteFile{CMS:2012fya}

\bibitem{CMS:2012awa}
CMS Collaboration,
  \href{http://cds.cern.ch/record/1460097}{CMS-PAS-TOP-12-001}.
\bibAnnoteFile{CMS:2012awa}

\bibitem{ATLAS:2012coa}
ATLAS Collaboration,
  \href{http://cds.cern.ch/record/1460441}{ATLAS-CONF-2012-095}.
\bibAnnoteFile{ATLAS:2012coa}

\bibitem{ATLAS:2013zzh}
ATLAS Collaboration,
  \href{http://cds.cern.ch/record/1547327}{ATLAS-CONF-2013-046}.
\bibAnnoteFile{ATLAS:2013zzh}

\bibitem{ATLAS:2013zzi}
ATLAS Collaboration,
  \href{http://cds.cern.ch/record/1562935}{ATLAS-CONF-2013-077}.
\bibAnnoteFile{ATLAS:2013zzi}

\bibitem{Tishchenko:2012ie}
MuLan Collaboration, \href{http://arxiv.org/abs/1211.0960}{{\tt arXiv:1211.0960
  [hep-ex]}}.
\bibAnnoteFile{Tishchenko:2012ie}

\bibitem{Bethke:2012jm}
S.~Bethke,
  \href{http://dx.doi.org/10.1016/j.nuclphysbps.2012.12.020}{\emph{Nucl. Phys.
  Proc. Suppl.} \textbf{234} (2013) 229},
  \href{http://arxiv.org/abs/1210.0325}{{\tt arXiv:1210.0325 [hep-ex]}}.
\bibAnnoteFile{Bethke:2012jm}

\bibitem{Kinoshita:1958ru}
T.~Kinoshita and A.~Sirlin,
  \href{http://dx.doi.org/10.1103/PhysRev.113.1652}{\emph{Phys. Rev.}
  \textbf{113} (1959) 1652}.
\bibAnnoteFile{Kinoshita:1958ru}

\bibitem{vanRitbergen:1999fi}
T.~van Ritbergen and R.~G. Stuart,
  \href{http://dx.doi.org/10.1016/S0550-3213(99)00572-6}{\emph{Nucl. Phys.}
  \textbf{B 564} (2000) 343}, \href{http://arxiv.org/abs/hep-ph/9904240}{{\tt
  arXiv:hep-ph/9904240}}.
\bibAnnoteFile{vanRitbergen:1999fi}

\bibitem{Hahn:2000kx}
T.~Hahn, \href{http://dx.doi.org/10.1016/S0010-4655(01)00290-9}{\emph{Comput.
  Phys. Commun.} \textbf{140} (2001) 418},
  \href{http://arxiv.org/abs/hep-ph/0012260}{{\tt arXiv:hep-ph/0012260}}.
\bibAnnoteFile{Hahn:2000kx}

\bibitem{Mertig:1998vk}
R.~Mertig and R.~Scharf,
  \href{http://dx.doi.org/10.1016/S0010-4655(98)00042-3}{\emph{Comput. Phys.
  Commun.} \textbf{111} (1998) 265},
  \href{http://arxiv.org/abs/hep-ph/9801383}{{\tt arXiv:hep-ph/9801383}}.
\bibAnnoteFile{Mertig:1998vk}

\bibitem{Tarasov:1997kx}
O.~Tarasov, \href{http://dx.doi.org/10.1016/S0550-3213(97)00376-3}{\emph{Nucl.
  Phys.} \textbf{B 502} (1997) 455},
  \href{http://arxiv.org/abs/hep-ph/9703319}{{\tt arXiv:hep-ph/9703319}}.
\bibAnnoteFile{Tarasov:1997kx}

\bibitem{Mertig:1990an}
R.~Mertig, M.~Bohm and A.~Denner,
  \href{http://dx.doi.org/10.1016/0010-4655(91)90130-D}{\emph{Comput. Phys.
  Commun.} \textbf{64} (1991) 345}.
\bibAnnoteFile{Mertig:1990an}

\bibitem{Awramik:2002vu}
M.~Awramik, M.~Czakon, A.~Onishchenko and O.~Veretin,
  \href{http://dx.doi.org/10.1103/PhysRevD.68.053004}{\emph{Phys. Rev.}
  \textbf{D 68} (2003) 053004}, \href{http://arxiv.org/abs/hep-ph/0209084}{{\tt
  arXiv:hep-ph/0209084}}.
\bibAnnoteFile{Awramik:2002vu}

\bibitem{Martin:2003qz}
S.~P. Martin, \href{http://dx.doi.org/10.1103/PhysRevD.68.075002}{\emph{Phys.
  Rev.} \textbf{D 68} (2003) 075002},
  \href{http://arxiv.org/abs/hep-ph/0307101}{{\tt arXiv:hep-ph/0307101}}.
\bibAnnoteFile{Martin:2003qz}

\bibitem{Martin:2005qm}
S.~P. Martin and D.~G. Robertson,
  \href{http://dx.doi.org/10.1016/j.cpc.2005.08.005}{\emph{Comput. Phys.
  Commun.} \textbf{174} (2006) 133},
  \href{http://arxiv.org/abs/hep-ph/0501132}{{\tt arXiv:hep-ph/0501132}}.
\bibAnnoteFile{Martin:2005qm}

\bibitem{Kniehl:2008cj}
B.~A. Kniehl and A.~Sirlin,
  \href{http://dx.doi.org/10.1103/PhysRevD.77.116012}{\emph{Phys. Rev.}
  \textbf{D 77} (2008) 116012}, \href{http://arxiv.org/abs/0801.0669}{{\tt
  arXiv:0801.0669 [hep-th]}}.
\bibAnnoteFile{Kniehl:2008cj}

\bibitem{Jegerlehner:2012kn}
F.~Jegerlehner, M.~Y. Kalmykov and B.~A. Kniehl,
  \href{http://dx.doi.org/10.1016/j.physletb.2013.04.012}{\emph{Phys. Lett.}
  \textbf{B 722} (2013) 123}, \href{http://arxiv.org/abs/1212.4319}{{\tt
  arXiv:1212.4319 [hep-ph]}}.
\bibAnnoteFile{Jegerlehner:2012kn}

\bibitem{Froggatt:1995rt}
C.~Froggatt and H.~B. Nielsen,
  \href{http://dx.doi.org/10.1016/0370-2693(95)01480-2}{\emph{Phys. Lett.}
  \textbf{B 368} (1996) 96}, \href{http://arxiv.org/abs/hep-ph/9511371}{{\tt
  arXiv:hep-ph/9511371}}.
\bibAnnoteFile{Froggatt:1995rt}

\bibitem{Froggatt:2001pa}
C.~Froggatt, H.~B. Nielsen and Y.~Takanishi,
  \href{http://dx.doi.org/10.1103/PhysRevD.64.113014}{\emph{Phys. Rev.}
  \textbf{D 64} (2001) 113014}, \href{http://arxiv.org/abs/hep-ph/0104161}{{\tt
  arXiv:hep-ph/0104161}}.
\bibAnnoteFile{Froggatt:2001pa}

\bibitem{Burgess:2001tj}
C.~Burgess, V.~Di~Clemente and J.~Espinosa, \emph{JHEP} \textbf{0201} (2002)
  041, \href{http://arxiv.org/abs/hep-ph/0201160}{{\tt arXiv:hep-ph/0201160}}.
\bibAnnoteFile{Burgess:2001tj}

\bibitem{Isidori:2007vm}
G.~Isidori, V.~S. Rychkov, A.~Strumia and N.~Tetradis,
  \href{http://dx.doi.org/10.1103/PhysRevD.77.025034}{\emph{Phys. Rev.}
  \textbf{D 77} (2008) 025034}, \href{http://arxiv.org/abs/0712.0242}{{\tt
  arXiv:0712.0242 [hep-ph]}}.
\bibAnnoteFile{Isidori:2007vm}

\bibitem{Bezrukov:2009db}
F.~Bezrukov and M.~Shaposhnikov,
  \href{http://dx.doi.org/10.1088/1126-6708/2009/07/089}{\emph{JHEP}
  \textbf{0907} (2009) 089}, \href{http://arxiv.org/abs/0904.1537}{{\tt
  arXiv:0904.1537 [hep-ph]}}.
\bibAnnoteFile{Bezrukov:2009db}

\bibitem{Shaposhnikov:2009pv}
M.~Shaposhnikov and C.~Wetterich,
  \href{http://dx.doi.org/10.1016/j.physletb.2009.12.022}{\emph{Phys. Lett.}
  \textbf{B 683} (2010) 196}, \href{http://arxiv.org/abs/0912.0208}{{\tt
  arXiv:0912.0208 [hep-th]}}.
\bibAnnoteFile{Shaposhnikov:2009pv}

\bibitem{Ford:1992pn}
C.~Ford, I.~Jack and D.~Jones,
  \href{http://dx.doi.org/10.1016/0550-3213(92)90165-8}{\emph{Nucl. Phys.}
  \textbf{B 387} (1992) 373}, \href{http://arxiv.org/abs/hep-ph/0111190}{{\tt
  arXiv:hep-ph/0111190}}.
\bibAnnoteFile{Ford:1992pn}

\bibitem{Feldstein:2006ce}
B.~Feldstein, L.~J. Hall and T.~Watari,
  \href{http://dx.doi.org/10.1103/PhysRevD.74.095011}{\emph{Phys. Rev.}
  \textbf{D 74} (2006) 095011}, \href{http://arxiv.org/abs/hep-ph/0608121}{{\tt
  arXiv:hep-ph/0608121}}.
\bibAnnoteFile{Feldstein:2006ce}

\bibitem{Agrawal:1997gf}
V.~Agrawal, S.~M. Barr, J.~F. Donoghue and D.~Seckel,
  \href{http://dx.doi.org/10.1103/PhysRevD.57.5480}{\emph{Phys. Rev.} \textbf{D
  57} (1998) 5480}, \href{http://arxiv.org/abs/hep-ph/9707380}{{\tt
  arXiv:hep-ph/9707380}}.
\bibAnnoteFile{Agrawal:1997gf}

\bibitem{Kobzarev:1974cp}
I.~Y. Kobzarev, L.~Okun and M.~Voloshin, \emph{Sov. J. Nucl. Phys.} \textbf{20}
  (1975) 644.
\bibAnnoteFile{Kobzarev:1974cp}

\bibitem{Coleman:1977py}
S.~R. Coleman, \href{http://dx.doi.org/10.1103/PhysRevD.15.2929,
  10.1103/PhysRevD.16.1248}{\emph{Phys. Rev.} \textbf{D 15} (1977) 2929}.
\bibAnnoteFile{Coleman:1977py}

\bibitem{Callan:1977pt}
C.~G.~J. Callan and S.~R. Coleman,
  \href{http://dx.doi.org/10.1103/PhysRevD.16.1762}{\emph{Phys. Rev.} \textbf{D
  16} (1977) 1762}.
\bibAnnoteFile{Callan:1977pt}

\bibitem{Coleman:1980aw}
S.~R. Coleman and F.~De~Luccia,
  \href{http://dx.doi.org/10.1103/PhysRevD.21.3305}{\emph{Phys. Rev.} \textbf{D
  21} (1980) 3305}.
\bibAnnoteFile{Coleman:1980aw}

\bibitem{Hebecker:2012qp}
A.~Hebecker, A.~K. Knochel and T.~Weigand,
  \href{http://dx.doi.org/10.1007/JHEP06(2012)093}{\emph{JHEP} \textbf{1206}
  (2012) 093}, \href{http://arxiv.org/abs/1204.2551}{{\tt arXiv:1204.2551
  [hep-th]}}.
\bibAnnoteFile{Hebecker:2012qp}

\bibitem{Redi:2012ad}
M.~Redi and A.~Strumia,
  \href{http://dx.doi.org/10.1007/JHEP11(2012)103}{\emph{JHEP} \textbf{1211}
  (2012) 103}, \href{http://arxiv.org/abs/1208.6013}{{\tt arXiv:1208.6013
  [hep-ph]}}.
\bibAnnoteFile{Redi:2012ad}

\bibitem{Fox:2002bu}
P.~J. Fox, A.~E. Nelson and N.~Weiner, \emph{JHEP} \textbf{0208} (2002) 035,
  \href{http://arxiv.org/abs/hep-ph/0206096}{{\tt arXiv:hep-ph/0206096}}.
\bibAnnoteFile{Fox:2002bu}

\bibitem{Benakli:2012cy}
K.~Benakli, M.~D. Goodsell and F.~Staub,
  \href{http://dx.doi.org/10.1007/JHEP06(2013)073}{\emph{JHEP} \textbf{1306}
  (2013) 073}, \href{http://arxiv.org/abs/1211.0552}{{\tt arXiv:1211.0552
  [hep-ph]}}.
\bibAnnoteFile{Benakli:2012cy}

\bibitem{Hall:2009nd}
L.~J. Hall and Y.~Nomura,
  \href{http://dx.doi.org/10.1007/JHEP03(2010)076}{\emph{JHEP} \textbf{1003}
  (2010) 076}, \href{http://arxiv.org/abs/0910.2235}{{\tt arXiv:0910.2235
  [hep-ph]}}.
\bibAnnoteFile{Hall:2009nd}

\bibitem{Giudice:2011cg}
G.~F. Giudice and A.~Strumia,
  \href{http://dx.doi.org/10.1016/j.nuclphysb.2012.01.001}{\emph{Nucl. Phys.}
  \textbf{B 858} (2012) 63}, \href{http://arxiv.org/abs/1108.6077}{{\tt
  arXiv:1108.6077 [hep-ph]}}.
\bibAnnoteFile{Giudice:2011cg}

\bibitem{Cabrera:2011bi}
M.~Cabrera, J.~Casas and A.~Delgado,
  \href{http://dx.doi.org/10.1103/PhysRevLett.108.021802}{\emph{Phys. Rev.
  Lett.} \textbf{108} (2012) 021802},
  \href{http://arxiv.org/abs/1108.3867}{{\tt arXiv:1108.3867 [hep-ph]}}.
\bibAnnoteFile{Cabrera:2011bi}

\bibitem{Arbey:2011ab}
A.~Arbey, M.~Battaglia, A.~Djouadi, F.~Mahmoudi and J.~Quevillon,
  \href{http://dx.doi.org/10.1016/j.physletb.2012.01.053}{\emph{Phys. Lett.}
  \textbf{B 708} (2012) 162}, \href{http://arxiv.org/abs/1112.3028}{{\tt
  arXiv:1112.3028 [hep-ph]}}.
\bibAnnoteFile{Arbey:2011ab}

\bibitem{Ibanez:2013gf}
L.~E. Ibanez and I.~Valenzuela,
  \href{http://dx.doi.org/10.1007/JHEP05(2013)064}{\emph{JHEP} \textbf{1305}
  (2013) 064}, \href{http://arxiv.org/abs/1301.5167}{{\tt arXiv:1301.5167
  [hep-ph]}}.
\bibAnnoteFile{Ibanez:2013gf}

\bibitem{Hebecker:2013lha}
A.~Hebecker, A.~K. Knochel and T.~Weigand,
  \href{http://dx.doi.org/10.1016/j.nuclphysb.2013.05.004}{\emph{Nucl. Phys.}
  \textbf{B 874} (2013) 1}, \href{http://arxiv.org/abs/1304.2767}{{\tt
  arXiv:1304.2767 [hep-th]}}.
\bibAnnoteFile{Hebecker:2013lha}

\bibitem{ArkaniHamed:2005yv}
N.~Arkani-Hamed, S.~Dimopoulos and S.~Kachru,
  \href{http://arxiv.org/abs/hep-th/0501082}{{\tt arXiv:hep-th/0501082}}.
\bibAnnoteFile{ArkaniHamed:2005yv}

\bibitem{Giudice:2006sn}
G.~Giudice and R.~Rattazzi,
  \href{http://dx.doi.org/10.1016/j.nuclphysb.2006.07.031}{\emph{Nucl. Phys.}
  \textbf{B 757} (2006) 19}, \href{http://arxiv.org/abs/hep-ph/0606105}{{\tt
  arXiv:hep-ph/0606105}}.
\bibAnnoteFile{Giudice:2006sn}

\bibitem{Bak:244579}
P.~Bak, C.~Tang and K.~Wiesenfeld,
  \href{http://dx.doi.org/10.1103/PhysRevLett.59.381}{\emph{Phys. Rev. Lett.}
  \textbf{59} (1987) 381}.
\bibAnnoteFile{Bak:244579}

\bibitem{Weinberg:1987dv}
S.~Weinberg, \href{http://dx.doi.org/10.1103/PhysRevLett.59.2607}{\emph{Phys.
  Rev. Lett.} \textbf{59} (1987) 2607}.
\bibAnnoteFile{Weinberg:1987dv}

\end{thebibliography}

\end{document}